ABSTRACT

Simulation of Micro-Void Development within Large Scale Additively Manufactured Polymer Composite Deposition Beads

Aigbe E. Awenlimobor, Ph.D.

Mentor: Douglas E. Smith, Ph.D.

A significant knowledge gap remains in fully understanding the process-structure-property relationship involved in extrusion-deposition additive manufacturing (EDAM) technology. Modeling the polymer melt flow extrusion-deposition process is crucial in understanding the development of the inherent bead microstructure. The current research presents a computational-based approach for investigating process-induced micro-voids and their impact on print properties. We develop a Generalized Newtonian Fluid (GNF) multiscale Finite Element Analysis (FEA) simulation tool to predict global and local flow fields during the polymer processing to investigate underlying mechanisms that may promote the micro-void development within the bead microstructure specifically the occurrence of low-pressure regions at sites of stress concentration such as at the tips of suspended fibers and at locations with abrupt changes in flow direction like the die-swell region just after the nozzle exit. We also investigate potential factors that may influence the development of these micro-voids, such as the particle interactions, the rheology of the suspension, the flow type, the size and geometry of the suspended particles, etc., which reveals new response behavior of suspended particles within the polymer melt. Additionally, we develop a novel technique to account for inter-fiber interactions by correlating the fiber-fiber coefficient of interaction with the effective fluid domain size that yields the same suspension steady-state fiber orientation. A novel means for determining the steady-state fiber orientation was also developed based on a Newton-Raphson approach using exact Jacobians. Furthermore, we present a method for quantifying and characterizing micro-voids within EDAM printed beads using X-ray micro-computed tomography imaging technique, which revealed a dominant heterogeneous mode of large-sized and irregular-shaped micro-void formation, especially at the fiber terminations, where the hydrostatic pressure drops significantly. The impact of micro-voids on effective material properties was also assessed based on a numerical homogenization scheme using realistic periodic representative volume elements (RVEs), which resulted in significant property losses. Overall, the research reveals a new mechanism of micro-void formation and potential ways of mitigating their formation. The direct implication of reduced bead porosity levels is the development of high-quality components for specialized applications, such as lightweight and high-strength integrity composites widely used in a variety of industries.

TABLE OF CONTENTS

LIST OF FIGURES

LIST OF TABLES

ACKNOWLEDGMENTS

I would like to express my deepest gratitude to my supervisor, Dr. Douglas E. Smith, for his invaluable guidance, support, and encouragement throughout this research journey. Your expertise and insights have been instrumental in shaping this dissertation, and I am profoundly grateful for your mentorship. I also wish to acknowledge the National Science Foundation (NSF) for providing the necessary grants (Award Number: 2055628) and also the financial support of Baylor University that made this research possible.

I also extend my heartfelt thanks to the members of my faculty committee, Dr. David A. Jack, Dr. Trevor J. Fleck, Dr. Min Y. Pack, and Dr. Emmanuel B. Agamloh, for their constructive feedback and unwavering support. Your contributions have significantly enriched my work, and I am thankful for your time and dedication.

To my parents, Monday and Evelyn Awenlimobor, and my brother Stanley Awenlimobor and his family, your love and support have been my anchor. Thank you for believing in me and providing the emotional and financial support necessary to complete this dissertation. Your encouragement has been a constant source of strength.

Lastly, to my friends, Emuobosan Enakerakpo, Idolor Ogheneovo, Musah Oshioke, Kenneth Ndukaife, and all my classmates, thank you for your understanding, patience, and encouragement. Your companionship and support have made this journey more enjoyable and less daunting. I am grateful for the moments of laughter and the words of motivation that have kept me going.

Thank you all for being a part of my journey...

# DEDICATION

*To "God Almighty" and to my beloved parents, "Evelyn and Monday Awenlimobor".*

## ATTRIBUTIONS

The individual contributions of various co-researchers and co-authors from collaborative research effort documented in select certain chapters are summarized below:

*Chapter Three*

*Microstructural Characterization of Large Area Additively Manufactured Polymer*

**Aigbe Awenlimobor:** Conceptualization, data curation, 3D image post processing (code generation) for region extraction, feature identification, formal analysis, result interpretation and visualization of the reconstructed 3D grayscale μ-CT voxel-data using MATLAB software, writing (All other sections).

**Neshat Sayah:** Printing of 13% CF/ABS single bead specimen from Baylor's 3D LAAM printer, μ-CT image scanning and acquisition of the printed bead specimen using NSI-X3000 X-Ray μ-CT system, 3D reconstruction of acquired 2D images using efX-CT NSI software, and writing (Sec. 3.1.1 & 3.1.2).

*Chapter Seven*

*2D Multi-Scale Extrusion-Deposition Polymer Composite Melt Flow Process Simulation*

**Aigbe Awenlimobor:** Conceptualization, data curation, single fiber evolution model development, multi-scale model integration and formal analysis, result interpretation and visualization, validation, writing (all sections).

**Zhaogui Wang:** 2D Planar deposition flow macro-scale model development and formal analysis and investigation.

*Chapter(s) Three - Seven*

**Aigbe Awenlimobor:** Conceptualization, model/algorithm development, code generation, formal analysis, writing (all sections), document review & editing, validation, visualization, validation, methodology, investigation, data curation.

**Dougals E. Smith:** Conceptualization, document review & editing, validation, supervision, software and resources provision, project administration, methodology, funding acquisition.

# CHAPTER ONE

# Introduction

## *1.1 Research Motivation and Objective*

Material extrusion additive manufacturing (MEX AM) technology offers numerous advantages compared to other subtractive technologies such as increased time and energy efficiency, high design flexibility and cost effectiveness. Although there has been remarkable progress in the advancement of MEX AM technology, there are yet aspects of the technology that are not completely understood such as the complex microstructural development within the parts printed from this technology especially the micro-void formation within the prints which directly affects the material behavior and in-service part performance. The ability to control the microstructure of the polymer composite prints during processing presents an advantage for significantly improving part performance, especially by minimizing the inherent micro-voids formation. The main objectives of the proposed research are (1) to understand the mode of formation and characteristics of micro-voids within LAAM printed beads and assess their impact on effective material properties, (2) to utilize computational finite element analysis (FEA) modelling technique to simulate flow processes in extrusion-deposition based additive manufacturing (EDAM) of particulate polymer composites to better understand the evolution of the inherent bead microstructure including the porosities and fibrous structure, and (3) to identify underlying mechanisms that are responsible for the development of process-induced micro-voids within the bead microstructure during polymer composite processing and to understand the various factors that may influence the void formation process. Micro-voids within printed beads are well known manufacturing defects that significantly impair the quality of fabricated components and could lead to component failure. The realization of these goals provides a means for leveraging our knowledge of suitable control parameters that would be tailored at mitigating the final part voidage levels. To the best of our knowledge, no known computational efforts currently exist in literature specifically aimed at understanding the development of 'process-induced' micro-voids within the microstructure of short fiber reinforced polymer (SFRP) composites including EDAM beads. Previous studies have utilized experimental methods to investigate potential sources and factors that influence final part voidage which does not address the actual dynamic mechanisms involved in the evolution of the complex microstructure during the polymer melt flow process in the extruder-nozzle that results in the development of micro-voids.

Presently, the widespread naïve perception on the main source of micro-voids within the prints is the mechanically entrapped air preexisting in the raw pellets prior to processing. However, literature has revealed a significant increase in the micro-void content from the raw pelletized feedstock to the final processed print beads without providing solid basis for the experimentally observed rise in void levels. By establishing a valid process-structure-property space map, this research is aimed at enhancing the in-service performance of additively manufactured (AM) SFRP composite parts by optimizing the inherent microstructural formations particularly the intra-bead voids or micro-voids existing within the composite bead. Our simulation is a first attempt at predicting the development of the local flow-field and the microstructural dynamics during EDAM polymer composite processing that helps provide new insight into the possible mechanisms that are potentially responsible for the observed rise in micro-void levels. For improved model accuracy, our investigation further explores various effects that assumptions employed when modeling the actual polymer melt flow extrusion-deposition process such as the shear-thinning fluid rheology, the inter and intra-fiber forces, confinement effects etc. Additionally, we carried out parametric studies on the sensitivity of the primary flow variable that determines the void formation and characteristics to various process parameters that could potentially be fine-tuned to minimize the final part voidage levels. Our hypothesis of the occurrence of low-pressure regions, especially at the tips of suspended fibers within the polymer melt that presents favorable sites for the nucleation of micro-void within the print beads are validated by comparing results of the local pressure response from our numerical simulation with experimental observations that reveal a high level of micro-voids nucleation at the tips of suspended fibers within typical EDAM printed polymer composite beads obtained from 3D image acquisition and analysis technique. These low-pressure regions may act as a sink that pulls pre-existing voids towards it causing coalescence/void growth and may likewise instigate the nucleation of voids from dissolved void forming species via certain void formation mechanisms. We further assessed the significant impact of these deleterious micro-voids on the resulting effective material properties of printed beads which further buttresses the importance of our current research.

## *1.2 Brief Introduction*

AM technology is a minimal wastage and cost saving technology with high manufacturing throughput, capable of producing large scale complicated geometries. For instance, the Thermwood Large Scale AM (LSAM) system is reported to have a printing capacity of 30m x 3m x 1.5m and a material output rate of 226kg/hr [1]. Extrusion-Deposition AM (EDAM) technology finds increasing application in various industries including automotive, aerospace, marine, space technology, renewable energy, housing, etc. particularly, in the fabrication of tooling and load-bearing components with complex intricate geometric structure and functionally graded materials (FGM) due to the inherently high design flexibility, speed, cost-effectiveness and large-scale capabilities. In pellet-based EDAM system, pelletized polymer composite feedstock material is melted as it is conveyed through the extruder screw before flowing through the contraction zone to the nozzle capillary and subsequently deposited onto a moving substrate where solidification takes place under atmospheric conditions. The shear thinning fluid rheology of the polymer melt usually results in shear rate dependent viscosity and viscoelastic local stiffness within the flow. Usually shear rates in excess of $5000s^{-1}$ can exist in the narrow annular clearance within the extruder due to the high rotational velocity of the screw [2] while shear rates typically below $300s^{-1}$ can be found in regions of the nozzle resulting from flow acceleration. The shear rate variability across the extruder further complicates the final bead microstructure. The pelletized polymer feedstock is often reinforced with chopped fibers for numerous advantages such as increased part dimensional stability (resulting from reduced coefficient of thermal expansion (CTE), improved stiffness, strength, flexure, and thermal conductivity of the part, and higher corrosion resistance. Despite these known advantages, studies have shown that the fiber inclusion in the polymer matrix introduces micro-voids within the bead microstructure [3]. The resulting microstructural constituents thus consist of reinforcing fibers, the polymer matrix, and microstructural voids. Voids within EDAM polymer parts typically exist in two different scales which include 1) The meso-scale voids or inter-layer voids and 2) the micro-scale or intra-layer micro-voids.

Meso-voids exist as narrow gaps between adjoining bead layers that align along the beads deposition direction and are prevalent in low fiber content composites with less deleterious effect to the mechanical properties of the part. However, the micro-voids that often segregate on the surface of individual fiber are predominant in high fiber content composites and pose as sites of stress concentration that effectively reduces the load bearing capacity of the composite part [3], [4]. For instance, interlaminar shear strength of composites have been reported to decrease by about 7% for each 1 % void up to a total void content of about 4% [5]. Similarly, the toughness of a polymer composite has been shown to reduce by as much as 15% for about 1.5- 3.5% micro-void contents [6]. The characterization of voids within the composite prints can provide useful information on the originating cause and type of voids. The inter-bead voids are usually prismatic shaped and are caused by weak inter layer adhesion, however the intra-bead voids have rather irregular shapes typically resembling a spheroid and involves a more complex microstructural formation process [7]. Inter-layer voids can be controlled somewhat with lateral bead space and post-deposition compaction (i.e., with a tamper or roller). Intra-layer micro-voids within the micro-structure of polymer composites are predominantly heterogenous in nature, existing predominantly at the interface of the fiber and matrix phase and are seldom homogenous in nature when nucleated in isolation within the matrix [3], [5]. Prior to processing, the raw pelletized feedstock are found to already contain certain quantity of voids or void forming species such as mechanically entrapped air during compounding process, and dissolved moisture/volatiles, residual solvents, etc. [5], [8], [9], [10] which are the known void sources within polymers. For instance, Vaxman et al. [5] recorded void volume fractions up to 6% from density measurement of unfilled Noryl extrudates. The pre-existing void (encapsulated air) in the precursor can be controlled by adequate venting measures [5], [11]. The void forming species (dissolved volatiles/moisture and residual solvents) are not voids in themselves but are sources that could promote void formation via a nucleation mechanism and are dependent on the material handling. Currently, it is not clear which of the composite's micro-constituents (i.e. polymer matrix, fiber or sizing agent) absorbs the highest proportion of the dissolved species, however the introduction of the fiber reinforcement within the polymers is found to exacerbate the observed void levels. Post extrusion, void contents up to 8, 20, and 35% void volume fractions were recorded in 10, 20, and 30 wt.% glass fiber filled Noryl extrudate respectively [5] . Two major mechanisms based on literature may contribute to the development of micro-voids within the polymer melt during EDAM processing namely:

1. Moisture/volatile absorption-desorption induced void nucleation mechanism which is based on "extension" of the classical theory of nucleation to polymeric flows [9], [12], [13], [14], [15].
2. Constrained volume contraction micro-void nucleation mechanism which results from uneven cooling across the extrudate due to thermal stratification from the core regions to the outer surface [5], [16], [17].

In both mechanisms, the formation of micro-voids within the molten polymer is by nucleation and growth. In the theoretical development of both mechanisms, a requirement for the nucleation of voids is the occurrence of sufficiently low localized pressure regions in the polymer melt below some reference value. The reference pressure could be the vapor pressure of the gaseous phase of the volatile content in the case of the former mechanism or simply the atmospheric pressure in the latter mechanism. In either case, knowledge of the local fluid pressure distribution amongst other process parameters like the temperature field, concentration gradient and distribution of dissolved void species etc., is important in determining the propensity for bubble entrapment and/or void nucleation in sites where they occur. Despite the challenges posed by voids ranging from reduced expectancy in material properties to increased property anisotropy, the research efforts towards investigating the defects in short fiber reinforced polymer composite parts is limited in comparison to the attention given to the studies of defects in long fiber consolidated polymer composite counterparts. Moreover, existing research on voids in the AM printing of chopped fiber polymer composites has placed more emphasis on the inter-bead voids or meso-voids that form between layers of deposited beads while studies on the more deleterious intra-bead voids that form within the complex fibrous microstructure has received very little focus.

Various factors have been experimentally found in the literature to influence the micro-void distribution within the composite beads such as the rheological properties of the polymer suspension, the extrusion operating conditions, the local fluid visco-elastic stiffness defined by the local resin richness or lack thereof, fiber orientation distribution, and the fiber's aspect ratio [5]. During solidification of the extrudate, the cooling rate and the fiber-matrix CTE mismatch were also observed to promote higher levels of void crystallization [5], [11]. At the terminal portion of the extrusion process where the polymer melt leaves the nozzle and where die expansion occurs, the micro-void content has been reported to increase significantly. However, in the filament feed and in the heating/extrusion zones of the extruder and nozzle, insignificant micro-void content were observed [18]. It is thus obvious from the preceding statements that there is overwhelming experimental evidence in literature that supports the existence and dependence of micro-voids within polymer composites beads on the various factors discussed above. Despite these known facts, there remains a significant knowledge gap in understanding the actual micro-voids formation and evolution process within the bead microstructures during EDAM polymer processing and also in establishing concrete relationships that may exist between void development and the prevailing process conditions and other relevant parameters. On the other hand, there has been significant progress and continuous improvement in modelling the evolution of other microstructural descriptor counterparts, mainly fiber orientation and distribution within the bead. One such analytical model used to predict the fiber orientation states is the Advani-Tuckers 2nd order fiber orientation tensor equation of state [19], [20] developed from the well-known Jeffery's equation [21]. Since introduced in short fiber polymer composites nearly 40 years ago, the orientation tensor approach has undergone various model improvement that more accurately simulate the momentum diffusion term and the appropriate 4th order orientation tensor closure approximation used in the model [22]. Alternatively, coupled, or uncoupled flow-fiber orientation numerical simulation models have also been developed to predict the flow field within the extruder-nozzle and the associated fiber orientation state. For instance, Finite Element Analysis (FEA) simulations was used independently by Heller et al. [23], Wang et al. [24] and Russell et al. [25] to simulate the flow of fiber filled polymer melt in a LSAM extruder nozzle to evaluate the orientation state of suspended short carbon fibers and the resulting thermo-mechanical properties. Likewise, coupled Smoothed Particle Hydrodynamic (SPH) and Discrete Element Method (DEM) numerical techniques were used by Yang et al. [26] to simulate the polymer deposition process of fiber filled polymer composite. While numerical models that accurately predict fiber orientation and distribution during polymer composite processing are important in estimating and controlling the average thermo-mechanical properties of prints, models that also describe the micro-void formation within the microstructure of the prints and predict their characteristics are likewise important owing to the significant impact they have on the microstructural properties and behavior of the printed parts. Currently, leading edge computational tools that simulate and characterize the development of the inherent bead microstructure during polymer composites processing are either inadequate or completely lacking; especially in their ability to relate the evolution of the inherent bead microstructure with the relevant process variables and flow parameters among other factors that may influence their development. In addressing the fundamental problem of the formation and growth process of microstructural voids within the printed beads, we develop a set of three hypotheses which are validated through a series of computational investigations presented in subsequent chapters.

Firstly, we hypothesize that besides the pre-existing voids present in the raw-pelletized feedstock that may subsist in the microstructure until the final stage of processing, two major mechanisms potentially promote pore formation within the microstructure of the polymer composite during processing at different regions of the EDAM extrusion-deposition process [5]. Within the extruder and nozzle, where the polymer melt temperature is relatively high, the moisture/volatile induced void nucleation mechanism is expected to be the driving mechanism. It is assumed that the polymer material (hydrophilic or hydrophobic) has some degree of void forming species such as moisture or dissolved additives/volatile content pre-existing in the raw pellets or absorbed during processing [5], [8], [11], [27]. Because of the multiphase constitution of the polymer melt flow, the mode of micro-void nucleation is predominantly of a heterogenous nature with the crystallization of a third phase at the fiber-matrix interface [3], [5], [13], [15]. Pores are predicted to nucleate in regions of the polymer melt with sufficiently low localized pressure below the moisture vapor pressure and at process temperatures above the glass transition temperature. More details on their model development can be found in [9], [12], [13], [14], [15]. Post-extrusion, when the visco-elastic polymer melt is ejected from the nozzle exit into the atmosphere, where die swell/expansion occurs and at the onset of solidification, the restrained volumetric shrinkage mechanism developed by Titomanlio et. al [16], [17] is expected to be the dominant mechanism responsible for the formation of voids since the former mechanism would only occur at process temperatures above the polymer melting or glass transition temperature [9], [27]. Constrained contraction of the inner core region of the extrudate due to uneven solidification resulting from temperature stratification across transverse sections of the extrudate may create a solidification front. When there is insufficient compensatory flow of polymer melt in the cavity in response to the pressure drop caused by densification, voids are expected to nucleate at regions where the cavity pressure $P$ drops below zero. The voids are likely to segregate at the interface of the fiber and matrix due to weak interfacial adhesion and owing to mismatch in the CTE between the microstructural constituents [5], [11]. Although, there is currently no satisfactory theoretical model for bubble nucleation in polymers [9], [27], these theories only provide a basis for our study of the flow-field pressure since, irrespective of the dominant void formation mechanism, it is evident that the occurrence of low local fluid pressure is a necessary requirement for void formation hence the fluid pressure is a primary variable. The low pressure most likely occurs at sharp transitions in the flow path geometry such as at the edge of a screw flight, at nozzle exit or cavitation at the fiber ends during flow acceleration. Most voids would nucleate at the fiber-matrix interface specifically close to the fiber's ends where the hydrostatic stress reaches a maximum on the fiber's surface [5], [28], [29]. According to Tekinalp et al. [3], the relative motion between the suspended fibers and the surrounding polymer flow is likely responsible for the high level of voids observed on the fiber matrix interface. In essence, the void distribution is expected to follow the local fiber concentration. The contributions to the overall void content from the failure of the sizing agent observed on the fiber-matrix interface is only minimal according to Vaxman et al. [5]. It is expected that the melt temperature of the sizing agent will be well above the operating temperature to avoid its failure during polymer melt processing. Moreover, the surface roughness of fiber fillers are typically on a nanoscale orders of magnitude less than the average fiber diameter making it less likely to entrap air. Additionally, homogenous mode bubble nucleation resulting from direct phase transformation of the dissolved void species due to boiling are known to only contribute a small fraction to the overall void content in polymer composites [5], [30]. Likewise, we can exclude the void nucleation mechanisms based on cavitation since the Reynolds number of polymer melt flow is negligible in the absence of inertia forces resulting in cavitation number much greater than unity.

Secondly, we postulate that void growth is governed by the difference in the internal pressure between the nucleated void and the surrounding pressure of the viscous polymer melt and by coalescence of smaller bubbles with larger ones driven by their pressure difference [5]. The void growth rate is expected to depend on the concentration of dissolved moisture/volatile, the molecular diffusivity, the visco-elastic non-Newtonian polymer melt rheology and the gradient profiles of the flow temperature and pressure [5]. The void size inevitably depends on the magnitude of local pressure drop in the gas bubble, the instant and streamline location in the viscous flow where the void nucleated during processing. For instance, voids that are nucleated early in the flow on streamline location with relatively low velocities will have sufficient time to grow and allow for diffusion of smaller gas bubbles along its travel path. Moreover, since the experimentally observed pore sizes are relatively small, voids likely form late in the extrusion process near the nozzle without sufficient time to grow [5], [18]. Additionally, the average fiber aspect ratio, its geometry and elastic/plastic properties may also influence the void formation and growth process [5], [11].

Accurate prediction of the local fluid pressure and consequently the likelihood of micro-void formation and growth depends on the level of sophistication and assumptions considered in the model to capture actual flow conditions in the extruder-nozzle. Such flow conditions include the shear-thinning fluid behavior that may be important in high shear regions of the polymer melt flow such as the lubrication flow region near the screw edge or flow acceleration region near the nozzle [31], the fiber geometry, inter-fiber hydrodynamic forces, wall effects, the intra fiber deformation forces (bending, buckling, & breakage), etc. Additionally, we can determine possible control variables that may mitigate the development of micro-voids during the polymer composite processing by carrying out detailed parametric studies to investigate the sensitivity of the pressure response to various flow parameters that may influence the micro-void formation identified in the second hypothesis.

### *1.3 Order of Dissertation*

Chapter Two presents summarily literature on extrusion-deposition additive manufacturing (AM) of short fiber polymer composites including a detailed background on process induced microstructural void formation. Known sources and causes of voids, their impact on properties and performance of printed parts and existing theoretical models for predicting their formation and growths are considered. The literature summary also provides a review of the current trend in analytical and numerical based methods for evaluating homogenized thermo-mechanical properties of short fiber composite materials. Lastly, the literature provides a review on multiscale simulation of the extrusion-deposition AM process.

In Chapter Three, detailed microstructural characterizations of a 13% carbon fiber filled ABS LSAM polymer composite bead specimen are performed using 3D X-ray micro computed tomography image acquisition and analysis to investigate the phenomenon of micro-void nucleation at the fiber/matrix interface, especially those that form at fiber tips. Since micro-voids within short fiber polymer composites beads produced by additive manufacturing (AM) technology are known to significantly impair quality and performance of printed parts, it is important to understand the formation behavior of these micro-voids. In-depth microstructural analysis and characterization of bead prints can provide useful insight into originating source and mode of the formation of these micro-voids during the polymer extrusion/deposition processing. The bead microstructure is characterized by using various metrics including the micro-constituents phase fractions and volume fractions of interest features, distribution of average micro-void size, average sphericity and average fiber orientation. To understand the impact of the final bead microstructural configuration on homogenized composite properties, the development of efficient and accurate material property predictive tool is very crucial. This predictive tool provides a reliable means for assessing the effectiveness of control measures in fine-tuning the microstructure of SFRP composite to meet desired performance requirement.

Chapter Four presents a finite element analysis (FEA) based numerical homogenization approach for evaluating the effective thermo-mechanical properties of LSAM particulate-filled composites using realistic periodic representative volume elements (RVEs) generated from reconstructed X-ray μ-CT image scans of the 3D printed bead. The chapter goes into detail on the process of determining a suitable RVE size from a single region of interest (ROI) extracted from the bead's volume based on some dispersion tolerance metric and presents a method for validation of the numerical procedure by benchmarking results of predicted effective quantities with the well-known Mori-Tanaka-Benveniste's analytical estimate [32]. Ultimately, Chapter Four aims to study the impact of the inherent micro-porosities on the resulting composite material behavior and investigate the effect of anisotropy due to spatial variation in the microstructure across the bead specimen on the computed composite homogenized properties. We expect a priori what is otherwise known from literature, that the inherent micro-voids would negatively affect the computed homogenized properties. It thus remains for us to present a detailed computational methodology which would be used to better understand the formation mechanism of these process-induced micro-void within the microstructure of polymer composite beads which make up the bulk of the remainder of the dissertation.

Simulating the flow behavior of fiber suspension during SFRP EDAM composite processing is a typical Fluid-Structure Interaction (FSI) problem that can provide useful insight into potential mechanisms responsible for the resulting microstructure of the SFRP composites particularly the deleterious micro-voids known to impair print quality. A common starting point for modelling dilute fiber suspensions has been to utilize the well-known Jeffery's analytical equations [21] which have been in existence since 1922. Although Jeffery-based models has gained popularity overtime in simulating the orientation dynamics of suspended particles in dilute viscous homogenous suspension, the model is seldom used in understanding the development of other microstructural formations such as the process induced micro-voids that form within polymer print beads. Chapter Five extends Jeffery's model to simulate particle behavior in a special class of homogenous Newtonian flows with combined extension and shear rate components typically found in axisymmetric EDAM nozzle flow contractions. The chapter also presents a method for optimization of Jeffery's pressure equation using exact gradients and Hessian to obtain the location within the fluid and the particle orientation at which the fiber surface pressure extremes reach a maximum.  The chapter further addresses some limitations of Jeffery's model. For instance, Jeffery's model is confined to simulating rigid ellipsoidal shaped particles in a viscous Newtonian homogenous flow and cannot model other phenomenon found in the actual extrusion-deposition polymer melt flow such as shear-thinning behavior of the polymer melt (that may be important in high shear lubrication flow regions such as near the screw edge or flow acceleration region near the nozzle), and cylindrical particle shapes with sharp geometrical transitions at the ends (which better characterizes the geometry of chopped fiber in polymer melt suspension). Moreover, other effects such the inter-fiber forces in concentrated or confined flows and the intra-fiber forces (e.g., Brownian effects, etc.) cannot be modeled with Jeffery's equation. The chapter presents a detailed methodology for the development of a FEA approach to simulate single particle motion in viscous suspension with GNF fluid rheology that can account for the various effects neglected by Jeffery's model and presents the results of various sensitivity analysis conducted considering other factors that may influences fiber surface pressure distribution like the particle aspect ratio, the initial particle orientation and the shear-extensional rate ratio. The FEA model is validated by comparing results of the Newtonian case with results obtained from the well-known Jeffery's analytical model. Preliminary findings from the study conducted in this chapter provide an improved understanding of key transport phenomenon related to physical processes involving FSI such as that which occurs within the flow-field developed during EDAM processing of SFRP composites. These results are expected to provide insight into important microstructural formations within the print beads.

Highly loaded fiber polymer suspension flows usually involve long and short-range hydrodynamic interaction forces between suspended particles. Unfortunately, Jeffery's equation is limited to simulating particle motion in dilute regime and does not account for momentum diffusion due to inter particle interaction in concentrated polymer suspension flow. Over the past four decades, more advanced macroscopic fiber orientation models have been developed that account for rotary diffusion due to fiber-fiber and fiber-matrix interaction such as the well-known Advani-Tucker's fiber orientation tensor evolution model. Unfortunately, these advanced models can only provide information about the overall transient fiber orientation state and cannot predict other field state information such as the local pressure distribution around suspended particles which may be useful in understanding the formation mechanism of other microstructural features like the inherent porosities. It is unrealistic to simulate the motion of every individual particle in the fiber suspension and their interaction with each other. Alternatively, we present a simpler and novel approach for accounting for rotary diffusion in our single particle FEA model which we explain in detail in Chapter Seven. The method ultimately relates the Folgar-Tucker's phenomenological interaction coefficient to the effective fluid domain radius of influence utilized in the single fiber FEA model. A crucial step in the methodology involves relating the interaction coefficient with the steady state fiber orientation tensor using one of the available advection-diffusion fiber orientation tensor evolution models.

Traditionally explicit numerical IVP-ODE transient solvers like the fourth order Runge-Kutta method are used to determine the steady-state fiber orientation. Chapter Six presents a computationally efficient and faster numerical method for determining the steady state fiber orientation for a range of diffusion interaction coefficients based on the Newton-Raphson iterative technique using exact derivatives of the second order fiber orientation tensor equations of state with respect to the independent components of the orientation tensor. The chapter considers various existing macroscopic-fiber orientation models and closure approximations to ensure robustness and reliability of the method and evaluates the performance and stability of the numerical scheme in determining physical solutions in complex flow fields. Validation of the derived exact partial derivatives of the fiber orientation tensor material derivative is performed by benchmarking with results of finite difference techniques. Fiber orientation is an important descriptor of the intrinsic microstructure of polymer composite materials and the ability to predict the orientation state accurately and efficiently is crucial in evaluating the bulk thermo-mechanical behavior and consequently performance of printed part.

As we previously established, simulating polymer melt flow during EDAM processing is crucial in understanding the underlying mechanisms responsible for the microstructural formation and associated properties of the print. The penultimate chapter of this dissertation presents a multi-scale computational FEA method that computes the global and local flow-field development within a typical EDAM polymer melt flow process particularly the fiber-induced local pressure fluctuations and orientation distribution across sections of the EDAM nozzle. On a macro-scale, the global flow field of a purely viscous, Newtonian planar polymer deposition flow through an EDAM nozzle is computed which provides input to a micro-scale model that simulates the evolution of a single ellipsoidal fiber along macro-model streamlines. The micro-scale single fiber evolution FEA model developed in Chapter Five serves as the micro-model in this multiscale simulation. Chapter Seven also presents a technique to account for rotary diffusivity due to short-range fiber-fiber interaction in the FEA simulation by determining an effective fluid domain size that is correlated with the interaction coefficient of the melt flow which yields an equivalent steady state orientation based on the Advani-Tuckers equation. For robustness of the solution, various possible motions of the fiber along individual EDAM flow paths from a given set of random initial fiber conditions are considered to determine pressure bounds on the fiber surface along each streamline. The chapter concludes by assessing the effect of shear thinning on the computed local flow-field responses. The simulation results of the pressure distribution around the surface of suspended fibers along streamlines of the EDAM flow-field are used to interpret the experimentally observed microstructural formations and characteristics specifically related to the micro-voids that form within typical EDAM printed beads.

The final chapter summarizes the various studies carried out in Chapter Three to Chapter Seven, and the various results and major conclusions of each chapter. Chapter Eight also proposes future extensions to the current research effort including various areas for model improvements, and other closely related research opportunities that can leverage the knowledge and outcome of the current research work.

# CHAPTER TWO

## Literature Review

### *2.1 Additive Manufacturing of SFRP Composites*

Additive Manufacturing (AM) finds increasing applications in the fabrication of tooling and end-use load-bearing components with complicated geometry and functionally graded materials due to the inherently high design flexibility and cost-effectiveness that characterize the technology. Although AM technology is generally characterized by low volume production rates due to associated low material throughput and high manufacturing cost compared to conventional techniques, it is however known to reduce tool design and production times and tooling cost compared to traditional tooling technologies [33]. Moreover, thermo-plastic based AM technology are much faster compared to the epoxy-based AM technology that generally requires extensive cure cycles with complex cure chemistries [34]. AM has been classified based on the processing state of matter into solid extrusion based, powder based or liquid based systems [35]. These systems are further subcategorized based on the various fabrication techniques employed. For instance, fused deposition modelling (FDM) technique is used in solid extrusion-based systems; selective laser sintering (SLS), electron beam melting (EBM), selective laser melting (SLM) and direct metal laser sintering (DMLS) are techniques used in powder-based systems, while stereolithography (SLA) is mainly used in liquid-based system [36]. FDM terminology and fused filament fabrication (FFF) are often used interchangeably, however both techniques differ slightly in the processing environmental conditions. While FDM takes place in a thermally controlled enclosure with limited envelope, FFF is conducted under atmospheric conditions [37]. Extensive literature review on AM technology can be found in [1], [34], [36], [38]. Additionally, AM of polymer composites can be classified based on the aspect ratio of the reinforcing agents broadly into discontinuous (short) or continuous (long) fiber composites [34]. Polymer composites have been manufactured using most of the available AM technologies listed above including FDM, SLS, and SLA. Of particularly interest to the current research is the FDM or extrusion-deposition AM (EDAM) technique. EDAM uses a thermoplastic as feedstock materials such as acrylonitrile butadiene styrene (ABS), polycarbonate (PC), polyactide (PLA), polyamide (PA), thermoplastic polyurethane (TPU), polyetherimide (PEI), polyethylene terephthalate (PET) and polyetheretherketone (PEEK). The feedstock material is heated into a viscoelastic polymeric melt state within a heating extruder chamber and the melted material is ejected through a nozzle/die under pressure which is then deposited as a bead onto a build surface to form the desired 3D geometry [39], [40], [41], [42]. Cooling and solidification of the polymer melt material follows immediately upon exposure of the processed material to the atmosphere at the nozzle exit. EDAM technology can be classified based on extrusion mechanism into three (3) broad categories namely the filament based, plunger or syringe based and the screw-based mechanism [37]. Overtime, EDAM technology has experienced significant scale-up in manufacturing capacity from desktop or small-scale AM (SSAM) to commercial medium- scale AM (MSAM) and large-scale AM (LSAM) systems which have higher material throughput and printing speed necessitated by industrial need in vast economic sectors including automotive, aerospace, renewable energy, defence etc and made possible by the utilization of pelletized feedstock. MSAM systems have build volumes ranging from 1 to 7m$^3$, extrusion nozzle exit diameters ranging from 0.1 to 4.0mm and deposition rates between 0.5 to 4.0 kg/h. Alternatively, LSAM systems have build volumes greater than $7m^3$, extrusion nozzle exit diameters ranging from 4.0 to 7.6mm and deposition rates between 4.0 to 50kg/h [43]. For example, the Oak Ridge National Laboratory (ORNL) Big Area Additive Manufacturing (BAAM) system has a build volume size of 6×2.5×1.8m and maximum material output of 45kg/h [1], [33]. Likewise, the Thermwood Corporation LSAM system has a larger build volume size reaching 30×3×1.5m and material output rates of up 227kg/h [44]. LSAM systems have less energy requirements than expected due to various factors considered in its design such as the elimination of the heating chamber typically found in SSAM systems, although measures taken to reduce cost and energy may result in a lowering of the print parts quality and introduction of defects such as warpage, delamination and cross-sectional tapering [43]. To combat these associated prints problems without necessarily modifying the manufacturing method or increasing production cost, the polymer feedstock materials in LSAM systems are usually reinforced with short glass fibers (GF) or carbon fibres (CF) to yield enhanced thermal-mechanical properties as compared with the neat polymers [42], [45], [46]. For example, Love et al. [46] showed that the addition of CF to ABS thermoplastic, significantly reduced the warping in manufactured parts by lowering the coefficient of thermal expansion. Tekinalp et al. [3] also showed an improvement in stiffness and strength along the print direction within the carbon fiber ABS (CF/ABS) composite compared to neat ABS due to the alignment of fibers along the print direction. Somireddy et al. [47] showed a significant improvement in flexural properties of the CF/ABS compared with the neat ABS. Their results showed tensile properties and Young Modulus of CF/ABS SFRP composites increased with increasing carbon fiber content up to a saturation limit of about 7.5%wt CF content, and included an accompanying decrease in toughness, yield strength and ductility [48]. Typical length scale of short fibres used in AM may range from that of milled fibres with length up to hundreds of microns to those of chopped fibres usually few millimetres long [34]. Maximum saturation limit of the fiber content and packing efficiency in a polymer composite material is limited by the average length and degree of alignment of the fiber reinforcement. High fiber loads up to 30% in polymer composites are achievable with highly aligned fibres or milled fibres with very small aspect ratio. Consequently, the fiber length is an important factor that affects the attendant SFRP composite properties. Unfortunately, the average fiber length used in SFRP composites (<150um) is well below the critical fiber length of 640um which has been deemed necessary for effective transfer of the fiber micro-constituent strength to the overall composite strength [34]. Moreover, long length discontinuous fibers are prone to excessive bending or breakage during polymer composite processing and may lead to clogging of the nozzle orifice. Although most fiber breakage occurs in regions of the screw extruder as compared to the extrusion-deposition event, the fiber length in both cases have been predicted to decrease exponentially with polymer processing time based on a kinetic model [49]. Although, there are obvious benefits resulting from the reinforcement of polymers with fibres, optimal material behaviour of manufactured composites is limited by the inherent complexities of the uncontrolled microstructure particularly the unwanted micro-voids and unpredictable distribution of the fiber orientation [1], [40], [50], [51]. The section following provides summarily literature on the voids in polymer composites manufactured from LSAM technology.

### *2.2 Voids in Large Area Additively Manufactured Polymer Composites*

Voids in short carbon fiber-reinforced composite AM printed parts can be categorized by their mechanism of formation into five distinct types: 1) raster gap voids 2) partial neck growth voids, 3) sub-perimeter voids, 4) intra-bead voids and 5) in-fill voids. More details on the description of each void types can be found in [52]. These voids occur at different levels of the multi-scale structure of a typical AM printing process. Figure 2.1 is a schematic showing the different levels of the multi-scale structure of a typical AM printing process which can be broken down into the macro-scale structure (cf. Figure 2.1a), the meso-scale and the micro-scale structure (cf. Figure 2.1b). Within the meso-structure, voids are broadly categorized into inter-layer voids and intra-layer (microstructural) voids (cf. Figure 2.1b). Inter-layer voids form gaps between the deposited beads that occur due to the rounded shape of the bead corners and insufficient bonding between beads of material during the 3D printing process [3], [53]. Conversely, intra-bead or microstructural voids (referred to here as micro-voids) develop within individual beads (cf. Figure 2.1c) where the micro-void size is typically much less than that of inter-layer voids. Yu et al. [30] found that the addition of carbon fibers into the polymer matrix increases the composite's viscosity, leading to microstructural voids forming within the AM part. Sayah et al. [40], [41] also identified the presence of micro-voids within the pellet's microstructure prior to the extrusion-deposition process as well as within the beads following deposition.

Although short fiber reinforcement of polymers offers various benefits such as enhanced thermal, mechanical, and anti-corrosion properties of the component parts with improved dimensional stability [3], however, the development of intra-bead micro-voids within the bead microstructure due to fiber reinforcement are known to alter the material behaviour that may impair the part performance and possibly result in compromised structural integrity. Intra-bead micro-voids also affect the capacitance and electrical permittivity of the short fiber polymer composite. Higher levels of voids within polymer composites can also increase moisture absorption during polymer composite processing and has been shown to degrade mechanical properties [10]. Inter-bead voids that form at the interstices of AM extrudate strands tend to orient in the print direction and can be as detrimental to the mechanical behaviour of the polymer composite as intra-bead micro-voids, especially those formed at the fiber-matrix interface which acts as sites of stress concentration that reduces the load bearing capacity of the polymer composite material [4], [34]. In both cases, voids are unwanted defects that arise within the structural fabric of the composite material due to unsuitable process conditions which should not be confused with micro-damage which is internal microstructural failures that occurs during composite loading [54].

EDAM process-induced intra-bead voids may originate during the flow of polymer melt through the extruder nozzle or during solidification when beads are deposited on the moving substrate. It has been experimentally observed that the micro-void content is highest when the polymer melt exits the nozzle during die-swell/expansion and subsequently decreases when the beads are deposited on the bed [18].

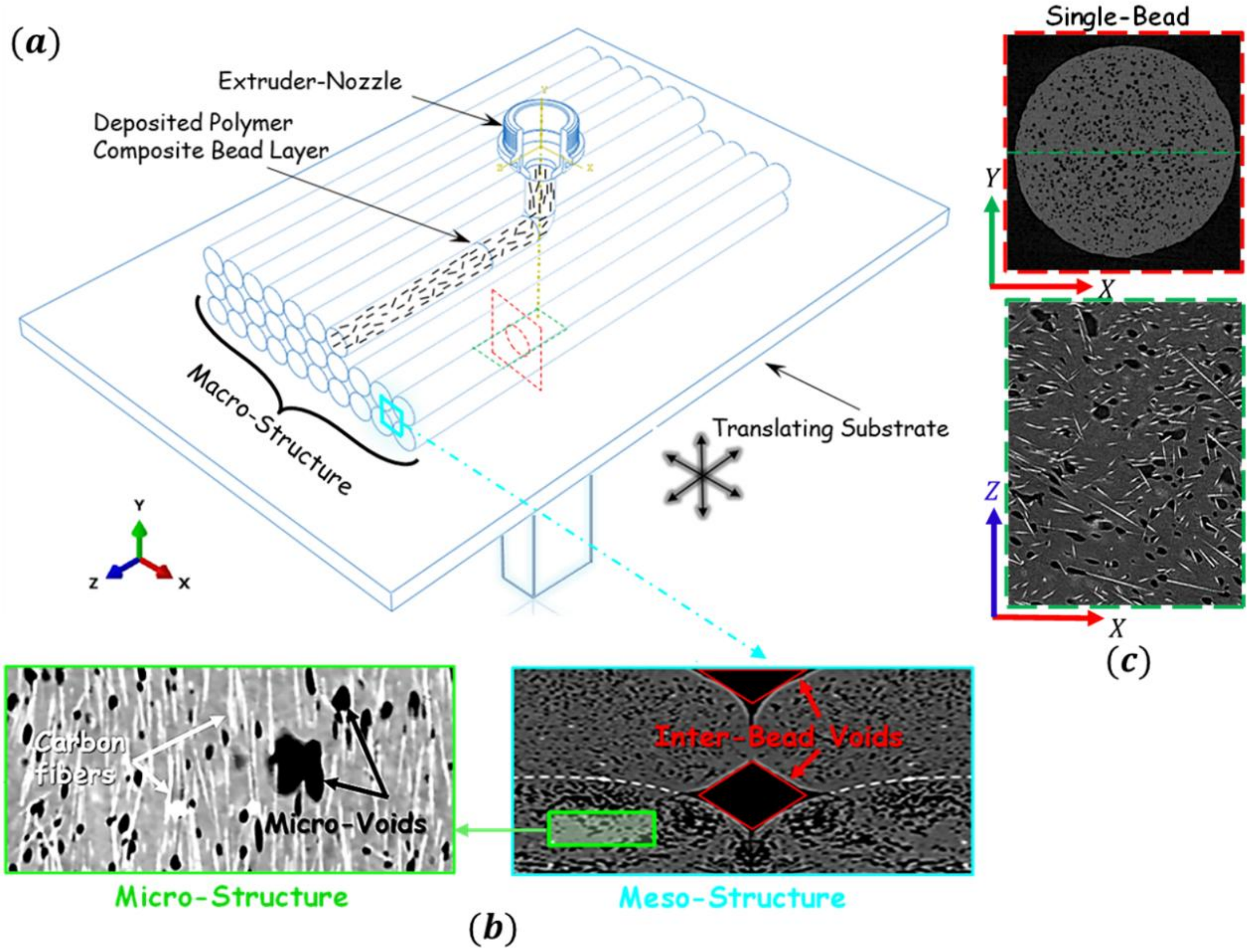

Figure 2.1: Schematic of the multi-scale structure of EDAM SFRP composite processing (a) macro-structure (b) meso-structure and micro-structure (c) cross section of a single bead layer showing the inherent micro-structure. (Image Credit: X-Ray μCT images provided by Dr. Neshat Sayah, Ph. D, Baylor University, 2024).

Various sources that may induce micro-void formation within EDAM beads during polymer composite extrusion/deposition processing include bubble encapsulation within the pellets during the compounding process (which can be reduced by adequate venting measures), absorption of moisture and gases including dissolution of chemical volatiles within the polymer melt, and development of internal stresses in excess of the intrinsic visco-elastic strength of the extrudate during cooling (often due to uneven contraction during solidification) [5], [8], [11]. Knowledge of the morphology of voids present within an EDAM printed sample can provide insight into the cause and type of voids present in the composite structure [7]. For instance, interlayer voids formed between adjoining strands having a flat-bar prismatic shape suggest weak interlayer adhesion as the responsible agent, while intra-bead micro-voids formed at the interface between the fiber and matrix constituents suggests a possible weak adhesive strength of the sizing agent. In the latter, ellipsoidal shaped voids and gas pockets have been shown to result from excessive temperature in the melt [7]. Narkis et al. [11] found that the contribution of the interfacial micro-void content due to fiber-matrix debonding compared to the overall void content within the extrudate was minimal. Interlayer voids can be somewhat managed by adjusting the lateral bead spacing and using post-deposition compaction methods, such as tampers or rollers. Among the two types of voids, those aligned in the loading direction are less harmful to the mechanical properties of the additively manufactured composite compared to the intralayer micro-voids within the bead structure. Additional factors that may influence the nucleation of micro-voids within polymers have been identified as well. For example, Vaxman et al. [5] and Sayah et al. [40] independently found that the print processing conditions like temperature, pressure, and flow rate affected the final void content of an extruded short fiber composite. During the process of extrudate solidification, factors like non-uniform and faster cooling rate, mismatch between the fiber and matrix thermal expansion coefficient and the die-swell at the nozzle exit of the free extrudate due to pressure difference upon atmospheric exposure were observed to promote voids [5], [11], [18].

Of particular interest in our work are contributing factors for void formation that relate to the reinforcing fiber constituent. Micro-void volume fraction has been experimentally observed to increase with increasing fiber concentration and fiber aspect ratio due to an increase in the effective viscosity of the polymer composite melt [5], [18], [48], [55]. At lower fiber volume fractions, interlayer voids between beads were experimentally observed to be more prevalent in the printed composite, however, when the fiber content is high, intra-bead micro-voids were found to more dominant where interlayer void content decreased due to reduced extrudate diameter resulting from lower die expansion and higher thermal conductivity [3], [4]. Additionally, Yang et al. [18] found that during polymer processing, the volume fraction of micro-voids is negligible within the extruder/nozzle, however the void content peaks when the polymer melt just exits the nozzle during die-swell and drops to a stable value upon bead deposition. The findings also revealed that an increase in the void content in the polymer composite at the nozzle exit where die swell occurs is a direct consequence of a decrease in the fiber volume fraction due to an overall decrease in the effective viscosity of the suspension. Sayah et al. [40] also showed that the degree of fiber misalignment in various regions of the extrudate correlate directly with the measured void volume fraction in these regions. Under favourable operating conditions (temperature, pressure and extrusion rate), micro-bubbles may form within the pure polymer matrix phase or at the interface of the fiber and matrix phase. Micro-voids that form at the fiber-matrix interface may be due in part to failure of the adhesive/sizing agent that results in fiber-matrix debonding. Of significance to our study is the observed higher likelihood of micro-voids occurring at the ends or tips of suspended fibers within the polymer composite beads with high fibre volume fractions [5], [11]. Although high fibre packing is found to reduce the potential of micro-voids nucleation at the interstices between fibres, the increased number of fibre ends are observed to provide favourable sites for void nucleation to occur. The air pockets or micro-voids that solidify at fiber terminations are typically characterized by melt pressure variations and micro-void growth/collapse or re-dissolution in the polymer matrix which is influenced by the difference between the internal micro-voids pressure and external pressure within the surrounding melt [11].

A relevant analogy can be drawn from micro-damage initiation as it relates to flow-induced micro-void formation where micro-damage nucleation sites tend to occur at stress concentrations occurring close to the ends of fibers [28], [29]. Known micro-damage nucleation mechanisms have been related to excessive interfacial shear stress at the fiber-matrix interface [29] and the maximum hydrostatic stress within the matrix [28], both of which occur at the fiber tips. In these cases, micro-crack initiation is shown to occur at the point where the maximum stress exceeds a critical value related to the intrinsic strength of the composite material such as the interfacial fiber-matrix bond strength or matrix fracture strength. Hu et al. [29] showed that fiber aspect ratio and orientation were significant microstructural parameters that influence the maximum stress at the fiber tips and consequently micro-crack initiation. Separate investigations conducted by Agyei et al. [56] and Hu et al. [29] showed that the local stiffness of the ductile fracture region within the matrix where micro-crack initiation and progression mainly occur depend on the stress concentration at fiber tips, the degree of fiber misalignment and the average tip distance between fibers within this region. In addition, computational studies performed by Awenlimobor et al. [57] indicated that the hydrostatic pressure within the fluid surrounding the fiber surface reaches an extreme value at the fiber tips where micro-voids typically occur, and the tip pressure depends on the fiber aspect ratio and orientation angle.

## *2.3 Micro-Void Nucleation Mechanisms*

Various mechanisms potentially responsible for void formation within the microstructure of short fiber polymer composite print beads have been investigated. Of the known mechanisms, the encapsulation of low molecular weight substances within the beads during compounding of the pelletized material and subsequent extrusion-deposition of the polymer melt in the EDAM nozzle extruder has been identified as one major cause of void formation [5]. Other process induced mechanisms include the uneven volumetric shrinkage mechanisms due to temperature stratification during solidification [5], [16], [17], volatile/moisture absorption and desorption induced void nucleation mechanism [9], [12], [13], [14], [15], and stress-instigated cracking mechanism [8]. All the listed mechanisms require a critical criterion to be satisfied for the onset of void initiation. For instance, volatile induced void formation requires the surrounding fluid pressure to drop below the polymer melts vapor and operating pressure. Likewise, volumetric shrinkage that leads to void nucleation results from insufficient compensatory flow of polymer melt once the cavity pressure drops below zero [16], [17]. These void formation mechanisms suggest that void nucleation within the microstructure of short fiber reinforced polymer composites during EDAM processing is to some extent dependent on the melt pressure field. These mechanisms do not act independently but often involves an interplay to achieve stable void development. Each could, however, be classified as a homogenous or heterogenous mechanism. In the former, micro-voids form within a single phase under critical conditions while in the latter class, micro-void formation occurs at the interface between two existing phases such as the interface between fiber and matrix [15]. The characteristics of a void may suggest the dominant mechanism responsible for their formation. For example, most micro-voids have been observed to be positioned at the ends of fibers, which suggests a heterogenous void formation process. Alternatively, the presence of isolated micro-voids within the matrix suggests the responsible mechanism is one of a homogenous nature. In subsection below, we present a summary of critical criterions of some known mechanisms necessary for void formation.

### *2.3.1 Moisture/Volatile Induced Mechanism*

The motivation for evaluating pressure on the fiber surface stems from classical nucleation theory that addresses void initiation and growth within a polymer melt investigated by Han and Han [12], Stewart [13], and Han [14], who also investigated the dynamics of void initiation in polymer melts under shear flow. Colton and Suh [15] distinguished between two mechanisms of nucleation which includes 1) homogenous classification involving the formation of a new stable phase in a primary phase with dissolved secondary components under critical conditions due to thermal fluctuations and molecular interaction, and 2) a heterogenous classification involving the crystallization of a third phase at the interface of two other phases, usually a liquid and a solid. Both forms of nucleation can coexist and occur concurrently under a mixed classification. However, in a system such as a colloidal solution, depending on the volume fraction of the suspension, a heterogenous nucleation is more likely to be dominant due to smaller activation energy barrier. The polymer composite material considered throughout this dissertation is composed of 13% filled carbon fiber filled Acrylonitrile Butadiene Styrene (13% CF/ABS) such that a heterogenous dominant mode of nucleation is expected to occur at the interface of the carbon fiber and polymer. Also, it is expected that the polymer material has some degree of absorbed moisture or dissolved additives/volatile. In the model development by Roychowdhury et al. [9], a necessary requirement for potential homogenous void nucleation is the occurrence of very low localized fluid pressure $P_L$ below the moisture vapor pressure $P_V$ i.e., $P_L < P_V$ at process temperature $\mathcal{T}_p$. The nucleation rate $J_n$ (i.e., $J_n \geq 1$ for void nucleation) as modified by Colton and Suh [15] in heterogenous systems is

$$J_n = N_v \sqrt{\frac{2\gamma_t}{\pi \widetilde{m}}} \exp\left[-\frac{16\pi\gamma_t^3}{3k_B \mathcal{T}(P_V - P_L)^2} S(\varphi)\right] \tag{2.1}$$

where $N_v$ is the number of molecules per unit volume of the volatile phase, $\widetilde{m}$ is the molecular mass of the volatile phase, $\gamma_t$ is the surface tension at characteristics temperature $\mathcal{T}$, and $k_B$ is the Boltzmann constant. In the above,

$$S(\varphi) = (1/4)(2 + \cos\varphi)(1 - \cos\varphi)^2 \tag{2.2}$$

where $\varphi$ is the wetting angle of the interface. Usually, the characteristics temperature of nucleation $\mathcal{T}_n$ stays well above the glass transition/melt temperature $\mathcal{T}_g/\mathcal{T}_m$ (i.e., $\mathcal{T}_n \sim \mathcal{T}_p \geq \mathcal{T}_g/\mathcal{T}_m$) and the phenomenon takes place almost instantaneously. Colton and Suh [15] determined the moisture vapor pressure from the moisture concentration distribution in the polymer using Henry's Law, $P_V = c/H_V$ where $c$ is the concentration and $H_V$ is Henry's constant for moisture in a polymer. Based on classical nucleation theory, the characteristics nucleation time $t_n$ is given by

$$t_n = r_c^2/D \tag{2.3}$$

where $D$ is the moisture diffusivity defined by $D = D_o e^{-A_E/\mathcal{T}}$; and $D_o$ is the moisture diffusion constant within the polymer, $A_E$ is an activation energy related material constant, and $\mathcal{T}$ is the temperature. $r_c$ is the critical radius on nucleation given by

$$r_c = 2\gamma_t/(P_V - P_L) \tag{2.4}$$

The simulated pressure response around suspended particles shows that the calculated localized fluid pressure $P_L$ may fall below processing pressure $P_\psi$ [57] which increases propensity for void nucleation at these sites. An additional requirement for void nucleation is that the void nucleation time $t_n$ must be less than the streamline deposition time $t_d$. i.e., $t_n < t_d$. Han and Han [12] showed that the classical nucleation theory under predicts the propensity for void nucleation in polymer solutions with significant proportion of dissolved volatile components. They observed nucleation at critical pressures $P_L$ above the vapor pressure $P_V$ and developed a more applicable model incorporating the Flory Huggins theorem to account for reduced entropies due to restrictions posed by macromolecules in the solvent yielding a nucleation rate of

$$J_n = [N_v][B_F]e^{(-\Delta F_p^*/nk_B\mathcal{T})} \tag{2.5}$$

where $B_F$ is the frequency factor given by

$$B_F = B_1[D(\mathcal{T})/4\pi r_c^2]\exp(-B_2/\mathcal{T}) \tag{2.6}$$

and $D(\mathcal{T})$ is the molecular diffusivity of the volatile phase which Han and Han [12] obtained using free volume theory of Vrentas and Duda given by

$$D(\mathcal{T}) = D_o(1 - 2\chi_F\vartheta_1)(1 - \vartheta_1)^2\exp(-E/R_G\mathcal{T})\exp\left(\varsigma\left(w_1\hat{V}_1^* + w_2\hat{V}_2^* q\right)/\hat{V}_{HF}^*\right) \tag{2.7}$$

The free energy for critical void nucleation in polymer solutions $\Delta F_p^*$ given by

$$\Delta F_p^* = (16/3)\pi\gamma_t^3(P_V - P_L)^2 - nk_B\mathcal{T}\left\{\ln\left(\vartheta_1\frac{P_G}{P_V}\right) + \vartheta_2\left(1 - \frac{V_1}{V_2}\right) + \chi_F\vartheta_2^2\right\} \tag{2.8}$$

In eqns. (2.6) through, (2.8) $B_1$ & $B_2$ are empirical constants dependent on the polymer solution, $w_i$, $\vartheta_i$ and $V_i$ are the weight fraction, volume fraction and molar volume of constituent $i$ respectively, subscript $i = 1$ for solvent and $i = 2$ for solute. In our material systems, the proportion of molar volume of the volatile phase in the polymer is much less than unity, i.e., $V_1/V_2 \ll 1$, $\varsigma$ is the free volume overlap factor, $q$ is the critical molar volume ratio of jumping units of solvent to jumping units of polymer solution, and $\hat{V}_{HF}^{*}$ is the average hole free volume per gram of mixture. $\chi_F$ is the Flory Huggins interaction parameter and $P_G/P_V$ defines the degree of saturation of the gas phase, $P_G$ being the pressure inside the critical bubble given as

$$P_G = (3/2)\rho_L \dot{r}_c^2 + 2\,\eta/r_c + 4\mu_0(\dot{r}_c/r_c) + P_L \tag{2.9}$$

were $\dot{r}_c$ is the growth rate at the onset of nucleation, $\rho_L$ is the liquid density and $\mu_0$ is the viscosity at zero shear. The surface tension $\gamma_t$ at the elevated temperature at which the polymers are processed is estimated using expression by Sugden [58] thus

$$\gamma_t(\mathcal{T}) = (P_a/\bar{V}(\mathcal{T}))^4 \tag{2.10}$$

where $P_a$ is the Parachor and $\bar{V}(\mathcal{T})$ is the molar volume of the liquid. The consequence of this is that the surface tension at an elevated temperature can be estimated with knowledge of the surface tension at a reference temperature through

$$\gamma_t(\mathcal{T}_2) = \gamma_t(\mathcal{T}_1)\left[\frac{\rho(\mathcal{T}_2)}{\rho(\mathcal{T}_1)}\right]^4 \tag{2.11}$$

and the contact angle can be obtained from equation by Girifalco and Good [58]

$$\cos\varphi = 2\hat{\varphi}\sqrt{\left({}^{\gamma_{t,s}}\!/\gamma_{t,l}\right)} - 1, \qquad \hat{\varphi} = 4\left[V_s^{-1/3} + V_l^{-1/3}\right]^{-1}, \qquad V = \frac{N_v}{\rho} \tag{2.12}$$

The details here provide a possible basis for estimating the potential for void nucleation within polymer composites during processing based on field solutions of the pressure response within the polymer melt flow and given a known amount of volatile content .

### 2.3.2 *Restrained Volumetric Shrinkage Mechanism*

The theory behind the constrained volume contraction void initiation mechanism has been developed by various researchers, including Titomanlio et. al. [16], [17]. The basic principle of void formation based on this mechanism is restricted contraction of the inner core region of the extrudate due to temperature stratification across a transverse section of the extrudate during solidification process which results in the creation of a tapered solidification front. Insufficient compensatory flow of polymer-melt from the extruder-nozzle in response to pressure drop in the enclosure created by the front (cf. Figure 2.2) due to densification would then lead to void formation. The melt viscosity and front geometry dictates the pressure drop within the cavity which is exacerbated near the front tip which further leads to void growth. Assuming a rigid, non-deformable solidification front an approximate expression that describes its geometry given as [16], [17]

$$\frac{s(z)}{R} = \left(1 - \frac{z}{L}\right)^{a} \tag{2.13}$$

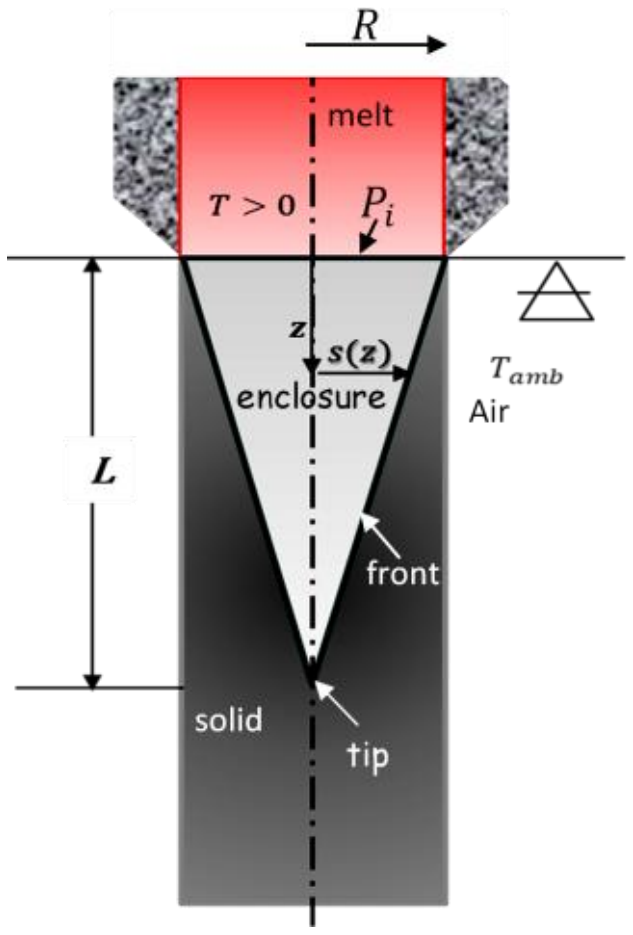

Figure 2.2: Schematic of the Polymer Extrusion at the Nozzle Exit showing Solidification Front

where $s(z)$ is the radial position of the front at any arbitrary distance $z$ from the nozzle of exit radius $R$, $a$ is a shape power index, and $L$ is the enclosure length which is proportional to the vertical velocity $\dot{z}$ of the solid extrudate given as

$$L = l\frac{\dot{z}R^2}{\alpha} \tag{2.14}$$

where $l$ is the dimensionless enclosure length, $\alpha$ is the thermal diffusion coefficient. Void nucleation takes place when the pressure $p\,(\zeta)$ at some arbitrary dimensionless axial distance $\zeta = (1 - z/L)$ within cavity drops below zero. i.e., $p\,(\zeta) \le 0$. Assuming the polymer melt rheology has a power law fluid definition and considering isothermal condition, the pressure $p(\zeta)$ within the cavity is given as

$$p(\zeta) = \left[p - C\frac{m\dot{z}^{n+1}R^{1-n}}{\alpha}\right](\zeta^{1-b} - 1) \tag{2.15}$$

where constants $C$ and $b$ are respectively given as

$$C = \left(\frac{3n+1}{n}\right)^n \frac{2\beta^n l}{b-1}, \qquad b = a(1+n) \tag{2.16}$$

$m$ is the power law viscosity coefficient and n is the power law viscosity exponent and β is the fractional volume contraction on solidification given as $\beta = (\rho_s - \rho_l)/\rho_l$. $\rho_l$ & $\rho_s$ are the densities of the polymer melt and solid extrudate phase respectively. The foregoing mechanisms presented above show that void nucleation within the microstructure of short fiber reinforced polymer composite during processing is to some extent dependent on the pressure field which is the hypothesis of the current study.

### *2.3.3 Stress Induced Mechanisms*

Various studies that investigate stress induced micro-void nucleation in polymers currently exists. Stresses within the polymer composite melt may arise from internal sources such as development of residual stresses within polymer undergoing restrained expansion or contraction or may arise from external sources when the polymeric composite is subjected to an applied load or imposed displacement. Eom et al. [8] studied the voids that form in a thermosetting polymer due to chemical shrinkage during cure that results in the buildup of internal tensile stresses which exceed the material strength. He proposed a critical value for the internal stress that determines the onset of void nucleation prior to gelation. In typical carbon fiber reinforced polymer deposition process, solidification occurs immediately following bead exposure to the atmosphere because of temperature differences. Depending on the thermal expansion coefficient, cooling rate and viscoelastic transformation process, volumetric shrinkage may occur giving rise to residual stresses within the substrate [4], [10], [59]. In thermosetting polymers, residual stress may also arise from excessive cure temperatures or poor heat transfer during curing typical in thick composites with associated thermal degradation [10]. Analogously, void formation would result within the bead microstructure when these residual stresses exceed the limit strength of the material. Micro-void nucleation due to complex microstructural behaviour of chopped fibre reinforced polymer composite materials under an external load have also been investigated by various researchers. Hu et al. [29], studied micro-voids that form at the ends of fiber due to shear stress concentration based on a shear-lag model that depended on the fiber's length and orientation. They found that shear stress reached a maximum at the fiber's end which may exceed the interfacial bonding strength between the fiber and matrix and likely result in micro-void nucleation at the interface. According to the shear lag model, the shear stress along the fiber length is given by:

$$\tau_m = \tau_Q \sinh \eta z - \sigma \sin \phi \cos \phi \sin \theta \,, \qquad \tau_Q = \tau_Q\left(G_m, E_m, E_f, l_f, r_f, v_f\right) \qquad (2.17)$$

where, $\tau_Q$ depends on the material properties and fiber geometry, $z$ is the transverse distance from the fiber center, $\sigma$ is the applied stress, $\phi$ defines the fiber orientation, $\theta$ is the polar angle measured from the plane normal to the fiber direction, $G_m$ is the matrix shear modulus, $E_m$ & $E_f$ are the Young modulus of the matrix and carbon fiber respectively, $r_f$ is the fibers radius, $l_f$ is the fiber's half length, and $v_f$ is the fibers volume fraction [35]. Additionally, they found that micro-voids potentially formed at regions with high agglomeration of fiber ends due to high stress concentration. Hanhan et al. [28] showed that the hydrostatic stress distribution in the matrix can be used to predict the probability and possible locations of void initiation within the composite microstructure. Experimentally, they showed that the location of fiber ends played an important role in determining where micro-voids form. By superposing the spatial locations of hydrostatic tensile stress extremes obtained from FEA simulations with the actual microstructural locations where the voids were observed to nucleate based on in-situ experimental data, they showed both locations correlated with each other. The hydrostatic stress in [51] was calculated as

$$\sigma_{hyd} = \frac{\sigma_{xx} + \sigma_{yy} + \sigma_{zz}}{3} \tag{2.18}$$

The various stress-induced micro-damage mechanisms presented here suggests that knowledge of the stress distribution of the composite polymer EDAM flow field can potentially provide information on the locations where voids are likely to nucleate. Based on Jeffery's assumption [21], the active stress on the surface of a fibre in viscous suspension is simply hydrostatic fluid pressure $\sigma_{hyd} = -p$. Accordingly, the focus in this research is on the pressure peak response on the surface of the fiber tip during its motion in homogenous viscous flow which provides a potential mechanism for micro-void segregation at the fiber terminations.

### *2.4 Computational Assessment of the Effective Thermo-Mechanical Properties of SFRP Composites*

The performance of randomly dispersed short fiber reinforced polymer (SFRP) composites depends on its microstructural characteristics such as the concentration, orientation and length distribution of the fiber reinforcement, the content, distribution and morphology of the inherent micro-voids, the fiber-matrix inter-layer adhesion, etc. [60]. Spatial variations in the heterogenous microstructure results in anisotropic macroscopic behavior of composite material. Property prediction of randomly distributed misaligned SFRP composite becomes more complicated with increased non-uniformity and anisotropy across the heterogenous composite microstructure.

Several computational micro-mechanics techniques have been developed by various researchers for estimating the effective material properties of SFRP as an alternative to experimental characterization which includes the analytical mean-field homogenization techniques, numerical modelling methods and the statistical continuum mechanics methods [61]. Analytical methods typically involve a two-step homogenization approach for property prediction of randomly misaligned SFRP composite. The first step determines the effective properties for the pseudo-grains of the decomposed RVE structure based on available analytical mean-field models for unidirectional aligned SFRP composite of uniform fiber length with isotropic microconstituent properties. The second then averages the predicted properties over the fiber orientation and length distribution amongst the pseudo-grains of the misaligned SFRP composite to account for spatial variations in the microstructural configuration using either the Voight or Reuss models [62], [63].

Numerous classical analytical micromechanics models for predicting approximate fourth-order elastic tensor of unidirectional SFRP composite with uniformly distributed fibers currently exists with varying degree of accuracy which are well documented in literature [60], [64], [65], [66], [67]. Analytical models have been derived from either variational or energy principles to provide solutions to the lower and upper bounds for composite stiffness. These includes first order bounds of Reuss [64] and Voight [68], second order bounds of Hashin and Shtrikman [69], Walpole [70], [71], [72], Willis [73] and Wu et al. [68], and third order bounds of Miller [74], Milton [75] and Beran et al. [76]. A family of models known as Eshelby's equivalent inclusion models have gained popularity over time. The original Eshelby's model was an exact solution to a single homogenous ellipsoidal inclusion. Since then, Eshelby's model has been extended to incorporate inhomogeneous inclusions with non-zero far-field strain including the effect of interactions between neighboring inclusions. The Mori-Tanaka approach [32] which modifies the dilute strain concentration tensor (Eshelby's tensor) to account for inter-particle interaction was extended to model short fiber composites by Taya et al. [77] and Taya M. [78]. The Halpin-Tsai empirical relations [75], [76] originally derived from the work of Hermans [79] and Hill [80] yielded pioneering solutions making it possible to directly derive the complete set of engineering constants for SFRP composites. More recently, Tandon and Weng [81] exploited the Mori-Tanaka's model to derive explicit solutions to the complete set of engineering constants particularly applicable to SFRP composites. Mori-Tanaka's assumption is only valid for low concentration particulate volume fractions and is a lower-bound solution to the composite stiffness. Lielen's model (double-inclusion model) [82] was developed for wide-range particulate volume fraction application by interpolating between the Hashin-Shtrikman-Willis composite stiffness bounds [69], [70], [71], [72], [73] using the inverse rule of mixture principle.

Without claim to completeness, other analytical micromechanics models developed to predict elastic properties of SFRP composite include the Cox-shear lag model [83], self-consistent method [84], the laminate analogy approach [60] etc. Bibliography on existing theoretical models for predicting other intrinsic quantities of unidirectional SFRP composites like the effective coefficient of thermal expansion (ECTE) and the effective thermal conductivity (ETC) are well documented [85], [86], [87]. Similar to the elasticity tensor, various solutions to the upper and lower bounds on the ECTE with differing levels of accuracy have been developed such as upper bound models of Voight [88] and Kerner [89], and lower bound models of Reuss [90] and Turner [91]. Other solutions to the limit bounds on the ECTE of transversely isotropic composites includes models of Van Fo Fy [92], Schapery [93] , Chamberlain [94] and Schneider [85], and Rosen and Hashin [95]. Analogically, the Mori-Tanaka's principle for predicting elasticity tensor of unidirectional SFRP composite has been extended by various researchers to predict the ECTE tensor [31], [96], [97], [98], [99], [100]. Other existing models for estimating the ECTE of SFRP composites includes but are not limited to the models of Chamis [101], Thomas [102] and Cribb [103] etc. Existing analytical models for predicting the ETC tensor of a unidirectional SFRP composite includes the equations of Halpin-Tsai [104], Nielsen [105], [106], [107], Nomura and Chou [108], Thornburg and Pears [109], Springer and Tsai [110] etc. The Giordano's approach for predicting the permittivity of unidirectional SFRP composites based on dielectric theory of inclusions has also been extended to estimate the thermal conductivity tensor [111]. Elasticity models based on Eshelby's theory of inclusion such as the Mori-Tanaka and Lielens' double inclusion models have likewise been extended to predict the thermal conductivity tensor of unidirectional SFRP composites which are known to yield more accurate results [100], [111], [112]. Traditionally, most mean-field theories used in the first homogenization step, such as the Mori-Tanaka-Benveniste formulations are limited to only two-phase composites. For multiphase heterogenous composites, such as one having inherent void inclusions or inhomogeneities, various studies [113], [114] have revealed higher levels of accuracy with multi-level homogenization schemes as compared to direct Voight averaging of the pseudo-grains obtained from RVE decomposition according to the different inclusion phases and characteristics. This usually involves a lower-level pre-homogenization of the matrix with embedded void phases or other inhomogeneities followed by an upper-level homogenization of the equivalent matrix with the filler reinforcements. Although the Mori-Tanaka model can be generalized for multiphase composites, Norris [115] has shown that the method does not always satisfy Hashin - Shtrikman and Hill - Hashin effective stiffness bounds. While theoretical asymptotic formulations based on multi-step and/or multi-level, mean-field homogenization approach are orders of magnitude faster and less computationally intensive in predicting effective properties of composites, they fall short in terms of accuracy when considering the interaction between inclusions or estimating the microscopic stresses associated with the particulates. This is especially true when analyzing composites with sharp phase property contrast or high inclusion aspect ratio and volume fraction [61], [116], [117]. Moreover, these approaches lack the capacity to accurately model geometric peculiarities of inclusions such as irregularities in particle morphology and characteristics, and the spatial variations in the distribution of microstructural features across the RVE which are typically found in actual SFRP composites [118].

With growing computational power and quantum processing speed, numerical boundary value problem (BVP) full-field methods, mainly finite element analysis (FEA) based homogenization methods have received more attention for estimating effective properties of SFRP composites. This is due in part to their high level of prediction accuracy and ability to model complex intricate microstructural geometric details associated with inclusions. Existing studies on numerical based homogenization methods for property prediction of random SFRP composites are predominantly based on computer-generated deterministic RVE volumes stochastically filled with particulates based on a statistical technique [119], [120], [121], [122], [123], [124]. Examples of SFRP composite elastic property numerical homogenization involving periodic deterministic RVEs generated from statistical based techniques (such as modified random sequential adsorption (RSA) showed good agreement with results obtained from analytical based methods include the studies of Berger et al. [125], Moussaddy et al. [126], Qi et al.[127], Mortazavi et al. [61], etc. These studies have shown that numerical based methods prevail in terms of accuracy over the analytical based methods when predicting properties of composites having inclusions with high aspect ratio and high-volume fraction [126], [128].

The continuum mechanics technique based on statistical correlation methods are known to perform poorly for property prediction of composites with non-spherical shaped inclusions. Several published works [119], [121], [128], [129] have revealed that the required RVE size and number of realizations, and the desired precision in predicting properties of heterogenous SFRP composites depend on several factors including the microstructural composition and concentrations, the microconstituents phase property contrasts, the morphology, characteristics and dispersion of inclusion phases and the evaluated quantity of interest. Given a desired level of accuracy and a reasonable RVE size, Kanit et al. [119] developed a method for determining the required number of deterministic RVE instances to predict the mean effective property of a random two-phase three-dimensional (3D) Voronoi mosaic SFRP composite with minimal dispersion in quantities. The method is independent of the choice of boundary conditions and particularly applies to predicting effective properties of large volumes with few realizations of reasonable sized RVEs.

More recently, accurate microstructural characterization has become possible with advancement in modern imaging techniques. Reconstruction of 3D voxelated grayscale radiographs obtained from X-ray micro-computed tomography (μ-CT) imaging technique has gained popularity for characterizing the microstructure of SFRP composites [28], [111], [118], [130], and has been used to generate realistic RVEs for more accurate micromechanical analysis. For instance, Guven et al. [118] generated realistic RVEs of various sizes from 3D X-ray μ-CT voxelated images which were then used to numerically evaluate the effective material properties of two-phase particulate filled polymer composite. His results were shown to be in close agreement with experimentally measured properties. Although their study was based on a two-phase SFRP composite, the method has been successfully extended to study the impact of micro-porosities or particulate inhomogeneities on predicted effective properties of multiphase particulate composites as well [113], [131]. While extensive studies have been performed that numerically assess the impact of porosities on the effective properties of SFRP composites using deterministic

RVEs [113], [116], to the best of the authors knowledge, no known studies currently exist, that conducted an assessment of porosities on the effective properties of SFRP composites utilizing realistic 3D X-ray μ-CT based RVEs. The closest study that utilized realistic RVEs was on nickel-reinforced alumina composites with roughly spherical shaped nickel particle reinforcement [131]. However, suspended particles typically found in AM manufactured SFRP composites are cylindrical shaped with high aspect ratio.

### *2.5 Overview of EDAM Process Simulation*

With computational advancement and sophistication, simulation of manufacturing flow processes has gained traction for providing in-depth understanding of the underlying physics responsible for process states to control and optimize the actual process and fine-tune the final print microstructure and effective properties to improve quality. Moreover, iterative design of manufacturing processes such as nozzle design, via experimental based approach could be very expensive compared to computational based methods. Additionally, in-process monitoring which may be intrusive and sensitive to disturbance is often limited by accessibility which complicates the measuring process [132]. Modelling of the flow process usually begins with identification of the manufacturing process and process variables including the feedstock material parameters (type and characteristics of reinforcement, matrix, and other additives). Three (3) major manufacturing methods are common to fabricate polymer composites which include: (1) short fiber suspension methods (2) squeeze flow methods (or advanced thermoplastic composites methods) and (3) porous media methods (or advanced thermoset composite methods). The current investigation focuses on the short fiber suspension methods which involve the transport of fiber filled polymer suspension into a mold cavity or through a die to form the composite.

It is helpful to further subdivide this method to include (a) injection molding, (b) compression molding and (c) extrusion processes. The reader is directed to [133] for further details on the process description, transport phenomena and applications. The extrusion and injection molding process have similar characteristics. However, the processing differs in that polymer composite melt flows into a closed cavity in injection molding to solidify into shape while the molten polymer is ejected through and shaped by a die into an open environment in extrusion process. Although injection molding is a well-established and widely used method to process large quantities of thermoplastic composites parts, the extrusion manufacturing method is a more promising technology due to the relatively lower material wastage, lower energy requirement and cost savings associated with the technology. Process modelling of extrusion-deposition AM (EDAM) method is more pertinent to the current scope of work.

Traditionally, polymer composite flow process modelling is often performed on a multi-scale level. At the macro-scale level, the length scale is typically on the order of the smallest part dimension (e.g. bead diameter) while at the microscale level, it is commonly on the order of the reinforcing particle's diameter. Coupling the macroscale and microscale physics is required to capture localized phenomena during the flow process. On the macroscale level, the primary aim is to relate the process/printing parameters to the flow or global deformation of the polymeric material. The key elements of a typical EDAM process (cf. Figure 2.3) include the feeding mechanism, heating and transport mechanism within the extruder-nozzle, bead deposition and road spreading/wetting process, inter-bead bond formation mechanism and bead cooling/solidification mechanism [134]. The choice of modelling approach depends on the phenomena being investigated and the associated process parameters that can be controlled. Various transport phenomena that influence the EDAM process/printing parameters includes the combined drag and pressure driven flow due to the turning action of the screw within the barrel (determining power and flowrate requirements), the contribution of the viscous dissipation in momentum transport to the overall energy transport (determining the systems heating or cooling requirements) and the phenomena of extrudate swell and melt fracture at the nozzle exit (affecting the bead shape and stability) [133].

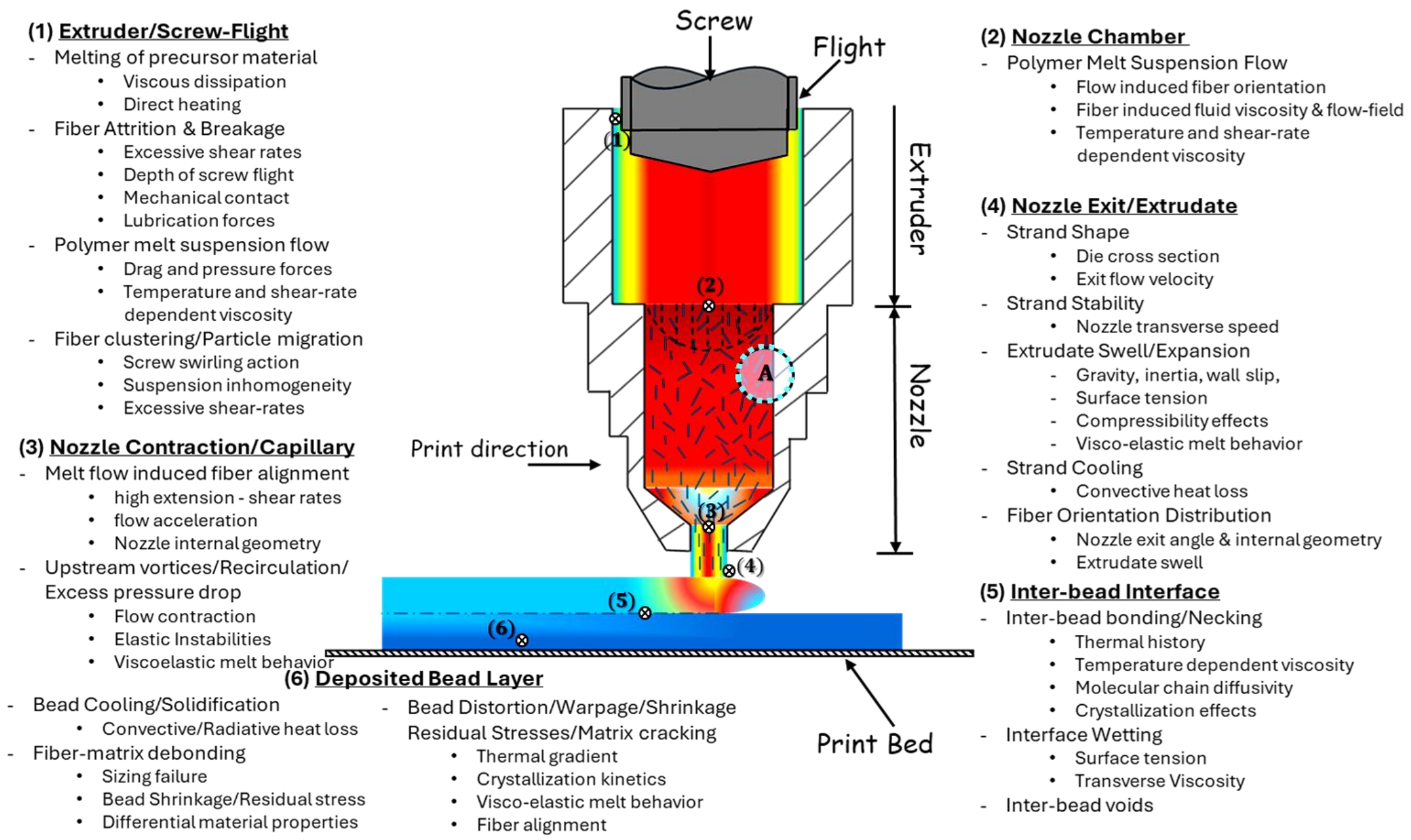

Figure 2.3: Aspects of EDAM Polymer Composite Processing Macro-Scale Level Modelling and typical outcomes of interest in the various regions of the polymer composite melt flow.

Modelling the flow of polymer melt during EDAM composite processing can provide useful information about underlying transport variables such as the velocity, pressure and temperature fields. The macroscale simulation of the melt flow process can be divided into two categories namely the extrusion flow through the extruder-nozzle and the deposition flow onto the substrate. As the heterogenous phase polymer melt flows through the extruder-nozzle, the orientation of the suspended fibres is determined by the flow-field. For highly loaded fibre suspension, the melt suspension viscosity and the flow-field are simultaneously influenced by the fiber spatial and orientation distributions. The mutually dependent phenomena necessitate a coupling of the flow-fibre orientation physics for more accurate process simulation. The rotational motion of the extruder screw induces a shearing flow between the screw flight and the barrel walls that results in high shear stresses. Typical swirling streamlines of non-Newtonian viscoelastic polymer melt flow within the extruder-nozzle obtained from a one-way coupled flow-fiber orientation FEA simulation can be found in [135]. Distinct regions of fiber clusters with directional alignment have been identified in printed bead samples [136] which are likely due to influence of flow swirling downstream the screw within the extruder-nozzle [137]. Fiber attrition and breakage have been observed to occur mostly at the feed and compression zone of the liquefier [138]. Fiber breakage is known to relate to the shear stresses that develop at the screw flight in the compression zone, as such the flight depth is a major design parameter to control fiber breakage. Deeper screw flight-depth characteristics of variable pitch screws can reduce shear stresses and resulting fiber breakage and vice versa for narrower flight-depth in standard screws [138]. On the contrary, higher shear stresses were found to improve fiber dispersion. Typical shear stress distribution across the screw flights in the metering zone for standard and variable-pitch screws of an extruder can be found in [138]. Euler buckling has been identified as the primary failure mode responsible for fiber breakage due to hydrodynamic forces acting on the fibers [49]. Assuming constant fiber diameter and a kinetic model, [49] derived analytical expression for the residual fiber length. Three relevant interactions were identified by [139] to contribute significantly to the fiber breakage mechanisms including the fiber–fiber interaction, fiber–wall interaction, and fiber–matrix interaction. Optimum screw design through simulation can be used to reduce fiber breakage and at the same time improve dispersion.

Suspended fibers in the polymer melt within an EDAM nozzle show a significant degree of alignment in the flow direction due to high shear stresses developed at the nozzle walls and high extension rates at the nozzle centre. Prior simulations revealed high levels of fiber alignment in the flow direction occur in the nozzle contraction region and nozzle capillary zones [24]. Flow vortices or recirculation are a result of abrupt nozzle contraction that are found at sharp corners and are found to be dependent on the visco-elastic properties of the polymer melt [140]. Mezi et al. [141] found that increased fluid shear thinning reduces the upstream vortices which influence the fiber orientation field and results in significant pressure drop. Moreover, the dominant shear induced normal stress difference at the nozzle contraction and die exit were shown to be primarily responsible for stable vortices and excess pressure drop [140]. The nozzle internal geometry and flow-field are thus key elements that determine the flow induced fiber orientation field. The flow aligned fiber orientation field in turn results in excessively high elongational viscosity within the nozzle due to the flow-fiber orientation coupling effect [142].

Cooling of the extrudate via convection begins once it exits the nozzle and is exposed to the ambient environment. Unexpectedly, melt flow simulations revealed that the rate of convective heat loss from the extrudate decreased with increased melt thermal conductivity within the nozzle [143]. As the bounding surfaces of the extrudate becomes unconfined and exposed to the open environment, the parabolic melt flow velocity profile is transformed to a plug-flow velocity profile and due to stress relaxation and release of elastically stored deformation energy, sudden expansion of the extrudate in the radial direction occurs, a phenomenon commonly referred to as extrudate swell/expansion [134], [143]. Extrudate swell ratios of polymeric composites have been predicted using coupled flow-fibre orientation simulation [23], [24], [143], [144] to be in the range of 1.05 to 1.3 [131]. The swell ratio is influenced by the viscoelastic properties of the polymer melt. Analytic approximations of the swell ratio developed by [145] are found to depend on the normal stress difference and shear stress at the nozzle wall. The effects of various factors on the extrudate swell have also been investigated by various researchers using simulation. For instance, [146] found that inertia and gravity effects which depends on the extrudate length significantly decreased swell ratio. Additionally, the effects of surface tension, wall slip, and pressure dependent-viscoelastic melt rheology were independently found to monotonically decrease swell ratios while compressibility effects resulted in an overall increase in the swell ratio. Flow-fibre orientation simulations also revealed that short fibres reinforcement reduced the extrudate swell ratio of polymer melt. However, the resulting fibre orientation distribution is seen to not be significantly affected by the extrudate swell phenomena [147].

As the bead is deposited onto the substrate or onto a previous bead layer, the polymer melt bends into a 90$^{o}$ shape. As a result of increasing radii from the bottom to the top across the bend, varying shear-rates develop which influences the fiber orientation state [1]. Various techniques have been employed to obtain the shape of the free surface of the deposited bead such as the shape optimization technique [144], the finite volume/front-tracking method [148], or the algebraic coupled level-set/volume-of-fluid method [149].

The shape minimization techniques can be subdivided into the zero-surface tension method and the zero-penetration method depending on the choice of boundary condition and solution variables [144]. Various process conditions including gravity, inertia, wall slip, surface tension, pressure dependent viscosity and compressibility were found to significantly affect the extrudate swell ratio of Newtonian fluid up to -90% / +50% [146]. Moreover, viscoelastic non-Newtonian polymeric melts have been found to have higher swell ratios compared to Newtonian fluids due to additional elastic effects [146]. Numerical simulation performed by [149] showed that the velocity ratio of the relative transverse velocity to nozzle exit flow velocity, and gap distance between the nozzle exit and substrate are parameters that influenced the bead shape. Low velocity ratio and small gap height resulted in better bead spreading and less circular and elongated cross section and vice versa [150]. In a different numerical study, the bead morphology, inter-bead distance and layer height were found to be important parameters in minimizing inter-bead void volumes [150].

Bead stability is another modeling aspect investigated by various researchers. For instance, Balani et al. [151] studied the effect of process parameters including the nozzle diameter, feed-rate and layer height which controlled flowrate, shear-rate and viscosity field, on extrudate deformation and inter-bead adhesion. Higher melt flow rates and higher shear rates were found to reduce the viscosity and cause low precision that resulted in excessive extrudate deformation due to a 'sharkskin' effect [152], [153]. High melt flow rates and small deposition times without provision for proper cooling of previous deposited bead layers can induce sagging due to gravity effects [1]. Consequently, excess deformation limits control of the resulting bead shape, bead surface roughness and print reliability [154], [151].

As hot extrudate is deposited on previous bead layer, surface wetting and reheating/remelting ensues at the bead interface which are two major factors responsible for effective interlayer bead adhesion [1]. The contact area between adjacent beads is determined by the wetting process. The dwell time ensures sufficient heating of the interface to allow adequate inter-molecular chain diffusion between adjacent beads through the interface thus ensuring proper fusion and inter-bead bond formation. A requirement for stronger inter-diffusion bond formation is that the temperature of the polymer melt is above the glass transition temperature [1], [134] which also reduce shape deformation and cracking [4]. Surface wetting depends on the melt viscosity transverse to the printing direction and the relative surface energies between the bead and the adjacent surface (i.e. surface tension) and these properties in turn depend on the fiber reinforcement [1], [134]. Analytical models developed to simulate the bead wetting process includes the Crockett model [155], [156], and the Frenkel-type energy-based model [157]. Bellini et al. [143] employed CFD to simulate the road spreading process as part of a complete EDAM process simulation. The bond formation or polymer sintering mechanism is described using the reptation theory [1], [144]. The process begins with the establishment of initial contact between adjacent beads, followed by the formation and growth of a neck at the bead-to-bead interface. Once a neck is formed, inter-diffusion of polymer chains across the neck takes place followed by a randomization of the polymer chains between the adjacent beads. Simultaneous cooling and phase transformation of the polymeric extrudate takes place due to convective heat loss during the wetting process which increases the melt viscosity.

Adequate chain diffusion needed for effective bond formation depends on the thermal history at the adjoining bead interface. Rapid crystallization of the viscoelastic polymeric melt may also retard the bond formation process due to excessive increase in the polymer viscosity. Various techniques for predicting the thermal history of the bead have been developed such as the 2D model of Thomas and Rodriguez [158], and the lumped capacity model of Belleheumer et al. [159]. A bead's thermal history depends on convective and radiative heat transfer across its surface and consequently the build environmental conditions that may or may not be controlled such as the air flow rate, and temperature. Heat dissipation from the print bead is facilitated by thermal conduction across the bead interfaces and through the conductive surface of the print bed. Phase transformation of the polymer melt from a viscous fluid to a viscoelastic solid during cooling results in an evolution of transient relaxation moduli of the polymer. The solidification process results in material shrinkage and the development of internal stresses due uneven cooling from temperature stratification in the radial direction coupled with the restraint posed by neighbouring beads. Additional stresses results from the anisotropic material behaviour due to the preferential alignment of the fiber reinforcement in the print direction and the crystallization effect in semi-crystalline polymers. Moreover, disparity in the coefficient of thermal expansion coefficient between the fiber and matrix constituents contributes to the internal stress development during cooling. Realistic simulation of the solidification process accounts for the various factors involved. One-dimensional steady state and 2D quasi steady state heat transfer analyses appear in literature to be insufficient in accurately simulating the heat transfer process and predicting the thermal history of the beads [160]. A comprehensive 3D analysis is necessary for a more accurate analysis of the bead cooling/solidification process. Typical temperature distribution across sections of a bead during the deposition process of fiber reinforced PEEK composite obtained from FEA simulation can be found in [161]. The heat transfer model was coupled with a non-isothermal dual crystallization kinetics model to predict the thermal history as well as the crystallization kinetics of the polymer composite and the influence on bond formation.

Numerous efforts have been made to simulate the solidification of prints to predict the resulting residual stress and part deformation including warpage and sagging [162], [163], [164], [165]. For example, Watanabe et al. [162] developed EDAM FEA process simulation models to predict the temperature distribution, deposited filament shape, and warp deformation of a two layer deposited polypropylene copolymer bead material. Their simulation results were shown to agree well with experiments.

Most deposition flow simulation models are based on numerous assumptions that oversimplify the actual solidification behavior. Currently, model improvements efforts are being made such as the development of realistic 3D models that accurately capture the necessary physics involved in the process such as the thermo-viscoelastic behavior of the polymer melt, the crystallization effects in semi-crystalline polymers and other non-linear effects [1], [161].

Much effort has been made to predict transport phenomena in EDAM polymer processing on a global level using macroscale simulation. However, microscale simulations are important to obtain a more accurate prediction of certain phenomena on a local level such as the local flow-field around suspended particles, the motion of suspended particles, the deformation of suspended particles and the rheology of the suspension. Theoretical analysis of single particle behavior in a viscous homogenous suspension is a well-known Fluid Structure Interaction (FSI) problem which has a variety of applications in key transport phenomena observed in physical rheological systems such as the movement of cells and platelets in blood plasma [166], the motion of reinforcing particles in fiber-filled polymer melt suspensions during polymer composite processing [1], proppants transport in fracturing fluids [167], migration of gaseous bubbles in quiescent viscous flows [168] etc. The rheology of particle suspensions is inherently complex due to a host of factors, including the presence of inter and intra particle forces arising from hydrodynamic interaction, contact collision between particles, confinement effect and particle deformability, Brownian disturbance, non-Newtonian viscoelastic fluid rheology, anisotropic particle geometry and concentration, and existence of various flow regimes within the system, etc. [22], [169], [170]. Various aspects of a typical microscale level simulation are depicted in Figure 2.4 below which shows typical localized transport phenomena investigated using on a microscale level such as the particle's dynamics and motion, particle's deformation and breakage, average suspension rheology, fiber-matrix debonding, etc. It also shows typical internal and external forces to be considered in a microscale simulation. The study of particle suspension dynamics often starts with the evaluation of single rigid spherical particle suspension under Newtonian simple shear flow which also provides insight into the rheology of dilute suspension [171], [172]. As an example, the dynamics of a single rigid ellipsoidal axisymmetric particle has been used extensively to investigate particle dynamics and flow-field structure of polymer composite melt flows during processing to assess their microstructure [141], [170], [173]. Theoretical studies on particle motion in a homogeneous viscous flow are commonly based on the assumptions of negligible inertia effects, Newtonian fluid rheology and non-deformable particle shape, conventionally referred to as “standard conditions” [174].

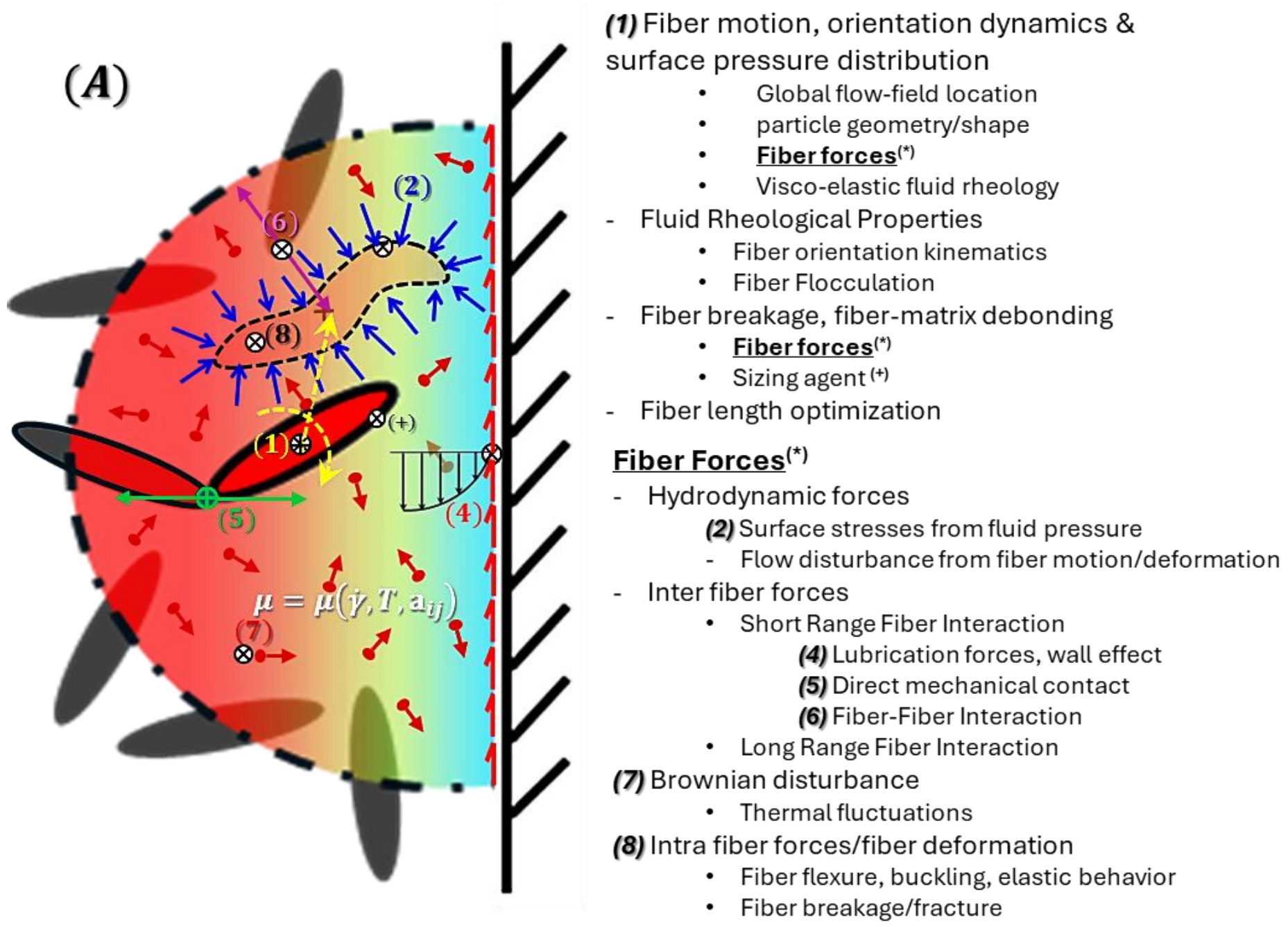

Figure 2.4: Aspects of EDAM Polymer Composite Processing Micro-Scale Level Modelling detailing local transport phenomena and forces considered in a typical model.

Pioneering works of Oberbeck [175], Edwardes [176] and Jeffery [21] evaluated the orbit of an ellipsoidal rigid particle suspended in a homogenous shear viscous flow, where particle motion was determined to be a function of initial condition which has been validated experimentally [177]. In other work, Bretherton showed that lateral positioning of spherical isotropic particles remains unchanged relative to their initial position in quiescent sedimentation or unidirectional shear viscous flow [178]. In addition, Cox [179] found that the orientation of transversely isotropic rigid particles in unconfined quiescent sedimentation would remain fixed at its initial value throughout its motion. These studies showed that under ‘standard conditions’, the motion and trajectory of a body of revolution depends on its initial conditions. For instance, the so called ‘degeneracy’ of Jeffery’s orbits is used to describe the indeterminacy of particle's motion in sheared viscous suspension whereby an axisymmetric particle may assume any of the infinitely possible metastable periodic orbits depending on its initial position. Experimental observations have revealed a tendency for suspended particles to eventually acquiesce to an equilibrium configuration within a finite timescale or equilibrium rate of approach irrespective of its initial configuration which is contrary to theoretical predictions based on "standard conditions" [174]. Jeffery [21] first suggested the possibility that spheroidal particles in a sheared viscous suspension with a theoretically indeterminate nature based on first order approximations, may eventually assume a  path of least energy dissipation. Taylor [180] was one of the earlier researchers to provide experimental basis for Jeffery's hypothesis and proposed that the higher order terms neglected in Jeffery's approximate equations were responsible for the observed departure in the actual particle's behavior from theoretical predictions. In a separate experimental study Saffmann et al. [181] showed that suspended particle's do not always settle in preferred configuration states. Saffmann determined that a non-Newtonian fluid viscosity not included in Jeffery's equations was primarily responsible for the observed discrepancy between theoretical predictions and actual particle's behavior. Other non-linear effects such as fluid and particle inertia, particle confinement and particle end effects were found to be insignificant within a finite timescale. Jeffery's equations are generally accepted to sufficiently predict a particle's kinematics in a dilute and semi-dilute viscous shear-thinning flows yielding only minor deviations from experimentally observed response [177], [178].

However, in the concentrated regime, predictions using Jeffery's model departure from experimental observations which become significant due to the combined effect of short range fiber interactions and shear-thinning fluid rheology neglected in Jeffery's model assumptions [182]. The effect of other rheological properties on the dynamics of a suspended particle such as higher order viscoelasticity fluid behavior (as found in actual FSI physical systems) have also been observed. An increase in the fluid elasticity results in a slow drift of prolate spheroids in sheared viscous suspension across spectrum of degenerate Jeffery orbits from a tumbling orbit to a log-rolling state and at drift rates proportional to the shear rate [183], [184]. Moreover, an excessive shear rate promoted particle realignment with the prevailing flow direction and the critical shear rate for flow realignment depended on particle aspect ratio and Ericksen's number.

More recently, computational models that account for particle inertia, non-Newtonian fluid rheology and/or shape deformability have emerged. These advanced models are often used to assess the departure of fiber kinematics based on each model consideration from related theoretical predictions based on "standard conditions". They are typically developed from analytically formulations based on variational principles or asymptotic series expansion about the limits of standard theoretical model assumptions [174], or they are developed from numerical based simulations [185]. Analytical models are computationally more efficient compared to numerical models, however analytical models are non-flexible, often restricted to predicting a specified set of outputs, and are less accurate due to oversimplification [132]. Models based on variational principles have been used to define limit bounds on the hydrodynamic drag coefficient of a spherical particle in GNF fluid subject to creeping flow [185]. Variational method has been successfully applied to obtain limit bounds solutions on the drag for spheres in GNF fluids for different viscosity models including the Newtonian model [186], power-law model [187], the Carreau model [188] and the Ellis model [189]. Variational method is more accurate for predicting hydrodynamic bounds in just Newtonian and power-law fluid models where limit bounds diverge with increasing shear-thinning [185]. Perturbation-based methods are generally used to compute solutions of fluid flow at relatively low Weissenberg number [190]. For instance, asymptotic perturbation about the leading order Newtonian fluid model has been used to evaluate the motion of transversely isotropic rigid particles in second-order viscoelastic fluid suspension [190], [191]. Consistent with experimental observations, at low shear rates, viscoelastic fluids cause suspended particles to slowly drift through various Jeffery's orbit until the attainment of an equilibrium orientation state in the flow vorticity direction. At higher shear rates, particles re-orient with the flow direction and their rotations are suppressed. Extension of Jeffery's theory to other particle shapes reveals that while prolate spheroids rotate towards a log-rolling position in the vorticity direction, oblate spheroids have an affinity for tumbling in the flow plane [192]. Deviations in particle shape from Jeffery's assumption of geometric asymmetry are found to produce significant changes in the particles motion. For general non-axisymmetric ellipsoids, Hinch and Leal [193] showed that particle motion is doubly periodic, consisting of a fast-tumbling motion around Jeffery's orbit and a slower drift representing a periodic change in Jeffery's orbit. On the contrary, application of the perturbation technique to investigate the effect of weakly shear-thinning fluid rheology on particles motion in unconfined sheared viscous suspension revealed that the degeneracy of Jeffery's orbit where unaffected by the non-Newtonian fluid rheology [194]. However, Jeffery's orbit and period were found to be instantaneously modified by the shear-thinning fluid behavior, and the quantitative modifications depended on the particle's initial conditions. Analytical based perturbation methods have also been used to study the effect of other 'non-Standard' Jeffery conditions on the configurational determinacy of suspended particles and effective fluid rheology of viscous flow suspension, such as particle and fluid inertia effects [174], [195], [196], contribution of Brownian disturbance [197], [198], [199], [200] and the effect of deformable particle shape [174], [201], [202]. As expected, the various phenomena investigated alter the dynamics, orbital configuration, and drift of Jeffery-like particles.

Numerical simulation techniques developed for particle motion are summarized in various review literature [22], [203], [204], [205]. Numerical method is tenable to increased model complexity and improved idealization with increased accuracy which comes at a high computational cost. Numerical based models are classified into mesh-free or particle-based method (PBM) and the traditional gridded continuum or element-based method (EBM) [203], [204]. PBM may be categorized based on physical or computational modelling. To avoid detraction from the primary focus of this dissertation, the reader is referred to existing review literature for more details [203], [204]. In PBM, the governing equations are discretized with moving sets of free particles retaining field-state information. PBM is a meshless, fully Lagrangian-based highly adaptive technique that allows for instantaneous tracking of individual particle response within a heterogenous multiphase system and capable of modeling flow fronts, free surfaces and accurately solving large deformation problems [206], [207], [208], [209], [210]. Examples of PBM include the explicit Smoothed Particle Hydrodynamic (SPH) and the Moving Particle Semi-Implicit (MPS) method and Discrete Element Method (DEM). The SPH method utilizes an explicit FSI coupling algorithm, while the MPS technique uses an implicit fully coupled FSI algorithm for improved prediction accuracy of the fiber and matrix motion and more accurate prediction of the suspension rheological properties which can help improve material properties [210]. Although PBM has been applied to evaluate complex single-phase flows with non-linear fluid rheology [211], [212], [213], [214], the behavior of suspended particles in non-linear suspension flow are seldom evaluated with this method. In DEM, the suspended fibers are represented as chains of discrete particles (either hard spheres, rods or ellipsoids connected by joints/linkages with predefined mechanical behavior) that interact through hydrodynamic forces, inter-particle forces (short and long range hydrodynamic forces, Brownian and colloidal force), and intra-particle forces (elastic, flexural forces etc.) and particle motion is computed by equilibrating the net force and torque on individual particles according to Newtons third Law [22], [205]. However, the fluid media in these simulations is modelled as a continuum governed by the Navier-Stokes flow equation. In DEM, one-way FSI coupling is often used to reduce computational cost. However for more accurate representation of the FSI interactions, back coupling is required to capture the effect of fluid hydrodynamics forces on particle dynamics and the resulting disturbance on the surrounding flow due to the fiber's motion. Typical DEM solution techniques include the Dynamic Numerical Simulation (DNS), Lattice Boltzmann Method (LBM), and particle Finite Element Analysis (pFEA). The representation of fiber particles as interconnected chains of discrete particles interlinked with joints having directional stiffness and failure property definition makes it possible to simulate fiber deformation and breakage at the joints. Applications of DEM to FSI problems are summarized in [22], [205]. DEM has been used extensively to study the behavior of single particles in Newtonian viscous suspension [215], [216], [217], [218], [219] and in non-linear viscous suspensions [220], [221]. Detailed bibliography on DEM based microscopic fiber suspension simulation can be found in Kugler et al. [22]. The literature presents different DEM model considerations including different types of particle discretization method, flow-field types, FSI coupling types and active fiber forces with regards to various quantities being investigated such as particle motion and deformation, suspension rheology, fiber breakage, optimum fiber length etc. (cf. Figure 2.4) PBM methods may be combined to simulate the EDAM process so as to exploit their advantages. For instance, the SPH method may be combined with the DEM method to simulate flows with moving boundaries/free surface, while capturing inter-particle interactions and large particle deformations [22], [26], [222].

EBM requires that the continuum domain be discretized into sub-domain units. EBM types include the Finite Element Method (FEM), the Finite Difference Technique (FDT), the Finite Volume Method (FVM) and the Boundary Element Method (BEM) [203], [205]. In EBM, individual domain units are interconnected via topological maps. EBM involves transformation of a complex Partial Differential Equation (PDE) into a system of algebraic equations with solutions computed at unit nodes, cells or elements level to yield an approximate general solution. EBM are well-established and highly evolved numerical techniques that are used extensively to solve Computational Fluid Dynamics (CFD) and FSI transport problems. However, because FSI problems often involve free surfaces, moving and/or deformable boundaries, and/or large deformation, the inherent complexities involved in remeshing, updating state variables and the errors introduced with excessive mesh distortion in EBM (even with the Arbitrary Lagrangian-Eulerian (ALE) technique) often makes EBM less attractive [203], [205]. On a single particle, physical modelling that balances the net hydrodynamic forces and couples on the surface of the particle is required to compute the particle's motion.

In complex FSI multi-particle suspension systems with a heterogenous distribution of suspended particles, it is customary to homogenize the multiphase continuum into a single uniform suspension phase having equivalent characteristics as the actual suspension using an averaging, smoothing or stochastic diffusing algorithm [205]. Coupling of the characteristic aggregate particles' state (orientation and spatial distribution) with the properties of the homogenized suspension is achieved using one of the available structure-based stress tensor rheological constitutive models [132], [133]. The evolution of the aggregate particles' average orientation dynamics can be computed using any of the available advection-diffusion moment tensor analytical models such as the Advani-Tucker's second order orientation equation of state [19], [22]. In BEM, solutions of state variables are computed only at the physical boundaries of the flow domain, hence reducing the problems dimensionality order as compared to other EBM. BEM simulations are thus faster, less computationally expensive and more accurate than other EBM. FDM and FVM has been used to compute flow field and fiber orientation dynamics in mold filling process [223], [224], [225]. BEM has been successfully implemented to study flow-field development of particulate suspension in viscous shear flow [226], [227], [228] and FEM has been used to study single particle behavior in linear viscous shear flow [57], [229], [230], [231]. Relevant to this study are the applications of EBM in non-linear single particle suspension. For example, 2D FEM has been used to simulate single rigid spheroidal particle behavior in dilute non-linear viscous shear flow [185], [231], [232], [233]. The studies showed that shear-thinning rheology only slightly affects the particle's kinematic, and this impact diminishes with increasing fiber slenderness. Moreover, increased shear-thinning was shown to significantly reduced the magnitude of the pressure distribution surrounding the particle surface while having a negligible effect on the surface pressure profile shape itself [185], [232]. Using a coupled FEM - Brownian dynamic simulation (BDS) based Langevin approach, Zhang et al. [234] simulated the effect of Brownian disturbance from surrounding fluid molecules on the motion of a single fiber, which was shown to be directly related to the magnitude of the Peclet number.

Macroscale level physics may vary depending on the manufacturing process and phenomena being investigated. Macroscale simulations are usually performed to investigate global phenomena and predict processing conditions and the global flow state such as the velocity, pressure and temperature distribution fields, flowrates, rate of heat transfer, etc. The heterogenous nature of fiber suspension involving a two-phase mixture often necessitates multiscale simulations to investigate localized phenomena such as the development of micro-voids, fiber orientation, fiber breakage, etc. Moreover, localized phenomena can in turn influence macroscopic behavior such as the suspension rheology. Multiscale simulation commonly involves coupling physics on two scales (i.e. macroscale and microscale) using constitutive equations [133].

Two phase short fiber suspension flow simulation can be classified into three types based on the method of representing the suspended solid fiber phase in the polymer melt mixture [132] which includes:

(a) Mathematical abstraction using analytical models such as the Folgar-Tucker's fiber orientation tensor model that predicts the transient fiber orientation tensor state based on the flow field velocity gradient, the particle's geometrical parameter and the fiber concentration accounted for in the phenomenological interaction coefficient used in the equation. Usually, the polymer composite melt is simulated using either any method such as EBM-FEM method [23], [24], [135], [237], or PBM-SPH method [207], [208]. Depending on the FSI coupling technique, either weak one-way or complete back-coupling model, the influence of the fibers on the polymer melt flow could be accounted for through the constitutive model used in the conservation equations that depends on the fiber orientation tensor state. This method is often used to study short fiber orientation evolution during EDAM processing.

(b) Discretization of the solid particle phase using either the PBM and/or EBM numerical approach. Here fiber motion and deformation such as fiber bending and breakage, etc. and its influence on the fluid's rheological properties and flow-field is simulated and visualized. This approach has been used to study fiber orientation evolution and nozzle clogging in EDAM processing. [26], [132] which is known to result from high degree of misaligned, long-length and cross-linked fibers in the nozzle contraction. Two discretization numerical approaches have been coupled together to simulate the matrix fluid phase and the fiber solid phase separately to exploit their unique advantages. For example [26] used a discrete SPH method to model the Newtonian incompressible polymer matrix phase and included bonded DEM particles to model suspended fiber particles making it possible to capture fiber motion, deformation and breakage in typical EDAM polymer composite melt flow processing.

(c) Phase homogenization method using an equivalent fluid phase mixture that combines properties of the pure polymer resin and the fiber inclusions. The homogenized fluid phase has been simulated using the MPS particle method where each particle is a composite material having equivalent physical properties of the resin and fiber phase present based on their weight fractions computed using the rule of mixture. The method has been used to simulate bead cooling during deposition and predict the evolution of deposited bead cross-section during solidification [238].

The choice of simulation method depends on the transport phenomena of the EDAM SFRP composite process being investigated, the desired degree of accuracy and the level of sophistication involved.

Evidently from the literature review, extensive efforts have been made to simulate various transport phenomenon associated with EDAM polymer composite process, however most simulation efforts have focused on global transport phenomenon which only requires macro-scale level modeling. Even when coupling the effect of the suspended fibers on the polymer deposition flow process, their influence is mostly used to study global transport phenomenon like prediction of global melt flow-field and fiber orientation distribution, extrudate swell and solidification behavior, bead deformation and shrinkage, inter-bead surface adhesion, etc. On the other hand, microscale level simulation has mostly been used to study particle motion and deformation in viscous homogenous flow suspension and evaluate the structure and rheology of dilute and semi-dilute suspension. However, there is little or no literature on multiscale level simulation used to study local transport phenomena in the actual EDAM polymer composite deposition flow process such as fiber breakage or the development of micro-voids within the polymer melt during processing. The current research is a first attempt that utilizes multiscale FEA based modelling approach to simulate particle motion along streamlines of EDAM polymer melt deposition flow process with an aim to investigate flow induced mechanisms that may be responsible for micro-void formation on the surface of suspended particles by studying the localized pressure distribution on the particles' surface. This has been discussed in detail in later chapters of this dissertation.

# CHAPTER THREE

## Microstructural Characterization of Large Area Additively Manufactured Polymer Composite Bead

Sections of this chapter are taken from: Awenlimobor, A., Sayah, N. and Smith, D.E., 2025. Micro-void nucleation at fiber-tips within the microstructure of additively manufactured polymer composites bead. *Composites Part A: Applied Science and Manufacturing*, *190*, p.108629.

Microstructural characterization of SFRP composites beads is crucial in understanding how the beads microstructure relates to the thermo-mechanical properties and part performance. These characterizations provide enhanced understanding of the effect of manufacturing process conditions on bead properties making it possible to optimize the bead microstructure and improve its microstructural properties and part performance. Techniques typically used to analyze the microstructure of polymer composites print beads include Optical Microscopy, Transmission Electron Microscopy (TEM), Raman Spectroscopy, Scanning Electron Microscopy (SEM). More recently, the advent of X-ray micro-computed tomography (μCT) imaging non-destructive analysis technique has led to higher resolution three-dimensional (3D) visualization and more accurate characterization of the microstructure of polymer composites at the micron scale as compared to 2D imaging techniques such as SEM [40], [50], [239]. μCT has been widely used to identify and characterize the microstructures of polymer composites including inherent micro-constituents' phases and contents and defects. Additionally, μCT can be used for in-situ real-time monitoring of processes at the micron-scale.

Extensive review literature on the study of micro-voids within EDAM printed SFRP composite beads was previously provided in detail in Section 2.2. Because intra-bead micro-voids are more prevalent in highly filled polymer composite beads and are known to be extremely detrimental to the composite part that cause significant property loss to materials as explained in the literature review and in [3], [4], [10], [34], it is useful to quantify and characterize them to gain fundamental insight into their formation mechanisms. From literature, it is important to study micro-voids with respect to various microstructural metrics that may provide better understanding of the micro-void development within complex microstructure of the print beads such as the proportion of the individual microconstituent phases, the average length, orientation and spatial distribution of fiber reinforcements, the spatial distribution and morphology of the inherent micro-voids and their interactions with other microconstituent phases, etc. Moreover, literature suggests a very high propensity for micro-void to segregate at the tips of fiber reinforcements especially in resin lean regions of the bead with markedly high fiber tips aggregation [5], [11], [29].

In this chapter, we aim to quantitatively characterize the micro-void content within an EDAM polymer composite bead microstructure with a focus on the relationship between micro-voids and fiber tips within the printed bead. In existing literature, the phenomenon of micro-voids nucleation at fibre tips have only previously been addressed from a qualitative perspective [5], [11]. To this end, the following experimentally examines microstructural formations of 13% carbon fiber filled ABS polymer composite EDAM beads using high resolution 3D X-Ray μ-CT imaging and computational methods for extracting quantifiable details from the μ-CT data.

### *3.1 3D Printing Process*

The Baylor University Large Scale AM (LSAM) system with a print volume of 48"× 48" × 6" high was used to produce short fiber polymer composite beads for characterizing microstructural voids in this study. The LSAM system is composed of a Strangpresse Model 19 single-screw extruder (Strangpresse, Youngstown, OH, USA) with three temperature control regions along its length and a nozzle diameter of 3.172 mm. PolyOne CF/ABS (Avient Corporation, Avon Lake, OH, USA) with 13% carbon fiber weight fraction was used as the LSAM feedstock. Pellets were dried in a convection oven at 80°C for twelve hours before the 3D extrusion/deposition processing was performed. Figure 1 (a-d) is a flow-chart that illustrates a typical polymer composite deposition of a single bead on a print bed and post-3D image acquisition and analysis of a cut section taken from a straight printed bead sample which used for our study. The bead sample was sufficiently long to ensure that a quasi-steady extrusion/deposition process was achieved. The EDAM internal nozzle geometry and printing process parameters appear in Figure 3.1(a) and Table 3.1.

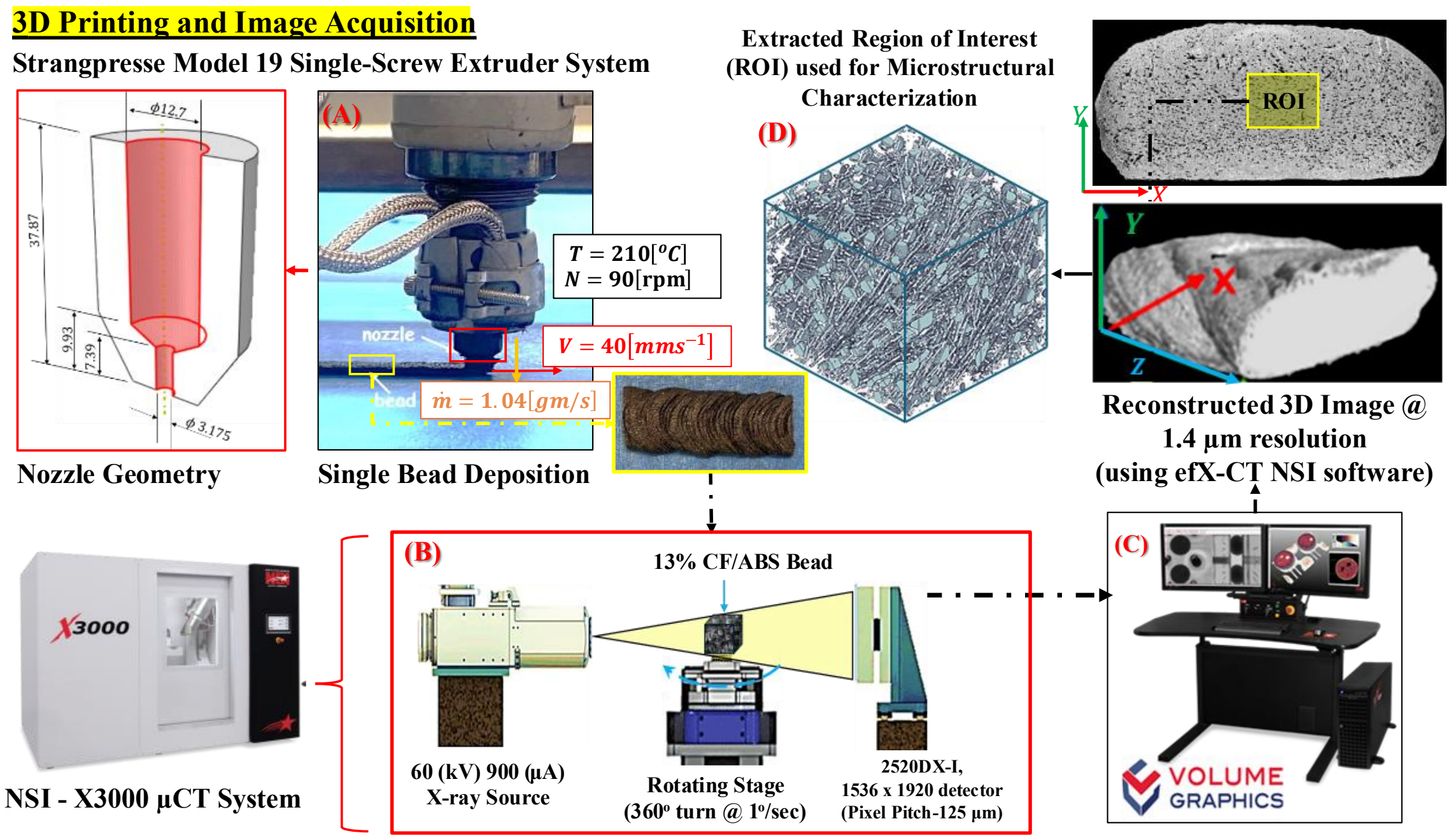

Figure 3.1: Flow-chart illustrating (a) typical LSAM bead printing (b) µ-CT image scanning and acquisition of the printed bead specimen using NSI-X3000 X-Ray µ-CT system (c) 3D reconstruction of acquired 2D images using efX-CT NSI software and (d) ROI extraction of the reconstructed 3D grayscale µ-CT voxel-data using MATLAB software.

Table 3.1: LSAM bead print process parameters

| Printing Process Parameters | Units | Value |
|---|---|---|
| Temperature | [°C] | 210 |
| Screw speed | [rpm] | 90 |
| Extruder mass flow rate | [gm/s] | 1.04 |
| Nozzle translation speed | [cm/min] | 240 |
| Nozzle diameter | [mm] | 3.17 |
| Nozzle height | [mm] | 1.20 |

### *3.2 µ-CT Image Acquisition Technique*

The North Star Imaging X3000, X-RAY µCT system (North Star Imaging, Rogers, MN, USA) was used to scan the CF/ABS deposited bead sample (cf. Figure 1b). µCT scans were performed at a resolution (voxel size) of 1.7 microns using an X-ray source at 60 kV and 900 µA to provide adequate contrast between the various constituent phases that compose the bead specimen. The sample was rotated 360 degrees in 1-degree increments, resulting in 2400 projections. The detector captured the transmitted X-ray signals, obtaining 2D attenuation distribution data. The acquired μCT scan data was then reconstructed using efX-CT software (North Star Imaging, Minnesota, USA), (cf. Figure 1c). During reconstruction, an outlier median filter preprocessing technique was used to reduce noise and improve the detection of boundaries between microstructural features such as voids and fibers within the ABS matrix.

### *3.3 μ-CT Image Data Post Processing*

μ-CT X-ray imaging techniques were used to generate 3D grayscale voxelated data based on density for a cube-shaped specimen with a side length of 0.35mm obtained from the CF/ABS bead where each voxel has a side length of 1.4μm (cf. Figure 1d). Unless stated otherwise, all post processing operations presented here are performed using built-in functions from MATLAB's (Mathworks, Natick, MA, USA) 3D image processing toolbox. The process used in this work for evaluating a Region of Interest (ROI) of a CF/ABS bead appears in Figures 3.2 and 3.3 which illustrate the description to follow where 'Seq.' refers to the event sequence for the image processing operation of interest. For each μ-CT dataset, grayscale data is classified into three groups using the '*imsegkmeans3*' statistical function to obtain binary data for each segment representing the different constituent phases that include ABS matrix, micro-void (air), and fiber inclusions as shown in the typical sample illustrated by Seq. #1 & #2 as shown in Figure 3.2. These images show a typical region of material from our LAAM bead appearing here for illustration of the imaging post processing analysis. The '*bwlabeln*' function is used to identify individual fiber or fiber clusters and void features by determining connected voxels having the same phase within each segment (cf. Figure 3.2, Seq. #3). Separation of the fiber clusters into individual pristine fibers is achieved by slightly adjusting the cluster's intensity value and filtering the data using the Hessian-based Frangi-Vesselness '*fibermetric*' function. The transformation of a typical fiber cluster into separate fiber vessels after grouping and filtering operation is depicted by Seq. #4a (cf. Figure 3.3). Subsequently, a skeletonization operation is performed to extract the individual vessel ribs using the '*bwskel*' function. Unfortunately, the filtering operation erodes the cluster data which results in the splitting of some ideally pristine fiber skeletons into broken fragments as can be seen in the resulting fragmented skeletal framework after Seq. #4b (cf. Figure 3.3). To resolve this, a custom algorithm is implemented that identifies and stitches together line fragments belonging to unique pristine fibers by matching orientation data of fragment pairs along their centroidal axes within proximity to each other and connecting missing voxels of nearby ends in the predetermined direction.

The resulting skeletal framework of pristine fibers after stitching end extension operation is shown in the image after Seq. #4c (cf. Figure 3.3). The stitching algorithm is limited by the efficacy of the built-in skeletonization function in obtaining a sufficiently smooth and central skeletal framework. Region property information including the centroid, orientation, and geometry data for individual line segments is obtained using the "*regionprops3*" function. After a successful stitching operation is complete, the endpoints of pristine fiber skeletons are obtained using the "*bwmorph3*" morphological operation function which are depicted by blue markers in the image after Seq. #4c (cf. Figure 3.3). We found that grayscale data erosion due to the filtering operation often results in shorter pristine fiber skeleton ribs that necessitated the development of an algorithm that extends the skeleton terminals along its principal direction to the edge of the fiber feature. Fiber end regions for fibers with clearly defined tips are defined by a 5-voxel unit radius around the end points of individual fiber skeleton within the fiber grayscale dataset. The fiber tip regions are depicted by the green regions in the superposed volumetric plot of the fiber cluster (gray) overlayed on its skeleton (red) after Seq. #4d (cf. Figure 3.3). For irregular fiber features having no unique tips and having a low aspect ratio (typically less than 3), the entire fragment is considered a tip.

By juxtaposing individual void features with fiber features and individual void features with fiber tip voxels through appropriate indexing operation, the fraction of voids by volume isolated within the ABS matrix and those in contact with fiber tips are, respectively, determined. The probability of a pristine fiber feature extending beyond the volumetric bounds of the cut-specimen is accounted for by excluding fiber tip regions within 5-voxel units of the volume bounding surface. Figure 3.3 shows the typical fiber tip regions (green) in a ROI with random pristine fiber samples (gray) after fiber tip exclusion zone definition (Seq. #5), and a volumetric superposed plot highlighting all relevant interest features with unique colormaps after the feature identification through indexing operation (Seq. #6) including pristine fiber regions (gray) with their associated fiber tips (green) and interacting tip voids (blue) together with a few samples of voids isolated within the matrix (red). The regional mean fiber orientation is obtained by averaging orientation data of fragmented segments of pristine fiber skeletons within the region and the splitting operation is achieved using the built-in '*spectralcluster*' function.

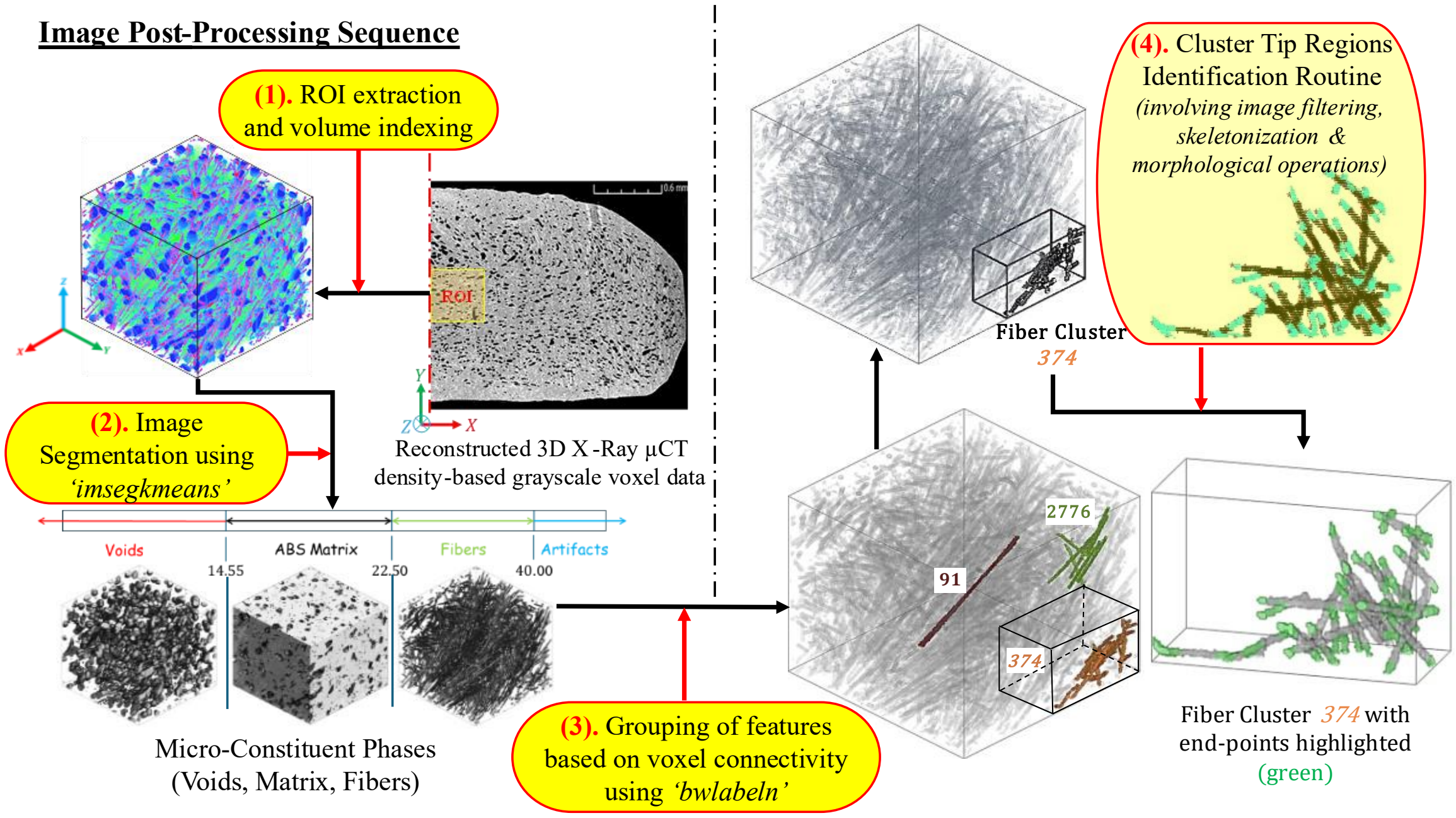

Figure 3.2: Flow-chart illustrating ROI extraction, grayscale thresholding based binary image segmentation of the bead sample, feature identification through grouping and filtering operation, and fiber cluster separation and tip region identification operation.

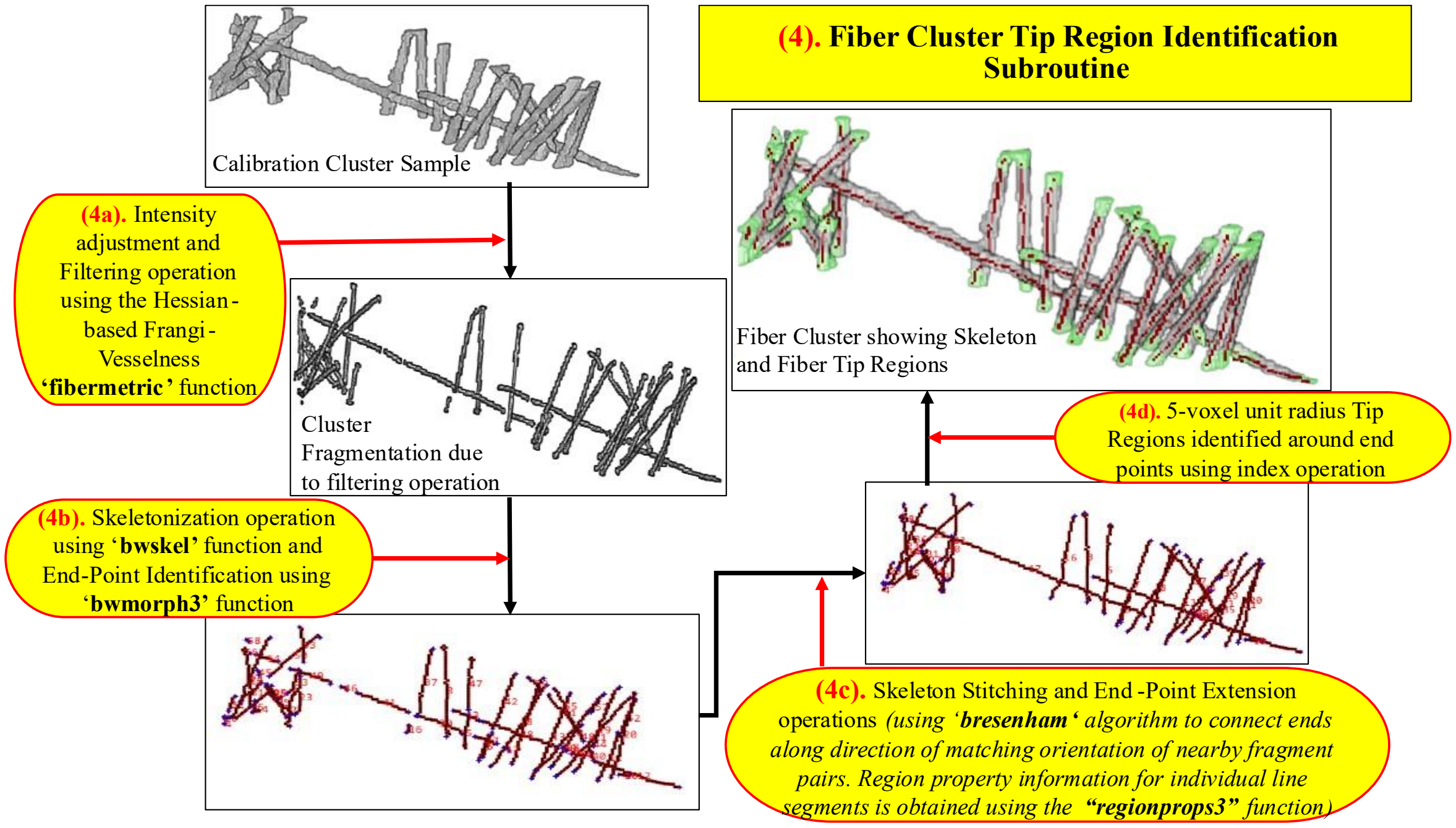

Figure 3.3: Flow-chart detailing the fiber cluster separation subroutine operation (cf. Figure 3.2, Seq. #4) including filtering, skeletonization, stitching and end-point extension and fiber tip region identification operations.

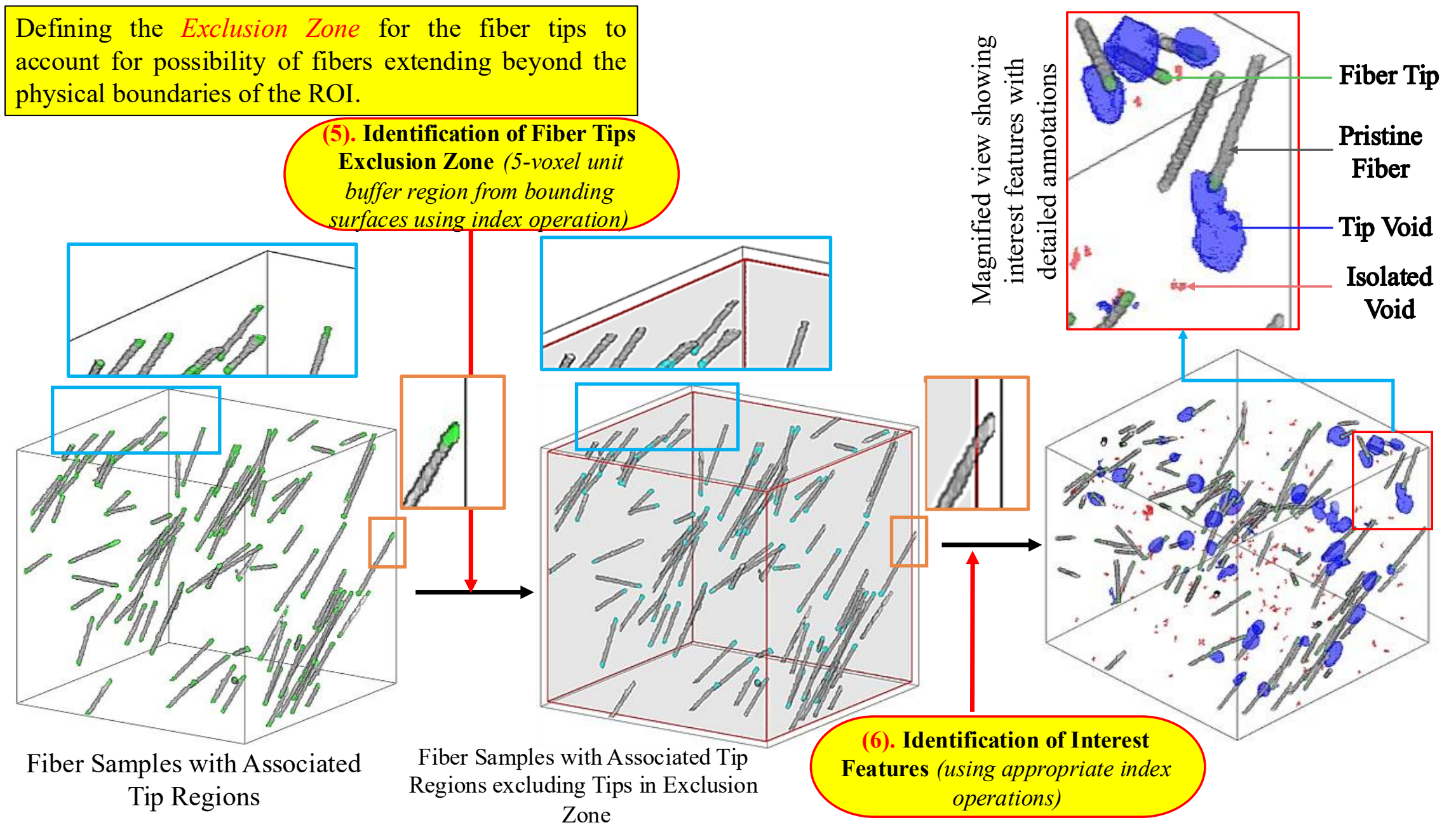

Figure 3.4: Flow-chart showing the fiber tip exclusion zone identification operation and the identification of interest features through indexing operation.

Figure 3.5a shows the resulting ROI highlighting relevant interest features contained within the volume after completion of the image post-processing process including the pristine fiber samples (gray), the voids interacting with fiber tips (blue) and the voids isolated within the matrix (red). For better visualization, Figure 3.5c shows a magnified cut section view extracted from the central region of the ROI (cf. Figure 3.5b) showing the relevant interest features within the volume. Of particular interest are the different classifications of the micro-void contained within the ROI volume determined from the image post-processing analysis. Figure 3.6a shows the volume content of homogenous micro-voids isolated within the matrix, while Figure 3.6b shows the content of heterogenous micro-voids touching fiber tips and Figure 3.6c shows the heterogenous micro-void content touching fiber but not fiber tips.

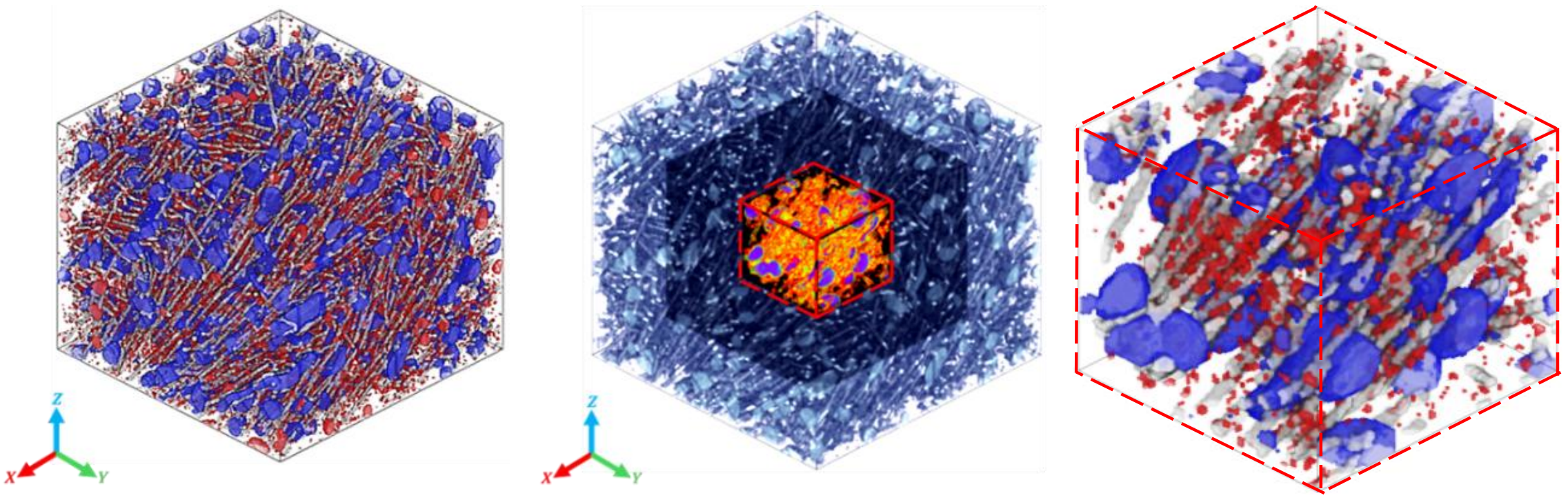

(a) (b) (c)

Figure 3.5: (a) ROI volume showing relevant interest features including pristine fiber samples (gray), micro-voids interacting with fiber tips (blue) and the micro-voids isolated within the matrix (red) (b) ROI volume highlighting a central region (c) magnified view of the central region extracted from the ROI volume (ROI Cubic Envelope Size:0.35mm x 0.35mm x 0.35mm).

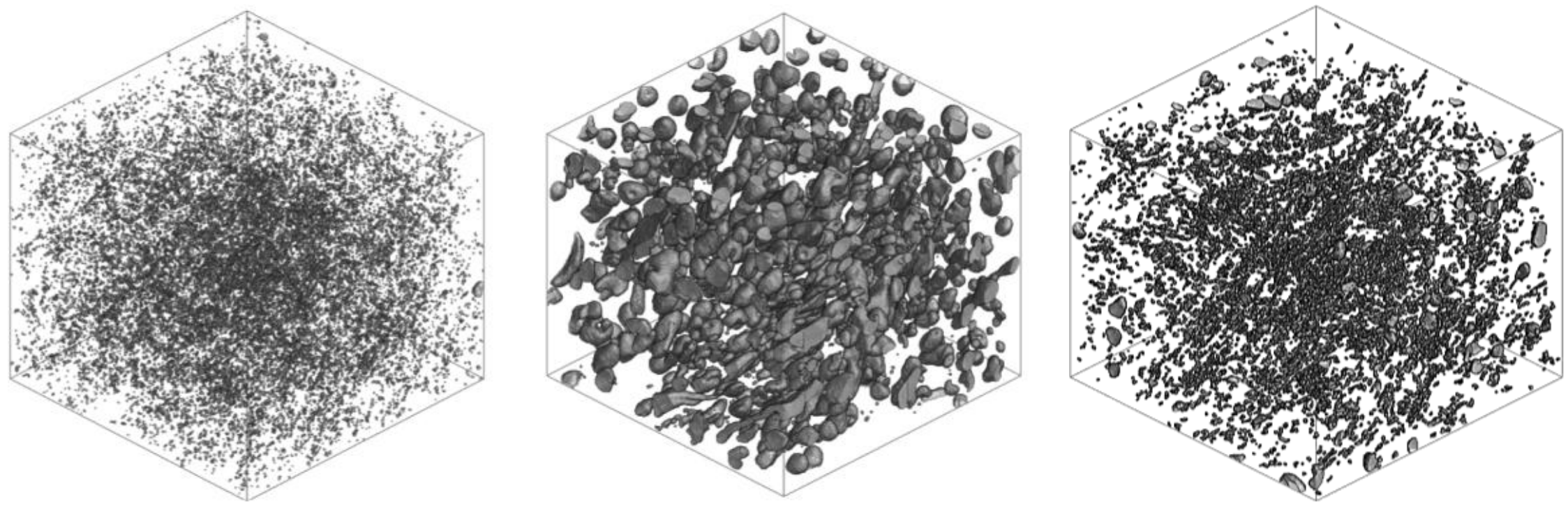

(a) (b) (c)

Figure 3.6: ROI volume showing (a) homogenous micro-voids isolated within the polymer matrix (b) heterogenous micro-voids with fiber tip interaction (b) heterogenous micro-voids without fiber tip interaction (Cubic Envelope Size:0.35mm x 0.35mm x 0.35mm).

Figure 3.7 presents tomography section slices of a typical microstructural region along the primary mid-planes of the 3D grayscale voxelated volume that shows the segmented microstructural features of interest including the micro-void regions in contact with a fiber tip (orange), the fiber regions (dark red), transition zones between two phases (bright greenish yellow) and the micro-void regions without fiber tip interaction (light blue spots).

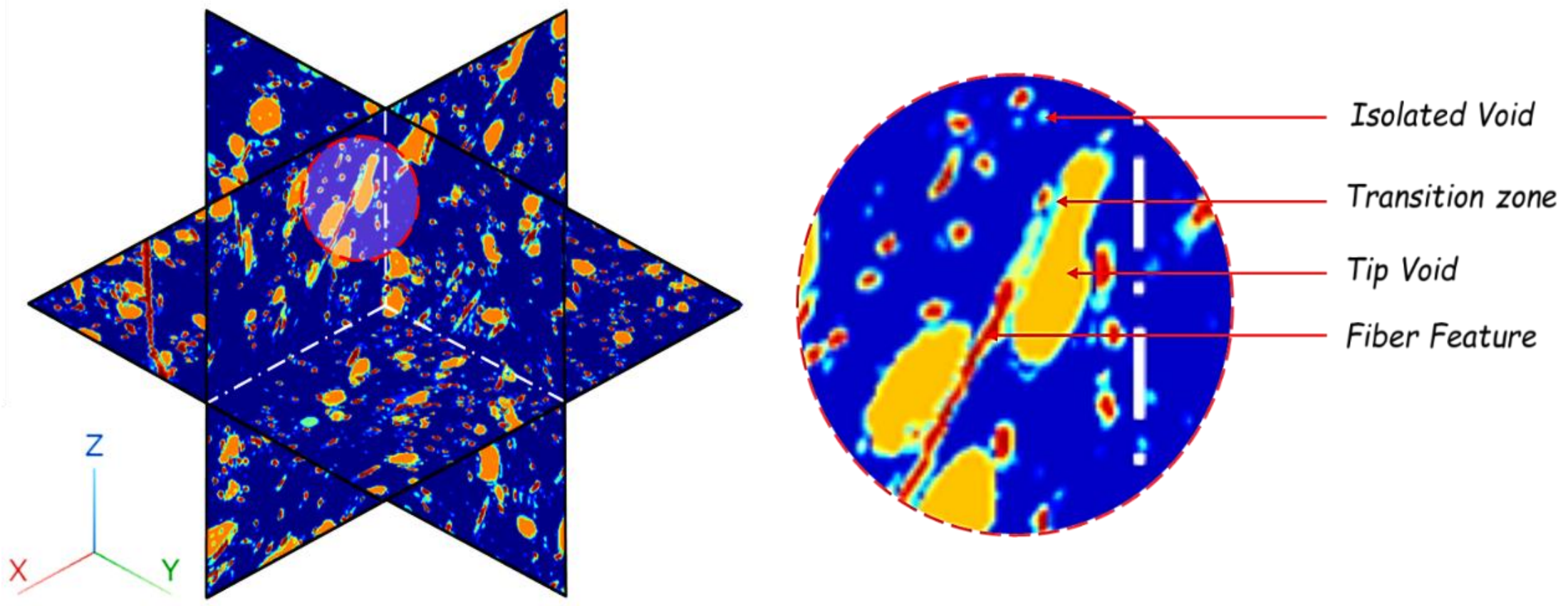

Figure 3.7: Volumetric tomography slice at the mid-planes of the 3D voxelated grayscale data.

### *3.4 μ-CT Data Analysis*

The microstructural features of the 13% CF/ABS polymer composite bead specimen obtained from the 3D μ-CT voxelated data post processing technique described above were analyzed for four (4) regions of interest (ROIs) within the LSAM printed bead as shown in Figure 3.8. The ROIs include ROI-I near the base of the bead near the build surface, ROI-II at the beads center, ROI-III near the free surface at the edge of the bead, and ROI-IV near the upper surface of the bead. These ROIs were chosen to provide a representative sampling of the bead cross section in regions that appear to have variations in microstructure. Each cubic ROI volume has a side length of 0.35mm and consists of 250 equal sized voxel cubes per side with each voxel unit having side length of 1.4μm, yielding a total of 15,625,000 voxels per ROI.

The microstructure within each of the four (4) ROIs is characterized by nine (9) metrics with regards to micro-void formation which includes (1) the volume fraction of each constituent phase (i.e., micro-void phase, $\vartheta_v$, fiber inclusions, $\vartheta_f$, and polymer matrix, $\vartheta_m$) (2) the fraction of micro-voids isolated within the polymer matrix $\vartheta_{vm}$ (or conversely, the fraction of those in contact with fibers $\vartheta_{vf} = 1 - \vartheta_{vm}$) (3) the fraction of micro-voids in contact with fiber tips $\vartheta_{vt}$ (4) the fraction of fibers having a tip in contact with a micro-void $\vartheta_{ft}$ (5) the average diameter of the micro-voids in contact with a fiber $d_{vf}$ (6) the average diameter of the micro-voids isolated within the matrix $d_{vm}$ (7) the average sphericity of micro-voids in contact with a fiber, $\Phi_{vf}$ (8) the average sphericity of the micro-voids isolated within the matrix $\Phi_{vm}$ and (9) the principal components of the region-averaged fiber orientation tensor $a_{ij}$.

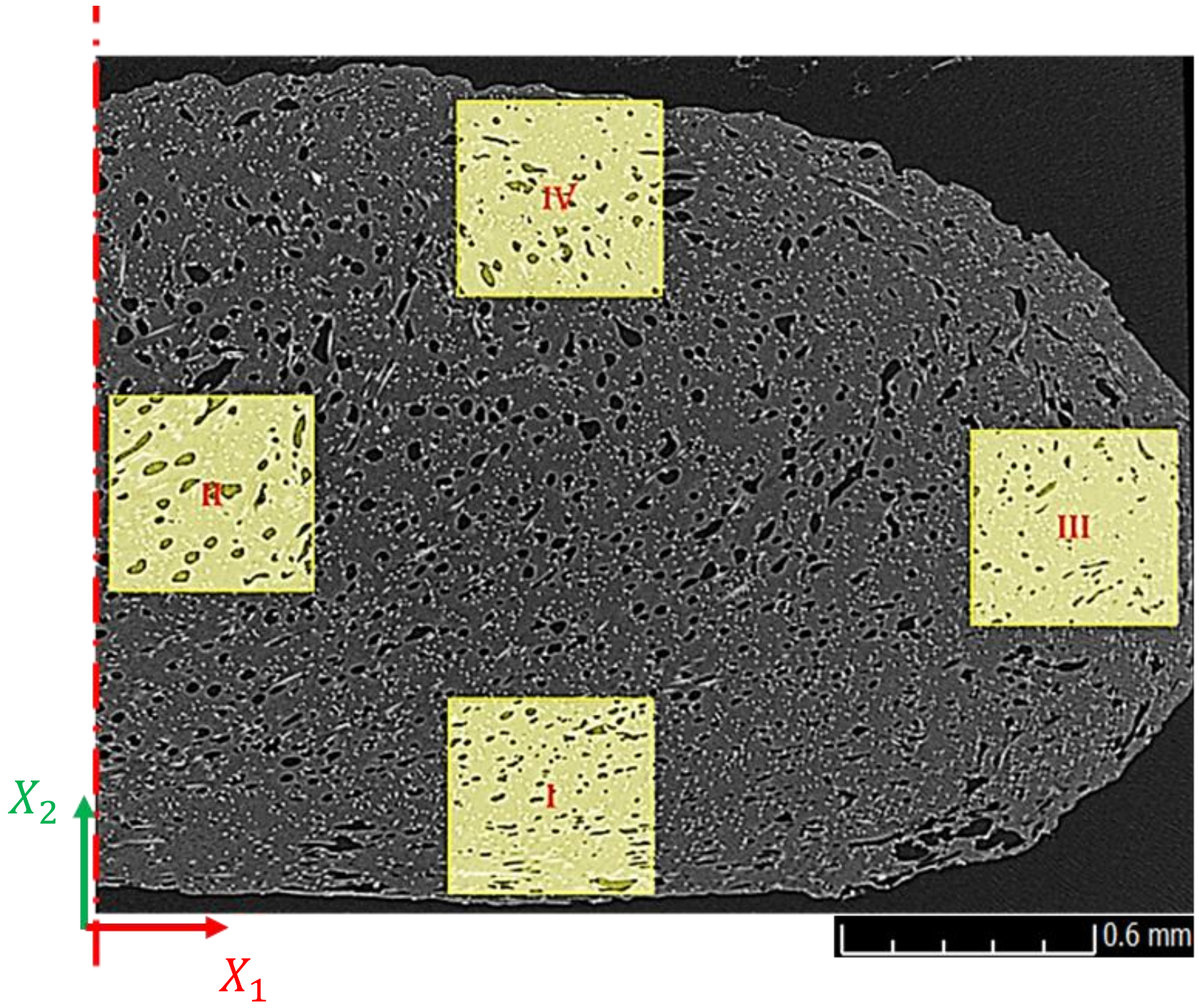

Figure 3.8: Regions of interest (ROIs) within the CF/ABS specimen from a polymer composite bead manufactured from Baylor's LSAM system.

### *3.5 Microstructure Assessment Metrics*

The volume fraction of the $p$-th constituent phase $\vartheta_p$ is simply the ratio of the volume of the $p$-th phase $V_p$ to the overall ROI volume $V$ written as

$$\vartheta_p = V_p/V \tag{3.1}$$

where subscript $(p)$ is either the void $(v)$, fiber $(f)$, or matrix $(m)$ phase. The volume fractions of the constituent phases should satisfy the conservation requirement (i.e. $\vartheta_m + \vartheta_v + \vartheta_f = 1$). The fraction $\vartheta_{vf}$ is defined as the ratio of the volume of micro-voids in contact with fiber $V_{vf}$ to the total micro-void volume within the ROI volume $V_v$. Likewise, the fraction $\vartheta_{ft}$ is defined as the ratio of fiber content with one or both tips in contact with one or more micro-voids $V_{ft}$ to the total fiber content in the ROI $V_f$. The fractions $\vartheta_{vf}$ and $\vartheta_{ft}$ are, respectively given as

$$\vartheta_{vf} = V_{vf}/V_v \qquad and\ \vartheta_{ft} = V_{ft}/V_f \tag{3.2}$$

It follows that the fraction of micro-voids isolated within the polymer matrix is given as $\vartheta_{vm} = 1 - \vartheta_{vf}$. In addition to quantifying the micro-void content, we also consider micro-void characteristics within the ROI volume including the micro-void average diameter $d_v$ and average sphericity $\Phi$. The micro-void equivalent diameter $d_v$ is from the diameter of a sphere with equal volume as the irregular shaped void as illustrated in Figure 3.9 and mathematically given as:

$$d_v = \sqrt[3]{{}^{6}/_{\pi}\, V_v} \tag{3.3}$$

The void sphericity $\Phi$ is a measure of the irregularity of the void shape and is computed based on the Wadell definition [240] as

$$\Phi = \frac{\sqrt[3]{36\pi V_v^2}}{A_v} \tag{3.4}$$

where $A_v$ and $V_v$ are the 3D voxelated boundary area and volume of each connected void region. The 3D boundary area is computed by isolating individual void features using a binary assignment and summing the total number of facially connected non-unity neighbor voxels to each bounding voxels of the individual void region. Figure 3.10 shows three (3) different representative micro-void features with different shapes and their sphericity values computed using eqn. *(3.4)*.

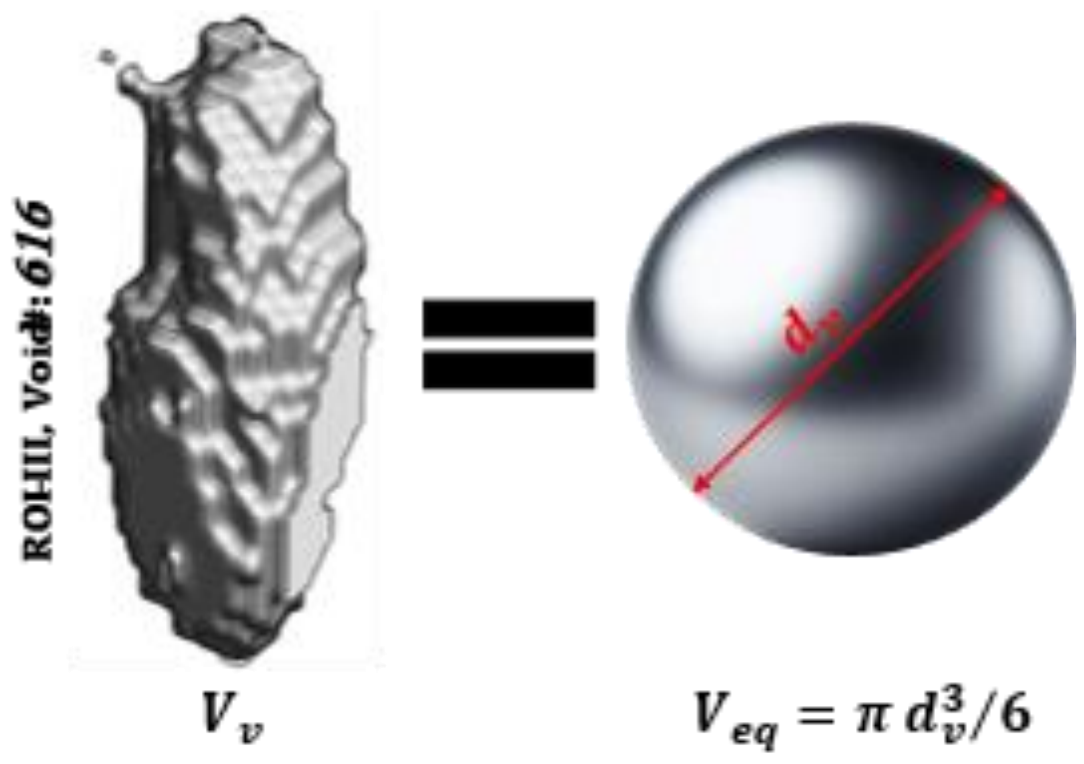

Figure 3.9: Visualization of a representative sphere element (right) with equal volume as an irregular shaped micro-void (left) used to determine equivalent void diameter and sphericity.

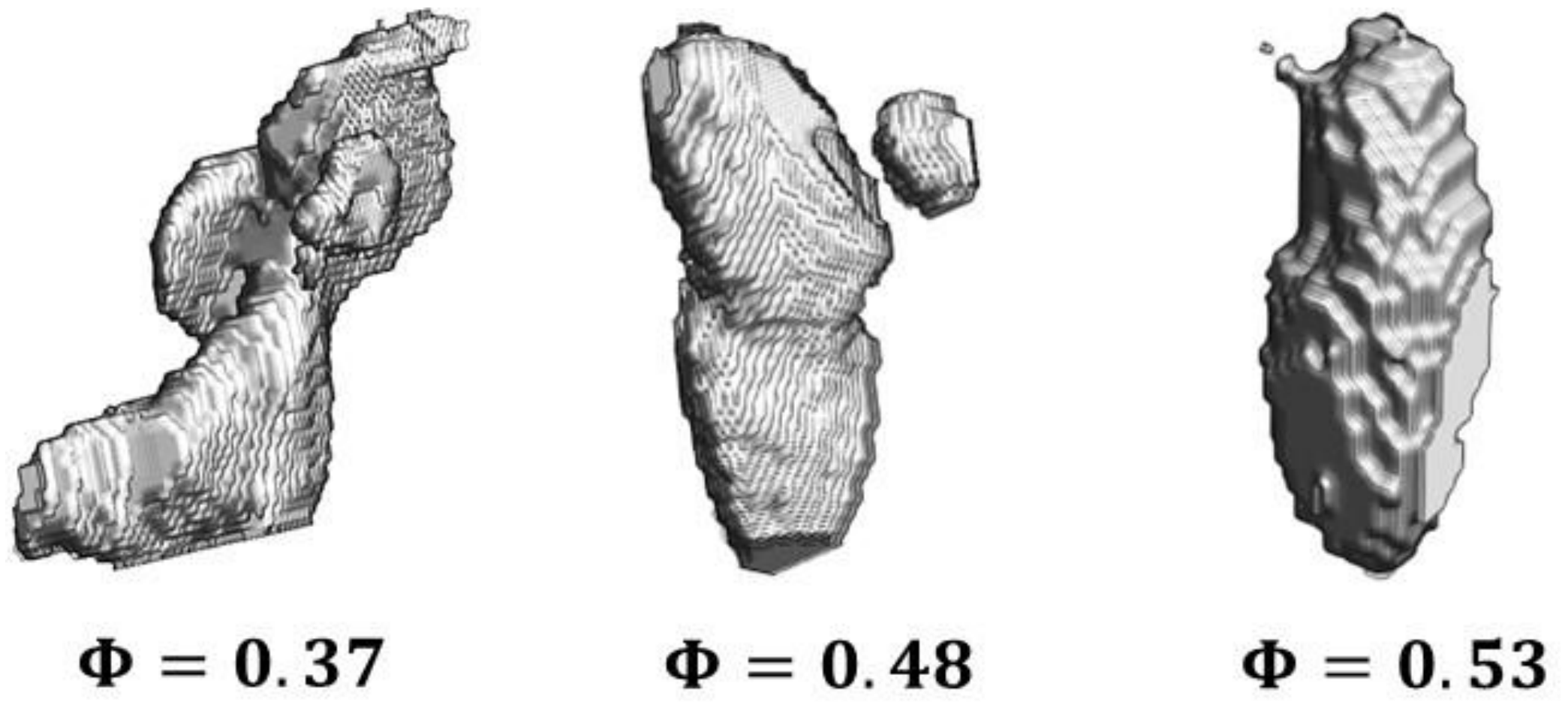

Figure 3.10: Typical micro-void features and their computed sphericity values.

The fiber orientation tensor $\mathrm{a}_{ij}$ for an ROI volume is determined from length weighted ensemble average of the dyadic products of individual fiber orientation vector $\underline{p}$ contained within the region volume written as

$$\mathrm{a}_{ij} = \frac{1}{n\bar{l}} \sum_{k=1}^{n} l_k p_i^k p_j^k \,, \qquad \bar{l} = \frac{1}{n} \sum_{k=1}^{n} l_k \tag{3.5}$$

where $l_k$ is the length of the $k^{th}$ fiber skeleton, $p_j^k$ is the $j^{th}$ orientation vector component of the $k^{th}$ fiber skeleton, and *n* is the total number of fiber skeletons contained within the ROI volume. The orientation vector $\underline{p}$ is given as

$$\underline{p} = [\cos\phi \sin\theta \quad \sin\phi \sin\theta \quad \cos\theta]^T \tag{3.6}$$

where the Euler angles $\phi$ and $\theta$ are shown in Figure 3.11. The normalization condition which relates the diagonal components of the orientation tensor requires that $\mathrm{a}_{ii} = 1$ where the repeated indices imply summation. The diagonal components of $\mathrm{a}_{ij}$ are used to describe the degree of fiber alignment with any of the orthogonal reference axis. In the current study, the z-axis indicates the print direction, and the y-axis is perpendicular to the print bed. The magnitude of the diagonal tensor components $\mathrm{a}_{kk}$ (no summation implied) ranges from 0 to 1, i.e. $(0 \leq \mathrm{a}_{kk} \leq 1)$, where a value of 1 indicates complete fiber alignment with the $k$-th coordinate direction and 0 indicates all fibers lie in a plane normal to the $k$-th coordinate direction.

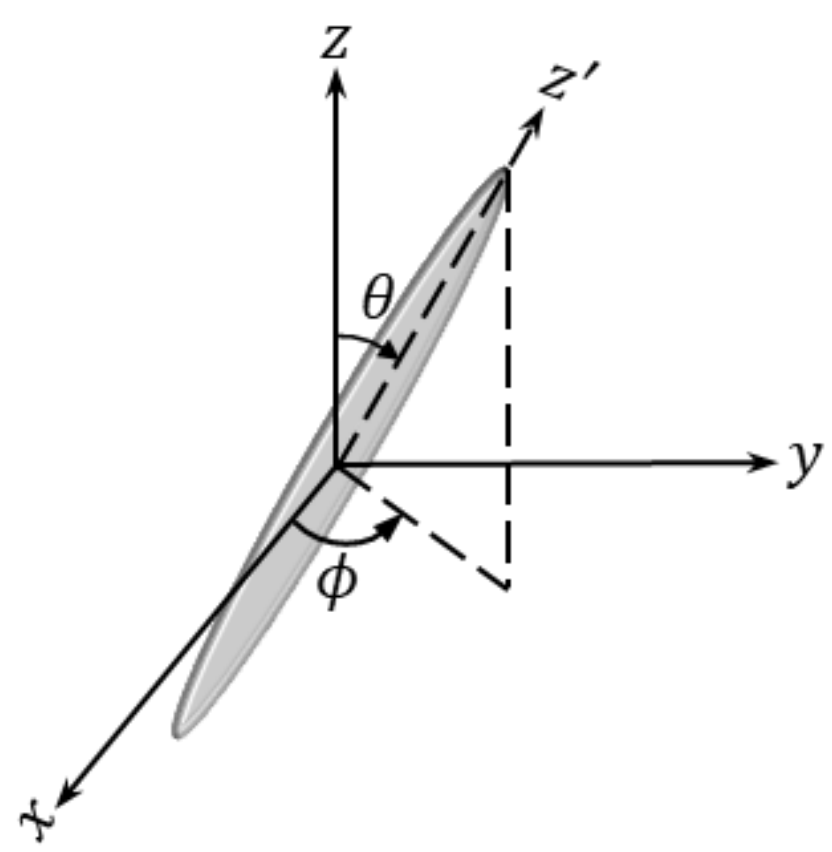

Figure 3.11: Showing the orientation of a single fiber with respect to the reference coordinate directions.

### *3.6 Results & Discussion*

Figure 3.12 shows microstructural features for the four (4) ROIs that result from the segmentation procedure described above in the methodology section which includes fibers (gray), voids touching fibers (blue), and voids not contacting a fiber tip (red). By visual inspection, it is apparent in Figure 3.12 there is a relatively high content of micro-voids touching fibers (blue) as compared to micro-voids not in contact with fiber tips (red). The fibers in each ROI can also be seen to be more aligned with the z-direction (print direction) and to a greater degree in ROI-III (cf. Figure 3.12c) with more densely packed and highly aligned pristine fiber striations as compared to other component directions. Moreover, the sizes of voids touching fiber tips (blue) can be seen to be relatively larger than that of other voids. We likewise observe more irregular and elongated shaped voids in ROI-III as compared to the other ROIs which have more spherical void shapes.

Table 3.2 contains values of the metrics defined above for assessing the microstructural features in each ROI. Calculated results reveal an average micro-void volume fraction $\vartheta_v$ near 11% with a standard deviation for $\vartheta_v$ less than 1% across all four ROIs. Within these four ROIs, the highest $\vartheta_v$ recorded is in ROI-II near the bead center while the lowest $\vartheta_v$ is in ROI-III near the edge of the bead. Most notable is that more than 89% of the micro-void volume (see e.g., $\vartheta_{vt}$ in Table 3.2) represents micro-voids that are in contact with a fiber tip in all four ROIs with as high as $\vartheta_{vt} = 95.7\%$ in ROI-III near the bead edge. In addition, the percentage of micro-void isolated within the matrix phase $\vartheta_{vm}$ is seen to increase with the overall void content in each ROI. The fraction of fiber skeletons having one or both tips in contact with one or more micro-voids (designated as $\vartheta_{ft}$) were on average observed to be higher in ROI-II and IV, nearing ~50%.

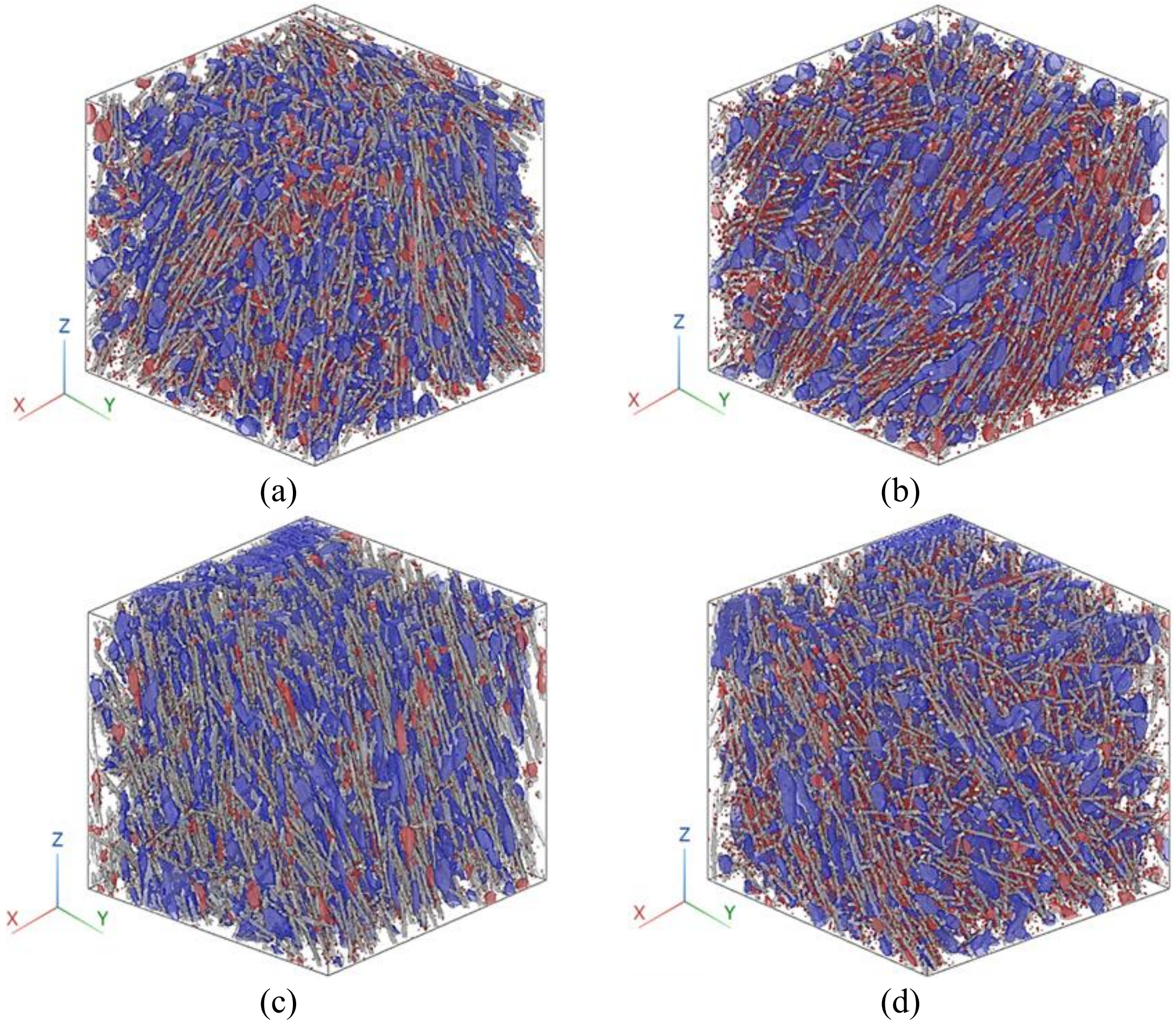

Figure 3.12: Segmented microstructural 3D image highlighting fiber features (gray), void with fiber tip interaction (blue), and voids without fiber tip interaction (red) for the different regions of interest of the CF/ABS bead (a) ROI-**I** (b) ROI-**II** (c) ROI-**III** (d) ROI-**IV.** (Cubic Envelope Size:0.35mm x 0.35mm x 0.35mm).

Table 3.2: Volume fractions of the microstructural features for the various regions of interest (ROIs) within the 13% CF/ABS EDAM printed bead.

| Symbol | Definition | ROI-I | ROI-II | ROI-III | ROI-IV |
|---|---|---|---|---|---|
| $\vartheta_v$ (%) | Void volume fraction | 10.70 | 12.27 | 10.06 | 11.17 |
| $\vartheta_f$ (%) | Fiber volume fraction | 6.96 | 6.65 | 7.25 | 7.53 |
| $\vartheta_{vm}$(%) | Fraction of voids isolated in matrix | 2.67 | 3.92 | 1.34 | 3.39 |
| $\vartheta_{vt}$ (%) | Fraction of voids touching fiber tip(s) | 90.17 | 89.68 | 95.70 | 91.39 |
| $\vartheta_{ft}$ (%) | Fraction of fibers skeleton with tip voids | 37.33 | 48.00 | 29.27 | 51.62 |

As may be expected, the average equivalent diameter of the voids contacting fiber(s) $d_{vf}$, were seen to be higher than the average equivalent diameter of micro-voids isolated within the matrix, $d_{vm}$ (cf. Table 3.3). The isolated micro-voids within the matrix on average had an equivalent diameter of $d_{vm} = 3.4\ \mu m$ with a standard deviation less than 0.4 $\mu m$ across all four ROIs. Alternatively, the equivalent diameter of the micro-voids in contact with fibers was on average seen to be higher in ROI-III - near the edge of the bead ($d_{vf} = 39.3\ \mu m$) followed by ROI-IV near the top surface of the bead ($d_{vf} = 35\ \mu m$) compared to the average equivalent diameter in other ROIs ($d_{vf} \approx 30\ \mu m$). Figure 3.13a-d shows the post-processed result of the heterogenous micro-voids in contact with fiber tips for the various ROI volumes. Evidently, ROI-III (cf. Figure 3.13c) is seen to have larger and more elongated micro-voids than other ROI's.

The distribution of equivalent diameter of micro-voids in contact with fibers in the various ROIs appear in Figure 3.14 along with fitted parameters for various 2-parameter probability distribution functions (pdf). In ROI-I (cf. Figure 3.14a) we found that the probability distribution of $d_{vf}$ is best represented by the Gamma pdf with a shape parameter $\alpha = 4.41$, and a scale parameter $\beta = 6.22$. Alternatively, the distributions in ROI-II & ROI-IV for $d_{vf}$ can best be fitted to a Weibull pdf (cf. Figure 3.14b & d) having a scale parameter of $\alpha = 32.27$ and a shape parameter $\beta = 2.71$ for ROI-II, and a scale parameter of $\alpha = 35.91$ with a shape parameter of $\beta = 2.07$ for ROI-IV. However, in ROI-III with larger sized voids, the distribution is best represented by a Lognormal pdf (cf. Figure 3.14c) having a location parameter $\alpha = 3.47$ and a scale parameter $\beta = 0.52$. The distributions of $d_{vf}$ in ROI-I, II, & IV are seen to peak near the mean value from either extremity of the histogram, however, the histogram of $d_{vf}$ in ROI-III is observed to skew to the right with larger void sizes.

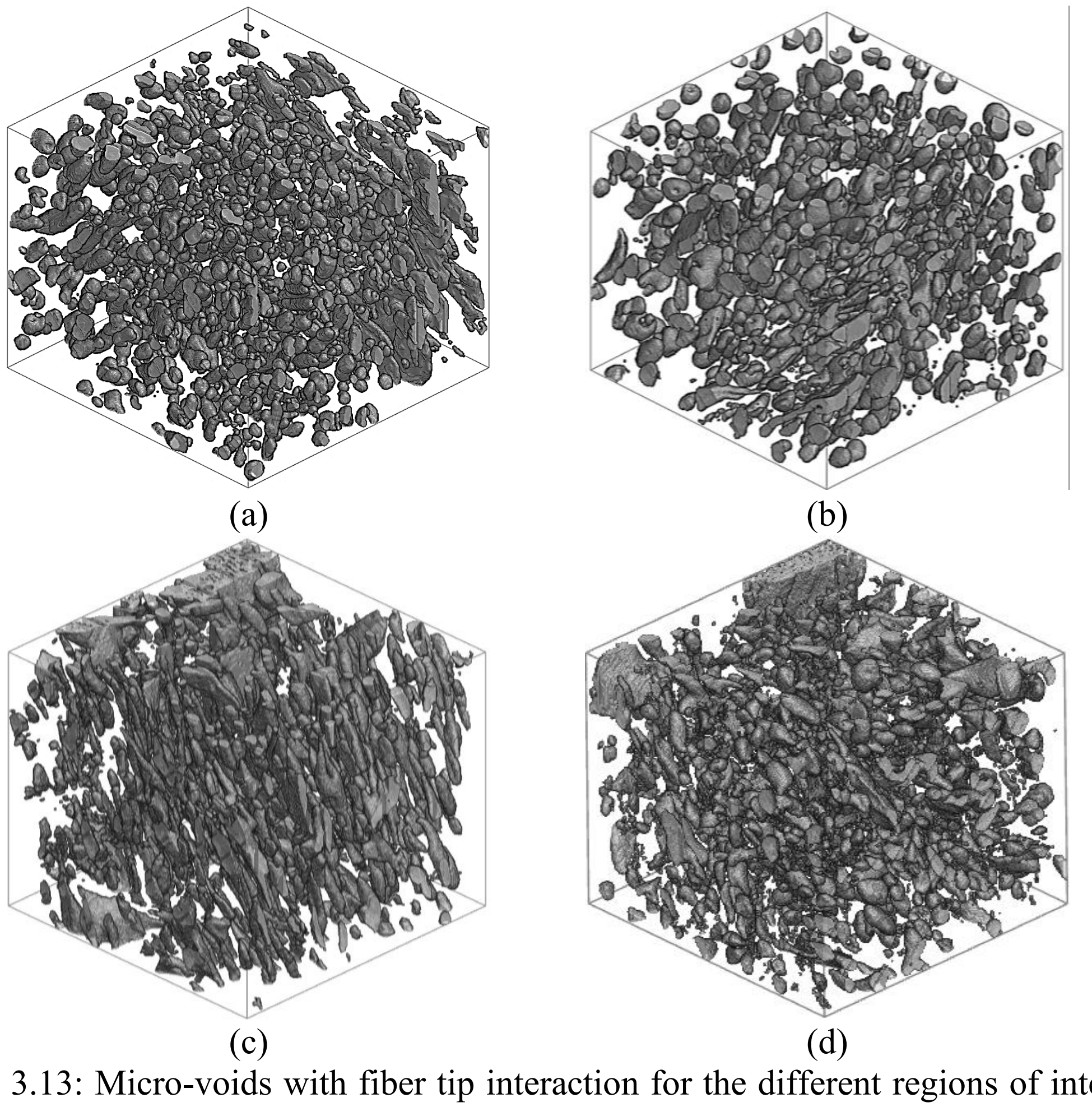

Figure 3.13: Micro-voids with fiber tip interaction for the different regions of interest of the CF/ABS bead (a) ROI-I (b) ROI-II (c) ROI-III (d) ROI-IV. (Cubic Envelope Size:0.35mm x 0.35mm x 0.35mm).

Table 3.3: Average diameter of the microstructural voids features with and without fiber interaction across all four (4) ROIs.

| Symbol | Definition | ROI-I | ROI-II | ROI-III | ROI-IV |
|---|---|---|---|---|---|
| $d_{vm}$ ($\mu m$) | Average diameter of voids isolated in matrix | 3.83 | 3.24 | 3.59 | 2.95 |
| $d_{vf}$ ($\mu m$) | Average diameter of voids touching fiber (s) | 30.96 | 29.35 | 39.26 | 34.85 |
| pdf | Probability Distribution Function Type | Gamma | Weibull | Lognormal | Weibull |
| $\alpha$ | Shape Parameter (Location for Lognormal) | 4.41 | 32.27 | 3.47 | 35.91 |
| $\beta$ | Scale Parameter | 6.22 | 2.71 | 0.52 | 2.07 |

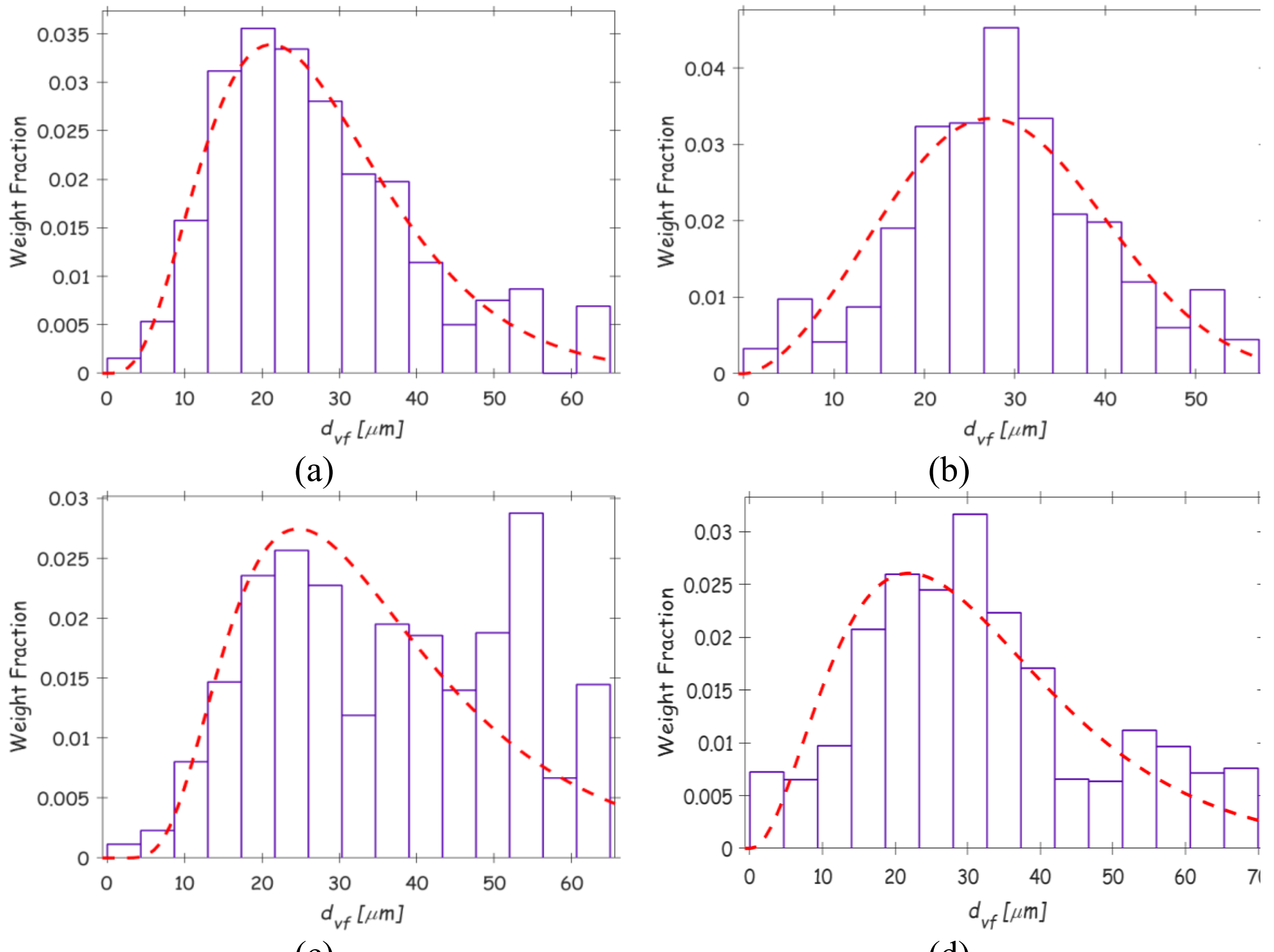

Figure 3.14: Distribution of the average void diameter across all regions of interest (ROI) of the 13% CF/ABS EDAM printed bead. (a) ROI–I (b) ROI-II (c) ROI-III (d) ROI-IV.

The isolated micro-voids have a higher overall mean sphericity value $\Phi_{vm} = 0.735$ with a standard deviation less than 0.01 compared to the overall mean sphericity value for the micro-voids in contact with fibers, $\Phi_{vf} = 0.6$ with a standard deviation less than 0.02.

The distribution of the sphericity for micro-voids in contact with fibers $\Phi_{vf}$ can be represented well with a Weibull probability distribution function in all four ROIs of the bead where all peaks are near the mean value as shown in Figure 3.15a-d. The parameters of the Weibull pdfs for the various ROIs appear in Table 3.4. Overall, the pdfs indicate that the bulk of the sphericity for micro-voids in contact with fibers are centered between about 0.62-0.67 across all ROIs.

Table 3.4: Average sphericity of the microstructural voids features with and without fiber interaction across all ROIs.

| Symbol | Definition | ROI-I | ROI-II | ROI-III | ROI-IV |
|---|---|---|---|---|---|
| $\Phi_{vm}$ | Average sphericity of voids isolated in matrix | 0.73 | 0.73 | 0.74 | 0.74 |
| $\Phi_{vf}$ | Average sphericity of voids touching fiber (s) | 0.58 | 0.60 | 0.59 | 0.62 |
| $\alpha$ | Weibull pdf Scale Parameter | 0.62 | 0.65 | 0.64 | 0.67 |
| $\beta$ | Weibull pdf Shape Parameter | 5.51 | 6.18 | 5.27 | 6.30 |

We see from the results of the ensemble average values of the fiber orientation principal components presented in Table 3.5 that there is higher degree of fiber alignment in the $z$ direction close to the bead edges (ROI-III) while the fibers near the bead's center (ROI-II) are more randomly oriented with a significant value of $a_{xx}$ which is in agreement with published data [40], [57]. Figure 3.16a-d shows the post-processed result of segmented fiber microstructural features for the various ROI volumes which provide visualization of the fiber orientation distribution within each ROI volume.

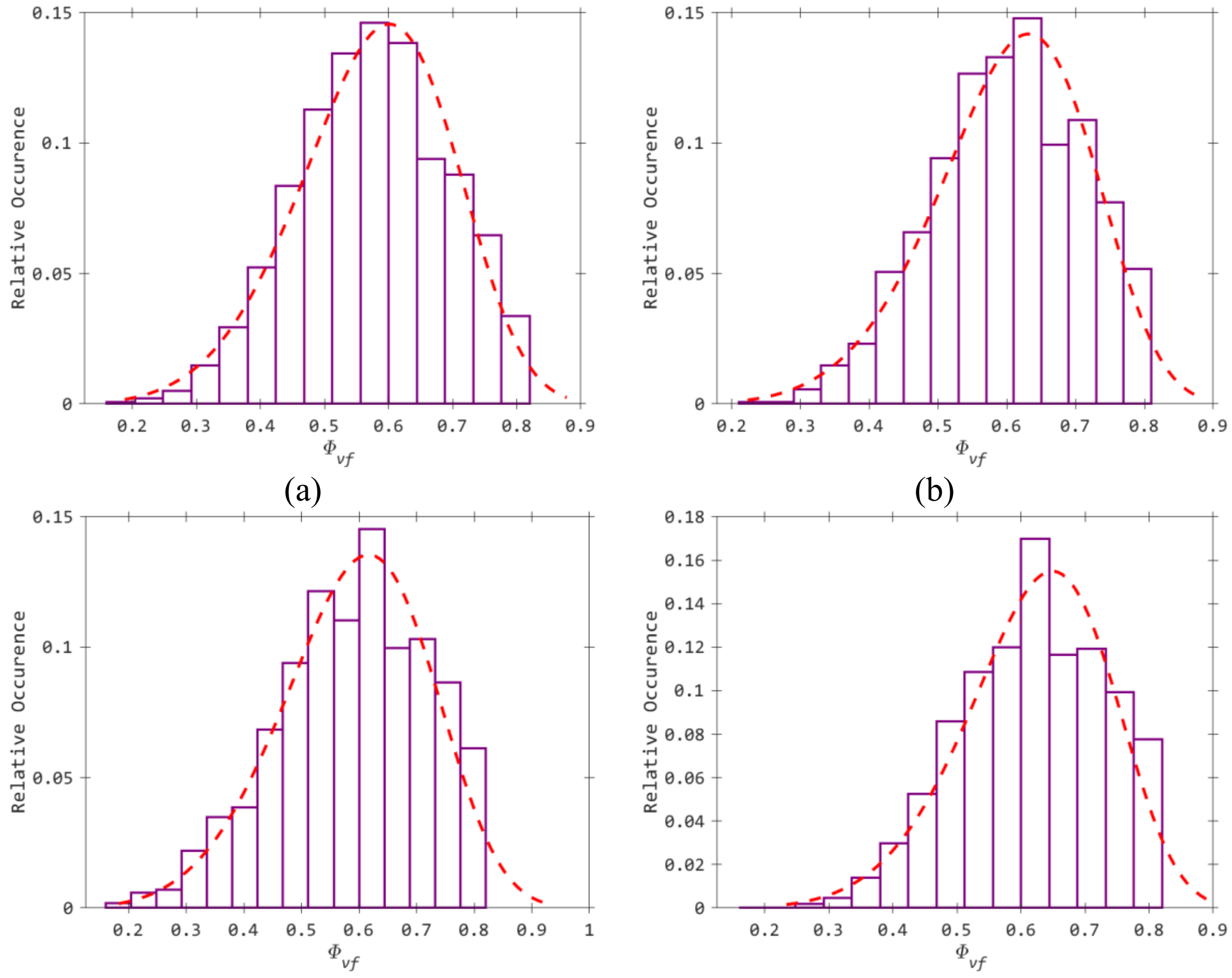

Figure 3.15: Distribution of the average void sphericity across all ROIs of the 13% CF/ABS EDAM printed bead. (a) ROI-I (b) ROI-II (c) ROI-III (d) ROI-IV.

Computed values of the $a_{zz}$ fiber orientation tensor component for all four (4) ROIs shows that ROI-I near the print bed (cf. Figure 3.17a) shows a more random orientation as evidenced by the $a_{zz}$ components as compared to other ROIs (cf. Figure 3.17c-d). This more random fiber arrangement can be seen in the magnified 3D sub-volume of the central microstructure of ROI-I (cf. Figure 3.18a) which also reveals a relatively low fiber volume fraction. In ROI-II close to the bead's center (cf. Figure 3.17b) and ROI-IV near the beads surface (cf. Figure 3.17d), the fibers are mostly either planarly or randomly oriented, although there is higher fiber alignment in the print direction for ROI-IV as compared to ROI-II. However, in ROI-III (cf. Figure 3.17b), the histogram of $a_{zz}$ is skewed to the right ($a_{zz} \rightarrow 1$) that indicates most of the fibers are highly aligned with the print direction. This high degree of fiber alignment is evident from the magnified 3D central sub-volume of the ROI-IV microstructure where the fibers are seen to be mostly oriented in the z-direction with a considerably high fiber volume fraction as compared to other ROIs (cf. Figure 3.18c).

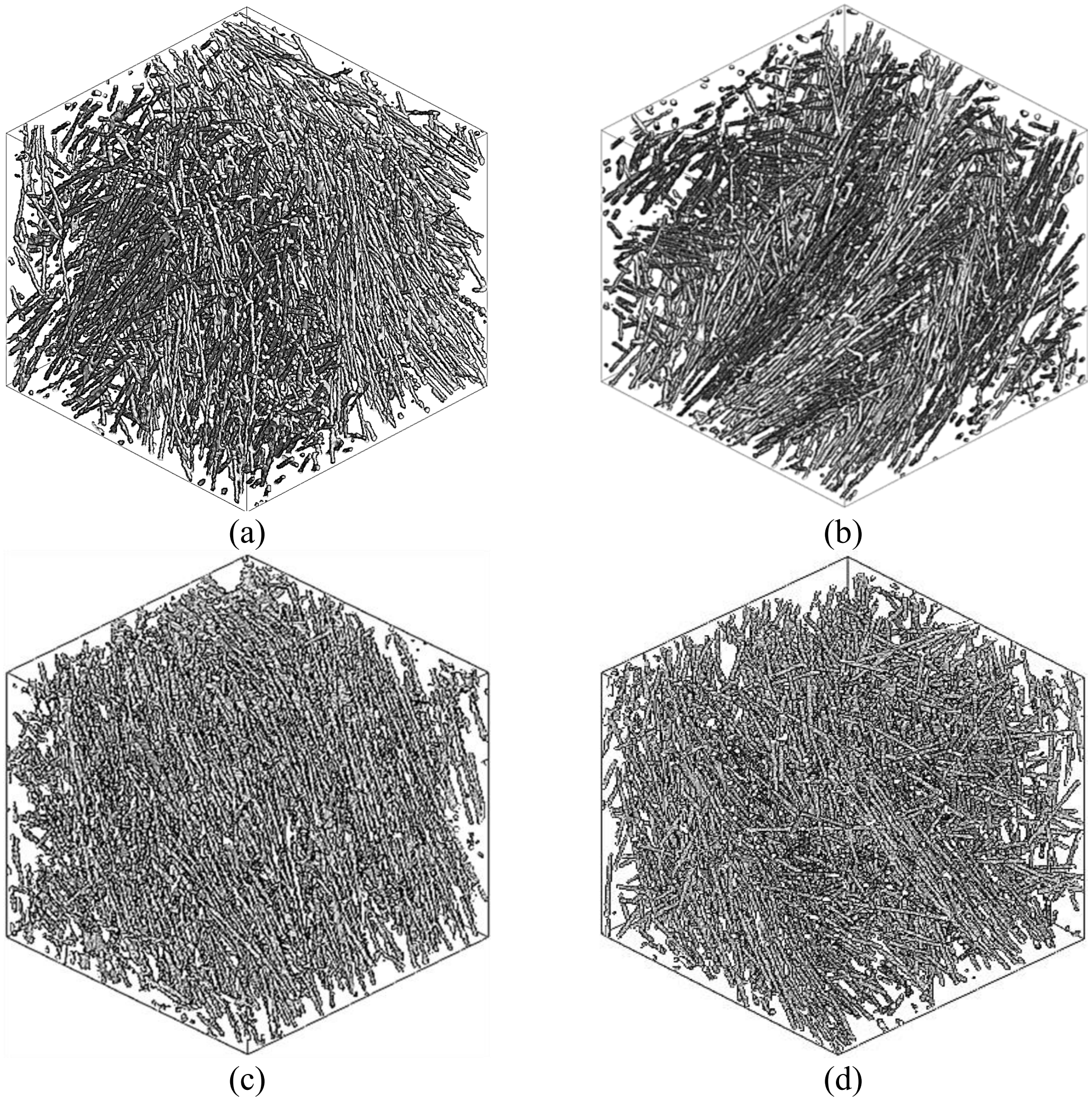

Figure 3.16: Fiber features within the microstructure of the different regions of interest of the CF/ABS bead (a) ROI-I (b) ROI-II (c) ROI-III (d) ROI-IV. (Cubic Envelope Size:0.35mm x 0.35mm x 0.35mm).

Table 3.5: Average values of the fiber orientation principal components in the various ROIs.

| | ROI-I | ROI-II | ROI-III | ROI-IV |
|---|---|---|---|---|
| $a_{xx}$ | 0.41 | 0.32 | 0.09 | 0.21 |
| $a_{yy}$ | 0.04 | 0.19 | 0.10 | 0.16 |
| $a_{zz}$ | 0.55 | 0.49 | 0.81 | 0.63 |

Consequently, the associated micro-void content with fiber tip contact is seen to be higher in ROI-III and IV which also has a relatively high degree of fiber alignment in the flow direction (z-direction); compared to the same in ROI-I and II which both have a higher degree of fiber alignment normal to the flow direction.

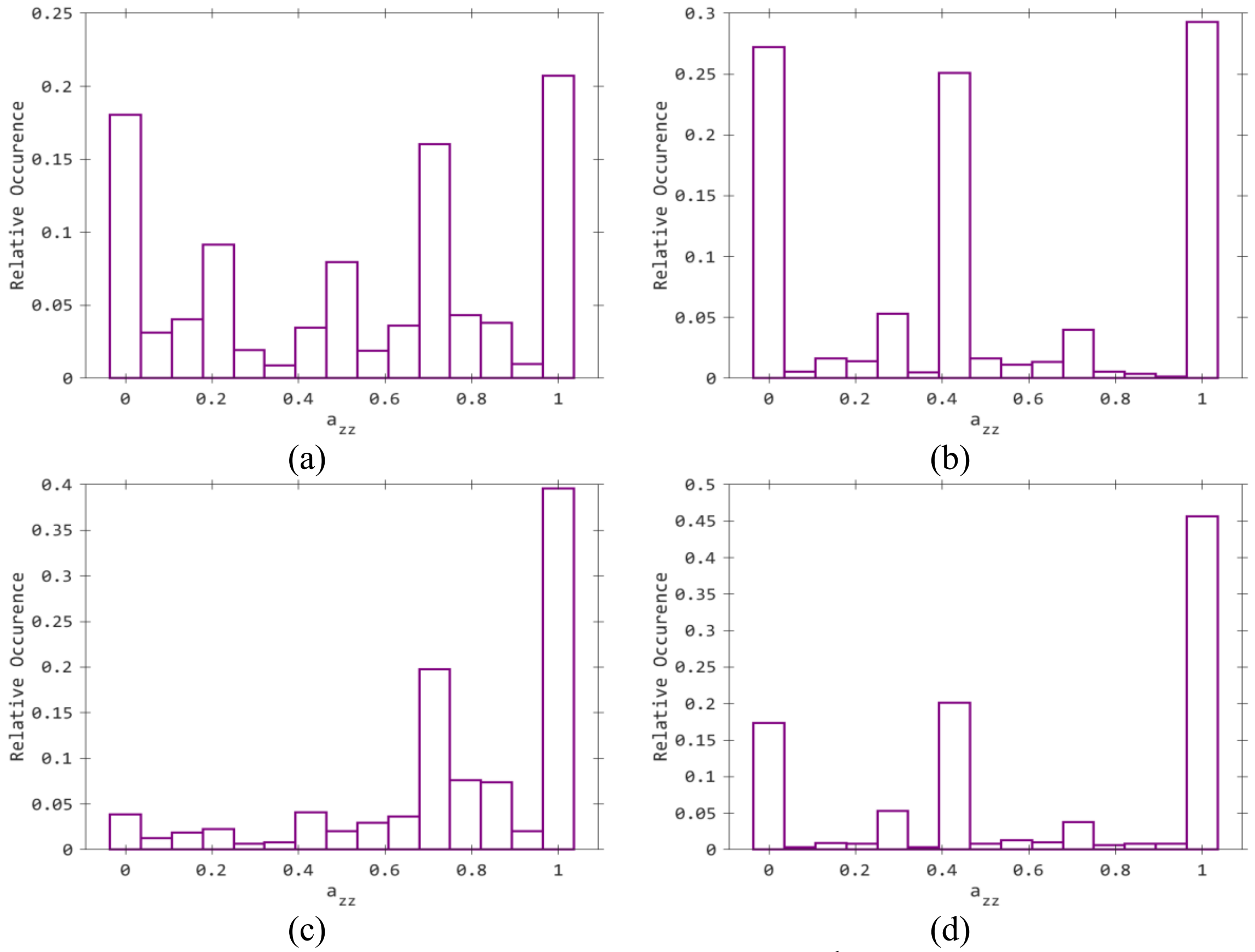

Figure 3.17: Distribution of the Azz component of the 2nd order fiber orientation tensor across all ROIs of the 13% CF/ABS EDAM printed bead. (a) ROI-I (b) ROI-II (c) ROI-III (d) ROI-IV.

These results suggest that a high degree of fiber alignment allows for a more compact arrangement of the fibers. Further, higher alignment appears to reduce the propensity for larger micro-void formation between fibers and hence the fiber tips provide more favorable sites for void formation [3]. We observe that fewer isolated voids (red) form between fibers in ROI-III (cf. Figure 3.18c) due to the highly compact fiber arrangement as compared to that in ROI-I (cf. Figure 3.18a) which has more isolated voids between fibers due to lower packing.

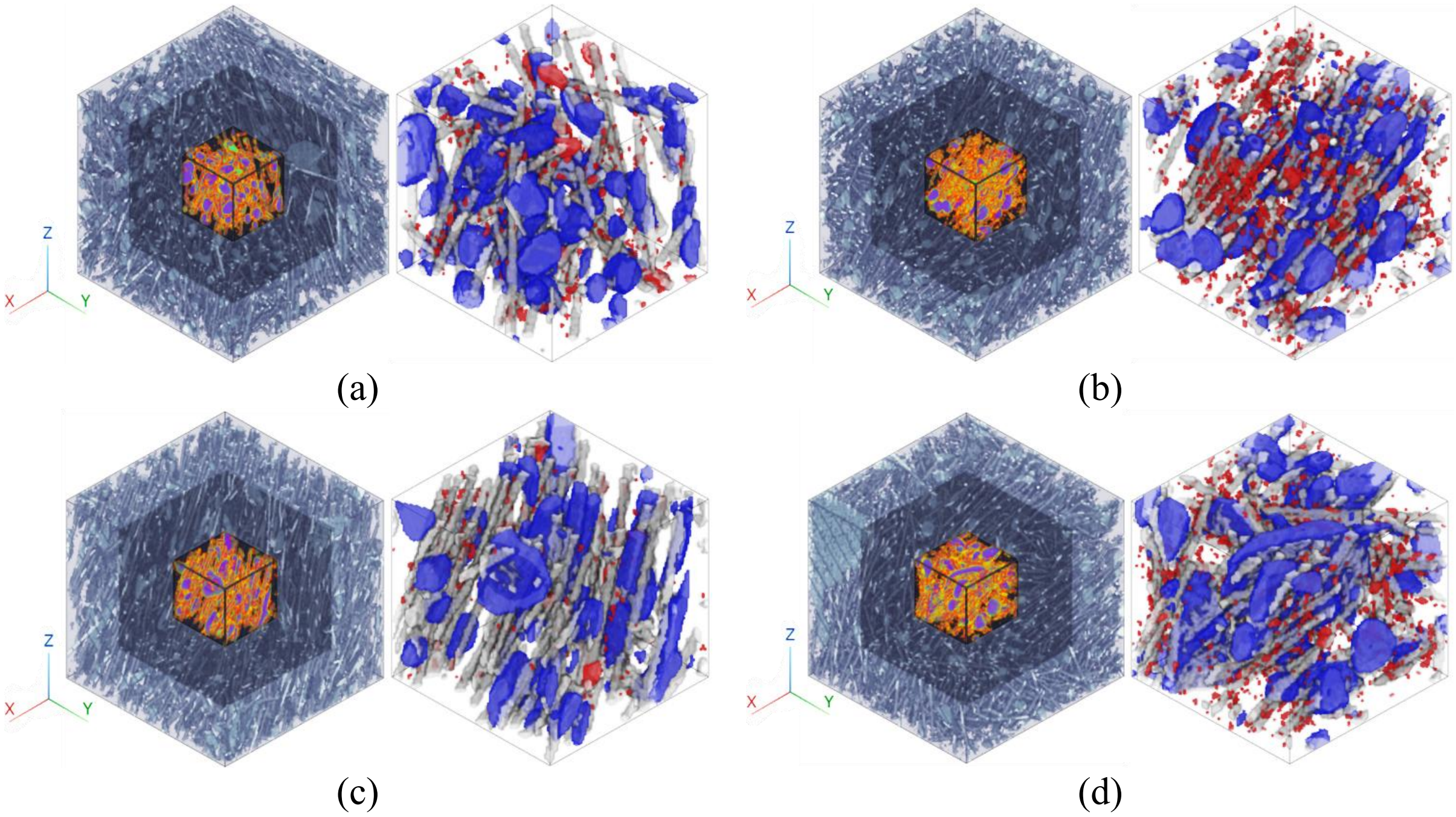

Figure 3.18: Volumetric plot showing extracted sub-volume view of the relevant microstructural features across all regions of interest (ROI) of the CF/ABS bead specimen. (a) ROI–I (b) ROI-II (c) ROI-III (d) ROI-IV.

### *3.7 Conclusion*

In conclusion, microstructural features including micro-void content, micro-void sphericity, and fiber orientation within a 13% CF/ABS bead specimen produced with EDAM has been evaluated using μ-CT scanning and image processing techniques. The results show an extremely high percentage of the micro-void contents form at the ends of the fibers, identified here as tip-voids. On average, the voids that nucleate at the fiber/matrix interface are relatively larger compared to those that are isolated within the ABS matrix (~9 times larger in diameter) and are also less spherical in shape. The homogenous micro-voids had an average equivalent diameter of 3.4μm and sphericity of 0.735 while the heterogenous micro-voids had an average equivalent diameter of 30μm and sphericity of 0.6. These observations are consistent with findings from literature [13], [15], [30]. Moreover, regions with a higher degree of fiber alignment with the flow direction have lower interstitial small-sized isolated voids possibly due to increased compactness. However, these regions have a higher micro-void content at the fiber tips (greater than 90% of the total void content) due to the increased number of fiber terminations which was also observation by Telkinalp et al. [3]. As we would see in Chapter Four, the inherent microstructural characteristics of the bead specimen affect the resulting thermo-mechanical properties and ultimately the part performance. Computational simulation studies that reveal mechanisms potentially responsible for the experimentally observed high volume content of micro-voids and the various factors that may influence their formation are presented in later chapters of this dissertation.

# CHAPTER FOUR

## Numerical Evaluation of the Effective Thermo-Mechanical Properties of Large Scale Additively Manufactured Short-Fiber Reinforced Polymer Composite

In this chapter, we evaluate the effective thermo-mechanical properties of 13% carbon fiber filled ABS (13% CF-ABS) SFRP composite manufactured via LAAM using the same regions of interest (ROI) presented in the previous chapter (Chapter Three). The goal of the SFRP composite assessment presented here is to understand the impact of micro-structural voids on the effective homogenized thermal and mechanical material properties. We employ a finite element based numerical homogenization approach using realistic representative volume elements (RVEs) developed from actual reconstructed 3D X-Ray μ-CT voxelated grayscale images of a 13% CF-ABS print bead specimen. Microstructural characterization of the printed bead specimen based on binary segmentation of the 3D grayscale voxelated data is performed to identify unique phases and microstructural features within the sample. Finite element models are defined based on the derived realistic RVEs to compute the effective thermo-mechanical properties at selected regions within a LAAM bead. To ensure domain continuity across the RVE boundaries, periodic bourndary conditions are prescribed on opposing boundary entities which ensures effective transfer of stress or heat flux across boundary surfaces. The effective elastic stiffness is derived from the homogenized macro-stresses and macro-strains under various prescribed load cases through a least-square linear regression fitting algorithm. Using the same finite element mesh, the homogenized thermal expansion coefficient (CTE) is computed from homogenized heat flux and temperature gradient obtained from steady-state heat transfer FE analysis based on the Fourier's law (in a manner similar to that in Wang [237]). Effective properties (i.e., elastic constants, CTE and thermal conductivity) computed using our numerical homogenization scheme are compared to results derived from analytical mean-field homogenization approach based on the Mori-Tanaka-Benveniste's formulation. The effects of the porosity on the effective properties are also quantified in the current assessment. Finally, a discrete minimization approach is developed to obtain a characteristic RVE instance from a given ROI volume with matching microstructural characteristics, and the effective thermo-mechanical properties across different regions of the LAAM printed bead specimen are computed and compared.

## *4.1 Computational Homogenization Methods*

In the current study, the effective properties for four (4) regions of interest (ROI) across the 13% CF/ABS bead specimen shown in Figure 3.8 are evaluated which includes: (a) ROI-I adjacent to the print bed, (b) ROI-II near the bead's center, (c) ROI-III along the outer edge of the bead and (d) ROI-IV adjacent to the top surface of the bead. The dimensions of each ROI are $0.35mm \times 0.35mm \times 0.35mm$. The CF/ABS bead was printed using a Strangpresse Model 19 single-screw extruder (Strangpresse, Youngstown, OH, USA) on a custom LAAM system at Baylor University  More details on the LAAM printing parameters and operating conditions can be found in [241].  The isotropic properties of the microconstituents of the 13% CF-ABS SFRP composite used in the homogenization analysis appearing in Table 4.1 are taken from [144]. The constituent properties shown are for a Tourayaca® T300 (Touray Industries, Tokyo, Japan) PAN based carbon fiber, and Lustran ® 433 ABS (INEOS Olefins & Polymers, London, UK) for the ABS polymer matrix. In Table 4.1, $E$ is the elastic modulus, $v$ is the poisson ratio, $\kappa$ is the thermal conductivity, $\rho$ is the density and $\zeta_s$ is the specific heat capacity.

Table 4.1: Average isotropic properties of the microconstituents of the 13% CF-ABS SFRP material

| | $E$ [GPa] | $v$ | $\alpha \times 10^{-6}$ [m/m − K] | $\kappa$ [W/m − K] | $\rho$ [g/cc] | $\zeta_s$ [J/kg · $K$] |
|---|---|---|---|---|---|---|
| Fiber | 230.0 | 0.20 | -0.61 | 3.060 | 1.76 | 777. |
| Matrix | 2.55 | 0.35 | 90.1 | 0.175 | 1.05 | 1865 |

## *4.2 Numerical FE Homogenization Method*

FEA models of different RVE's were generated from reconstructed 3D X-ray $\mu$-CT voxel-based radiographs of each of the ROIs in Figure 3.8 and Figure 4.1a. Binary segmentation of each ROI volume into the three (3) microstructural constituents (matrix, fiber and voids) was achieved via grayscale data thresholding with detailed procedures provided in [241]. Sufficient image resolution to accurately capture the microstructural features was performed using a voxel cube with side length of $1.4\mu m$ yielding a total of 250 voxels per length or 15,625,000 voxels per ROI. The FEA models for the elasticity analyses were generated directly using the scripting interface of Abaqus/Standard (Abaqus 2023, Simulia, Dassault Systemes, Waltham, Massachusetts) where each voxel within the RVE is represented as a 3D solid 8-node fully integrated iso-parametric continuum brick element (C3D8). A separate element set is created for each segmented microstructural constituent. For the heat transfer analysis, diffusive-C3D8 elements (i.e. DC3D8) are employed instead based on the same mesh as that for the elasticity analysis. Relevant material property definitions for the individual microstructural phases are also created and assigned to their respective material sections through the section assignment input syntax. Figure 4.1 (a-b) shows a sample RVE block extracted from the ROI-II (cf. Figure 3.8) which highlights the three microstructural phases including fibers (gray), micro-voids (red), and the ABS matrix (transparent volume). Figure 4.1 (c) shows the resulting FEA model created from directly importing the segmented voxelated data which illustrates the three microstructural phases appearing on the outer RVE surfaces. Figure 4.1 (d–f) shows the individual element sets of the each microstructural phases including the ABS matrix (cf. Figure 4.1 (d)), the inherent micro-voids (cf. Figure 4.1 (e)) and the fiber reinforcements (Figure 4.1 (f)). Although, the stepped-like surface of micro-features within the RVE generated with voxel data are likely to induce stress concentrations, Guven et al [118] has shown that the impact on evaluated effective properties for small displacement analysis are only minimal.

This study compares results for three RVE sizes using ROI-II as a case study. The smallest sized RVE (RVE-I) has a cube side lengths of 70μm with 125,000 elements and 125 RVE realizations (cf. Figure 4.2 (a)) within ROI-II, while the mid-sized RVE (RVE-II) has a side-lengths of 116.2μm with 571,787 elements and 27 RVE realizations (cf. Figure 4.2 (b)) within ROI-II. The largest RVE (RVE-III) has side-lengths of 175μm, with a total of 1,953,125 elements and 8 RVE realizations within ROI-II (cf. Figure 4.2 (c)). With the foregoing RVE classification, we define a numbering scheme for identifying a particular realization - Z, belonging to a RVE type - Y within a ROI volume X is defined as ROI-X, RVE-Y, #Z. Complete adhesion between the filler and matrix constituent is assumed. In all cases, the element side length equals the length of the voxel cube of $1.4\mu m$ which is one-fifth the average fiber diameter of $7.0\mu m$.

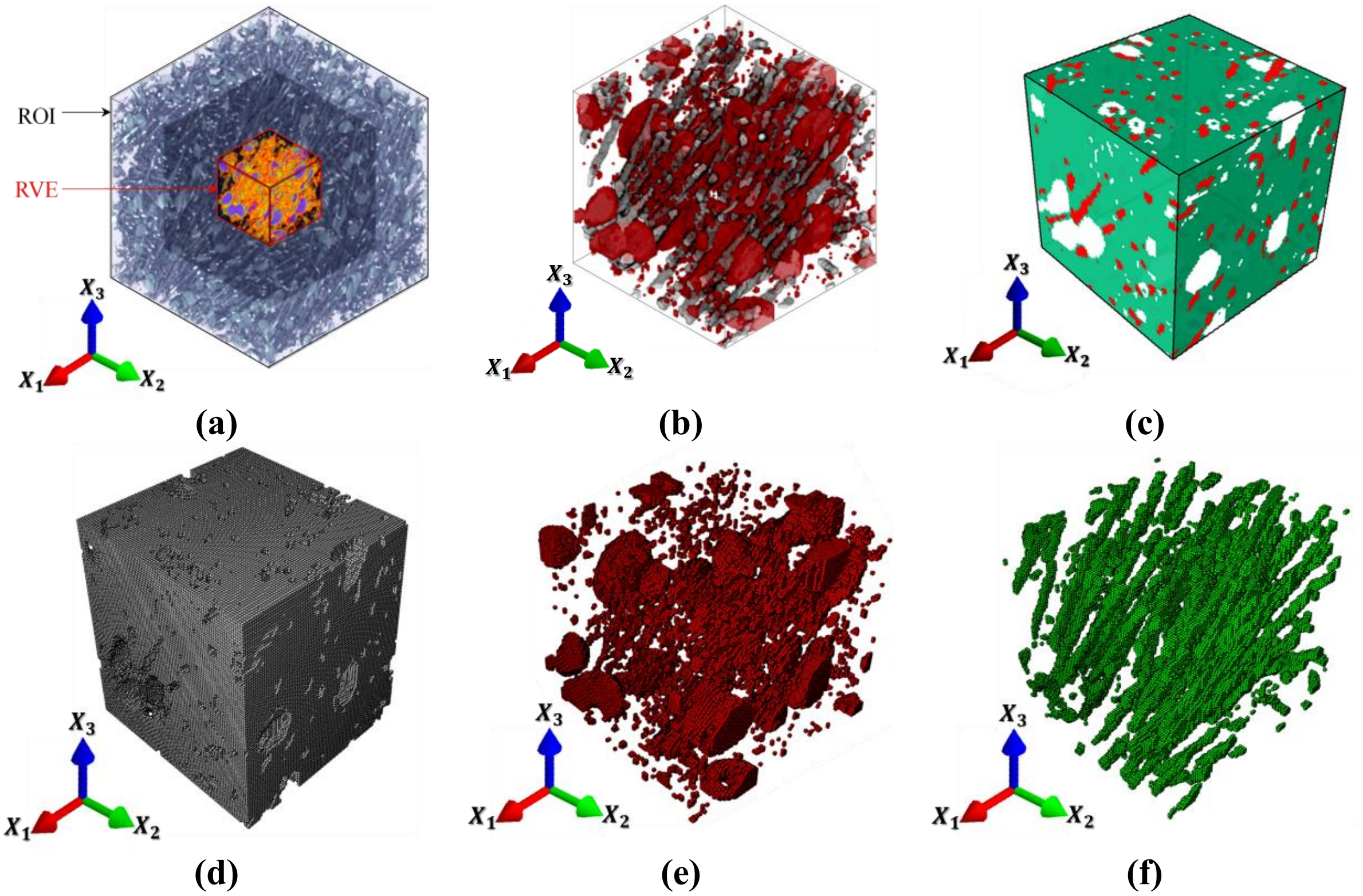

Figure 4.1: (a) Reconstructed $\mu CT$ 3D-scans of ROI-II highlighting mid-block RVE volume (b) magnified view of the mid-block RVE showing segmented micro-constituents, (c) FEA model imported from $\mu CT$ voxel data of the RVE, (d) ABS matrix FEA elements (e) micro-voids FEA elements and (f) fiber reinforcement FEA elements.

Periodic boundary conditions (PBC) on the cubic RVE outer surfaces are defined according to formulations in [242], [243], [244] for the stress and thermal RVE simulations as

$$u_i^{F_k^+} - u_i^{F_k^-} = \sum_j c_F \hat{\varepsilon}_{ij} \Delta x_j^k \qquad \theta^{F_k^+} - \theta^{F_k^-} = \sum_j c_F \widehat{\nabla}_j \theta \Delta x_j^k \qquad c_F = \delta_{jk} \tag{4.1}$$

$$u_i^{\Sigma_{kn}^+} - u_i^{\Sigma_{kn}^-} = \sum_j c_\Sigma \hat{\varepsilon}_{ij} \Delta x_j^k \qquad \theta^{\Sigma_{kn}^+} - \theta^{\Sigma_{kn}^-} = \sum_j c_\Sigma \widehat{\nabla}_j \theta \Delta x_j^k \qquad c_\Sigma = -\left(-e_{jkm}\right)^n \tag{4.2}$$

$$u_i^{V_k^+} - u_i^{V_k^-} = \sum_j c_V \hat{\varepsilon}_{ij} \Delta x_j^k \qquad \theta^{V_k^+} - \theta^{V_k^-} = \sum_j c_V \widehat{\nabla}_j \theta \Delta x_j^k \qquad c_V = (-1)^{\delta_{jk}} \tag{4.3}$$

where $u_i^{N_k}$ is the $i^{th}$ component of displacement and $\theta^{N_k}$ is the temperature, both applied on the $k^{th}$directional identifier on opposing entities $N$ of the periodic structure including faces ($N = F^{\pm}$), $n^{th}$ edges ($N = \Sigma_n^{\pm}$) and vertices ($N = V^{\pm}$). $\delta_{ij}$ is the Kronecker delta and $e_{ijk}$ is the Levi-Civita permutation tensor. The quantity $\hat{\varepsilon}_{ij}$ is the average macro-strain tensor over the periodic RVE. $\Delta x_j^k$ is the projection of the $j^{th}$ dimension of the RVE along the $k^{th}$direction and for our orthogonal shaped RVE, $\Delta x_j^k = \Delta L_j$ (cf. Figure 4). Indices $i, j, k \in \{1,2,3\}$ identify the cartesian degrees of freedoms and $m = k\delta_{jk} + (1 - \delta_{jk})(6 - j - k)$ in Eqn. (4.2). The PBC multi-point constraints (MPC) on opposing entities are defined in the model via the Abaqus equation input syntax.

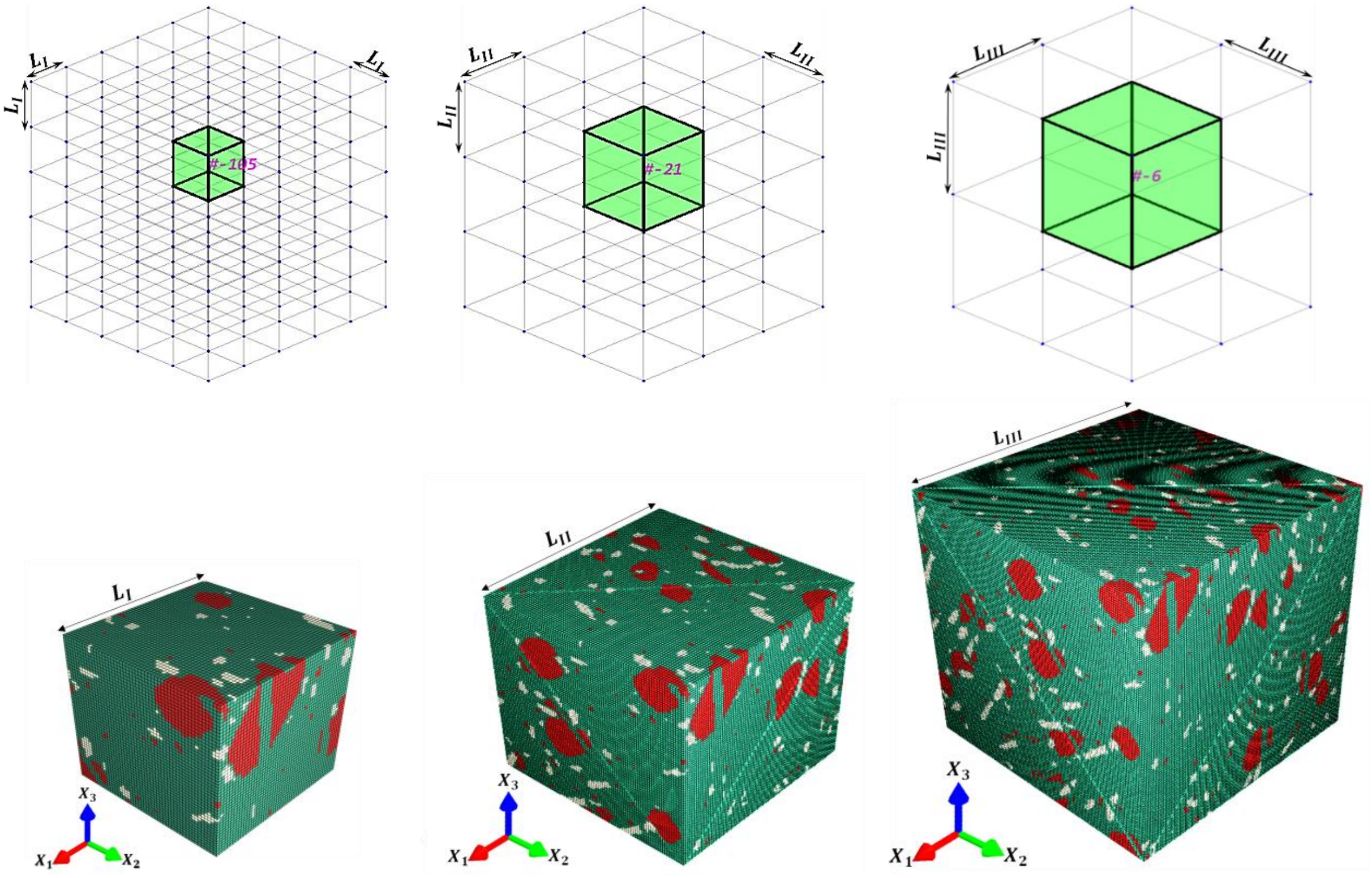

**(a)** **(b)** **(c)**

Figure 4.2: ROI-II partitioning into (a) RVE - I: 125 realizations with 125,000 elements per realization with $L_I = 70\ \mu m$ (b) RVE - II: 27 realizations with 571,787 elements per realization with $L_{II} = 116.2\ \mu m$ (c) RVE - III: 8 realizations with 1,953,125 elements per realization with $L_{III} = 175\ \mu$.

To avoid redundancy, edges and vertices are excluded from the facet node set definitions, and vertices are excluded from the edge node set definitions. The prescribed macro-strains $\hat{\varepsilon}_{ij}$ or temperature gradient $\widehat{\nabla}_j\theta$ are imposed through an extra set of dummy nodes that are coupled to degree of freedom MPC nodes on opposing PBC entities ($N^{\pm}$) with a respective displacement or temperature magnitude equal to the RHS value of the constraint equation.

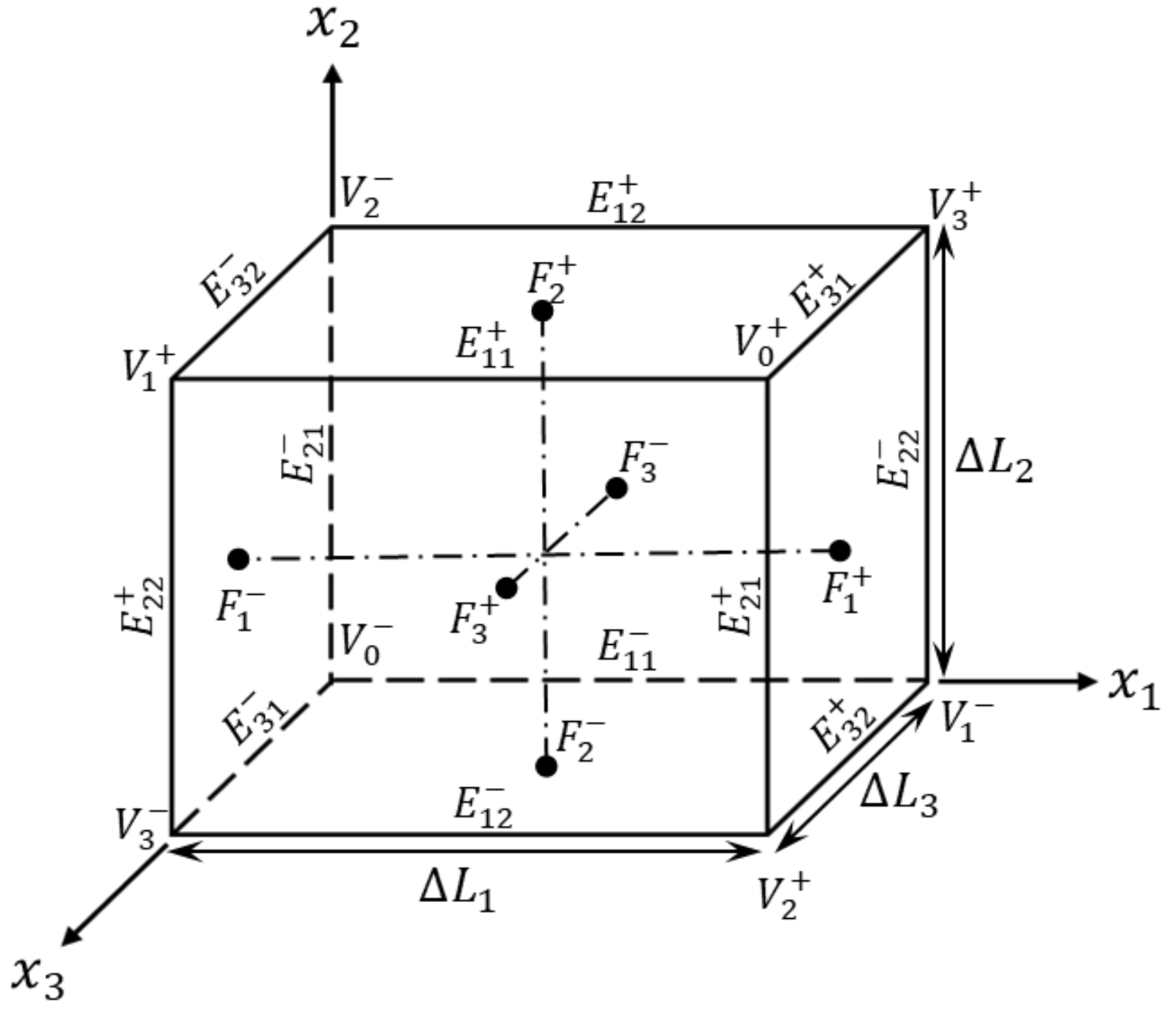

Figure 4.3: Single periodic RVE structure showing definitions of entities and coordinate directions used in the PBC formulations (Eqn. (4.1)-(4.3))

### *4.2.1 Evaluating the Effective Engineering Constants.*

For the stress analysis, six load cases are employed with permutation indices $kl$ $(11, 22, 33, 23, 13, 12)$ which are applied through the displacement PBC equations (cf. Eqns. 1-3). For load case $kl$, the applied strain $\hat{\varepsilon}_{ij}^{kl}$ may be written as

$$\hat{\varepsilon}_{ij}^{kl} = \frac{\epsilon}{2}\left(1 + \delta_{ij}\right)\delta_{ik}\delta_{lj} \tag{4.4}$$

where $\epsilon$ is the magnitude of the imposed strain, and repeated indices do not imply summation. We assume $\epsilon = 0.25$ in all elasticity analyses. The homogenized equivalent macro-stresses $\hat{\sigma}_{ij}$ and macro-strains $\hat{\varepsilon}_{ij}$ of the heterogenous RVE volume $\Omega$ is evaluated through volume averaging, satisfying the Hill-Mandel condition of equivalent strain energy [245], [246] between the idealized homogenized volume with macro-stress $\hat{\sigma}_{ij}$ and macro-strain $\hat{\varepsilon}_{ij}$ and heterogenous volume with stress $\sigma_{ij}(x)$ and strain $\varepsilon_{ij}(x)$ (both a function of material point $\boldsymbol{x}$ within the domain $\Omega$ of the RVE) given as

$$\frac{1}{2}\int_{\Omega} \sigma_{ij}(\boldsymbol{x})\varepsilon_{ij}(\boldsymbol{x})d\Omega = \frac{1}{2}\hat{\sigma}_{ij}\hat{\varepsilon}_{ij}\Omega \tag{4.5}$$

The equivalent macro-stresses $\hat{\sigma}_{ij}$ and macro-strains $\hat{\varepsilon}_{ij}$ derived from Eqn. (4.5) may be written as

$$\hat{\sigma}_{ij} = \frac{1}{\Omega}\int_{\Omega} \sigma_{ij}(\boldsymbol{x})d\Omega\,, \qquad \hat{\varepsilon}_{ij} = \frac{1}{\Omega}\int_{\Omega} \varepsilon_{ij}(\boldsymbol{x})d\Omega\,, \qquad i,j = 1,2,3\;;\; x \in \Omega \tag{4.6}$$

The effective elastic constant tensor $\hat{C}_{ijkl}$ is obtained from the homogenized quantities according to the constitutive relation

$$\hat{\sigma}_{ij} = \hat{C}_{ijkl}\hat{\varepsilon}_{kl}, \qquad i,j,k,l = 1,2,3 \tag{4.7}$$

which is conveniently written in contracted notation as

$$\hat{\sigma}_{m} = \hat{C}_{mn}\hat{\varepsilon}_{n}, \qquad m = f_1(i,j), \qquad n = f_1(k,l) \tag{4.8}$$

where$f_1(i,j)$ is a linear indices transformation functions given as

$$f_1(i,j) = r\delta_{ij} + \left(1 - \delta_{ij}\right)(9 - i - j) \tag{4.9}$$

The material constant $\hat{C}_{mn}$ is computed from a least-square linear regression fitting algorithm that minimizes the relative error in the components $\hat{\sigma}_m$ of the stress tensor of Eqn. (4.8). Let the $i^{th}$ component of the stress and strain tensors for $j^{th}$ load case be denoted as $\widehat{\Psi}_{ij} = \hat{\sigma}_i^j$ and $\widehat{\Xi}_{ij} = \hat{\varepsilon}_i^j$, respectively, We define $\widehat{D}_{ij}$ such that $\widehat{D}_{ij} = \widehat{\Xi}_{ik}\widehat{\Xi}_{jk}$ and the block diagonal matrices $\widehat{P}_{ijkl}$ and $\widehat{Q}_{ijkl}$, respectively as

$$\widehat{P}_{ijkl} = \widehat{D}_{ij}\delta_{kl}, \qquad and, \qquad \widehat{Q}_{ijkl} = \widehat{\Xi}_{ij}\delta_{kl} \tag{4.10}$$

The material constant $\hat{C}_{mn}$ is then computed from the linear algebraic expression given as

$$\widehat{P}_{ijkl}\hat{C}_{kl} = \hat{b}_{ij}, \qquad \hat{b}_{ij} = \widehat{Q}_{iljk}\widehat{\Psi}_{kl} \tag{4.11}$$

For simplicity the expression of Eqn. (4.11) can represented in the reduced order form as

$$\widehat{P}_{mn}\hat{C}_n = \hat{b}_m, \qquad m = f_2(i,j), \qquad n = f_2(k,l) \tag{4.12}$$

The linear indices transformation functions $f_2(i,j)$ that appear in Eqn. (4.12) is given as

$$f_2(i,j) = i + 6(j-1) \tag{4.13}$$

The reduced order tensor $\widehat{P}_{mn}$ is a 36 x 36 matrix and the tensors $\hat{C}_n$ and $\hat{b}_m$ are 36 x 1 vectors. Depending on the requirements of material property behavior desired from the least square fitting of the elastic constant $\hat{C}_n$, such as matrix and material symmetry, orthogonality, isotropy etc., constraints that define linear relationships between components $\hat{C}_n$ are imposed with the constraint matrix $X_{vn}$ through the linear relation

$$X_{vn}\hat{C}_n = 0, \qquad n = 1 \dots 36 \tag{4.14}$$

In the current study, only two material requirements are needed for a complete definition of the constraint matrix $X_{vn}$ which includes the condition of matrix symmetry and material orthogonality defined through sets of linear equation constraint submatrices $X'_{rn}$ and $X''_{sn}$, respectively such that

$$X_{vn} = \begin{bmatrix} X'_{rn} \\ X''_{sn} \end{bmatrix} \tag{4.15}$$

The necessary condition of matrix symmetry $\hat{C}_{ij} = \hat{C}_{ji}$ requires 15 essential constraints, and thus $X'_{rn}$ is a 15 x 36 sized submatrix such that:

$$X'_{rn} = \delta_{pn} - \delta_{qn}$$
$$r = f_3(i,j)\,, \qquad p = f_2(i,j), \qquad q = f_2(j,i), \qquad i = 1 \cdots 5, \qquad j = i + 1 \cdots 6 \tag{4.16}$$

where the linear indices transformation functions $f_3(i,j)$ is given as

$$f_3(i,j) = j - .5(i^2 - 11i + 12) \tag{4.17}$$

Further conditions for material orthotropy having three orthogonal planes of symmetry require the definition of 12 additional constraints that set the shear-extension (9) and biplanar shear-shear (3) coupling terms of the elastic constants to zero, i.e.,

$$\hat{C}_{ij} = 0$$
$$i = 1 \ldots 5, \qquad j = \begin{cases} 4 \ldots 6 & i \leq 3 \\ i + 1 \ldots 6 & i > 3 \end{cases} \tag{4.18}$$

The constraint submatrix $X''_{sn}$ is thus a 12 x 36 matrix given as

$$X''_{sn} = \delta_{np}, \qquad s = f_4(i,j), \qquad p = f_2(i,j) \tag{4.19}$$

The linear indices transformation functions $f_4(i,j)$ that appear in Eqn. (4.19) is given as

$$f_4(i,j) = \begin{cases} 3i + j - 6 & i \leq 3 \\ i + j + 1 & i > 3 \end{cases} \tag{4.20}$$

A Lagrange multiplier method is used to combine constitutive relation of Eqn. (4.12) and the constraint equations in Eqn. (4.14) to obtain the linear algebraic systems of equation given as [247]

$$\begin{bmatrix} \hat{P}_{mn} & X_{nv} \\ X_{vn} & 0_{vv} \end{bmatrix} \begin{bmatrix} \hat{C}_n \\ \lambda_v \end{bmatrix} = \begin{bmatrix} \hat{b}_m \\ 0_v \end{bmatrix} \tag{4.21}$$

where $\lambda_v$ is the vector of the Lagrange multipliers with one component for each imposed constraint. The accuracy of the regression fit is assessed by the coefficient of determination for each $j^{th}$ load case $R_j^2$ given as

$$R_j^2 = 1 - St^j/Sr^j \tag{4.22}$$

where,

$$St^j = \sum_{\forall i}\left[\hat{\sigma}_i^j - \hat{C}_{ik}\hat{\varepsilon}_k^j\right]^2, \qquad Sr^j = \sum_{\forall i}\left[\hat{\sigma}_i^j - \langle\sigma\rangle^j\right]^2, \qquad \langle\sigma\rangle^j = \frac{1}{n}\sum_{\forall i}\hat{\sigma}_i^j \tag{4.23}$$

The computed solution of $\hat{C}_n$ transformed to the 6 x 6 matrix form $\hat{C}_{mn}$ and inverted yields the effective compliance matrix $\hat{S}_{mn}$from which we can derive the 9 independent engineering constants according to

$$\hat{S}_{mn} = \left[\hat{C}^{-1}\right]_{mn} = \begin{bmatrix} \frac{1}{E_{11}} & -\frac{\nu_{21}}{E_{22}} & -\frac{\nu_{31}}{E_{33}} & & & \\ -\frac{\nu_{12}}{E_{11}} & \frac{1}{E_{22}} & -\frac{\nu_{32}}{E_{33}} & & & \\ -\frac{\nu_{13}}{E_{11}} & -\frac{\nu_{23}}{E_{22}} & \frac{1}{E_{33}} & & & \\ & & & \frac{1}{G_{23}} & & \\ & & & & \frac{1}{G_{13}} & \\ & & & & & \frac{1}{G_{12}} \end{bmatrix} \tag{4.24}$$

### *4.2.2 Evaluating the Effective Coefficient of Thermal Expansion.*

The evaluation of the effective CTE in our approach is based on the Duhamel-Neumann law [248], [249] where the fundamental constitutive expression that relates the effective stress $\hat{\sigma}_{ij}$ to the effective total strain $\hat{\varepsilon}_{kl}$ for a structure subjected to a temperature change $\Delta\theta$ is given as

$$\hat{\sigma}_{ij} = \hat{C}_{klij}[\hat{\varepsilon}_{kl} - \hat{\alpha}_{kl}\Delta\theta] \tag{4.25}$$

where $\hat{C}_{klij}$ is the effective elastic stiffness tensor of the homogenized material and $\hat{\alpha}_{kl}$ is the effective CTE tensor. In contracted notation, Eqn. 22 becomes

$$\hat{\sigma}_m = \hat{C}_{nm}[\hat{\varepsilon}_n - \hat{\alpha}_n\Delta\theta] \tag{4.26}$$

In our analysis, we assume $\Delta\theta = 65^oC$. In evaluating the effective CTE, the total strain $\varepsilon_{ij}$ in the sets of defined PBC constraints in Eqn. (4.1)-(4.3) above is set to zero, giving $\hat{\varepsilon}_{kl} = 0$. Eqn. (4.26) reduces to

$$\hat{\sigma}_m = -\hat{C}_{nm}\hat{\alpha}_n\Delta\theta \tag{4.27}$$

From the results of the homogenized equivalent macro-stresses $\hat{\sigma}_m$ derived from the thermal expansion analysis based on an applied temperature difference $\Delta\theta$ over the entire RVE volume, the effective CTE can be computed by rearranging Eqn. (4.27) above as

$$\hat{\alpha}_n = -\hat{S}_{nm}\hat{\sigma}_m/\Delta\theta \tag{4.28}$$

### *4.2.3 Evaluating the Effective Thermal Conductivity.*

A heat transfer analysis is performed on the RVE to compute the effective thermal conductivity using three thermal load cases with permutation indices $kk$ $(11, 22, 33)$ applied through the temperature PBC constraints in Eqns. 1-3. These thermal loads are defined to produce orthogonal temperature gradients such that for case $k$ we define

$$\hat{\nabla}_j^k\theta = \delta_{jk}\Theta \tag{4.29}$$

where $\Theta$ is the magnitude of the imposed temperature difference across the RVE in the $x_j$ direction. In our analyses, we impose $\Theta = 100^oC$. The general heat conduction energy conservation equation at an arbitrary material point within the RVE volume is given as

$$\nabla_i\kappa_{ij}\nabla_j\theta + \dot{q}_v = \rho c_p\dot{\theta} \tag{4.30}$$

where $\nabla_i$ is the gradient operator vector, $\kappa_{ij}$ is the thermal conductivity of the material, $\theta$ is the temperature, $\dot{q}_v$ is the rate of internal heat generation within the material, and $\rho$ and $c_p$ are the density and specific heat capacity respectively, with all quantities being evaluated at specified material point within the RVE volume. In the FEA analysis for evaluating the TC tensor $\hat{\kappa}_{ij}$, we assumed steady state ($\dot{\theta} = 0$) and there is no internal heat generation ($\dot{q}_v = 0$). The resulting temperature field distribution is the solution to

$$\nabla_i \kappa_{ij} \nabla_j \theta \;= 0 \tag{4.31}$$

From Fourier's law of steady state heat conduction, the heat flux $q_i$ at any material point within the conducting medium is

$$q_i = -\kappa_{ij} \nabla_j \theta \tag{4.32}$$

Integrating Eqn. (4.31) over the domain $\Omega$, applying the Gauss divergence theorem, and making appropriate substitution for the heat flux $q_i$ from Eqn. (4.32) yields

$$\int_{\Omega} \nabla_i \kappa_{ij} \nabla_j \theta \; d\Omega = \int_{\Lambda} \kappa_{ij} \nabla_j \theta \, \vec{\mathrm{n}}_i \; d\Lambda = -\int_{\Lambda} q_i \, \vec{\mathrm{n}}_i \; d\Lambda = 0 \tag{4.33}$$

The Hill-Mandel condition of equivalent thermal dissipation between the assumed homogenous and heterogenous regions yields [116]

$$-\hat{q}_k \widehat{\nabla}_k \theta = -\{\widehat{q_k \nabla_k}\}\theta = -\frac{1}{\Omega} \int_{\Omega} q_k(x) \nabla_k \theta(x) d\Omega \tag{4.34}$$

Applying the same macrohomogeneity principle as was done with the elasticity solution in Section 1.1, we obtain spatial averages of the local heat flux $\hat{q}_i$ and temperature gradient $\widehat{\nabla}_i \theta$ given as [112], [129], [250]

$$\hat{q}_i = \frac{1}{\Omega} \int_{\Omega} q_i d\Omega \,, \qquad \widehat{\nabla}_i \theta = \frac{1}{\Omega} \int_{\Omega} \nabla_i \theta d\Omega \tag{4.35}$$

The effective TC tensor $\hat{\kappa}_{ij}$ of the RVE volume is obtained from the Fourier's law of steady heat conduction given as

$$\hat{q}_i^k = -\hat{\kappa}_{ij}\hat{\nabla}_j^k\theta \tag{4.36}$$

For simplicity, let the $i^{th}$ component of the equivalent homogenized heat flux vector $\hat{q}_i^k$ and temperature gradient $\hat{\nabla}_j^k\theta$ for the $k^{th}$ load case be denoted as $\hat{Q}_{ik} = \hat{q}_i^k$ and $\widehat{\nabla T}_{jk} = \hat{\nabla}_j^k\theta$, respectively, such that Eqn. (4.36) can be rewritten as

$$\hat{Q}_{ik} = -\hat{\kappa}_{ij}\widehat{\nabla T}_{jk} \tag{4.37}$$

Rearranging Eqn. (4.37) yields the effective ETC tensor $\hat{\kappa}_{ij}$ given as

$$\hat{\kappa}_{ij} = -\hat{Q}_{ik}\left[\widehat{\nabla T}^{\,-1}\right]_{jk} \tag{4.38}$$

### *4.3 Analytical Mean-Field Homogenization Method*

Analytical methods for computing thermal-elastic properties of discontinuous fiber reinforced composites with suspended fibers of random orientation first appeared in Advani and Tucker's [19], [62] two step micromechanics homogenization evaluations. In their approach, the first step computes average properties of decomposed pseudo-grains having unidirectional, uniform length fibers using a mean-field theory (see e.g., [64]) or numerical FE analysis, and a second step that performs an orientation and/or a fiber length averaging of the aggregates using either the Voight's or Reuss' assumption. This widely accepted approach for computing elastic material properties from orientation tensor data account for the randomly dispersed spatially varying fiber orientation and length distributions in the heterogenous macro-scale volume of fiber-filled polymer composites. In the following, we present the Mori-Tanaka-Benveniste's analytical mean field homogenization approach for computing effective material properties, i.e. components of the 4th order elastic stiffness tensor $\hat{C}_{ijkl}$, the 2nd order coefficient of thermal expansion (CTE), $\hat{\alpha}_{ij}$ and the 2nd order thermal conductivity tensor $\hat{\kappa}_{ij}$.

### *4.3.1 Effective Engineering Elastic Constants – Analytical Approach.*

Various empirical micromechanics models have been developed to predict the elastic properties of unidirectionally aligned discontinuous short fiber reinforced polymer composite such as those briefly discussed above and summarized in [64]. One such model which we use to validate results from our numerical calculations is the Eshelby based – Mori-Tanaka formulations for calculating the effective composite moduli. The general equation for the mean-field homogenized composite stiffness $C_{ijkl}$ is given as

$$C_{ijkl} = \left[\vartheta_m C^m_{ijuv} + \sum_p \vartheta_p C^p_{ijrs} B^p_{rsuv}\right] [\tilde{B}^{-1}]_{uvkl},$$
$$\tilde{B}_{ijkl} = \left[\vartheta_m \delta_{ijkl} + \sum_p \vartheta_p B^p_{ijkl}\right] \tag{4.39}$$

where $C^m_{ijkl}$ and $C^p_{ijkl}$ are the isotropic matrix and particulate (either fiber or void) stiffness tensors, respectively. The particle's strain concentration tensor $B^p_{ij}$ according to the Mori-Tanaka model corresponds to the Hashin-Shtrikman-Willis lower bounds solution for the stiffness tensor computed as

$$[B^{-1}]^p_{ijkl} = \delta_{ijkl} + \Pi^m_{ijrs}\left([C^{-1}]^m_{rsuv} C^p_{uvkl} - \delta_{rskl}\right) \tag{4.40}$$

where $\Pi_{ijkl}$ is the Eshelby's elasticity tensor appearing in APPENDIX A. We assume the micro-void inclusions are spherical shaped with unity aspect ratio. Length averaging of the stiffness tensor is first performed using the length distribution of the inclusions within the composite as

$$C_{ijkl} = \sum_{\forall n} w^n C^n_{ijkl}(\hat{a}^n_r) \tag{4.41}$$

where $w^n$ is the weight fraction of the $n^{th}$ pseudo-grain of an inclusion phase with average aspect ratio $\hat{a}^n_r$. Subsequently, the orientation average of the fourth order transversely isotropic elastic stiffness tensor $\hat{C}_{ijkl}$ is computed using the fourth-order fiber orientation averaging scheme [62]

$$\begin{aligned}\hat{C}_{ijkl} = \beta_1 \mathrm{a}_{ijkl} + \beta_2\left(\mathrm{a}_{ij}\delta_{kl} + \mathrm{a}_{kl}\delta_{ij}\right) + \beta_3\left(\mathrm{a}_{ik}\delta_{jl} + \mathrm{a}_{il}\delta_{jk} + \mathrm{a}_{jk}\delta_{il} + \mathrm{a}_{jl}\delta_{ik}\right) \\ + \beta_4\left(\delta_{ij}\delta_{kl}\right) + \beta_5\left(\delta_{ik}\delta_{jl} + \delta_{il}\delta_{jk}\right)\end{aligned} \tag{4.42}$$

where $\mathrm{a}_{ijkl}$ is the 4th order orientation tensor of an inclusion phase computed using any of the suitable closure approximations detailed in [19], [62], [251], and the $\beta_i$ , $i$ = 1,..,5 are invariants of the unidirectional transversely isotropic elasticity tensor $C_{mn}$ (written in contracted notation)

$$\begin{aligned}\beta_1 = C_{11} + C_{22} - 2C_{12} - 4C_{66}, \;\; \beta_2 = C_{12} - C_{23}, \;\; \beta_3 = C_{66} + \frac{1}{2}(C_{23} - C_{22}) \\ \beta_4 = C_{23}, \qquad \beta_5 = \frac{1}{2}(C_{22} - C_{23})\end{aligned} \tag{4.43}$$

The elastic engineering constants are computed from the orthotropic stiffness matrix of the contracted form of Eqn. (4.42) based on Eqn. (4.24).

### *4.3.2 Effective Coefficient of Thermal Expansion – Analytical Approach.*

Similar to the approach presented above for the analytical elastic stiffness prediction, a two-step homogenization approach is employed to analytically evaluate the effective CTE tensor for the discontinuous fiber reinforced polymer composite with voids considered in this work. Analytical orientation averaged effective CTE tensor $\hat{\alpha}_{ij}$ for a discontinuous fiber reinforced composite having randomly oriented cylindrical fiber inclusions may be computed from [62], [144], [247]

$$\hat{C}_{ijkl}\,\hat{\alpha}_{kl} = \widehat{[C:\alpha]}_{ij} \quad \text{where} \quad [C:\alpha]_{ij} = C_{ijkl}\alpha_{kl} \tag{4.44}$$

and $\hat{C}_{ijkl}$ is given in Eqn. 42. The orientation average for tensor product $\widehat{[C:\alpha]}_{ij}$ may be calculated using the second-order orientation averaging scheme from Advani & Tucker [19], [62] given as

$$\widehat{[C:\alpha]}_{ij} = \gamma_1 \mathrm{a}_{ij} + \gamma_2 \delta_{ij} \tag{4.45}$$

where $\gamma_1$ and $\gamma_2$ are, respectively, the invariants of the tensor product $[C:\alpha]_{ij}$ obtained from the double contraction of the transversely isotropic elasticity tensor $C_{ijkl}$ and CTE tensor $\alpha_{kl}$ for a unidirectional discontinuous fiber composite given as

$$\gamma_1 = [C:\alpha]_{11} - [C:\alpha]_{22} \quad \text{and} \quad \gamma_2 = [C:\alpha]_{22} \tag{4.46}$$

For consistency, we instead employ the Mori-Tanaka-Benveniste's equation for estimating the tensor product $C_{ijkl}\alpha_{kl}$ of unidirectional short fiber reinforced polymer composite with isotropic constituents given as [31], [96], [97], [98]

$$[C:\alpha]_{ij} = C_{ijkl}\alpha_{kl} = \left[\vartheta_m C^m_{uvkl}\alpha_{kl} + \sum_p \vartheta_p C^p_{uvrs} B^p_{rskl} \alpha^p_{kl}\right] [\tilde{B}^{-1}]_{uvij} \tag{4.47}$$

where terms given here appear in Section 2.2.1. Length averaging of the tensor product $[C:\alpha]_{ij}$ is performed prior to orientation averaging on the right-hand side of Eqn. 44 according to

$$[C:\alpha]_{ij} = \sum_{\forall n} w^n [C:\alpha]^n_{ij}(\hat{a}^n_r) \tag{4.48}$$

### 4.3.3 Effective Thermal Conductivity – Analytical Approach.

The Mori-Tanaka's model presented above for predicting the homogenized 4th order elasticity tensor for unidirectional particulate composite has been extended [100], [111], [112] to estimate other 2nd order tensor properties of the composite material including the thermal conductivity tensor. The resulting expression for evaluating the composite's thermal conductivity tensor $\kappa_{ij}$ is as

$$\kappa_{ij} = \left[\vartheta_m \kappa_{is}^m + \sum_{\forall p} \vartheta_p \kappa_{ir}^p A_{rs}^p\right] \left[\tilde{A}^{-1}\right]_{sj} \quad \text{with} \quad \tilde{A}_{ij} = \vartheta_m \delta_{ij} + \sum_{\forall p} \vartheta_p A_{ij}^p \qquad (4.49)$$

where $\kappa_{is}^m$ and $\kappa_{ir}^p$ are the isotropic matrix and particulate (either fiber or void) thermal conductivity tensors. The intensity-concentration tensor $A_{ij}^p$ that couples the mean temperature gradients between the particulate inclusions, and the isotropic matrix corresponds to the lower bound solution based on Hashin-Shtrikman-Willis single variational principle and is computed as

$$[A^{-1}]_{ij}^p = \delta_{ij} + \mathcal{K}_{ir}\left\{[\kappa^{-1}]_{rs}^m \kappa_{sj}^p - \delta_{rj}\right\} \qquad (4.50)$$

In Eqns. (4.49)-(4.50) above, $\mathcal{K}_{ij}$ is the Eshelby's thermal conductance tensor having only non-zero diagonal components, which for spheroidal inclusions, is given as

$$\mathcal{K}_{22} = \mathcal{K}_{33} = 0.5 a_r \chi_r^3 \{a_r \chi_r^{-1} - \ln[a_r + \chi_r^{-1}]\}, \qquad \mathcal{K}_{11} = 1 - 2\mathcal{K}_{22} \qquad (4.51)$$

In our calculations, we assume the void particles are spherically shaped for simplicity. In this case, the Eshelby's tensor is simply one-third the identity tensor, i.e. $\mathcal{K}_{ij} = {}^1\!/_3\, \delta_{ij}$. Similar to the procedure adopted in elasticity stiffness tensor homogenization, length averaging of the computed transversely isotropic thermal conductivity tensor using the length distribution of the fiber inclusion phase within the SFRP composite is performed on a weight-based averaging scheme according to

$$\kappa_{ij} = \sum_{\forall n} w^n \kappa_{ij}^n (\hat{a}_r^n) \tag{4.52}$$

The invariants of $\kappa_{ij}$ i.e. $\psi_i$ are computed in the usual manner as

$$\psi_1 = \kappa_{11} - \kappa_{22}, \qquad \psi_2 = \kappa_{22} \tag{4.53}$$

Subsequently, the second order orientation averaged thermal conductivity tensor $\hat{\kappa}_{ij}$ for the misaligned fiber reinforced composite is calculated as [62], [111]

$$\hat{\kappa}_{ij} = \psi_1 \mathrm{a}_{ij} + \psi_2 \delta_{ij} \tag{4.54}$$

### *4.4 Density and Specific Heat Estimation*

We also need the average density $\hat{\rho}$ and specific heat capacity $\hat{\zeta}_s$ for the composite material which is computed here using the basic rule of mixture equation since these scalar quantities do not depend on the distribution of fiber orientation or fiber length within the RVE. The average density $\hat{\rho}$ for $p$ particulate phases is given as

$$\hat{\rho} = \vartheta_m \rho_m + \sum_{\forall p} \vartheta_p \rho_p \tag{4.55}$$

where $\rho_m$ and $\rho_p$ are the matrix and particulate (either fiber or void) isotropic density. Likewise, the average specific heat capacity $\hat{\zeta}_s$ for $p$ particulate phases is given as

$$\hat{\zeta}_s = \vartheta_m \zeta_s^m + \sum_{\forall p} \vartheta_p \zeta_s^p \tag{4.56}$$

where $\zeta_s^m$ and $\zeta_s^p$ are the matrix and particulates (fiber and void) isotropic specific heat capacity values.

### *4.5 Evaluating the Magnitude of the Effective Quantities*

The magnitude of the effective quantities is used here to define equivalent isotropic properties of the RVE for the elastic modulus, the CTE and the TC. This provides a convenient means for quantifying the dispersion in the effective quantity and for comparing the numerical FE results with the analytical MT estimates. The effective elastic modulus $E^{eff}$ magnitude and effective Poisson ratio $\nu^{eff}$ magnitude are given as [252]

$$\frac{1}{E^{eff}} = \frac{1}{3G^{eff}} + \frac{1}{9K^{eff}}, \qquad \nu^{eff} = \frac{3K^{eff} - 2G^{eff}}{6K^{eff} + 2G^{eff}} \tag{4.57}$$

where $K^{eff}$ is the apparent effective bulk modulus defined as the average between the Voight upper $K_V$ and Reuss lower $K_R$ first order bounds on the bulk modulus and is given as

$$K^{eff} = \frac{1}{2}[K_V + K_R], \qquad \frac{1}{K_R} = \hat{S}_{iijj}, \qquad K_V = \frac{1}{9}\hat{C}_{iijj} \tag{4.58}$$

The effective shear modulus magnitude $G^{eff}$ is obtained from the average of the Voight upper $G_V$ and Reuss lower $G_R$ first order bounds on the shear modulus and is given as

$$G^{eff} = \frac{1}{2}[G_V + G_R], \qquad \frac{1}{G_R} = \frac{2}{5}\left[\hat{S}_{ijij} - \frac{1}{3}\hat{S}_{iijj}\right], \qquad G_V = \frac{1}{10}\left[\hat{C}_{ijij} - \frac{1}{3}\hat{C}_{iijj}\right] \tag{4.59}$$

Likewise, the apparent effective CTE magnitude $\alpha^{eff}$ is computed from the Hill's average of the Voight lower bound $\alpha_V$ and Reuss upper bound $\alpha_R$ values of the CTE tensor [253]. i.e.

$$\alpha^{eff} = \frac{1}{2}[\alpha_V + \alpha_R], \qquad \alpha_V = \frac{1}{3}\hat{\alpha}_{ii}, \qquad \alpha_V = \frac{1}{9K_V}\hat{C}_{iikl}\hat{\alpha}_{kl} \tag{4.60}$$

where summation convention is implied. The apparent effective TC magnitude is given as [116], [119]:

$$\kappa^{eff} = \frac{1}{3}\hat{\kappa}_{ii} \tag{4.61}$$

## *4.6 Results & Discussion*

The results of the thermo-mechanical properties including the engineering constants, effective CTE and TC obtained from both numerical and analytical homogenization approaches are presented in the following. The influence of the inherent microstructural porosity on the resulting thermo-mechanical properties are also evaluated and quantitative assessment of the macroscale property anisotropy due to spatial variation of the microstructural configurations across various ROIs are presented. The measure of dispersion in the computed effective properties for each ROI is quantified using the coefficient of variation statistical parameter given as

$$\xi = \sigma/\mu \tag{4.62}$$

where $\xi$ assesses the ability of selected RVE size in representing the ROI volume. The parameters $\sigma$ and $\mu$ are, respectively, the standard deviation and mean value of the reported effective quantity from the different RVE's of the ROI volume. Minimization of the $\xi$ value is desirable when determining a suitable RVE size in accurately predicting a composite property $Z$. From this point onward, the effective composite properties are reported in their normalized form with respect to the equivalent properties of the isotropic matrix phase $Z_m$ and are distinguished from actual dimensional quantities by an overbar accent. i.e. $\bar{Z} = Z/Z_m$.

### *4.6.1 Thermo-Mechanical Property Estimates for ROI-II*

The first set of results presented here are solutions obtained for the three RVE sizes within ROI-II appearing in Fig. 3 based on numerical FE homogenization method presented in Section 4.2. Numerically computed quantities are compared to results obtained from corresponding analytical approach from Section 4.3. The effect of the inherent micro-voids on the resulting effective thermo-mechanical properties is also quantified. Average microstructural properties of ROI–II follow from results provided in [241] and include the fiber volume fraction ${\vartheta_f}^{\mathrm{II}} = 6.65\%$, the void volume fraction $\vartheta_v^{\mathrm{II}} = 12.27\%$, and the region averaged 2[nd] order fiber orientation tensor $\hat{a}_{ij}^{\mathrm{II}}$ given as [241]

$$\hat{a}_{ij}^{\mathrm{II}} = \langle p_i p_j \rangle = \begin{bmatrix} 0.32 & 0.04 & 0.12 \\ 0.04 & 0.17 & -0.02 \\ 0.12 & -0.02 & 0.51 \end{bmatrix} \tag{4.63}$$

From the region averaged orientation $\hat{a}_{ij}^{\mathrm{II}}$, we observe higher fiber alignment in the print direction ($x_3$-direction) with less fiber alignement in the $x_1$-direction parallel to the build surface. The distribution of the fiber aspect ratio for the ROI-II region appears in Figure 4.4 where it is fitted to a Weibull function given as

$$w_f = \frac{\phi_1}{{\phi_2}^{\phi_1}} \hat{a}_r^{(\phi_1 - 1)} e^{-(\hat{a}_r/\phi_2)^{\phi_1}} \tag{4.64}$$

with a shape parameter $\phi_1 = 22.72$ and a scale parameter $\phi_2 = 1.65$. The weighted average aspect ratio for this region is $\hat{a}_r^{\mathrm{II}} = 20.31$. The average fiber aspect ratio is limited by the ROI envelope which may under-represent the specimens true mean aspect ratio. Partitioning of the ROI into several RVE realizations further limits the average fiber aspect ratio. The mean fiber aspect ratio $\hat{a}_r$ and fiber aspect ratio coefficient of variation $\xi_r$ from the complete sets of realizations of each RVE sizes (RVE-I, RVE-II, and RVE-III) of ROI-II are presented in Table 4.2. These microstructural characteristics for ROI-II are used to compute the thermo-mechanical properties of the region.

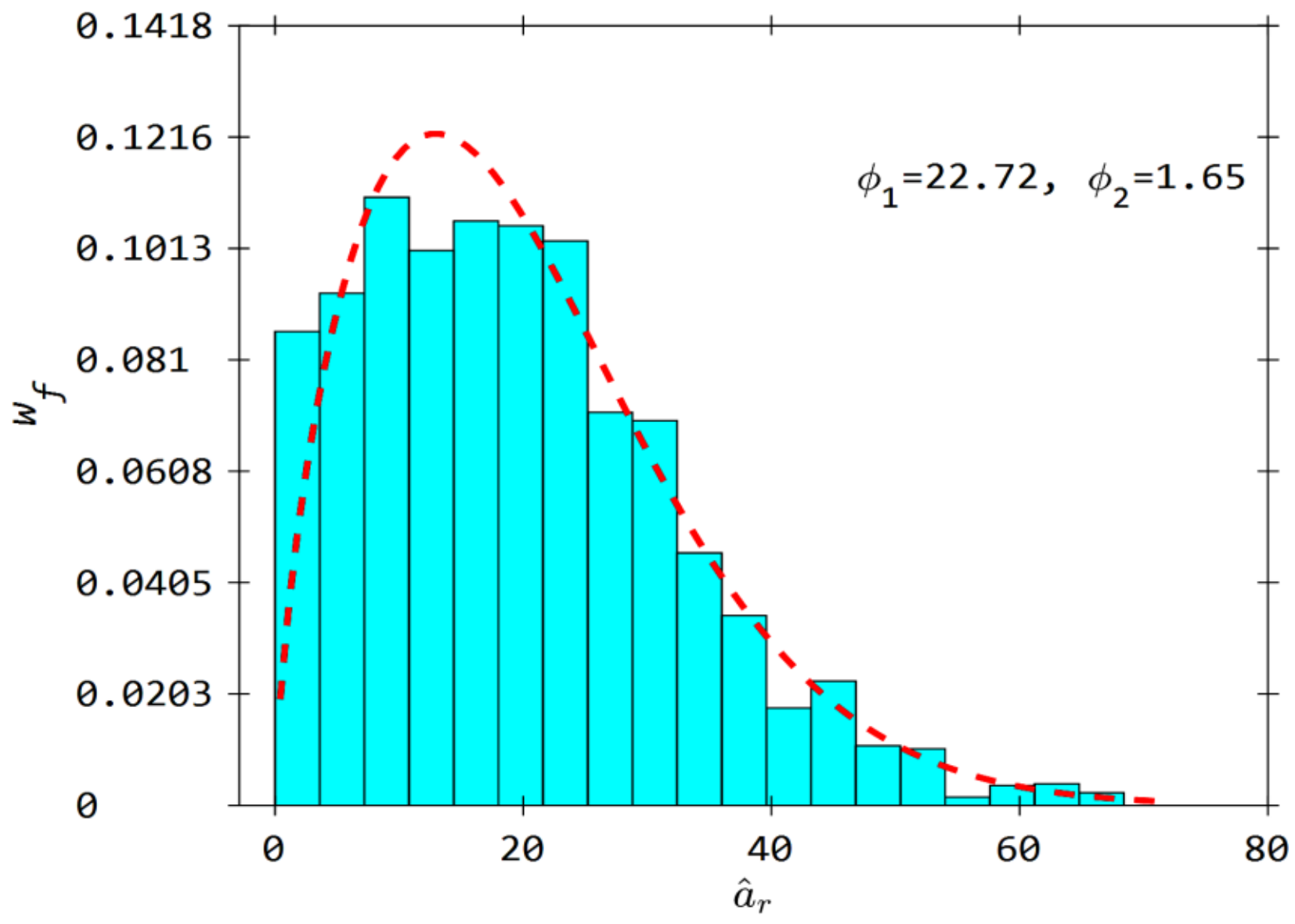

Figure 4.4: Fiber Aspect Ratio Distribution across extracted for ROI-II

Table 4.2: Mean fiber aspect ratio $\hat{a}_r$ and fiber length coefficient of variation $\xi_r$ from realizations of (a) RVE-I, (b) RVE-II, (c) RVE-III from ROI-II.

| | (a) | (b) | (c) |
|---|---|---|---|
| $\hat{a}_r$ | 8.58 | 12.38 | 15.69 |
| $\xi_r$ [%] | 10.50 | 7.19 | 7.01 |

*4.6.1.1 Effective Elastic Stiffness and Engineering Constants.* Typical displacement contours of the single RVE-II, #14 instance extracted from partitioning of ROI-II as shown in Figure 4.2 appear in Figure 4.5 (a-c). In this example, the RVE is subjected to tensile and shear loading corresponding to a total of six load cases as described in Section 4.2.1. The loads are prescribed on the bounding entities of the RVE through periodic boundary conditions as described in Section 4.2. The single step static stress analysis considers the effect of all three microstructural constituents including the ABS matrix, fiber and micro-

void inclusion (cf. Figure 4.1) on the stiffness behavior of the RVE using the micro-constituents material properties presented in Table 4.1. The homogenized stresses and strains obtained from the six different load cases are used to compute the effective stiffness using procedures detailed in Section 4.2.1. Figure 4.5a and Figure 4.5b show the uneven displacement on opposing faces of the RVE in response to the tensile loading which is due to the anisotropy introduced by the non-uniform spatial and orientation distribution of the fiber and void inclusions within the RVE. Additionally, the distinct features of the inherent microstructure are reflected by the local deformations (undulations) on the boundaries of the RVE (cf. Figure 4.5).

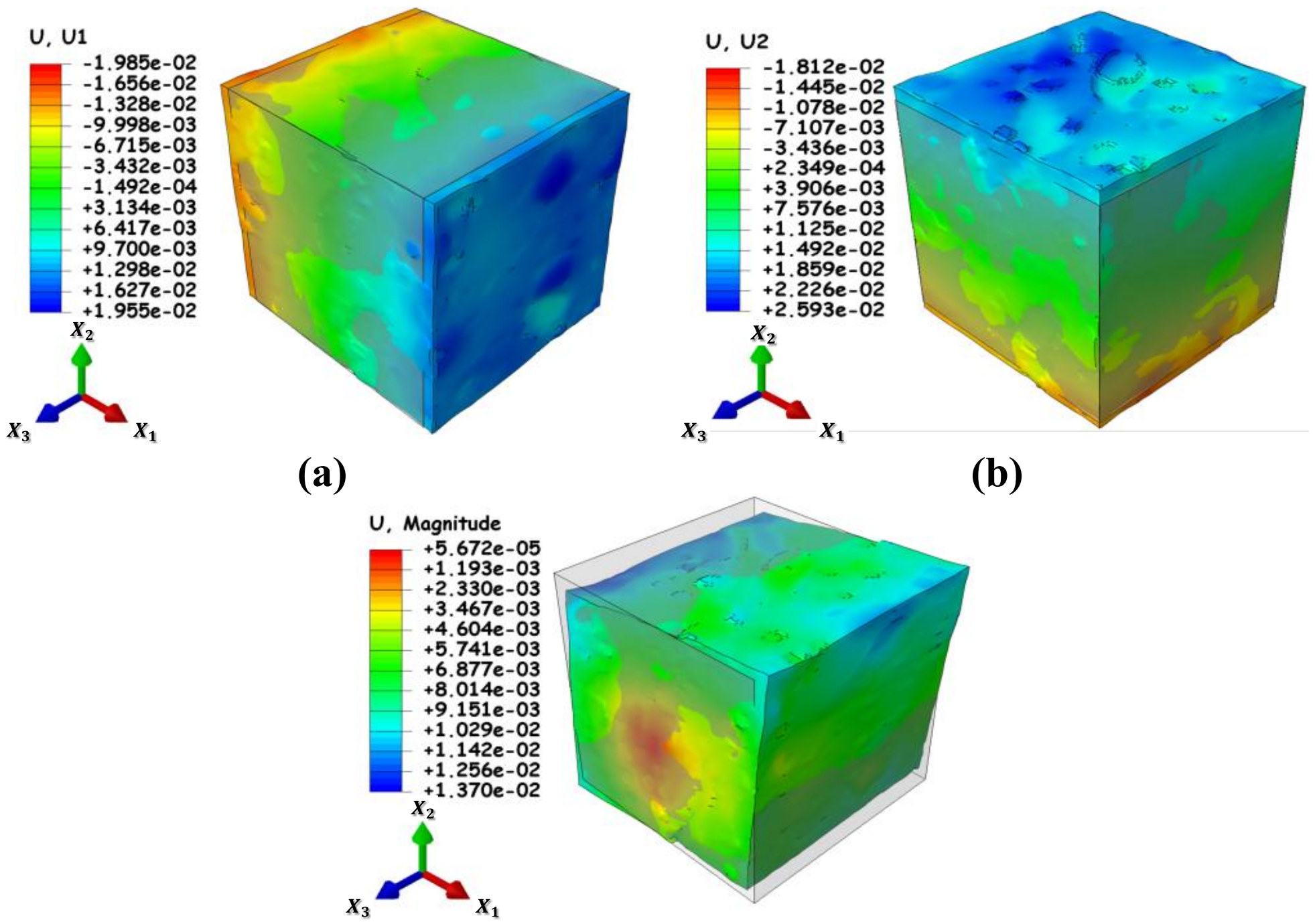

Figure 4.5: Isometric view of the deformation contours of RVE-II, #14 from ROI-II overlayed over the undeformed volume under different loading (a) tensile response in $x_1$-direction (b) tensile response in $x_2$-direction (c) shear response in $x_1x_2$-plane.

Figure 4.6 (a-c) presents contours of displacement magnitude on opposing bounding faces of the RVE in the direction of the applied load for $x_1$ and $x_2$ tensile deformations and the $x_1$-$x_2$ shear deformation load cases which verifies our correct implementation of the periodic constraints on the boundaries of the RVE. This is evident from similar topographical features on opposing bounding faces of the RVE as shown in Figure 4.6. The periodic constraints enforce domain continuity without overlapping or separation among neighboring RVE boundaries that ensure effective transfer of loads between adjacent RVE boundaries.

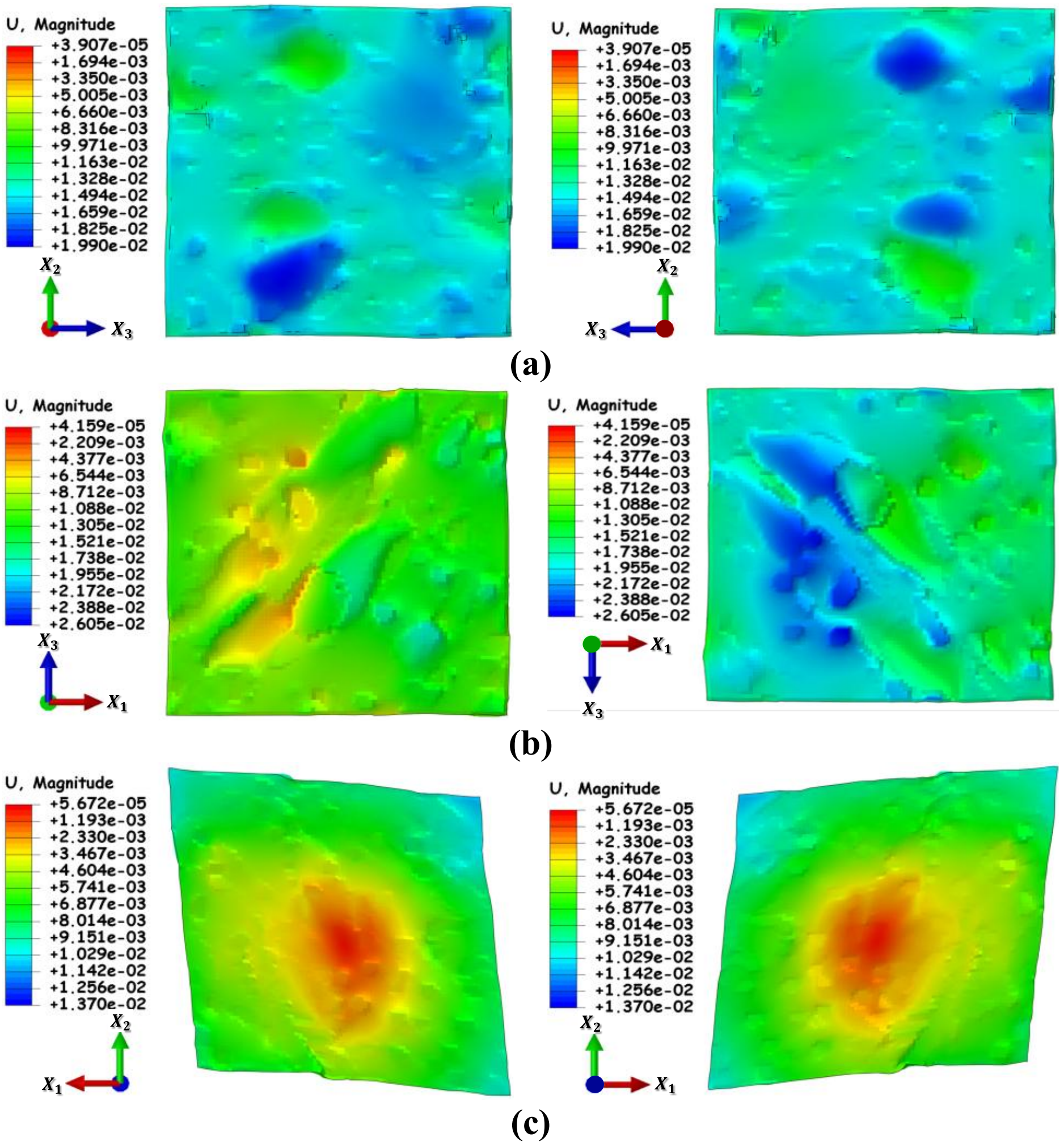

Figure 4.6: Deformation contours of RVE-II, #14 from ROI-II showing topography of opposing bounding face pairs to validate the implementation of the PBC for loading in (a) $x_1$-direction (b) $x_2$-direction (c) $x_1x_2$-plane. The images show RVE bounding faces in the negative (upper row) and positive (lower row) direction (or normal in case of shear) to the direction of the applied loading.

The homogenized stresses and strains from the six (6) different load cases are used to compute the effective stiffness and compliance of the RVE volume as described in Section 4.2.1 which appear in Table 4.3 and Table 4.4. The results are presented for both (a) non-porous microstructure without considering the micro-void inclusions (i.e. micro-void constituent properties are replaced by ABS properties) and (b) porous microstructure considering the micro-void inclusions. Table 4.3 below provides the computed mean elastic stiffness $\bar{C}_{mn}$ derived from the orthotropic regression fitting procedure in Section 4.2.1 for the complete sets of 27 RVE-II realizations of ROI-II (cf. Figure 4.2b) which are compared to the Mori-Tanaka's mean-field estimate using the length distribution in Figure 4.4 and the region averaged orientation tensor of ROI-II appearing in Eqn. (4.63). Values in Table 3 show a good agreement between results obtained from our computational approach and those from the related analytical method. The results of the stiffness tensor in Table 4.3 and elastic moduli in Table 4.4 show that the 13% CF/ABS bead RVE specimen exhibits a somewhat transversely isotropic behavior along the $x_3$-plane. These results show that the largest component of the stiffness tensor $\bar{C}_{33}$ (cf. Table 4.3) and the largest elastic moduli $\bar{E}_{33}$ (cf. Table 4.4) both coincide with the largest average fiber orientation tensor component $\hat{a}_{33}$ showing that material stiffness increases with increasing degree of fiber alignment, as expected. Further, the results of Table 4.3 and Table 4.4 confirm that there is a clear reduction in elastic stiffness when micro-void inclusions are considered [100], [113], [131]. In addition, regression coefficients from the computational homogenization least squares approach given in Section 4.2 for ROI-II data indicate a moderately high accuracy of the regression fitting procedure (i.e., $\langle R^2 \rangle > 0.80$) from the set of realizations of the various RVE sizes reported in Table 4.5. This along with the relatively low standard deviations $\sigma_{std}$ also appearing in Table 4.5 show that the computed homogenized elastic constants provide a good fit to the data with minor variation. Although Table 4.4 shows that the effective moduli increase with increasing RVE size, the rate of increase between consecutive RVE's is seen to decrease by an average of -22.8% and -44.3% for the elastic and shear moduli respectively of the non-porous microstructure and by an average of -42.7% and -42.9% for the elastic and shear moduli respectively of the porous microstructure. Additionally, Table 4.5 shows that the average regression fitting coefficients is seen to decrease with increasing RVE size.

Table 4.3: Computed values of average elastic stiffness $\bar{C}_{mn}$ for ROI-II region obtained from numerical (FE) homogenization approach based on RVE-II and the orientation averaged Mori-Tanaka (MT) method for 13% CF/ABS SFRP composite (components not shown indicate a zero value).

| | non-porous | porous |
|---|---|---|
| FE | $\begin{bmatrix} 1.34 & 1.20 & 1.33 & & & \\ 1.20 & 1.28 & 1.21 & & & \\ 1.33 & 1.21 & 1.51 & & & \\ & & & 1.51 & & \\ & & & & 1.79 & \\ & & & & & 1.49 \end{bmatrix}$ | $\begin{bmatrix} 0.98 & 0.75 & 0.89 & & & \\ 0.75 & 0.89 & 0.77 & & & \\ 0.89 & 0.77 & 1.15 & & & \\ & & & 1.21 & & \\ & & & & 1.48 & \\ & & & & & 1.18 \end{bmatrix}$ |
| MT | $\begin{bmatrix} 1.36 & 1.18 & 1.32 & & & \\ 1.18 & 1.21 & 1.21 & & & \\ 1.32 & 1.21 & 1.61 & & & \\ & & & 1.44 & & \\ & & & & 1.71 & \\ & & & & & 1.37 \end{bmatrix}$ | $\begin{bmatrix} 0.99 & 0.74 & 0.83 & & & \\ 0.74 & 0.87 & 0.75 & & & \\ 0.83 & 0.75 & 1.21 & & & \\ & & & 1.20 & & \\ & & & & 1.45 & \\ & & & & & 1.13 \end{bmatrix}$ |

Table 4.4: Mean values of engineering elastic constants for ROI-II computed from the numerical FE homogenization schemes for all RVE cases (i.e., RVE-I, II and III having 125, 27, and 8 realizations, respectively) with (a) non-porous microstructure (b) porous microstructure.

| Cases | RVE | $\bar{E}_{11}$ | $\bar{E}_{22}$ | $\bar{E}_{33}$ | $\bar{G}_{23}$ | $\bar{G}_{13}$ | $\bar{G}_{12}$ | $\bar{\nu}_{23}$ | $\bar{\nu}_{13}$ | $\bar{\nu}_{12}$ |
|---|---|---|---|---|---|---|---|---|---|---|
| | I | *1.39* | *1.38* | *1.58* | *1.45* | *1.67* | *1.42* | *0.82* | *0.93* | *0.97* |
| (a) | II | *1.42* | *1.41* | *1.65* | *1.51* | *1.79* | *1.49* | *0.78* | *0.93* | *0.98* |
| | III | *1.45* | *1.43* | *1.70* | *1.55* | *1.87* | *1.52* | *0.76* | *0.94* | *0.97* |
| | I | *1.08* | *1.04* | *1.27* | *1.15* | *1.34* | *1.11* | *0.72* | *0.86* | *0.90* |
| (b) | II | *1.12* | *1.07* | *1.35* | *1.21* | *1.48* | *1.18* | *0.78* | *0.93* | *0.98* |
| | III | *1.14* | *1.09* | *1.40* | *1.26* | *1.56* | *1.21* | *0.76* | *0.94* | *0.97* |

Table 4.5: Mean values and standard deviation of the coefficient of determination $R^2$ for ROI-II computed from the least square regression fitting procedure for all RVE cases (i.e., RVE-I, II and III).

| Cases | non-porous | | | porous | | |
|---|---|---|---|---|---|---|
| RVE | I | II | III | I | II | III |
| $\langle R^2 \rangle$ | 0.92 | 0.90 | 0.88 | 0.87 | 0.84 | 0.81 |
| $\sigma_{std}$ | 0.05 | 0.05 | 0.05 | 0.08 | 0.07 | 0.08 |

Figure 4.7 presents box plots showing the mean value ('x' marker), the interquartile range (box) consisting of the lower quartile, median line and upper quartile, the extremums of the data range (whiskers), and outliers (isolated dots) for the predicted engineering constants for each set of realizations RVE-I, RVE-II, and RVE-III of ROI-II appearing in Figure 3. The results show an overall reduction in the variability of computed properties with increasing RVE size. Mean values tend to converge as the RVE size is increased for all engineering constants which agrees with the work of Kanit et al. [119]. Their work showed that small but reasonable sized RVEs with a sufficient number of realizations can accurately predict effective properties as would larger sized RVE with smaller number of realizations. We do, however, observe some deviation in computed engineering constants for the smallest RVE (i.e., RVE-I). Although micro-voids appear to reduce the elastic moduli for all RVE cases (i.e., up to 24% reduction observed), the mean Poisson ratios computed from RVE - II and III are unaffected by micro-void inclusions (cf. Figure 4.7b and Figure 4.7c, respectively) contrary to that shown in RVE-I results (cf. Figure 4.7a) which shows a significant effect of voids on the Poisson ratio with deviations as high as 11.6%. This suggests that RVE-I is insufficient in accurately predicting the elastic modulus of the CF/ABS composite. Partitioning of the ROI into smaller RVE volumes results in increased variability in average microstructural characteristics across the RVE realizations which potentially leads to increased dispersion in the predicted effective properties of the ROI volume as observed from the box plots of Figure 4.7.

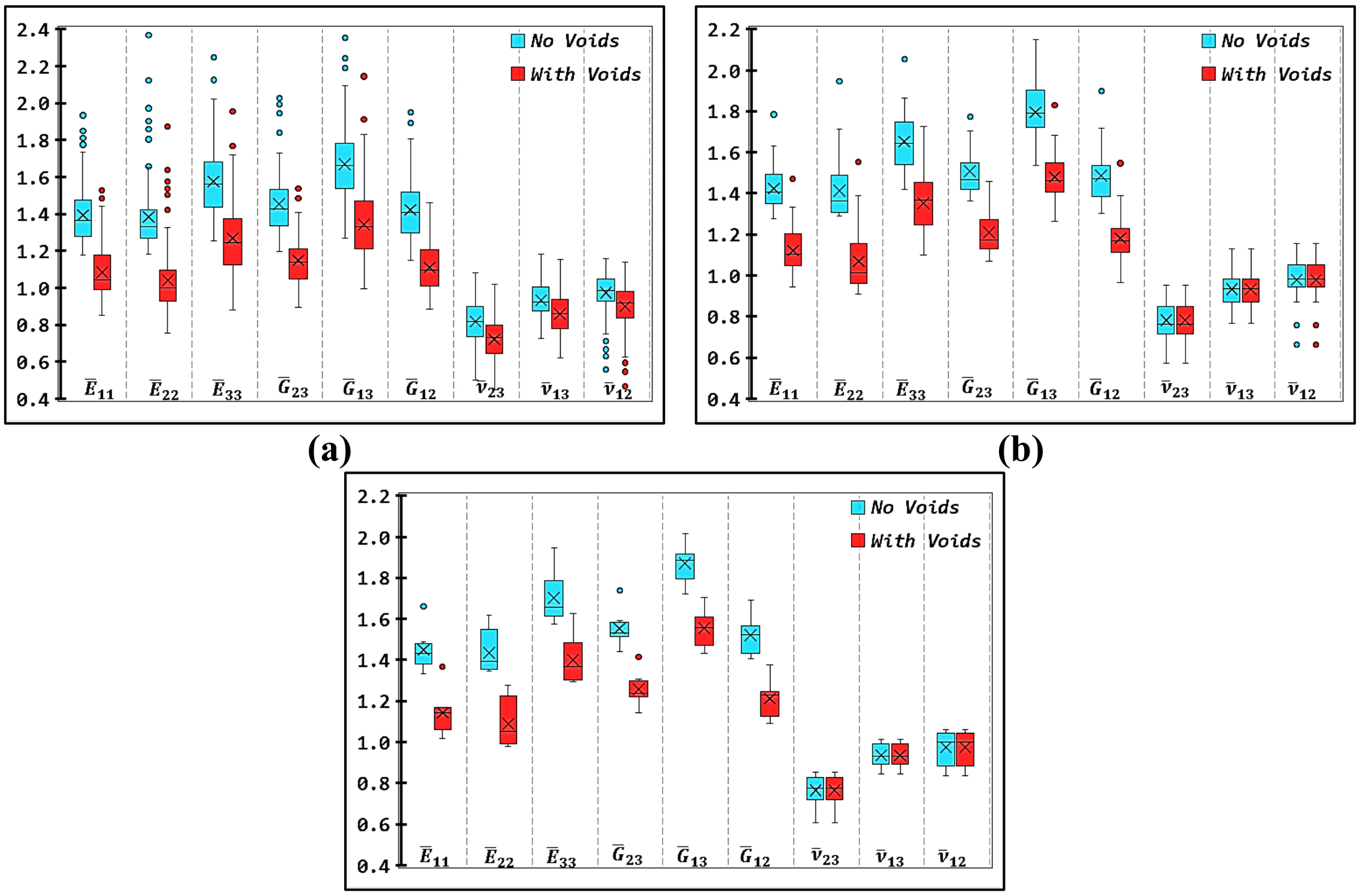

Figure 4.7: Box plot showing the mean values, interquartile intervals, and outliers of the computed normalized engineering constants from RVE realization datasets for (a) RVE – I (b) RVE-II and (c) RVE-III. Results are shown for analysis cases with micro-voids (red) and without micro-voids (blue) present within the bead's microstructure.

Measured microstructural features within ROI-II including fiber length distribution, fiber orientation tensor values, fiber volume fraction and micro-voids are used to compute and compare the mean effective elastic modulus $\overline{E}^{eff}$ obtained from both Mori-Tanaka's mean-field and the numerical FE-based homogenization approaches using the method of Hills [252] according to Eqn. (4.57). The relative error in predicted values between both methods presented in Table 4.6 show that the Mori-Tanaka estimates are reasonably comparable to the numerical predictions with errors that are mostly below 10% for the elastic moduli. Further, these results indicate a good degree of accuracy which improves with increasing RVE size. Likewise, the predicted effective elastic modulus for the porous case ($E^{eff}$ ~ 1.235) is comparable to the mean values obtained from tensile test experiment by Russell T. [254], for the same 13% CF/ABS test sample ($E_{LT}^{eff}$ ~ 1.221) and the predicted effective shear modulus ($G^{eff}$ ~ 1.286) aligns closely with the flexural test experimental mean values ($G_{LT}^{eff}$ ~ 1.295) for the same test sample [254]. The higher errors may stem from the original Mori-Tanaka-Benveniste model which was formulated for two-phase composites with ellipsoidal inclusions as Norris A. N [115] has shown that the model's extension to multiphase inclusions may perform poorly and may violate the Hashin - Shtrikman stiffness bounds. Moreover, the Mori-Tanaka's predictions have been reported by Mortazavi et al., [61] and Breuer et al., [117] to deviate significantly from numerical estimates with increasing aspect ratio and volume fraction of the fiber inclusions.

Table 4.6: Relative error $\bar{\epsilon}$ [%] in the predicted effective elastic modulus magnitude, $\bar{E}^{eff}$ of ROI-II between Mori-Tanaka's analytical model and numerical FE homogenization schemes for all RVE cases (I, II &III).

| $\bar{E}^{eff}$ | non-porous | | | porous | | |
|---|---|---|---|---|---|---|
| | I | II | III | I | II | III |
| FE | *1.46* | *1.52* | *1.56* | *1.13* | *1.20* | *1.23* |
| MT | *1.31* | *1.41* | *1.47* | *1.08* | *1.16* | *1.22* |
| $\bar{\epsilon}$ [%] | *10.21* | *8.20* | *5.86* | *4.49* | *3.38* | *1.49* |

The coefficient of variation $\xi$ computed with Eqn. (4.62) for each RVE computation of ROI-II appears in Table 4.7. These results show that the $\xi$ value decreases with increasing RVE size by roughly -45% from RVE-I to RVE-II and -40% from RVE-II to RVE-III for the numerical FE calculation. Additionally, the $\xi$ values are relatively higher for the porous RVE structures compared to the corresponding non-porous RVE due to the increased variability introduced by the non-uniform distribution of the inherent micro-voids. The $\xi$ values computed from the analytical MT approach are relatively lower than the numerical FE counterparts for all RVE sizes. The $\xi$ value of RVE-II in both cases is less than 5%. Although RVE-III yields slightly more accurate results than the others considered, the computational requirements are excessive, and the gains of higher accuracy do not justify the higher computational cost.

Table 4.7: Coefficient of Variation $\xi$ [%] in the effective elastic modulus magnitude, $\bar{E}^{eff}$ for all RVE cases (I, II & III) of ROI-II and for both Mori-Tanaka's analytical model and numerical FE homogenization schemes.

| $\xi$ | non-porous | | | porous | | |
|---|---|---|---|---|---|---|
| | I | II | III | I | II | III |
| FE | *7.13* | *3.91* | *2.37* | *7.78* | *4.23* | *2.47* |
| MT | *3.54* | *2.18* | *2.12* | *4.17* | *2.26* | *2.39* |

*4.6.1.2 Effective Coefficient of Thermal Expansion*. Effective CTE are presented here where we compare results from the numerical FE homogenization approach (cf. Section 4.2.2) to those computed using Mori-Tanaka's mean-field method in Section 4.3.2. Typical displacement contours from the numerical FE simulation are plotted in Figure 4.8 for RVE-II, realization #14 subjected to a uniform thermal load of $\Delta\theta = 65^0\mathrm{C}$. As stated earlier, PBC constraints are imposed on the bounding entities of the RVE with zero applied strain (cf. Section 4.2.2).

The displacement contour of Figure 4.8 shows the thermal response of the porous RVE microstructure consisting the ABS matrix, fiber and micro-void inclusion (cf. Figure 4.1) based on the micro-constituents material properties presented in Table 4.1. The stresses computed from the single-step static stress analysis are used to obtain the effective

CTE from Eqn. (4.28) using the stiffness tensor for the corresponding RVE instances computed in the preceding section (i.e. Section 4.6.1.1).

Figure 4.8 shows non-uniform deformation of the RVE in all directions due to the anisotropic nature of the inherent microstructure. Additionally, there are distinct local deformations at the RVE boundaries which are likely introduced by the property contrast of the various micro-constituents at the RVE boundaries. Mean values of the diagonal components of the effective CTE tensor for the three RVE sizes shown in Figure 4.2, including both non-porous and porous microstructure of the CF/ABS SFRP bead are presented in Table 4.8. The maximum observed discrepancy in the effective CTE component values between RVE sizes is seen to drop from 6.3% between RVE-I and II to 3.3% between RVE-II and III. The CTE component values in Table 4.8 is observed to be consistently lower for the $\bar{\alpha}_{33}$ component with the highest degree of fiber alignment ($\hat{a}_{33}$) and the CTE is highest for the $\bar{\alpha}_{22}$ component with the lowest degree of fiber alignment, i.e. $\hat{a}_{22}$ (cf. Eqn. (4.63)) for all RVE cases considered. Additionally, the inherent micro-voids are seen to only slightly reduce the CTE component values.

Box plots appearing in Figure 4.9 indicate that the degree of variation in effective CTE decreases with increasing RVE size. We record higher variances in the predicted quantities with the smallest RVE (RVE-I) having more outliers outside the interquartile range (cf. Figure 4.9a) compared to largest RVE (RVE-III) with shorter error-bars and minimal variation in predicted effective CTE values. From these results, the presence of micro-voids within the bead microstructure only slightly reduces the predicted effective CTE values in all RVE cases (i.e., less than 5%). The estimated effects of micro-voids on the volumetric effective CTE values $\alpha_V$ are much lower (i.e., less than 3.25%). These trends agree well with results provided by others (see e.g., [98], [131], [255]). The Mori-Tanaka's (MT) mean-field estimates of the effective CTE magnitude, $\bar{\alpha}^{eff}$ (cf. Eqn. (4.57)-(4.59)) agree well with the numerical FE predictions with a maximum observed discrepancy of about 5.5% for the non-porous composite and 7.0% for the porous composite (cf. Table 4.9)

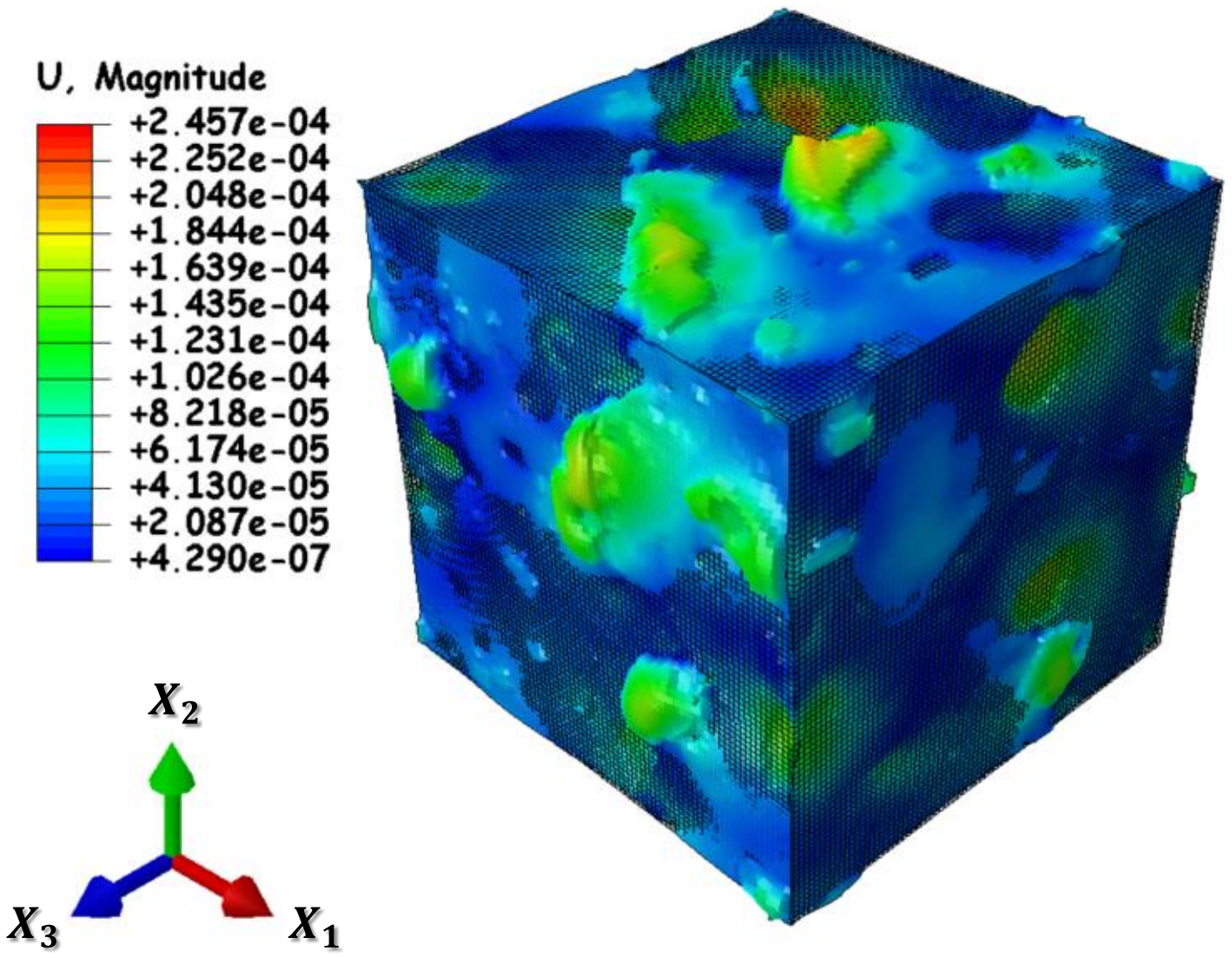

Figure 4.8: Deformation contour of RVE-II, #14 from ROI-II region overlayed on the undeformed mesh geometry and showing the bulk response of the volume under thermal load of $\Delta\theta = 65^0C$.

Table 4.8: Mean values of the diagonal components of the effective CTE tensor for all RVE cases (I, II, and III) of ROI-II computed using numerical FE homogenization.

| RVE | non-porous | | | porous | | |
|---|---|---|---|---|---|---|
| case | $\bar{\alpha}_{11}$ | $\bar{\alpha}_{22}$ | $\bar{\alpha}_{33}$ | $\bar{\alpha}_{11}$ | $\bar{\alpha}_{22}$ | $\bar{\alpha}_{33}$ |
| I | *0.83* | *0.91* | *0.70* | *0.81* | *0.89* | *0.67* |
| II | *0.82* | *0.90* | *0.66* | *0.79* | *0.88* | *0.63* |
| III | *0.81* | *0.89* | *0.64* | *0.79* | *0.87* | *0.61* |

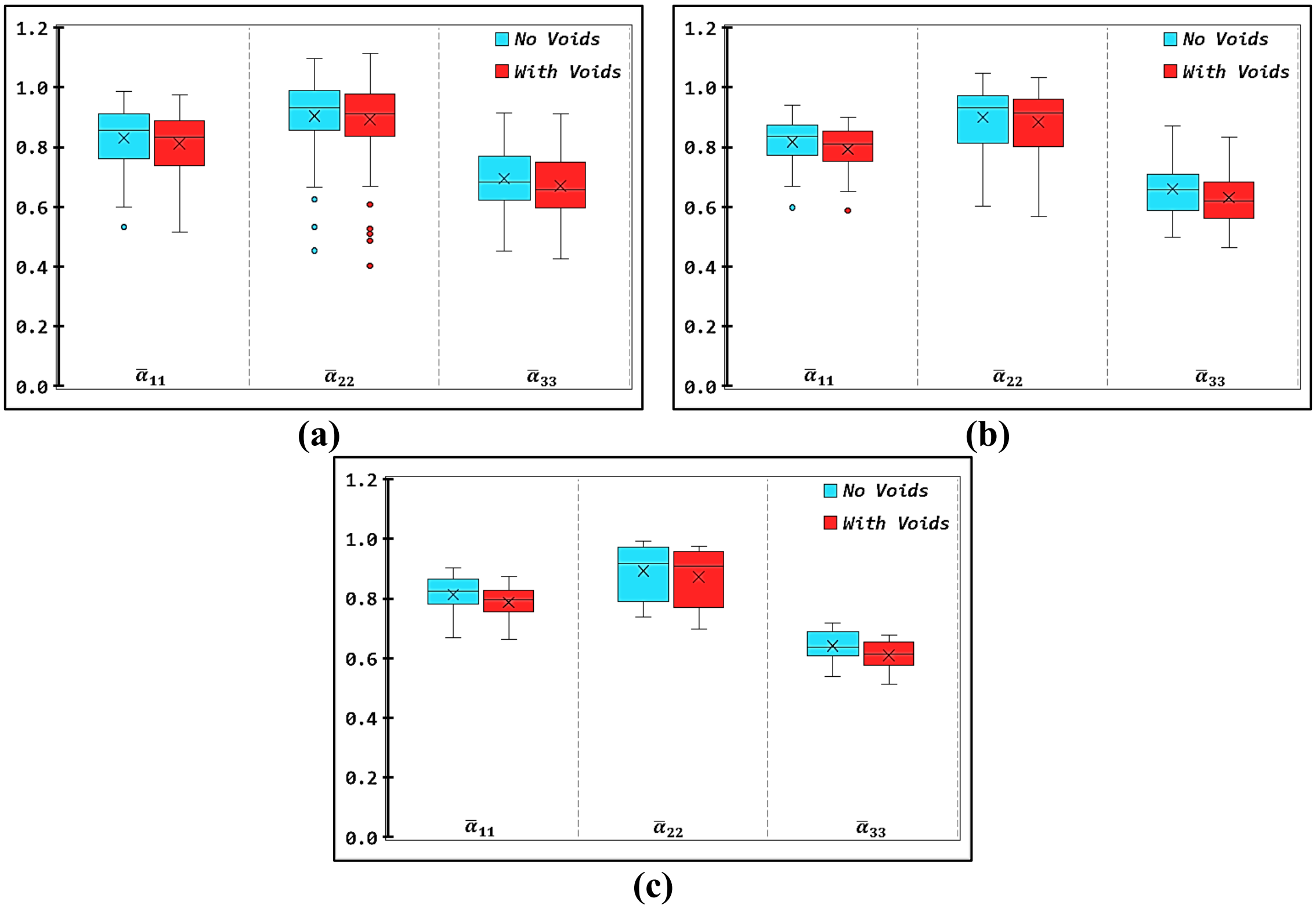

Figure 4.9: Box plot showing the mean values, interquartile intervals, and outliers of the computed components of the normalized effective CTE tensor from the RVE realization datasets for (a) RVE – I (b) RVE-II (c) RVE-III. Results are shown for analysis case with micro-voids (red) and without micro-voids (blue) present within the bead's microstructure.

Note that the analytical MT calculated CTE values depend on the accuracy of the calculated stiffness tensor $\bar{C}_{mn}$ in Table 4.3 and appear to improve with increasing RVE size dropping to about 2.0% for the non-porous composite and 4.6% for the porous composite. Similar to the numerical FE results, porosity appears to the same effect on the analytical MT results as that seen with the numerical FE data where effective CTE are reduced less than 1% when micro-voids are considered.

Table 4.9: Computed normalized $\bar{\alpha}^{eff}$ and the associated relative error $\bar{\epsilon}$ [%] in the computed effective CTE magnitude of ROI-II between the numerical FE homogenized and Mori-Tanaka's (MT) analytical results for all RVE sizes (RVE - I, II, and III).

| | non-porous | | | porous | | |
|---|---|---|---|---|---|---|
| $\bar{\alpha}^{eff}$ | I | II | III | I | II | III |
| FE | *0.81* | *0.79* | *0.78* | *0.79* | *0.76* | *0.75* |
| MT | *0.85* | *0.82* | *0.79* | *0.84* | *0.81* | *0.79* |
| $\bar{\epsilon}$ [%] | *5.45* | *3.85* | *2.00* | *7.02* | *6.07* | *4.59* |

Based on the results of the calculated coefficient of variation $\xi$ for the effective CTE magnitude $\bar{\alpha}^{eff}$ computed from the different realizations of the various RVEs of ROI-II presented in Table 4.10 the $\xi$ values are much lower compared to corresponding $\xi$ values computed for the effective modulus (cf. Table 4.7), with $\xi \leq 4\%$ for the non-porous RVE and $\xi \leq 5.5\%$ for the porous RVE. Like the effective modulus, the $\xi$ values for the effective CTE are relatively higher for the porous RVE compared to the non-porous RVE. Additionally, the $\xi$ values from the numerical FE approach are relatively higher than the analytical MT values for RVE-I and RVE-II but relatively lower for RVE-III.

Table 4.10: Coefficient of variation $\xi$ [%] in the normalized effective CTE magnitude ($\bar{\alpha}^{eff}$) for all RVE cases (I, II and III) of ROI-II and for Mori-Tanaka's analytical model and numerical FE homogenization schemes.

| $\xi$[%] | non-porous | | | porous | | |
|---|---|---|---|---|---|---|
| | I | II | III | I | II | III |
| FE | *4.00* | *2.28* | *1.12* | *5.36* | *3.16* | *1.61* |
| MT | *2.89* | *1.86* | *1.91* | *3.23* | *2.03* | *1.99* |

The influence of micro-voids on the effective CTE is shown to be minimal in these results. The impact studies may thus be more relevant to the effective composite stiffness and thermal conductivity.

*4.6.1.3 Effective Thermal Conductivity.* The results of the effective TC of the 13% CF/ABS composite based on the numerical and analytical FE homogenization evaluation procedures described in Sections 4.2.3 and 4.3.3 respectively are reported here. We first consider a non-porous analysis of RVE-II realization #14 from ROI-II subjected to thermal gradient along each of the three principal references axes. The thermal gradients are prescribed on the bounding entities of the RVE through the set of PBC equations given in Eqn.(4.1)-(4.3). The material properties for the micro-constituents used for the analysis are also provided in Table 4.1. Single-step steady-state heat transfer analysis is carried out and the output of the homogenized heat fluxes and temperature gradients computed from the FE analysis for the three different load cases are used to compute the effective thermal conductivity tensor using the numerical FE homogenization procedures detailed in Section 4.2.3.

Temperature contours of three separate simulations appear in Figure 4.10. The plots show a non-uniform distribution of the temperature gradients along the principal coordinate axes due to the microstructural heterogeneity across the composite sample coupled with the relatively high contrast in the isotropic thermal conductivity between fiber and matrix phases ($\bar{\kappa}^f \sim 17.5$). Table 4.11 shows the principal components of the predicted effective TC tensor for all three RVE sizes appearing in Figure 4.2, for both non-porous and porous composite microstructures. The effective mean component values of TC from all RVEs considered are seen to be within close range of each other (ranging from 0% to a maximum of 1.5% discrepancy between consecutive RVE values). The micro-voids are seen to reduce the computed effective TC values by 10% - 12%. The maximum component of the conductivity tensor (i.e. $\bar{\kappa}_{33}$) is observed to coincide with the component of maximum average fiber orientation (i.e. $\hat{a}_{33}$).

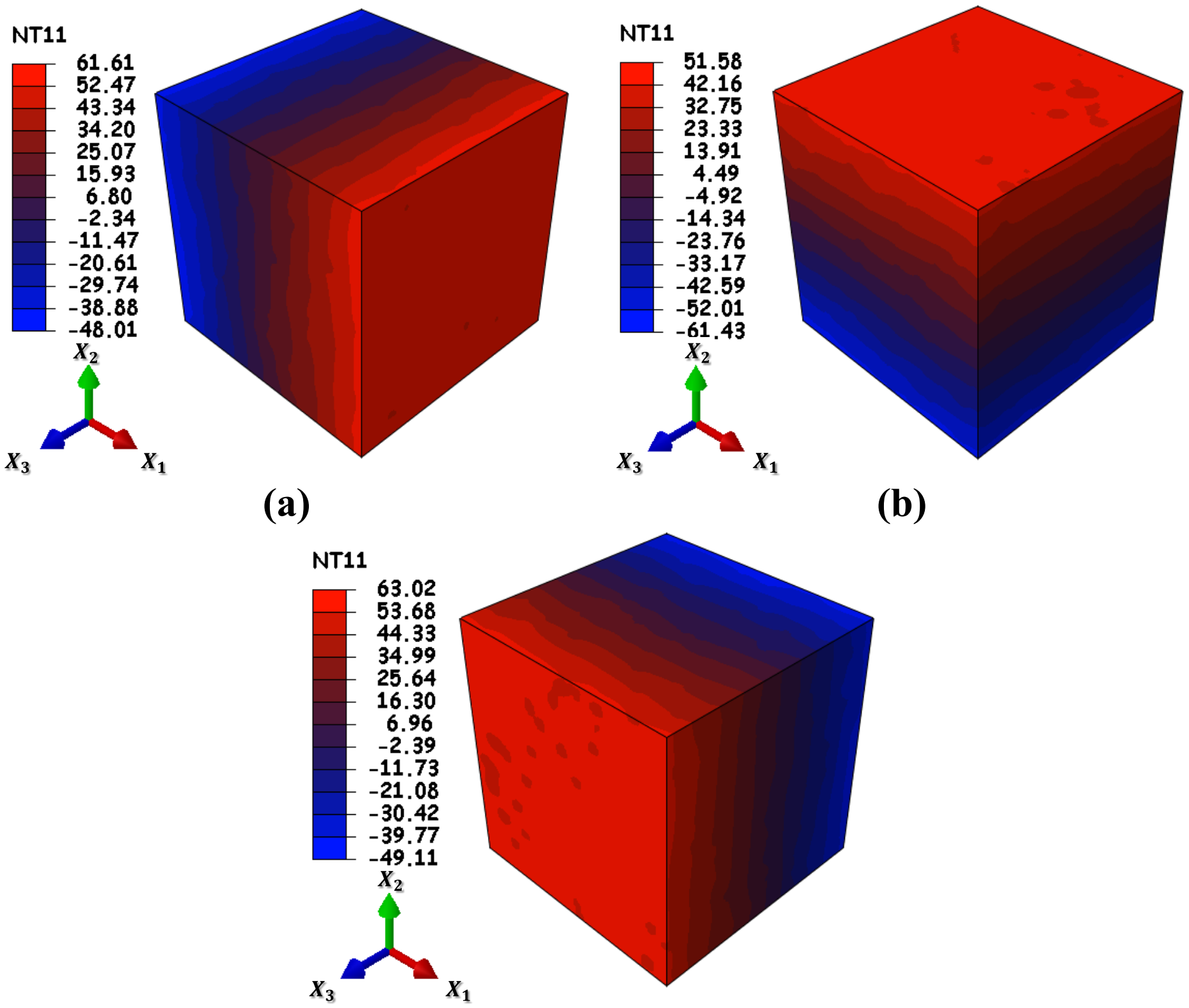

**(c)**

Figure 4.10: Temperature contours of RVE-II, realization #14 from ROI-II for three thermal load conditions: (a) thermal gradient in $x_1$-direction (b) thermal gradient in $x_2$-direction (c) thermal gradient in $x_3$-direction. NT11 is the temperature in [$^0C$].

Table 4.11: Mean values of effective TCs for all RVE cases (RVE-I, II, and III) of ROI-II computed with the numerical FE homogenization scheme.

| RVE | non-porous | | | porous | | |
|---|---|---|---|---|---|---|
| type | $\bar{\kappa}_{11}$ | $\bar{\kappa}_{22}$ | $\bar{\kappa}_{33}$ | $\bar{\kappa}_{11}$ | $\bar{\kappa}_{22}$ | $\bar{\kappa}_{33}$ |
| I | *1.39* | *1.32* | *1.47* | *1.22* | *1.15* | *1.31* |
| II | *1.40* | *1.33* | *1.49* | *1.24* | *1.16* | *1.33* |
| III | *1.41* | *1.33* | *1.50* | *1.24* | *1.16* | *1.35* |

From the error-bar plots of Figure 4.11 we see a clear reduction in the dispersion of quantities as the RVE size increases from RVE-I (cf. Figure 4.11a) to RVE-III (cf. Figure 4.11c). The error-bar shrinks considerably for largest RVE case, i.e. RVE-III, although the mean values for all three RVE cases are within close range to each other with maximum observed discrepancy in all quantities between consecutive RVE sizes for both porous and non-porous microstructural considerations dropping from about 1.8% between RVE-I and II to a value of only 0.8% between RVE II and III. The results of the effective TC magnitude $\bar{\kappa}^{eff}$ presented in Table 4.12 below shows very good agreement between the values obtained from both numerical (FE) homogenization and the Mori-Tanaka (MT) analytical methods with a maximum discrepancy of only 3.5% recorded for the smallest sized RVE (RVE-I) which reduces with increasing RVE size to about 1% for RVE-III.

Although there is minimal differences in computed effective TC values among the various RVE sizes in Table 4.11, results of the dispersion in the measured effective TC appearing in Table 4.13 shows that the coefficient of variation $\xi$ decreases with increasing RVE size as observed with other effective quantities and falls below 3% for both the porous and non-porous RVE-II micro-structures, computed from both numerical FE and analytical MT methods. The $\xi$ values for the effective TC are seen to be relatively lower than the values computed for the effective moduli and in some cases slightly higher than those computed for the effective ETC. This is because determination of sufficient RVE size is known to be dependent on the property being evaluated [119]. Although we observe less dispersion in the computed effective TC quantities for RVE-III ($\xi < 2\%$), the gain in accuracy does not measure up to the added cost of computation due to the increased RVE size and associated mesh points.

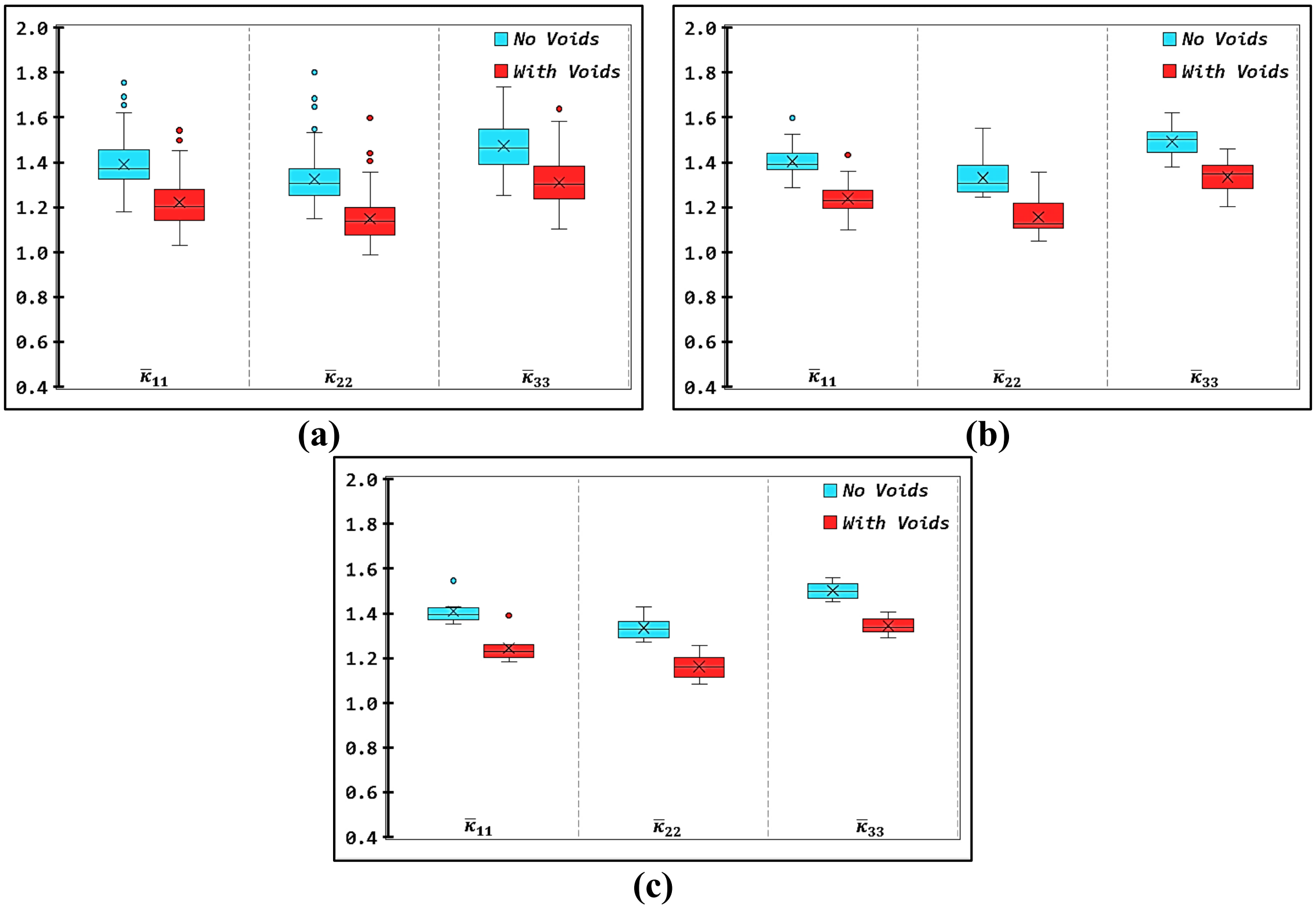

Figure 4.11: Box plot showing the mean values ('x' markers), median values (horizontal lines) interquartile intervals (box limits), extremum (whiskers) and outliers (isolated points) of the diagonal components of the normalized effective TC tensor from the RVE realizations for (a) RVE–I (b) RVE-II and (c) RVE-III. Results are shown for analysis case with micro-voids (red) and without micro-voids (blue) present within the bead's microstructure.

Table 4.12: Relative error $\bar{\epsilon}$ [%] in the predicted effective TC magnitude ($\bar{\kappa}^{eff}$) of ROI-II between the numerical FE homogenization scheme and Mori-Tanaka's (MT) analytical result for all RVE sizes (RVE - I, II, and III).

| $\bar{\kappa}^{eff}$ | non-porous | | | porous | | |
|---|---|---|---|---|---|---|
| | I | II | III | I | II | III |
| FE | *1.40* | *1.41* | *1.42* | *1.23* | *1.24* | *1.25* |
| MT | *1.35* | *1.38* | *1.40* | *1.18* | *1.22* | *1.24* |
| $\bar{\epsilon}$ [%] | *3.48* | *1.79* | *0.98* | *3.46* | *1.95* | *1.16* |

Table 4.13: Coefficient of variation $\xi$ [%] in the predicted normalized apparent ETC ($\bar{\kappa}^{eff}$) of ROI-II for both Mori-Tanaka's analytical model and numerical FE homogenization schemes and for all RVE cases (I, II and III).

| $\xi$ | non-porous | | | porous | | |
|---|---|---|---|---|---|---|
| | I | II | III | I | II | III |
| FE | *5.63* | *2.92* | *1.78* | *5.49* | *2.80* | *1.83* |
| MT | *4.85* | *2.59* | *1.83* | *4.86* | *2.47* | *1.93* |

### *4.6.2 Thermo-Mechanical Properties Over the Bead Cross Section*

To better understand the influence of the spatial variation in the bead microstructure on the effective thermo-mechanical properties across the LAAM bead, analysis is performed by selecting a single characteristic RVE instance of type II (see e.g., RVE-II in Figure 4.2) from all four ROIs in Figure 3.8 (taken from [241]). In Section 4.6.1 it was established that RVE-II provides reasonable values of effective properties of the 13% CF/ABS SFRP as compared to accepted analytical calculations. We assume that if the average value of a mathematical descriptor that defines the microstructural characteristics of a RVE instance chosen from a particular ROI volume deviate minimally from the corresponding value of the overall ROI volume, then the average effective properties of the characteristic RVE instance should also deviate minimally from the overall ROI volume. We define the deviation $\delta Z_j$ in the effective property $Z$ in terms of the effective property of the $j^{th}$ RVE realization $Z_j^{RVE}$ and that of its ROI $Z^{ROI}$ as

$$\delta Z_j = \frac{Z_j^{RVE} - Z^{ROI}}{Z^{ROI}}, \qquad Z = \vartheta_f,\ \ \vartheta_v,\ \ \bar{E}^{eff},\ \ \bar{\alpha}^{eff},\ \ \bar{\kappa}^{eff} \tag{4.65}$$

We also consider a measure of the deviation in the average fiber orientation $\delta p_j$ written in terms of the average fiber orientation of the $j^{th}$ RVE realization $\underline{\hat{p}}_j^{RVE}$ and that of its ROI $\underline{\hat{p}}^{ROI}$ as

$$\delta p_j = 1 - \cos \delta\varphi_j = 1 - \frac{\underline{\hat{p}}_j^{RVE} \cdot \underline{\hat{p}}^{ROI}}{\left|\underline{\hat{p}}_j^{RVE}\right| \left|\underline{\hat{p}}^{ROI}\right|} \tag{4.66}$$

$\delta p_j$ can be physically interpreted as a measure of the angular deviation between average fiber orientation vector of the $j^{th}$ RVE, $\underline{\hat{p}}_j^{RVE}$ and the average fiber orientation vector of its associated ROI, $\underline{\hat{p}}^{ROI}$. A value of $\delta p_j = 0$ implies $\underline{\hat{p}}_j^{RVE}$ and $\underline{\hat{p}}^{ROI}$ are perfectly aligned while $\delta p_j = 1$ implies both vectors are completely misaligned. As would become evident in the sections to follow, we show that the deviation in the computed effective properties between an RVE realization and its associated ROI diminishes with the deviation in the microstructural characteristics. The purpose of establishing a valid relationship between the ROI/RVE effective properties and the inherent microstructural characteristics is to determine a means for estimating the effective properties of the ROI using an optimum RVE realization with matching microstructural characteristics as its associated ROI volume. For this exercise, ROI-II region and RVE-II realizations are utilized. The 3D regression plots of Figure 4.12 show the relationship between the deviation in effective thermal-mechanical properties with the deviations in the volume fractions and average fiber orientation for the various non-porous RVE-II realizations within ROI-II. In Figure 4.12 (a), the deviation in the effective modulus $\delta \bar{E}^{eff}$ has a minimum value of $\delta \bar{E}_{24}^{eff} = 0.03\%$ and a maximum $\delta \bar{E}_{15}^{eff} = 8.12\%$. We note that RVE-II, realization #24 has a fiber volume fraction $\delta \vartheta_{f_{24}} = 0.13\%$ corresponding to the minimum $\delta \vartheta_f$ for ROI-II while $\delta \vartheta_{f_{15}} = 16.68\%$ corresponding to the maximum $\delta \vartheta_f$ for ROI-II . In addition, $\delta p_{24} = 0.004$ (or $\delta \varphi_{24} = 5.33^o$) for the RVE-II (cf. Eqn. (4.66)), realization #24 which has the minimum $\delta \bar{E}^{eff}$. Note, however, that $\delta \varphi_{24}$ is not the minimum value of $\delta \varphi$ over the complete set of RVE-II realizations in ROI-II which occurs at realization #7 with a value of $\delta \varphi_{min} = 1.46^o$.Further, $\delta \varphi$ reaches a maximum value of $\delta \varphi_{max} = 16.20^o$ for realization #16. To determine the RVE realization having the smallest deviation overall, we define $\delta \Upsilon^{eff} = [\, \delta \bar{E}^{eff} + \delta \bar{\alpha}^{eff} + \delta \bar{\kappa}^{eff}]/3$ which has a minimum value for RVE-II, realization #24 which is also has a minimum value of $\delta \vartheta_f$. The associated values of the deviation in the effective properties (cf. Figure 4.12) are well below 1% (i.e., $\delta \bar{\alpha}_{24}^{eff} =$ $0.33\%,\ \delta\bar{\kappa}_{24}^{eff} = 0.11\%$). The relationships for the regression lines in for the Figure 4.12 are given as

$$0.5210 - 2.7837\delta\vartheta_f = 1.2299 - 96.3324\delta p = 0.5159 - 6.6355\delta\bar{E}^{eff} \quad (4.67)$$
$$0.5111 - 2.7862\delta\vartheta_f = 1.2400 - 99.0200\delta p = 0.5060 - 8.6749\delta\bar{\alpha}^{eff} \quad (4.68)$$
$$0.5144 - 2.7883\delta\vartheta_f = 1.2542 - 100.4985\delta p = 0.5093 + 11.8825\delta\bar{\kappa}^{eff} \quad (4.69)$$

At the minimum point of the regression lines for the deviation in all effective properties where $\delta\bar{E}^{eff} = \delta\bar{\alpha}^{eff} = \delta\bar{\kappa}^{eff} = 0$ (cf. Figure 4.12), the deviation in the fiber volume fraction $\delta\vartheta_f$ approaches zero ($\delta\vartheta_f = 0.18\%$), however the deviation in the average fiber orientation is small but not zero (i.e., $\delta p = 0.007$ or $\delta\varphi = 6.98^o$). RVE-II, realization #24 has the closest matching microstructural characteristics to the minimum point of the regression line amongst other instances in ROI-II volume. Minimization of $\delta\vartheta_f$ takes precedence to minimizing $\delta p$ or $\delta\varphi$ in selecting an appropriate RVE instance from a given ROI based on the foregoing arguments. .On the contrary, minimizing $\delta p$ prior to minimizing $\delta\vartheta_f$, gives a non-optimal RVE instance (RVE-II, #7) that yields unnacceptable objective function values, i.e. at $\delta p_{min} = 0.0003$, or $\delta\varphi_{min} = 1.46^o, \delta\vartheta_f = 15.95\%$, $\delta\bar{E}_7^{eff} = 5.69\%$, $\delta\bar{\alpha}_7^{eff} = 3.01\%$, and $\delta\bar{\kappa}_7^{eff} = 5.03\%$ which shows significant deviations in the microstructural characteristics between RVE-II, realization #7 and the entire ROI-II region. Moreover, the effective properties considered here depend strongly on the fiber volume fraction and very weakly on the average fiber orientation as noted elsewhere [117], [128]. Accordingly, we select characteristic RVE-II realizations from the all ROI volumes following the approach of matching RVE-ROI microstructural characteristics and present values of the deviation in the microstructural characteristics over the ROI in Figure 4.13.

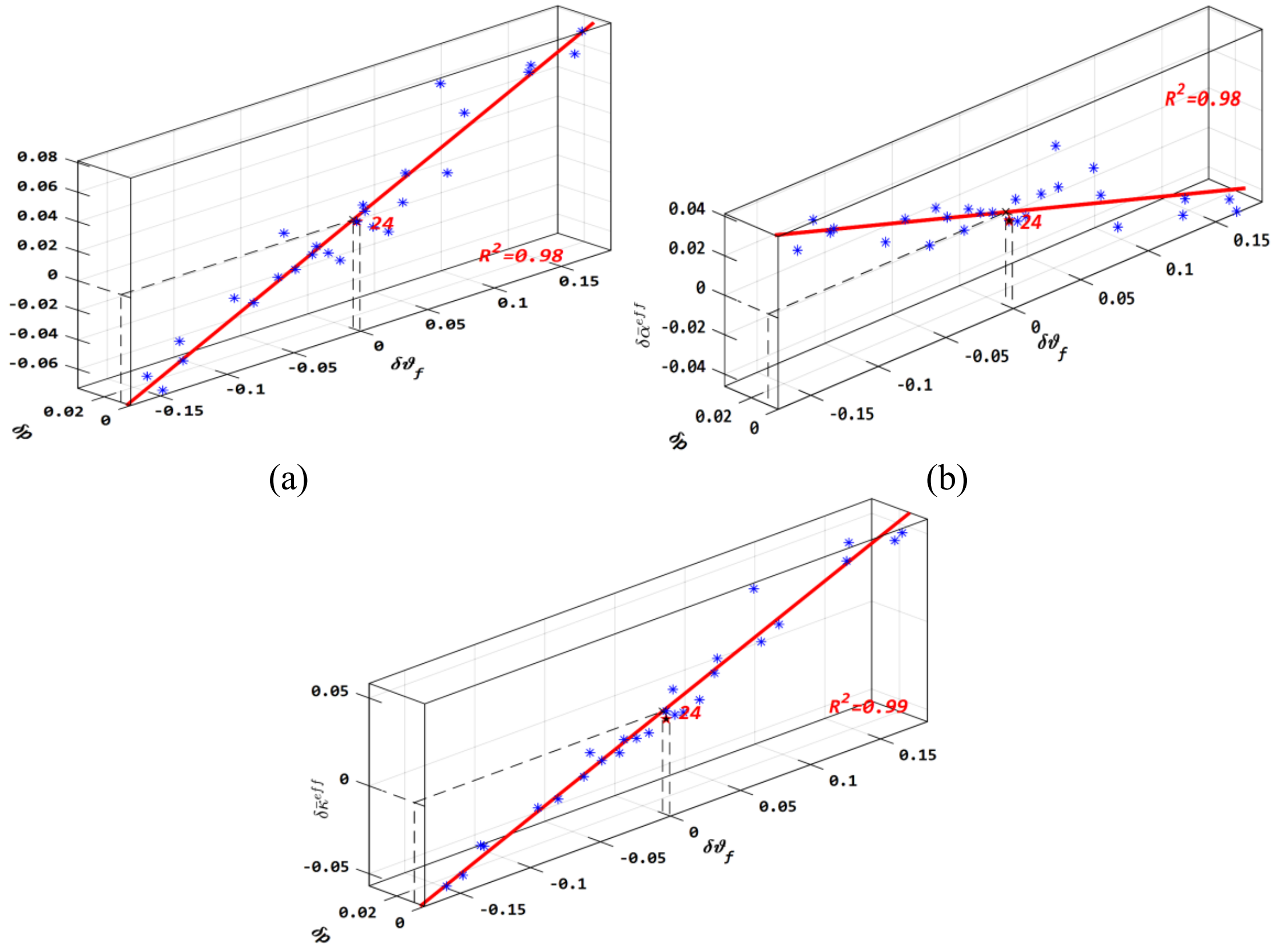

(c)

Figure 4.12: 3D correlation plot between the distribution of the deviation in the fiber fraction $\delta\vartheta_f$ and orientation vector $\delta\rho$ versus the deviation in the (a) apparent effective modulus, $\delta\bar{E}^{eff}$ (b) apparent ECTE, $\delta\bar{\alpha}^{eff}$ and (c) apparent ETC, $\delta\bar{\kappa}^{eff}$, for the various RVE-II instances of ROI-II.

The minization sequence is important for RVE realizations with marginally close minimum $\delta\vartheta_f$ values but distant $\delta p$ values. For instance, in ROI-IV (cf. Figure 4.13d), RVE-II, #3 and RVE-II, #16 both have $\delta\vartheta_f$ values less than 0.5% (i.e. 0.33% and 0.15% respectively), however RVE-II, #3 has a value of $\delta p_3 = 0.0024$ or $\delta\varphi_3 = 3.98^o$ while RVE-II, #16 has a value of $\delta p_{16} = 0.0194$ or $\delta\varphi_{16} = 11.31^o$ which indicates high misalignment between the RVE-II, #16 and ROI-IV average fiber orientation. Consequently, RVE-II, #3 with a reasonbly low $\delta\varphi$ is selected. The selected RVE-II realizations for the each ROI volume are (a) ROI-I, RVE-II, realization #14 (b) ROI-II, RVE-II, realization #24 (c) ROI-III,

RVE-II, realization #23 (d) ROI-IV, RVE-II, realization #3. The above RVE-II realizations are selected from their respctive ROI since they have nearly matching microstructural characteristics as their respectic ROI volumes.

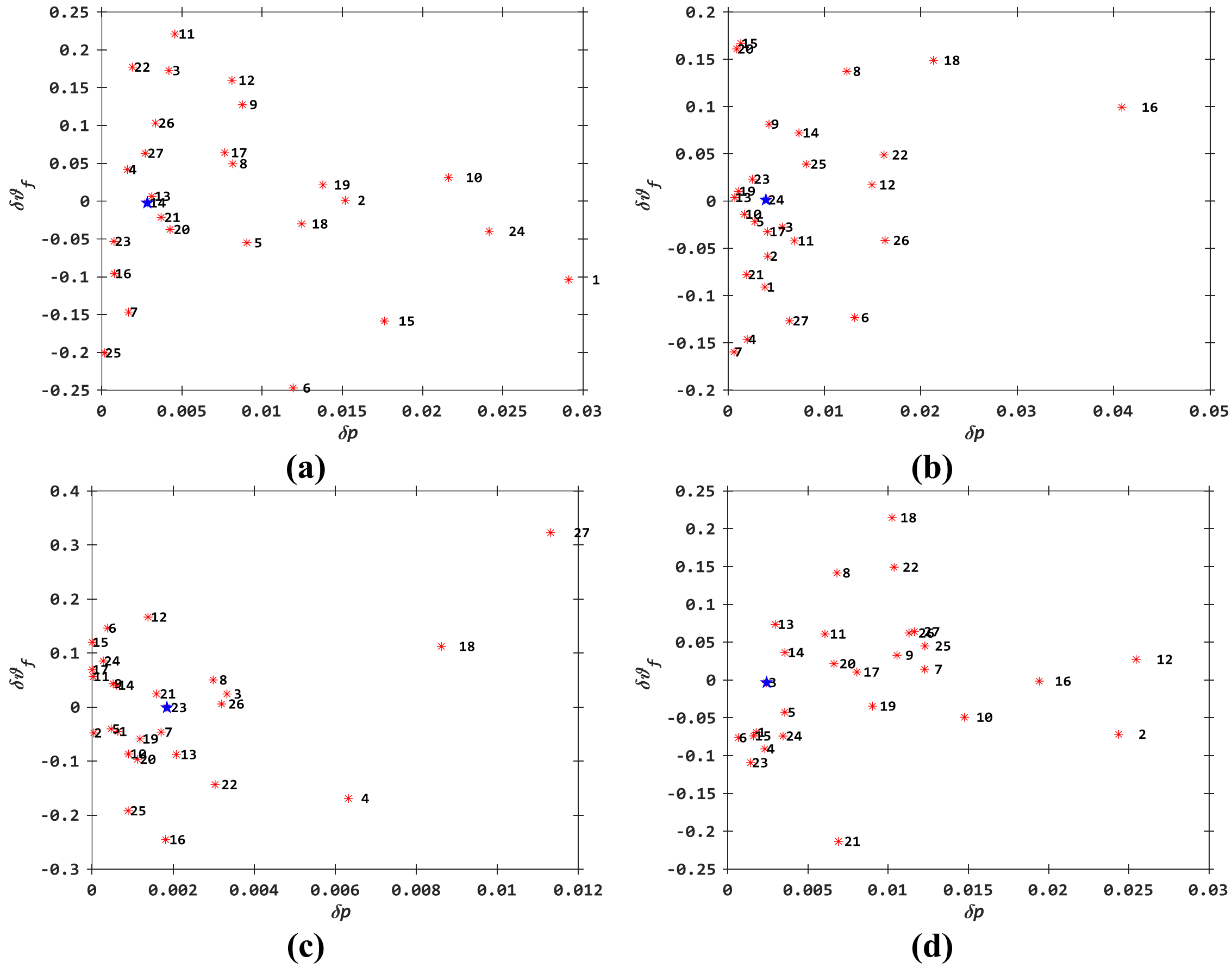

Figure 4.13: Deviation in the fiber volume fraction and average fiber orientation for the RVE-II realizations of each ROI: (a) ROI-I, (b) ROI-II, (c) ROI-III, and (d) ROI-IV for the non-porous RVE microstructure.

The resulting deviations in the computed microstructural parameters for the RVE realizations in Figure 14 are presented in Table 4.14. For all selected cases, the minimum deviation in the fiber volume fraction $\delta\vartheta_f$ is seen to be below 0.35% and the minimum deviation in the average fiber orientation vector $\delta\varphi$ is seen to be below $5.5^o$.

Table 4.14: Deviation in microstructural parameter values for the non-porous RVEs of each ROI: (a) ROI-I, RVE-II, realization #14 (b) ROI-II, RVE-II, realization #24, (c) ROI- III, RVE-II, realization #23, and (d) ROI- IV, RVE-II, realization #3.

| | (a) | (b) | (c) | (d) |
|---|---|---|---|---|
| $\delta\vartheta_f$ [%] | -0.23 | 0.13 | -0.10 | -0.33 |
| $\delta\vartheta_v$ [%] | 3.73 | -10.26 | 18.92 | -9.31 |
| $\delta\varphi$ [deg] | 4.45 | 5.33 | 3.86 | 3.56 |

Figure 4.14 shows the microstructure of selected RVE-II realizations from the four ROI volumes obtained from 3D X-ray μCT imaging and subsequent image processing given in [241]. The estimated average fiber volume fractions $\vartheta_f$ and components of the second order fiber orientation tensor for the various ROI - RVE's are presented in Table 3.5. Visual inspection of the microstructures given in Figure 4.14 support the reported values in Table 3.5. For example, ROI-III, RVE-II, realization #23 and ROI- IV, RVE-II, realization #3 appear more densely packed than the other RVE volumes, hence their high fiber volume fractions. Likewise, ROI- III, RVE-II, realization #23 is seen to have a relatively high alignment in the $x_3$-direction appearing almost vertical, as compared to other RVE realizations shown.

Effective properties of the RVE realizations from each ROI shown in Figure 4.14 for the non-porous microstructure computed using the FE homogenization procedure in Section 4.2 are presented in Table 4.16. The results of the predicted quantities computed based on selected RVE-II realizations are expected to be within range of the predicted effective properties for their associated ROI volumes computed from the complete set of RVE-II realizations since the RVE/ROI deviations in the relevant microstructural properties (cf. Table 4.14) are seen to be as low as reasonably possible (i.e. $|\delta\vartheta_f| \leq 0.33\%$, $\delta\varphi \leq 5.33^0$). The reported effective quantities for the non-porous microstructure of the various single characteristics RVE-II instances from their respective ROIs in Table 4.16 were found to be comparable with average values of the effective quantities obtained from the overall set of realizations of RVE-II for each ROI with a maximum observed discrepancy of only 1.7%. As expected, the results show that the effective modulus $\bar{E}^{eff}$ and thermal conductivity $\bar{\kappa}^{eff}$ increase with increasing fiber volume fraction and increasing degree of fiber alignment with the print direction (z-axis) across the different bead regions. Conversely, the effective CTE $\bar{\alpha}^{eff}$ is seen to decrease with increasing fiber volume fraction and increasing fiber alignment with the print direction.

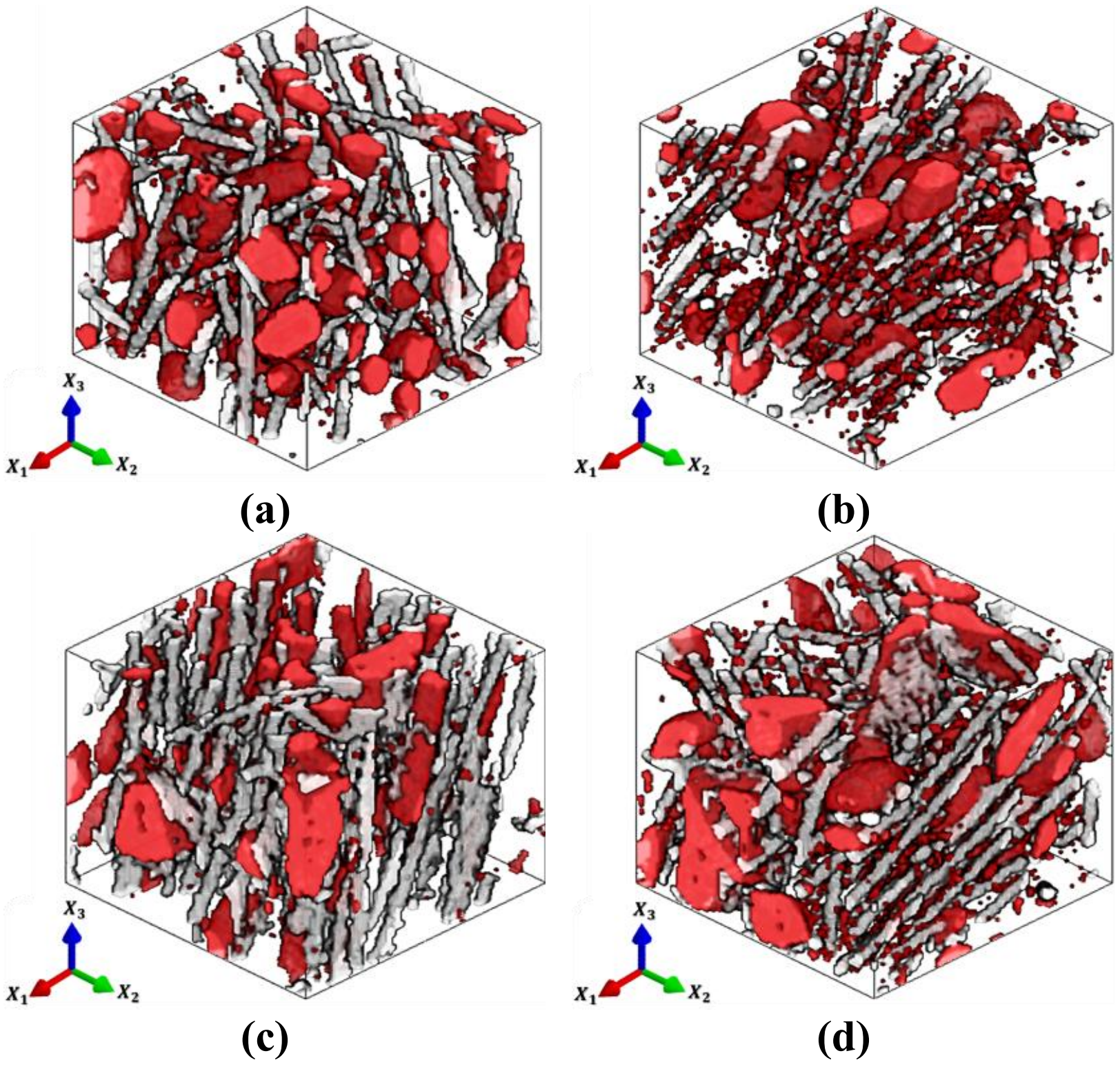

**(a)** **(b)** **(c)** **(d)**

Figure 4.14: 3D-μCT volumes showing microstructure (fiber - gray, voids – red) of select RVE-II realizations of each ROI volume: (a) ROI-I, RVE-II, realization #14 (b) ROI- II, RVE-II, realization #24, (c) ROI- III, RVE-II, realization #23, and (d) ROI- IV, RVE-II, realization #3

Table 4.15: Average values of the fiber volume fraction and orientation tensor components for the select RVE-II realizations: (a) ROI-I, RVE-II, realization #14 (b) ROI- II, RVE-II, realization #24, (c) ROI- III, RVE-II, realization #23, and (d) ROI- IV, RVE-II, realization #3

| Parameters | (a) | (b) | (c) | (d) |
|---|---|---|---|---|
| $\vartheta_f$ [%] | *6.95* | *6.66* | *7.24* | *7.51* |
| $\vartheta_v$ [%] | *11.10* | *11.01* | *11.96* | *10.13* |
| $\langle p_1 p_1 \rangle$ | *0.34* | *0.25* | *0.08* | *0.22* |
| $\langle p_2 p_2 \rangle$ | *0.04* | *0.23* | *0.08* | *0.20* |
| $\langle p_3 p_3 \rangle$ | *0.62* | *0.52* | *0.84* | *0.58* |

Table 4.16: (a) Computed values of effective thermo-mechanical properties for selected non-porous RVE-II realizations: (a) ROI-I, RVE-II, realization #14 (b) ROI-II, RVE-II, realization #24, (c) ROI- III, RVE-II, realization #23, and (d) ROI- IV, RVE-II, realization #3.

| | (a) | (b) | (c) | (d) |
|---|---|---|---|---|
| $\bar{E}^{eff}$ | *1.54* | 1.52 | *1.55* | *1.60* |
| $\bar{\alpha}^{eff}$ | *0.76* | 0.79 | *0.77* | *0.74* |
| $\bar{\kappa}^{eff}$ | *1.43* | 1.41 | *1.45* | *1.47* |

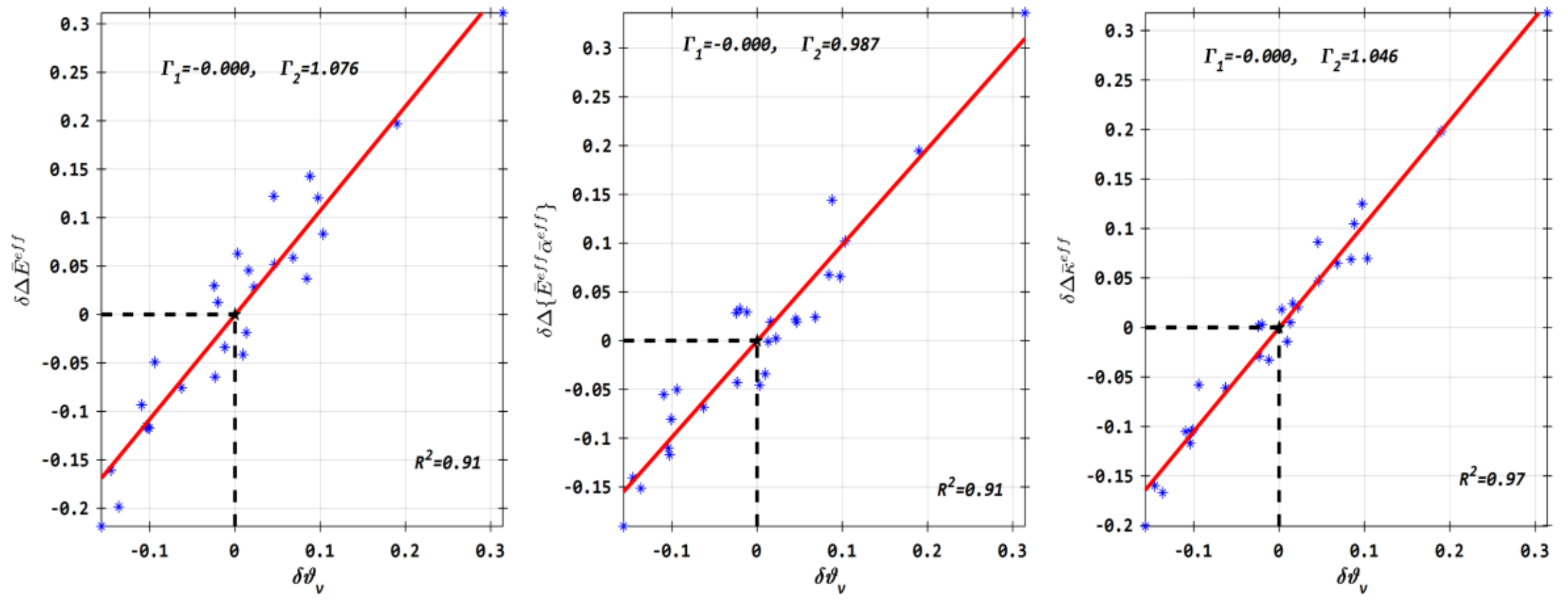

(a) (b) (c)

Figure 4.15: Correlation between the deviation in the (a) net effective modulus (b) net product of the effective modulus and thermal expansion coefficient (c) net thermal conductivity; versus the deviation in the void fraction for the various RVE-II instances of ROI-II volume.

Once the effective properties of the different non-porous ROIs have been computed using relevant RVE instances, the effect of the inherent porosity on the properties can be approximated using established linear relationships that correlate the deviation in the effective properties difference, $\delta\Delta Z$ due to the inherent porosity to the deviation in volume fraction of the porosity, $\delta\vartheta_v$ between the selected RVE instances and their respective ROI volumes. i.e.

$$\delta\Delta Z_j = \frac{\Delta Z_j^{RVE} - \Delta Z^{ROI}}{\Delta Z^{ROI}}, \qquad \Delta Z = \Delta\bar{E}^{eff},\ \Delta\bar{\alpha}^{eff},\ \Delta\bar{\kappa}^{eff} \tag{4.70}$$

where $\Delta Z = Z^{porous} - Z^{non-porous}$ is the difference in the magnitude of effective property $Z$ between the porous and non-porous microstructure. The inherent porosity is assumed to impact primarily the integrity of the polymer matrix and the contribution of its properties to the overall behavior of the composite material. Accordingly, developed linear relationship, $\delta\Delta Z = \Gamma_1 + \Gamma_2\delta\vartheta_v$ in Figure 4.15 (a-c) for ROI-II would apply to other ROI regions across the entire bead section. The ECTE values are weighted with their corresponding elastic modulus values when developing it's linear relationship. Given the difference between the porous and non porous effective properties for the characteristic RVE-II instances of the various ROI volumes, $\Delta Z^{RVE}$ (cf. Table 4.17), we can backtrack the associated difference between the porous and non porous effective properties of the various ROI volumes, $\Delta Z^{ROI}$ from Eqn. (4.65) using the linear relationships presented in Figure 4.15 (a-c). Consequently, given $\Delta Z^{ROI}$ and the effective properties of the non-porous microstructure of the various ROI volumes $Z^{ROI}$ (cf. Table 4.16), we compute the approximate effective properties of the porous microstructure of the various ROI volumes (cf. Table 4.18). Again, the evaluated effective quantities for the porous microstructure presented in Table 4.18 based on volume information of the various single characteristics RVE-II instances from their respective ROI were found to match closely with the mean values of the effective quantities obtained from the complete set of RVE-II realizations for each ROI with a maximum observed discrepancy of 2.5%. As expected, the porosity is observed to reduce the effective properties in the different ROI regions of the bead.

Table 4.17: (a) Estimated values of difference in the effective thermo-mechanical properties ($\Delta Z^{RVE}$) between the porous and non-porous microstructure for the selected RVE-II instances of the various ROI volumes (a) ROI-I, RVE-II, #14 (b) ROI-II, RVE-II, #24, (c) ROI- III, RVE-II, #23, and (d) ROI- IV, RVE-II, #3.

| | **(a)** | **(b)** | **(c)** | **(d)** |
|---|---|---|---|---|
| $\Delta\bar{E}^{eff}$ | *0.31* | *0.29* | *0.40* | *0.29* |
| $\Delta\bar{\alpha}^{eff}$ | *0.04* | *0.02* | *0.02* | *0.05* |
| $\Delta\bar{\kappa}^{eff}$ | *0.15* | *0.15* | *0.17* | *0.14* |

Table 4.18: (a) Approximate values of effective thermo-mechanical properties for the porous microstructure of the various ROI volumes (a) ROI-I, RVE-II, #14 (b) ROI- II, RVE-II, #24, (c) ROI- III, RVE-II, #23, and (d) ROI- IV, RVE-II, #3.

| | (a) | (b) | (c) | (d) |
|---|---|---|---|---|
| $\bar{E}^{eff}$ | *1.24* | *1.20* | *1.23* | *1.27* |
| $\bar{\alpha}^{eff}$ | *0.72* | *0.76* | *0.76* | *0.68* |
| $\bar{\kappa}^{eff}$ | *1.28* | *1.24* | *1.31* | *1.31* |

### *4.6.3 Effective Property Correlation Studies*

To better understand the variation of the effective properties across the 13% CF/ABS bead appearing in Figure 3.8, we compute outputs using the numerical homogenization procedure in Section 4.2 for all RVE-II realizations. Renderings of these realizations appear for the various ROI volumes in Figure 4.16. Correlations between computed quantities across the bead cross section for each ROI volume are presented here. Figure 4.17 presents the least squares linear fitted curve effective elastic modulus $\bar{E}^{eff}$ (cf. Eqn. (4.71)) as a function of average fiber volume fraction $\vartheta_f$ over all ROIs. Similarly, Figure 4.17b shows the correlation between the $\bar{E}_{33}$ elastic modulus (cf. Eqn. (4.72)) and $\hat{a}_{33}$ average fiber orientation tensor component in the print direction obtained from datasets for the non-porous RVE-II realizations for the various ROIs.

$$\bar{E}^{eff} = 0.9063 + 9.1362\, \vartheta_f \tag{4.71}$$
$$\bar{E}_{33} = 0.5536 + 2.4565\, \hat{a}_{33} \tag{4.72}$$

Results in Figure 4.17 indicate a good correlation exists between the fiber volume fraction and effective elastic modulus with a correlation coefficient $R^2 = 0.94$ which implies that the fiber volume fraction is an important microstructural parameter for predicting the modulus of SFRP composites, as expected. Similarly, there is a reasonable correlation between the $\bar{E}_{33}$ elastic modulus and $\hat{a}_{33}$ average fiber orientation tensor component with a correlation coefficient $R^2 = 0.75$, which suggests that the degree of fiber alignment with the print direction is a key indicator of this effective elastic property for our SFRP composite.

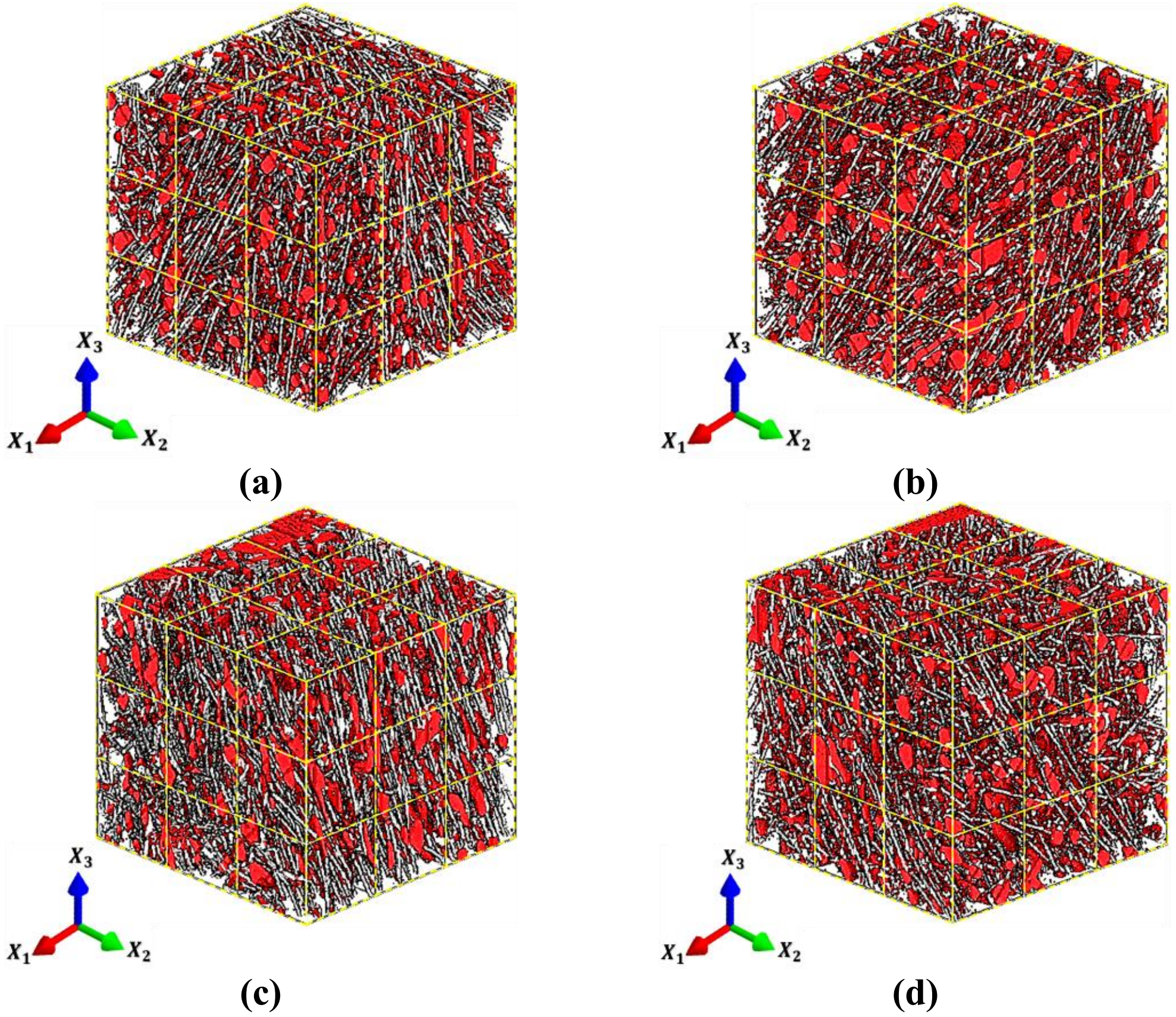

Figure 4.16: 3D-μCT volume view showing internal microstructure (fiber -gray, voids – red) of the various ROI volumes and the RVE-II instances for each: (a) ROI-I, (b) ROI- II, (c) ROI- III (d) ROI- IV

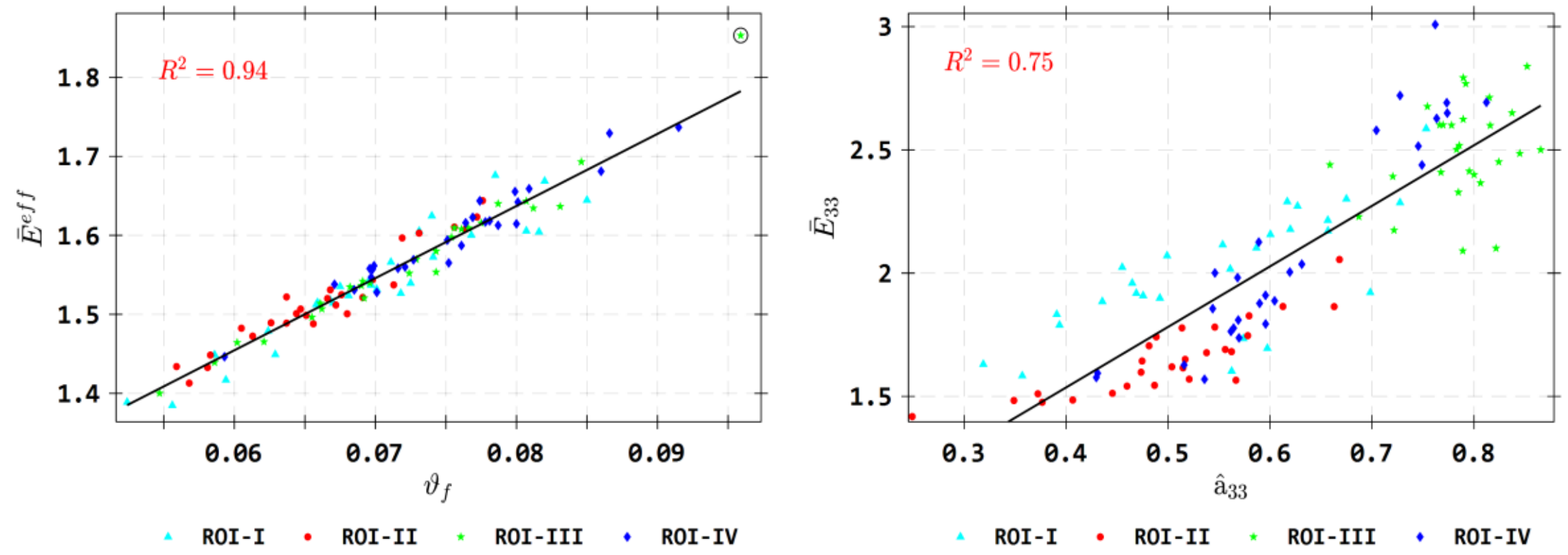

**(a)** **(b)**

Figure 4.17: Linear correlation plots between the (a) effective elastic modulus $\bar{E}^{eff}$ and the fiber volume fraction $\vartheta_f$ (b) elastic modulus component $\bar{E}_{33}$ and the average fiber orientation tensor component $\hat{a}_{33}$; for non-porous RVE-II instances of the various ROIs.

Linear correlations between effective CTE $\bar{\alpha}^{eff}$ and the average fiber volume fraction $\vartheta_f$ (cf. Eqn. (4.73)) for the non-porous RVE-II realizations for the various ROIs appears in Figure 4.18a which reveals an inverse relation exists where a higher volume fraction of particles yields a lower effective CTE, also as expected given the relatively low CTE of the carbon fiber material as expected (see e.g., [98], [131], [255]).

$$\bar{\alpha}^{eff} = 0.9918 - 3.1942\, \vartheta_f \quad (4.73)$$
$$\bar{\alpha}_{33} = 1.0445 - 0.8400\, \hat{a}_{33} \quad (4.74)$$

These results show a good correlation exists between $\bar{\alpha}^{eff}$ and $\vartheta_f$ with a correlation coefficient $R^2 = 0.81$. Similarly, a good correlation between the average fiber orientation component in the print direction $\hat{a}_{33}$ and the $\bar{\alpha}_{33}$ component (cf. Eqn. (4.74)) of effective CTE tensor ($R^2 = 0.85$) is seen. The degree of fiber alignment in the print direction is observed to vary inversely with the $\bar{\alpha}_{33}$ component of the effective CTE tensor along the same direction. Higher fiber alignment tends to allow for a higher packing density, giving more weight to the suspended carbon fiber properties.

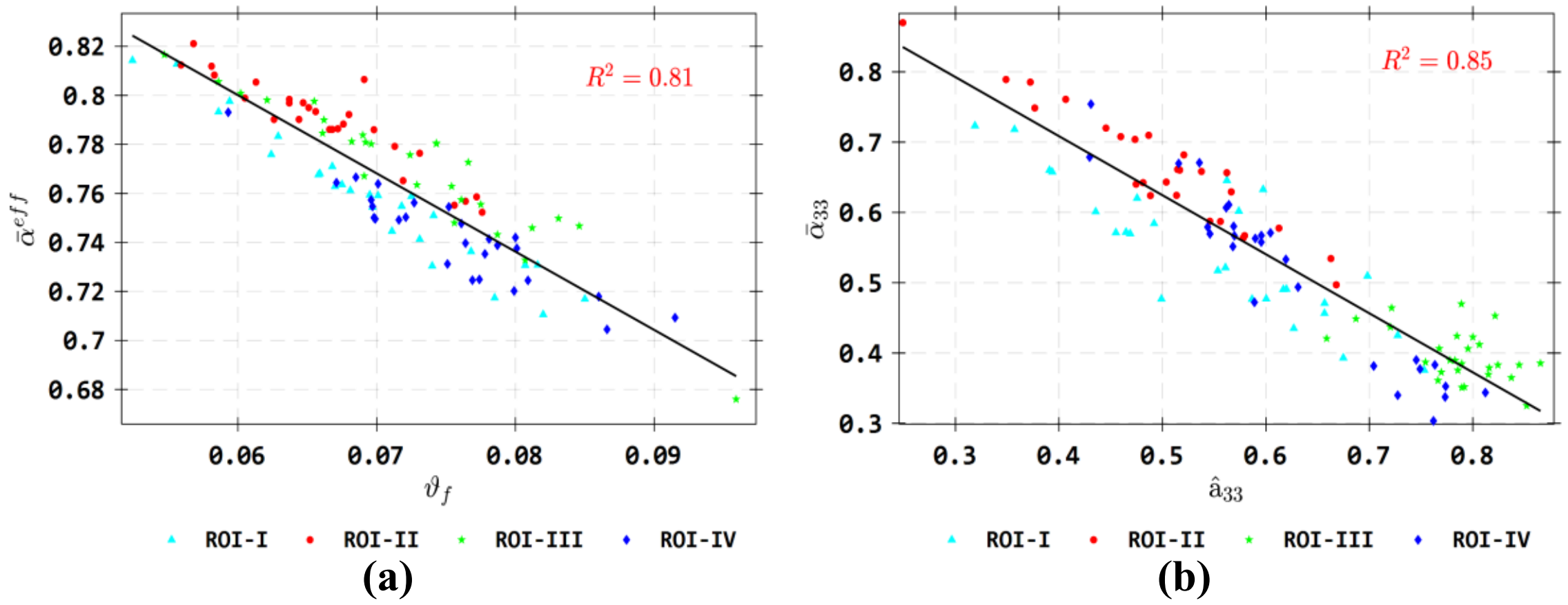

Figure 4.18: Linear correlation plots between (a) the apparent ECTE magnitude $\bar{\alpha}^{eff}$ with the fiber volume fraction $\vartheta_f$ (b) the ECTE tensor component $\bar{\alpha}_{33}$ and the average fiber orientation tensor component $\hat{a}_{33}$; for non-porous RVE-II instances of the various ROIs.

A near perfect correlation ($R^2 = 1.0$) is seen between fiber volume fraction $\vartheta_f$ and apparent effective TC magnitude $\bar{\kappa}^{eff}$ (cf. Eqn. (4.75)) for the non-porous RVE-II realizations from various ROIs. Conversely, a very weak correlation ($R^2 = 0.66$) is to occur between the average fiber orientation component in the print direction $\hat{a}_{33}$ and the $\bar{\kappa}_{33}$ component (cf. Eqn. (4.76)) of the effective TC tensor.

$$\bar{\kappa}^{eff} = 0.9548 + 6.8198\,\vartheta_f \tag{4.75}$$
$$\bar{\kappa}_{33} = 0.9815 + 0.8079\,\hat{a}_{33} \tag{4.76}$$

This implies that for the two-phase SFRP composite, the apparent effective TC $\bar{\kappa}^{eff}$ has a strong linear dependence on the volume fraction of the fiber reinforcement as was also identified in Tian et. al [250].

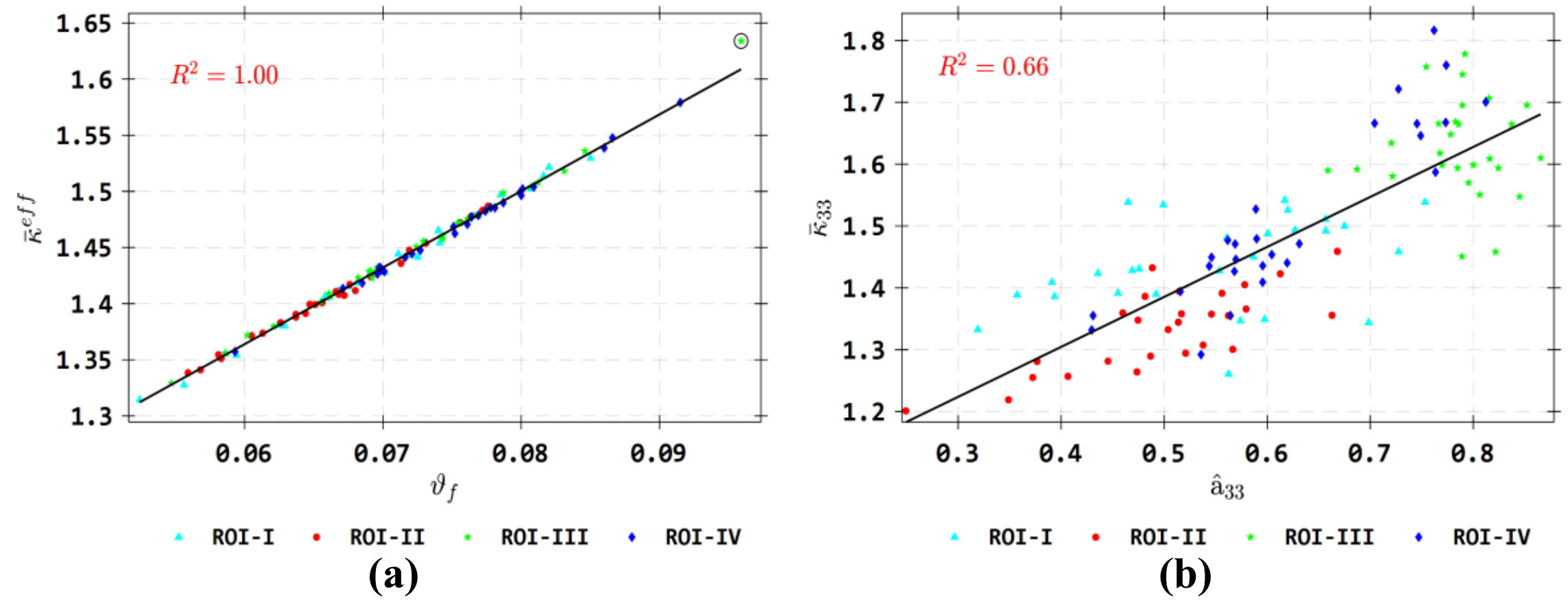

Figure 4.19: Linear correlation plots between (a) the apparent ETC magnitude $\bar{\boldsymbol{\kappa}}^{eff}$ and the fiber volume fraction $\vartheta_f$ (b) the ETC tensor component $\bar{\alpha}_{33}$ and the average fiber orientation tensor component $\hat{a}_{33}$; for non-porous RVE-II instances of the various ROIs.

## *4.7 Conclusion*

Numerical FE homogenization method was developed and applied to evaluate effective thermo-mechanical properties of a 13% CF-ABS SFRP composite where the material's microstructure was obtained using X-ray μ-CT characterization techniques that provided 3D voxel based realistic RVEs Sensitivity analysis was performed to determine a suitable computationally efficient RVE based on three RVE sizes and realization sets that yield effective properties with minimum dispersion in predicted quantities (less than 5%). Predicted effective properties obtained from our numerical FE approach were shown to agree well with estimated properties based on the Mori-Tanaka mean-field homogenizations technique. Assessment of the influence of micro-voids on the material behavior of the SFRP composite indicated an overall reduction in equivalent properties of the composite, as expected, where the effect of porosity on the effective CTE were minimal. Parameter dependent studies showed that the computed effective properties had a linear correlation with the fiber volume fraction and the average fiber orientation within the bead specimen consistent with literature [117], [127], [128]. The effective modulus and thermal conductivity were observed to vary proportionally with the fiber volume fraction and degree of fiber alignment with the print direction. Conversely the effective thermal expansion coefficient was observed to vary inversely with fiber volume fraction and degree of fiber alignment with the print direction. For increased simplicity and computational efficiency, a microstructure-based minimization approach that involves selection of a single characteristic RVE instance from a given realization set with nearly matching microstructural properties as the overall parent ROI region was used to obtain quick estimates of the effective thermo-mechanical properties across regions of the non-porous printed bead strand and with very small prediction error tolerance. Overall, the effective modulus and thermal conductivity were predicted to be higher at the edges and top surface of the print bead where the volume fraction and degree of fiber alignment with the print direction are seen to be highest, and the properties were lower in regions closer to the bead center with less densely packed and more randomly oriented fibrous microstructure. The opposite behavior was observed for the thermal expansion coefficient across the bead sections.

# CHAPTER FIVE

## Simulating Particle Motion in Viscous Homogenous Suspension Flow

Sections of this chapter are taken from: Awenlimobor, A. and Smith, D.E., 2024. Effect of shear-thinning rheology on the dynamics and pressure distribution of a single rigid ellipsoidal particle in viscous fluid flow. *Physics of Fluids*, *36*(12).

From the numerical evaluation performed in the preceding chapter and in line with numerous literature [113], [116], [131], the bead microstructure including the fibrous and porous structure significantly affect the resulting effective material properties and print quality. As such, understanding the mechanisms that are responsible for the development of the bead's microstructure, especially the micro-void formation mechanisms, is crucial. Presently, there is limited understanding on the known cause of micro-voids and the factors responsible for their formation in SFRP composites. Not until recently has there been increasing research interest in understanding mechanisms responsible for process-induced micro-void formation using computational-based simulation approach [57], [235], [241], [256]. Simulating the EDAM polymer composite melt flow-field process can provide valuable insight into potential mechanisms responsible for the microstructural development within the print beads, especially the micro-voids formation. Since intra-bead void nucleation is a localized (microscale) transport phenomena (occurring on the order of the fiber dimension) and is known to be heterogenous in nature forming at the particle-fluid interface, coupled multiscale simulation is required to accurately study this phenomenon. Previous studies [5], [9], [12], [13], [14], [15], [16], [17] have revealed that the local surrounding fluid pressure is a relevant process variable that significantly influences the nucleation of micro-voids in polymeric materials which itself depends on the particle dynamics. While previous numerical studies on fiber suspension flows have mainly focused on the particle dynamics that are mostly based on linear shear flow, the local flow-field surrounding the particle, including the velocity and pressure distribution has received very little attention. Moreover, existing studies that also investigated the pressure field surrounding a particle are based on flow analysis around fixed particle in space [185], [221], [257] that do not consider the influence of the particle's dynamics on the pressure distribution.

As in many previous works on short fiber composites, it is helpful to consider Jeffery [21] when studying the dynamics of short fiber suspensions. Jeffery's equation has been widely used to evaluate particle dynamics in viscous, low Reynolds number, Newtonian fluid flow. Most physical processes involving the flow of particle suspension like EDAM polymer composites melt flow process possess non-linear suspension rheology and contains arbitrary shaped deformable particles with complex hydrodynamic interactions which are unaccounted for in Jeffery's model. Recently, Jeffery's equation has been extended to capture various effects neglected by the assumptions made in his initial work such as the influence of fiber's shape and symmetry [258], [259], the effect of a fiber's flexibility/deformability [201], [202], [217], [260], the influence of neighboring particles in a concentrated suspension [19], [261], the effect of a non-Newtonian and visco-elastic fluid rheology [190], [194], [262], etc. In these prior studies, model advancement and application of Jeffery's equations has been primarily focused on particle dynamics, and has yet to be employed to better understand micro-void formation within print beads.

Moreover, the various extensions to Jeffery's equation have not specifically addressed the flow-field velocity and pressure surrounding the fiber surface during its motion.

The current chapter presents the 3D FEM model development which is used to investigate the effect of non-standard Jeffery's condition including the effect of generalized Newtonian fluid (GNF) rheology on the dynamics and surface pressure distribution of a single particle suspended in viscous homogenous flows. Firstly, we explore the effect of various factors such as the fibers geometric aspect ratio and initial fiber angle on the single particle motion and surface pressure distribution for a single particle suspended in Newtonian homogenous flow-field using Jeffery's equation. Typical size of particles encountered during Extrusion Deposition Additive Manufacturing (EDAM) polymer composite processing are on average hundreds of microns in magnitude depending on the particles concentration and system's scale, usually around $50 - 100\mu m$ for small scale EDAM systems and $\sim 300\mu m$ for large scale EDAM systems [263]. The rotary Peclet number that characterizes these polymeric melt flow through an EDAM nozzle are orders of magnitude high (i.e. $Pe_r \gg 1$). Brownian effects arising from particle interaction with the surrounding fluid molecules are thus insignificant and have been ignored in the current investigation since the hydrodynamic forces are expected to dominate the particle's motion. Jeffery's equations are a good starting point for studying particles behavior in these Newtonian flows. More rigorous stochastic statistical analysis accounting for Brownian disturbance such as that conducted by Leal et al. [264] and Zhang et al. [234] is a relevant study for future consideration. The generalized Newtonian FEA single fiber motion model development is a non-linear extension to the Newtonian formulations of Zhang et al. [230], [234], [265] and Awenlimobor et al. [57], [235] assuming a power-law non-Newtonian fluid behavior for fiber suspension rheology. A two (2) stage Newton Raphson numerical algorithm is used in our simulation, firstly to solve for the steady-state flow-field velocities and pressure distribution within the flow domain and secondly to compute the resulting translational and rotational velocities of the rigid spheroidal particle during its motion in various homogenous flow fields by equilibrating the net force and couple acting on the particles surface and the fiber's instantaneous positions and orientations are updated using a numerical ordinary differential equation (ODE) solution technique. FEA model validation is achieved by comparing steady state responses at a single time step of the quasi-transient analysis of a single particle motion along Jeffery's orbit obtained from a custom-built FEA simulation with results obtained Jeffery's Equations. Likewise, the behavior of the particle (kinematics and surface pressure response) in various Newtonian homogenous flow fields are benchmarked for both Jeffery's Model and FEA simulation.

Finally, we investigate the resulting effect of particle shape and the shear-thinning fluid rheology on the particle's dynamics and evolution of the pressure distribution response on the fibers' surface in the various homogenous flow fields using our validated FEA model. These findings are particularly useful in controlling process parameters to optimize the microstructure of particulate polymer composites to improve print properties.

### *5.1 Single Particle Motion Computational Models*

The subsequent sections provide in detail the methods used for predicting the behavior of a single three-dimensional (3D) rigid ellipsoidal particle suspended in Newtonian and non-Newtonian viscous homogenous shear-extension flows. The first section presents Jeffery's formulation for the flow-field development around an ellipsoid and explicit derivations for the particle motion (angular velocities and orientation angles) in a special class of linear homogenous flow with combined extension and shear rate velocity gradient components that idealizes typical flow conditions found in various sections of an EDAM extruder-nozzle. The second section details the FEA model development for obtaining particle angular velocities, orientation angles and field velocities and pressure distribution surrounding a particle suspended in non-linear creeping shear flow with a power-law fluid definition. Subsequent sections present results of the model validation by comparing the evolution of the particle's angular velocities and surface pressure distribution obtained from both Jeffery's analytical equations and FEA numerical model for different Newtonian flow cases and particle aspect ratio. Except stated otherwise, we consider a geometric aspect ratio of $r_e = 6$ for the prolate spheroid, a consistency index of $m = 1\ Pa \cdot s^n$ for the power-law fluid or a viscosity of $\mu_1 = 1\ Pa \cdot s$ for Newtonian fluid, and shear rate of $\dot{\gamma} = 1\ s^{-1}$ for the various flow cases.

## *5.2 Standard Jeffery Analytical Model*

Jeffery [21] derived analytical equations for the motion of a single 3D ellipsoidal particle suspended in a Newtonian homogenous viscous creeping flow by linearization of the Navier Stokes equations assuming a zero Reynolds number. The following includes a summary of Jeffery's particle-fluid interaction dynamics model where he obtained expressions for the velocity and pressure field within the fluid surrounding the particle. The equations for the pressure and velocity within a Newtonian fluid having viscosity $\mu_1$ are respectively given as

$$p = p_0 + 2\mu_1 \Lambda_{ij}^{III} \nabla_{X_i} \nabla_{X_j} \Omega \qquad (5.1)$$

and

$$\dot{X}_i = \dot{X}_i^{\infty} + \nabla_{X_i} \Lambda_j^I \chi_j + \epsilon_{ijk} \nabla_{X_j} \Lambda_{km}^{II} X_m + \Lambda_{jk}^{III} X_k \nabla_{X_i} \nabla_{X_j} \Omega - \Lambda_{ij}^{III} \nabla_{X_j} \Omega \qquad (5.2)$$

where the position vector $\underline{X}$, gradient operator $\underline{\nabla}$ and integral function $\underline{\chi}$ are given respectively as

$$\underline{X} = [X_1 \quad X_2 \quad X_3]^T, \qquad \underline{\nabla}_{\underline{X}} = [\partial/\partial X_1 \quad \partial/\partial X_2 \quad \partial/\partial X_3]^T, \qquad \underline{\chi} = [\chi_1 \quad \chi_2 \quad \chi_3]^T \quad (5.3)$$

In the above, the Laplace function $\Omega$ is defined in terms of the independent position vector variables $\underline{X}$ and $\lambda$ as

$$\Omega = \Omega(\underline{X}, \lambda) = \int_{\lambda}^{\infty} \frac{1}{\Delta} \left\{ \sum_{j=1}^{3} \frac{X_j^2}{И_j^2 + \lambda} - 1 \right\} d\lambda, \qquad \Delta^2 = \prod_{j=1}^{3} \left( И_j^2 + \lambda \right) \qquad (5.4)$$

where $\lambda$ is an arbitrary offset distance from the particle's surface obtained from the positive real roots of

$$\sum_{j=1}^{3} \frac{X_j^2}{И_j^2 + \lambda} = 1, \qquad \lambda \geq 0 \qquad (5.5)$$

The undisturbed fluid velocity $\dot{X}_i^{\infty}$ in eqn. *(5.2)* above is given as

$$\dot{X}_i^{\infty} = L_{ij} X_j \qquad (5.6)$$

where $L_{ij}$ is the velocity gradient tensor. The constant-coefficient tensors $\Lambda_i^I, \Lambda_{ij}^{II}$ & $\Lambda_{ij}^{III}$ that appear in eqns. *(5.1)* - *(5.2)* above are given as

$$\underline{\Lambda}^I = \begin{bmatrix} R \\ S \\ T \end{bmatrix}, \qquad \underline{\underline{\Lambda}}^{II} = \begin{bmatrix} U & & \\ & V & \\ & & W \end{bmatrix}, \qquad \underline{\underline{\Lambda}}^{III} = \begin{bmatrix} A & H & G' \\ H' & B & F \\ G & F' & C \end{bmatrix} \qquad (5.7)$$

where expressions for the components shown here are given in APPENDIX B **(B.1)**. The terms in $\Lambda_{ij}^{III}$ are simply the stresslet and torque acting on the rigid ellipsoid suspended in linear ambient flow-field [266]. The tensors $\Lambda_i^I, \Lambda_{ij}^{II}$ & $\Lambda_{ij}^{III}$ are functions of the symmetric rate of deformation tensor $\Gamma_{ij}$ and the antisymmetric vorticity tensor $\Xi_{ij} = \epsilon_{imn}\Xi_m\delta_{nj}$ obtained by decomposing the velocity gradient tensor $L_{ij}$ according to

$$L_{ij} = \nabla_{X_j}\dot{X}_i = \Gamma_{ij} + \Xi_{ij}, \qquad \Gamma_{ij} = \frac{1}{2}\left[L_{ij} + L_{ji}\right], \qquad \Xi_{ij} = \frac{1}{2}\left[L_{ij} - L_{ji}\right] \tag{5.8}$$

The velocity gradient $L_{ij}$ is given with respect to the particle's local coordinate axis and is thus a function of the independent particle orientation angle vector $\underline{\Theta} = [\phi \quad \theta \quad \psi]^T$ obtained by a transformation operation according to

$$L_{ij} = Z_{X_{mi}} \mathcal{L}_{mn}\, Z_{X_{nj}} \tag{5.9}$$

where $L_{ij}$ is the velocity gradient in the global reference frame axis. The transformation tensor $Z_{\theta_{ij}}$ is given in terms of the Euler angles as:

$$Z_{X_{ij}} = \Pi_{mi}^{(1)}\Pi_{nm}^{(2)}\Pi_{jn}^{(3)} \tag{5.10}$$

where,

$$\Pi_{ij}^{(k)} = \delta_{in}\delta_{jn} + (1 - \delta_{in})(1 - \delta_{jn})\left[\delta_{ij}\cos\Theta_k + (j - i)\sin\Theta_k\right], \qquad n = 2 + -1^k \tag{5.11}$$

At the particle's surface, the field velocity is given by

$$\dot{X}_i^p = \dot{X}_i\,\big|_{\lambda=0} = \epsilon_{ijk}\dot{\Psi}_j X_k \tag{5.12}$$

The particle's angular velocity $\dot{\Psi}_i$ in the local reference frame is given by the expression.

$$\dot{\Psi}_i = \Xi_i + M_i D_i \tag{5.13}$$

where no summation is implied by repeated indices and $\Xi_i$ is the vorticity vector, $D_k$ contains non-diagonal terms of the symmetric rate of deformation tensor $\Gamma_{ij}$, i.e.

$$D_k = \Gamma_{ij}, \qquad i \neq j \neq k$$

and the constant coefficient matrix $M_k$ is defined as

$$M_k = \frac{И_i^2 - И_j^2}{И_i^2 + И_j^2}, \qquad i \neq j \neq k \qquad (5.14)$$

The angular velocities in the global reference coordinate axis $\underline{\dot{\Theta}}$ based on Euler's definition are obtained by the transformation operation

$$Z_{\Theta_{ij}} \dot{\Theta}_j = \dot{\Psi}_i \qquad (5.15)$$

where the transformation operator $\underline{\underline{Z}}_\Theta$ is given as (cf. Figure 5.1a) for the Euler definition of orientation angles

$$\underline{\underline{Z}}_\Theta = \begin{bmatrix} \cos\theta & 0 & 1 \\ -sin\,\theta\cos\psi & \sin\psi & 0 \\ \sin\theta\sin\psi & \cos\psi & 0 \end{bmatrix} \qquad (5.16)$$

Figure 5.1a illustrates the ellipsoidal particle of interest suspended in simple shear flow as shown. The normal and shear stress components at any point in the flow field may be evaluated for incompressible fluid as

$$\sigma_{ij} = -p\delta_{ij} + \mu_1 \left[ \nabla_{X_i} \dot{X}_j + \nabla_{X_j} \dot{X}_i \right] \qquad (5.17)$$

On the particle's surface, the stress reduces to $\sigma_{ij} = -p\delta_{ij}$ implying that the only active stresses on the particle's surface are the hydrostatic pressure acting normal to the surface.

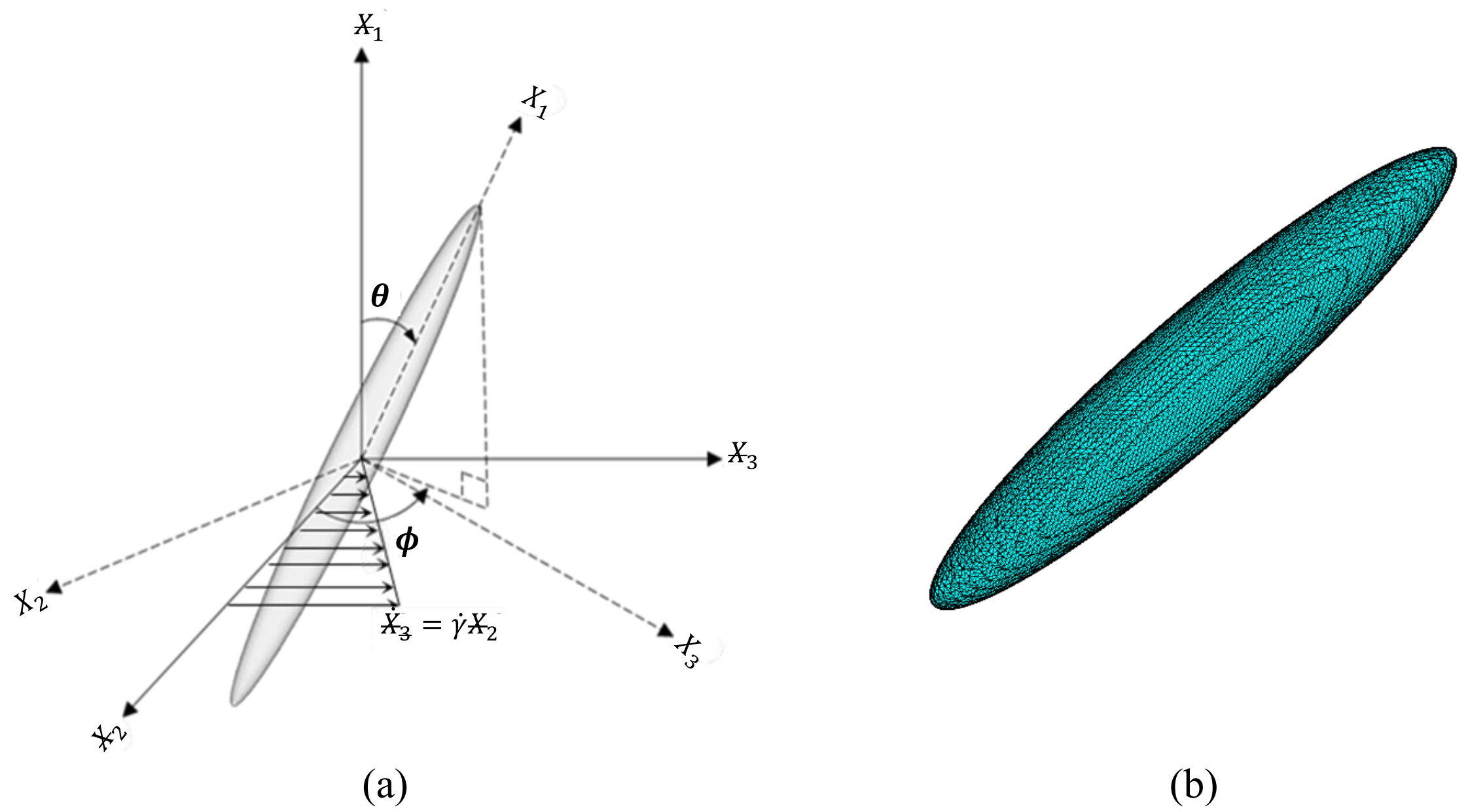

Figure 5.1: (a) Fiber orientation angles definition (b) Mesh refinement on the fiber surface.

The two-dimensional (2D) contraction of Jeffery's expression for the field velocities and pressure surrounding a rigid particle of elliptical shape in planar homogenous flow field can be found in APPENDIX B (B.2). Our main interest here is to evaluate the motion, and surface pressure and velocity of the ellipsoidal inclusion using Jeffery's equations given above. To compute surface pressure and velocity distribution on the particle surface, the ellipsoidal surface is discretized using MATLAB's inbuilt PDE modeller (MathWorks, Natick, MA, USA) where vertices were imposed at ends of the ellipsoid to enable the calculation of particle tip pressure (cf. Figure 5.1b). At the mesh points, the flow-field pressure and velocities are evaluated using eqns. *(5.1)*-*(5.2)* respectively. The degree of mesh refinement is critical to obtaining accurate pressure extremities and locations on the particle surface. A 4$^{th}$ order explicit Runge-Kutta ordinary differential equation (ODE) technique is used to numerically integrate the particle's angular velocities (cf. eqn. *(5.13)*) with time to obtain solutions of the particle orientation angles, and the associated field state (pressure and velocities on each node of the particle surface) based on Jeffery's model equations.

### *5.3 Optimization of Jeffery's Pressure*

The current objective is to minimize the pressure $p$ on the surface of the fiber and about all possible orientation configuration defined by the surface of the unit sphere S such that:

$$\oint d\underline{\rho} = \int_{\phi=0}^{2\pi} \int_{\theta=0}^{\pi} \sin\theta \, d\theta \, d\phi \tag{5.18}$$

The minimization objective function is thus the pressure $p$ given in eqn. *(5.1)* above which we can rewrite as a function of the dependent variables $\underline{\Theta}$, and $\underline{X}$, such that:

$$p(\underline{X}, \underline{\Theta}) = p_0 + \varpi_j Д_j \tag{5.19}$$

and the constraints are thus the surface of the ellipsoidal fiber, and the surface of the unit sphere S that defines all possible fiber orientation, i.e.,

$$\sum_{j=1}^{3} \frac{X_j^2}{И_j^2} - 1 = 0 \qquad \text{on the fiber's surface} \tag{5.20}$$

$$0 \leq \phi \leq 2\pi, \quad 0 \leq \theta \leq \pi \qquad \text{on the surface of the unit sphere S} \tag{5.21}$$

The constant coefficient vector $\underline{\varpi}$ is given as:

$$\underline{\varpi} = [A \quad B \quad C \quad F + F' \quad G + G' \quad H + H'] \tag{5.22}$$

Definition of the constants $A, B, C, F, G, H$ have been provided in APPENDIX B (B.1) and are functions of the components of the deformation tensor $\Gamma_{ij}$, and the vorticity tensor $\Xi_{ij}$ given in eqn. *(5.8)* above. The vector $Д_n$ in form contracted notation contains components of the hessian of the Laplace function Ω that appear in the pressure equation (cf. eqn. (5.19)) given as

$$Д_n = \nabla_{X_i}\nabla_{X_j}\Omega, \qquad n = f(i,j) \tag{5.23}$$

where the function $f(i,j)$ is given as

$$f(i,j) = i\delta_{ij} + \left(1-\delta_{ij}\right)(9-i-j) \tag{5.24}$$

Moreover, the Laplace equation must be satisfied. i.e.,

$$\nabla^2\Omega = \nabla_{X_k}\nabla_{X_k}\Omega = 0 \tag{5.25}$$

Derivation of the exact gradient and hessian of the pressure used in the optimization process are provided in detail in the succeeding section.

### 5.3.1 *Obtaining Exact Derivatives of Jeffery's Pressure.*

The optimization operation requires the gradient and hessian of the pressure which we obtain explicitly by taking first and second derivatives of the pressure equation with respect to the independent variable vectors $\underline{X}$ and $\underline{\Theta}$. i.e., the gradient of the pressure is given as

$$\underline{\nabla}p = \begin{bmatrix}\underline{\nabla}_{\underline{X}} \\ \underline{\nabla}_{\underline{\Theta}}\end{bmatrix} p \tag{5.26}$$

where

$$\nabla_{X_i}p = \varpi_j \nabla_{X_i}Д_j, \qquad \nabla_{\Theta_i}p = Д_j\nabla_{\Theta_i}\varpi_j, \qquad i = 1-3, \qquad j = 1-6 \tag{5.27}$$

and the hessian is given as

$$\underline{\underline{\nabla}}^2 p = \begin{bmatrix}\underline{\nabla}_{\underline{X}}\underline{\nabla}_{\underline{X}}^T & \underline{\nabla}_{\underline{X}}\underline{\nabla}_{\underline{\Theta}}^T \\ \underline{\nabla}_{\underline{\Theta}}\underline{\nabla}_{\underline{X}}^T & \underline{\nabla}_{\underline{\Theta}}\underline{\nabla}_{\underline{\Theta}}^T\end{bmatrix} p \tag{5.28}$$

Since the hessian is symmetric, the relevant components of $\underline{\underline{\nabla}}^2 p$ are given as

$$\nabla_{X_i}\nabla_{X_j}p = \varpi_k\left[\nabla_{X_i}\nabla_{X_j}Д_k\right], \qquad \nabla_{X_i}\nabla_{\Theta_j}p = \left[\nabla_{X_i}Д_k\right]\left[\nabla_{\Theta_j}\varpi_k\right], \tag{5.29}$$

$$\nabla_{\Theta_i}\nabla_{\Theta_j}p = Д_k\left[\nabla_{\Theta_i}\nabla_{\Theta_j}\varpi_k\right], \qquad i,j = 1-3, \qquad k = 1-6$$

The derivative operators are distributive over the differentiable elements and sub-elements of the constant coefficient vector $\underline{\varpi}$ and can be assembled from the derivatives of its individual components. Typical first and second order derivatives of the constants in $\underline{\varpi}$ are presented in eqn. (5.30) below from which the others can be surmised.

$$\begin{gathered}\nabla_{\Theta_i}\varpi_1 = \nabla_{\Theta_i}A = \frac{1}{6}\left\{\frac{2Ч''_{1_0}\nabla_{\Theta_i}\Gamma_{11} - Ч''_{2_0}\nabla_{\Theta_i}\Gamma_{22} - Ч''_{3_0}\nabla_{\Theta_i}\Gamma_{33}}{Ч''_{2_0}Ч''_{3_0} + Ч''_{3_0}Ч''_{1_0} + Ч''_{1_0}Ч''_{2_0}}\right\} \\ \nabla_{\Theta_i}\nabla_{\Theta_j}\varpi_1 = \nabla_{\Theta_i}\nabla_{\Theta_j}A = \frac{1}{6}\left\{\frac{2Ч''_{1_0}\nabla_{\Theta_i}\nabla_{\Theta_j}\Gamma_{11} - Ч''_{2_0}\nabla_{\Theta_i}\nabla_{\Theta_j}\Gamma_{22} - Ч''_{3_0}\nabla_{\Theta_i}\nabla_{\Theta_j}\Gamma_{33}}{Ч''_{2_0}Ч''_{3_0} + Ч''_{3_0}Ч''_{1_0} + Ч''_{1_0}Ч''_{2_0}}\right\}\end{gathered} \tag{5.30}$$

The components of $\underline{\varpi}$ are functions of the components of the deformation rate tensor $\Gamma_{ij}$ and the vorticity tensor $\Xi_{ij}$ which are obtained from the decomposition of the velocity gradient $L_{ij}$ in the local fiber reference frame according to the transformation operation of eqn. *(5.8)* and is thus function of the fiber orientation angles $\underline{\Theta}$. i.e.

$$\langle \nabla_{\Theta_k}\Gamma_{ij},\ \nabla_{\Theta_k}\Xi_{ij}\rangle = \frac{1}{2}\left[\nabla_{\Theta_k}L_{ij} \pm \nabla_{\Theta_k}L_{ji}\right] \tag{5.31}$$

Likewise, the second derivatives can be written as

$$\langle \nabla_{\Theta_l}\nabla_{\Theta_k}\Gamma_{ij},\ \nabla_{\Theta_l}\nabla_{\Theta_k}\Xi_{ij}\rangle = \frac{1}{2}\left[\nabla_{\Theta_l}\nabla_{\Theta_k}L_{ij} \pm \nabla_{\Theta_l}\nabla_{\Theta_k}L_{ji}\right] \tag{5.32}$$

where the operator $\nabla_{\Theta_k} = \partial\,/\partial\Theta_k$. The first derivative of the velocity gradient tensor with respect to k^th component of $\underline{\Theta}$ in the fibers local coordinate axis is obtained by the product rule and expressed in indicial notation as

$$\nabla_{\Theta_k}L_{ij} = \nabla_{\Theta_k}Z_{X_{mi}}Ƚ_{mn}\,Z_{X_{nj}} + Z_{X_{mi}}Ƚ_{mn}\,\nabla_{\Theta_k}Z_{X_{nj}} \tag{5.33}$$

The derivative of the transformation tensor $Z_{X_{ij}}$ with respect to $\underline{\Theta}$ , i.e. $\nabla_{\Theta_k}Z_{X_{ij}}$ is a third order tensor given as

$$\nabla_{\Theta_k} Z_{X_{ij}} = \delta_{k1} \nabla \Pi_{mi}^{(1)} \Pi_{nm}^{(2)} \Pi_{jn}^{(3)} + \delta_{k2} \Pi_{mi}^{(1)} \nabla \Pi_{nm}^{(2)} \Pi_{jn}^{(3)} + \delta_{k3} \Pi_{mi}^{(1)} \Pi_{nm}^{(2)} \nabla \Pi_{jn}^{(3)} \tag{5.34}$$

The derivative $\partial Z_{X_{ij}} / \partial \Theta_k$ is trivial. Since $\Pi_{ij}^{(k)}$ is conveniently represented in indicial notation as given in eqn. *(5.11)*, it is easy to differentiate $\Pi_{ij}^{(k)}$ with respect to $\Theta_k$,. i.e.

$$\nabla \Pi_{ij}^{(k)} = (1 - \delta_{in})(1 - \delta_{jn})\left[-\delta_{ij} \sin \Theta_k + (j - i) \cos \Theta_k\right] \tag{5.35}$$

Following from eqn. (5.33) above, the second derivative of the velocity gradient $L_{ij}$ with respect to $\underline{\Theta}$ in the fibers local coordinate axis via product rule is given as

$$\begin{aligned} \nabla_{\Theta_l} \nabla_{\Theta_k} L_{ij} = {} & \nabla_{\Theta_l} \nabla_{\Theta_k} Z_{X_{mi}} \L_{mn} Z_{X_{nj}} + \nabla_{\Theta_k} Z_{X_{mi}} \L_{mn} \nabla_{\Theta_l} Z_{X_{nj}} \\ & + \nabla_{\Theta_l} Z_{X_{mi}} \L_{mn} \nabla_{\Theta_k} Z_{X_{nj}} + Z_{X_{mi}} \L_{mn} \nabla_{\Theta_l} \nabla_{\Theta_k} Z_{X_{nj}} \end{aligned} \tag{5.36}$$

The second derivative of the transformation tensor $Z_{X_{ij}}$ with respect to $\underline{\Theta}$ , i.e. $\nabla_{\Theta_l} \nabla_{\Theta_k} Z_{X_{ij}}$ that appear in eqn. (5.36) above is a fourth order tensor given as

$$\begin{aligned} \nabla_{\Theta_l} \nabla_{\Theta_k} Z_{X_{ij}} = {} & \delta_{l1} \delta_{k1} \nabla^2 \Pi_{mi}^{(1)} \Pi_{nm}^{(2)} \Pi_{jn}^{(3)} + \delta_{l2} \delta_{k1} \nabla \Pi_{mi}^{(1)} \nabla \Pi_{nm}^{(2)} \Pi_{jn}^{(3)} \\ & + \delta_{l3} \delta_{k1} \nabla \Pi_{mi}^{(1)} \Pi_{nm}^{(2)} \nabla \Pi_{jn}^{(3)} + \delta_{l1} \delta_{k2} \nabla \Pi_{mi}^{(1)} \nabla \Pi_{nm}^{(2)} \Pi_{jn}^{(3)} \\ & + \delta_{l2} \delta_{k2} \Pi_{mi}^{(1)} \nabla^2 \Pi_{nm}^{(2)} \Pi_{jn}^{(3)} + \delta_{l3} \delta_{k2} \Pi_{mi}^{(1)} \nabla \Pi_{nm}^{(2)} \nabla \Pi_{jn}^{(3)} \\ & + \cdots \delta_{l1} \delta_{k3} \nabla \Pi_{mi}^{(1)} \Pi_{nm}^{(2)} \nabla \Pi_{jn}^{(3)} + \delta_{l2} \delta_{k3} \Pi_{mi}^{(1)} \nabla \Pi_{nm}^{(2)} \nabla \Pi_{jn}^{(3)} \\ & + \delta_{l3} \delta_{k3} \Pi_{mi}^{(1)} \Pi_{nm}^{(2)} \nabla^2 \Pi_{jn}^{(3)} \end{aligned} \tag{5.37}$$

From eqn. (5.35) above, we can conveniently obtain second derivatives of $\Pi_{ij}^{(k)}$ with respect to $\underline{\Theta}$. i.e.

$$\nabla^2 \Pi_{ij}^{(k)} = -(1 - \delta_{in})(1 - \delta_{jn})\left[\delta_{ij} \cos \Theta_k + (j - i) \sin \Theta_k\right] \tag{5.38}$$

The $\underline{\Theta} -$ derivatives of the fiber angular velocities with respect to its local coordinate axis are linear superposition of the derivatives of the individual terms in eqn. *(5.13)* and given as

$$\nabla_{\Theta_i}\dot{\Psi}_j = \nabla_{\Theta_i}\Xi_j + M_{jk}\nabla_{\Theta_i}D_k \quad \Big| \quad \nabla_{\Theta_m}\nabla_{\Theta_n}\dot{\Psi}_j = \nabla_{\Theta_m}\nabla_{\Theta_n}\Xi_j + M_{jk}\nabla_{\Theta_m}\nabla_{\Theta_n}D_k \tag{5.39}$$

where the operator $\underline{\nabla}_{\underline{\Theta}}^{(n)}$ is distributive over the components of $\Xi_j$ and $D_j$ as in the usual manner. For instance,

$$\begin{aligned} \nabla_{\Theta_i}\Xi_1 &= \nabla_{\Theta_i}\xi, \qquad \nabla_{\Theta_i}D_1 = \nabla_{\Theta_i}\Gamma_{23} \\ \nabla_{\Theta_m}\nabla_{\Theta_n}\Xi_1 &= \nabla_{\Theta_m}\nabla_{\Theta_n}\xi, \qquad \nabla_{\Theta_m}\nabla_{\Theta_n}D_1 = \nabla_{\Theta_m}\nabla_{\Theta_n}\Gamma_{23} \end{aligned} \tag{5.40}$$

The first derivatives of the coefficient vector $\underline{Д}$ containing terms of the derivatives Laplace function $\underline{\nabla}_{\underline{X}}^2\Omega$ with respect to $\underline{X}$ is given as

$$\nabla_{X_m}Д_n = \nabla_{X_m}\nabla_{X_j}\nabla_{X_k}\Omega, \qquad n = f(j,k) \tag{5.41}$$

Similarly, the second derivative of the coefficient vector $\underline{Д}$ with respect to $\underline{X}$ is a third order tensor $\nabla_{X_r}\nabla_{X_s}Д_m$ given as

$$\nabla_{X_r}\nabla_{X_s}Д_n = \nabla_{X_r}\nabla_{X_s}\nabla_{X_j}\nabla_{X_k}\Omega, \qquad n = f(j,k) \tag{5.42}$$

The terms of the higher order derivatives of the Laplace function Ω found in expressions for $\nabla_{X_m}Д_n$ and $\nabla_{X_r}\nabla_{X_s}Д_n$ in eqns. (5.41) - (5.42) above are given in the next section below.

### *5.3.2 Obtaining higher-order derivatives of the Laplace Function Ω.*

The Laplace function $\Omega$ found in Jeffery's equations for the field velocities and pressure distribution is an integral function in terms of position descriptor variable $\lambda$ and is given as

$$\Omega = \int_{\lambda}^{\infty} f(\underline{X},\lambda)\,d\lambda, \qquad f(\underline{X},\lambda) = \frac{1}{\Delta}\left[\sum_{j=1}^{3}\frac{X_j^2}{И_j^2+\lambda} - 1\right] \tag{5.43}$$

To obtain derivatives of Ω, the well-known Leibnitz integral theorem finds particular use in differentiating definite integral functions with limits that are function of the differentiable variable. For instance, the first-order partial derivative of the Laplace function with respect to $X_j$ using the Leibnitz theorem can be evaluated from the expression in eqn. (5.44) below:

$$\frac{d\Omega}{dX_j} = \int_{\lambda}^{\infty} \frac{\partial}{\partial X_j}\{f(\underline{X},\lambda)\}\, d\lambda + f(\underline{X},\infty)\frac{d\infty}{dX_j} - f(\underline{X},\lambda)\frac{d\lambda}{dX_j} \tag{5.44}$$

By definition, $f(\underline{X},\lambda) = 0$ since

$$\left[\sum_{j=1}^{3}\frac{X_j^2}{И_j^2+\lambda} - 1\right] = 0, \qquad and, \qquad \frac{d\infty}{dX_j} = 0 \tag{5.45}$$

Therefore

$$\frac{d\Omega}{dX_j} = \int_{\lambda}^{\infty} \frac{\partial}{\partial X_j}\{f(\underline{X},\lambda)\}\, d\lambda = 2X_j \int_{\lambda}^{\infty} \frac{1}{\Delta}\left[\frac{d\lambda}{И_j^2+\lambda}\right] = 2Ч_j X_j \tag{5.46}$$

In eqn. (5.46) above, repeated indices do not imply summation. For subsequent higher order derivatives of $\Omega$, it is important to make some necessary definitions. Firstly, we define $Ҏ^{(n)}$ such that:

$$\frac{1}{Ҏ^{(n)}} = \sum_{j=1}^{3}\frac{X_j^2}{\left(И_j^2+\lambda\right)^n}, \qquad Ҏ^{(2)} = \frac{1}{2}X_j \nabla_{X_j}\lambda \tag{5.47}$$

Also, we define $¥^{(n)}$ such that:

$$\frac{1}{¥^{(n)}} = \sum_{j=1}^{3}\frac{1}{\left(И_j^2+\lambda\right)^n}, \qquad ¥ = ¥^{(1)} \tag{5.48}$$

Additionally, the first and second derivatives of $\lambda$ with respect to components of $\underline{X}$ and its permutations are important in concisely obtaining higher order derivatives of $\Omega$. By differentiating $\Delta f(\underline{X},\lambda)$, and making necessary substitutions, we obtain for the first derivatives of $\lambda$ thus:

$$\nabla_{X_j}\lambda = \frac{2X_j}{И_j^2 + \lambda}Ҏ^{(2)} \tag{5.49}$$

Similarly, the second derivatives of $\lambda$ with respect to $X_j$ are given as

$$\begin{aligned} \nabla_{X_j}^2\lambda &= \frac{1}{X_j}\nabla_{X_j}\lambda - \left[2\frac{1}{И_j^2 + \lambda} - 2\frac{Ҏ^{(2)}}{Ҏ^{(3)}}\right]\nabla_{X_j}\lambda^2 \\ \nabla_{X_i}\nabla_{X_j}\lambda &= -\left[\frac{1}{И_i^2 + \lambda} + \frac{1}{И_j^2 + \lambda} - 2\frac{Ҏ^{(2)}}{Ҏ^{(3)}}\right]\nabla_{X_i}\lambda\,\nabla_{X_j}\lambda \end{aligned} \tag{5.50}$$

With the above definitions we can concisely present expressions for typical forms of the second-order partial derivatives of $\Omega$ with respect to permutations of $X_j$ vector using Leibnitz integral theorem thus from which derivatives with respect to other permutations of the differentiable variables are implicit.

$$\nabla_{X_j}^2\Omega = 2Ч_j - \frac{1}{Ҏ^{(2)}\Delta}\left[\nabla_{X_j}\lambda\right]^2, \qquad \nabla_{X_i}\nabla_{X_j}\Omega = -\frac{1}{Ҏ^{(2)}\Delta}\nabla_{X_i}\lambda\,\nabla_{X_j}\lambda \tag{5.51}$$

Likewise, third-order partial derivatives of $\Omega$ with respect to permutations of components of $X_j$ are conveniently presented in eqns. (5.52) - (5.54) from which components of $\nabla_{X_m}Д_n$ can be deduced.

$$\nabla_{X_j}^3\Omega = -3\frac{1}{X_j}\frac{1}{Ҏ^{(2)}\Delta}\left[\nabla_{X_j}\lambda\right]^2 + \frac{1}{Ҏ^{(2)}\Delta}\left[\frac{1}{2¥} + 3\frac{1}{И_j^2 + \lambda} - 2\frac{Ҏ^{(2)}}{Ҏ^{(3)}}\right]\left[\nabla_{X_j}\lambda\right]^3 \tag{5.52}$$

$$\nabla_{X_i}^2\nabla_{X_j}\Omega = -\frac{1}{X_i}\frac{1}{Ҏ^{(2)}\Delta}\nabla_{X_i}\lambda\,\nabla_{X_j}\lambda + \frac{1}{Ҏ^{(2)}\Delta}\left[\frac{1}{2¥} + 2\frac{1}{И_i^2 + \lambda} + \frac{1}{И_j^2 + \lambda} - 2\frac{Ҏ^{(2)}}{Ҏ^{(3)}}\right]\left[\nabla_{X_i}\lambda\right]^2\nabla_{X_j}\lambda \tag{5.53}$$

$$\nabla_{X_i}\nabla_{X_j}\nabla_{X_k}\Omega = +\frac{1}{Ҏ^{(2)}\Delta}\left[\frac{1}{2¥} + \frac{1}{И_i^2 + \lambda} + \frac{1}{И_j^2 + \lambda} + \frac{1}{И_k^2 + \lambda} - 2\frac{Ҏ^{(2)}}{Ҏ^{(3)}}\right]\nabla_{X_i}\lambda\,\nabla_{X_j}\lambda\nabla_{X_k}\lambda \tag{5.54}$$

Additionally, fourth – order partial derivatives of $\Omega$ with respect to permutations of components of $X_j$ are given in eqns. (5.55) - (5.58) from which components of $\nabla_{X_r}\nabla_{X_s}Д_n$, can be deduced.

$$
\begin{aligned}
\nabla^4_{X_j}\Omega = \frac{1}{\text{Ƥ}^{(2)}\Delta}\Bigg\{ & 15\frac{1}{X_j^2}\left[\nabla_{X_j}\lambda\right]^2 - 36\left[\frac{1}{И_j^2+\lambda} - \frac{\text{Ƥ}^{(2)}}{\text{Ƥ}^{(3)}}\right]\frac{1}{X_j}\left[\nabla_{X_j}\lambda\right]^3 \\
& + \Bigg[-\frac{1}{4¥^2} - \frac{1}{2¥^{(2)}} - 3\frac{1}{¥}\frac{\text{Ƥ}^{(2)}}{\text{Ƥ}^{(3)}} + 2\frac{1}{¥}\frac{1}{И_j^2+\lambda} + 16\frac{1}{\left(И_j^2+\lambda\right)^2} - 36\frac{1}{И_j^2+\lambda}\frac{\text{Ƥ}^{(2)}}{\text{Ƥ}^{(3)}} \\
& + 12\left[\frac{\text{Ƥ}^2}{\text{Ƥ}^{(3)}}\right]^2 + 6\frac{\text{Ƥ}^{(2)}}{\text{Ƥ}^{(4)}}\Bigg]\left[\nabla_{X_j}\lambda\right]^4\Bigg\}
\end{aligned} \tag{5.55}
$$

$$
\begin{aligned}
\nabla^3_{X_i}\nabla_{X_j}\Omega = \frac{1}{\text{Ƥ}^{(2)}\Delta}\Bigg\{ & 3\left[\frac{1}{2¥} + 2\frac{1}{И_i^2+\lambda} + \frac{1}{И_j^2+\lambda} - 2\frac{\text{Ƥ}^{(2)}}{\text{Ƥ}^{(3)}}\right]\frac{1}{X_i}\left[\nabla_{X_i}\lambda\right]^2\nabla_{X_j}\lambda \\
& + \Bigg[-\frac{1}{4¥^2} - \frac{1}{2¥^{(2)}} - 3\frac{1}{¥}\frac{1}{И_i^2+\lambda} - \frac{1}{¥}\frac{1}{И_j^2+\lambda} + 3\frac{1}{¥}\frac{\text{Ƥ}^{(2)}}{\text{Ƥ}^{(3)}} - 12\frac{1}{(И_i^2+\lambda)^2} \\
& - 2\frac{1}{\left(И_j^2+\lambda\right)^2} - 6\frac{1}{И_i^2+\lambda}\frac{1}{И_j^2+\lambda} + 18\frac{1}{И_i^2+\lambda}\frac{\text{Ƥ}^{(2)}}{\text{Ƥ}^{(3)}} + 6\frac{1}{И_j^2+\lambda}\frac{\text{Ƥ}^{(2)}}{\text{Ƥ}^{(3)}} \\
& - 12\left[\frac{\text{Ƥ}^{(2)}}{\text{Ƥ}^{(3)}}\right]^2 + 6\frac{\text{Ƥ}^{(2)}}{\text{Ƥ}^{(4)}}\Bigg]\left[\nabla_{X_i}\lambda\right]^3\nabla_{X_j}\lambda\Bigg\}
\end{aligned} \tag{5.56}
$$

$$
\begin{aligned}
\nabla^2_{X_i}\nabla^2_{X_j}\Omega = \frac{1}{\text{Ƥ}^{(2)}\Delta}\Bigg\{ & -\frac{1}{X_i}\frac{1}{X_j}\nabla_{X_i}\lambda\,\nabla_{X_j}\lambda + \left[\frac{1}{2¥} + \frac{1}{И_i^2+\lambda} + 2\frac{1}{И_j^2+\lambda} - 2\frac{\text{Ƥ}^{(2)}}{\text{Ƥ}^{(3)}}\right]\frac{1}{X_i}\nabla_{X_i}\lambda\left[\nabla_{X_j}\lambda\right]^2 \\
& + \left[\frac{1}{2¥} + 2\frac{1}{И_i^2+\lambda} + \frac{1}{И_j^2+\lambda} - 2\frac{\text{Ƥ}^{(2)}}{\text{Ƥ}^{(3)}}\right]\frac{1}{X_j}\left[\nabla_{X_i}\lambda\right]^2\nabla_{X_j}\lambda \\
& + \Bigg[-\frac{1}{4¥^2} - \frac{1}{2¥^{(2)}} - 2\frac{1}{¥}\frac{1}{И_i^2+\lambda} - 2\frac{1}{¥}\frac{1}{И_j^2+\lambda} - 6\frac{1}{(И_i^2+\lambda)^2} - 6\frac{1}{\left(И_j^2+\lambda\right)^2} \\
& - 8\frac{1}{И_i^2+\lambda}\frac{1}{И_j^2+\lambda} + 3\frac{1}{¥}\frac{\text{Ƥ}^{(2)}}{\text{Ƥ}^{(3)}} + 12\frac{1}{И_i^2+\lambda}\frac{\text{Ƥ}^{(2)}}{\text{Ƥ}^{(3)}} + 12\frac{1}{И_j^2+\lambda}\frac{\text{Ƥ}^{(2)}}{\text{Ƥ}^{(3)}} - 12\left[\frac{\text{Ƥ}^{(2)}}{\text{Ƥ}^{(3)}}\right]^2 \\
& + 6\frac{\text{Ƥ}^{(2)}}{\text{Ƥ}^{(4)}}\Bigg]\left[\nabla_{X_i}\lambda\right]^2\left[\nabla_{X_j}\lambda\right]^2\Bigg\}
\end{aligned} \tag{5.57}
$$

$$
\begin{aligned}
\nabla^2_{X_i}\nabla_{X_j}\nabla_{X_k}\Omega = \frac{1}{\text{Ƥ}^{(2)}\Delta}\Bigg\{ & \left[\frac{3}{2}\frac{1}{¥} - 2\frac{\text{Ƥ}^{(2)}}{\text{Ƥ}^{(3)}}\right]\frac{1}{X_i}\nabla_{X_i}\lambda\,\nabla_{X_j}\lambda\nabla_{X_k}\lambda \\
& + \Bigg[-\frac{9}{4}\frac{1}{¥^2} - \frac{3}{2}\frac{1}{¥^{(2)}} - 3\frac{1}{¥}\frac{1}{И_i^2+\lambda} - 2\frac{1}{(И_i^2+\lambda)^2} + 6\frac{1}{И_i^2+\lambda}\frac{\text{Ƥ}^{(2)}}{\text{Ƥ}^{(3)}} + 9\frac{1}{¥}\frac{\text{Ƥ}^{(2)}}{\text{Ƥ}^{(3)}} \\
& - 12\left[\frac{\text{Ƥ}^{(2)}}{\text{Ƥ}^{(3)}}\right]^2 + 6\frac{\text{Ƥ}^{(2)}}{\text{Ƥ}^{(4)}}\Bigg]\left[\nabla_{X_i}\lambda\right]^2\nabla_{X_j}\lambda\nabla_{X_k}\lambda\Bigg\}
\end{aligned} \tag{5.58}
$$

Lastly, the derivatives of the variables $\chi_i$ with respect to components of the position vector $\chi_j$ that appear in Jeffery's expressions for the field velocities are given in eqns. (5.59)-(5.60)below, where repeated indices do not imply summation.

$$\nabla_{X_j}\chi_j = -2\frac{\text{Ҏ}^{(2)}}{\Delta^3}\prod_{j=1}^{3}\text{И}_j = \text{Ж} \tag{5.59}$$

$$\nabla_{X_j}\chi_i = \text{Ч}_i'X_k + \text{Ж}\frac{(\text{И}_i^2+\lambda)}{(\text{И}_j^2+\lambda)}\frac{X_j}{X_i}, \qquad k = 6 - i - j \mid i \neq j \tag{5.60}$$

### *5.3.3 Finite Difference Validation of the Gradient and Hessian of Jeffery's Pressure.*

The derived gradients and hessian of the pressure obtained in the preceding sections are validated by simple finite difference approximations for an arbitrary design variable vector $\underline{\text{ɣ}} = [\underline{X} \quad \underline{\Theta}]^T$. Given a small perturbation ϱ the central finite difference (FD) gradient of the objective function $p(\text{ɣ}_j)$ can be approximated as

$$\nabla_k p = \frac{p(\text{ɣ}_j + \varrho\delta_{jk}\text{ɣ}_j) - p(\text{ɣ}_j - \varrho\delta_{jk}\text{ɣ}_j)}{2\varrho\text{ɣ}_k} + O(\varrho^2) \tag{5.61}$$

Likewise, the hessian approximation of $p(\text{ɣ}_j)$ is obtained via the same central finite difference method. i.e.

$$\nabla_{jk}^2 p = \frac{\nabla_j p(\text{ɣ}_i + \varrho\delta_{ik}\text{ɣ}_i) - \nabla_j p(\text{ɣ}_i - \varrho\delta_{ik}\text{ɣ}_i)}{2\varrho\text{ɣ}_k} + O(\varrho^2) \tag{5.62}$$

In the eqns. (5.61) - (5.62) above, there is no summation over repeated indices. The metric adopted to assess the accuracy of the derived gradient and hessian tensors is the Frobenius norm of the difference between the exact values and finite difference approximations. i.e., the error of the gradient, $\text{ç}^{(1)}$ and the error of the hessian, $\text{ç}^{(2)}$ are estimated according to the respective the expressions in eqn. (5.63) below:

$$\text{ç}^{(1)} = \left\|\underline{\nabla}p^{exact} - \underline{\nabla}p^{FD}\right\|_2, \qquad \text{ç}^{(2)} = \left\|\underline{\underline{\nabla}}^2 p^{exact} - \underline{\underline{\nabla}}^2 p^{FD}\right\|_2 \tag{5.63}$$

Given a random fiber orientation state $\underline{\Theta}^i$ and any arbitrary spatial position $\underline{X}^i$ at an instant $t_i$ within the flow domain such that $\lambda \geq 0$, say,

$$\underline{\Theta}^i = [\pi/4 \quad -\pi/3 \quad 2\pi/5]^T, \qquad \underline{X}^i = [5.45 \quad 0.85 \quad 0.25]^T \tag{5.64}$$

and considering a flow field with a fluid viscosity $\mu = 1\ Pa.s$, and random velocity gradient $\underline{\underline{L}}$ say

$$\underline{\underline{L}} = \begin{bmatrix} 0.8143 & 0.3500 & 0.6160 \\ 0.2435 & 0.1966 & 0.4733 \\ 0.9293 & 0.2511 & 0.3517 \end{bmatrix}$$

Based on expression of eqn. (5.63) above, assuming $\varrho = 10^{-4}$, we obtain for the error estimates of the gradient and hessian i.e., $ç^{(1)}$ & $ç^{(2)}$ the following respective values

$$ç^{(1)} = 9.0047 \times 10^{-8}, \qquad ç^{(2)} = 7.7596 \times 10^{-7}$$

### *5.4 Homogenous Flow Considerations*

Various homogenous flows similar to those used in short fiber composite fiber orientation simulations [267] are considered here which serve as input for our particle motion studies. These homogenous flows also serve as a basis for understanding the flow fields development in common extrusion-deposition additive manufacturing (EDAM) polymer composite processing that involves a combination of shearing and extensional components within the flow (cf. APPENDIX B, B.3). The following flows are considered in this study:

(i) Simple Shear flow (SS), i.e., $L_{23} = \dot{\gamma}$

(ii) Two Stretching/Shearing flows (SUA), including simple shear in $X_2X_3$ plane superimposed with uniaxial elongation in the $X_3$-direction, i.e., $L_{11} = L_{22} = -\dot{\varepsilon},\ L_{33} = 2\dot{\varepsilon}\ , L_{23} = \dot{\gamma}$. Two cases are considered, balanced shear/stretch, $\dot{\gamma}/\dot{\varepsilon} = 10$, and dominant stretch, $\dot{\gamma}/\dot{\varepsilon} = 1$

(iii) Uniaxial Elongation flow (UA), in the $X_3$ direction, i.e., $L_{11} = L_{22} = -\dot{\varepsilon},\ L_{33} = 2\dot{\varepsilon}$

(iv) Biaxial Elongation (BA), flow in the $X_2 - X_3$ plane, i.e., $L_{11} = -2\dot{\varepsilon},\ L_{22} = L_{33} = \dot{\varepsilon}$

(v) Two shear/planar-elongation flows (PST), including simple shear in $X_2 - X_3$ plane superimposed on planar elongation in $X_1 - X_3$ plane, i.e., $L_{11} = -\dot{\varepsilon}, L_{33} = \dot{\varepsilon}, L_{23} = \dot{\gamma}$. Two cases are considered including balanced shear-planar elongation with $\dot{\gamma}/\dot{\varepsilon} = 10$, and dominant planar elongation with $\dot{\gamma}/\dot{\varepsilon} = 1$.

(vi) Balanced shear/bi-axial elongation flow (SBA), simple shear in the $X_2 - X_3$ plane superimposed on biaxial elongation, i.e., $L_{33} = \dot{\varepsilon},\ L_{22} = \dot{\varepsilon}, L_{23} = \dot{\gamma}, L_{11} = -2\dot{\varepsilon}$. Two cases are considered which include $\dot{\gamma}/\dot{\varepsilon} = 1$ and $\dot{\gamma}/\dot{\varepsilon} = 10$

(vii) Triaxial Elongation flow (TA), i.e., $L_{11} = L_{22} = L_{33} = \dot{\varepsilon}$

(viii) Balanced shear/tri-axial elongation flow (STA), including simple shear in the $X_2 - X_3$ plane superimposed on biaxial elongation, i.e., $L_{11} = L_{22} = L_{33} = \dot{\varepsilon}, L_{23} = \dot{\gamma}$, Two cases are considered i.e. $\dot{\gamma}/\dot{\varepsilon} = 1$, and $\dot{\gamma}/\dot{\varepsilon} = 10$

Classification of the various combined homogenous flow regimes based on the flow parameter $\bar{\nu}$ (cf. APPENDIX B, B.3) is given in Table 5.1 below

Table 5.1: Flow parameter values $\bar{\nu}$ for the combined homogenous flow types

| $\dot{\gamma}/\dot{\varepsilon}$ | SUA | PST | SBA | STA |
|---|---|---|---|---|
| 1 | 0.5657 | 0.3820 | 0.5657 | 0.4514 |
| 10 | 0.0283 | 0.0098 | 0.0283 | 0.0146 |

For visualization purposes and to better interpret the results that follows in later section, typical flow streamlines around a particle suspended in the mixed mode flow conditions are presented in Figure 5.2. In all flow types, simple shear is applied in the $X_2 - X_3$ plane and the particle is initially oriented in the $X_2$ direction. The SUA flow (cf. Figure 5.2a) tends to orient the particle such that its major axis aligns with the $X_3$ direction of stretching, thus mitigating the tumbling motion in the $X_2 - X_3$ shear plane that occurs under simple shear flow alone. The inward flow in the y-direction initially accelerates the particle, aiding the tumbling motion into the direction of applied extension.  High shear to extension rate dominance is thus required to prevent the particle from stalling in the $X_3$ direction. In the PST flow type shown in Figure 5.2b, the $X_1$ direction inward flow tends to constrain particle tumbling motion in the $X_2 - X_3$ shear plane and promotes preferential alignment of the particle in the z-direction and there is no flow in the y-direction that influence the particles initial motion. Unlike the SUA flow condition, in the SBA flow regime (cf. Figure 5.2c), the $X_1$ direction inward flow limits particle tumbling motion in the $X_2 - X_3$ shear plane without promoting directional preference for the particle alignment in the shear plane. Hence there is no tendency for particle stall to occur irrespective of the shear-extension rate dominance. Since the STA flow type has equal applied extension in all principal directions, the deviator of the velocity gradient has no principal component, and the particle's behavior under this flow type is similar to that under simple shear flow.

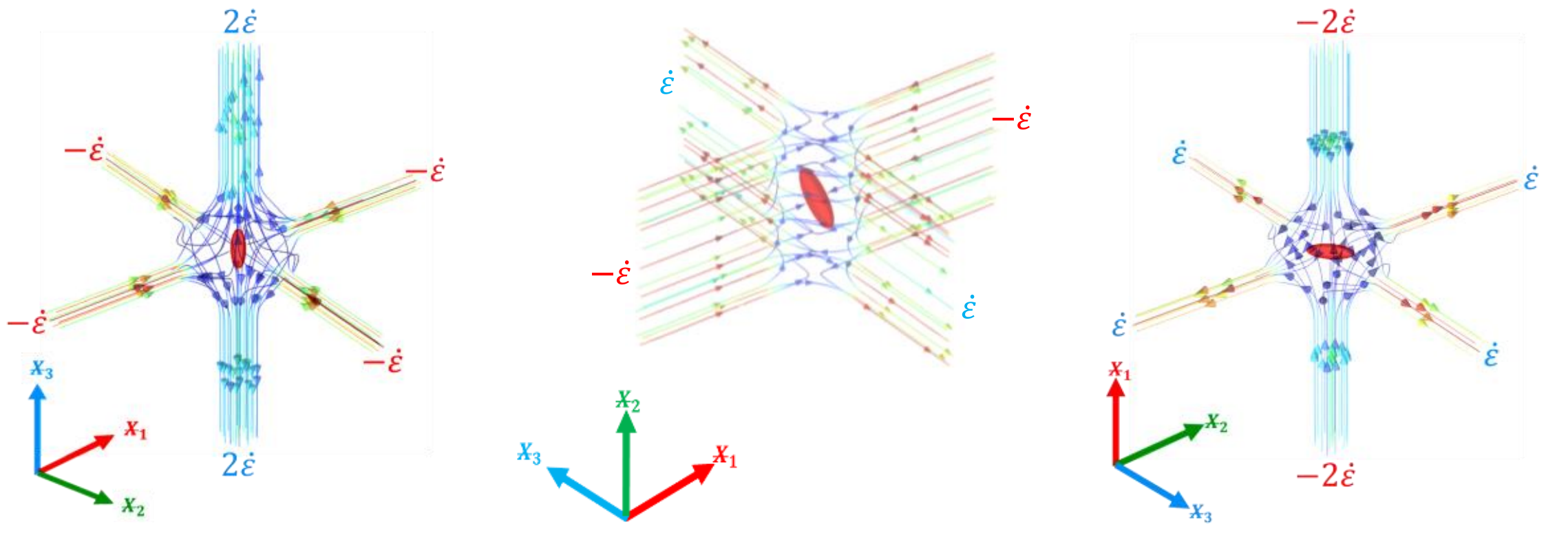

(a) (b) (c)

Figure 5.2: Visualization of the suspended particle in the combined shearing in $X_2 - X_3$ plane and (a) uniaxial elongation (SUA), (b) planar stretching (PST), and (c) biaxial elongation (SBA) flow conditions.

For the case of an axisymmetric ellipsoidal particle suspended in unconfined simple shear flow (see type (i) flow above) with velocity gradient $Ł_{23} = \dot{\gamma}$, Jeffery [21] derived analytical expressions for the particle's angular velocities given as

$$\dot{\phi}(t) = \frac{\dot{\gamma}}{2}[\kappa \cos 2\phi + 1], \qquad \dot{\theta}(t) = \frac{\dot{\gamma}}{2}\frac{(\kappa \sin 2\phi)\sqrt{(\kappa \cos 2\phi + 1)\varsigma^2(1+\kappa)}}{[(\kappa \cos 2\phi + 1) + \varsigma^2(1+\kappa)]}, \qquad (5.65)$$
$$\dot{\psi}(t) = -\frac{\dot{\gamma}}{2}(\kappa \cos 2\phi)\cos\theta$$

where the precession $\dot{\phi}$ is observed to be independent of $\theta$ and $\varsigma$ is the orbit constant. By integrating the angular velocities, Jeffery further obtained expressions for the corresponding particle orientation angles which may be written as

$$\phi(t) = \tan^{-1}\left\{\sqrt{\frac{1+\kappa}{1-\kappa}}\tan\left[\frac{\dot{\gamma}}{2}\sqrt{1-\kappa^2}\,t\right]\right\}, \quad \theta(t) = \tan^{-1}\left\{\varsigma\frac{\sqrt{1+\kappa}}{\sqrt{\kappa\cos 2\phi + 1}}\right\}, \qquad (5.66)$$
$$\psi(t) = \int_0^t \left(\frac{\dot{\gamma}}{2} - \dot{\phi}\right)\cos\theta \, dt$$

where $\dot{\gamma}$ is the shear-rate, $\kappa$ is a shape factor given as $\kappa = ({r_e}^2 - 1)/({r_e}^2 + 1)$. The orbit constant of integration $\varsigma$ can be shown to become $\varsigma = \tan\theta_0$ when $\phi_0 = 0$ and $\theta_0 \leq \theta \leq \tan^{-1}\{r_e\varsigma\}$[21]. For in-plane particle rotation, $\varsigma = +\infty$ such that $\theta = \pi/2$, $\psi = 0, \dot{\psi} = \dot{\theta} = 0$. Yamane et al. [205] provides a general equation for calculating the orbital constant $\varsigma$ as a function of the orientation vector $\rho_i$ given as

$$\varsigma^2 = \frac{1}{{r_e}^2}\left(\frac{\rho_1}{\rho_3}\right)^2 + \left(\frac{\rho_2}{\rho_3}\right)^2 \qquad (5.67)$$

The corresponding period for the in-plane particle tumbling motion in simple shear flow about the ellipsoid's polar axis is

$$\tau_1 = \frac{4\pi}{\dot{\gamma}\sqrt{1-\kappa^2}} \qquad (5.68)$$

As the ellipsoid rotates in the $X_2 - X_3$ plane of shear flow, $\dot{\phi}$ reaches a maximum value when the particle is oriented normal to the principal direction of the fluid motion, i.e., at $\phi = n\pi,\ |n| \geq 0$ (cf. Figure 5.1a), and attains a minimum value when it aligns in the flow direction i.e., at $\phi = n\pi/2\,, |n| \geq 1$ [261]. The limit of the precession is thus $0 \leq \dot{\phi} \leq \dot{\gamma}$ for ellipsoidal particles and $\dot{\phi} = \dot{\gamma}/2$ for spherical particles. The extremum of the nutation $\dot{\theta}$ occurs when $\phi = Re\{.5\cos^{-1} q\}$, where $q$ is the solution to the cubic equation defined as

$$\{q : \kappa^2 q^3 + 3\kappa(\beta + 1)q^2 + (\kappa^2 + 2\beta + 2)q + \kappa(1 - \beta) = 0\}, \qquad \beta = \varsigma^2(1 + \kappa) \tag{5.69}$$

The nutation ranges between $-\dot{\gamma}/4 \leq \dot{\theta} \leq \dot{\gamma}/4$ for spheroidal particles, and it is critical for rodlike particles when $\varsigma = 1/\sqrt{2}$, and for disc-like particles when $\varsigma = +\infty$. It attains a value of $\dot{\theta} = 0$ for spherical particles. Likewise, the particle spin rate, $\dot{\psi}$ reaches a minimum at $\phi = n\pi,\ n \geq 0$, and a maximum value at $\phi = .5\cos^{-1}\left\{\left[-(3\beta + 4) \pm \sqrt{\beta(9\beta + 8)}\right]/4\kappa\right\}$. The spin-rate ranges between $-\dot{\gamma}/2 \leq \dot{\psi} \leq \dot{\gamma}/2$ and it is critical for rod-like particles when $\varsigma = 0$ and for disc-shaped particle when $\varsigma = +\infty$. We now consider a more complicated flow condition and derive expressions for the case of an axisymmetric particle suspended in combined elongation and shear flow, i.e., flow types (ii, v, vi, & viii) given above following similar procedures adopted by Jeffery [21] for the case of simple shear flow. Consider a flow with velocity gradient of the form

$$\underline{\underline{L}} = \begin{bmatrix} \dot{\varepsilon}_1 & 0 & 0 \\ 0 & \dot{\varepsilon}_2 & 0 \\ 0 & \dot{\gamma} & \dot{\varepsilon}_3 \end{bmatrix} \tag{5.70}$$

where the $trace\left(\underline{\underline{L}}\right) = 0$, i.e., $\dot{\varepsilon}_1 + \dot{\varepsilon}_2 + \dot{\varepsilon}_3 = 0$. It can be shown that the angular velocities of a particle for this $\underline{\underline{L}}$ may be written as

$$\begin{bmatrix} \dot{\phi} \\ \dot{\theta} \\ \dot{\psi} \end{bmatrix} = \begin{bmatrix} \frac{\dot{\gamma}}{2} + \frac{\kappa}{2}\{\dot{\gamma}\cos 2\phi - [\dot{\varepsilon}_2 - \dot{\varepsilon}_3]\sin 2\phi\} \\ \frac{\kappa}{4}\{\dot{\gamma}\sin 2\phi + [\dot{\varepsilon}_2 - \dot{\varepsilon}_3]\cos 2\phi - [2\dot{\varepsilon}_1 - \dot{\varepsilon}_2 - \dot{\varepsilon}_3]\}\sin 2\theta \\ -\frac{\kappa}{2}\{\dot{\gamma}\cos 2\phi - [\dot{\varepsilon}_2 - \dot{\varepsilon}_3]\sin 2\phi\}\cos\theta \end{bmatrix} \tag{5.71}$$

where the in-plane angular velocity reduces to

$$\dot{\phi} = \frac{d\phi}{dt} = \frac{1}{2}\{\dot{\gamma}(1 + \kappa\cos 2\phi) - [\dot{\varepsilon}_2 - \dot{\varepsilon}_3]\kappa\sin 2\phi\} \tag{5.72}$$

By integrating $\dot{\phi}$ in eqn. *(5.72)(5.71)*, we obtain an expression for the in-plane orientation angle $\phi$ in these flow-types with characteristics velocity gradient $\underline{\underline{L}}$ given as

$$\tan\phi = \frac{k\kappa}{\kappa - 1}\tan\left[\tan^{-1}\frac{1}{k}\left[\frac{\dot{\varepsilon}_2 - \dot{\varepsilon}_3}{\dot{\gamma}} + \frac{\kappa - 1}{\kappa}\tan\phi_0\right] - \frac{1}{2}k\kappa\dot{\gamma}t\right] - \frac{\kappa}{\kappa - 1}\frac{\dot{\varepsilon}_2 - \dot{\varepsilon}_3}{\dot{\gamma}} \tag{5.73}$$

where,

$$k = \sqrt{\frac{1}{\kappa^2} - \frac{\dot{\varepsilon}_2 - \dot{\varepsilon}_3}{\dot{\gamma}}^2 - 1} \tag{5.74}$$

If the initial orientation $\phi_0 = 0$, then eqn. *(5.73)* reduces to

$$\tan\phi = -\frac{\kappa}{\kappa - 1}\left[\frac{k^2 + \left[\frac{\dot{\varepsilon}_2 - \dot{\varepsilon}_3}{\dot{\gamma}}\right]^2}{k\cot[.5k\kappa\dot{\gamma}t] + \frac{\dot{\varepsilon}_2 - \dot{\varepsilon}_3}{\dot{\gamma}}}\right] \tag{5.75}$$

By integrating $\dot{\theta}$ in eqn. *(5.71)*, we can directly obtain an expression for $\theta$ as

$$\tan\theta = \left[\frac{\frac{1}{\kappa} + \cos 2\phi_0 - \frac{[\dot{\varepsilon}_2 - \dot{\varepsilon}_3]}{\dot{\gamma}}\sin 2\phi_0}{\frac{1}{\kappa} + \cos 2\phi - \frac{[\dot{\varepsilon}_2 - \dot{\varepsilon}_3]}{\dot{\gamma}}\sin 2\phi}\right]^{1/2} \tan\theta_0\, e^{-\kappa/2[2\dot{\varepsilon}_1 - \dot{\varepsilon}_2 - \dot{\varepsilon}_3]t} \tag{5.76}$$

It can be shown that for the special case of initial polar orientation angle $\phi_0 = 0$, then eqn. *(5.76)* reduces to

$$\tan\theta = \left[\frac{\frac{1}{\kappa}+1}{\frac{1}{\kappa}+\cos 2\phi - \frac{[\dot{\varepsilon}_2-\dot{\varepsilon}_3]}{\dot{\gamma}}\sin 2\phi}\right]^{1/2} \tan\theta_0\, e^{-\kappa/2[2\dot{\varepsilon}_1-\dot{\varepsilon}_2-\dot{\varepsilon}_3]t} \tag{5.77}$$

Further, the spin $\psi(t)$ for these flow conditions may be written in integral form as

$$\psi(t) = \int_0^t \left(\frac{\dot{\gamma}}{2} - \dot{\phi}\right)\cos\theta\, dt \tag{5.78}$$

The quarter-period of rotation may be derived from eqn. *(5.75)* by finding the pole of the above expression of $\tan\phi$ as

$$\tau_1^{0.25} = \frac{2}{k\kappa\dot{\gamma}}\left[\pi - \tan^{-1}\left[\frac{k\dot{\gamma}}{\dot{\varepsilon}_2-\dot{\varepsilon}_3}\right]\right] \tag{5.79}$$

The period for a complete tumbling motion in this flow type is obtained by finding the zero of $\tan\phi$ in eqn. *(5.75)* above which is given as

$$\tau_1 = \frac{4\pi}{k\kappa\dot{\gamma}} \tag{5.80}$$

When $(\dot{\varepsilon}_2-\dot{\varepsilon}_3)/\dot{\gamma} = 0$, the flow is essentially simple shear, and the period is as given in eqn. *(5.68)* above. The particle stalls when $k^2 \leq 0$, i.e., when

$$\frac{\dot{\varepsilon}_2-\dot{\varepsilon}_3}{\dot{\gamma}} \geq \frac{\sqrt{1-\kappa^2}}{\kappa} \tag{5.81}$$

and the stall angle $\phi_s$ is derived by equating $\dot{\phi} = 0$ (cf. eqn. *(5.73)*) to obtain

$$\tan 2\phi_s = \left[\frac{\dot{\varepsilon}_2-\dot{\varepsilon}_3}{\dot{\gamma}} \pm i\frac{k}{\kappa}\right] \Big/ \left[\frac{\dot{\varepsilon}_2-\dot{\varepsilon}_3}{\dot{\gamma}}^2 - \frac{1}{\kappa^2}\right], \qquad \phi_s = \begin{cases}\phi_s + \pi/2, & \phi_s < 0 \\ \phi_s, & \phi_s \geq 0\end{cases} \tag{5.82}$$

Correspondingly, given a stall angle tolerance $\phi_{tol}$, the time for particle stall is obtained by equating eqn. (5.75) and (5.82), i.e. $t_s\ :\ \phi(t_s) = \phi_s - \phi_{tol}$. When eqn. *(5.82)* is satisfied ($k = 0$), the stall angle may be shown to be

$$\phi_{onset} = \tan^{-1} r_e \tag{5.83}$$

The particle orientation at stall for the special class of homogenous flows (described as ii, v, vi, and viii above) can be obtained by using Newton-Raphson numerical iterative process to zero the angular velocities thus

$$\underline{\Theta}_s^{\rho+} = \underline{\Theta}_s^{\rho-} - \underline{\underline{J}}_{\Theta_1}{}^{-1}\underline{\dot{\Theta}}^{\rho} \tag{5.84}$$

where $\underline{\Theta}_s^{\rho-} = [\phi_s \quad \theta_s]^T$, $\underline{\dot{\Theta}}^{\rho} = [\dot{\phi} \quad \dot{\theta}]^T$, and the Jacobian $\underline{\underline{J}}_{\Theta}$ is given as

$$\underline{\underline{J}}_{\Theta_1} = \begin{bmatrix} -4\dot{\theta}\operatorname{cosec}2\theta - \kappa[2\dot{\varepsilon}_1 - \dot{\varepsilon}_2 - \dot{\varepsilon}_3] & 0 \\ \left\{\dot{\phi} - \dfrac{\dot{\gamma}}{2}\right\}\sin 2\theta & 2\dot{\theta}\cot 2\theta \end{bmatrix} \tag{5.85}$$

For particle motion in more general class of Newtonian homogenous flows with velocity gradient $\underline{\underline{L}}$ the stall angle can be obtained using the Newton-Raphson procedure in APPENDIX B (B.4).

Jeffery's model derivations are limited to the standard assumption of single rigid ellipsoidal shaped particle suspended in Newtonian viscous linear homogenous flows. Practically speaking, the pressure driven flow of polymer melt through EDAM nozzle contraction during material processing is more accurately characterized by a quadratic ambient flow-field such as given in Lubansky et al. [268]. As such, development of a more realistic solution would involve a velocity gradient with higher order polynomial terms which is a relevant direction for future studies. For more general conditions, it is common to employ the Finite Element Analysis (FEA) which are not bound by the limitations of the Jeffery's model and can include inter and intra fibre forces, non-ellipsoidal fibre shape, non-Newtonian visco-elastic fluid rheology, confinement flows, and other deviations from standard conditions. Moving beyond Jeffery's model assumptions may result in a preferred particle configuration that is independent of its initial orientation and may cause the particle to align with the flow or vorticity direction [180], [181], [182]. In the sections following we describe an FEA modelling approach that may be used to investigate the effect of Generalized Newtonian Fluid (GNF) rheology on the particle dynamics and surface pressure response.

### *5.5 FEA Single Particle Model with GNF Rheology*

In the FEA model analysis present here, we simulate the motion of a single rigid spheroidal particle suspended in homogenous viscous flow with GNF rheology. The flow domain $\vartheta$ for the single particle micromodel analysis is shown in Figure 5.3a. The model extends the Newtonian fluid single fiber model developed by Zhang et. al. [230], [234], [265] and implemented by Awenlimobor et al.,[232], [233] to simulate GNF flow. In this approach, the governing equations are based on the Stokes assumption of creeping, incompressible, isothermal, steady state, low Reynolds number viscous flow where the mass and momentum conservation equations may be written as

$$\nabla_{X_i}\dot{X}_i = 0 \tag{5.86}$$

$$\nabla_{X_i}\sigma_{ij} + f_j = 0 \tag{5.87}$$

In the above, $\nabla_{X_i}$ is the gradient operator, $\dot{X}_i$ is the flow velocity vector, $f_j$ is the body force vector, and $\sigma_{ij}$ is the Cauchy stress tensor given as

$$\sigma_{ij} = \tau_{ij} - p\delta_{ij} \tag{5.88}$$

In eqn. *(5.88)*, $p$ is the hydrostatic fluid pressure, $\delta_{ij}$ is the kronecker delta, and $\tau_{ij}$ is the deviatoric stress tensor defined in terms of the strain rate tensor $\dot{\gamma}_{ij}$ by the constitutive relation

$$\tau_{ij} = 2\mu(\dot{\gamma})\dot{\gamma}_{ij} \tag{5.89}$$

where the viscosity $\mu$ is a function of the strain rate magnitude $\dot{\gamma} = \sqrt{2\dot{\gamma}_{ij}\dot{\gamma}_{ji}}$. The simulations presented below solve eqns. (5.86)-(5.89) for quasi-steady velocity and pressure within the fluid domain surrounding the ellipsoidal inclusion using our custom finite element analysis (FEA) program developed in MATLAB. We assume a non-porous particle surface with zero slip allowance and velocity boundary conditions are prescribed with respect to the particle's local coordinate reference axes.

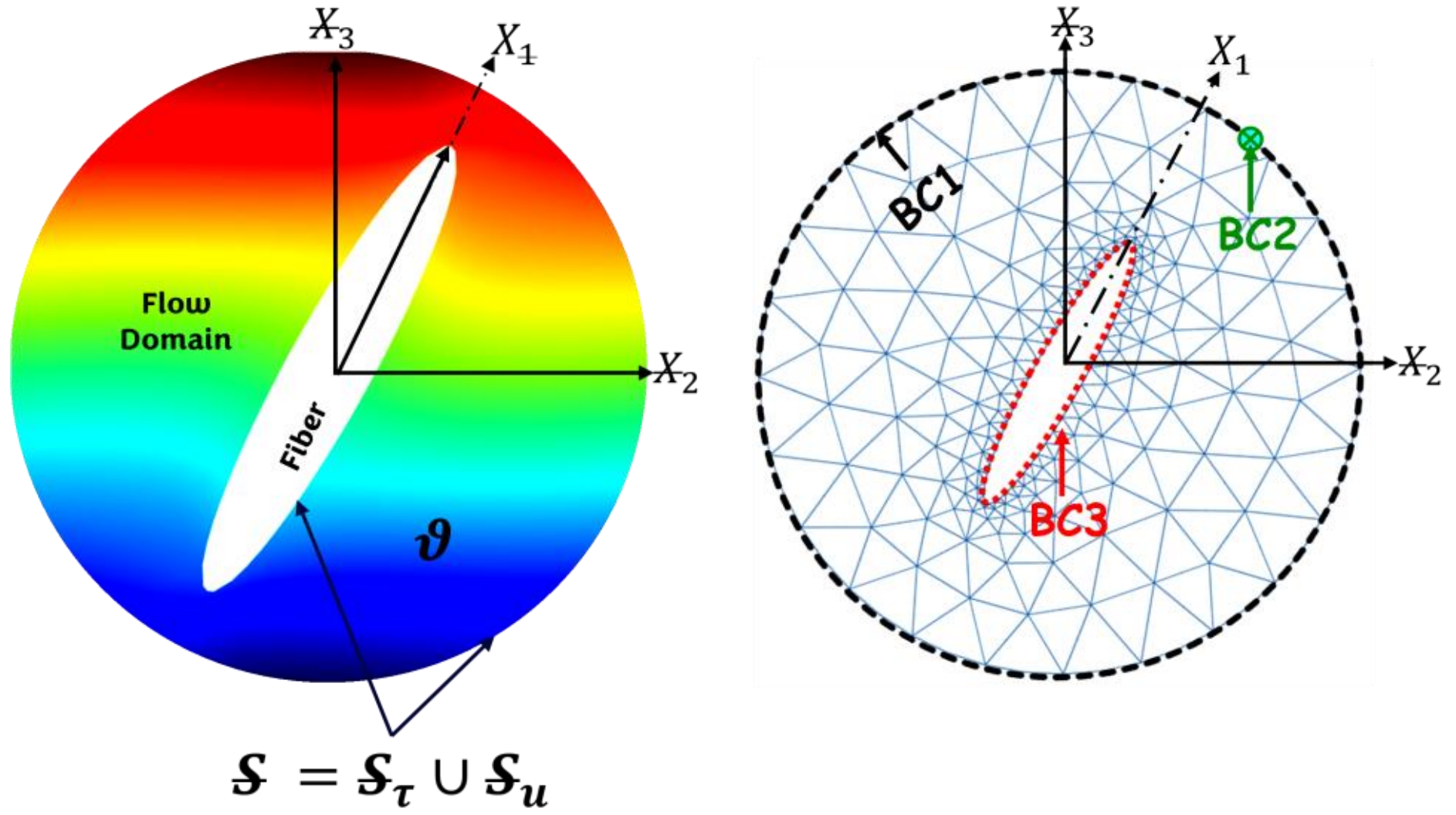

(a) (b)

Figure 5.3: FEA model showing (a) flow domain (b) prescribed boundary conditions.

Similar to previous single particle Newtonian fluid analyses [57], the velocities and velocity gradients of the prevailing flow are used to compute the far-field velocities on the fluid domain boundary $\dot{X}_i^{BC1}$ (cf. Figure 5.3b) of the micromodel as

$$\dot{X}_i^{BC1} = \dot{X}_i^{\infty} = Z_{X_{ji}}\dot{\mathcal{X}}_j^{\psi} + Z_{X_{mi}}Ł_{mn}\, Z_{X_{nj}}\Delta X_j^{BC1} \tag{5.90}$$

where $Z_{X_{ij}}$ is the local to global transformation tensor, $\dot{\mathcal{X}}_j^{\psi}$ is the flow-field velocity vector, $Ł_{mn}$ is the velocity gradient tensor in global reference frame and $\Delta \mathcal{X}_j$ is the position vector with respect to the particle's center. In 2D, $Z_{X_{ij}}$ is simply

$$Z_{X_{ij}} = \delta_{ij}\cos\phi - Б_{ij}\sin\phi \tag{5.91}$$

where $Б_{ij} = i - j$. Again, referring to Figure 5.3b, the velocity on the particle's surface $\dot{X}_i^{BC3}$ is computed from the particle's center translational and rotational velocities assuming rigid body motion which is written with respect to the particle's local reference axis as

$$\dot{X}_i^{BC3} = \dot{X}_i^{p} = Z_{X_{ji}}\dot{\mathcal{X}}_j^{c} + \epsilon_{ijk}Z_{\Theta_{jn}}\dot{\Theta}_n\Delta X_k^{BC3} \tag{5.92}$$

where $\dot{X}_i^c$ is the particle's center translational velocity vector and $\dot{\Theta}_i$ is the particle's angular velocity vector. In 2D, eqn. (5.92) above can be simplified to

$$\dot{X}_i^{BC3} = \dot{X}_i^{p} = Z_{X_{ji}}\dot{\mathcal{X}}_j^{c} + \dot{\phi}Б_{ji}\Delta X_j^{BC3}, \tag{5.93}$$

A pressure point constraint $p_{BC2}$ is imposed at a node on the far-field fluid domain (see, e.g., BC2 in Figure 5.3b) with a magnitude equal to the prescribed static fluid pressure $p_0$, i.e.

$$p_{BC2} = p_0 \tag{5.94}$$

We define a fluid domain size factor $\mathbb{m} = d_f/2И_3$ [57] (where $d_f$ is the diameter of the flow domain and $И_3$ is the major axis length of the particle). The flow domain size thus increases linearly with the size of the particle. In our analysis, we utilize a factor of $\mathbb{m} = 10$ which is determined to be sufficiently large to yield accurate results. The fluid domain discretization for the base case having a particle geometric aspect ratio $r_e = 6$ appears in Figure 5.4a&b where an increasing mesh density is used near the particle and particles tip.

All FEA simulations are performed with a 10-node quadratic, iso-parametric tetrahedral serendipity element as shown in Figure 5.4c.

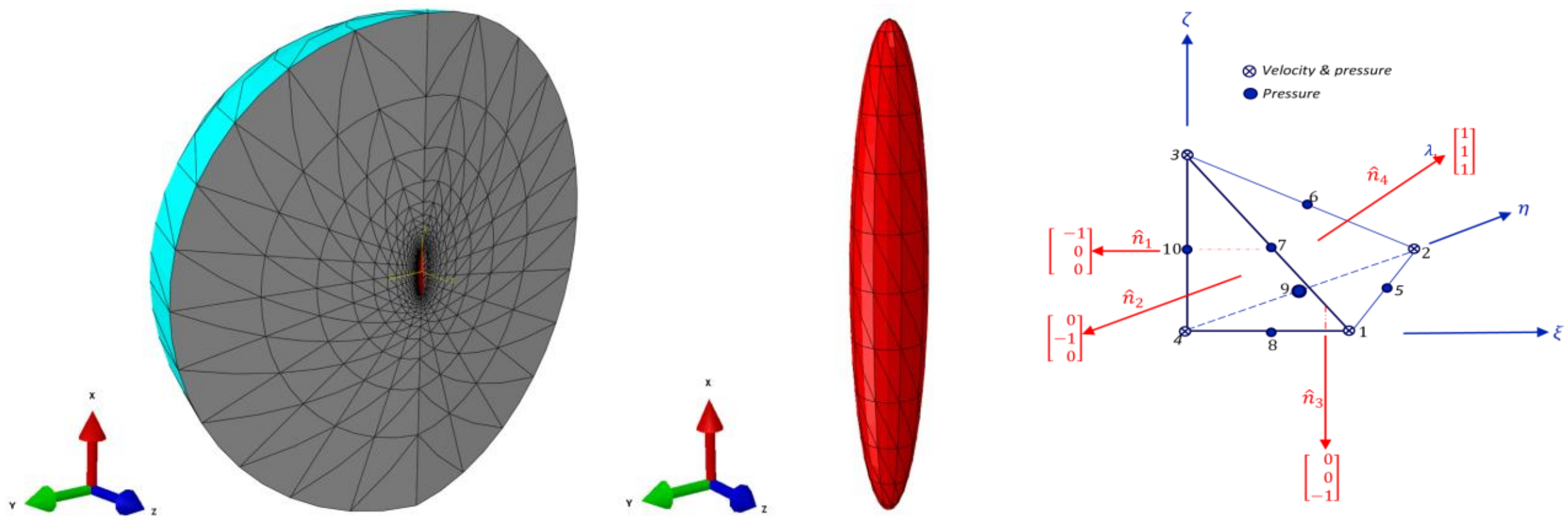

(a) (b) (c)

Figure 5.4: 3D Single suspended particle finite element model (a) Fluid domain discretization (b) magnified view of the domain mesh on the surface of the rigid particle (c) element selection with active degrees of freedom.

For the two-dimensional (2D) single fiber simulation, discretization of the micro-model fluid domain is achieved using a radial seed of 60-unit cells with a unidirectional geometric bias of 1.1 and circumferential seed of 60-unit cells resulting in a total of 1800 triangular elements as shown in Figure 5.5a. We employ a 6-node quadratic, iso-parametric triangle serendipity element (cf. Figure 5.5b) which has been found to give accurate results for low Reynolds number fluid flow problems [269].

For 2D sensitivity analysis involving very large fibers aspect ratio $r_e = 30$, the MATLAB inbuilt PDE modeler is used to discretize the fluid domain with increasing mesh density towards the fiber and fibers tip as shown in Figure 5.6a & b. The weak form of the nonlinear finite element equations may be transformed in the usual manner to a system of algebraic equations written in terms of the solution variable vector $\underline{u}$ and the global system residual vector $\underline{\Sigma}$ as

$$\underline{\Sigma} = \underline{\underline{K}}\left(\underline{u}\right)\underline{u} - \underline{f} \tag{5.95}$$

where $\underline{\underline{K}}$ is the global system 'stiffness' matrix, $\underline{u} = [\underline{v} \quad \underline{p}]^T$ is the primary solution vector containing nodal velocities $\underline{v}$ and pressures $\underline{p}$ and $\underline{f}$ is the secondary variable vector containing the associated nodal reaction forces and flow rates.

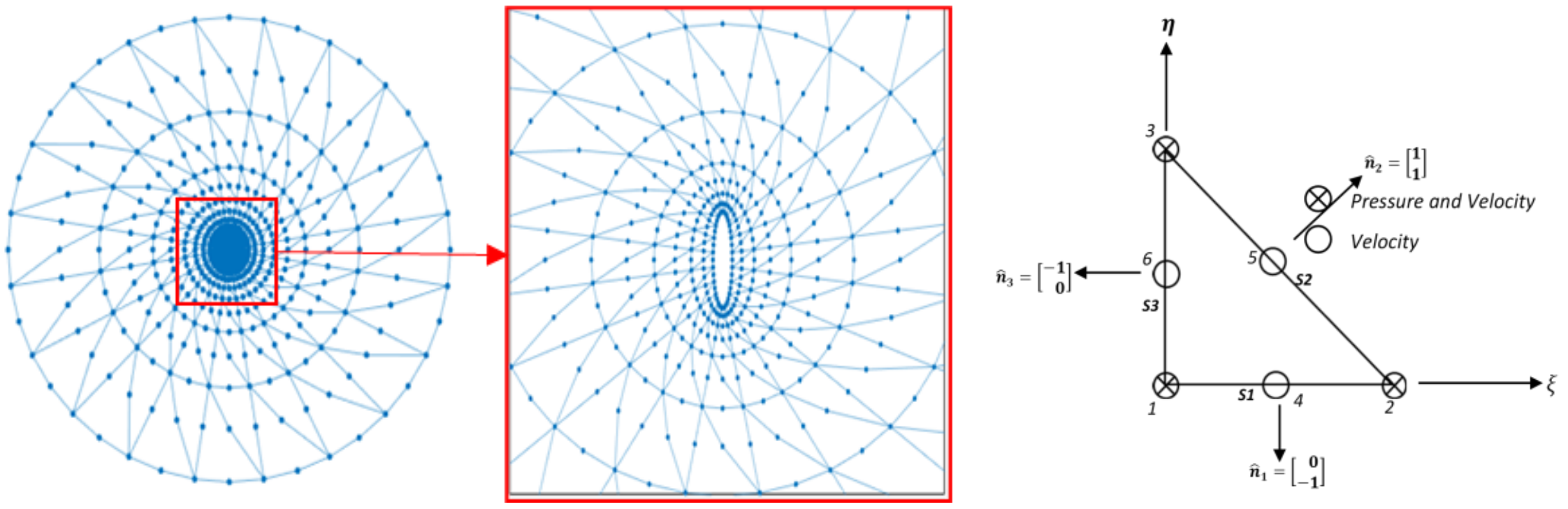

(a) (b)

Figure 5.5: 2D Single suspended particle finite element model (a) Fluid domain discretization (b) element selection with active degrees of freedom.

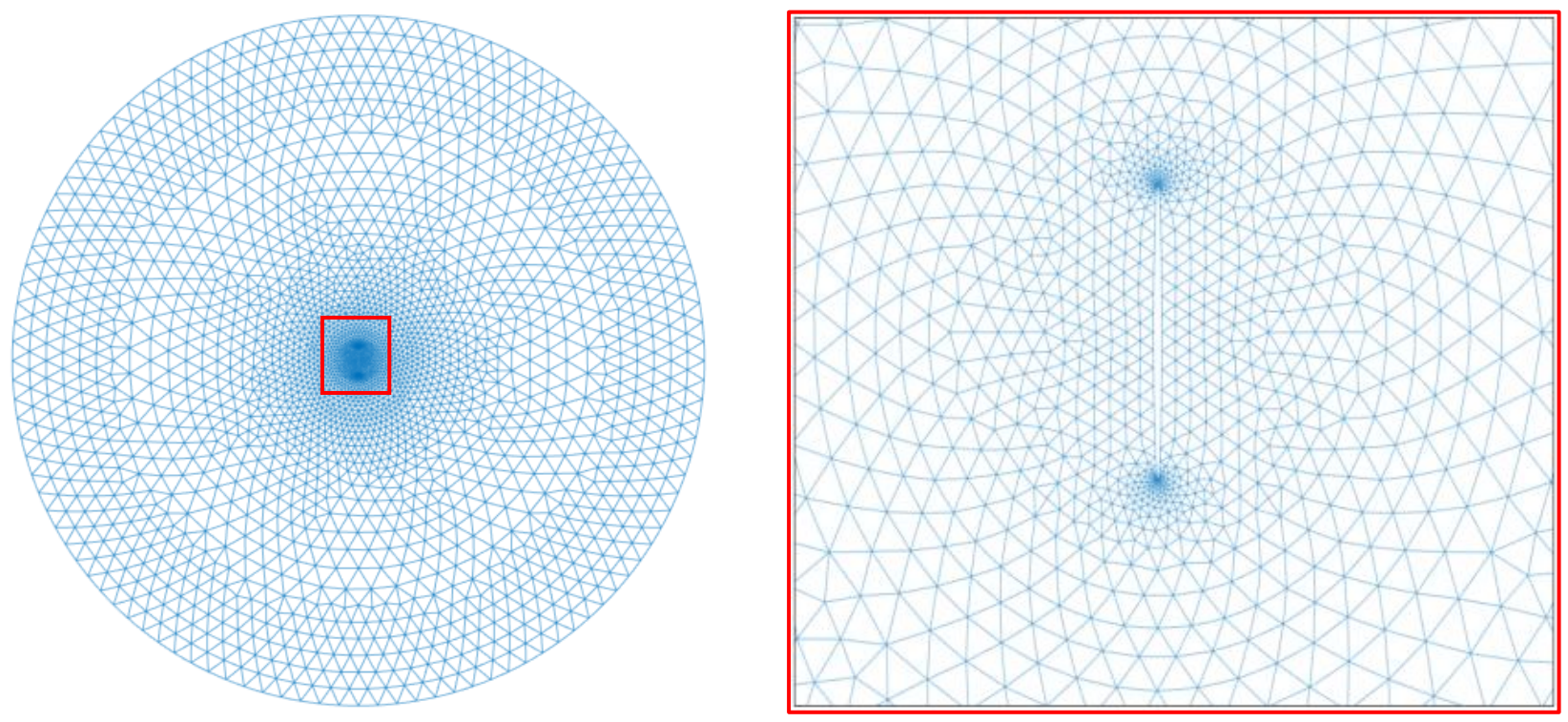

(a) (b)

Figure 5.6: (a) Fluid domain discretization (b) magnified view of the domain mesh around the rigid fiber

To simplify the solution procedure, the global system matrix is partitioned into essential $'e'$ (known) and free $'f'$ (unknown) degrees of freedom (dofs) as

$$\underline{\Sigma} = \begin{Bmatrix} \underline{\Sigma}_f \\ \underline{\Sigma}_e \end{Bmatrix} = \begin{Bmatrix} \underline{\underline{K}}_{ff} & \underline{\underline{K}}_{fe} \\ \underline{\underline{K}}_{ef} & \underline{\underline{K}}_{ee} \end{Bmatrix} \begin{Bmatrix} \underline{u}_f \\ \underline{u}_e \end{Bmatrix} - \begin{Bmatrix} \underline{f}_f \\ \underline{f}_e + \underline{\mathrm{g}}_e \end{Bmatrix} \tag{5.96}$$

where $\underline{u}_f$ & $\underline{g}_e$ are the unknown quantities to be computed in the finite element analysis. The unknown free velocity and pressure dofs in $\underline{u}_f$ are computed via a Newton Raphson iterative algorithm by zeroing the free residual vector $\underline{\Sigma}_f$. i.e $\underline{u}_f$ is iteratively updated until it approaches the actual solution according to

$$\underline{u}_f{}^{+} = \underline{u}_f{}^{-} - \underline{\underline{J}}_{ff}{}^{-1} \underline{\Sigma}_f \tag{5.97}$$

In the above, the Tangent Stiffness Matrix (TSM) or Jacobian $\underline{\underline{J}}_{ff}$ is obtained by differentiating the free residual vector $\underline{\Sigma}_f$ defined in eqn. (5.96) with respect to the free degrees of freedom $\underline{u}_f$ to obtain

$$\underline{\underline{J}}_{ff} = \frac{\partial \underline{\Sigma}_f}{\partial \underline{u}_f} = \frac{\partial \underline{\underline{K}}_{ff}}{\partial \underline{u}_f} \underline{u}_f + \frac{\partial \underline{\underline{K}}_{fe}}{\partial \underline{u}_f} \underline{u}_e + \underline{\underline{K}}_{ff} - \frac{\partial \underline{f}_f}{\partial \underline{u}_f} \tag{5.98}$$

For the linear system, i.e. $\underline{\underline{K}} \neq \underline{\underline{K}}\left(\underline{u}\right)$,

$$\underline{u}_f = \underline{\underline{K}}_{ff}{}^{-1} \left( \underline{f}_f - \underline{\underline{K}}_{fe} \underline{u}_e \right) \tag{5.99}$$

The unknown reactions forces and flow rates at the essential dofs in $\underline{g}_e$ are computed by setting the essential residual vector $\underline{\Sigma}_e = 0$ (cf. eqn. (5.96)) to obtain as

$$\underline{g}_e = \underline{\underline{K}}_{ef} \underline{u}_f + \underline{\underline{K}}_{ee} \underline{u}_e - \underline{f}_e \tag{5.100}$$

The global residual vector and Jacobian are assembled from individual element residual $\underline{\Sigma}^e$ and element tangent stiffness matrices $\underline{\underline{J}}^e$ in the usual manner. The element residual vector $\underline{\Sigma}^e$ is written in terms of the FEA integral equations as

$$\underline{\Sigma}^e = \begin{Bmatrix} \underline{\Sigma}_1^e \\ \underline{\Sigma}_2^e \end{Bmatrix} = \begin{Bmatrix} \int_{\vartheta^e} \underline{\omega}_1 \left(\underline{\nabla} \cdot \underline{v}\right) d\vartheta \\ \int_{\vartheta^e} \left(\underline{\underline{\nabla}}_S \cdot \underline{\omega}_2\right)^T \mu(\dot{\gamma}) \underline{\underline{C}}_o \left(\underline{\underline{\nabla}}_S \cdot \underline{v}\right) d\vartheta - \int_{\vartheta^e} \underline{p} \left(\underline{\nabla} \cdot \underline{\omega}_2\right) d\vartheta - \int_{\vartheta^e} \rho \underline{\omega}_2^T \underline{f} d\vartheta - \int_{S_T^e} \underline{\omega}_2^T \underline{\bar{t}}\, dS \end{Bmatrix} \tag{5.101}$$

where $\underline{\Sigma}_1^e$ & $\underline{\Sigma}_2^e$ are element residual vectors derived from mass and momentum conservation, respectively, $\underline{\omega}_1$ and $\underline{\omega}_2$ are the arbitrary FEA weighting functions on the continuity and momentum equation, respectively, $\underline{\nabla}$ and $\underline{\underline{\nabla}}_S$ are the gradient vector and symmetric gradient matrix operator, respectively, defined in [270], $\underline{p}$ and $\underline{v}$ are the pressure and velocity field variables, $\rho$ is the fluid density, $\mu(\dot{\gamma})$ is the non-Newtonian fluid viscosity, $\underline{\underline{C}}_o$ is a constant coefficient matrix, $\underline{\bar{t}}$ and $\underline{f}$ are the surface traction and the body force vectors, and $S_T^e$ and $\vartheta^e$ are the element surface and interior domains of integration, respectively. The element TSM $\underline{\underline{J}}^e$ is obtained by differentiating the element residual vector $\underline{\Sigma}^e$ with respect to the element solution variables $\underline{u}^e$ which contains $\underline{p}^e$ and $\underline{v}^e$ , i.e., $\underline{u}^e = \left[\underline{v}^e \quad \underline{p}^e\right]^T$ and

$$\underline{\underline{J}}^e = \frac{\partial \underline{\Sigma}^e}{\partial \underline{u}^e} = \frac{\partial}{\partial \underline{u}^e} \begin{Bmatrix} \underline{\Sigma}_1^e \\ \underline{\Sigma}_2^e \end{Bmatrix}, \qquad \underline{\underline{J}}^{eT} = \begin{bmatrix} \frac{\partial}{\partial \underline{v}^e} & \frac{\partial}{\partial \underline{p}^e} \end{bmatrix}^T \begin{Bmatrix} \underline{\Sigma}_1^e \\ \underline{\Sigma}_2^e \end{Bmatrix}^T \tag{5.102}$$

First order Façade derivatives are used to approximate the tangent stiffness matrix according to

$$\frac{\partial \underline{\Sigma}}{\partial \underline{u}} \Delta \underline{u} = \underline{\Sigma}\left(\underline{u} + \Delta \underline{u}\right) - \underline{\Sigma}\left(\underline{u}\right), \qquad \underline{\Sigma} = \underline{\Sigma}\left(\underline{u}\right) \tag{5.103}$$

which we apply to the continuity residual term $\underline{\Sigma}_1^e$ to obtain derivatives with respect to the velocity and pressure as

$$\frac{d\underline{\Sigma}_1^e}{d\underline{v}}\Delta\underline{v} = \int_{\vartheta^e} \underline{\omega}_1\left(\underline{\nabla}\cdot\Delta\underline{v}\right) d\vartheta, \qquad \frac{d\underline{\Sigma}_1^e}{d\underline{p}}\Delta\underline{p} = 0 \tag{5.104}$$

Similarly, derivatives of the momentum conservation term with respect to the solution variables after algebraic manipulations are, respectively, given as

$$\frac{d\underline{\Sigma}_2^e}{d\underline{v}}\Delta\underline{v} = \int_{\vartheta^e} \left(\underline{\underline{\nabla}}_s\underline{\omega}_2\right)^T \mu\underline{\underline{C}}_o\underline{\underline{\nabla}}_s\Delta\underline{v}d\vartheta + \int_{\vartheta^e} \frac{1}{\mu^2}\frac{1}{\dot{\gamma}}\frac{\partial\mu}{\partial\dot{\gamma}}\left[\left(\underline{\underline{\nabla}}_s\underline{\omega}_2\right)^T \mu\underline{\underline{C}}_o\underline{\underline{\nabla}}_s\underline{v}\right]\left[\left(\underline{\underline{\nabla}}_s\underline{v}\right)^T \mu\underline{\underline{C}}_o\underline{\underline{\nabla}}_s\Delta\underline{v}\right] d\vartheta \tag{5.105}$$

$$\frac{d\underline{\Sigma}_2^e}{d\underline{p}}\Delta\underline{p} = -\int_{\vartheta^e} \left(\underline{\nabla}\cdot\underline{\omega}_2\right)\Delta\underline{p}\, d\vartheta \tag{5.106}$$

It follows that the Galerkin formulation written as the element residual vector $\underline{\Sigma}^e$ and tangent stiffness matrix $\underline{\underline{J}}^e$ in tensorial representation are given respectively as

$$\underline{\Sigma}^e = \begin{Bmatrix} \int_{\vartheta^e} \underline{\underline{B}}_s^{e^T}\mu(\dot{\gamma})\underline{\underline{C}}_o\underline{\underline{B}}_s^e d\vartheta & -\int_{\vartheta^e} \underline{B}^{e^T}\underline{\Phi}^e d\vartheta \\ -\int_{\vartheta^e} \underline{\Phi}^{e^T}\underline{B}^e d\vartheta & \underline{\underline{0}} \end{Bmatrix} \begin{Bmatrix} \underline{v}^e \\ \underline{p}^e \end{Bmatrix} - \begin{Bmatrix} \int_{\vartheta^e} \rho\underline{\underline{N}}^{e^T}\underline{f}\, d\vartheta + \int_{S_{\mathrm{T}}^e} \underline{\underline{N}}^{e^T}\underline{\bar{t}}\, dS \\ \underline{0} \end{Bmatrix} \tag{5.107}$$

and

$$\underline{\underline{J}}^e = \frac{d\underline{\Sigma}^e}{d\underline{u}^e} = \tag{5.108}$$

$$\begin{Bmatrix} \int_{\vartheta^e} \underline{\underline{B}}_s^{e^T}\mu\underline{\underline{C}}_o\underline{\underline{B}}_s^e d\vartheta + \int_{\vartheta^e} \frac{1}{\mu^2}\frac{1}{\dot{\gamma}}\frac{\partial\mu}{\partial\dot{\gamma}}\left(\underline{\underline{B}}_s^{e^T}\mu\underline{\underline{C}}_o\underline{\underline{B}}_s^e\underline{v}^e\right)\left(\underline{v}^{e^T}\underline{\underline{B}}_s^{e^T}\mu(\dot{\gamma})\underline{\underline{C}}_o^T\underline{\underline{B}}_s^e\right) d\vartheta & -\int_{\vartheta^e} \underline{B}^{e^T}\underline{\Phi}^e d\vartheta \\ -\int_{\vartheta^e} \underline{\Phi}^{e^T}\underline{B}^e d\vartheta & \underline{\underline{0}} \end{Bmatrix}$$

where

$\underline{\Phi}^e$ and $\underline{\underline{N}}^e$ are the pressure and velocity interpolation functions, respectively,
$\underline{B}^e$ and $\underline{\underline{B}}_s^e$ are 'strain' displacement matrices
$\underline{v}^e$ and $\underline{p}^e$ are respectively the velocities and pressures degrees-of-freedom (dof) at the respective element nodes
$S^e$ and $\vartheta^e$ are the element boundary surfaces and domain of integration, respectively.

In eqn. (5.108), $\dot{\gamma}$ is the scalar magnitude of the strain rate tensor $\underline{\underline{\dot{\gamma}}}$ which may be written in terms of FEA quantities as

$$\dot{\gamma} = \sqrt{\frac{1}{2}\underline{\underline{\dot{\gamma}}}:\underline{\underline{\dot{\gamma}}}} = \sqrt{\left(\underline{\underline{\nabla_s v}}\right)^T \underline{\underline{C_o}} \left(\underline{\underline{\nabla_s v}}\right)}, \qquad \dot{\gamma} = \sqrt{\underline{v}^{e^T} \underline{\underline{B_s^e}}^T \mu(\dot{\gamma}) \underline{\underline{C_o}}\, \underline{\underline{B_s^e}}\, \underline{v}^e} \tag{5.109}$$

In this work, we consider the non-Newtonian viscosity $\mu(\dot{\gamma})$ as that of a power-law shear-thinning fluid given as

$$\mu = m\dot{\gamma}^{n-1} \tag{5.110}$$

where $m$ is the flow consistency coefficient in $Pa \cdot s^n$ and $n$ is the power-law index, and $\dot{\gamma}$ is the scalar magnitude of the deformation tensor $\dot{\gamma}_{ij}$. In the second integral of the momentum equation Jacobian in eqn. (5.108) above, it is convenient to introduce a variable $\alpha = 1/(\mu^2\dot{\gamma})\,(\partial\mu/\partial\dot{\gamma})$ to simplify the expression and make it generally applicable to other GNF fluids. It follows that $\alpha$ can be written for the power-law fluid as

$$\alpha = \frac{1}{\mu^2}\frac{1}{\dot{\gamma}}\frac{\partial\mu}{\partial\dot{\gamma}} = \frac{1}{\mu\dot{\gamma}^2}(n-1) \tag{5.111}$$

Alternatively, for a Carreau-Yasuda fluid, the expression for $\mu$ and $\alpha$ are, respectively,

$$\frac{\mu - \mu_\infty}{\mu_0 - \mu_\infty} = \{1 + (\lambda\dot{\gamma})^{\mathrm{a}}\}^{(n-1)/\mathrm{a}} \quad and \quad \alpha = \frac{1}{\mu^2}\frac{1}{\dot{\gamma}}\frac{\partial\mu}{\partial\dot{\gamma}} = \frac{1}{\dot{\gamma}^2}\frac{\mu - \mu_\infty}{\mu^2}\left\{\frac{n-1}{1 + (\lambda\dot{\gamma})^{-\mathrm{a}}}\right\} \tag{5.112}$$

where, $\mu_0$ is the zero-shear viscosity, $\mu_\infty$ is an infinite-shear viscosity, $\lambda$ is a time constant, and a is a fitting parameter.

### *5.6 Single Particle Motion with GNF Rheology*

In our numerical approach, the particle's motion is computed based on an appropriate explicit numerical ordinary differential equation solution technique by calculating its linear and rotational velocities that results in a zero net hydrodynamic force and torque acting on the particle's surface. Again, we adopt Newton-Raphson's iterative method to determine the nonlinear solution of particle's translational and rotational velocities as

$$\underline{\dot{Y}}^{+} = \underline{\dot{Y}}^{-} - \underline{\underline{J_H}}^{-1}\underline{\Sigma_H} \tag{5.113}$$

where $\underline{\dot{Y}}$ contains the particle's linear velocities $\underline{\dot{X}}^c$ and rotational velocity $\underline{\dot{\Psi}}$, i.e., $\underline{\dot{Y}} = \left[\underline{\dot{X}}^c \quad \underline{\dot{\Psi}}\right]^T$ and $\underline{\Sigma_H}$ is the particle hydrodynamic residual vector which is composed of the particle's hydrodynamic forces $\underline{F_H}$ and couple $\underline{Q_H}$ , i.e., $\underline{\Sigma_H} = \left[\underline{F_H} \quad \underline{Q_H}\right]^T$ as a function of the particle's velocity, i.e., $\underline{\Sigma_H} = \underline{\Sigma_H}\left(\underline{\dot{Y}}\right)$. Since calculations are performed with respect to the particle's local reference frame, the particle's velocity vector is transformed to global coordinate system according to the eqn. (5.114)

$$\underline{\dot{\not{Y}}} = \underline{\underline{Z_{\dot{Y}}}}\,\underline{\dot{Y}} \tag{5.114}$$

where variables on the global reference frame are accented by a strikethrough and the particle's velocity transformation tensor $\underline{\underline{Z_{\dot{Y}}}}$ 3D and 2D are respectively given by

$$\underline{\underline{Z_{\dot{Y}}}} = \begin{bmatrix} \underline{\underline{Z_X}} & \underline{\underline{0}} \\ \underline{\underline{0}}^T & \underline{\underline{Z_\Theta^{-1}}} \end{bmatrix}, \qquad \underline{\underline{Z_{\dot{Y}}}} = \begin{bmatrix} \underline{\underline{Z_X}} & 0 \\ 0^T & 1 \end{bmatrix} \tag{5.115}$$

We calculate the net hydrodynamic force vector $\underline{F_H}$ and couple $\underline{Q_H}$ on the particle by vector summation of the nodal reactions forces and torques on the particle surface as

$$\underline{F_H} = -\sum_{k}^{n_k} \underline{g_e^{(k)}}, \qquad \underline{Q_H} = -\sum_{k}^{n_k} \underline{\Delta X^{(k)}} \times \underline{g_e^{(k)}} \tag{5.116}$$

where $\underline{\Delta X^{(k)}}$, and $\underline{g_e^{(k)}}$ are the position vector and the nodal reaction force vector at the $k^{\text{th}}$ node on the particle surface ($BC3$), respectively, and $n_k$ is the total number of nodes on $BC3$. The particle hydrodynamic Jacobian $\underline{\underline{J_H}}$ in eqn. (5.113) above is obtained by differentiating the components of the particle hydrodynamic residual vector $\underline{\Sigma}_H$ with respect to components of the particle's velocity vector $\underline{\dot{Y}}$ as

$$\underline{\underline{J}}_H = \frac{\partial \underline{\Sigma}_H}{\partial \underline{\dot{Y}}} = \frac{\partial}{\partial \underline{\dot{Y}}}\left[\underline{F}_H \quad \underline{Q}_H\right]^T = \left[-\sum_k^{n_k} \frac{\partial \underline{g}_e^{(k)}}{\partial \underline{\dot{Y}}} \quad -\sum_k^{n_k} \underline{\Delta X}^{(k)} \times \frac{\partial \underline{g}_e^{(k)}}{\partial \underline{\dot{Y}}}\right]^T \tag{5.117}$$

Differentiating the global system FEA residual vector $\underline{\Sigma}$ in eqn. *(5.96)* with respect to the particle velocity vector $\underline{\dot{Y}}$ we obtain the derivative of the nodal reaction force vector $d\underline{g}_e/d\underline{\dot{Y}}$ in eqn. (5.117) as

$$\frac{d\underline{g}_e}{d\underline{\dot{Y}}} = \left\{\frac{\partial \underline{\underline{K}}_{ef}}{\partial \underline{u}_e}\underline{u}_f + \frac{\partial \underline{\underline{K}}_{ee}}{\partial \underline{u}_e}\underline{u}_e + 2\underline{\underline{K}}_{ee} - \frac{d\underline{f}_e}{d\underline{u}_e}\right\}\frac{d\underline{u}_e}{d\underline{\dot{Y}}} + \left\{\frac{\partial \underline{\underline{K}}_{ef}}{\partial \underline{\overline{u}}_f}\underline{u}_f + \frac{\partial \underline{\underline{K}}_{ee}}{\partial \underline{\overline{u}}_f}\underline{u}_e + 2\underline{\underline{K}}_{ef} - \frac{d\underline{f}_e}{d\underline{\overline{u}}_f}\right\}\frac{d\underline{u}_f}{d\underline{\dot{Y}}} \tag{5.118}$$

where the derivative $d\underline{u}_f/d\underline{\dot{Y}}$ is written in terms of the derivative $d\underline{u}_e/d\underline{\dot{Y}}$ as

$$\frac{d\underline{u}_f}{d\underline{\dot{Y}}} = -\left\{\frac{\partial \underline{\underline{K}}_{ff}}{\partial \underline{\overline{u}}_f}\underline{u}_f + \frac{\partial \underline{\underline{K}}_{fe}}{\partial \underline{\overline{u}}_f}\underline{u}_e + 2\underline{\underline{K}}_{ff} - \frac{d\underline{f}_f}{d\underline{\overline{u}}_f}\right\}^{-1}\left\{\frac{\partial \underline{\underline{K}}_{ff}}{\partial \underline{u}_e}\underline{u}_f + \frac{\partial \underline{\underline{K}}_{fe}}{\partial \underline{u}_e}\underline{u}_e + 2\underline{\underline{K}}_{fe} - \frac{d\underline{f}_f}{d\underline{u}_e}\right\}\frac{d\underline{u}_e}{d\underline{\dot{Y}}} \tag{5.119}$$

To obtain the FEA model derivatives in the above, we differentiate the global FEA system residual $\underline{\Sigma}$ in eqn. *(5.96)* with respect to the solution variable $\underline{u}$ to obtain the global FEA system Jacobian $\underline{\underline{J}}$ as

$$\underline{\underline{J}} = \frac{d\underline{\Sigma}}{d\underline{u}} = \begin{Bmatrix} \underline{\underline{J}}_{ff} & \underline{\underline{J}}_{fe} \\ \underline{\underline{J}}_{ef} & \underline{\underline{J}}_{ee} \end{Bmatrix}$$
$$= \begin{Bmatrix} \left\{\frac{\partial \underline{\underline{K}}_{ff}}{\partial \underline{u}_f}\underline{u}_f + \frac{\partial \underline{\underline{K}}_{fe}}{\partial \underline{u}_f}\underline{u}_e + \underline{\underline{K}}_{ff} - \frac{\partial \underline{f}_f}{\partial \underline{u}_f}\right\} & \left\{\frac{\partial \underline{\underline{K}}_{ff}}{\partial \underline{u}_e}\underline{u}_f + \frac{\partial \underline{\underline{K}}_{fe}}{\partial \underline{u}_e}\underline{u}_e + \underline{\underline{K}}_{fe} - \frac{\partial \underline{f}_f}{\partial \underline{u}_e}\right\} \\ \left\{\frac{\partial \underline{\underline{K}}_{ef}}{\partial \underline{u}_f}\underline{u}_f + \frac{\partial \underline{\underline{K}}_{ee}}{\partial \underline{u}_f}\underline{u}_e + \underline{\underline{K}}_{ef} - \frac{\partial \underline{f}_e}{\partial \underline{u}_f}\right\} & \left\{\frac{\partial \underline{\underline{K}}_{ef}}{\partial \underline{u}_e}\underline{u}_f + \frac{\partial \underline{\underline{K}}_{ee}}{\partial \underline{u}_e}\underline{u}_e + \underline{\underline{K}}_{ee} - \frac{\partial \underline{f}_e}{\partial \underline{u}_e}\right\} \end{Bmatrix} \tag{5.120}$$

where eqn. (5.120) has been expanded to include all free and essential degrees of freedom in $\underline{u} = \{\underline{u}_f \quad \underline{u}_e\}^T$. In addition, the nodal reaction force vector derivative $d\underline{g}_e/d\underline{\dot{Y}}$ in eqn. (5.118) is written in terms of the submatrices of the global FEA system Jacobian $\underline{\underline{J}}$ as

$$\frac{d\underline{g}_e}{d\underline{\dot{Y}}} = \left\{\underline{\underline{K}}_{ee} + \underline{\underline{J}}_{ee}\right\}\frac{d\underline{u}_e}{d\underline{\dot{X}}} + \left\{\underline{\underline{K}}_{ef} + \underline{\underline{J}}_{ef}\right\}\frac{d\underline{u}_f}{d\underline{\dot{Y}}} \tag{5.121}$$

Likewise, the derivative $d\underline{u}_f/d\underline{\dot{Y}}$ in eqn. (5.119) is also written in terms of the submatrices of the global system Jacobian $\underline{\underline{J}}$ as

$$\frac{d\underline{u}_f}{d\underline{\dot{Y}}} = -\left\{\underline{\underline{K}}_{ff} + \underline{\underline{J}}_{ff}\right\}^{-1}\left\{\underline{\underline{K}}_{fe} + \underline{\underline{J}}_{fe}\right\}\frac{d\underline{u}_e}{d\underline{\dot{Y}}} \tag{5.122}$$

Again, for the linear consideration i.e. $\underline{\underline{K}} \neq \underline{\underline{K}}\left(\underline{u}\right)$, $\underline{\underline{J}} = 0$ the derivative of the nodal reaction force vector derivative $d\underline{g}_e/d\underline{\dot{Y}}$ in eqn. (5.118) reduces to

$$\frac{\partial \underline{g}_e}{\partial \underline{\dot{Y}}} = \left(\underline{\underline{K}}_{ee} - \underline{\underline{K}}_{ef}\underline{\underline{K}}_{ff}^{-1}\,\underline{\underline{K}}_{fe}\right)\frac{\partial \underline{u}_e}{\partial \underline{\dot{Y}}} \tag{5.123}$$

Given the initial condition of the particle, $\underline{Y}^{j-1}$ at any instant with an associated velocity $\underline{\dot{Y}}^{j-1}$ at each *j^th^* time step, we update particle's position and orientation $\underline{Y}^{j}$ using on an explicit fourth order Runge-Kutta method. i.e.

$$\underline{Y}^{j} = \underline{Y}^{j-1} + \frac{\Delta t}{6}\left[\underline{\mathcal{K}}_1^{j-1} + 2\underline{\mathcal{K}}_2^{j-1} + 2\underline{\mathcal{K}}_3^{j-1} + \underline{\mathcal{K}}_4^{j-1}\right] \tag{5.124}$$

where

$$\begin{aligned} &\underline{\mathcal{K}}_1^{j-1} = f_Y\left(t^{j-1}, \underline{Y}^{j-1}\right) = \underline{\dot{Y}}^{j-1},\ \underline{\mathcal{K}}_2^{j-1} = f_Y\left(t^{j-1} + {}^{\Delta t}/_2, \underline{Y}^{j-1} + {}^{\Delta t}/_2\,\underline{\mathcal{K}}_1^{j-1}\right) \\ &\underline{\mathcal{K}}_3^{j-1} = f_Y\left(t^{j-1} + {}^{\Delta t}/_2, \underline{Y}^{j-1} + {}^{\Delta t}/_2\,\underline{\mathcal{K}}_2^{j-1}\right),\ \underline{\mathcal{K}}_4^{j-1} = f_Y\left(t^{j-1} + \Delta t, \underline{Y}^{j-1} + \Delta t\,\underline{\mathcal{K}}_3^{j-1}\right) \end{aligned} \tag{5.125}$$

and the function $f_Y$ is used to evaluate the particles velocities $\underline{\dot{Y}}$ at time $t$ and position $\underline{Y}$

### *5.7 Cylindrical Particle Geometry*

In reality, the geometry of pristine particle consolidations present within a typical polymer composite bead are not ellipsoidal in shape with smooth edges but are better represented by cylindrical particles. Moreover, the chopped ends of the particles reinforcement do not possess a clearly defined tip as the ellipsoid but are characterized by sharp geometrical transitions at the particle terminations that likely result in pressure singularities. Unfortunately, besides other drawbacks, Jeffery's model equations are only applicable to ellipsoidal shaped particles and cannot model arbitrary shaped particles, however, our FEA simulation has the advantage of modelling complex particle shapes. To investigate the existence of exacerbated pressure extremes at the particle ends, we consider a cylindrical shaped particle in our FEA simulation choosing a cylindrical aspect ratio that yields the equivalent hydrodynamic ellipsoidal aspect ratio for the base case (i.e. $r_e = 6$). We develop a fluid domain mesh using ABAQUS Std. (Simulia ABAQUS, Dassault Systemes SE, Velizy-Villacoublay, France) for the single cylinder suspension using similar fluid domain size ratio, 10 times the cylinder length as shown in Figure 5.7a below. Mesh refinement zone close the cylinder surface is defined to accurately capture the field response on the particles surface (cf. Figure 5.7b). As would become evident in subsequent chapters, the pressure at the particles tip is dependent on the tip curvature and aspect ratio. With ellipsoidal shaped particles, both geometric attributes are interdependent and cannot be decoupled which limits our understanding of the individual contribution of both attributes to the surface pressure at the particles tip. With cylindrical shaped particles, however, we can independently study the individual contribution of both geometric attributes to the tip pressure response. In our analysis, we consider different end conditions (i.e. edge curvature radii - $r_\kappa$) as shown in Figure 5.7c ranging from small fillet radius to perfectly hemispherical (i.e. $0.05 \leq \bar{r}_\kappa \leq 0.5$) where $\bar{r}_\kappa = r_\kappa / И_1$ is the normalized curvature radius and $И_1$ is the cylinder diameter. By adjusting the cylindrical height while maintaining a constant diameter, $И_1$ we determine their respective cylinders geometric aspect ratio that are hydrodynamically equivalent to the ellipsoidal aspect ratio of $r_e = 6$ by matching output of the dynamic response. After careful determination of cylinder appropriate heights, the results of the evolution of the angular velocity along Jeffery's orbit for the various cylindrical particles with different end conditions are benchmarked with angular velocity of the ellipsoidal particle with aspect ratio of $r_e = 6$ using the same fluid viscosity and shear rate. To study the independent effect of the cylinder aspect ratio on the pressure response while maintaining a constant particle end curvature, the cylinder with the hemispherical end was chosen. For objectivity, the aspect ratio was varied by adjusting the length of the straight section of the cylinder while retaining a constant mesh for the hemispherical curved surface. Moreover, since only the cylinder with the hemispherical end has clearly defined unique tips where the surface pressure extremes are expected to occur, it provides a biased means for studying the decoupled effect of aspect ratio on the tip pressure response compared to the ellipsoid. In our investigation, we consider five (5) aspect ratios for the cylinder with the hemispherical end ranging from $7.0 \leq r_c \leq 7.4$ in steps of 0.1.

### *5.8 Validation of FEA Model Development*

To validate our FEA model-based particle motion simulations to calculations performed with Jeffery's equations, we first define the particle surface pressure $\overline{p}$ in dimensionless form as

$$\overline{p} = \frac{p - p_0}{\mu_1 \dot{\gamma}_c} \tag{5.126}$$

where $\dot{\gamma}_c$ is a characteristic strain rate of the flow-field. For a given $\mu_1$ and $\dot{\gamma}_c$, $\overline{p}$ is evaluated from eqn. (5.126) where $p$ is computed from Jeffery's model (cf. eqn. (5.1)) and similarly from the nodal pressure solution of the FEA model described above.

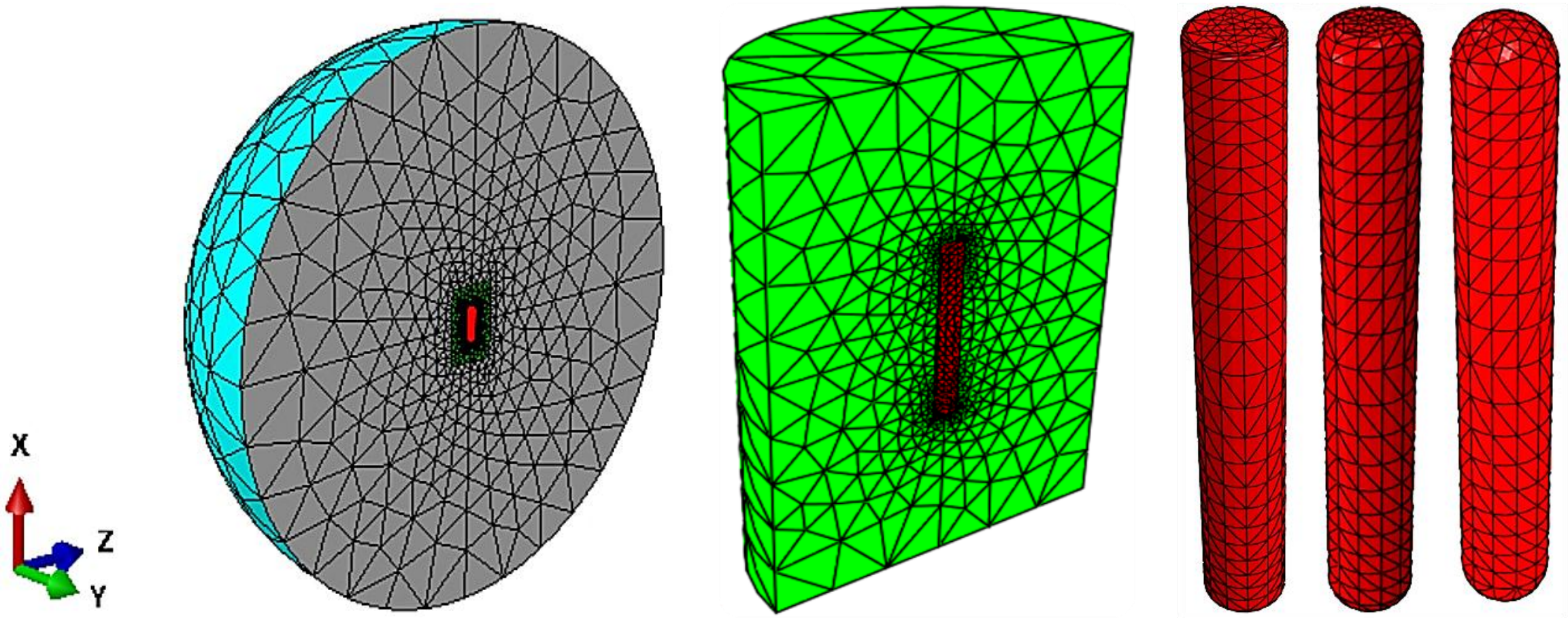

(a) (b) (c)

Figure 5.7: (a) Fluid domain discretization around the cylindrical particle (b) mesh refinement around the cylindrical particle surface (c) cylindrical particle with different end conditions (edge curvature radius).

Likewise, the flow-field velocity magnitude is normalized with respect to the tangential velocity at the particle's tip is given as

$$\bar{v} = |\underline{\dot{X}}|/|\underline{\dot{X}}^t|, \qquad \underline{\dot{X}}^t = \underline{\Theta} \times \underline{X}^t \tag{5.127}$$

where $\underline{X}^t$ is the position vector at particle's tip defined by the major axis length. To ensure consistency between the Jeffery's model equations and Finite Element Analysis (FEA) simulation results, we consider the particle's motion and surface pressure distribution for the case of a single rigid ellipsoidal particle suspended in viscous homogenous Newtonian (i.e., power-law index $n$ = 1) flow. The FEA model is shown to exactly match Jeffery's results for a range of particle aspect ratios including $r_e = 1, 2, 3, 6,$ and 10 (cf. Figure 5.8a for $\dot{\phi}$ and Figure 5.8b for $\bar{P}$). $\bar{P}$ is the dimensionless pressure at the particle's tip.

Additionally, Jeffery's orbit exactly matches our FEA results for the various flow conditions described above as shown in Figure 5.8c&d which show components of the particle unit vector $\rho_i$, and maximum and minimum normalized surface pressure $\tilde{p}$. Results in Figure 5.8a & b are for one period of Jeffery's orbit, however, given that values at the end point exactly match within 0.25%, we expect the accuracy of our numerical approach to remain as particle rotations continue.

The FEA results of the particle's angular velocity and tip surface pressure for the suspended rigid particle motion in simple shear flow in both 2D and 3D space are compared and validated against their corresponding reference counterparts computed from Jeffery's analytic equations as shown in Figure 5.9. We likewise observe very good agreement in the responses obtained from both FEA and Jeffery's solution irrespective of the dimensional space. While the particle's in-plane angular velocity is unaffected by the dimensional space evident from the overlapping curves in Figure 5.9a, the same is not the case for the particle's surface pressure as the pressure response magnitude is observed to reduce significantly with reduction in dimensionality of the computational space as can be observed from Figure 5.9b.

Figure 5.10 shows the evolution of the maximum (red trend) and minimum (cyan trend) pressure on the fiber surface over the tumbling period from Jeffery's (black markers) and FEA simulation (colored trends) in both 2D (dashed trend) and 3D (continuous trend) space. The location of the pressure extremes varies from point to point on the fiber's surface during its motion along Jeffery's orbit. As a result, the pressure extreme depends on the mesh refinement on the fiber surface which results in minor discrepancies observed between the extreme pressure profiles obtained from Jeffery's exact solution and FEA simulation in Figure 5.10.

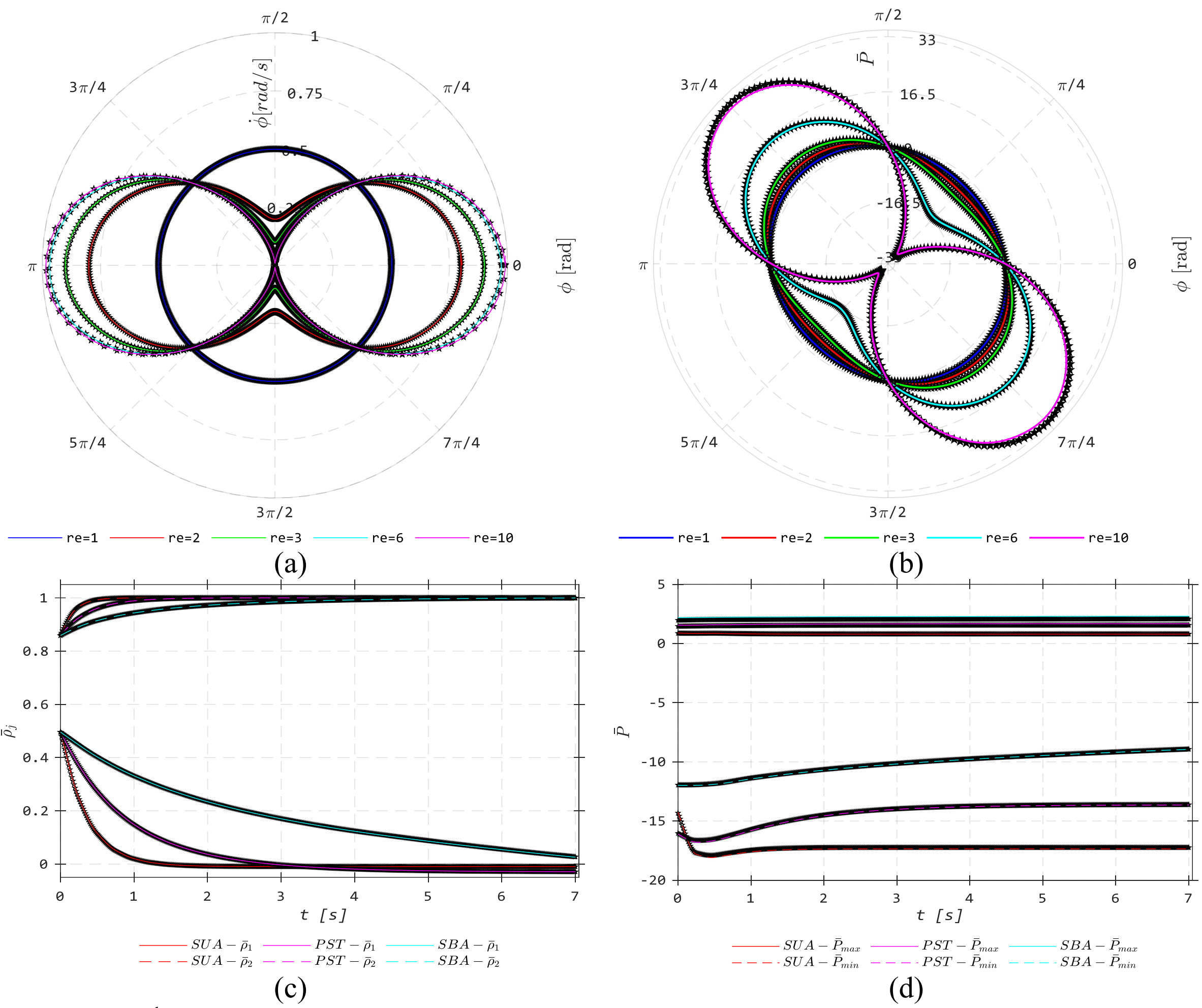

Figure 5.8[1]: FEA (colored lines) and Jeffery's results (black markers) of the evolution of the particle's (a) angular velocity, & (b) tip pressure, in simple shear flow for particle tumbling in the shear plane with different aspect ratios $1 \leq r_e \leq 10$; (c) orientation components, and (d) minimum (dashed) and maximum (continuous) surface pressure for particle with initial orientation, $\phi^0 = \pi/3, \theta^0 = 11\pi/24, \psi^0 = 0$ suspended in different combined flow types - SUA, PST and SBA with $\dot{\gamma}/\dot{\varepsilon} = 1$.

[1] Results of the 3rd component of the particle's orientation vector (i.e. $\bar{\rho}_3$) is implicit given the normalization condition $\bar{\rho}_i\bar{\rho}_i = 1$.

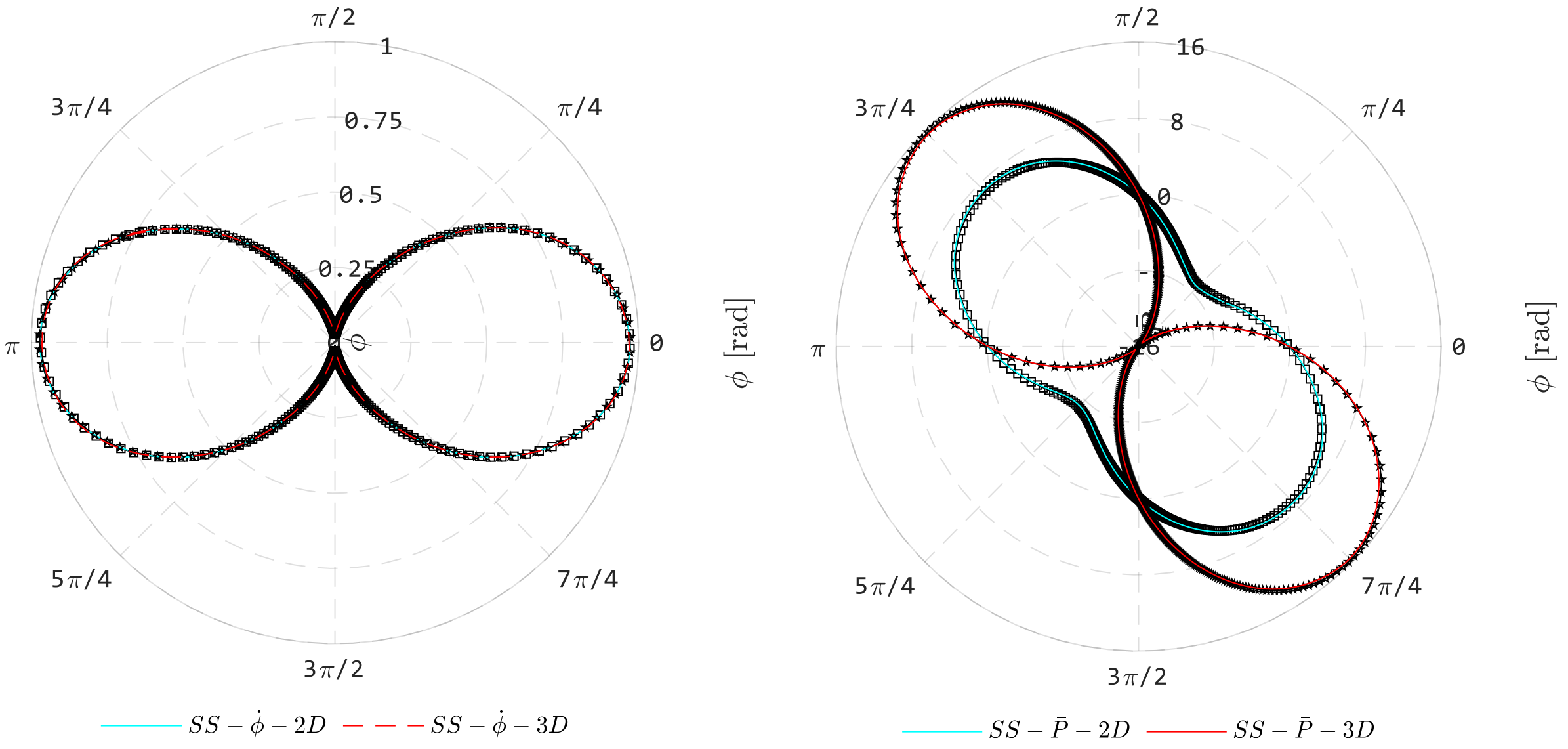

(a) (b)

Figure 5.9: FEA (colored lines) and Jeffery's results (black markers) of the evolution of the particle's (a) in-plane angular velocity (b) tip pressure, in simple shear flow and for in-plane tumbling of the particle in both 2D space (cyan curve) and 3D space (red curve) with $r_e = 6$.

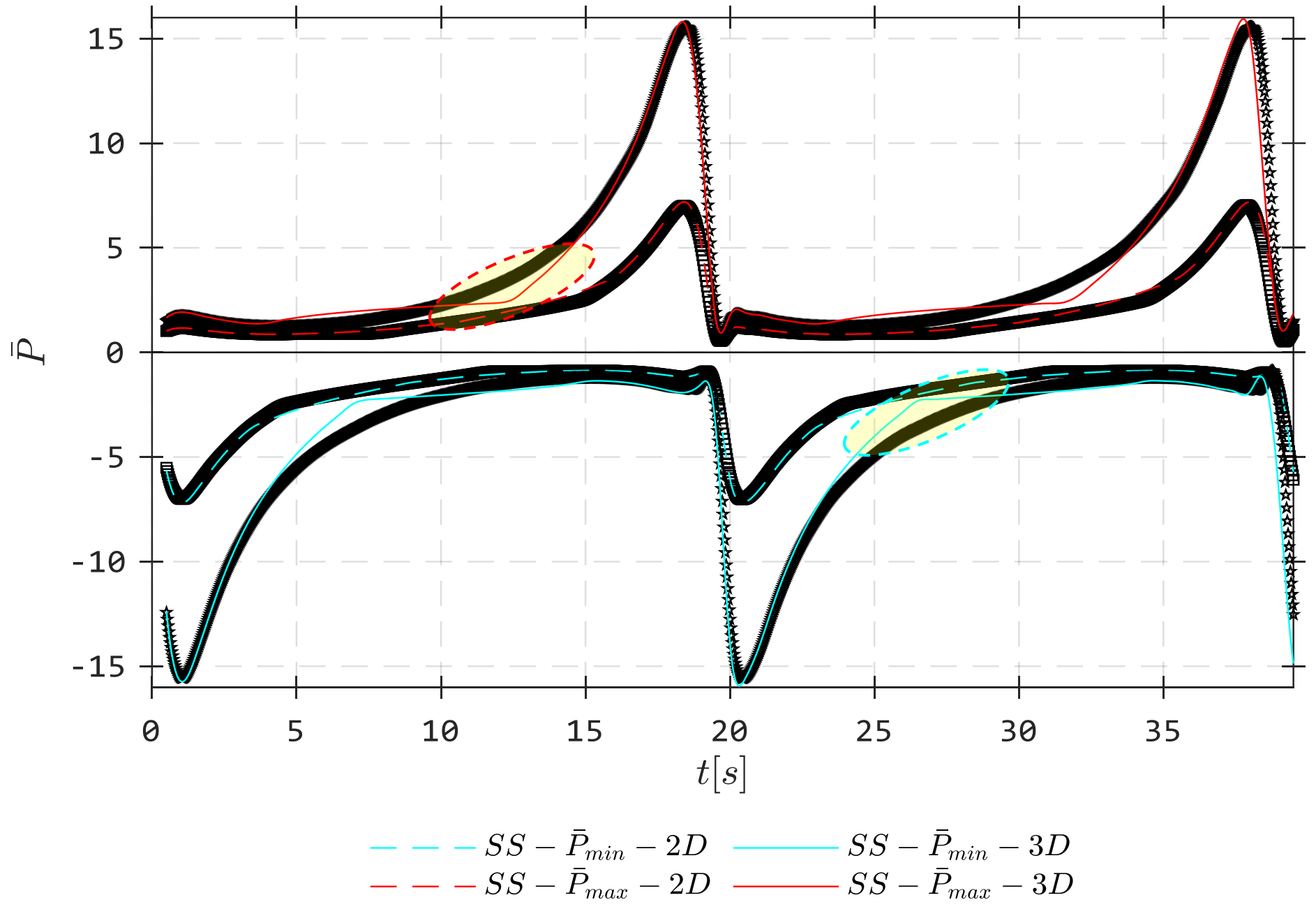

Figure 5.10: FEA (colored lines) and Jeffery's results (black markers) of the evolution of the maximum (red lines) and minimum (cyan lines) surface pressure for particle tumbling in in 2D (dashed line) and 3D (continuous line) space.

The GNF power law FEA model development is validated by benchmarking pressure response obtained from the custom-built MATLAB FEA simulation for a single steady state condition and fiber configuration with outputs obtained from a similar simulation developed with COMSOL Multiphysics software (COMSOL, Inc., Burlington, MA, USA) using same model input. A fibers geometric aspect ratio $r_e = 6$ is used for the validation exercise, and a simple shear flow field with a shear rate of $\dot{\gamma} = 1\ s^{-1}$ is imposed. We consider two different power law fluid definition with different flow behavior index $n$, for the first case (a) $n = 0.2$, and for the second case (b) $n = 1.$, both cases having a consistency coefficient $m = 1\ Pa \cdot s^n$. An initial fiber configuration corresponding to an orientation $\phi_0 = -0.7762\ rad$ and angular velocity of $\dot{\phi}_0 = -0.5087\ rads^{-1}$ has been used for the steady state analysis which is where the first minimum pressure peak occurs on the fibers surface during its evolution along Jeffery's orbit. The result of the pressure distribution for both cases presented in Figure 5.11 and the pressure extremes on the fiber's surface in

Table 5.2 shows there is good agreement between COMSOL simulation and inbuilt MATLAB FEA simulations. We observe a maximum discrepancy in pressure limits of about 6%.

The 3D FEA formulations are direct extensions in dimensionality to the 2D FEA model and the results of the computed responses are likewise expected to agree with results obtained from COMSOL Multiphysics.

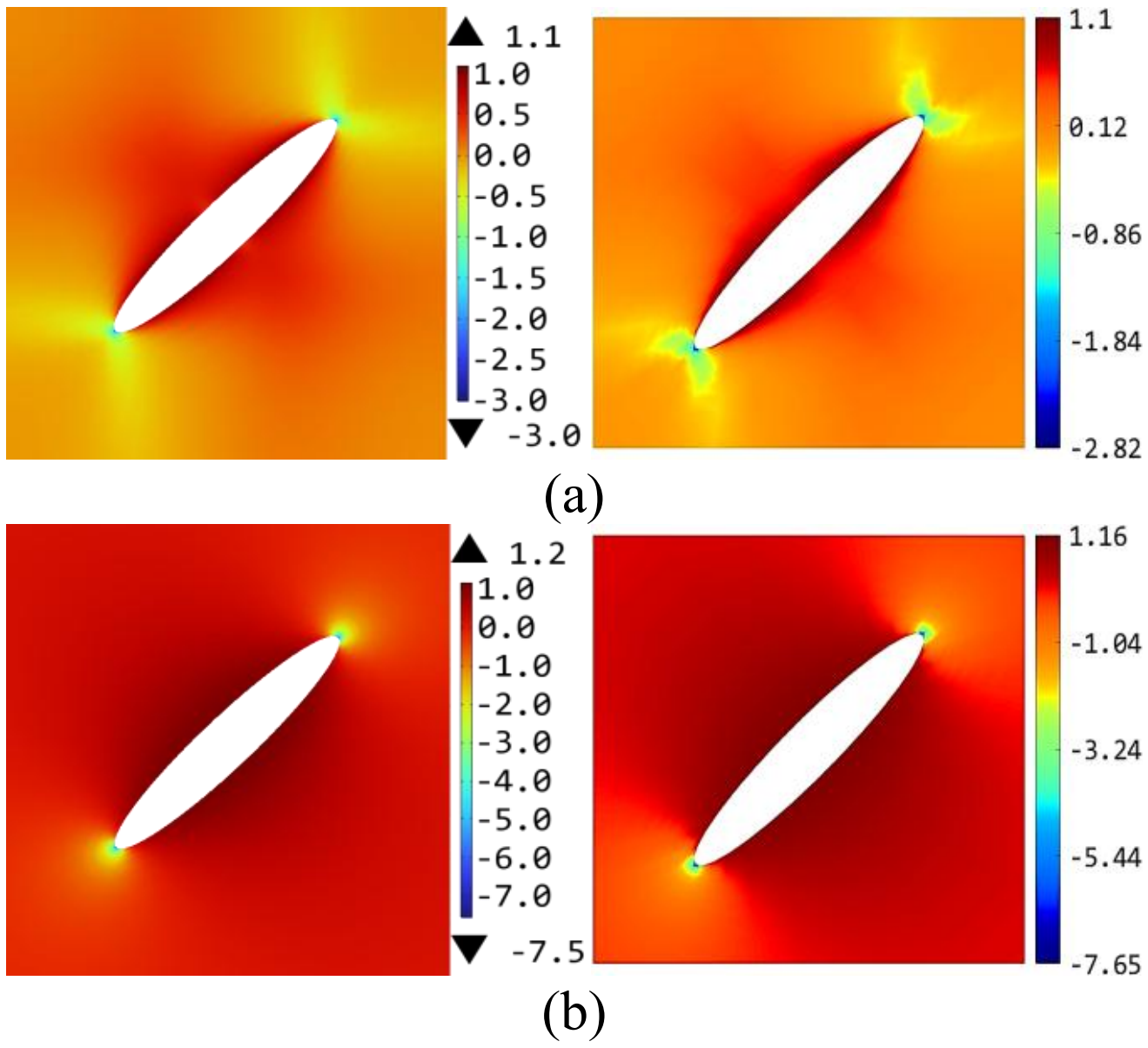

Figure 5.11: Figure showing pressure distribution around the fiber for power law index corresponding to (a) $n = 0.2$, (b) $n = 1.0$, for COMSOL (left of each case) and MATLAB (right of each case) simulations.

Table 5.2: Table comparing results of minimum and maximum fiber surface pressure obtained from both COMSOL and inbuilt MATLAB FEA simulations or both cases of power law indices (i.e., $n = 0.2$, & $n = 1.0$)

| | $n = 0.2$ | | $n = 1.0$ | |
|---|---|---|---|---|
| | $P_{min}$ | $P_{max}$ | $P_{min}$ | $P_{max}$ |
| MATLAB | -2.83 | 1.11 | -7.65 | 1.17 |
| COMSOL | -3.01 | 1.10 | -7.55 | 1.17 |

### *5.9 Validation of Jeffery's Pressure Optimization Scheme*

The implementation of the optimization scheme to obtain the minimum surface pressure on the particles surface using exact derivatives of the Jeffery's pressure equation is validated by comparing outputs of optimum spatial location $X_j^{opt}$ and fiber orientation, $\phi_{opt}$ where the minimum particle surface pressure, $\bar{P}_{min}$ occurs with outputs obtained from the IVP-RK4 method, $\phi_{RK4}$ for the different homogenous flow cases and for $\dot{\gamma}/\dot{\varepsilon} = 1$ as shown in Table 5.3 below. The peak pressure magnitude on the particle's surface occurs when the particle tumbles in the shear plane. The results of the in-plane particle orientation angle at minimum pressure is seen to match closely from both numerical methods and this fiber orientation at the optimum location is seen to correspond to the principal flow direction, $\phi_{prin}$ (cf. Table 5.3). Moreover, the optimum location on the particle's surface, $X_j^{opt}$ shows that the minimum pressure occurs at the particle's tip.

Table 5.3: Results of the in-plane particle orientation angle, the location of peak minimum pressure occurrence on the particles surface and the associated minimum pressure for the different homogenous flow conditions ($\dot{\gamma}/\dot{\varepsilon} = 1$ ).

| | $\phi_{RK4}$ | $\phi_{opt}$ | $\phi_{prin}$ | $X_1^{opt}$ $[mm]$ | $X_2^{opt}[mm]$ | $X_3^{opt}[mm]$ | $\bar{P}_{min}$ |
|---|---|---|---|---|---|---|---|
| SS | 0.7863 | 0.7853 | 0.7854 | 0.0600 | 0.0000 | 0.0000 | -15.7059 |
| SUA | 1.3842 | 1.4110 | 1.4099 | 0.0600 | 0.0000 | 0.0000 | -18.1227 |
| UA | | | ***No minimum found | | | | 9.0701 |
| BA | | | ***No minimum found | | | | -9.0701 |
| PST | 1.1776 | 1.1781 | 1.1781 | 0.0600 | 0.0000 | 0.0000 | -16.9603 |
| SBA | 0.7977 | 0.7854 | 0.7854 | 0.0600 | 0.0000 | 0.0000 | -13.0741 |
| TA | | | ***No minimum found | | | | 0.0000 |
| STA | 0.7864 | 0.7853 | 0.7854 | 0.0600 | 0.0000 | 0.0000 | -5.9363 |

### *5.10 Results and Discussion*

The Results and Discussion section is divided into two sub-sections. The first sub-section presents particles behavior (orientation dynamics and surface pressure distribution) in a Newtonian fluid, considering the various homogenous flows described above and the effect of geometric aspect ratio and particles initial orientation on the particles motion and evolution of the surface pressure. The subsequent sub-section presents in detail the effect of shear-thinning power-law fluid rheology on the particles behavior in the various combined homogenous flows and for different shear-to-extension rate ratio ($\dot{\gamma}/\dot{\varepsilon} = 1$ and $10$). The section also presents the results of sensitivity studies on the influence of the ellipsoidal aspect ratio and initial particle orientation on the particles behavior in non-Newtonian simple shear flow.

### *5.10.1 Particle Motion in Newtonian Homogenous Flows*

*5.10.1.1 Effect of Particle Aspect Ratio and Flow Conditions.* The 2D FEA sensitivity analysis on the particle's geometric aspect ratio $r_e$ carried out showed that the magnitude of $r_e$ varies directly with the max and min pressures on the particle's surface as it rotates through Jeffery's orbit in simple shear flow. Figure 5.12 illustrates that the minimum pressure on the particle surface drops as the shape of the ellipsoid oblates from a prolate spheroid to a perfect sphere at which point there are no noticeable pressure peaks on the particle surface during its evolution, as expected. A closer inspection of the pressure contour plots appearing in Figure 5.13 shows the location of minimum pressure on the particle surface and that these low-pressure sites occur at the particle tip. The shear rate magnitude and Newtonian viscosity is observed to influence computed pressure response as that for particle aspect ratio, i.e., higher shear rate and viscosity result in a higher peak pressure at sites where they occur on the particle surface as shown in Figure 5.14 and Figure 5.15. These factors (particle aspect ratio, viscosity, and flow shear rate), however, affect Jeffery's period differently. By mere inspection of the definition of Jeffery's tumbling period (cf. eqn. (5.68)), the period is observed to vary directly with aspect ratio (i.e., implying faster tumbling for shorter particles) the reverse is the case with the shear rate magnitude which varies inversely with the period as higher shear rate results in higher particle angular velocities, as predicted by Jeffery. However, Jeffery's period is unaffected by the viscosity magnitude. In summary, higher geometric aspect ratios, shear rate magnitude and viscosity result in lower particle surface pressure drop for suspended particles in simple shear flow.

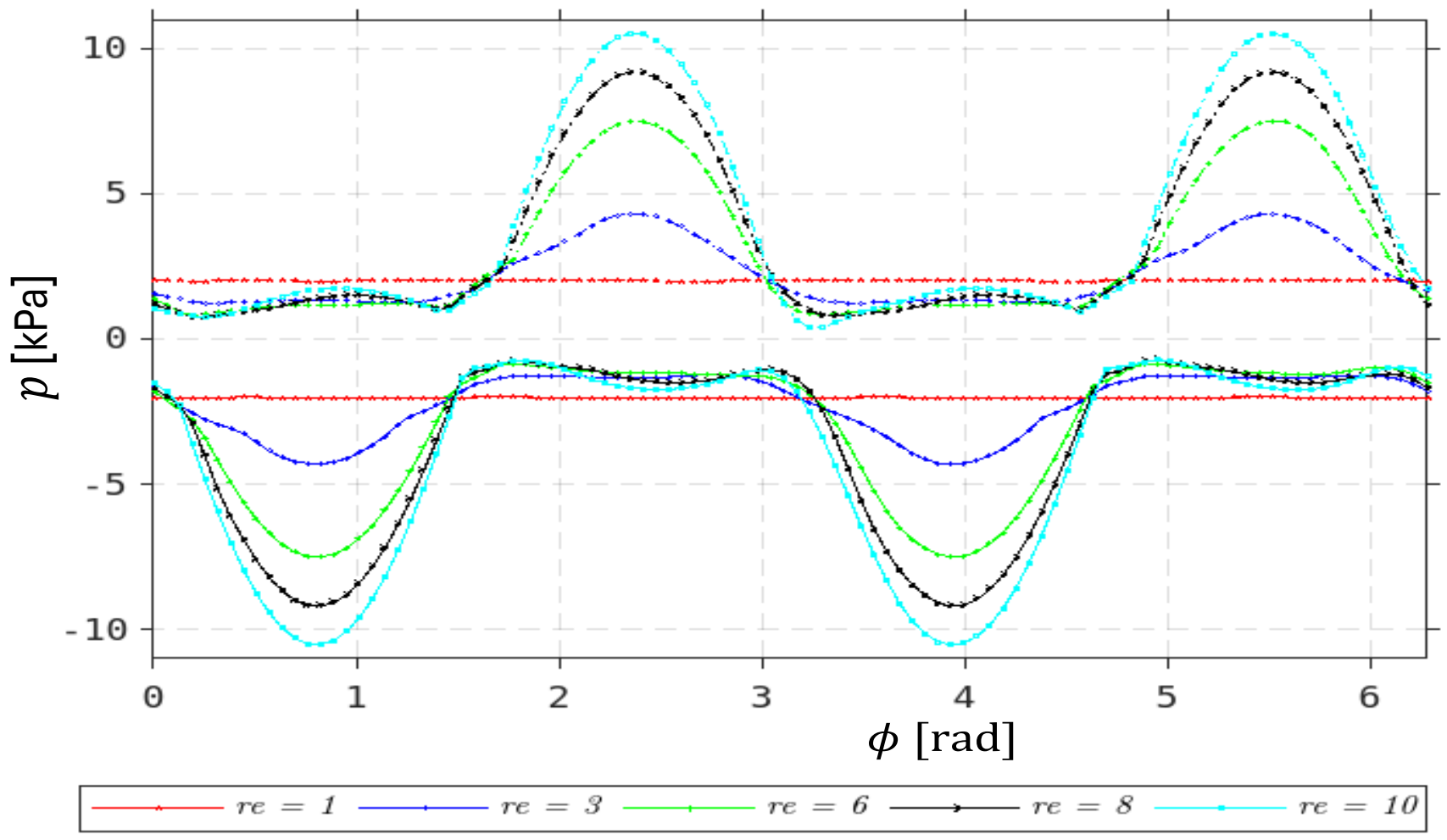

Figure 5.12: Maximum (upper curves $p > 0$) and minimum (lower curves $p < 0$) particle surface pressures for various aspect ratio in simple shear flow ($\dot{\gamma} = 1 s^{-1}$).

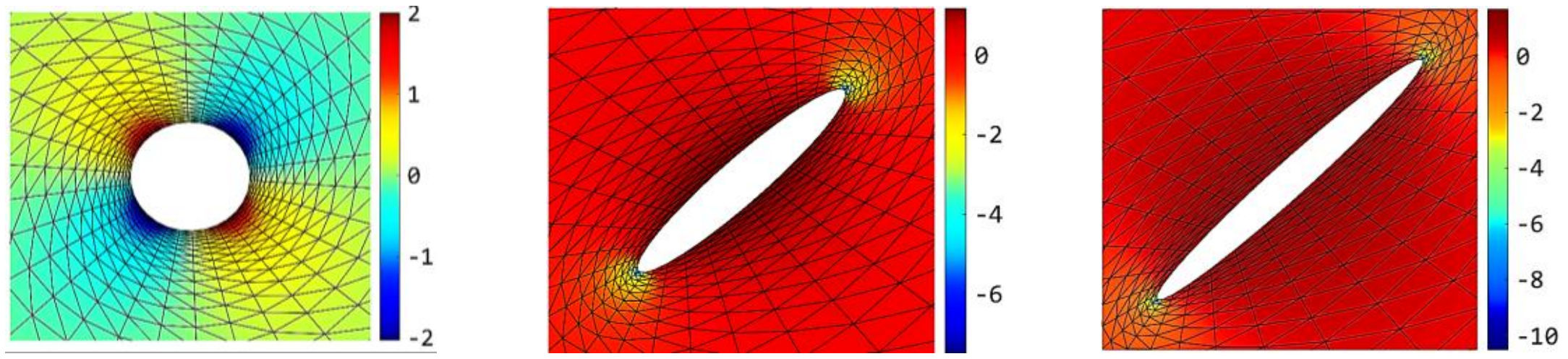

(a) (b) (c)

Figure 5.13: Pressure distribution around particle's surface for at the point of minimum pressure drop for `different particle's aspect ratio (a) $r_e = 1$ (b) $r_e = 6$ (c) $r_e = 10$.

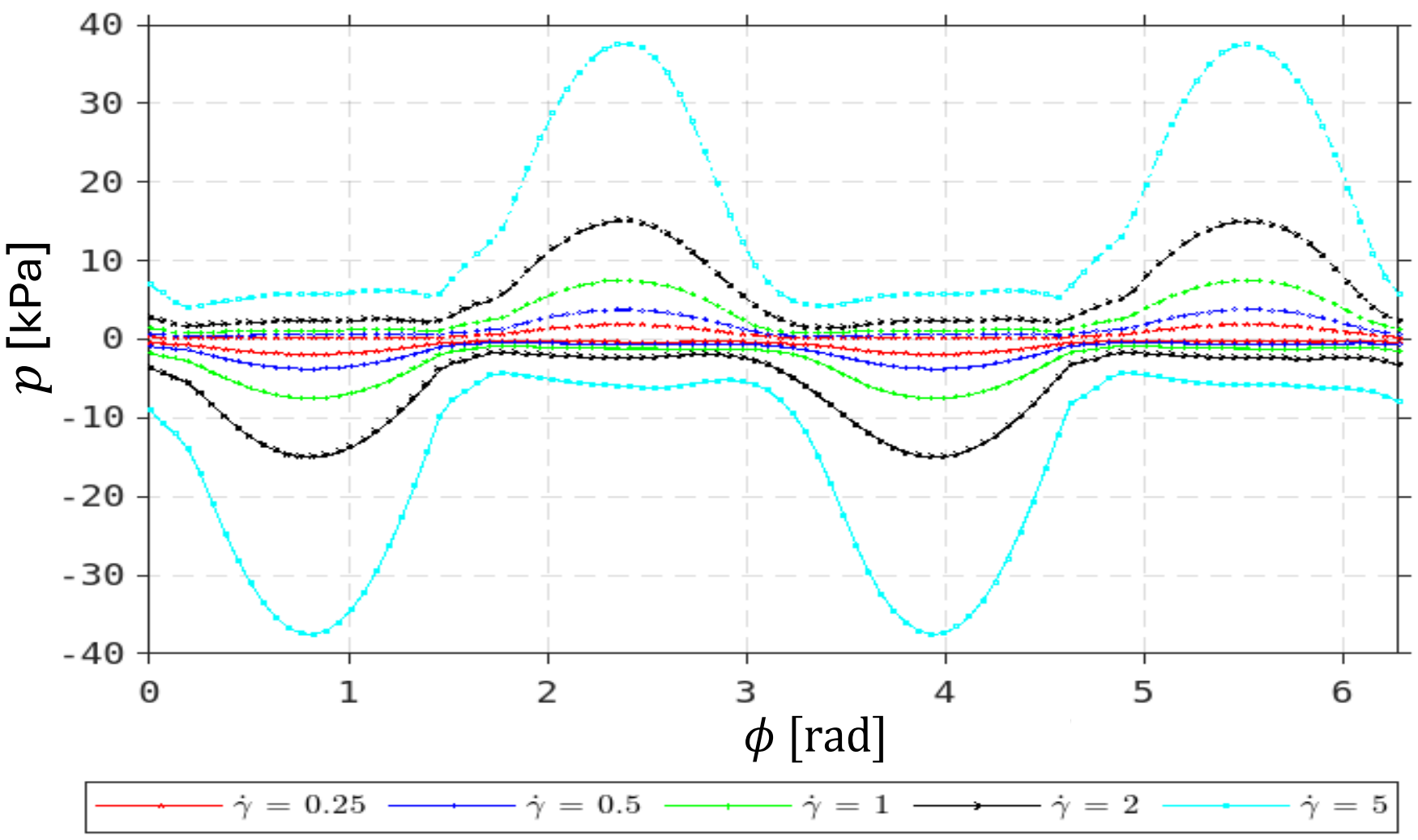

Figure 5.14: Maximum (upper curves $p > 0$) and minimum (lower curves $p < 0$) particle surface pressures for various shear rate values in simple shear flow ($r_e = 6$). The units for $\dot{\gamma}$ are $s^{-1}$.

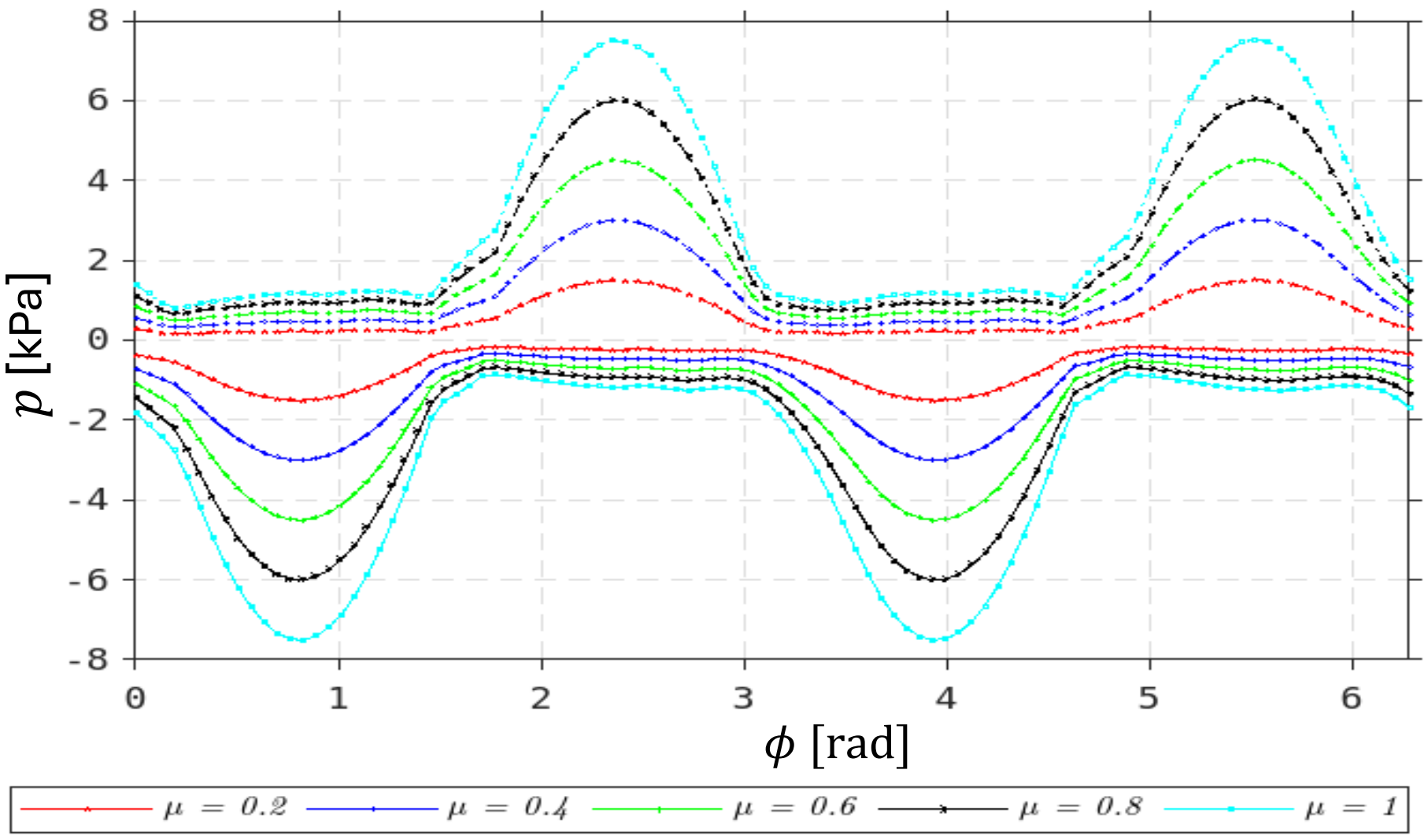

Figure 5.15: Maximum (upper curves $p > 0$) and minimum (lower curves $p < 0$) particle surface pressure limits for various Newtonian viscosities in simple shear flow ($r_e = 6$). The units for $\mu$ are $Pa \cdot s$.

From the foregoing preliminary pressure sensitivity 2D Newtonian studies, we see that the peak pressure extreme on the surface of a particle suspended in Newtonian purely viscous simple shear flow is influenced by the fluid viscosity $\mu_1$, the magnitude of the shear rate $\dot{\gamma}$, and the particle aspect ratio $r_e$. For completeness, we further explore 3D particle behavior in Newtonian purely viscous flow using Jeffery's equations. For a given aspect ratio, the net pressure $p - p_0$, computed from eqn. *(5.1)* is seen to have a linear dependence on the Newtonian viscosity $\mu_1$ and shear rate $\dot{\gamma}$, i.e. $(p - p_0)/\mu_1\dot{\gamma}$ is constant. However, as $r_e$ increases, so does the extreme tip pressure. Figure 5.8b shows that the particle's tip pressure magnitude is proportional to the $r_e$ of the ellipsoidal particle, which is likely due to the increased particle length, the reduced particle tip curvature which occurs as $r_e$ $is$ increased, or both. From eqns. *(5.71)* & *(5.72)*, it can be shown that the particle's tip pressure extremes occur at an orientation angle of $\phi = \pm\pi/4$ when the angular velocity $\dot{\phi} = \dot{\gamma}/2$ which also corresponds to the principal flow directions for simple shear flow.

Further, at the position where the particle's precession approaches extremum at $\phi = n\pi/2, |n| \geq 0$, the particles tip pressure goes to zero irrespective of the geometric aspect ratio. Figure 5.16 shows the pressure distribution on the surface of rigids spheroidal particles at the location of orbital minimum surface pressure extreme for different aspect ratios and for particle motion in the plane of shear flow. It is evident that the minimum pressure on the particles surface occurs at the particle tips and the pressure peak magnitudes increases with the geometric aspect ratio.

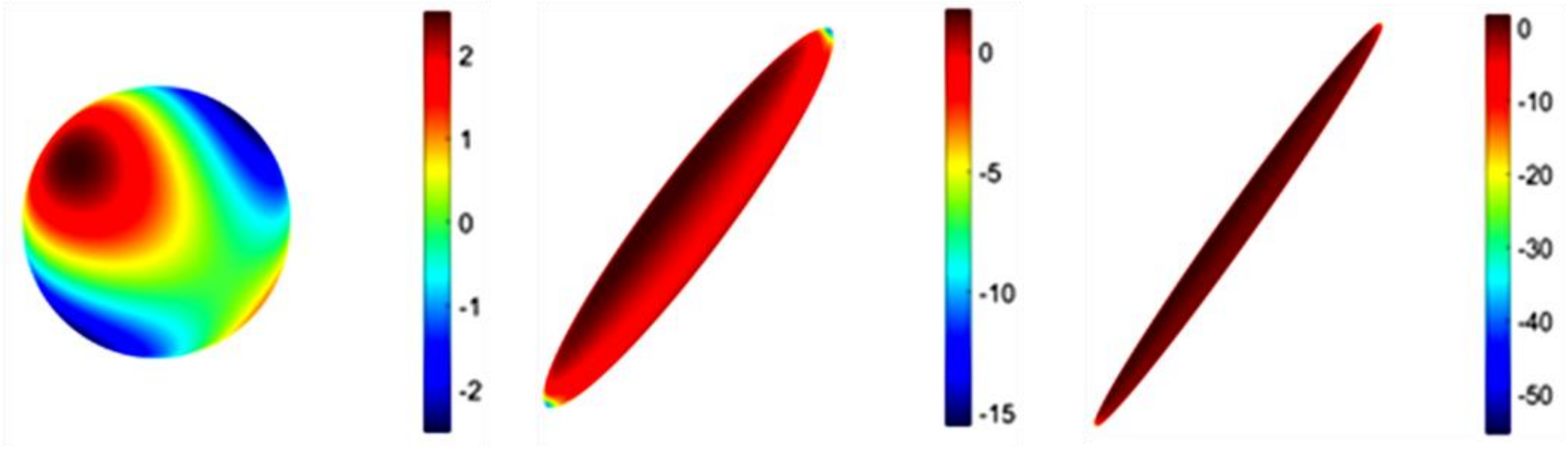

(a) (b) (c)

Figure 5.16: Pressure Distribution around the particle surface at the point of minimum peak pressure occurrence ($\phi = \pi/4$) for different aspect ratio (a) $r_e = 1$ (b) $r_e = 6$ (c) $r_e = 15$.

With increased ellipsoidal aspect ratio, the curvature radius at the particle's tip reduces. It is important to understand the relation of the tip pressure magnitude with the tip geometry (i.e. the curvature radius, $r_\kappa = 1/r_e$) and with the relative positioning of the tip in the constant velocity gradient flow-field (defined by the particles geometric parameter, $\kappa$). Figure 5.17a shows the relationship between the spheroidal orbital minimum tip pressure, $\overline{P}_{min,\kappa}$ normalized with respect to the spherical reference values, $\overline{P}_{min,0}$, (i.e. $\kappa = 0$) and the curvature radius for a prolate spheroid with unity minor axis length. This relationship obtained through a typical curve fitting procedure can be represented by eqn.

(5.128). The Newtonian orbital minimum tip pressure ratio is seen to decrease exponentially with increasing tip curvature radius as

$$\overline{P}_{min,\kappa}/\overline{P}_{min,0} = 0.63 + 0.39 r_\kappa^{-1.53} - 4.81 \exp(14.47 r_\kappa) \tag{5.128}$$

Alternatively, the Newtonian orbital minimum tip pressure ratio can be represented in terms of the geometric parameter $\kappa$ as shown in Figure 5.17b and can be written as

$$\overline{P}_{min,\kappa}/\overline{P}_{min,0} = 1.87\kappa + 10.74\kappa^{19.56} + 0.82 \exp(4.54\kappa^{56.62}) \tag{5.129}$$

Figure 5.17b shows that as $\kappa$ tends to unity approaching a slender rod, the particle tip orbital minimum pressure goes to infinity. Note that the mean aspect ratio of short fiber fillers experimentally measured in 13% CF/ABS large scale EDAM printed bead were found to be about $r_e = 45,\ \kappa = 0.999$ [271], [272], that would theoretically yield high pressure spikes at the particle tips in the polymer suspension during polymer composite processing based on Jeffery's model assumption, which have been suggested by Awenlimobor et al. [57] to be potentially responsible for micro-void nucleation at the fiber tips.

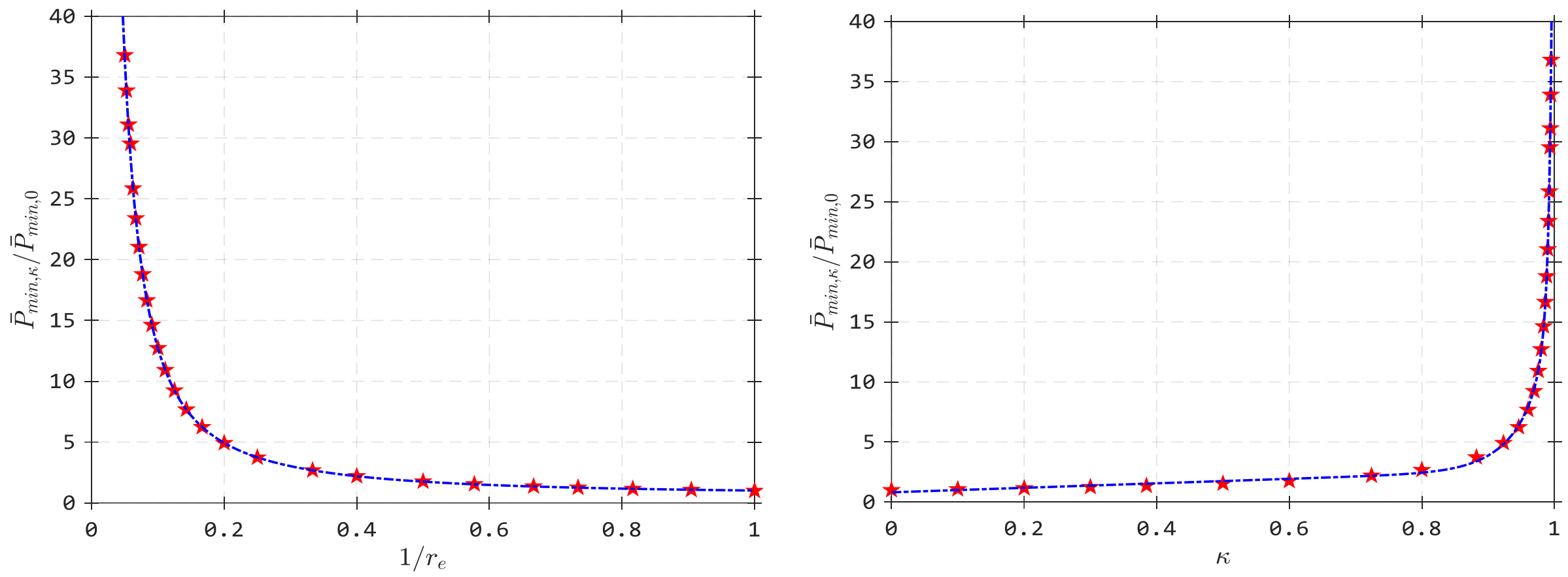

(a) (b)

Figure 5.17: Relationship between the particle's orbital minimum pressure normalized with respect to the minimum surface pressure on a sphere in Newtonian fluid flow as a function of (a) radius of curvature ($r_\kappa$), and (b) geometric parameter $\kappa$. Results are shown for a particle tumbling in simple shear flow with $\mu_1 = 1\ Pa \cdot s$ and $\dot{\gamma} = 1s^{-1}$ .

*5.10.1.2 Effect of Initial Particle Orientation.* In Figure 5.18a, we present the particle's motion in simple shear flow for various initial particle azimuth angle $\theta_0 = 2\pi/24 \leq \theta \leq 11\pi/24$ ($\phi_0 = 0$) based on Jeffery's solution given above. As expected, the particle's motion is periodic, and the period is the same for all orbits. The orbit becomes narrower as we increase the initial out-of-plane orientation angle which reduces the effective aspect ratio (seen as that projected to the shear plane), resulting in lower peak pressure extremes. Figure 5.18b shows that the angle at which the particle pressure extreme occurs shifts as the particle is oriented further out of the shear plane. Eventually, setting the initial out of plane orientation to zero would lead to the particle spinning about its axis in a log-rolling position with near-zero surface pressure due to negligible disturbance velocity. The phase diagrams (cf. Figure 5.18c&d) reveals a symmetric behavior in particle dynamics. As the particle moves further out of plane (i.e. $\varsigma \rightarrow 0$), the location of the tip pressure extremes converges towards the location of minimum precession at $\phi = \pm\pi/2$, but as the particle moves towards the shear-plane, the pressure extreme locations coincide with the direction of the principal axis of the flow ($\phi = \pm\pi/4$). Figure 5.19 shows the particle's configuration at the location of minimum particle tip pressure along select Jeffery's orbits with various initial azimuth angle $\theta_0$. For the particle tumbling in the shear plane of the flow ($\theta_0 = -\pi/2$) we see that the particle's orientation coincides with the principal direction of the flow ($\phi = \pi/4$) but as it moves further out of plane, the peak pressure location moves closer towards the upper limit of azimuthal inclination for each orbit (i.e. $\phi \rightarrow \pi/2$).

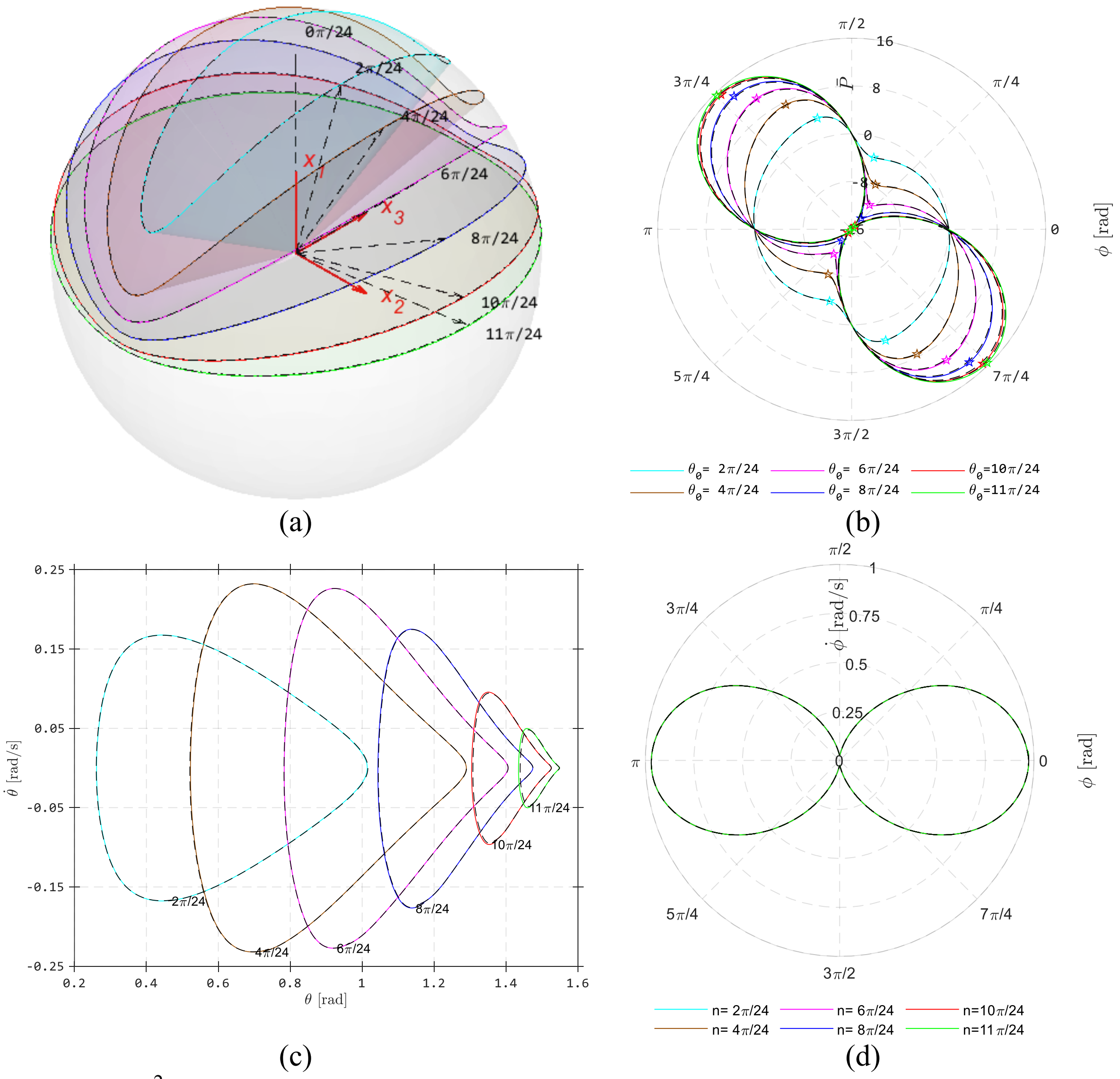

Figure 5.18[2]: Results for different initial particle orientation showing (a) Jeffery's orbits (b) particle tip pressure evolution where the asterisk (*) indicates location of the tip pressure extreme (c) phase diagram of azimuth angle $\theta$ vs nutation $\dot{\theta}$ (d) polar plot of the precession $\dot{\phi}$ vs polar angle $\phi$. Results are shown for $-2\pi/24 \leq \theta_0 \leq -12\pi/24$ and for simple shear flow with $\mu_1 = 1\ Pa \cdot s$ and $\dot{\gamma} = 1s^{-1}, r_e = 6$ .

[2] The results presented in Figure 5.18a-d are also validated with both FEA and Jeffery's analytical calculation. The black dashed lines are results obtained from Jeffery's equation and the continuous colored lines are results from FEA computations.

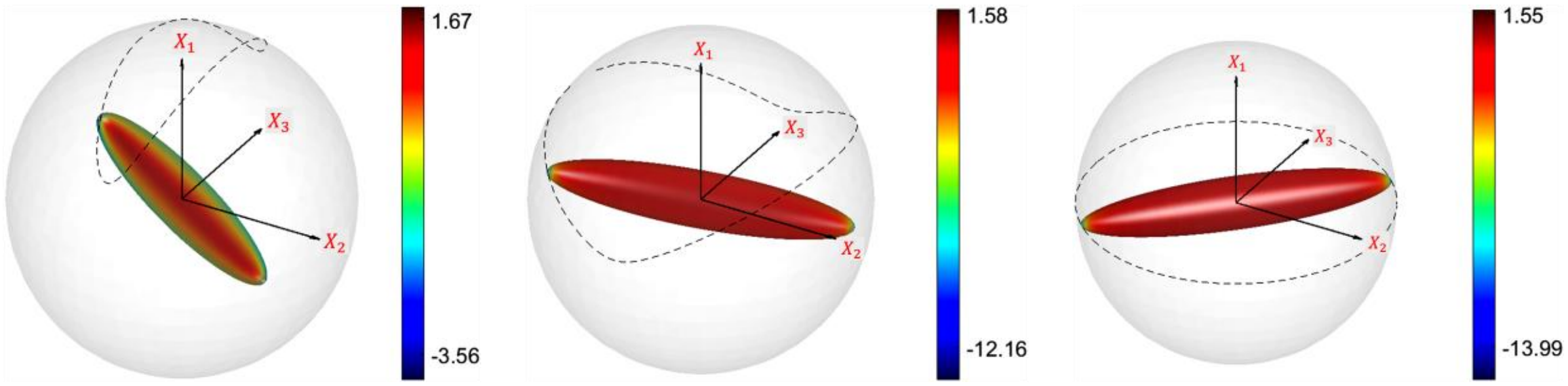

(a) (b) (c)

Figure 5.19: Spatial configuration of the particle at the point of minimum pressure occurrence and for various initial azimuthal angle $\theta_0$ of (a) $\theta_0 = 2\pi/24$, (a) $\theta_0 = 8\pi/24$, (a) $\theta_0 = 12\pi/24$. Results are shown for $\mu_1 = 1\ Pa \cdot s$ and $\dot{\gamma} = 1s^{-1},\ r_e = 6$.

Figure 5.20a shows a nearly linear relationship between the particle's orbital minimum tip pressure and the polar angle location along the corresponding Jeffery's orbits. As noted above, when the particle is tumbling in shear plane (i.e., $\varsigma = +\infty$), the location of the particle's surface extreme pressure coincides with the ellipsoidal tip location. However, as the particle becomes oriented more out-of-plane (i.e. $\varsigma \to 0$), the location of minimum pressure on the particle surface at the orientation of peak pressure occurrence is slightly shifted away from the tip down the leeward side trailing the flow. Figure 5.20b shows the difference between the minimum pressure on the fibers surface and tip pressure ($\delta\bar{P}$) at the instant when the peak occurs along Jeffery's orbit. The result shows that a higher initial out of plane orientation leads to greater deviation of the fiber tip pressure from its surface pressure extreme magnitude.

The particle orbital maximum nutation $\dot{\theta}$ itself peaks at a Jeffery's orbit that passes through ($\phi, \theta = \pm\pi/4$) irrespective of the aspect ratio. In Figure 5.21a, the continuous lines trace the paths of orbital maximum nutation across the degenerate spectrum of Jeffery's orbit for different aspect ratios, and the dashes lines are the Jeffery's orbit that cuts across the location of peak nutation for different ellipsoidal aspect ratios. From Figure 5.21b, the peak nutation across the spectrum of Jeffery's orbit is observed to increase with the aspect ratio and approaches the critical value at $\dot{\theta} = \dot{\gamma}/4$.

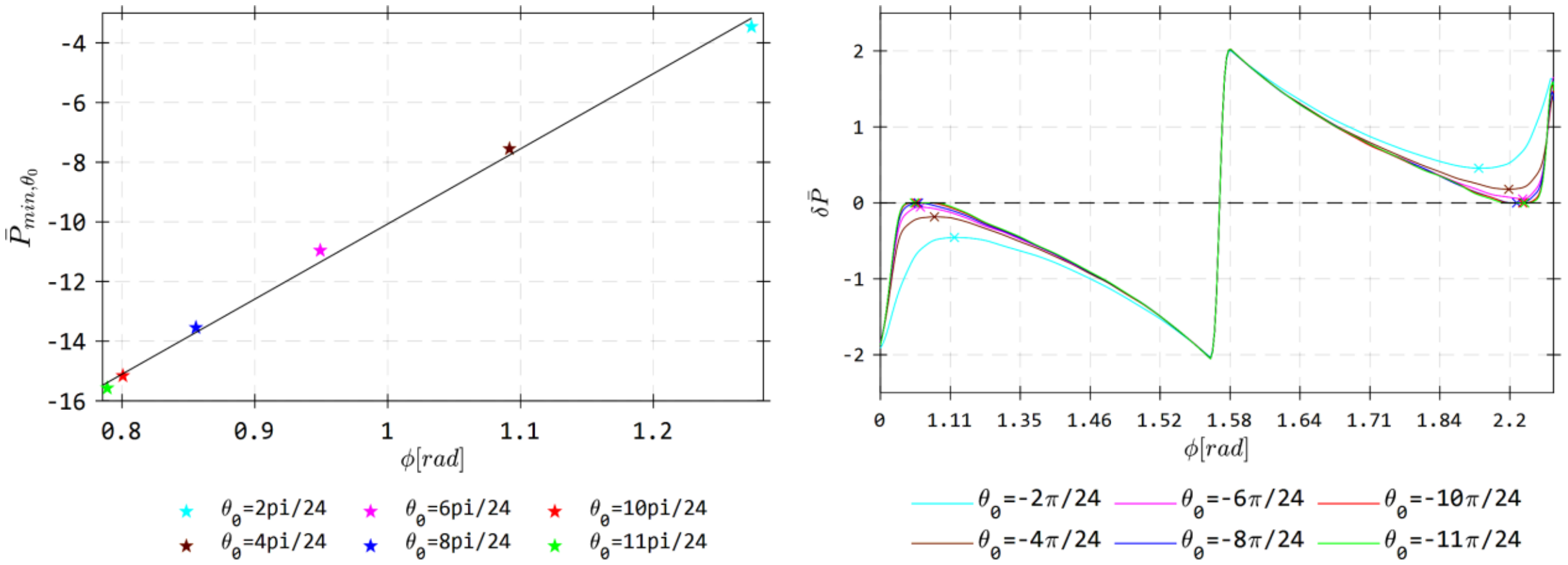

(a) (b)

Figure 5.20: Tip pressure results (a) Orbital minimum particle tip pressure versus polar angle, and (b) difference in the instantaneous particle tip pressure and actual surface pressure extremum, for different Jeffery's orbit and for $\mu_1 = 1\ Pa \cdot s$ and $\dot{\gamma} = 1s^{-1}$, $r_e = 6$.

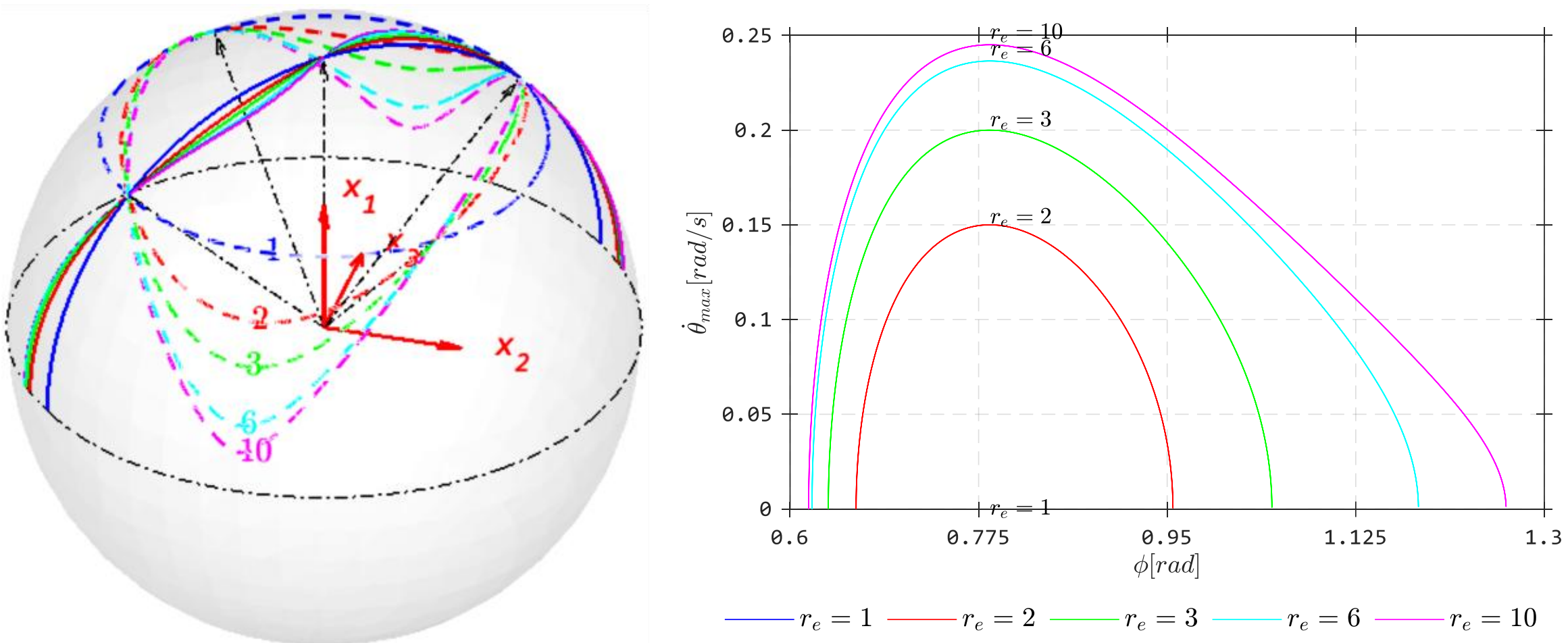

(a) (b)

Figure 5.21: Out-of-plane Jeffery orbits (a) the path of orbital maximum nutation across degenerate spectrum of Jeffery's orbit for aspect ratios of 1, 2, 3, 6, and 10 (continuous lines) and critical Jeffery's orbit at which the orbital maximum nutation attains peak magnitude for the same aspect ratios (dashed lines). (b) phase plot of the orbital maximum nutation across degenerate spectrum of Jeffery's orbit for different aspect ratios.

*5.10.1.3 Effect of Particle Shape.* As we earlier stated, a drawback of Jeffery's equation is its inability to model arbitrary shaped particles with different end conditions which can be accounted for in our FEA simulation. Most chopped fibers used to reinforce polymer composites are cylindrical shaped with different end conditions. Cylindrical shaped particles allow us to study the decoupled effect of edge curvature radius and aspect ratio on the particle surface pressure response. We present results of the cylindrical particle's responses (cf. Figure 5.22) for different end curvature radius ranging from $0.05 \leq \bar{r}_\kappa \leq 0.5$ for a cylinder of diameter $И_1 = 1.0$, individually calibrated to be hydrodynamically equivalent to an ellipsoid with an aspect ratio $r_e = 6$. The fibers orientation angle, rotational velocity, and surface pressure were computed using flow parameters of $\mu_1 = 1\ \mathrm{Pa.s}$, $\dot{\gamma} = 1\ \mathrm{s}^{-1}$. Figure 5.22a shows that the evolution of the cylinders' angular velocity for the different edge curvature cases with different geometric aspect ratios (colored lines) are perfectly superposed on the angular velocity profile of the ellipsoid with $r_e = 6$ (black dotted line). The corresponding cylindrical geometric aspect ratio, $r_c$ for the different end cases is seen to vary inversely with the edge curvature radius, $\bar{r}_\kappa$ (cf. Figure 5.22c). For an equivalent ellipsoidal aspect ratio of $r_e = 6$, the cylindrical geometric aspect ratio $r_c$ was approximated as a cubic function of the tip curvature $\bar{r}_\kappa$ according to

$$r_c = 7.806 - 1.282\,\bar{r}_\kappa - 1.463\bar{r}_\kappa^2 + 3.859\bar{r}_\kappa^3 \tag{5.130}$$

Expectedly, the pressure extremes on the particle's surface are observed to increase with decreasing edge curvature radius (cf. Figure 5.22b). The minimum surface pressure is observed to drop in magnitude from a value of $\bar{P}_{min} = -19.41$ when $\bar{r}_\kappa = 0.05$ to about $\bar{P}_{min} = -8.68$ when $\bar{r}_\kappa = 0.5$. Recalling that for the ellipsoidal particle of dynamically equivalent aspect ratio, the orbital minimum pressure occurs at a value of $\overline{P}_{min} = -15.71$. It follows that unlike the ellipsoidal particle, the cylindrical hydrodynamic equivalent could have higher or lower pressure magnitudes at the ends depending on the edge curvature radius. As the particle tumbles in and out of alignment with the principal direction of the flow, the minimum and maximum pressure cycles through mesh-points along the surface of the particle in the plane of the shear flow. Figure 5.22d shows the pressure distribution on the cylinder particle surface at the instant of orbital minimum surface pressure for different edge curvatures. The instantaneous pressure extremes occur at the terminations of the curved section at the cylinder's end in comparison to the ellipsoid where the instantaneous pressure extremes occur at its vertices.

Although the aspect ratio of the cylinder is slightly adjusted for each end curvature cases to hydrodynamically match the aspect ratio of the reference ellipsoid aspect ratio, we argue that the observed change in the pressure extreme magnitude is mainly a result of the change in the edge curvature radius rather than the aspect ratio. To validate this, we perturb the geometric aspect ratio of the cylinder by about 6 % (i.e. $7.0 \leq r_c \leq 7.4$) for a constant edge curvature, $\overline{r}_\kappa \leq 0.5$ (hemispherical end case) similar to the range of adjustment in aspect ratio obtained for different curvature cases in Figure 5.22b. For objectivity, we adjust only the length of the straight section of the cylinder to ensure the curvature and mesh integrity of the curved section is unaffected which is where we expect pressure extreme would occur.

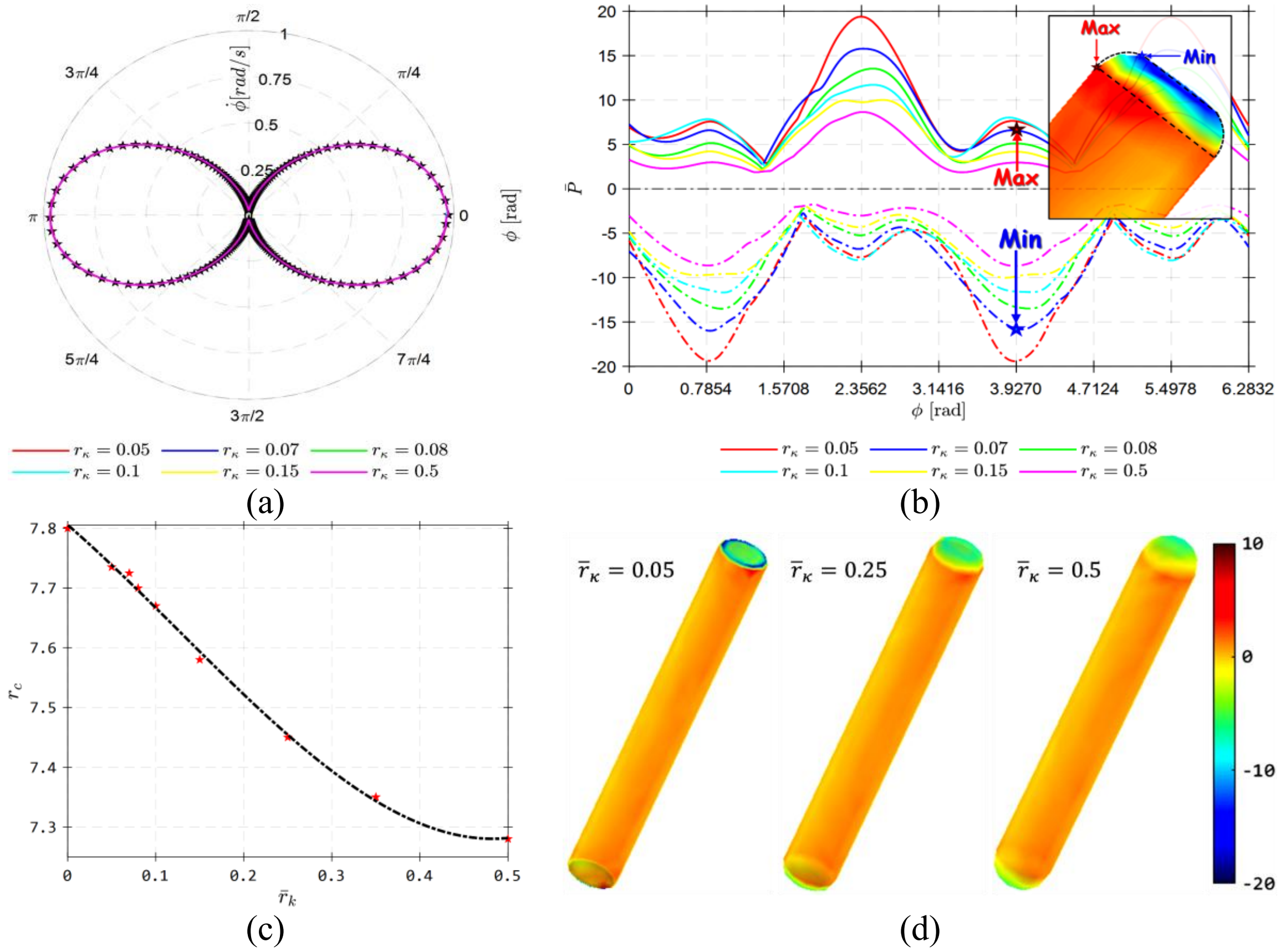

Figure 5.22: Evolution of (a) the angular velocity (b) pressure extremes; on the surface of the cylindrical particle for different curvature radius $0.05 \le \bar{r}_\kappa \le 0.5$. (c) Relation between the cylindrical geometric aspect ratio and edge curvature radius, and (d) Pressure distribution on the surface of the particle at the instant of orbital minimum surface pressure occurrence for different cylinder edge curvatures. Results are shown for cylinders with hydrodynamic equivalent ellipsoidal aspect ratio of $r_e = 6$ tumbling in simple shear flow ($\mu_1 = 1\ Pa.s,\ \dot{\gamma} = 1\ s^{-1}$).

Figure 5.23a&b shows that the angular velocity and pressure extremes on the cylinder surface are not significantly affected by the perturbation in the aspect ratio. This is because the flow-field is symmetric and of a constant velocity gradient, and the perturbation in cylinder aspect ratio only slightly and linearly perturbs the disturbance velocity $\delta \dot{X}_i^d$ on the surface of the particle and the corresponding pressure field such that $\delta \dot{X}_i^d = \mathbb{E}_{ik} \delta X_k$ where $\mathbb{E}_{ik} = \epsilon_{ijk} \Psi_j - L_{ik}$ is constant.

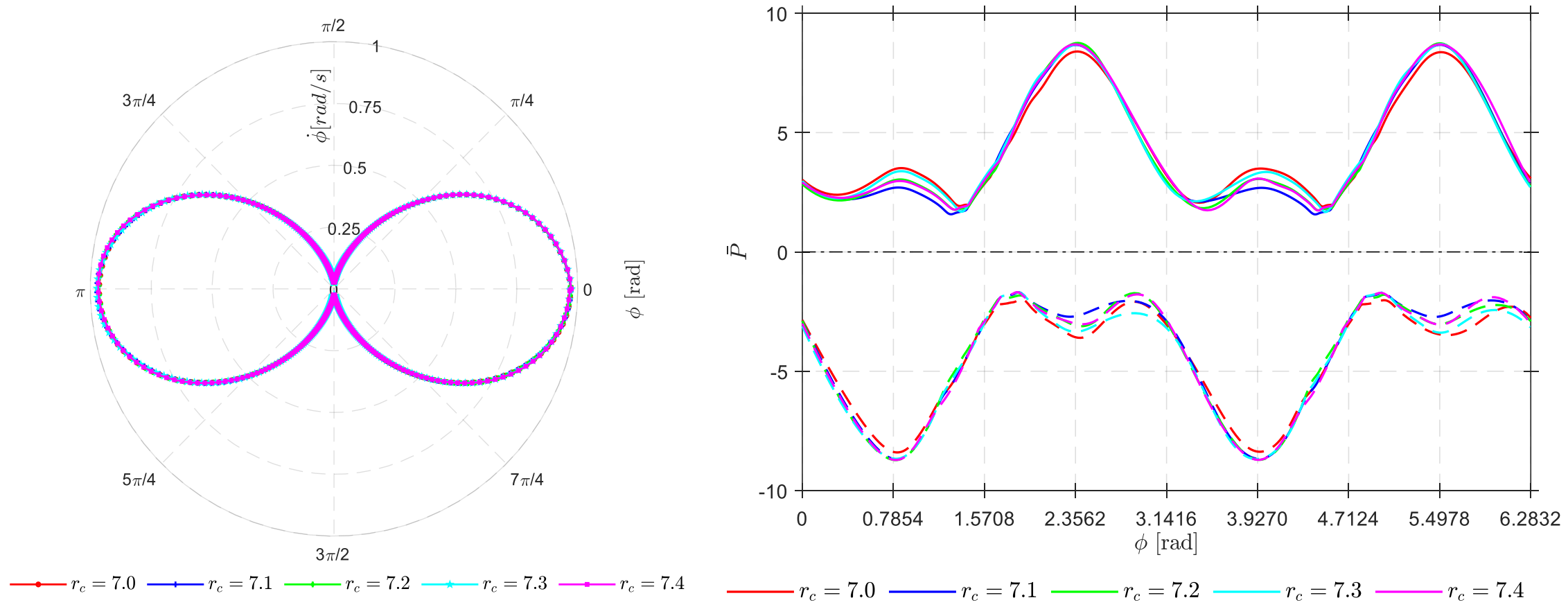

(a) (b)

Figure 5.23: Evolution of (a) the angular velocity (b) pressure extremes; on the surface of the cylindrical particle for cylinder with different geometric aspect ratio $7.0 \leq r_c \leq 7.4$ and with constant end curvature radius, $\bar{r}_\kappa \leq 0.5$. Results are shown for cylindrical particles tumbling in simple shear flow ($\mu_1 = 1\ Pa.s,\ \dot{\gamma} = 1\ s^{-1}$).

*5.10.1.4 Effect of Flow-type & Elongational ratio (Steady Homogenous Flows).* For the investigation of the behaviour of single rigid spheroidal particle suspended in Newtonian homogenous flows, Jeffery's equations are sufficient and computationally more efficient than our numerical solutions. The basic homogenous flows discussed in the methodology section above that consider various combinations of stretching and shearing rate are expected in polymer composite melt flow applications such as material extrusion/deposition additive manufacturing (see e.g., Awenlimobor et al. [57]). In all Newtonian flow analyses considered here, we employ an aspect ratio of $\mathrm{r_e} = 6$, a viscosity of $\mu_1 = 1\ \mathrm{Pa \cdot s}$ and a shear rate of $\dot{\gamma} = 1\ \mathrm{s}^{-1}$ where applicable. The particle is initially oriented in the $\underline{\mathrm{X}}_2$-direction (i.e. $\phi^0 = 0, \theta^0 = -\pi/2, \psi^0 = 0$) and rotates in the $\underline{\mathrm{X}}_2 - \underline{\mathrm{X}}_3$ shear plane.

Figure 5.24 shows the calculated particle in-plane angular velocity ($\phi$) and particle tip pressure ($\bar{P}$) in the various homogenous flows for two cases of shear-to-extension rate ratio ($\dot{\gamma}/\dot{\varepsilon}$) where applicable. Here we use the overbar to indicate a dimensionless pressure as in eqns. 75 and 76. In the planar extensional flows (i.e. UA, BA, & TA flows), we observe an absence of particle motion, however, the particle begins to rotate with the introduction of a non-zero shear velocity gradient component (cf. Figure 5.24a). In the extension-shear SUA flow (i.e., $\dot{\gamma}/\dot{\varepsilon} = 1$), the particle is initially accelerated by the combined action of the inward flow in the $\underline{X}_2$-direction and the shear flow in the $\underline{X}_2 - \underline{X}_3$ plane. The particle eventually stalls at $\phi_s = 1.58$ rad as it aligns with the $\underline{X}_3$-direction due to the applied stretching and relatively low shear rate. In the PST flow case, there is no flow in the $\underline{X}_2$-direction that influences the initial particle motion, however the inflow in the $\underline{X}_1$-direction keeps the particle motion in the $\underline{X}_2 - \underline{X}_3$ shear plane. Like the SUA flow case, the applied stretching and relatively high extensional dominance causes the particle to stall at $\phi_s = 1.60$ rad as it turns to align in the $\underline{X}_3$-direction. The SUA and PST mixed mode flow types are asymmetric in the $\underline{X}_2 - \underline{X}_3$ plane. In the SBA flow regime, the inward flow in the $\underline{X}_1$-direction prevents out-of-plane motion of the particle, and there is no provision for preferential orientation in the $\underline{X}_2 - \underline{X}_3$ plane due to uniform stretching in the $\underline{X}_2 - \underline{X}_3$-shear plane. As a result, the particle tumbles continuously. The STA and SS flow types are essentially similar in terms of their influence on the particle's behavior. The only difference observed between these flow types is in the calculated particle tip pressure. At the onset of particle motion at $\phi^0 = 0$ the net pressure at the particle tip is zero ($\bar{P} = 0$) for cases with no net flow in the $\underline{X}_2$-direction. However, the particle tip has a net positive pressure ($\bar{P} = +9.07/+8.71$) for the UA/SUA flows due to the inflow in the $\underline{X}_2$-direction, and the outflow in the $\underline{X}_2$-direction creates a net negative pressure on the particles tip ($\bar{P} = -9.07/-8.71$) for the BA/SBA cases. As the shear flow induces particle rotation, the tip pressure drops gradually until it reaches a minimum, at which point the particles orientation coincides with a principal flow direction (cf. Figure 5.24b).

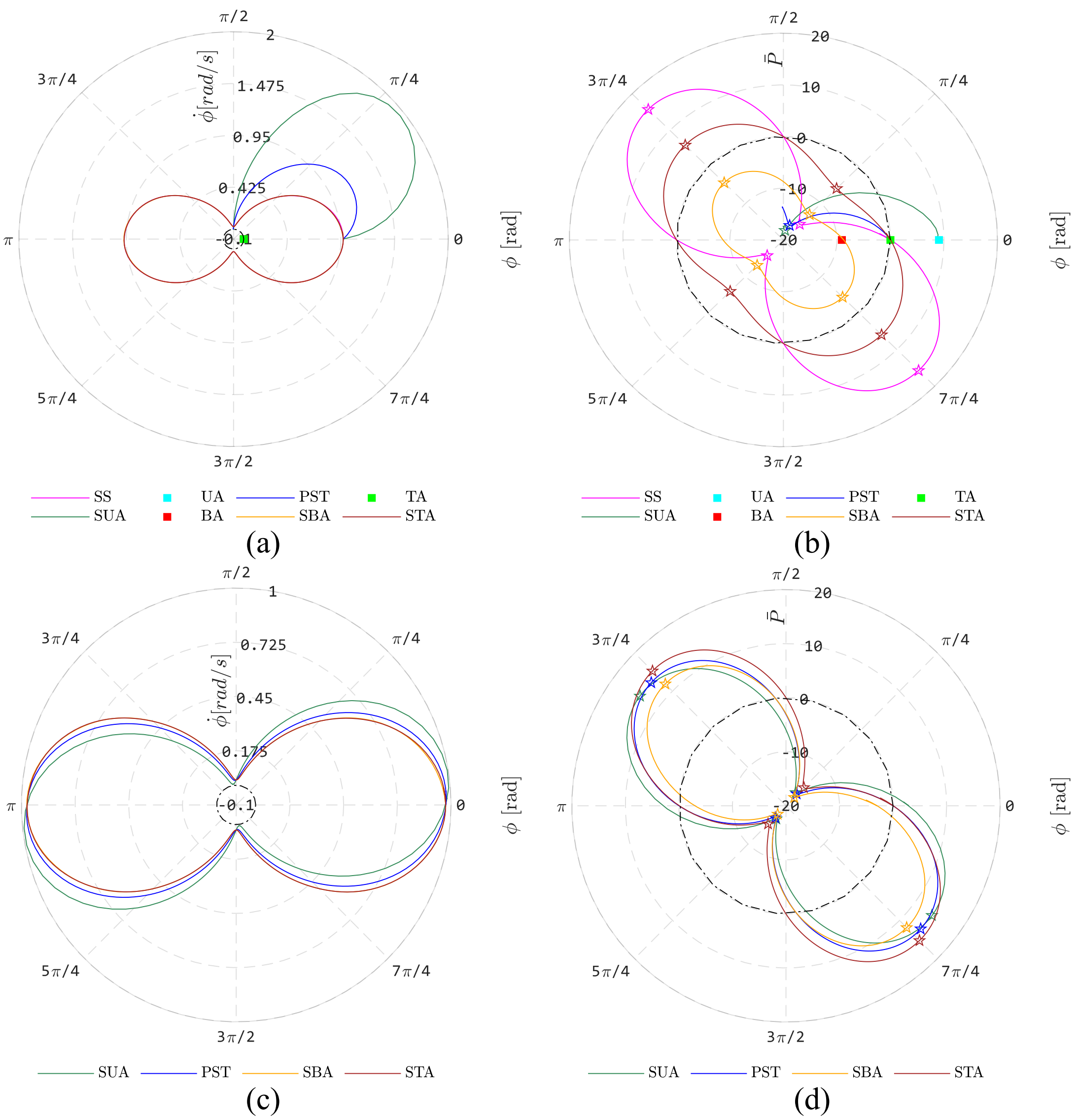

Figure 5.24: Polar plot of the evolution of the particle's (a) precession $\dot{\gamma}/\dot{\varepsilon} = 1$ (where applicable) (b) tip pressure $\dot{\gamma}/\dot{\varepsilon} = 1$ (c)precession $\dot{\gamma}/\dot{\varepsilon} = 10$ (d)tip pressure $\dot{\gamma}/\dot{\varepsilon} = 10$ for particle in the various homogenous flow types. In all cases, $\dot{\gamma} = 1\ s^{-1}$, $\mu_1 = 1\ Pa \cdot s$.

In an event where the particle does not stall, the pressure on the particle tip fluctuates between its minimum and maximum limits at locations where its orientations coincide with the principal flow directions. For the axisymmetric flows, the particle tip pressure extremes occur at $\phi = \pm\pi/4$, while for the SUA asymmetric flow (i.e., $\dot{\gamma}/\dot{\varepsilon} = 1$), this occurs at $\phi = +1.41$ rad. Alternatively, for the PST asymmetric flow, the pressure extreme occurs at $\phi = +1.18$ rad. Cessation of the particles motion under the combined SUA and PST flow conditions is lifted once the conditions of eqn. *(5.81)* are violated, i.e. when $\dot{\gamma}/\dot{\varepsilon} \geq 3\kappa/\sqrt{1-\kappa^2}$ for the SUA flow condition and $\dot{\gamma}/\dot{\varepsilon} \geq \kappa/\sqrt{1-\kappa^2}$ for the PST flow conditions. In the current study where we assumed $\kappa = .9459$, the particle does not stall when $\dot{\gamma}/\dot{\varepsilon} \geq 8.75$ for SUA flow condition and when $\dot{\gamma}/\dot{\varepsilon} \geq 2.92$ for the PST flow condition. With increased shear strain rate (i.e., for $\dot{\gamma}/\dot{\varepsilon} = 10$), the particle rotates periodically for all combined flow conditions (cf. Figure 5.24c). Since $\dot{\varepsilon}_2 = \dot{\varepsilon}_3 = \dot{\varepsilon}$ , for the axisymmetric combined flow cases, the particle does not stall regardless of the magnitude of $\dot{\gamma}/\dot{\varepsilon}$. One exception is seen for ellipsoidal particles with small but finite thickness such as in the case of a thin rod when $\kappa \to 1$or in the case of a circular disc when $\kappa \to 0$, both of which are degenerate cases as described by Jeffery [21]. As the shear rate increases, the asymmetric flows become more symmetrical and the particle's surface pressure magnitudes are increased (cf. Figure 5.24d). Additionally, increased shear rate also moves the orientation where tip pressure extremes occur (i.e. at the point where it coincides with the principal flow directions). For example, in the SUA flow case, the orientation where pressure extremes occurs are at $\phi = -0.640, +0.931$ rad while the same occurs at $\phi = -0.736, +0.835$ rad for the PST flow case. Figure 5.25a&b shows that the particles orbital minimum surface pressure and corresponding orientation approaches a stable equilibrium value with increasing shear rates. Particle motion analyses show that cessation of the rotation depends on the value of $\dot{\gamma}/\dot{\varepsilon}$, i.e. for the SUA and PST flows as shown in Figure 5.26. The tumbling period is seen to asymptote from either direction to the orientation where conditions for the onset of particle stall is satisfied which is seen to occur at a limit stall angle of approximately $\phi_p = 1.72$ rad. To the left of the red-dashed vertical limit lines in Figure 5.26a, or beneath the red-dashed horizontal line in Figure 5.26b, defining the asymptote events, the particle would stall, however the reverse situation is expected beyond these limits.

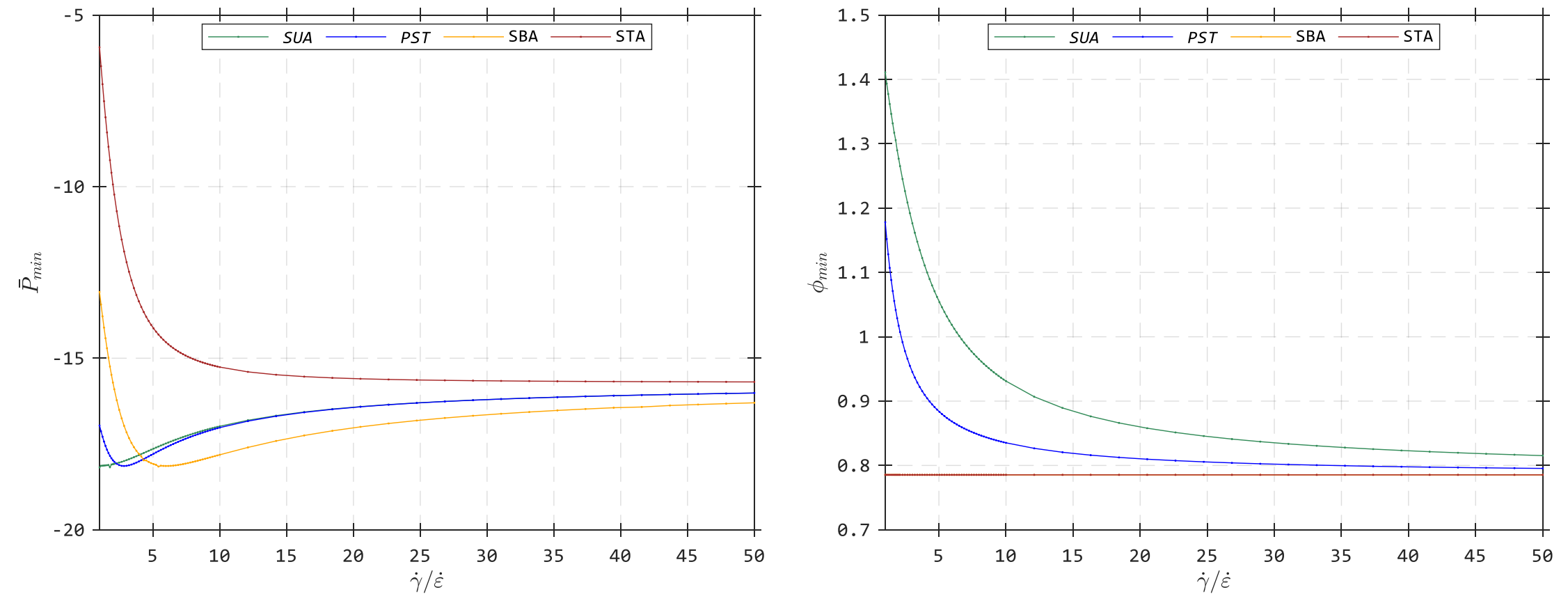

(a) (b)

Figure 5.25: (a) In-plane particle orientation angle at the instant of peak minimum pressure occurrence on the particles surface and (b) corresponding minimum pressure, for different shear dominance factor $\dot{\gamma}/\dot{\varepsilon}$ and for the for the combined homogenous flow conditions.

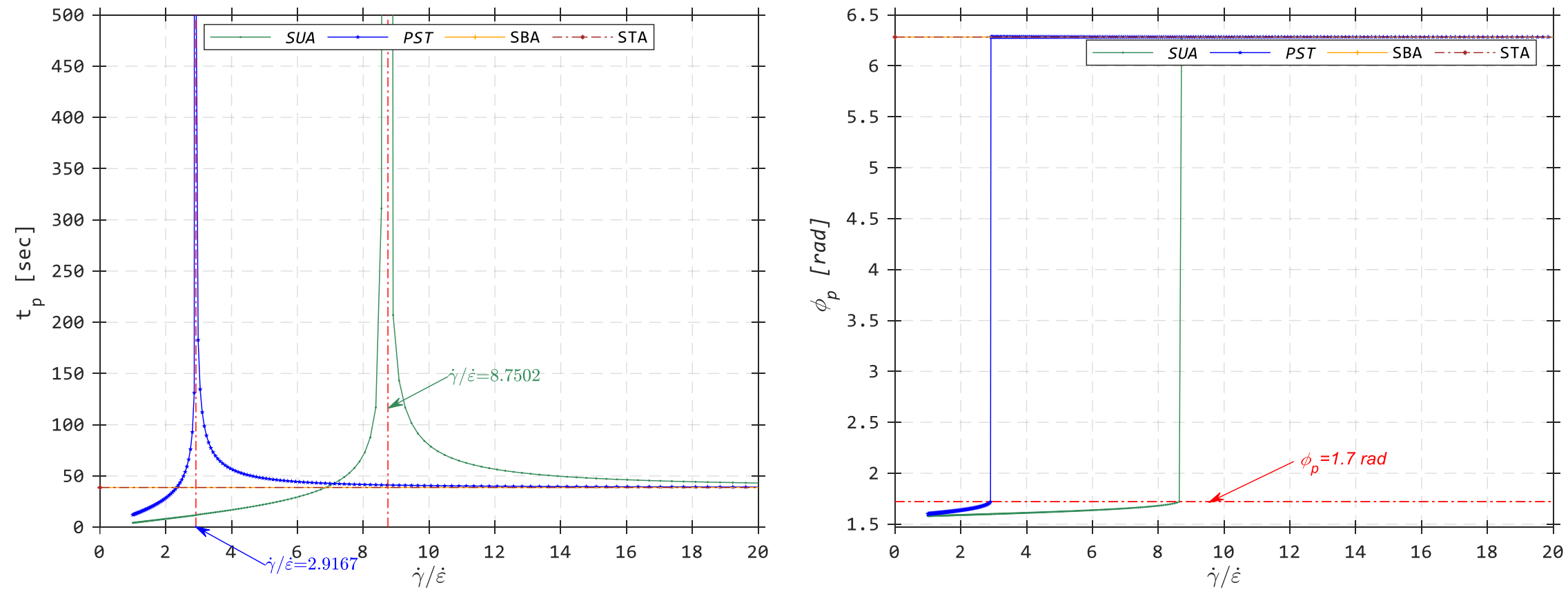

(a) (b)

Figure 5.26: Particle motion analysis (a) particle tumbling period (or stall time where applicable) and (b) corresponding particle rotation angle, for different shear dominance factor $\dot{\gamma}/\dot{\varepsilon}$ and for the for the combined homogenous flow conditions.

*5.10.1.5 Effect of Flow-type & Elongational ratio (Unsteady Center Gated Disk Flow).* The results of the pressure field around the fiber surface for this flow type are applicable to understanding the micro-void formations in injection molding of polymer composite materials. The flow conditions necessitate a negative pressure gradient along the radial flow direction. The flow characteristics involve a combination of spatially varying shear and planar elongation deformation rates. Closer to the plates, the flow is dominated by shear while regions closer to the centerline of the axisymmetric flow, are dominated by extensional flow conditions and the transition zones involves a combination of both shear and extensional velocity gradient driven flow [147]. The shear dominance increases with layer height. Numerical and experimental studies have shown that within a thin inner layer lining the wall, fibers are randomly oriented in the flow plane, and within thicker outer shells but close to the wall, the fibers are mostly aligned with the flow direction in the shear plane. In the core regions near the centerline where extensional flow is dominant, the fibers are mostly aligned in the direction perpendicular to the shear plane and the fiber orientation distribution in the transition regions are indeterminate [147], [273]. The calculations here are performed based on a half gap height (half disk spacing) h of $1.5\text{mm}$ and fiber dimensions of $60\mu\text{m}$ major axis length and $10\mu\text{m}$ minor axis radius for the ellipsoid. An inlet radius of $r_1 = h$ and outlet radius of $r_2 = 30h$ is assumed.

The results of the minimum pressure on the fibers surface over the possible range of fiber orientation configuration described by the surface of the unit sphere at the flow outlet and at layer heights $X_{0_3} = 0., .5, 1$ and the associated fiber orientation where the peak occurs which has been obtained from the optimization analysis are compared to results of the minimum pressure obtained by multidimensional grid analysis based on a discretization of 100 elements in the radial axis $\mathcal{X}_{0_r}$, 96 elements along the longitudinal axis $\phi$ and 48 elements along the latitudinal axis $\theta$ and computed at the fibers tip where it has been pre-determined to occur from the optimization analysis at the different layer heights. Figure 5.27 also shows that the direction at which the fiber orientation at peak pressure magnitude aligns with one of the principal directions of the flow at the associated spatial position. The pressure magnitudes are highest at the wall lining and the centerline of the flow (about -10.36), and lower at intermediate regions (about -9.63 at $\mathcal{X}_{0_3} = .25$). Figure 5.28a-c shows the evolution of the fiber orientation for a fiber with initial random orientation state at different layer heights including the flow centerline $\mathcal{X}_{0_3} = 0.$, the inner wall lining of the disk $\mathcal{X}_{0_3} = 1.$, and at an intermediate region $\mathcal{X}_{0_3} = .5$.

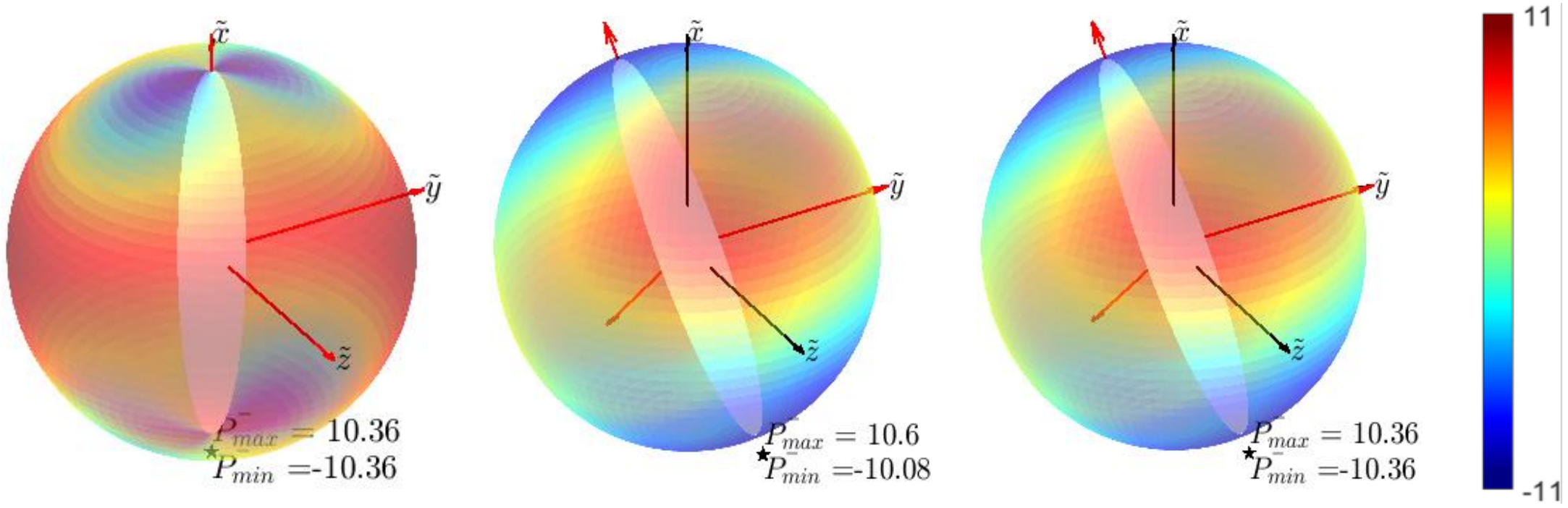

(a) (b) (c)

Figure 5.27: Distribution of the fiber surface "dimensionless" pressure over all possible orientation at the instant of peak minimum pressure occurrence on the fiber surface at the flow outlet at $\mathcal{X}_{0_r} = .25$ and for different layer heights (a) $\mathcal{X}_{0_3} = 0.$ (b) $\mathcal{X}_{0_3} = .5$ and (c) $\mathcal{X}_{0_3} = 1.$

The results show that at the centerline characterized by stretch dominant flow (cf. Figure 5.28a) the fiber abruptly reorients almost parallel to the transverse flow direction and stalls while at intermediate region with mixed stretching and shearing flow, the fibers gradually re-orients in the flow plane more favorably to the flow direction. At the walls, where shear is dominant, the fiber gradually but continuously tends to align with the flow direction (cf. Figure 5.28c) in line with the conclusion of Ferec at al. [147]. From the evolution of the fibers tip pressure magnitude in Figure 5.28d, at distance away from the flow inlet, the pressure magnitude increases from the walls to the centerline of the flow. The fibers trajectories are relatively longer at higher vertical distance from the centerline.

Figure 5.29a shows the distribution of the minimum fiber tip pressure over all possible direction at various layer heights. The values are seen to be relatively less severe than those observed in the shear dominant homogenous flow discussed in earlier sections. The corresponding instantaneous directions of peak minimum fiber tip pressure for the various layer heights occur in the shear plane and lie between the flow axis direction and an azimuthal inclination of $\theta_0 = -\pi/4$ which are the in-plane principal directions for pure elongation and pure shear respectively. The deviation of the instantaneous fiber orientation vector from the corresponding instantaneous direction of minimum fiber tip pressure would influence the extreme pressure magnitude at the fiber tips. For the case of a fiber initially orientated randomly at the flow inlet, Figure 5.29b shows the cosine of angle between the instantaneous fiber orientation and the corresponding unit direction of instantaneous minimum fiber tip pressure for varying layer heights. The result shows minimal deviation angle at the flow centerline and gradually increasing deviation angle with increasing layer height and radial distances.

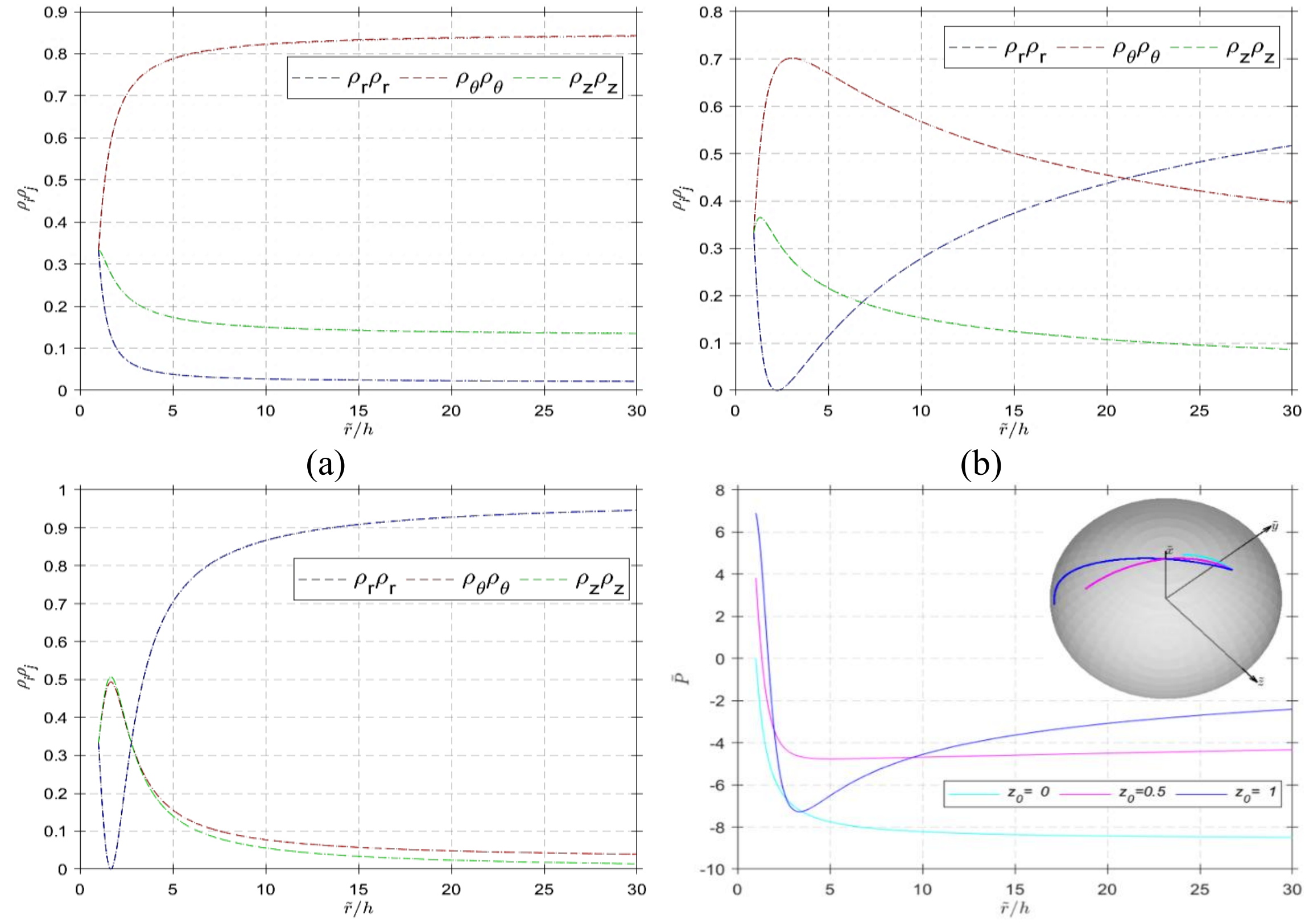

(c)

Figure 5.28: Evolution of the fiber orientation component at different layer heights (a) $X_{0_3} = 0.$ (b) $X_{0_3} = .5$ and (c) $X_{0_3} = 1$ for a fiber with initial random orientation at $r_0 = h$ (d) Evolution of the fibers tip pressure at the different layer heights (i.e. $X_{0_3} = 0., .5, 1.$).

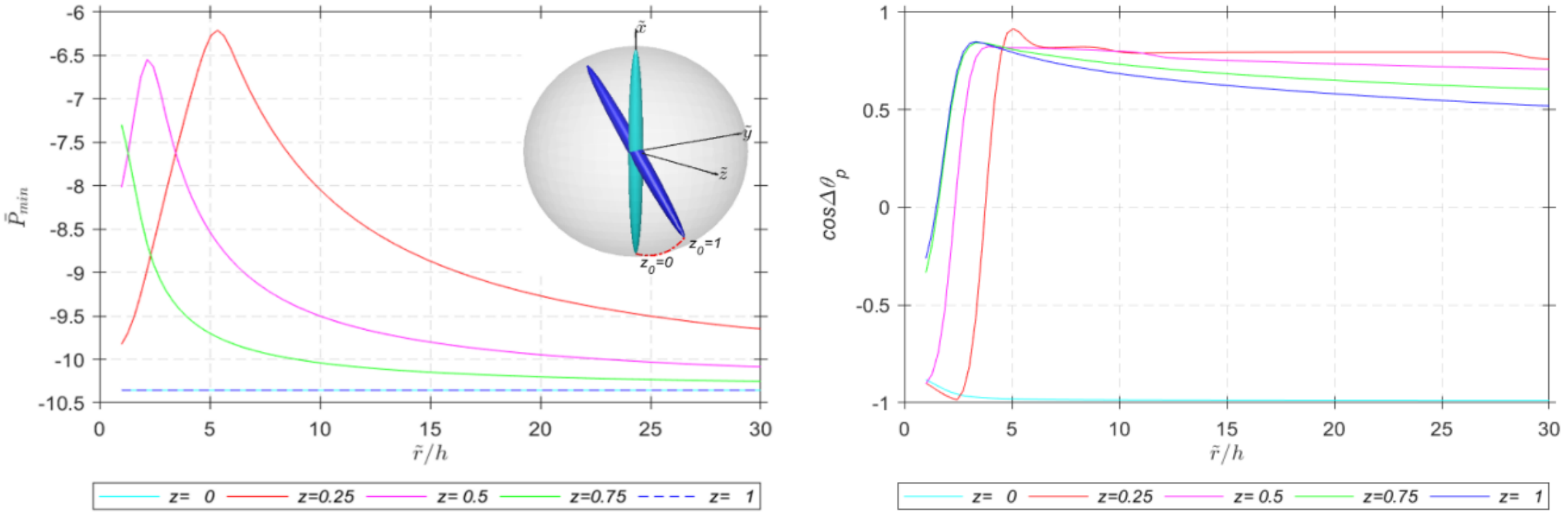

(a) (b)

Figure 5.29: (a) Distribution of the instantaneous minimum fiber tip pressure over all possible fiber configuration (b) Evolution of the cosine of angle between the fiber orientation vector and direction of peak minimum pressure for a fiber with initial random orientation.

### *5.10.2 Particle Motion in Non-Newtonian Homogenous Flows*

The results presented above focused on a single rigid ellipsoidal particle in various combined extensional and shear Newtonian homogenous flows that are considered typical of those in an EDAM nozzle during polymer composite processing. It is well understood, however, that thermo-plastic polymer materials are inherently non-Newtonian. Moreover, the addition of filler reinforcements to polymers are known to increase the melt viscosity and the shear-thinning fluid behavior in the nozzle. Additionally, high shear regions of complex flows such as the lubrication zone near the screw edge or regions of flow acceleration near the nozzle are known to result in flow segregation of highly shear-thinning polymer melt suspension into resin lean highly viscous domains and resin rich low-viscosity domains. As such understanding the particle behavior in shear-thinning fluid within various flow regimes is important in understanding microstructural development within polymer composite beads. The sections to follow present results obtained with the nonlinear FEA modeling approach presented above which considers a non-Newtonian shear-thinning power-law fluid rheology.

*5.10.2.1 Effect of Flow-type & Elongational ratio.* In this section, we consider the response of a single 3D ellipsoidal particle in simple homogeneous power-law fluid flows computed using the FEA method described above. The results presented in Figure 15 are for an ellipsoid with geometric ratio $r_e = 6$ rotating in a power-law fluid with a flow shear rate of $\dot{\gamma} = 1\ s^{-1}$ and power-law indices ranging from 0.2 to 1.0.

Figure 5.30a shows that the shear-thinning behavior has a slight influence on the particle's dynamic motion as reduction in the power-law index slows down the particle. The limits of the particle's in-plane angular velocity are observed to increase with increasing power-law index. Further, Figure 5.30b shows that the particle surface pressure extremes increase with decreased shear-thinning. Additionally, it is interesting to note that even though the orbit formed from particle tumbling in the shear-plane appears to exhibit little noticeable difference due to shear-thinning, Figure 5.31a shows that the tumbling period significantly increases with increasing shear-thinning.

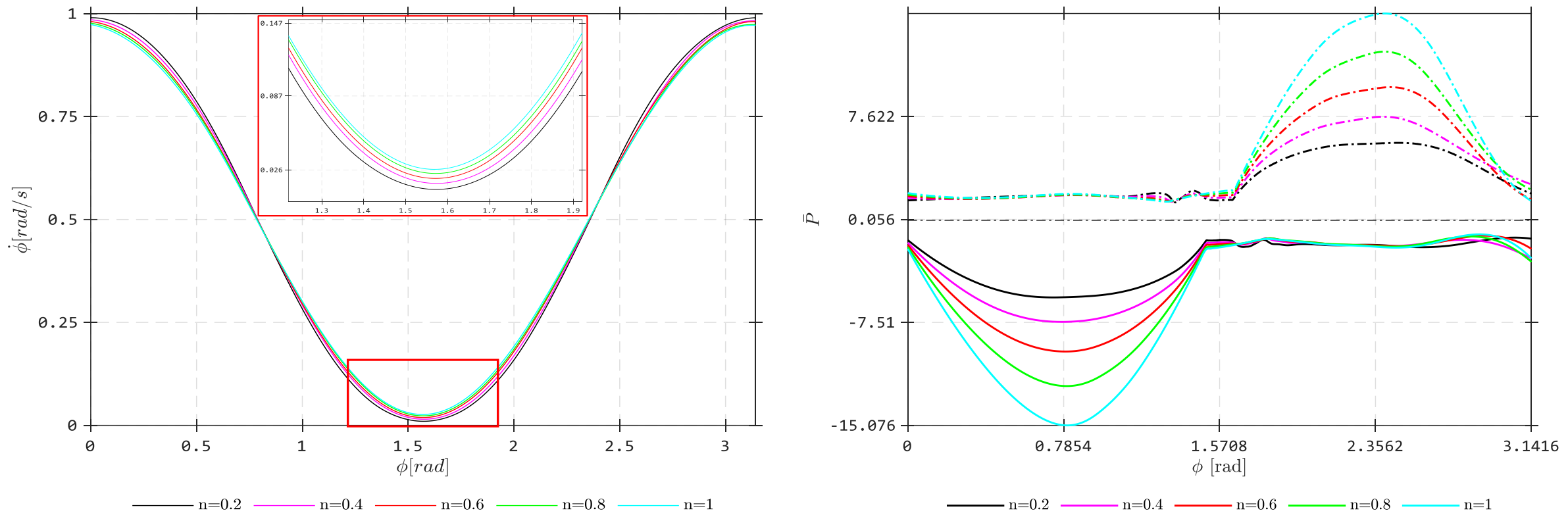

(a) (b)

Figure 5.30: FEA computed shear-thinning response of (a) particle polar angle $\phi$ vs precession $\dot{\phi}$ and (b) surface pressure extremes for particle motion in simple shear flow. Results are shown, for $r_e = 6, 0.2 \leq n \leq 0.8, m = 1\ Pa \cdot s^n$, $\dot{\gamma} = 1s^{-1}$ and $\phi^0 = 0, \theta^0 = -\pi/2, \psi^0 = 0$.

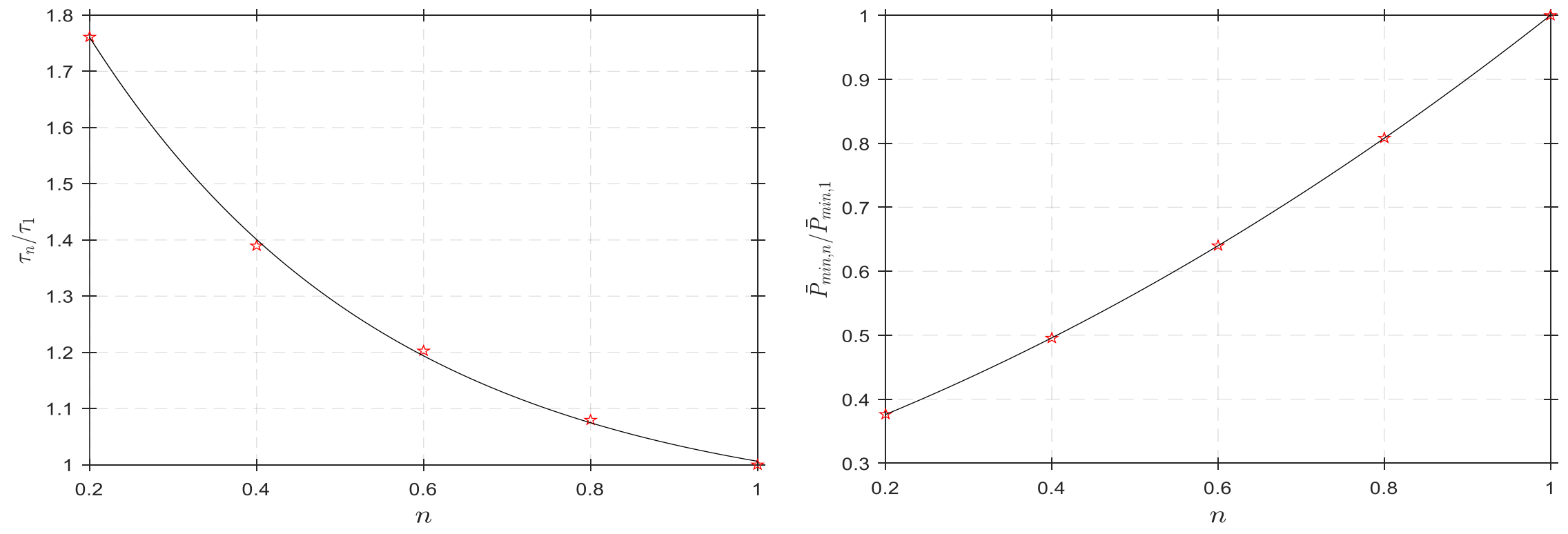

(a) (b)

Figure 5.31: Non-Newtonian to Newtonian ratio of the (a) particle's in-plane tumbling period (b) Orbital minimum particle tip pressure. Results are shown for $r_e = 6, 0.2 \leq n \leq 0.8, m = 1\ Pa \cdot s^n, \dot{\gamma} = 1s^{-1}$ and $\phi^0 = 0, \theta^0 = -\pi/2, \psi^0 = 0$.

The relationship between the particle tumbling period $\tau_n$ and the power-law index $n$ under simple shear flow conditions was determined through a typical curve fitting procedure to follow

$$\tau_n = \tau_1(0.9135 + 1.4724e^{-2.7645n}) \qquad (5.131)$$

where $\tau_n$ is the tumbling period in a shear-thinning fluid with power-law index $n$ and $\tau_1$ is the particle tumbling period for the Newtonian case, i.e. when $n = 1$. Figure 5.31b shows that the orbital minimum particle tip pressure has a quadratic variation with the flow behavior index as described as

$$\bar{P}_{min,n} = \bar{P}_{min,1}(0.28 + 0.42n + 0.30n^2) \qquad (5.132)$$

which implies that the shear-thinning effect on particle pressure distribution can be interpreted as having the same effect as would a modification of the Newtonian viscosity, agreeing with the findings of Ji et al.[221] and Awenlimobor et al. [232].

Figure 5.32 shows the pressure field around the ellipsoidal particle at various instants during the particle tumbling motion in the plane of the shear flow. The contours show an intensification of the pressure on the particle surface as the power-law index increases from $n = 0.2$ to $n = 1.0$. The pressure intensification is observed to be higher at orientations of peak orbital pressure extreme magnitudes (i.e. at $\phi = \pm\pi/4$). These observations can be explained from the plot of the disturbance in the velocity $\dot{X}_i^d$ [194] around the surface of the particle due to the particles motion defined as the difference between the flow-field velocity and free stream velocity, i.e. $\dot{X}_i^d = \dot{X}_i - \dot{X}_i^\infty$ (cf. Figure 5.33). We observe a higher magnitude of the velocity disturbance around same location on the particles surface where pressure extremes are observed to occur (i.e. at the particle tips). Likewise, the intensity of the disturbance is seen to increase with increasing power-law index and the magnification is higher at critical orientation angles where the orbital peak pressure extremes occur during alignment with the principal flow directions (i.e. at $\phi = \pm\pi/4$). The lower pressure intensities are thus a result of lower disturbance in the velocity field around the particle caused by the deceleration of the particles motion in the shear-thinning fluid.

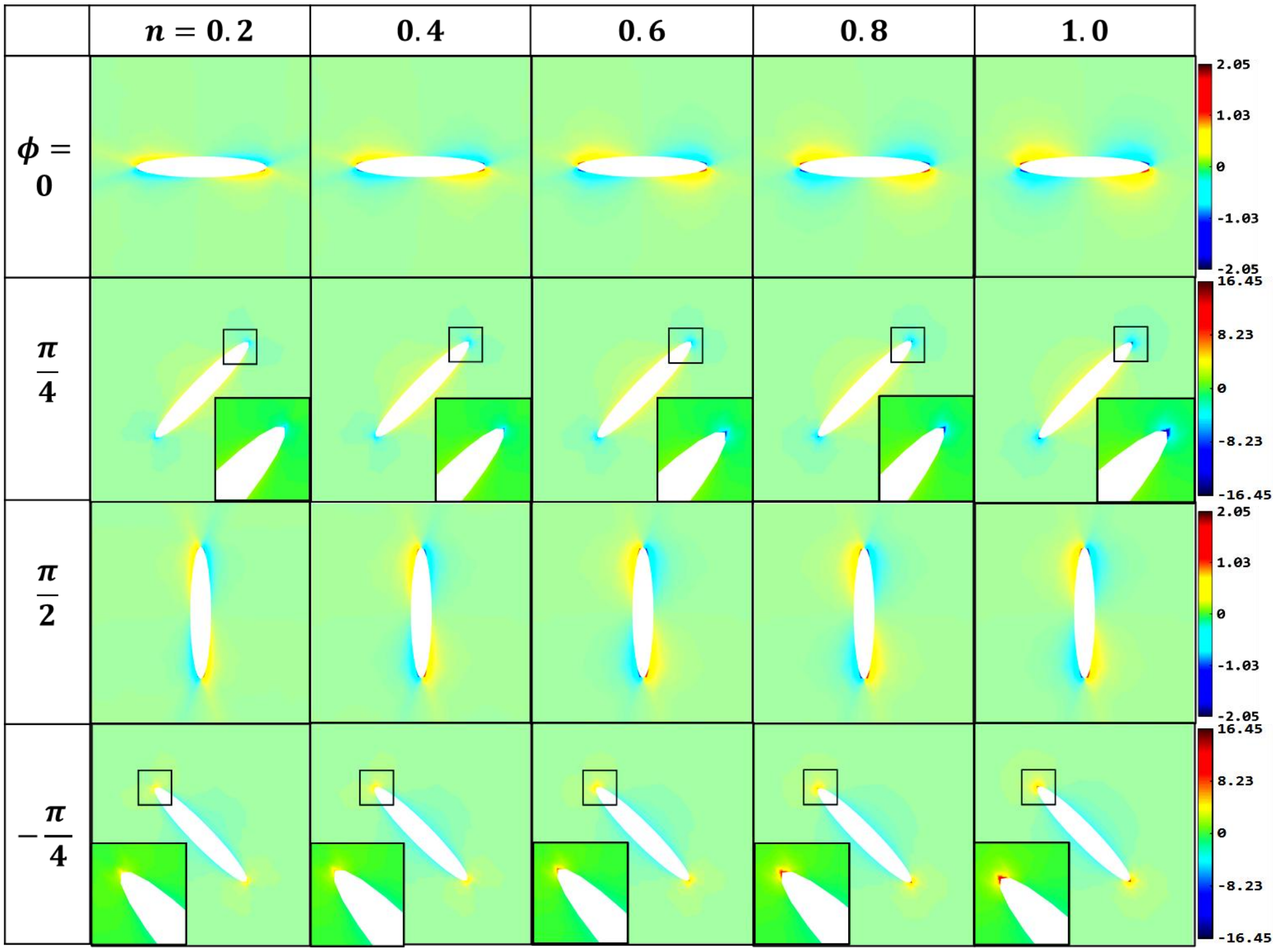

Figure 5.32: Mid - sectional plot of the pressure distribution around the ellipsoidal particle for at different instants during the particle's in-plane tumbling motion ($\phi = 0, \pi/4, \pi/4, \pi/4$) and for different power-law indices ($0.2 \leq n \leq 0.8$). Results are shown for $r_e = 6$, $m = 1\ Pa \cdot s^n$, $\dot{\gamma} = 1s^{-1}$ and $\phi^0 = 0, \theta^0 = -\pi/2, \psi^0 = 0$.

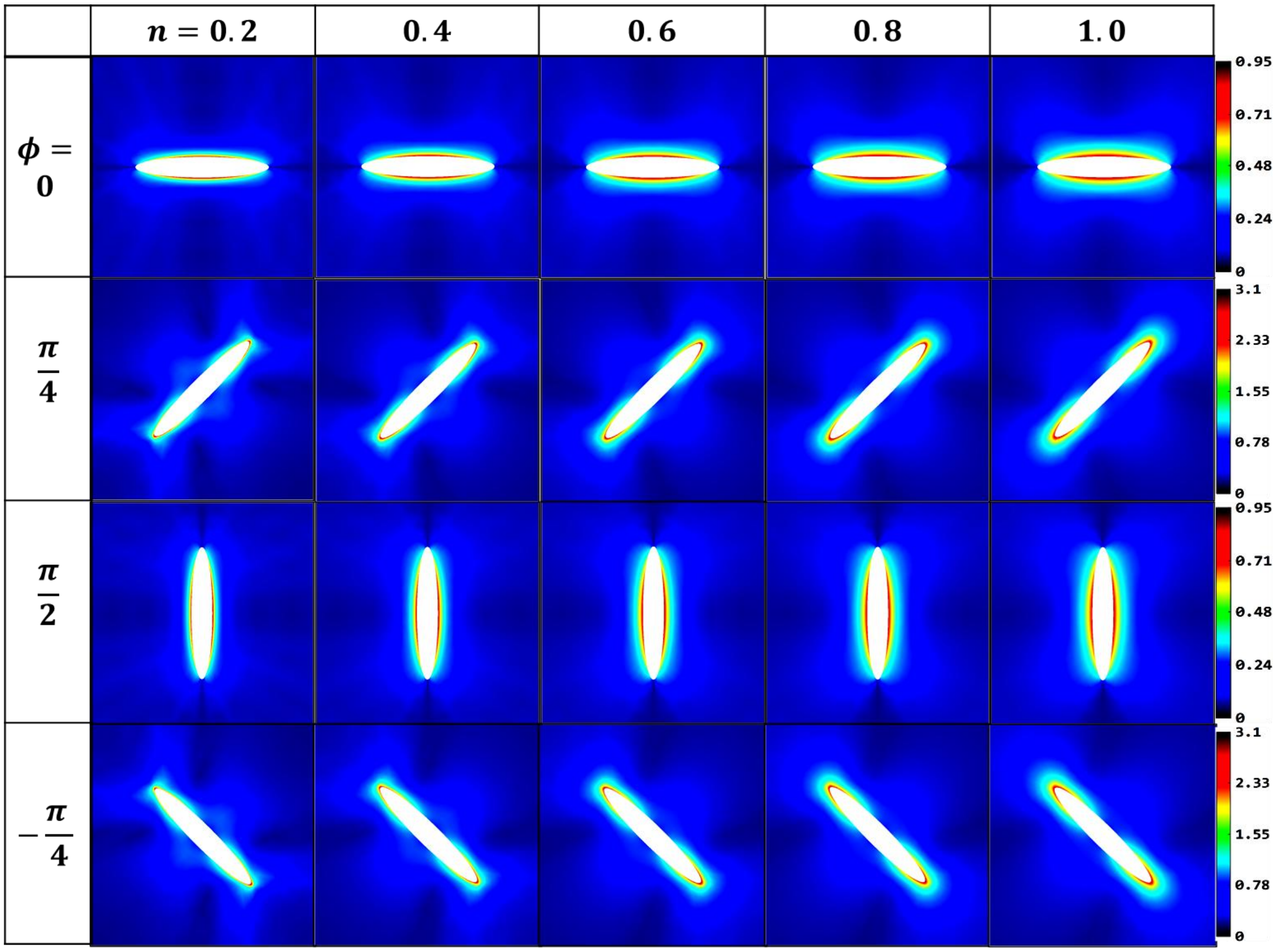

Figure 5.33: Mid - sectional plot of the disturbance velocity around the ellipsoidal particle for at different instants during the particle's in-plane tumbling motion ($\phi = 0, \pi/4, \pi/4, \pi/4$) and for different power-law indices ($0.2 \leq n \leq 0.8$). Results are shown for $r_e = 6, m = 1\ Pa \cdot s^n, \dot{\gamma} = 1s^{-1}$ and $\phi^0 = 0, \theta^0 = -\pi/2, \psi^0 = 0$.

Figure 5.34a-d shows the computed results of the single rigid ellipsoidal particle in combined shear and uniaxial extension (SUA) flow type with a power-law index $n$ ranging from 0.2 to 1 while considering two shear-extension rate ratios (i.e., $\dot{\gamma}/\dot{\varepsilon} = 1$ and 10). Figure 5.34a and Table 5.4 shows that the particle stalls in the SUA flow with $\dot{\gamma}/\dot{\varepsilon} = 1$) and the shear-thinning fluid behavior slightly increases particle rotation speed and shortens the trajectory which is evident from the slight reduction in the time to particle stall and the stall angle with decreasing power-law index. Figure 5.34b shows that the shear-thinning fluid reduces the magnitude of the particle surface pressure extremes in the SUA flow,

however, the shear-thinning rheology does not affect the orbital angle location where the minimum peak magnitude pressure occurs (i.e. at $\phi = +1.41$ rad). In the shear dominant flow condition when $\dot{\gamma}/\dot{\varepsilon} = 10$, the particle tumbles periodically under slightly non-Newtonian rheological fluid behavior ($n \geq 0.8$), however further reduction in the power-law index ($n < 0.8$) causes the particle to eventually stall in a preferred orientation along the direction of stretching (cf. Figure 5.34c). This implies that the conditions for particle stall in a shear-thinning fluid is dependent on the competing influence of the shear-extensional rate factor and the intensity of the shear-thinning fluid behavior.

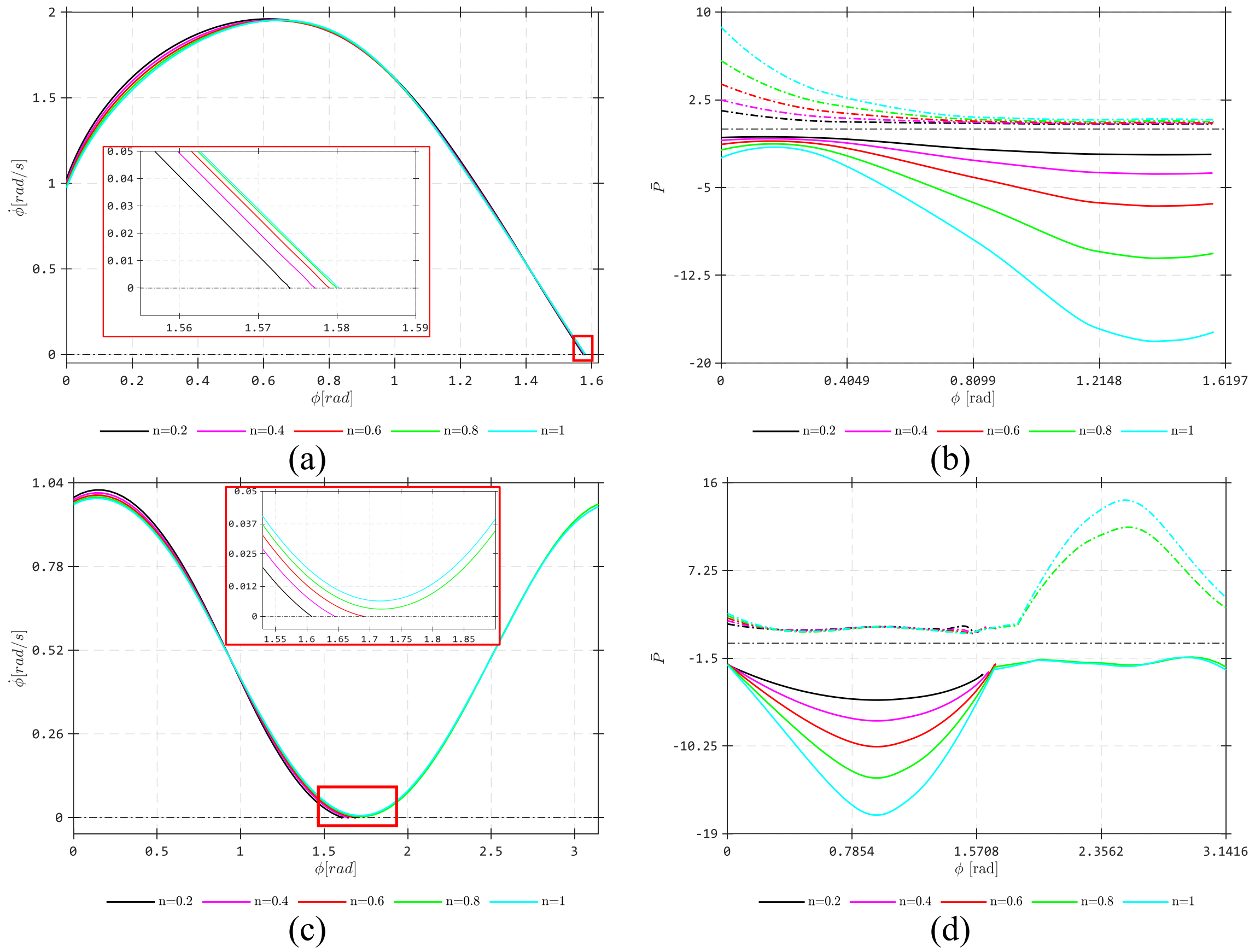

Figure 5.34: Phase diagram of the particles polar angle $\phi$ vs (a) precession $\dot{\phi}$ - $\dot{\gamma}/\dot{\varepsilon} = 1$ (b) surface pressure maximum (dashed) and minimum (continuous) - $\dot{\gamma}/\dot{\varepsilon} = 1$ and (c) precession $\dot{\phi}$ - $\dot{\gamma}/\dot{\varepsilon} = 10$, (d) surface pressure maximum (dashed) and minimum (continuous) - $\dot{\gamma}/\dot{\varepsilon} = 10$, for particle motion in combined shear and uniaxial extension

(SUA) flow. Results are shown, for $0.2 \leq n \leq 0.8$, $m = 1\ Pa \cdot s^n$, $\dot{\gamma} = 1s^{-1}$ and $\phi^0 = 0, \theta^0 = -\pi/2, \psi^0 = 0$.

Table 5.4: Particle stall time $\tau_s$ and particle stall angle $\phi_s$ for single ellipsoidal particle motion in SUA shear-thinning flow for different flow behavior index $0.2 \leq n \leq 1.0$ with $m = 1\ Pa \cdot s^n$, $\dot{\gamma} = 1s^{-1}$ and $\dot{\gamma}/\dot{\varepsilon} = 1$.

| $n$ | 0.2 | 0.4 | 0.6 | 0.8 | 1.0 |
|---|---|---|---|---|---|
| $\tau_s$ | 3.922 | 3.982 | 4.012 | 4.032 | 4.032 |
| $\phi_s$ | 1.574 | 1.577 | 1.579 | 1.580 | 1.580 |

Table 5.5 shows that the particle stall time ($\tau_s$) and stall angle ($\phi_s$) when $n < 0.8$, and half period ($\tau_n^{0.5}$) for the cases where the particle tumbles periodically (i.e. when $n \geq 0.8$). As expected, at the location of the orbital extreme pressure magnitude where the particle orientation coincides with the principal flow direction (at $\phi = +0.931, +2.502$ rad), the surface extreme pressure magnitudes are observed to decrease with the intensity of the shear-thinning fluid rheology. The pressure fluctuations on the particle's tip as it tumbles continuously in the shear dominant flow or the local pressure that subsist at particle's tip as it stalls in the extension dominant flow condition are important in understanding the final microstructural formations within printed polymer composite beads [57].

Table 5.5: Half-period/stall time (where applicable) $\tau_s$ and stall angle $\phi_s$ (where applicable) for single ellipsoidal particle motion in SUA shear-thinning flow for different flow behavior index $0.2 \leq n \leq 1.0$ with $m = 1\ Pa \cdot s^n$, $\dot{\gamma} = 1s^{-1}$ and $\dot{\gamma}/\dot{\varepsilon} = 10$.

| $n$ | 0.2 | 0.4 | 0.6 | 0.8 | 1.0 |
|---|---|---|---|---|---|
| $\tau_n^{0.5}$ or $\tau_s$ | 31.930 | 39.777 | 65.146 | 59.070 | 40.156 |
| $\phi_s$ | 1.607 | 1.644 | 1.689 | - | - |

In the combined shearing/planar stretching (PST) flow, the shear-thinning fluid rheology does not deter the particle's acquiescence into preferred orientation state under the extension-rate dominant flow condition (i.e. $\dot{\gamma}/\dot{\varepsilon} = 1$). However, the shear-thinning is observed to decelerate the particles motion, prolong the stall event and extend the particles trajectory to stall contrary to what was observed in the SUA flow. Figure 5.35a reveals a slight reduction in the peak in-plane angular velocity with decreasing power-law index and Table 5.6 shows that the stall time and stall angle both of which increase with increased shear-thinning. The particle tip pressure magnitudes are nonetheless observed to decrease with increased shear-thinning as expected (cf. Figure 5.35b). The particle in-plane orientation at the location of orbital minimum surface pressure (i.e. at $\phi = +1.18$) is unaltered by the shear-thinning effect.

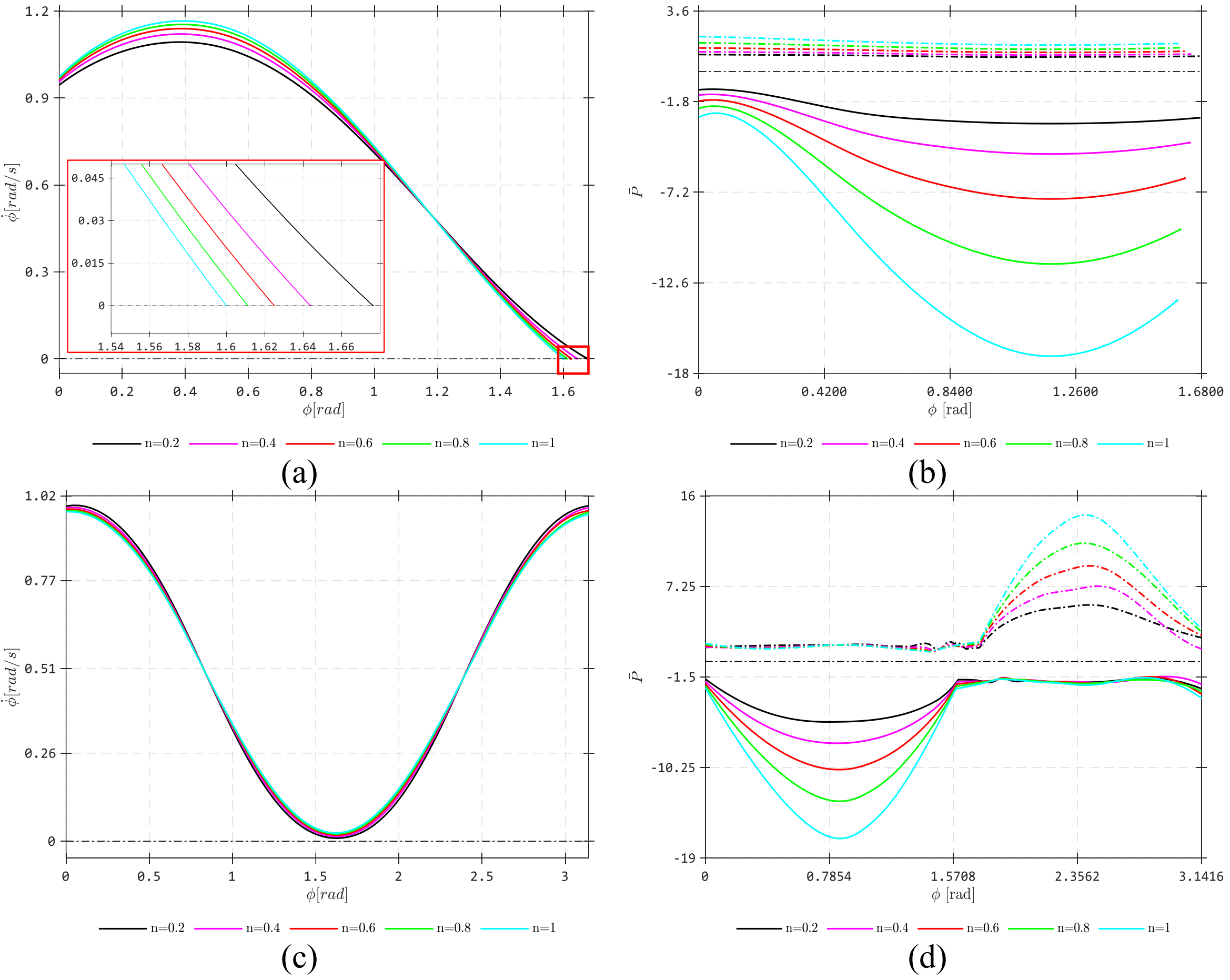

Figure 5.35: Phase diagram of the particles polar angle $\phi$ vs (a) precession $\dot{\phi}$ - $\dot{\gamma}/\dot{\varepsilon} = 1$ (b) surface pressure maximum (dashed) and minimum (continuous) - $\dot{\gamma}/\dot{\varepsilon} = 1$ and (c) precession $\dot{\phi}$ - $\dot{\gamma}/\dot{\varepsilon} = 10$, (d) surface pressure maximum (dashed) and minimum (continuous) - $\dot{\gamma}/\dot{\varepsilon} = 10$, for particle motion in combined shear and planar stretching (PST) flow. Results are shown, for $0.2 \leq n \leq 0.8$, $m = 1\,Pa \cdot s^n$, $\dot{\gamma} = 1s^{-1}$ and $\phi^0 = 0, \theta^0 = -\pi/2\,, \psi^0 = 0$.

The shear-thinning effect does not stall the particle under the shear-rate dominant condition (i.e. when $\dot{\gamma}/\dot{\varepsilon} = 10$) in the PST flow contrary to what was observed in the SUA flow. However, at the local minimum of the particle's angular velocity evolution curve when its deceleration approaches zero (cf. Figure 5.35c), the increased shear-thinning effect is observed to further decelerate particle motion and bring it closer to stall condition.

Table 5.6: Particle stall time $\tau_s$ and particle stall angle $\phi_s$ for single ellipsoidal particle motion in PST shear-thinning flow for different flow behavior index $0.2 \leq n \leq 1.0$ with $m = 1\ Pa \cdot s^n$, $\dot{\gamma} = 1s^{-1}$ and $\dot{\gamma}/\dot{\varepsilon} = 1$.

| $n$ | 0.2 | 0.4 | 0.6 | 0.8 | 1.0 |
|---|---|---|---|---|---|
| $\tau_s$ | 13.337 | 11.796 | 11.026 | 10.515 | 10.135 |
| $\phi_s$ | 1.676 | 1.644 | 1.625 | 1.611 | 1.600 |

Table 5.7 shows that the particles tumbling period increases with decreasing power-law index indicating the deceleration of the particle rotation with increased shear-thinning. The sustained particle motion allows for continuous fluctuations between particle surface pressure extremes at the particle tip. As would be expected, the pressure magnitudes are observed to decrease with increased shear-thinning (cf. Figure 5.35d). Further, the in-plane orientation at the orbital location of particle surface tip pressure extremum (i.e. at $\phi = +0.835, +2.406\text{rad}$) is unaltered by the shear-thinning effect.

Table 5.7: Half-period for single ellipsoidal particle motion in PST shear-thinning flow for different flow behavior index $0.2 \leq n \leq 1.0$ with $m = 1\ Pa \cdot s^n$, $\dot{\gamma} = 1s^{-1}$ and $\dot{\gamma}/\dot{\varepsilon} = 10$.

| $n$ | 0.2 | 0.4 | 0.6 | 0.8 | 1.0 |
|---|---|---|---|---|---|
| $\tau_n^{0.5}$ | 36.181 | 28.045 | 24.169 | 21.839 | 20.280 |

Under the balanced shear and bi-axial elongation (SBA) flow condition, inward flow normal to the shear plane coupled with uniform stretching along the shear plane promotes particle in-plane tumbling motion. Under this flow condition, the particle does not stall irrespective of the magnitude of the extension rate. However, while the increased shear-thinning is observed to accelerate the particles motion when $\dot{\gamma}/\dot{\varepsilon} = 1$, it is shown to slightly decelerate the particles motion under a higher shear rate i.e. $\dot{\gamma}/\dot{\varepsilon} = 10$ (cf. Figure 5.36a & c). When $\dot{\gamma}/\dot{\varepsilon} = 1$ the limits of particle in-plane angular velocity are observed to decrease with increased shear-thinning and vice versa when $\dot{\gamma}/\dot{\varepsilon} = 10$. The shear-thinning effect decreases the particle tumbling period when $\dot{\gamma}/\dot{\varepsilon} = 1$ and increases the period when $\dot{\gamma}/\dot{\varepsilon} = 10$ (cf. Table 5.8). Under a lower shear rate ($\dot{\gamma}/\dot{\varepsilon} = 1$), there are no noticeable peaks in the evolution of the particle maximum surface pressure, contrary to what is observed when $\dot{\gamma}/\dot{\varepsilon} = 10$. As would be expected, the particle surface pressure extremes are observed to decrease with increased shear-thinning and the location of orbital minimum surface pressure at $\phi = \pm\pi/4$ is unaffected by the shear-thinning rheology (cf. Figure 5.36b & d).

Observation of particle behavior in the flow types considered here as applied to polymer melt flow conditions during EDAM processing suggests that the shear-thinning effect increases the particle stall tendency closer to the EDAM nozzle center where a higher extension rate dominance is seen. Shear-thinning is seen here to have a similar effect as decreasing the shear-to-extension rate ($\dot{\gamma}/\dot{\varepsilon}$), thus shifting the boundaries of the extension dominant region outward (cf. APPENDIX B, B.3). Irrespective of the flow regime, the shear-thinning rheology reduces the pressure magnitude which has a similar effect to reducing the viscosity magnitude in a Newtonian fluid.

Additionally, in high shear dominant flow regions of the EDAM nozzle, the shear-thinning effect is generally expected to slow down the particles motion, while close to the nozzle center, dominated by high extension-rate, the particle's stall event is expected to be promoted by shear-thinning effects.

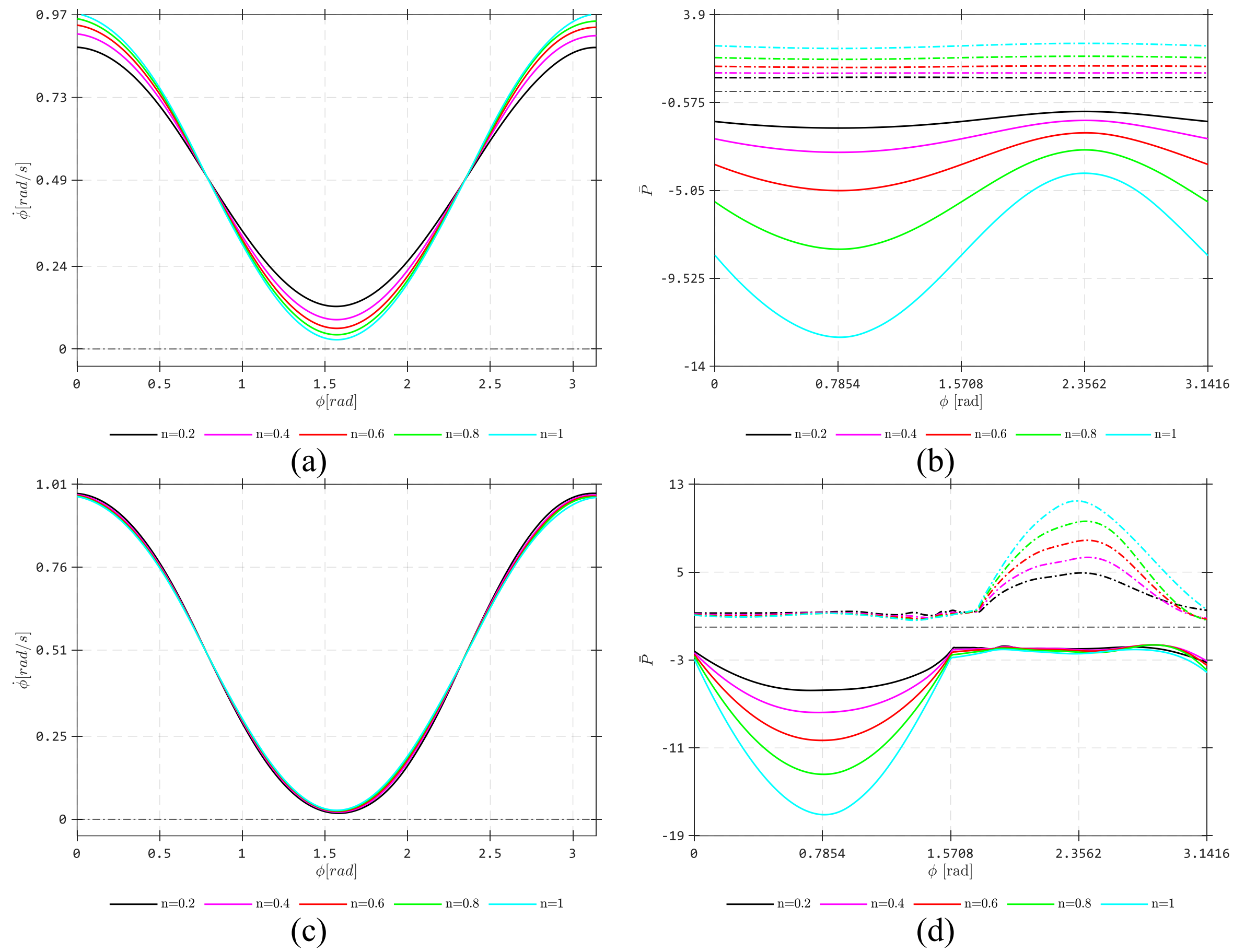

Figure 5.36: Phase plots of the particles polar angle $\phi$ vs (a) precession $\dot{\phi}$ - $\dot{\gamma}/\dot{\varepsilon} = 1$ (b) surface pressure maximum (dashed) and minimum (continuous) - $\dot{\gamma}/\dot{\varepsilon} = 1$ and (c) precession $\dot{\phi}$ - $\dot{\gamma}/\dot{\varepsilon} = 10$, (d) surface pressure maximum (dashed) and minimum (continuous) - $\dot{\gamma}/\dot{\varepsilon} = 10$, for particle motion in combined shear and biaxial extension (SBA) flow. Results are shown, for $0.2 \leq n \leq 0.8$, $m = 1\ Pa \cdot s^n$, $\dot{\gamma} = 1 s^{-1}$ and $\phi^0 = 0, \theta^0 = -\pi/2, \psi^0 = 0$.

Table 5.8: Half tumbling period $\tau_n^{0.5}$ for single ellipsoidal particle motion in SBA shear-thinning flow for different flow behavior index $0.2 \leq n \leq 1.0$ and different shear to extension rate ratio ($\dot{\gamma}/\dot{\varepsilon}$) with $m = 1\ Pa \cdot s^n$, $\dot{\gamma} = 1 s^{-1}$.

| $\tau_n^{0.5}$ | | $n$ | | | | |
|---|---|---|---|---|---|---|
| | | 0.2 | 0.4 | 0.6 | 0.8 | 1.0 |
| $\dot{\gamma}/\dot{\varepsilon}$ | 1 | 9.558 | 11.273 | 13.266 | 15.799 | 19.453 |
| | 10 | 25.265 | 22.650 | 21.155 | 20.138 | 19.423 |

*5.10.2.2* *Effect of Initial Particle Orientation.* In earlier sections we showed that the pressure magnitudes on the surface of a particle suspended in a Newtonian simple shear flow reduces as the orbit constant $\varsigma$ (cf. eqn. (5.65) & (5.66)) goes from $\varsigma = +\infty$ where the particle is tumbling in the shear plane to $\varsigma = 0$ where the particle is spinning about its axis perpendicular to the shear plane. It was also shown that the tumbling period was unaffected by Jeffery's orbit. The effect of shear-thinning rheology on the particle motion for various Jeffery orbits are presented in this section. We consider particle motion in simple shear flow with shear rate of $\dot{\gamma} = 1\mathrm{s}^{-1}$ and for a GNF power-law fluid rheology with a power-law index of $\mathrm{n} = 0.5$ and a consistency index of $\mathrm{m} = 1\ \mathrm{Pa} \cdot \mathrm{s}^{\mathrm{n}}$. The same geometric aspect ratio of $\mathrm{r_e} = 6$ as was previously used is considered here.

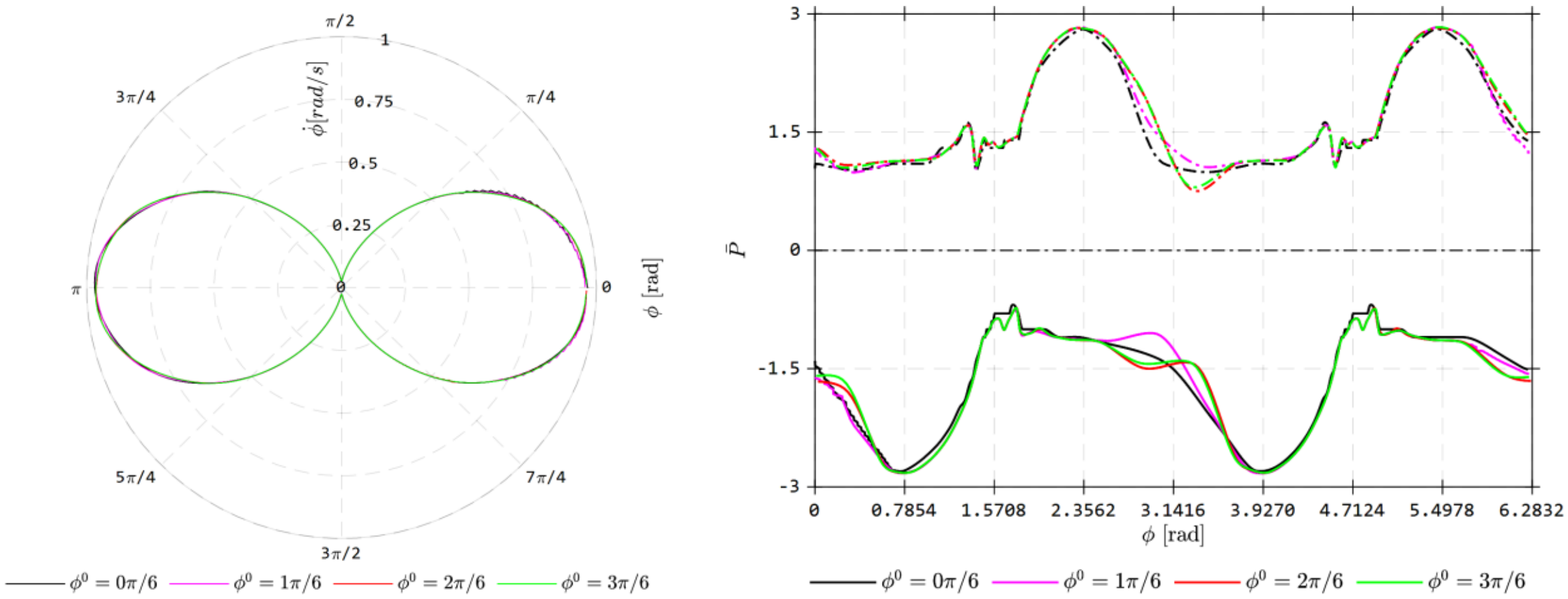

(a) (b)

Figure 5.37: Figure showing results of (a) evolution of the fibers angular velocity in space along Jeffery's orbit (b) evolution of fibers limit surface pressure in space along Jeffery's orbit. Results are presented for different fiber initial orientation ( $0 \leq \phi^0 \leq \pi/2$ ) and for flow shear rate $\dot{\gamma} = 1\ s^{-1}$, fiber aspect ratio $r_e = 6$, and flow behavior index $n = 0.2$.

The 2D sensitivity analysis on the fibers' initial condition showed that the angular velocity of the fiber is unaltered by the initial condition (cf. Figure 5.37a), nor is its limit pressure peaks on the fiber's surface affected (cf. Figure 5.37b) in a shear-thinning fluid with strong non-Newtonian characteristics (flow behavior index $n = 0.2$). In the 3D analysis, the Jeffery's orbits are altered slightly by the shear-thinning fluid which occurs to a greater extent as the fiber is oriented further out of the shear plane (i.e., as $\varsigma = +\infty$ is moved to $\varsigma = 0$) as shown in Figure 5.38a. The initial particle polar angle on a particular Newtonian Jeffery's orbit is observed to also modify the particle trajectory. Figure 5.38a and b also show that trajectory of the particle motion in an orbit with an initial azimuth angle of $\theta^0 = 2\pi/24$ with two initial starting positions at the vertices of the Newtonian conical orbit. With an initial starting position at the vertex of the directrix of the Newtonian conical orbit on the major axis (at $\phi^0 = \pi/2$), the particle path is seen to dilate outwardly defined by the outer curve (dashed cyan line) from the Newtonian orbit (continuous black line). However, starting the particle from the co-vertex of the directrix of the Newtonian orbit on the minor axis (i.e. $\phi^0 = 0$), the orbit constricts inwardly defined by the inner curve (continuous cyan line). Both curves clearly illustrate the extent of deviation in the particle path from the Newtonian orbit and that for a given power-law index and set of flow parameters. The fluid shear-thinning is seen to influence the particles motion similar to elongating or shortening the particle, depending on the initial position on the orbit. This observed behavior is consistent with conclusions by Abtahi et al.[194].

The fluid shear-thinning is seen to have a more profound effect on the surface pressure of particles on Jeffery orbit closer to the shear plane ($\varsigma \to +\infty$) as compared to orbits farther out of plane (i.e. close to $\varsigma \to 0$). The net pressure drop ($\delta\bar{P}$) due to the shear-thinning effect.t is seen to be proportional to the magnitude of the particle surface pressure as shown in Figure 5.38c. Likewise, the net pressure drop of particle tip pressure is seen to depend on its initial starting position as is evident from the net pressure curves shown for each initial polar angle on the orbit farthest from the shear plane ($\theta^0 = 2\pi/24$), i.e. dashed cyan line for $\phi^0 = 0$ and continuous cyan line for $\phi^0 = \pi/2$ .

As expected, the particle dynamics are also affected by the shear-thinning rheology. The envelope of the phase diagram of the particle's nutation (cf. Figure 5.38d) contract inwardly from the Newtonian envelope due to the shear-thinning effect irrespective of the initial position on the orbit. The shear-thinning rheology appears to have less effect on the particle's precession as the Jeffery's orbit is oriented further out of plane, i.e. when $\varsigma \to 0$ (cf. Figure 5.38e), however, this effect on the particle's nutation is more profound as $\varsigma \to 0$. Although, the particle's period of tumbling is independent on the Jeffery's orbit in Newtonian flow, the tumbling period is observed to be influenced by the Jeffery's orbit under shear-thinning flow conditions. Figure 5.39a shows the relationship between the tumbling period $\tau_{0.5}$ and the initial azimuth angle, $\theta_0$ for the particle motion in non-Newtonian power-law fluid, with flow behaviour index of $n = 0.5$. The relationship in Figure 5.39a can be described as

$$\tau_{0.5} = \tau_1\left(1.2976 - 0.7358e^{-3.84958\theta_o}\right) \tag{5.133}$$

which has been obtained using a typical curve fitting procedure. Overall, the shear-thinning fluid rheology slows down a particle's motion which occurs to a greater degree as the tumbling orbit approaches the shear plane (i.e. $\varsigma \to +\infty$). Additionally, the reduction in the minimum surface pressure magnitudes due to shear-thinning becomes more significant as $\varsigma \to +\infty$ and vice-versa. The relationship between the particles orbital minimum tip pressure $\bar{P}_{min}$ and the initial particles out-of-plane orientation $\theta^0$ appearing in Figure 5.39b clearly shows a gradual widening of the gap between the Newtonian and non-Newtonian pressure profiles.

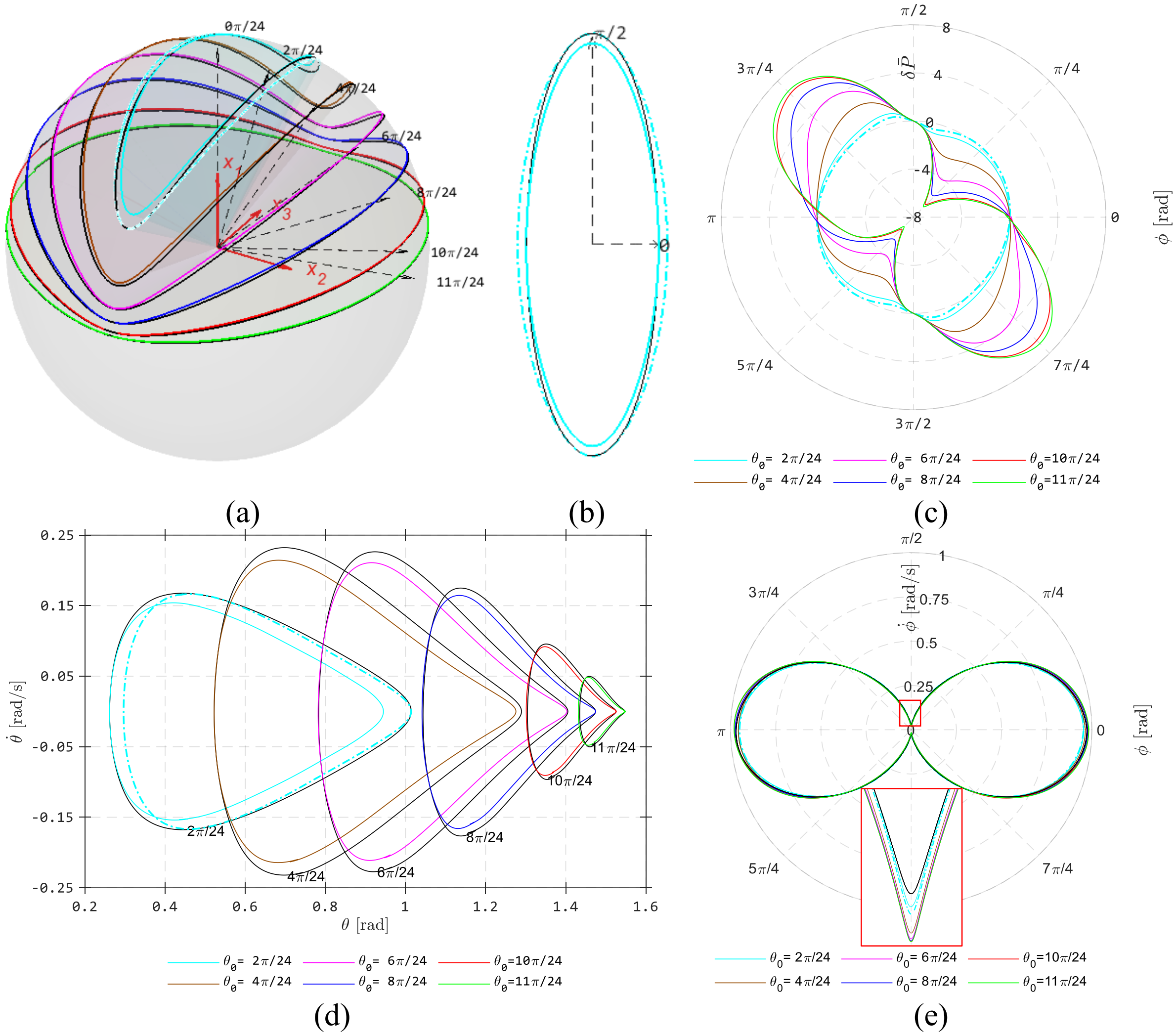

Figure 5.38: Effect of fluid shear-thinning on Jeffery's orbit: (a) particle's orbits in simple shear flow (b) dilated orbit (dashed cyan line $\phi^0 = \pi/2$), constricted orbit (continuous cyan line, $\phi^0 = 0$) and Newtonian orbit (black line) for $\theta^0 = 2\pi/24$ (c) difference in particle tip pressure between NT and GNF fluid (d) phase diagram of azimuth angle $\theta$ vs nutation $\dot{\theta}$ (e) polar plot of precession $\dot{\phi}$ vs polar angle $\phi$, for different initial particle orientation between $-2\pi/24 \leq \theta^0 \leq -12\pi/24$, $\phi^0 = 0$, $\psi^0 = 0$ and for NT fluid (dashed) and GNF power-law fluids (continuous) with $m = 1\ Pa \cdot s^n, n = 0.5$.

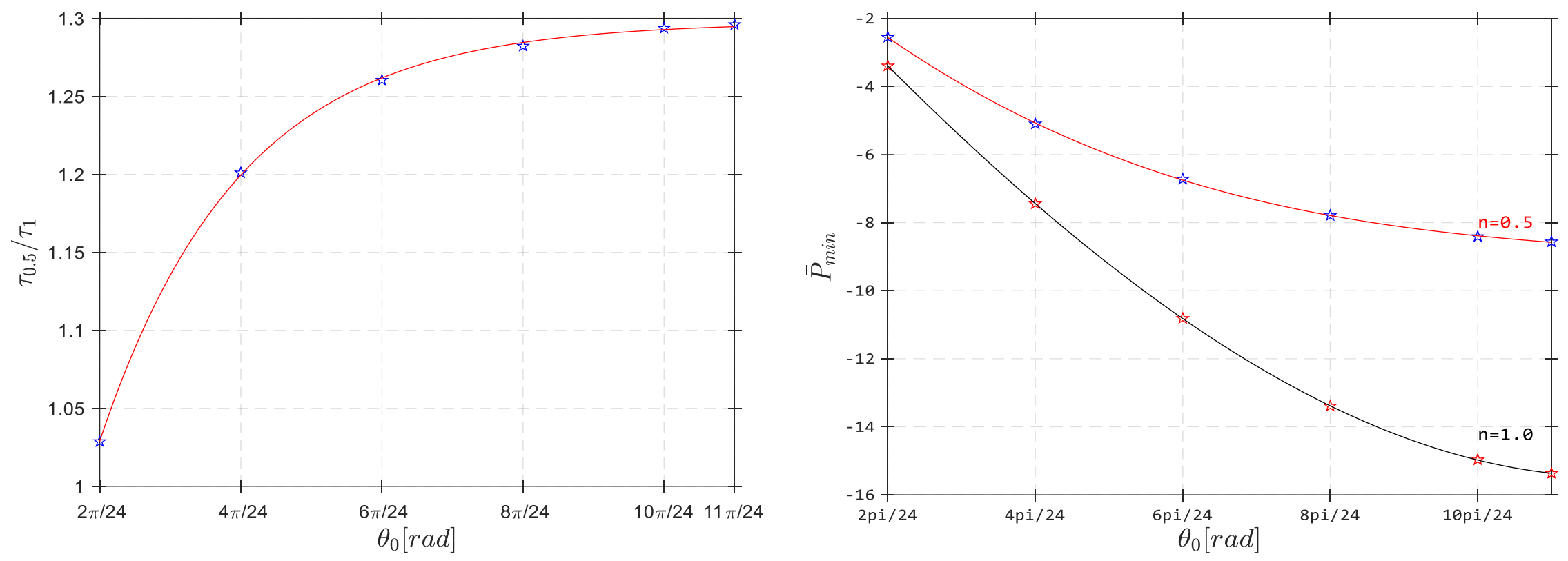

Figure 5.39: Effect of shear-thinning comparing (a) the non-Newtonian to Newtonian tumbling period $\tau_{0.5}/\tau_1$, and (b) the non-Newtonian (red line) and Newtonian (black line) particles orbital minimum tip pressure $\bar{P}_{min}$, versus the initial azimuth angle $\theta_0$, considering GNF power-law fluid, with with $m = 1\ Pa \cdot s^n, n = 0.5$ and initial orbit position $\phi^0 = 0$.

#### *5.10.2.3 Effect of Geometric Aspect Ratio.*

For completeness, we now consider the effect of the geometric aspect ratio on particle behaviour in shear-thinning simple shear flow making comparisons to the behaviour in a Newtonian fluid. The result for the evolution of the 2D rigid ellipsoidal fiber along Jeffery's orbit in viscous fiber suspension simple shear flow with shear thinning fluid rheology having flow behavior index ranging from 0.2 to 1.0 are presented in Figure 5.40 below for two (2) cases of fibers geometric aspect ratio, i.e., a prolate spheroid with geometric ratio $r_e = 6$ and a slender fiber with geometric ratio $r_e = 30$. For the first case, a shear rate of $\dot{\gamma} = 1\ s^{-1}$ is used however to reduce the orbit period for the case with high aspect, a shear rate of $\dot{\gamma} = 3\ s^{-1}$ was used given the definition of the Jeffery's orbit period (cf. eqn. (5.68)). For objectivity, the normalized quantities of the fiber's response are reported, i.e. $\bar{\dot{\phi}} = \dot{\phi}/\dot{\gamma}$ for the angular velocity and $\bar{p} = (p - p_0)/\mu\dot{\gamma}$ for the surface pressure. The results in Figure 5.40a & b show that the shear-thinning effect on the particles dynamics becomes more pronounced with increasing fibers aspect ratio. Figure 5.41a shows that the shear-thinning slightly slows down the particle motion and to a greater extent for higher aspect ratio particles. Likewise, the minimum and maximum pressure peaks on the fiber's surface are observed to increase proportionally with the flow behavior index for both fiber aspect ratio cases (cf. Figure 5.40c&d). Figure 5.41b shows the decline rate in the magnitude of the orbital peak pressure minimum with the power law index is non-linear and greater for the higher aspect ratio 2D particle compared to the lower aspect ratio particle.

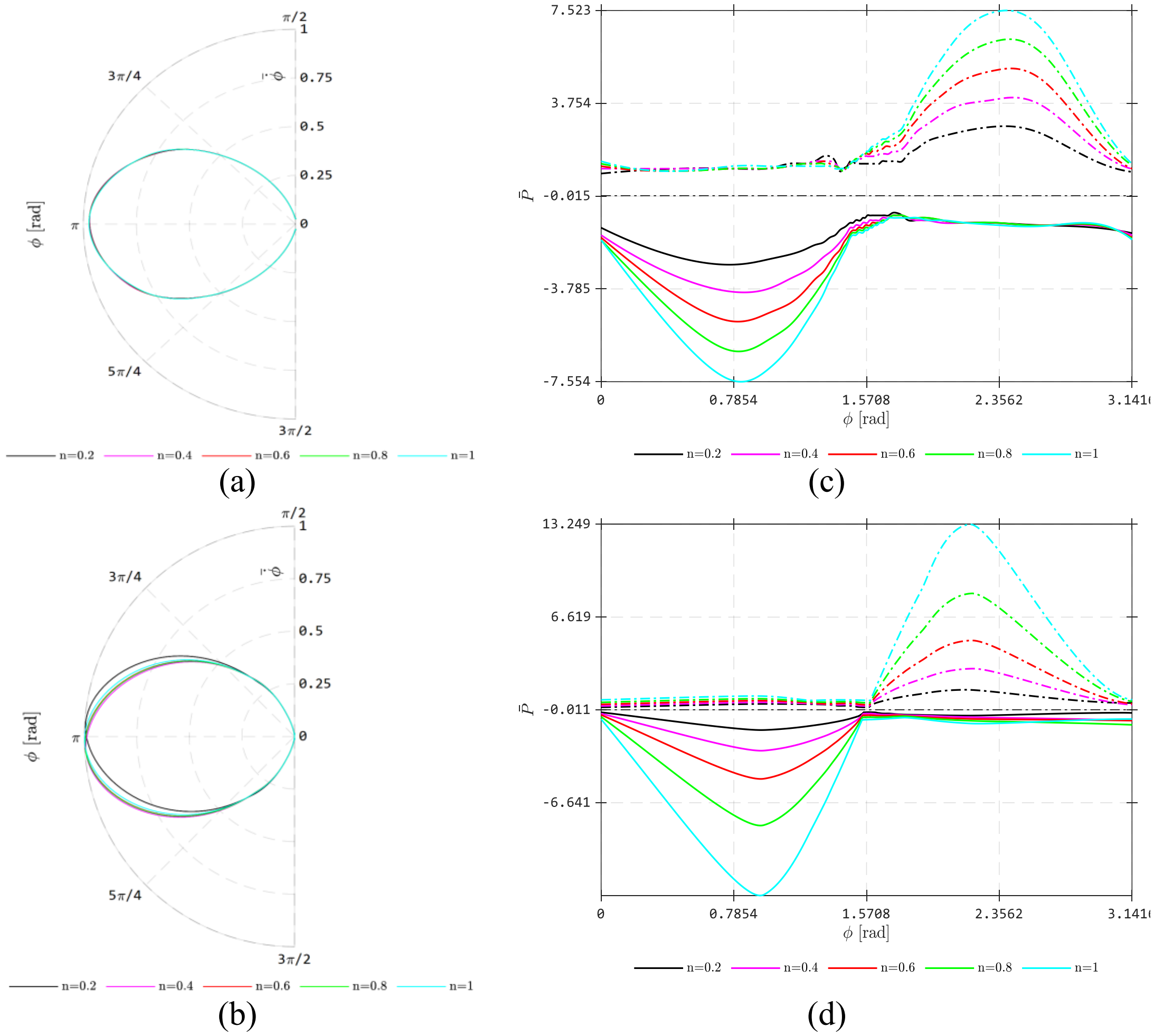

Figure 5.40: Figure showing results of evolution of fibers angular velocity normalized with shear rate (a) fiber with aspect ratio $r_e = 6$ (b) fiber with aspect ratio $r_e = 30$. Also shown are results of evolution of fibers limit surface pressure in time along Jeffery's orbit for (c) fiber with aspect ratio $r_e = 6$ (d) fiber with aspect ratio $r_e = 30$.

Likewise, the 3D sensitivity study on the influence of the particle geometric aspect ratio on its field state shows that the aspect ratio significantly influences the observed particle kinematic behaviour and the surface pressure distribution in Newtonian shear flow. The 3D studies allow us to study the combined effect of shear-thinning fluid rheology, initial out of plane orientation and aspect ratio on the particle's behaviour.

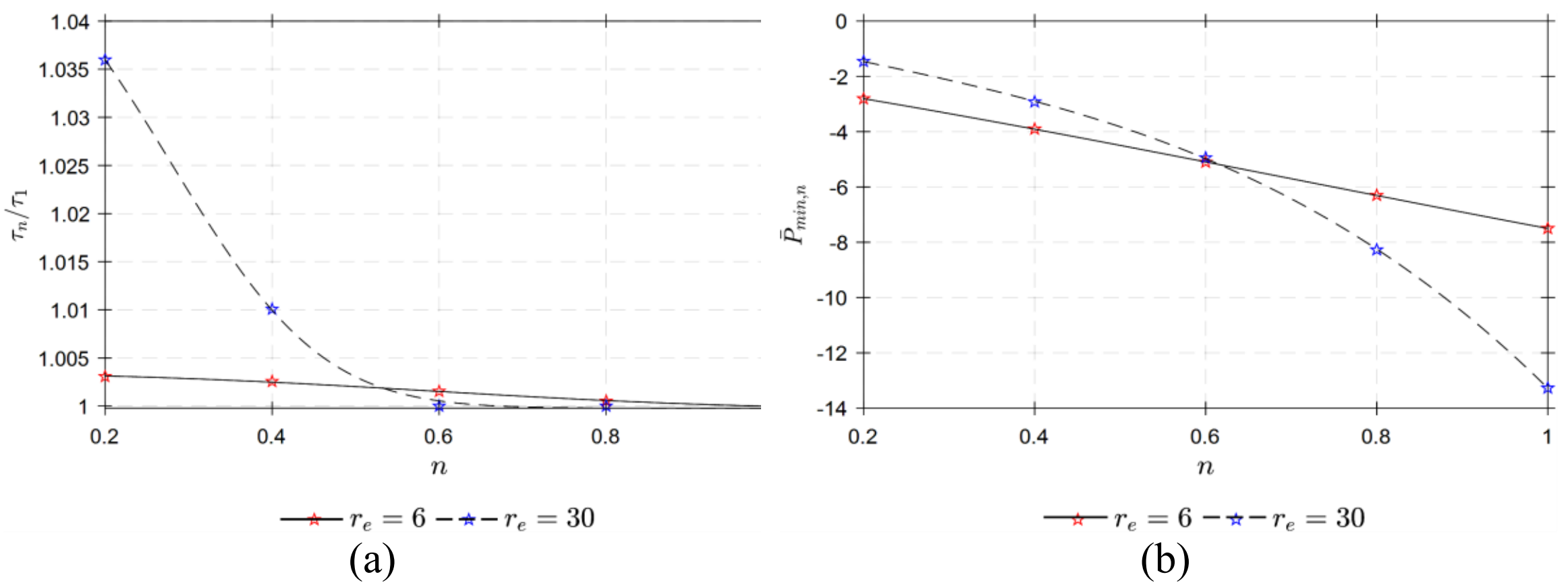

Figure 5.41: 2D results showing (a) the ratio of the non-Newtonian to Newtonian period, (b) the orbital peak pressure minimum, for different power law indices $n = 0.2 - 1.0$, and for different fiber aspect ratio $r_e = 6$, and $r_e = 30$.

Previous studies showed that the shear-thinning effect on the particle's orbit are magnified with increasing initial out of plane orientation $\theta^0$ [194]. As such we consider Jeffery's orbit with initial particle orientation of $\phi^0 = 0, \theta^0 = 2\pi/24$, and $\psi^0 = 0$. Figure 5.42 shows the deviation in particle trajectories, pressure and dynamic responses between the shear-thinning and Newtonian fluid for various particle aspect ratios. For spherical shaped particles, shear-thinning has no significant effect on the particles orbit, or the evolution of the particle's surface pressure and dynamic responses. However, as the particle aspect ratio increases up to $r_e = 6$, we observe considerable deviation in the particle trajectory (cf. Figure 5.42a) consistent with the findings of Abtahi et al. [194]. Similar to results that appear above, the particle trajectory is elongated or constricted depending on the initial starting position on a particular Newtonian Jeffery's orbit. With a further increase in the particle's slenderness, i.e. as $\kappa \rightarrow 1$, modification of the particle's trajectory due to shear-thinning becomes negligible as was also observed by Ferec et al. [231].

Likewise, the impact of shear-thinning on particle angular velocities is initially observed to increase with increasing aspect ratio (cf. Figure 5.42c&d). The non-linear effects, however, gradually decline with further increase in ellipsoid's slenderness. The shear-thinning behaviour is observed to slightly decrease the particles orbit period with slight increase in the aspect ratio. Further increases in the particle's slenderness, however, results in the shear-thinning behaviour prolonging the tumbling period. At lower aspect ratios, the pressure drag which does not depend on the local viscosity dominates the hydrodynamic resistance, however, with longer particles, the skin friction drag becomes significant due to the increased surface area and change in apparent viscosity [185]. Since a decrease in the apparent viscosity is known to slow down particle motion, we experience longer tumbling periods with considerable increase in the particle aspect ratio (cf. Figure 5.43a). The shear-thinning effect on the pressure response however continues to increase with the particle length (cf. Figure 5.42b & Figure 5.43b) which can be attributed to the hydrostatic stress intensification at the particle's tip arising from the increased particle length and/or the related decrease in the tip curvature.

Since typical EDAM printed fiber-filled polymer composites are known to have very high aspect ratios $r_e > 45$ [271], [272], the shear-thinning rheology is expected to have negligible effects on particle angular velocity and trajectory. However, we expect the non-Newtonian fluid to slow down the particles kinematics and reduce the surface pressure distribution.

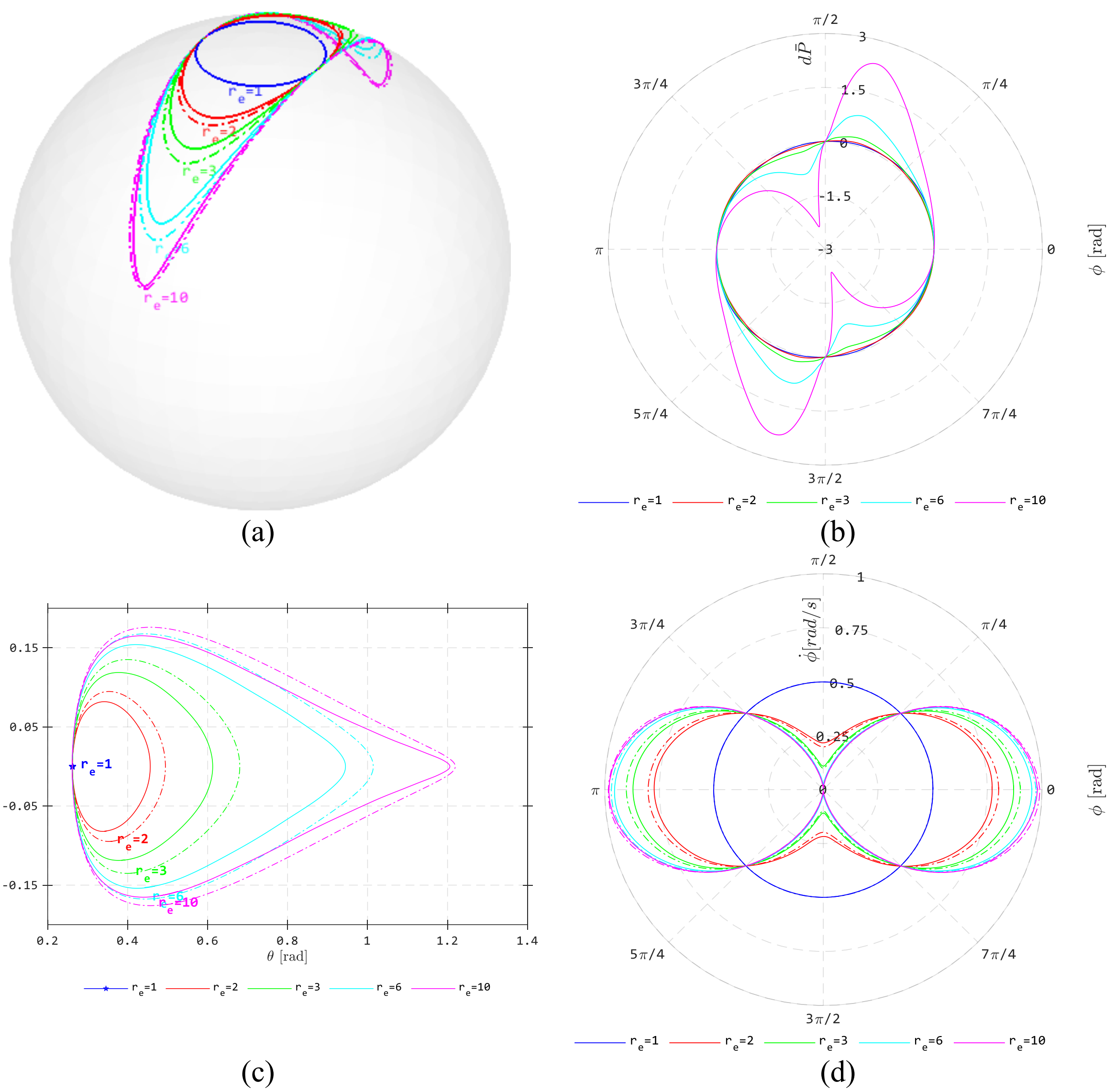

Figure 5.42: Showing (a) particle's orbits in simple shear flow (b) difference in particle tip pressure evolution between NT and GNF fluid (c) phase diagram of azimuth angle $\theta$ vs nutation $\dot{\theta}$ (d) polar plot of the polar angle $\phi$ vs precession $\dot{\phi}$, for different particle aspects $r_e$ and for NT fluid (dashed) and GNF power-law fluids (continuous) with $m = 1\ Pa \cdot s^n, n = 0.5$. Initial particle orientation is $\phi^0 = 0, \theta^0 = 2\pi/24,\ \psi^0 = 0$.

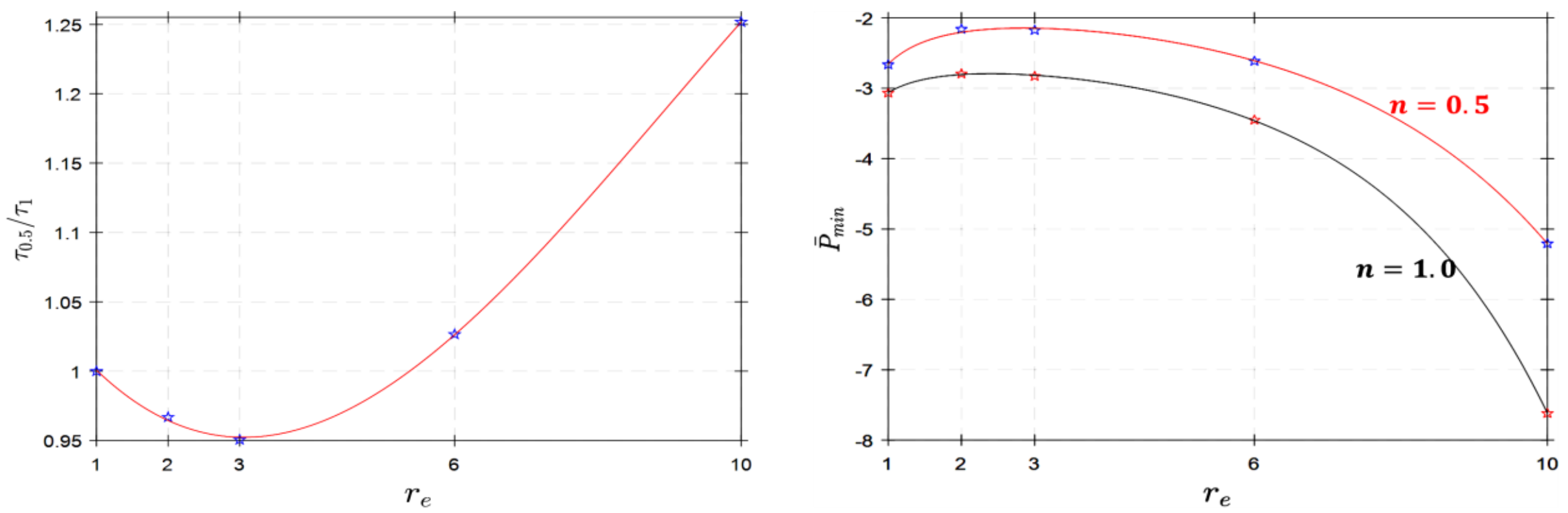

(a) (b)

Figure 5.43: Effect of shear-thinning comparing (a) the non-Newtonian to Newtonian tumbling period $\tau_{0.5}/\tau_1$, and (b) the non-Newtonian (red line) and Newtonian (black line) particles orbital minimum tip pressure $\bar{P}_{min}$, versus the fiber aspect ratio $r_e$, considering GNF power-law fluid, with with $m = 1\,Pa \cdot s^n, n = 0.5$ and initial particle orientation $\phi^0 = 0, \theta^0 = 2\pi/24, \psi^0 = 0$.

*5.10.2.4 Effective viscosity of shear-thinning suspension.* The flow behavior index of the shear-thinning fluid has an effect analogous to the influence of a Newtonian fluid viscosity on the pressure response on the fiber surface. Figure 5.15 shows the variation of the 2D fiber surface limit pressure response with the Newtonian viscosity $\mu_1$ (or consistency coefficient m for power law fluid with behavior index of n = 1). We earlier observed that the pressure magnitude on the fibers surface increases with increasing Newtonian viscosity like the influence of the flow behavior index on the pressure response (Figure 5.40b). This suggests the occurrence of regions of low and high viscosities extremes on the fibers surface during the fiber tumbling motion within the non-Newtonian fluid.

Figure 5.44 shows extracted data points (blue dots) of the instantaneous shear viscosity and shear rate scalar magnitude on the fibers surface over the complete period of fiber tumbling motion and for a power law index n = 0.2. The average viscosity $\eta_1$ and the viscosity corresponding to the average shear rate magnitude $\eta_2$ on the fibers surface over the entire period are also shown. From both values, $\eta_2$ is observed to be a better representation (definition) of the 'effective'mean viscosity on the fibers surface with an order of magnitude like the flow viscosity due to the imposed shear rate on the flow-field.

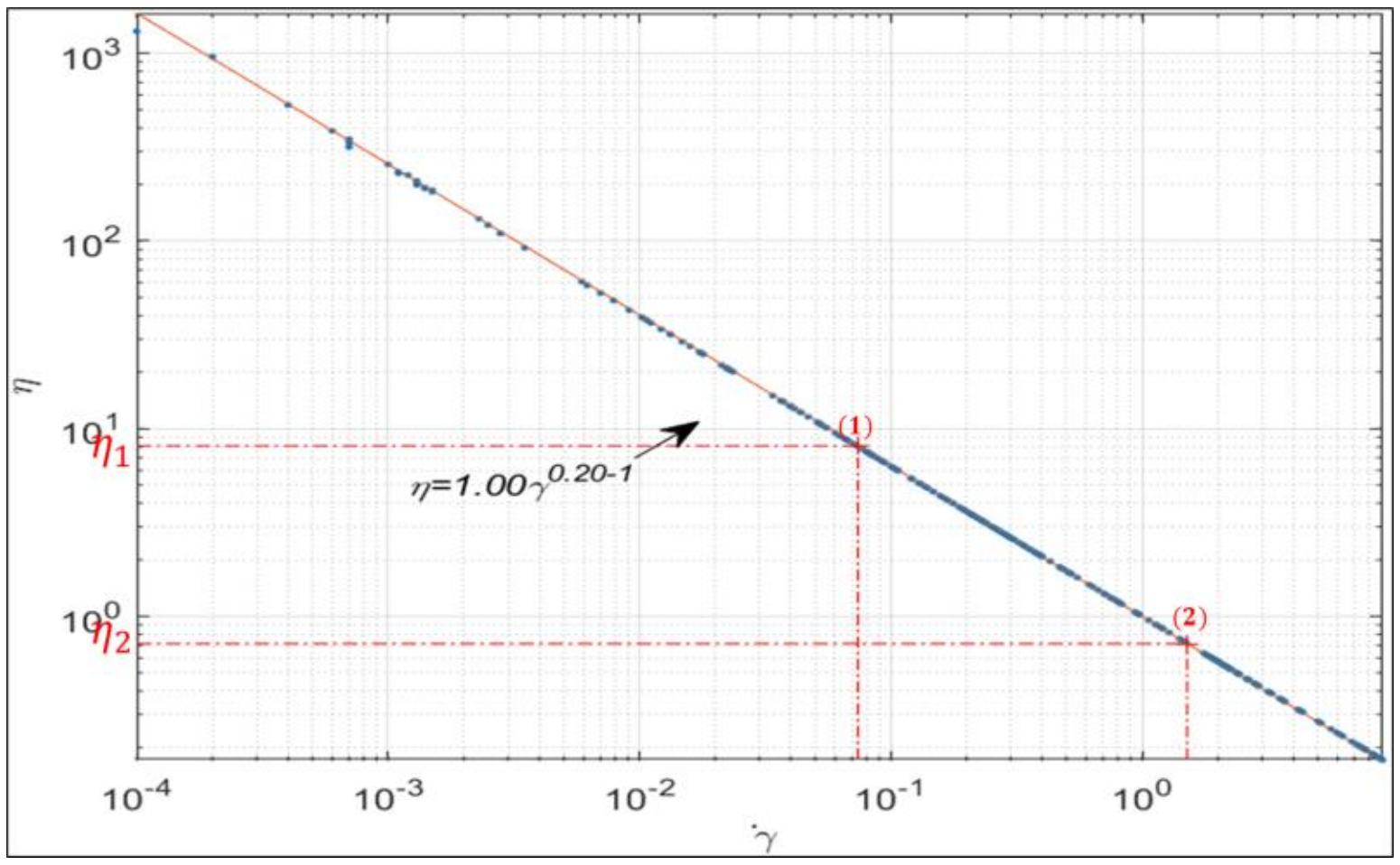

Figure 5.44: Scatter plot of the shear rates and resulting shear viscosities on the 2D fibers surface over the complete period of fiber tumbling motion. Indicated on the plot are the mean value points (1 & 2) of the shear rate and viscosities.

To gain a better understanding on the dynamics of the shear viscosities on the fiber surface during its tumbling motion and its influence on the fibers surface limit pressures, we present transient profiles of the evolution of the effective mean shear viscosity $\eta_2$ and the corresponding viscosity limits at each time interval for different flow behavior index (cf. Figure 5.45). From the profiles, we see that although the limits of shear rates magnitudes and resulting viscosities increase with decreasing flow behavior index, the effective mean viscosity $\eta_2$ on the fibers surface only slightly shifts below the actual farfield viscosity $\eta_0$. Following the effect of the Newtonian viscosity observed on the pressure limits on the fiber surface (cf. Figure 5.44a), we can infer in a qualitative sense that the decreasing trend in the effective mean viscosities with decreasing flow behavior index observed over the tumbling period in Figure 5.45a-d above are responsible for the low pressure limit magnitudes on the fiber surface.

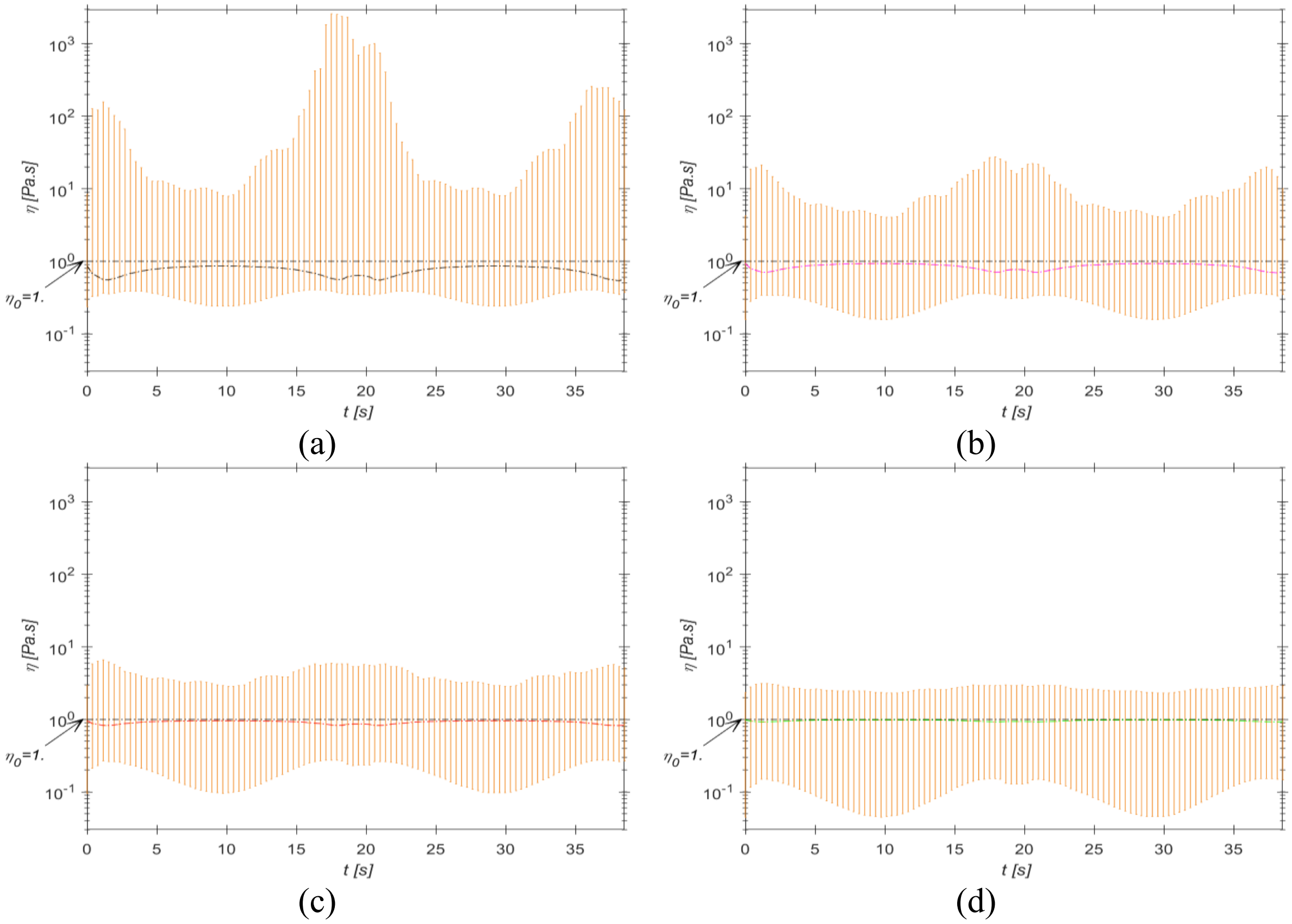

Figure 5.45[3]: Figure showing time evolution of the average and limits of the non-Newtonian viscosity on the fiber's surface for different flow behavior index (a) $n = 0.2$, (b) $n = 0.4$, (c) $n = 0.6$, (d) $n = 0.8$. For the Newtonian case, all upper & lower limits of the viscosities and effective mean viscosity transient profile all coincide with the far-field viscosity at $\eta_0$=1.

## *5.11 Conclusion*

In conclusion, a non-linear FEM numerical approach has been implemented to investigate the effects of shear-thinning fluid rheology in combination with other factors including the particles aspect ratio and initial particle orientation on the dynamics and surface pressure distribution on a particle suspended in viscous homogenous flow. The particles behavior in a special class of homogenous flows that typifies conditions found in melt flow regions of the of an extrusion nozzle during polymer composite additive manufacturing processing is also studied.

[3] The plots indicate viscosity limits on the fiber surface are exacerbated as the flow behavior index decreases due to the power law relationship. i.e., $\lim_{\substack{n \to 0 \\ \dot{\gamma} \to \infty}} \eta \to 0$ and $\lim_{\substack{n \to 0 \\ \dot{\gamma} \to 0}} \eta \to \infty$.

In the Newtonian flow, the ellipsoidal particle stalls in extension dominant asymmetric flow regimes but tumbles periodically in axisymmetric flows irrespective of the magnitude of the extension rate. The stall event in asymmetric flows is dictated by the shear-to-extension rate ratio. Increased shear dominance increases flow symmetry and tendency for continuous and periodic particle tumbling. The tumbling period in the asymmetric flows is expectedly dependent on the shear-to-extension rate ratio. The tumbling period increases asymptotically with increasing extension dominance until the conditions for stall based on Jeffery's equation are satisfied. On the other hand, the evolution time to particle stall is shown to increase asymptotically with increased shear dominance until the conditions for stall are violated. With sustained particle motion, the particle tip pressure fluctuates between extremums at the instant where its orientation aligns with the principal flow directions. An increase in the ellipsoidal particle aspect ratio was shown to affect the particles dynamics and increase the tumbling period. It also was shown to exacerbate the pressure extremes at the particle tip which could be caused by the increased aspect ratio alone, or the related reduction in tip curvature, or both. With a narrowing of Jeffery's orbit as the particle tumbles further out of the shear plane, the particle surface pressure extremes are observed to decrease and the surface location of the pressure extreme further deviates from the particle's tip location. The orbital peak particle tip pressure magnitude follows a somewhat linear relationship with the polar location on the orbit across spectrum of degenerate Jeffery's orbit.

Although in the 2D study, the particle's dynamics is unaffected by the shear-thinning fluid, the behavior of the suspended particle in the 3D study is shown to be affected by the shear-thinning fluid rheology. In the axisymmetric flows where the particle motion ensues periodically, the shear-thinning fluid rheology slows down the particles motion and increases the tumbling period. Cessation of particle motion (i.e., a stall condition) in the asymmetric homogenous flows is shown to be dictated by a competing influence of the shear-thinning intensity and shear to extension rate dominance. The shear-thinning was found to have an effect similar to decreasing the shear-rate dominance of the prevailing flow on the particles motion. Irrespective of the homogenous flow type, the magnitude of the particle surface pressure distribution was observed to significantly decrease with increased shear-thinning intensity due to an accompanying decrease in the effective viscosity of the fluid around the particle surface. The orbital location at which the pressure magnitude extremes on the particles surface are, however, unaffected by the shear-thinning rheology. On the shear-plane, shear-thinning rheology has no noticeable effect on the particles' orbit, however, with a narrowing of the Jeffery orbit as we move further out of plane, the particle's trajectory deviates further from the Newtonian reference path. The shear-thinning rheology may either constrict or dilate the Newtonian orbit depending on the initial starting location of the particle on the orbit. The elongation of the particle's motion and the lowering of the pressure on the surface of the particle by the shear-thinning effect is augmented with widening of Jeffery's orbit as the particle tumbles closer to the shear plane. For spherical particles, the shear-thinning fluid has no significant effect on the dynamics or surface pressure distribution, but with increased aspect ratio, modification of the particle's trajectory and dynamics due to the non-linear effects becomes significant until a critical point, where the non-linear effects are reversed. With excessive particle slenderness, the impact of the shear-thinning fluid on the particle's trajectory and dynamics diminishes. On the contrary, the effects of the shear-thinning on lowering of the particle surface pressure magnitude is proportionally elevated with increasing aspect ratio.

# CHAPTER SIX

## Determination of Steady State Fiber Orientation State based on an Exact Jacobian Newton – Raphson Numerical Scheme

Mechanical performance of printed parts depends on the inherent process-induced microstructural properties. Of the different bead microstructural descriptors for short fiber reinforced composite polymer, the fiber orientation is a key parameter that is useful for accurately predicting the material behavior. As such, appropriate macroscopic modelling of the average fiber orientation distribution is crucial in evaluating these properties. Various analytical models have been developed over time for estimating the average flow induced fiber orientation during polymer composite processing. Traditionally, Jeffery's equation [21] has been used to simulate fiber orientation evolution in dilute fiber suspension. However, Jeffery's model fails when considering semi-dilute or concentrated fiber suspension or confinement lubrication flows involving long and short-range hydrodynamic fiber interaction forces [182]. This has led to the development of more advanced advection-diffusion macroscopic fiber orientation evolution models that account for the neglected effects of momentum diffusion due to inter fiber interactions in concentrated suspensions such as the Folgar-Tucker model [261], [274] or the various variants of the Advani-Tuckers even-order moment tensor model [19], [22]. The preceding chapter (Chapter Five) presented extensive study on the behavior of single particle in dilute viscous homogenous suspension flow with GNF fluid rheology without considering the effects of rotary diffusion due to hydrodynamic fiber interaction forces.

As would appear in subsequent chapter, a novel numerical approach that relates the momentum diffusion phenomenological interaction coefficient and the effective fluid domain radius of influence utilized in the single fiber model is developed to capture the effect of fiber-fiber interaction. The first step involves relating the interaction coefficient with the steady state fiber orientation tensor using one of the available advection-diffusion fiber orientation tensor evolution models. In a separate study involving a series of single fiber motion FEA simulations, a relation between the fiber stall orientation angles and the effective fluid domain size that results in cessation of the fibers rotary motion due flow disturbance introduced by the hydrodynamic interaction between the fiber wall and the adjacent far-field flow is established. Subsequently, the relationship between the steady state fiber orientation and the fiber interaction coefficient is determined from the established correlations. It is thus apparent that a numerical method for determining the steady state fiber orientation for a range of diffusion interaction coefficients using any of the available fiber orientation evolution models which vary in degree of prediction accuracy is pertinent to the obtaining the relevant relationship which is the focus of the current chapter. The steady state values of orientation tensors in homogeneous flows have traditionally been computed with time evolution numerical IVP-ODE techniques such as the 4$^{th}$ order Runge-Kutta method or predictor-corrector methods (see e.g.,[19], [275]). However, when the objective is to obtain a steady-state orientation state, the transient solution process can require excessive computations for orientation tensor solution that are not needed. Here we present a computationally efficient approach based on Newton-Raphson iteration for determining the steady state values of orientation tensors. This approach zeros the time derivative of the 2$^{nd}$ order orientation tensor using explicit derivatives of the 2$^{nd}$ order orientation tensor rate of change equation with respect to its 2$^{nd}$ order orientation tensor components. Here we consider various fiber orientation models and closure approximations and compare their performance in homogeneous flow fields. We benchmark the results of the explicit derivatives with those obtained using finite differences to ensure accuracy and our Newton-Raphson iteration results are compared with time evolution numerical solutions in terms of both accuracy and efficiency.

Table 6.1. List of abbreviations and acronyms used in this chapter

| Acronym | Definition | Acronym | Definition |
|---|---|---|---|
| *Acronyms related to Fiber Orientation Tensor Models* | | | |
| *JF* | Jeffery | *SOK* | Slow Orientation Kinematics |
| *FT* | Folgar- Tucker | *IRD* | Isotropic Rotary Diffusion |
| *AT* | Advani - Tucker | *ARD* | Anisotropic Rotary Diffusion |
| *HD* | Hydrodynamic AT component | *iARD* | Improved ARD |
| *PT* | Phelps-Tucker | *pARD* | Principal spatial tensor ARD |
| *SRF* | Strain Reduction Factor | *mARD* | Moldflow ARD |
| *RSC* | Reduced Strain Closure | *WPT* | Wang-Phelps-Tucker |
| *RPR* | Retarding Principal Rate | *NEM-MS* | Nematic 'Onsager' potential of non-Brownian Maier-Saupe form |
| *IOK* | Intrinsic Orientation Kinematics | *Dz* | Thickness moment of interaction coefficient |
| *Acronyms related to 4$^{th}$ Order Tensor Closure Approximations* | | | |
| *DFC* | Distribution function calculation | *ORF* | Orthotropic fitted closure |
| *HOF* | Herman's orientation factor ($f$) | *ORL* | ORF low fiber-fiber interaction coefficient |
| *HYB* | Hybrid closure | *ORW* | ORF for wide range of fiber-fiber interaction coefficient values with 2$^{nd}$ order orientation tensor eigenvalue polynomial expansions |
| *HYB1* | HYB based on HOF calculated from 2$^{nd}$ order orientation tensor | *ORW3* | ORW with 3$^{rd}$ order orientation tensor eigenvalue polynomial expansions. |
| *HYB2* | HYB based on HOF calculated from invariants of the 2$^{nd}$ order orientation Tensor | *NAT* | Natural closure |
| *HL* | Hinch & Leal closures | *NAT1* | Extended quadratic fit NAT closure |
| *ISO* | Isotropic closure | *NAT2* | Exact midpoint fit NAT closure |
| *LIN* | Linear closure | *IBOF* | Invariant Based Optimal Fitting |
| *QDR* | Quadratic closure | *EBOF* | Eigenvalue Based Optimal Fitting |
| *HL1* | HL first composite closure | *RE* | Rational Elliptical closure |
| *HL2* | HL second composite closure | *WTZ* | Wetzel RE closure |
| *SF2* | HL closure for strong flows | *LAR* | Least Absolute Residuals regression |
| *IBF* | Invariant based fitted closure | *LAR4* | 4th order optimal fitting LAR closure |
| *EBF* | Eigenvalue based fitted closure | *LAR32* | Rational LAR closure with 3rd order numerator and 2nd order denominator optimal fitting |
| *ORT* | Orthotropic linear closure | *FFLAR4* | Flow-Field LAR4 closure |
| *ORS* | Orthotropic smooth closure | *VST* | Verweyst closure |

## *6.1 Representation of Fiber Orientation State*

Figure *6.1* shows an orientation state of a prolate ellipsoid including the local coordinate systems $x_1'$, $x_2'$, $x_3'$ and Euler angles definition where the orientation vector $\underline{\rho}$ lies along the major axis of the ellipsoid.

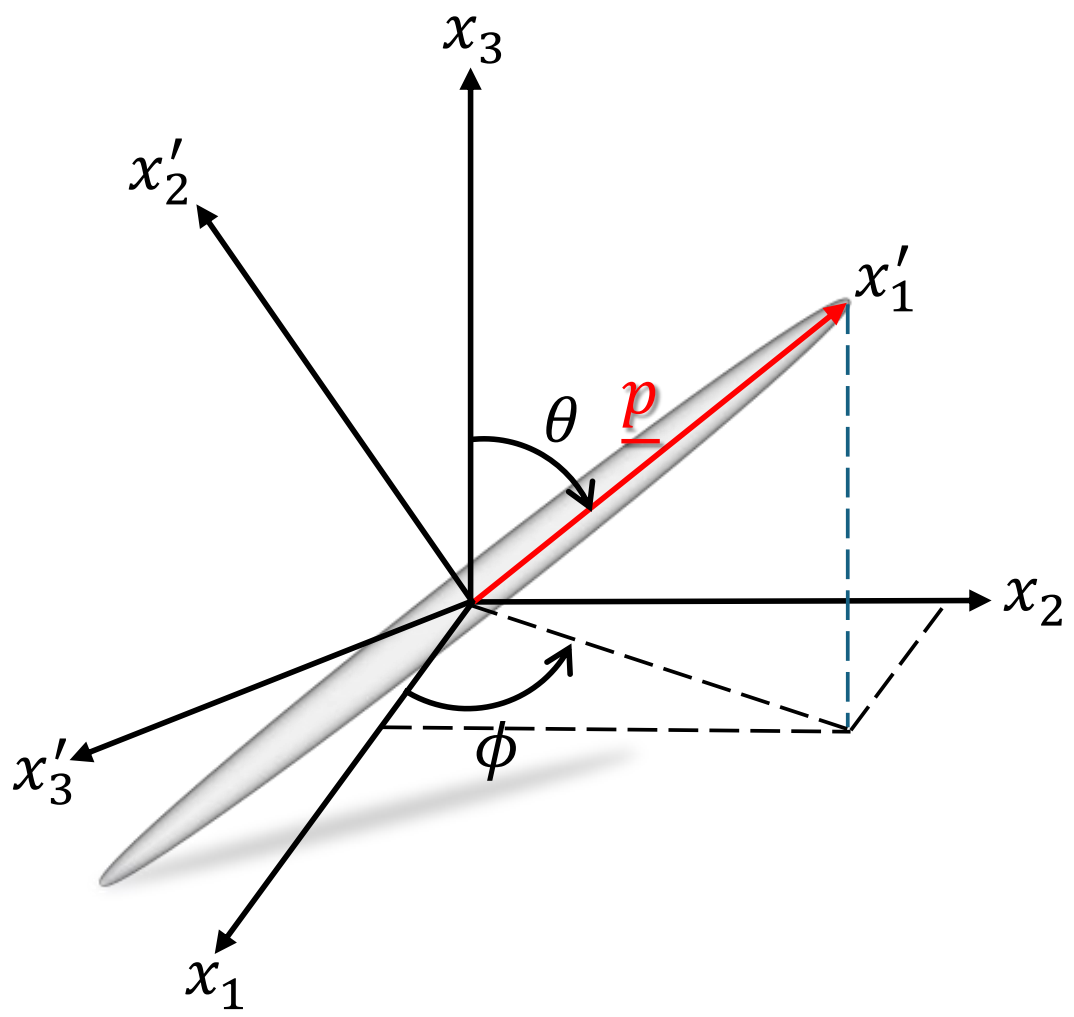

Figure 6.1. Single 'rigid' ellipsoidal fiber orientation

The orientation state of a fiber as shown in Figure *6.1* can be described by the unit vector $\underline{\rho}$ associated as [19]

$$\underline{\rho} = [\sin\theta\cos\phi \quad \sin\theta\sin\phi \quad \cos\theta]^T \tag{6.1}$$

The orientation of the unit vector $\underline{\rho}$ is commonly described in terms of the probability distribution function (PDF) $\psi\left(\underline{\rho}\right)$ over all possible directions of $\underline{\rho}$. It can be shown that $\psi\left(\underline{\rho}\right) = \psi\left(-\underline{\rho}\right)$ satisfies the normalization condition

$$\oint \psi\left(\underline{\rho}\right) d\underline{\rho} = \int_{\theta=0}^{\pi} \int_{\phi=0}^{2\pi} \psi(\theta,\phi)\sin\theta\, d\theta d\phi = 1 \tag{6.2}$$

The PDF $\psi\left(\underline{\rho}\right)$ also satisfies the continuity condition [19]

$$\frac{D\psi}{Dt} = -\frac{\partial}{\partial\underline{\rho}}\left(\psi\dot{\underline{\rho}}\right) \tag{6.3}$$

A set of higher-order, even order tensors, can also be used to describe the orientation state of a single fiber or fiber suspension such as the 2nd and 4th order orientation tensors, given respectively as

$$\mathrm{a}_{ij} = \oint \rho_i\rho_j\psi\left(\underline{\rho}\right)d\underline{\rho} \qquad and \quad \mathrm{a}_{ijkl} = \oint \rho_i\rho_j\rho_k\rho_l\psi\left(\underline{\rho}\right)d\underline{\rho} \tag{6.4}$$

Orientation tensors in Equation (6.4) are completely symmetric i.e.

$$\begin{gathered} \mathrm{a}_{ij} = \mathrm{a}_{ji} \\ \mathrm{a}_{ijkl} = \mathrm{a}_{jikl} = \mathrm{a}_{kijl} = \mathrm{a}_{lijk} = \mathrm{a}_{ikjl} = \mathrm{a}_{iljk} = \cdots, \qquad 24\,permutations \end{gathered} \tag{6.5}$$

and a normalization condition requires that

$$\mathrm{a}_{ii} = 1, \qquad \mathrm{a}_{ijkk} = \mathrm{a}_{ij} \tag{6.6}$$

Consequently, there are only 5 independent components in $\mathrm{a}_{ij}$ and 9 independent components in $\mathrm{a}_{ijkl}$.

### *6.2 Determination of Preferred Orientation*

The focus of this chapter is to develop a numerical approach for determining steady state values of the orientation vector $\rho_i$ and orientation tensor components $\mathrm{a}_{ij}$ that compute a zero rate of change of each corresponding evolution equation using the Newton-Raphson. The rate of change equation (often referred to as the residual, $\Sigma_i$) for both $\rho_i$ and $\mathrm{a}_{ij}$ are set to zero, respectively, as

$$\Sigma_i = \frac{D\rho_i}{Dt} = 0 \qquad and \qquad \Sigma_{ij} = \frac{D\mathrm{a}_{ij}}{Dt} = 0 \tag{6.7}$$

where $i, j \in 1,2,3$ and summation convention is implied unless otherwise stated. Newton-Raphson iteration computes the root (in our case the orientation state) as [276]

$$\rho_i^+ = \rho_i^- - [J_{ij}(\rho_i^-)]^{-1}\Sigma_j(\rho_i^-) \tag{6.8}$$

for the orientation vector and

$$\mathrm{a}_{ij}{}^+ = \mathrm{a}_{ij}{}^- - [J_{mnij}(\mathrm{a}_{ij}{}^-)]^{-1}\,\Sigma_{mn}\left(\mathrm{a}_{ij}{}^-\right) \tag{6.9}$$

for the 2nd order orientation tensor where the $(+)$ and $(-)$ superscripts denote the current and previous iteration, respectively. The implication of Equations (6.8) and (6.9) is the need to compute the Jacobians $J_{ij}$ and $J_{mnij}$ which are the derivatives of the orientation vector and tensor, respectively, with respect to its components that define them, i.e.,

$$J_{ij} = \frac{\partial \Sigma_i}{\partial \rho_j} \qquad and \quad J_{mnij} = \frac{\partial \Sigma_{mn}}{\partial \mathrm{a}_{ij}} \tag{6.10}$$

The above computations, including the 2nd and 4th order tensor inverse calculations, are conveniently performed using contracted tensor notation. Since there are only 5 independent components of $\mathrm{a}_{ij}$ (see discussion below), the residual $\Sigma_{mn}$ may be represented in contracted form as the vector $\Sigma_r$ (and similarly, the tensor $\mathrm{a}_{ij}$ may be represented as the vector $\mathrm{a}_r$), and the Jacobian $J_{mnij}$ as a matrix $J_{rs}$. The contracted notation is related to the tensor components through the index mapping

$$r(m,n) = n - \frac{1}{2}(m-1)(m-6), \qquad for\, n = m \ldots 3, \qquad for\, m = 1, 2 \tag{6.11}$$

It follows that Equation (6.9) in contracted notation becomes

$$\mathrm{a}_r{}^+ = \mathrm{a}_r{}^- - [J_{rs}(\mathrm{a}_r{}^-)]^{-1}\Sigma_r(\mathrm{a}_r{}^-) \tag{6.12}$$

In the following sections, we present existing models for rate of change equations of the orientation vector $\rho_i$ and tensor $\mathrm{a}_{ij}$ based those considered in the review by Kugler et al. [22]. Equating each rate of change equation to zero yields the Newton-Raphson residual in Equation (6.7) which is then differentiated to obtain expressions for the associated Jacobian.

## *6.3 Macroscopic Fiber Orientation Modelling*

Macroscopic fiber orientation modelling is usually required in polymers processing to predict the bulk response of chopped fibers in composites parts and ultimately determine part performance. The choice of macroscopic model depends on various processing parameters such as the concentration of fiber suspension, flow type and strength, fiber geometry and volume fraction, material rheology, etc. Our algorithm is based on fiber orientation modelling in the dilute and concentrated regime given the classification of fiber suspension concentration presented in detail in the literature review.

### *6.3.1 Fiber Orientation Modelling in the Dilute Regime*

Jeffery's hydrodynamic model for the motion of a single rigid ellipsoidal particle in an incompressible Newtonian viscous fluid flow field forms the basis for evaluating fiber orientation for dilute suspensions[4]. Jeffery assumed that an ellipsoidal particle is convected with the bulk motion of the undisturbed surrounding fluid where the rate of change of $\underline{\rho}$ written in component form is given as [21], [22], [278].

$$\dot{\rho}_i = \omega_{ij}\rho_j + \xi\left(\dot{\gamma}_{ij}\rho_j - \dot{\gamma}_{kl}\rho_k\rho_l\rho_i\right) \tag{6.13}$$

where $\omega_{ij}$ and $\dot{\gamma}_{ij}$ are the anti-symmetric and symmetric decomposition of the rate of deformation tensor $L_{ij}$ given respectively as

[4] Dilute suspension is a heterogenous suspension where the average interparticle spacing is relatively large such that there is no restriction on the fibers motion due to hydrodynamic forces or mechanical contact. More details on fiber suspension concentration regimes can be found in [62], [277]

$$\omega_{ij} = \frac{1}{2}\left(\frac{\partial v_i}{\partial x_j} - \frac{\partial v_j}{\partial x_i}\right) \quad and \quad \dot{\gamma}_{ij} = \frac{1}{2}\left(\frac{\partial v_i}{\partial x_j} + \frac{\partial v_j}{\partial x_i}\right) \tag{6.14}$$

such that $L_{ij} = \partial v_i/\partial x_j = \dot{\gamma}_{ij} + \omega_{ij}$, and $\xi$ is a particle shape parameter given as $\xi = (r_e^2 - 1)/(r_e^2 + 1)$ for a particle aspect geometric ratio $r_e$. We define components $\Sigma_i$ of the Newton-Raphson residual in Equation (6.8) for computing the steady-state orientation of the vector $\underline{\rho}$ in Equation (9) as

$$\Sigma_i^{JF} = \dot{\rho}_i \tag{6.15}$$

where the superscript $JF$ denotes the 'Jeffery' type residual. The Jacobian $J_{mn}^{JF} = {\partial \Sigma_m^{JF}}/{\partial \rho_n}$ is obtained by taking derivatives of the components $\Sigma_m^{JF}$ with respect to $\rho_n$, i.e.

$$J_{mn}^{JF} = \frac{\partial \dot{\rho}_m}{\partial \rho_n} = \omega_{mj}\delta_{jn} + \xi\left[\dot{\gamma}_{mj}\delta_{jn} - \dot{\gamma}_{kl}(\delta_{kn}\rho_l\rho_m + \rho_k\delta_{ln}\rho_m + \rho_k\rho_l\delta_{mn})\right] \tag{6.16}$$

where we note that derivatives of $\underline{\rho}$ with respect to itself form the identity tensor., i.e., $\partial\rho_i/\partial\rho_j = \delta_{ij}$. Equation (6.16) contains only 2 independent components of the orientation vector $\Sigma_i^{JF}$ making it possible to represent the components of $J_{mn}^{JF}$ as a $2 \times 2$ matrix. Although Jeffery's model for dilute suspensions has limited direct application since the effect of momentum diffusion due to fiber-fiber interaction is ignored, however, the NR method can be used to find the steady state solution for flows with high extension to shear rate ratio, such as those considered in [233]. Moreover, Jeffery's assumption ignores the flexural and fracture behavior of the fiber and assumes no slip contact between the fiber and surrounding fluid. Studies have shown that fibers suspended in industrial polymer melt flows tend to orient relative to the flow field [181] which is not captured in Jeffery's work. As a result, enhancements to the Jeffery's single fiber model have been made to better capture the bulk behavior of fibers in semi-dilute and concentrated suspension which include fiber-fiber interaction.

### *6.3.2 Fiber Orientation Modelling in Semi-dilute and Concentrated Regime*

In the concentrated regime, the average interparticle spacing is very small (orders of magnitude less than the smallest particle dimension) such that the fiber motion is affected by hydrodynamic forces and possible direct mechanical contact with other particles. Definitions for the semi-dilute and concentrated fiber suspension regimes can be found in [62], [277]. The effect of the interparticle interaction on the single particle's motion is typically modeled by including a momentum diffusion term in Eqn. 13. Here we consider various diffusion models which are commonly used in commercial SFRP simulations having fiber suspensions in the concentrated regime. In the sections following, the NR method for resolving the steady state fiber orientation is developed for multiple 2$^{nd}$ order fiber orientation tensor and rotary diffusion models and include several 4$^{th}$ order closure approximations.

*6.3.2.1 The Folgar-Tucker Diffusion Model.* Although theoretically feasible, it is computationally expensive and nearly impractical to simulate the behavior of every individual particle in an industrially relevant short fiber polymer suspension flow. A model that accounts for the effect of momentum diffusion due to short- and long-range fiber-fiber interaction in non-dilute fiber suspensions was first proposed by Folgar-Tucker model [261], [274] by incorporating a rotary diffusion term into Jeffery's single fiber model as.

$$\dot{\rho}_i^{FT} = \dot{\rho}_i^{JF} - D_r \frac{1}{\psi} \frac{\partial \psi}{\partial \rho_i} \tag{6.17}$$

where $D_r$ is the rotary diffusivity constant that introduces a Brownian diffusion effect among contacting particles. For long slender particles (i.e., $\xi \approx 1$) Folgar and Tucker set $D_r = C_I \dot{\gamma}$ where $C_I$ is a phenomenological interaction coefficient and $\dot{\gamma} = \sqrt{2\dot{\gamma}_{ij}\dot{\gamma}_{ji}}$ is the scalar magnitude of the strain rate tensor $\dot{\gamma}_{ij}$. The orientation PDF $\psi$ (ODF) in Equation (6.17) defines the probability of a given fiber has a particular orientation (see [19], [261], [274]). The time evolution of the ODF is defined by the Fokker-Plank equation for probability distribution function (PDF) of fiber orientation. Conventionally a numerical method such as finite volume (cf. Bay [279]) and more recently a more computationally efficient exact spherical harmonics method (cf. Montgomery-Smith et al. [280]) has been used to solve the Folgar-Tuckers (FT) equation of change for fiber orientation, but these methods have yet to see significant application in molding or extrusion processes. Advani and Tucker [19] proposed a tensorial representation of fiber orientation evolution based on the series expansion of even order moments of the ODF. Their approach simplifies computations which has led to the widespread use of orientation tensors as the preferred method of evaluating fiber orientation for short fiber polymer composites. Advani and Tucker developed an equation for the rate of change of $\mathrm{a}_{ij}$, identified here and below as the orientation material derivative tensor $\dot{\mathrm{a}}_{ij}$, in terms of the 2nd and 4th order tensor as

$$\frac{D\mathrm{a}_{ij}^{AT}}{Dt} = \dot{\mathrm{a}}_{ij}^{AT} = \dot{\mathrm{a}}_{ij}^{HD} + \dot{\mathrm{a}}_{ij}^{IRD} \tag{6.18}$$

where $\dot{\mathrm{a}}_{ij}^{HD}$ is the hydrodynamic component derived from Jeffery's equation (cf. Eqn. 11) given as

$$\dot{a}_{ij}^{HD} = -\left(\omega_{ik}a_{kj} - a_{ik}\omega_{kj}\right) + \xi\left(\dot{\gamma}_{ik}a_{kj} - a_{ik}\dot{\gamma}_{kj} - 2\dot{\gamma}_{kl}a_{ijkl}\right) \tag{6.19}$$

and $\dot{a}_{mn}^{IRD}$ is adds isotropic rotary diffusion as

$$\dot{a}_{ij}^{IRD} = 2D_r\left(\delta_{ij} - \alpha a_{ij}\right) \tag{6.20}$$

In the above, $\alpha$ is a dimension factor (i.e., $\alpha = 3$ for 3D orientation and $\alpha = 2$ for 2D planar orientation). It follows that the residual in Equation (6.8) for the Advani-Tucker model is

$$\Sigma_{mn}^{AT} = \dot{a}_{ij}^{AT} \tag{6.21}$$

The associated Jacobian $J_{mnij}^{FT}$ for the Advani-Tucker model is obtained by differentiating $\Sigma_{mn}^{AT}$with-respect-to the components of $a_{ij}$ yielding

$$J_{mnij}^{AT} = \frac{\partial \Sigma_{mn}^{AT}}{\partial a_{ij}} = \frac{\partial \dot{a}_{mn}^{HD}}{\partial a_{ij}} + \frac{\partial \dot{a}_{mn}^{IRD}}{\partial a_{ij}} \tag{6.22}$$

Given that the derivative of $a_{ij}$with respect itself is

$$\frac{\partial a_{mn}}{\partial a_{ij}} = \delta_{mi}\delta_{nj} \tag{6.23}$$

the derivative terms on the right-hand side of Equation (6.22) may be written as

$$\frac{\partial \dot{a}_{mn}^{HD}}{\partial a_{ij}} = (-\omega_{mk} + \xi\dot{\gamma}_{mk})\delta_{ki}\delta_{nj} + (\omega_{kn} + \xi\dot{\gamma}_{kn})\delta_{mi}\delta_{kj} - 2\xi\dot{\gamma}_{kl}\frac{\partial a_{mnkl}}{\partial a_{ij}} \tag{6.24}$$

$$\frac{\partial \dot{a}_{mn}^{IRD}}{\partial a_{ij}} = -2D_r\alpha\delta_{mi}\delta_{nj} \tag{6.25}$$

Expressions for derivatives of $a_{mnkl}$ (i.e., ${\partial a_{mnkl}}/{\partial a_{ij}}$) in Equation (6.24) and others using well known closure approximations are presented in sections to follow. Since there are only 5 independent components of the 2nd order tensor $a_{ij}$, in contracted notation, we can represent the residual $\Sigma_{mn}^{FT}$ as a vector $\Sigma_{r}^{FT}$ and the Jacobian $J_{mnij}^{FT}$ as a matrix $J_{rs}^{FT}$. Any reordering convention could be used. Here we employ.

$$r(m,n) = n - \frac{1}{2}(m-1)(m-6), \qquad for\ n = m \dots 3, \qquad for\ m = 1, 2 \tag{6.26}$$

The Advani-Tucker's $n^{th}$ order orientation evolution model is less accurate depending on the order of the tensor and thus requires a closure approximation. Due to experimentally observed disparity in the fiber orientation kinematics compared to those computed from the Advani and Tucker orientation model [19], more recent modifications have been proposed in an effort to reduce the rate of alignment in polymer melt flow.

*6.3.2.2 Strain Reduction Factor (SRF) Model.* The SRF model was developed by Huynh [281] to reduce the rate of alignment as compared to the AT model by introducing a strain reduction factor ꝁ $(0 < ꝁ < 1)$ into Equation 16 (i.e., multiplying the right-hand-side of Equation 16 by ꝁ) to slow down the orientation kinetics as observed experimentally. He based his premise on a reduced bulk strain of fiber clusters in a concentrated suspension flow. Although the predictions of the steady state orientation based on this model for simple shear flow with suitable determination of ꝁ matched experimental results [282], however it gave an initial overshoot at small strain. The residual and Jacobian in this case is just a multiplication of ꝁ with that previously obtained for the FT model.

$$\Sigma_{mn}^{SRF} = ꝁ\Sigma_{mn}^{AT}, \qquad J_{mnij}^{FT} = ꝁJ_{mnij}^{AT}, \qquad 0 < ꝁ < 1 \tag{6.27}$$

Steady state orientations predicted by the SRF model have been experimentally validated [9], however, it tends to overshoot measured alignments in the initial transient. The SRF model also does not satisfy the rheological test of material objectivity and, therefore, cannot be applied to complex flows.

*6.3.2.3 Reduced Strain Closure (RSC) Model.* Wang et al. [282] addressed material objectivity with their reduced strain closure (RSC) model which applies a reduction factor only to the evolution rate of the spectral decomposed principal directions $\underline{\underline{\Phi}}$ of the orientation tensor $\dot{\underline{\Lambda}}$, without modifying the rate of rotation $\underline{\underline{\dot{\Phi}}}$ given as

$$\dot{\Lambda}_i^{RSC} = \kappa\dot{\Lambda}_i^{AT}, \qquad \dot{\Phi}_{ij}^{RSC} = \dot{\Phi}_{ij}^{AT}, \qquad \mathrm{a}_{mn}|\mathrm{a}_{mn} = \Lambda_i \Phi_{mi}\Phi_{ni} \tag{6.28}$$

where | denotes the mathematical abbreviation for 'such that', and repeated indices imply summation. It follows that the material derivative for the RSC model may be written as [282]

$$\dot{\mathrm{a}}_{mn}^{RSC} = \dot{\mathrm{a}}_{mn}^{AT} - (1-\kappa)\dot{\mathrm{a}}_{mn}^{\Delta AT} \tag{6.29}$$

where the kernel of $\dot{\mathrm{a}}_{mn}^{\Delta AT}$ defines the part of the standard AT model affected by the modification factor, $\kappa$ to yield the RSC model is given as

$$\dot{\mathrm{a}}_{mn}^{\Delta AT} = \dot{\Lambda}_i^{AT}\Phi_{mi}\Phi_{ni} = 2\xi\dot{\gamma}_{kl}(\mathbb{L}_{mnkl} - \mathbb{M}_{mnrs}\mathrm{a}_{rskl}) + \dot{\mathrm{a}}_{mn}^{IRD} \tag{6.30}$$

where

$$\mathbb{L}_{mnkl} = \dot{\Lambda}_i\Phi_{mi}\Phi_{ni}\Phi_{ki}\Phi_{li}, \qquad \mathbb{M}_{mnkl} = \Phi_{mi}\Phi_{ni}\Phi_{ki}\Phi_{li} \tag{6.31}$$

The Newton-Raphson residual for the RSC model is $\Sigma_{mn}^{RSC} = \dot{\mathrm{a}}_{mn}^{RSC}$ and the Jacobian is obtained by taking partial derivatives of Equation 31 as

$$J_{mnij}^{RSC} = J_{mnij}^{FT} - (1-\kappa)\frac{\partial\dot{\mathrm{a}}_{mn}^{\Delta AT}}{\partial\mathrm{a}_{ij}} \tag{6.32}$$

where

$$\frac{\partial\dot{\mathrm{a}}_{mn}^{\Delta AT}}{\partial\mathrm{a}_{ij}} = 2\xi\dot{\gamma}_{kl}\frac{\partial}{\partial\mathrm{a}_{ij}}\{\mathbb{L}_{mnkl} - \mathbb{M}_{mnrs}\mathrm{a}_{rskl}\} + \frac{\partial\dot{\mathrm{a}}_{mn}^{IRD}}{\partial\mathrm{a}_{ij}} \tag{6.33}$$

which can be re-written as

$$\frac{\partial \dot{\mathrm{a}}_{mn}^{\Delta AT}}{\partial \mathrm{a}_{ij}} = 2\xi\dot{\gamma}_{kl}\left[\frac{\partial \mathbb{L}_{mnkl}}{\partial \mathrm{a}_{ij}} - \mathrm{a}_{rskl}\frac{\partial \mathbb{M}_{mnrs}}{\partial \mathrm{a}_{ij}} - \mathbb{M}_{mnrs}\frac{\partial \mathrm{a}_{rskl}}{\partial \mathrm{a}_{ij}}\right] + \frac{\partial \dot{\mathrm{a}}_{mn}^{IRD}}{\partial \mathrm{a}_{ij}} \tag{6.34}$$

In the above, the derivatives of 4th order tensors $\mathbb{M}_{mnkl}$ and $\mathbb{L}_{mnkl}$ are obtained by differentiating equations 33, respectively, as

$$\begin{aligned}\frac{\partial \mathbb{M}_{mnkl}}{\partial \mathrm{a}_{rs}} &= \frac{\partial}{\partial \mathrm{a}_{rs}}\{\Phi_{mi}\Phi_{ni}\Phi_{ki}\Phi_{li}\} \\ &= \Phi_{ni}\Phi_{ki}\Phi_{li}\frac{\partial \Phi_{mi}}{\partial \mathrm{a}_{rs}} + \Phi_{mi}\Phi_{ki}\Phi_{li}\frac{\partial \Phi_{ni}}{\partial \mathrm{a}_{rs}} + \Phi_{mi}\Phi_{ni}\Phi_{li}\frac{\partial \Phi_{ki}}{\partial \mathrm{a}_{rs}} \\ &\quad + \Phi_{mi}\Phi_{ni}\Phi_{ki}\frac{\partial \Phi_{li}}{\partial \mathrm{a}_{rs}}\end{aligned} \tag{6.35}$$

and

$$\frac{\partial \mathbb{L}_{mnkl}}{\partial \mathrm{a}_{rs}} = \Phi_{mi}\Phi_{ni}\Phi_{ki}\Phi_{li}\frac{\partial \dot{\Lambda}_i}{\partial \mathrm{a}_{rs}} + \dot{\Lambda}_i\frac{\partial}{\partial \mathrm{a}_{rs}}\{\Phi_{mi}\Phi_{ni}\Phi_{ki}\Phi_{li}\} \tag{6.36}$$

where the derivatives of the eigenvalues $\Lambda_i$ and eigenvectors $\Phi_{jk}$ appearing in (cf. APPENDIX C, C.1) [283] have been used to simplify the final expression.

*6.3.2.4 Retarding Principal Rate (RPR) Model.* Tseng et al. [284], [285] developed a Retarding Principal Rate (RPR) model, which, like the RSC model, reduce fiber orientation kinetics. Their approach is based on a coaxial modification to the principal directions of the orientation tensor evolution rate via a nonlinear correlation. The material derivative tensor $\dot{\mathrm{a}}_{\mathrm{mn}}^{\mathrm{X}}$ of any standard model X (where X can define the AT, RSC, or other orientation tensor evolution model), is linearly combined with the RPR correction to slow down the orientation response rate. i.e.

$$\dot{\mathrm{a}}_{mn}^{X-RPR} = \dot{\mathrm{a}}_{mn}^{X} + \dot{\mathrm{a}}_{mn}^{RPR} \tag{6.37}$$

where the RPR correction $\dot{\mathrm{a}}_{mn}^{RPR}$ is written in terms of its eigenvalue derivative matrix $\dot{\Lambda}_{kl}^{IOK}$ as

$$\dot{a}_{mn}^{RPR} = -\Phi_{mk}\dot{\Lambda}_{kl}^{IOK}\Phi_{nl}, \qquad \dot{\Lambda}_{kl}^{IOK} = \dot{\Lambda}_{kl}^{IOK}\left(\underline{\dot{\Lambda}}^{X}\right) \tag{6.38}$$

where $\dot{\Lambda}_{kl}^{IOK}$ is the correction to the principal values of the orientation tensor material derivative based on the IOK assumption (see e.g., Eqn. (6.42) below). The superscript on $\dot{\Lambda}_{kl}^{IOK}$ indicates the intrinsic orientation kinetics (IOK) assumption. Given that the correction is coaxial, the rotation tensor growth rate is unaffected and is obtained from the spectral decomposition of $a_{mn}$, where $\underline{\underline{\Phi}}$ satisfies

$$\Lambda_{mn} = \Phi_{km}a_{kl}\Phi_{ln} \tag{6.39}$$

The columns of the eigenmatrix obtained in this way are reordered in descending order with respect to the magnitude of the eigenvalues i.e.,

$$\Phi_{ij} = \left\{ \Phi_{ij} \mid \Lambda_j = \Lambda_{jj}, \qquad \Lambda_1 \geq \Lambda_2 \geq \Lambda_3 \right\} \tag{6.40}$$

where | denotes the mathematical abbreviation for 'such that' and the repeated indices in $\Lambda_{jj}$ here and in the following, do not imply summation but rather represents the $jth$ diagonal component of $\Lambda_{ij}$. The growth rate of the principal eigenvalues of the standard model – X, i.e. $\dot{\Lambda}_{kl}^{X}$ is obtained from

$$\dot{\Lambda}_{kl}^{X} = \Phi_{km}\dot{a}_{kl}^{X}\Phi_{ln}, \qquad \dot{\Lambda}_{k}^{X} = \dot{\Lambda}_{kk}^{X} \tag{6.41}$$

The diagonal tensor $\dot{\Lambda}_{kk}^{IOK}$ is defined by a 2-parameter non-linear correlation to the principal values of the standard model $\dot{\Lambda}_{kl}^{X}$ such that

$$\dot{\Lambda}_{kk}^{IOK} = \dot{\Lambda}_{k}^{IOK} = \alpha\left[\dot{\Lambda}_{k}^{X} - \beta\left(\left\{\dot{\Lambda}_{k}^{X}\right\}^{2} + 2\dot{\Lambda}_{l}^{X}\dot{\Lambda}_{m}^{X}\right)\right], \qquad \dot{\Lambda}_{kl}^{IOK}\Big|_{k\neq l} = 0 \tag{6.42}$$

For an RPR corrected model, the NR residual $\Sigma_{mn}^{X-RPR}$ is simply the material derivative

$$\Sigma_{mn}^{X-RPR} = \dot{a}_{mn}^{X-RPR} \tag{6.43}$$

and the Jacobian $J_{mnij}^{X-RPR}$ is given as

$$J_{mnij}^{X-RPR} = \frac{\partial \dot{\mathrm{a}}_{mn}^{X}}{\partial \mathrm{a}_{ij}} + \frac{\partial \dot{\mathrm{a}}_{mn}^{RPR}}{\partial \mathrm{a}_{ij}} \tag{6.44}$$

The partial derivative of the RPR correction term $\dot{\mathrm{a}}_{mn}^{RPR}$ is

$$\frac{\partial \dot{\mathrm{a}}_{mn}^{RPR}}{\partial \mathrm{a}_{ij}} = -\left\{\frac{\partial \Phi_{mk}}{\partial \mathrm{a}_{ij}} \dot{\Lambda}_{kl}^{IOK} \Phi_{nl} + \Phi_{mk} \frac{\partial \dot{\Lambda}_{kl}^{IOK}}{\partial \mathrm{a}_{ij}} \Phi_{nl} + \Phi_{mk} \dot{\Lambda}_{kl}^{IOK} \frac{\partial \Phi_{nl}}{\partial \mathrm{a}_{ij}}\right\} \tag{6.45}$$

and the partial derivative of the modified growth rate of eigenvalues tensor $\dot{\Lambda}_{kl}^{IOK}$ is obtained by taking the derivative of Equation (6.42) as

$$\frac{\partial \dot{\Lambda}_{kk}^{IOK}}{\partial \mathrm{a}_{ij}} = \frac{\partial \dot{\lambda}_{k}^{IOK}}{\partial \mathrm{a}_{ij}} = \alpha \left[\frac{\partial \dot{\lambda}_{k}^{X}}{\partial \mathrm{a}_{ij}} - 2\beta \left(\dot{\lambda}_{k}^{X} \frac{\partial \dot{\lambda}_{k}^{X}}{\partial \mathrm{a}_{ij}} + \frac{\partial \dot{\lambda}_{l}^{X}}{\partial \mathrm{a}_{ij}} \dot{\lambda}_{m}^{X} + \dot{\lambda}_{l}^{X} \frac{\partial \dot{\lambda}_{m}^{X}}{\partial \mathrm{a}_{ij}}\right)\right],$$

$$\left.\frac{\partial \dot{\Lambda}_{kl}^{IOK}}{\partial \mathrm{a}_{ij}}\right|_{k \neq l} = 0 \tag{6.46}$$

where

$$\frac{\partial \dot{\Lambda}_{kl}^{X}}{\partial \mathrm{a}_{ij}} = \left\{\frac{\partial \Phi_{km}}{\partial \mathrm{a}_{ij}} \dot{\mathrm{a}}_{kl}^{X} \Phi_{ln} + \Phi_{km} \frac{\partial \dot{\mathrm{a}}_{kl}^{X}}{\partial \mathrm{a}_{ij}} \Phi_{ln} + \Phi_{km} \dot{\mathrm{a}}_{kl}^{X} \frac{\partial \Phi_{ln}}{\partial \mathrm{a}_{ij}}\right\}, \qquad \frac{\partial \dot{\lambda}_{k}^{X}}{\partial \mathrm{a}_{ij}} = \frac{\partial \dot{\Lambda}_{kk}^{X}}{\partial \mathrm{a}_{ij}} \tag{6.47}$$

In the above, $\partial \dot{\mathrm{a}}_{mn}^{X} / \partial \mathrm{a}_{ij}$ is the partial derivative of material derivative $\dot{\mathrm{a}}_{mn}^{X}$ with respect to the components of $\mathrm{a}_{ij}$ obtained a priori, and the partial derivatives of the eigenvector matrix with respect to the same $\left(i.e., \partial \Phi_{mn} / \partial \mathrm{a}_{ij}\right)$ can be obtained as shown in Appendix A (cf. [283]). Prediction of fiber orientation with IRD-based models have been validated experimentally for short-fiber/thermoplastic composites (SFT) with fiber length typically in the range of 0.2mm to 0.4mm [286]. However, long-fiber/thermoplastic composites (LFT) (e.g., with fiber lengths above 10mm), IRD models tend to be less accurate with decreased alignment. Modifications to the rotary diffusion term have been proposed for improving the accuracy when predicting the components of the orientation tensor. Ranganathan et al. [287] assumed an isotropic rotary diffusivity that inversely varies with the degree of alignment of the orientation tensor which was implemented using a phenomenological interaction parameter that depends on the reciprocal of the inter fiber spacing. The applicability of their model is limited to the transient orientation state while being well suited for long range fiber-fiber interaction. Unfortunately, their model is known to be unsuitable for LFTs steady state orientation prediction as with other IRD models likely due to its isotropic diffusivity.

*6.3.2.5 Anisotropic Rotary Diffusion (ARD) Models.* Various ARD models and modified forms have been developed based on the definition of the spatial diffusion tensor. Most models utilize the moment-tensor form for the ARD representation developed by Phelps and Tucker [286]. Fan et al. [262] and Phan-Thien [288] were the first to propose an anisotropic rotary diffusion (ARD) moment-tensor model by replacing the FT scalar phenomenological interaction parameter with a second order anisotropic rotary diffusion tensor. In a similar manner, Koch [289] developed an ARD model suited for semi-dilute suspensions with an anisotropic spatial diffusion tensor that depends on the orientation state and the rate of deformation tensor. However, their model was based on the more complicated PDF form for the orientation tensor representation rather than the moment-tensor form and proved ineffective in LFT modelling. Phelps et al. [286] extended the work of Fan [262] and Phan-Thien et al. [288] by developing a more general moment-tensor anisotropic diffusion model that depends on the spatial diffusion and orientation tensors.

The derivation of the spatial diffusion tensor was written as a function of the orientation state and rate of deformation tensor in a similar manner to that proposed by Hand [290]. Phelps's model showed remarkable improvements in predicting orientation states of LFTs. The general expression for the 2$^{nd}$ order orientation tensor evolution rate is a linear combination of the Jeffery's-based model and the and the rotary diffusion term given as

$$\dot{a}_{mn}^{PT} = \dot{a}_{mn}^{HD} + \dot{a}_{mn}^{ARD} \tag{6.48}$$

where the rotary diffusion term $\dot{a}_{mn}^{ARD}$ is defined in terms of the spatial diffusion coefficient and the orientation state given as

$$\dot{a}_{mn}^{ARD} = \dot{\gamma}[2\mathbb{C}_{mn} - 2\mathbb{C}_{rs}\delta_{rs}a_{mn} - 5(\mathbb{C}_{mk}a_{kn} + a_{mk}\mathbb{C}_{kn}) + 10a_{mnkl}\mathbb{C}_{kl}] \tag{6.49}$$

and $\underline{\underline{\mathbb{C}}}$ is the spatial diffusion tensor. Based on this model, the NR residual and Jacobian are respectively given as

$$\Sigma_{mn}^{PT} = \dot{a}_{mn}^{PT} \tag{6.50}$$

$$J_{mnij}^{PT} = \frac{\partial \dot{a}_{mn}^{PT}}{\partial a_{ij}} = \frac{\partial \dot{a}_{mn}^{HD}}{\partial a_{ij}} + \frac{\partial \dot{a}_{mn}^{ARD}}{\partial a_{ij}} \tag{6.51}$$

where the derivative of the rotary diffusion ($ARD$) term is obtained using product rule as

$$\begin{aligned}\frac{\partial \dot{a}_{mn}^{ARD}}{\partial a_{ij}} = \dot{\gamma}\Bigg[2\frac{\partial \mathbb{C}_{mn}}{\partial a_{ij}} - 2\left(\frac{\partial \mathbb{C}_{rs}}{\partial a_{ij}}\delta_{rs}a_{mn} + \mathbb{C}_{rs}\delta_{rs}\delta_{mi}\delta_{nj}\right) + \cdots \\ - 5\left(\frac{\partial \mathbb{C}_{mk}}{\partial a_{ij}}a_{kn} + \mathbb{C}_{mk}\delta_{ki}\delta_{nj} + \delta_{mi}\delta_{kj}\mathbb{C}_{kn} + a_{mk}\frac{\partial \mathbb{C}_{kn}}{\partial a_{ij}}\right) + \cdots \\ + 10\left(\frac{\partial a_{mnkl}}{\partial a_{ij}}\mathbb{C}_{kl} + a_{mnkl}\frac{\partial \mathbb{C}_{kl}}{\partial a_{ij}}\right)\Bigg]\end{aligned} \tag{6.52}$$

In the Mold-Flow ARD (mARD) model developed by Bakharev [291], the Phelps and Tucker's rotary diffusion ARD expression is truncated to include just the linear terms. i.e.

$$\dot{a}_{mn}^{mARD} = \dot{\gamma}[2\mathbb{C}_{mn} - 2\mathbb{C}_{kl}\delta_{kl}a_{mn}] \tag{6.53}$$

$$\frac{\partial \dot{a}_{mn}^{mARD}}{\partial a_{ij}} = \dot{\gamma}\left[2\frac{\partial \mathbb{C}_{mn}}{\partial a_{ij}} - 2\left(\frac{\partial \mathbb{C}_{kl}}{\partial a_{ij}}\delta_{kl}a_{mn} + \mathbb{C}_{kl}\delta_{kl}\delta_{mi}\delta_{nj}\right)\right] \tag{6.54}$$

The corresponding evolution rate equation for the 2nd order orientation tensor based on mARD model is given as

$$\dot{a}_{mn}^{mPT} = \dot{a}_{mn}^{HD} + \dot{a}_{mn}^{mARD} \tag{6.55}$$

Various forms of the spatial diffusion coefficient $\mathbb{C}_{mn}$ used in the ARD model (indicated by a superscript in the following) have been developed. The Phelps and Tucker [286] representation $\mathbb{C}_{mn}^{PT}$ based on a modification to Hand's anisotropic tensor [290] is given as a function of the rate of deformation tensor $\dot{\gamma}_{ij}$ and orientation state $a_{ij}$ as

$$\mathbb{C}_{mn}^{PT} = b_1\delta_{mn} + b_2 a_{mn} + b_3 a_{mk} a_{nk} + \frac{b_4}{\dot{\gamma}}\dot{\gamma}_{mn} + \frac{b_5}{\dot{\gamma}^2}\dot{\gamma}_{mk}\dot{\gamma}_{nk} \tag{6.56}$$

where the $b_i$ are dimensionless constants are determined experimentally. The derivative of the $\mathbb{C}_{mn}^{PT}$ with respect to $a_{ij}$ that is substituted into Eqn. 56 is

$$\frac{\partial \mathbb{C}_{mn}^{PT}}{\partial a_{ij}} = b_2\delta_{mi}\delta_{nj} + b_3\left(\delta_{mi}\delta_{kj}a_{nk} + a_{mk}\delta_{ni}\delta_{kj}\right) \tag{6.57}$$

The sensitivity of the numerical stability of the model response on the PT model parameters $b_i$ coupled with the complicated process involved in their determination are major limitations to the application of this model. Tseng et al. [292] developed an improved anisotropic rotary diffusion model (iARD) with a two-parameter spatial diffusion tensor representation written in terms of $\dot{\gamma}_{ij}$ and its scalar magnitude $\dot{\gamma}$ that couples the effect of fiber-matrix interaction and fiber-fiber interaction given as

$$\mathbb{C}_{mn}^{iARD} = C_I\left(\delta_{mn} - 4C_M\frac{\dot{\gamma}_{mk}\dot{\gamma}_{nk}}{\dot{\gamma}^2}\right) \tag{6.58}$$

where $C_I$ and $C_M$ are the fiber-fiber and fiber-matrix interaction parameters respectively. An alternate form of Eqn. 60 is given as

$$\mathbb{C}_{mn}^{iARD} = C_I\left(\delta_{mn} - C_M\tilde{L}_{mn}\right), \qquad \tilde{L}_{mn} = (L_{mk}L_{nk})/(L_{rs}L_{rs}) \tag{6.59}$$

where $L_{mn}$ is the rate of deformation tensor. The derivative $\mathbb{C}_{mn}^{iARD}$with respect to $\mathrm{a}_{ij}$ is simply zero, i.e.,

$$\frac{\partial \mathbb{C}_{mn}^{iARD}}{\partial \mathrm{a}_{ij}} = 0 \tag{6.60}$$

Due to the lack of material objectivity of the rate of deformation tensor $L_{mn}$, in Eqn. 61, Tseng et al. [293] developed an improved objective principal spatial tensor ARD model (pARD) that is coaxial with the orientation tensor given as

$$\mathbb{C}_{mn}^{pARD} = \left\{ C_I \Phi_{mk} \mathbb{D}_{kl} \Phi_{nl}, \qquad \underline{\underline{\Phi}} \mid \mathrm{a}_{mn} = \Phi_{mk} \bar{\mathrm{a}}_{kl} \Phi_{nl} \right\} \tag{6.61}$$

where the tensor $\mathbb{D}_{kl}$ contains only diagonal terms and its trace is unity. i.e.

$$\mathbb{D}_{kk} = 1, \qquad \mathbb{D}_{kl}|_{k \neq l} = 0 \tag{6.62}$$

The derivative of $\mathbb{C}_{mn}^{pARD}$ with respect to the 2nd order orientation tensor for substitution into Eqn. 56 is

$$\frac{\partial \mathbb{C}_{mn}^{pARD}}{\partial \mathrm{a}_{ij}} = C_I \left\{ \frac{\partial \Phi_{mk}}{\partial \mathrm{a}_{ij}} \mathbb{D}_{kl} \Phi_{nl} + \Phi_{mk} \mathbb{D}_{kl} \frac{\partial \Phi_{nl}}{\partial \mathrm{a}_{ij}} \right\} \tag{6.63}$$

Another ARD model form suggested by Wang [294] called the WPT model involved truncating the terms of the PT model to obtain

$$\mathbb{C}_{mn}^{WPT} = b_1 \delta_{mn} + b_3 \mathrm{a}_{mk} \mathrm{a}_{nk} \tag{6.64}$$

Falvoro et al. [278] provided an alternative form of $\mathbb{C}_{mn}^{WPT}$where he replaced the coefficients in Eqn. 66 with a weighted superposition of the interaction coefficient $C_I$ written as

$$\mathbb{C}_{mn}^{WPT} = C_I\big((1 - w)\delta_{mn} + w \mathrm{a}_{mk} \mathrm{a}_{nk}\big) \tag{6.65}$$

where $w$ is the weighting factor. The derivative of $\mathbb{C}_{mn}^{pARD}$ with respect to $\mathrm{a}_{ij}$ for substitution into Eqn. 56 is

$$\frac{\partial \mathbb{C}_{mn}^{WPT}}{\partial \mathrm{a}_{ij}} = wC_I(\delta_{mk}\mathrm{a}_{nk} + \mathrm{a}_{mk}\delta_{nk}) \tag{6.66}$$

Lastly, we consider the $D_z$ ARD model (cf. Falvoro et al. [278]) developed by Moldflow for simulating 2.5D flow processes. Their model is defined in terms of the interaction coefficient $C_I$, a moment of interaction thickness parameter $D_z$, and the unit normal to the mold surface $\hat{n}$. The expression for $\mathbb{C}_{mn}$ for the $D_z$ ARD model is

$$\mathbb{C}_{mn}^{Dz} = C_I^{Dz}(\delta_{mn} - (1 - D_z)\hat{n}_m\hat{n}_n) \tag{6.67}$$

and

$$\frac{\partial \mathbb{C}_{mn}^{Dz}}{\partial \mathrm{a}_{ij}} = 0 \tag{6.68}$$

*6.3.2.6 Nematic Potential (NEM) Model.* Latz et al. [295] developed a fully coupled flow-orientation model for concentrated suspensions utilizing a 2-parameter nematic potential effective collision ARD (NEM) model that couples the phenomenological effect of the momentum diffusion due to fiber-fiber interaction and a topological interaction effect of diffusion due to an exclusion volume mechanism. The material derivative of the 2nd order orientation tensor based on the nematic diffusion model is

$$\dot{\mathrm{a}}_{mn}^{nem} = \dot{\mathrm{a}}_{mn}^{HD} + \dot{\mathrm{a}}_{mn}^{IRD-MS} \tag{6.69}$$

where

$$\dot{\mathrm{a}}_{mn}^{IRD-MS} = \dot{\gamma}[C_I(\delta_{mn} - \alpha\mathrm{a}_{mn}) + U_0(\mathrm{a}_{mk}\mathrm{a}_{kn} - \mathrm{a}_{kl}\mathrm{a}_{mnkl})] \tag{6.70}$$

In the above, $U_0$ is the 'Onsager' nematic topological interaction coefficient of the Maier-Saupe potential. Typically, for stability, $U_0 \leq 4C_I$ for 2D analysis and $U_0 > 8C_I$ for 3D analysis. The NR residual and Jacobian are, respectively, given as

$$\Sigma_{mn}^{nem} = \dot{\mathrm{a}}_{mn}^{nem} \tag{6.71}$$

$$J_{mnij}^{nem} = \frac{\partial \dot{\text{a}}_{mn}^{HD}}{\partial \text{a}_{ij}} + \frac{\partial \dot{\text{a}}_{mn}^{IRD-MS}}{\partial \text{a}_{ij}} \tag{6.72}$$

where the derivative of the nematic diffusion term is obtained from Eqn. 72 as

$$\begin{aligned} \frac{\partial \dot{\text{a}}_{mn}^{IRD-MS}}{\partial \text{a}_{ij}} = \dot{\gamma} \Bigg[ & -C_I \alpha \delta_{mi} \delta_{nj} \\ & + U_0 \left( \delta_{mi} \delta_{kj} \text{a}_{kn} + \text{a}_{mk} \delta_{ki} \delta_{nj} - \delta_{ki} \delta_{lj} \text{a}_{mnkl} - \text{a}_{kl} \frac{\partial \text{a}_{mnkl}}{\partial \text{a}_{ij}} \right) \Bigg] \end{aligned} \tag{6.73}$$

Latz et al. [295] found the influence of the topological interaction on the fiber orientation state to be flow dependent having a significant effect on channel and contraction flows and a relatively lesser influence for flow around cylinders.

Kugler et al. [22] , Favaloro et al. [278], Agboola et al. [296] and Park and Park [297] present detailed review and comparison of existing fiber orientation models. The foregoing ARD models find useful application in polymer composite industry and have been incorporated in mold-filling flow computations in injection molding process simulations [223], [224], [298], [299], [300], [301], [302]. Like with most commercial software used in simulation of the injection molding process such as Autodesk Moldflow and Moldex3D, it is common to combine multiple diffusion models when predicting the orientation state with the goal of improved accuracy. One such combination is the ARD-RSC models whose material time derivative is expressed as [293]

$$\dot{\text{a}}_{mn}^{pARD-RSC} = \dot{\text{a}}_{mn}^{RSC} - \text{ķ}\dot{\text{a}}_{mn}^{IRD} + \dot{\text{a}}_{mn}^{ARD} + \dot{\text{a}}_{mn}^{\Delta RSC} \tag{6.74}$$

where

$$\dot{\text{a}}_{mn}^{\Delta RSC} = -2\dot{\gamma}(1 - \text{ķ})[\mathbb{M}_{mnkl} - \delta_{kl}\text{a}_{mn} - 5(\mathbb{L}_{mnkl} - \mathbb{M}_{mnrs}\text{a}_{rskl})]\mathbb{C}_{kl} \tag{6.75}$$

and $\dot{\text{a}}_{mn}^{IRD}$, $\dot{\text{a}}_{mn}^{RSC}$, $\dot{\text{a}}_{mn}^{ARD}$ are given in Eqns. (6.20), (6.29), and (6.49) respectively. In this case the NR residual $\Sigma_{mn}^{pARD-RSC}$ is the material derivative $\dot{\text{a}}_{mn}^{pARD-RSC}$, i.e.,

$$\Sigma_{mn}^{pARD-RSC} = \dot{\mathrm{a}}_{mn}^{pARD-RSC} \tag{6.76}$$

and the Jacobian is obtained by taking partial derivative of Eqn. 76 with respect to $\mathrm{a}_{ij}$ as

$$J_{mnij}^{pARD-RSC} = \frac{\partial \dot{\mathrm{a}}_{mn}^{RSC}}{\partial \mathrm{a}_{ij}} - \mathrm{k}\frac{\partial \dot{\mathrm{a}}_{mn}^{IRD}}{\partial \mathrm{a}_{ij}} + \frac{\partial \dot{\mathrm{a}}_{mn}^{ARD}}{\partial \mathrm{a}_{ij}} + \frac{\partial \dot{\mathrm{a}}_{mn}^{\Delta RSC}}{\partial \mathrm{a}_{ij}} \tag{6.77}$$

where

$$\frac{\partial \dot{\mathrm{a}}_{mn}^{\Delta RSC}}{\partial \mathrm{a}_{ij}} = -2\dot{\gamma}(1-\mathrm{k})\left\{ \begin{aligned} &\left[\frac{\partial \mathbb{M}_{mnkl}}{\partial \mathrm{a}_{ij}} - \delta_{kl}\delta_{mi}\delta_{nj} - 5\frac{\partial}{\partial \mathrm{a}_{ij}}\{\mathbb{L}_{mnkl} - \mathbb{M}_{mnrs}\mathrm{a}_{rskl}\}\right]\mathbb{C}_{kl} \\ &+[\mathbb{M}_{mnkl} - \delta_{kl}\mathrm{a}_{mn} - 5(\mathbb{L}_{mnkl} - \mathbb{M}_{mnrs}\mathrm{a}_{rskl})]\frac{\partial \mathbb{C}_{kl}}{\partial \mathrm{a}_{ij}} \end{aligned} \right\} \tag{6.78}$$

The partial derivative terms on the right-hand side (R.H.S) of Eqn. (6.78) appear in the previous sections above.

### *6.3.3 Closure Approximations and Their Explicit Derivatives*

All fiber orientation tensor evolution equations include a tensor of higher order that must be evaluated or approximated. The common approach is to introduce a closure approximation that defines a higher order fiber orientation tensor as a function of lower order information. Closure approximations are either based on an analytical expression or formed through a curve fitting process. The non-fitted analytical closure approximations include the those constructed from combinations of lower order orientation and tensors and include the linear (LIN), quadratic (QDR), and the hybrid (HYB) closures. The general class of Hinch and Leal's [303] composite closure approximations precontracted with the deformation rate tensor also falls under this category. Fitted closures are derived in terms of independent constants which are obtained using test data from solutions to the PDF for multiple standard flow conditions. The fitted closures include Eigenvalue-Based Fitted (EBF) and Invariant-Based Fitted (IBF) closures [304], [305], [306], [307]. Higher order optimal fitted closures such as the Eigenvalue-Based Optimal Fitted (EBOF) and Invariant-Based Optimal Fitted (IBOF) closures [304], [305], [306], [307] which are extensions to the EBF and IBF closures respectively have also been developed for improved model accuracy. More details on different closure approximations and their associated advantages and disadvantages is discussed in later sections and also in [22], [251], [267], [275]. Other fitted closure approximations include the neural network based closures by Jack et al. [308], [309] and the 6th order Invariant based orthotropic fitted closure by Jack and Smith [20], [307]. Derivatives of the orientation tensor closure approximation are used in the Newton-Raphson iteration method to compute the steady-state fiber orientation tensor state. These derivatives used in this study appear below.

*6.3.3.1 Quadratic Closure Approximation*. The quadratic closure $\tilde{a}_{ijkl}$ was introduced by Doi [310] and Lipscomb [311] and is defined as the dyadic product of the 2nd order orientation tensor $a_{ij}$ as

$$\tilde{a}_{ijkl} = a_{ij}a_{kl} \tag{6.79}$$

The derivative of $\tilde{a}_{ijkl}$ with respect to $a_{mn}$ is

$$\frac{\partial \tilde{a}_{ijkl}}{\partial a_{mn}} = \frac{\partial a_{ij}}{\partial a_{mn}} a_{kl} + a_{ij}\frac{\partial a_{kl}}{\partial a_{mn}} = \delta_{im}\delta_{jn}a_{kl} + a_{ij}\delta_{km}\delta_{ln} \tag{6.80}$$

The QDR closure inherently lacks symmetry but does preserve the required symmetry of the computed $a_{mn}$ when employed in the Advani-Tucker equation (cf. Eqn. (6.18)).

*6.3.3.2 Linear Closure Approximation*. The linear (LIN) closure approximation of $\hat{a}_{ijkl}$ first proposed by Hand [290] uses all possible products of the 2nd order orientation tensor $a_{ij}$ and the identity tensor $\delta_{ij}$ as

$$\hat{a}_{ijkl} = -h_1(\delta_{ij}\delta_{kl} + \delta_{ik}\delta_{jl} + \delta_{il}\delta_{jk}) + h_2(a_{ij}\delta_{kl} + a_{ik}\delta_{jl} + a_{il}\delta_{jk} + \delta_{ij}a_{kl} + \delta_{ik}a_{jl} + \delta_{il}a_{jk}) \quad (6.81)$$

where $h_1$ and $h_2$ are numerical factors which vary based on spatial dimensionality and given in Table 6.2. It follows that the derivative of $\hat{a}_{ijkl}$ with respect to $a_{mn}$ is

$$\frac{\partial \hat{a}_{ijkl}}{\partial a_{mn}} = h_2(\delta_{im}\delta_{jn}\delta_{kl} + \delta_{im}\delta_{kn}\delta_{jl} + \delta_{im}\delta_{ln}\delta_{jk} + \delta_{im}\delta_{jn}\delta_{kl} + \delta_{im}\delta_{kn}\delta_{jl} + \delta_{im}\delta_{ln}\delta_{jk}) \quad (6.82)$$

Table 6.2: Numerical factors of the linear closure

| | *Solid* (3*D*) | *Planar* (2*D*) |
|---|---|---|
| $h_1$ | 1/35 | 1/24 |
| $h_2$ | 1/7 | 1/6 |

The LIN closure is exact for random orientation distribution while the QDR closures are exact for a uniaxially aligned fiber orientation.

*6.3.3.3 Hybrid Closure Approximation*. The hybrid closure approximation $a_{ijkl}$ is a weighted combination of the $\hat{a}_{ijkl}$ and $\tilde{a}_{ijkl}$ defined above written in terms of Herman's Orientation factor f as [19]

$$a_{ijkl} = f\tilde{a}_{ijkl} + (1 - f)\hat{a}_{ijkl} \quad (6.83)$$

Advani & Tucker [19] proposed an expression for $f$ as an invariant of the orientation state given as $f = a_f a_{ij} a_{ji} - b_f$, where $a_f$ and $b_f$ are constants that depends on the spatial dimension as shown in Table 6.3.

Table 6.3: Constants of the hybrid closure

| | $Solid\ (3D)$ | $Planar\ (2D)$ |
|---|---|---|
| $a_f$ | $3/2$ | $2$ |
| $b_f$ | $1/2$ | $1$ |

The derivative of the hybrid closure approximation $\mathrm{a}_{ijkl}$ with respect to components of the 2nd order tensor $\mathrm{a}_{mn}$ is

$$\frac{\partial \mathrm{a}_{ijkl}}{\partial \mathrm{a}_{mn}} = f\frac{\partial \tilde{\mathrm{a}}_{ijkl}}{\partial \mathrm{a}_{mn}} + (1-f)\frac{\partial \hat{\mathrm{a}}_{ijkl}}{\partial \mathrm{a}_{mn}} + \frac{\partial f}{\partial \mathrm{a}_{mn}}\left(\tilde{\mathrm{a}}_{ijkl} - \hat{\mathrm{a}}_{ijkl}\right) \tag{6.84}$$

where

$$\frac{\partial f}{\partial \mathrm{a}_{mn}} = a_f\left(\delta_{im}\delta_{jn}\mathrm{a}_{ji} + \mathrm{a}_{ij}\delta_{jm}\delta_{in}\right) \tag{6.85}$$

An alternative form of $f$ proposed by Advani & Tucker [19] is

$$f = 1 - \alpha^{\alpha} e_{ijk}\mathrm{a}_{i1}\mathrm{a}_{j2}\mathrm{a}_{k3} \tag{6.86}$$

where it follows that

$$\frac{\partial f}{\partial \mathrm{a}_{mn}} = -\alpha^{\alpha} e_{ijk}\left\{\delta_{im}\delta_{1n}\mathrm{a}_{j2}\mathrm{a}_{k3} + \mathrm{a}_{i1}\delta_{jm}\delta_{2n}\mathrm{a}_{k3} + \mathrm{a}_{i1}\mathrm{a}_{j2}\delta_{km}\delta_{3n}\right\} \tag{6.87}$$

The hybrid model is observed to perform better during the transient orientation prediction; however, the hybrid closure was shown to over-predict the steady state fiber alignment compared with simulation results based on the more accurate orientation distribution function which are, however, computationally expensive since they require a finite difference grid in space and time.

*6.3.3.4 Hinch and Leal Closure Approximation.* Hinch and Leal [303] developed numerous composite closure approximations for the 4th order orientation tensor in precontracted forms with the deformation rate tensor. The accuracy of their predictions was shown to depend on the flow field and flow magnitude considered. Hinch and Leal closure approximations are not explicit expressions of the 4th order orientation tensor $a_{ijkl}$ but are instead written in contracted form with the deformation rate tensor i.e., $\gamma_{kl}a_{ijkl}$. Advani and Tucker provides a general explicit expression of $a_{ijkl}$ (cf. Equation (6.88)) summarizing all Hinch and Leal closures as

$$\begin{aligned} a_{ijkl} = {} & \beta_1(\delta_{ij}\delta_{kl}) + \beta_2(\delta_{ik}\delta_{jl} + \delta_{il}\delta_{jk}) + \beta_3(\delta_{ij}a_{kl} + a_{ij}\delta_{kl}) \\ & + \beta_4(a_{ik}\delta_{jl} + \cdots. + a_{jl}\delta_{ik} + a_{il}\delta_{jk} + a_{jk}\delta_{il}) + \beta_5(a_{ij}a_{kl}) \\ & + \beta_6(a_{ik}a_{jl} + a_{il}a_{jk}) + \cdots. + \beta_7(\delta_{ij}a_{km}a_{ml} + a_{im}a_{mj}\delta_{kl}) \\ & + \beta_8(a_{im}a_{mj}a_{kn}a_{nl}) \end{aligned} \tag{6.88}$$

where the $\beta_i$ are defined in Table 6.4. The partial derivative of the above expression with respect to $a_{rs}$ is

$$\begin{aligned} \frac{\partial a_{ijkl}}{\partial a_{rs}} = {} & \left[\frac{\partial \beta_1}{\partial a_{rs}}(\delta_{ij}\delta_{kl}) + \frac{\partial \beta_2}{\partial a_{rs}}(\delta_{ik}\delta_{jl} + \delta_{il}\delta_{jk}) + \frac{\partial \beta_3}{\partial a_{rs}}(\delta_{ij}a_{kl} + a_{ij}\delta_{kl}) \right. \\ & + \frac{\partial \beta_4}{\partial a_{rs}}(a_{ik}\delta_{jl} + a_{jl}\delta_{ik} + a_{il}\delta_{jk} + a_{jk}\delta_{il}) + \frac{\partial \beta_5}{\partial a_{rs}}(a_{ij}a_{kl}) \\ & + \frac{\partial \beta_6}{\partial a_{rs}}(a_{ik}a_{jl} + a_{il}a_{jk}) + \frac{\partial \beta_7}{\partial a_{rs}}(\delta_{ij}a_{km}a_{ml} + a_{im}a_{mj}\delta_{kl}) \\ & \left. + \frac{\partial \beta_8}{\partial a_{rs}}(a_{im}a_{mj}a_{kn}a_{nl})\right] \\ & + \Big[\beta_3(\delta_{ij}\delta_{kr}\delta_{ls} + \delta_{ir}\delta_{js}\delta_{kl}) \\ & + \beta_4(\delta_{ir}\delta_{ks}\delta_{jl} + \delta_{jr}\delta_{ls}\delta_{ik} + \delta_{ir}\delta_{ls}\delta_{jk} + \delta_{jr}\delta_{ks}\delta_{il}) \\ & + \beta_5(\delta_{ir}\delta_{js}a_{kl} + a_{ij}\delta_{kr}\delta_{ls}) \\ & + \beta_6(\delta_{ir}\delta_{ks}a_{jl} + a_{ik}\delta_{jr}\delta_{ls} + \delta_{ir}\delta_{ls}a_{jk} + a_{il}\delta_{jr}\delta_{ks}) \\ & + \beta_7(\delta_{ij}\delta_{kr}\delta_{ms}a_{ml} + \delta_{ij}a_{km}\delta_{mr}\delta_{ls} + \delta_{ir}\delta_{ms}a_{mj}\delta_{kl} + a_{im}\delta_{mr}\delta_{js}\delta_{kl}) \\ & + \beta_8(\delta_{ir}\delta_{ms}a_{mj}a_{kn}a_{nl} + a_{im}\delta_{mr}\delta_{js}a_{kn}a_{nl} + a_{im}a_{mj}\delta_{kr}\delta_{ns}a_{nl} \\ & + a_{im}a_{mj}a_{kn}\delta_{nr}\delta_{ls})\Big] \end{aligned} \tag{6.89}$$

Mullens [275] provided a summary Table (cf. Table 6.4) for the $\beta_i$ factors of the Hinch and Leal closures subdivided into weak flow (WF - Isotropic, Linear and Quadratic), strong flow (SF), and Hinch and Leal composite flows (HL – HL1 and HL2) closure forms.

Table 6.4: Summary of the Hinch and Leal closure $\beta_i$ factors for the different flow classifications

| | | $\beta_1$ | $\beta_2$ | $\beta_3$ | $\beta_4$ | $\beta_5$ | $\beta_6$ | $\beta_7$ | $\beta_8$ |
|---|---|---|---|---|---|---|---|---|---|
| WF | ISO | $\frac{1}{15}$ | $\frac{1}{15}$ | … | … | … | … | … | … |
| | LIN | $-\frac{1}{35}$ | $-\frac{1}{35}$ | $\frac{1}{7}$ | $\frac{1}{7}$ | … | … | … | … |
| | QDR | … | … | … | … | 1 | … | … | … |
| SF | SF2 | … | … | … | … | 1 | 1 | … | $-\frac{2}{\langle \mathrm{a}^2 \rangle}$ |
| HL | HL1 | … | … | $\frac{2}{5}$ | … | $-\frac{1}{5}$ | $\frac{3}{5}$ | $-\frac{2}{5}$ | … |
| | HL2 | $\frac{26}{315}\alpha$ | $\frac{26}{315}\alpha$ | $\frac{16}{63}\alpha$ | $-\frac{4}{21}\alpha$ | 1 | 1 | … | $-\frac{2}{\langle \mathrm{a}^2 \rangle}$ |

where the parameters $\langle \mathrm{a}^2 \rangle$ and $\alpha$ are respectively

$$\langle \mathrm{a}^2 \rangle = \mathrm{a}_{ij}\mathrm{a}_{ji}, \qquad \alpha = \exp\left[2\frac{1-3\langle \mathrm{a}^2 \rangle}{1-\langle \mathrm{a}^2 \rangle}\right] \tag{6.90}$$

and the partial derivatives are respectively given as

$$\begin{gathered} \frac{\partial \langle \mathrm{a}^2 \rangle}{\partial \mathrm{a}_{rs}} = \delta_{ir}\delta_{js}\mathrm{a}_{ji} + \mathrm{a}_{ij}\delta_{jr}\delta_{is} \\ \frac{\partial \alpha}{\partial \mathrm{a}_{rs}} = -\frac{4\alpha}{(1-\langle \mathrm{a}^2 \rangle)^2}\frac{\partial \langle \mathrm{a}^2 \rangle}{\partial \mathrm{a}_{rs}}, \quad \frac{\partial}{\partial \mathrm{a}_{rs}}\left\{\frac{k}{\langle \mathrm{a}^2 \rangle}\right\} = -\frac{k}{\langle \mathrm{a}^2 \rangle^2}\frac{\partial \langle \mathrm{a}^2 \rangle}{\partial \mathrm{a}_{rs}} \end{gathered} \tag{6.91}$$

*6.3.3.5 Eigenvalue based Fitted (EBF) Closure Approximations.* More recently, fitted closure approximations have emerged including the Eigenvalue-Based Fitted (EBF) closures that require a principal axis transformation which ensures material objectivity. Orthotropic closure approximations for the 4$^{\text{th}}$ order tensor implicitly impose objectivity such that the approximation is independent of the coordinate frame selection. Indeed, the Orthotropic Smooth (ORS), Orthotropic Fitted (ORF) and ORF for low fiber-fiber interaction coefficient (ORL) closures are included in the EBF family of closures originally developed by Cintra and Tucker [267] based on the assumption of coincident principal axis of the 2nd and 4th order tensor. The ORF has been shown to have better performance compared to non-fitted closure approximations, however, the ORF behaved poorly for flows with very low interaction coefficients and sometimes gave non-physical oscillations similar to that seen with the Hinch and Leal closure (HL2) under the same condition. The ORL behaves better for low interaction coefficient in simple shear flow, but overpredicts flow direction alignment and is unstable for radial diverging flows. Due to the inherent symmetries of $a_{ijkl}$, its components can be represented in (6 x 6) contracted notation as

$$A_{rs} = \mathrm{a}_{ijkl} \tag{6.92}$$

where the index of the contracted notation is related to the index notation according to

$$r = \begin{cases} i = j & for,\ \delta_{ij} = 1 \\ (9 - i - j) & for,\ \delta_{ij} = 0 \end{cases} \text{ and } s = \begin{cases} k = l & for,\ \delta_{kl} = 1 \\ (9 - k - l) & for,\ \delta_{kl} = 0 \end{cases} \tag{6.93}$$

It follows that the derivative of the 4th order tensor with respect to $\mathrm{a}_{mn}$ in contracted notation are

$$\frac{\partial A_{rs}}{\partial \mathrm{a}_{mn}} = \frac{\partial \mathrm{a}_{ijkl}}{\partial \mathrm{a}_{mn}} \tag{6.94}$$

Symmetry of the 4th order tensor requires $\mathrm{a}_{ijkl} = \mathrm{a}_{klij}$ which implies that $A_{rs} = A_{sr}$ . The contracted tensor $A_{rs}$ transformed to the principal axes has the orthotropic form $\bar{A}_{rs}$ given as

$$\underline{\underline{\bar{A}}} = \begin{bmatrix} \bar{A}_{11} & \bar{A}_{12} & \bar{A}_{13} & & & \\ \bar{A}_{21} & \bar{A}_{22} & \bar{A}_{23} & & & \\ \bar{A}_{31} & \bar{A}_{32} & \bar{A}_{33} & & & \\ & & & \bar{A}_{44} & & \\ & & & & \bar{A}_{55} & \\ & & & & & \bar{A}_{66} \end{bmatrix} \tag{6.95}$$

The contracted tensor transforms from its principal reference frame to the original coordinate axes according to

$$A_{rs} = \gamma_{ri}\gamma_{sj}\bar{A}_{ij} \tag{6.96}$$

The 6x6 transformation matrix $\gamma_{ij}$ is given as $\gamma_{ij} = \mathfrak{m}_{im}\, \mho_{mn}\, \mathfrak{m}_{nj}^{-1}$, where $\mathfrak{m}_{ij} = k\delta_{ij},\ k = \begin{cases} 1, & i \leq 3 \\ 2, & i > 3 \end{cases}$ and $\mho_{mn}$ is the contracted form of $\mho_{ijkl}$ where $\mho_{ijkl} = \Phi_{ik}\Phi_{jl} + (1 - \delta_{kl})\Phi_{jk}\Phi_{il}$. The modal matrix $\Phi_{ij}$ whose *k*th column are the corresponding eigenvectors $\underline{x}^k$ of eigenvalues $\Lambda_k = \Lambda_{kk}$ (no summation implied) is obtained from the spectral decomposition of $\mathrm{a}_{ij}$ as

$$\underline{\underline{\Phi}} \mid \mathrm{a}_{mn} = \Phi_{mk}\Lambda_{kl}\Phi_{nl} \tag{6.97}$$

A more direct way is to transform the 4th order orientation tensor given in its principal reference frame as $\bar{\mathrm{a}}_{mnpq}$ to that in the original reference frame $\mathrm{a}_{ijkl}$ according to

$$\mathrm{a}_{ijkl} = \Phi_{im}\Phi_{jn}\Phi_{kp}\Phi_{lq}\bar{\mathrm{a}}_{mnpq} \tag{6.98}$$

where $\bar{\mathrm{a}}_{mnpq}$ is reconstructed can from the contracted form $\bar{A}_{rs}$. Using the product rule

$$\begin{aligned} \frac{\partial \mathrm{a}_{ijkl}}{\partial \mathrm{a}_{rs}} &= q_{im}q_{jn}q_{kp}q_{lq}\frac{\partial \bar{a}_{mnpq}}{\partial \mathrm{a}_{rs}} + \cdots \\ &\quad + \left(\frac{\partial q_{im}}{\partial \mathrm{a}_{rs}}q_{jn}q_{kp}q_{lq} + q_{im}\frac{\partial q_{jn}}{\partial \mathrm{a}_{rs}}q_{kp}q_{lq} + q_{im}q_{jn}\frac{\partial q_{kp}}{\partial \mathrm{a}_{rs}}q_{lq} \right. \\ &\quad \left. + q_{im}q_{jn}q_{kp}\frac{\partial q_{lq}}{\partial \mathrm{a}_{rs}}\right)\bar{a}_{mnpq} \end{aligned} \tag{6.99}$$

Derivative of the eigenvector tensor $\underline{\underline{\Phi}}$ appears in [283] (cf. APPENDIX C, C.1). Symmetry requirements of the transformed orthotropic tensor reduces the total number of independent non-zero components to 9, and additional special symmetry properties of the exact 4th order tensor requires that $\bar{\mathrm{a}}_{ijkl} = \bar{\mathrm{a}}_{kjil} = \bar{\mathrm{a}}_{ljki} = \bar{\mathrm{a}}_{ikjl} = \bar{\mathrm{a}}_{ilkj}$ reducing the non-zero independent components in $A_{ij}$ to the 6 diagonal terms. i.e.,

$$\bar{A}_{ij} = \bar{A}_{kk} \quad k = 9 - i - j\,, \qquad i \neq j \tag{6.100}$$

where repeated indices for $\bar{A}_{kk}$ does not imply summation. The normalization property $\mathrm{a}_{ijkk} = \mathrm{a}_{ij}$ of the exact 4th order tensor further requires that

$$\begin{bmatrix} \bar{A}_{44} \\ \bar{A}_{55} \\ \bar{A}_{66} \end{bmatrix} = \underline{\underline{ß}}^{-1} \left\{ \begin{bmatrix} \Lambda_1 \\ \Lambda_2 \\ \Lambda_3 \end{bmatrix} - \begin{bmatrix} \bar{A}_{11} \\ \bar{A}_{22} \\ \bar{A}_{33} \end{bmatrix} \right\} \tag{6.101}$$

where $\Lambda_i$ are the eigenvalues of $\mathrm{a}_{ij}$, with $\sum_i \Lambda_i = 1$ and $ß_{ij} = 1 - \delta_{ij}$. Based on the foregoing conditions, the only three surviving non-zero independent terms are $\bar{A}_{11}$, $\bar{A}_{22}$ and $\bar{A}_{33}$. The general form for orthotropic closure is to express the three surviving non-zero independent components ($\bar{A}_{11}$, $\bar{A}_{22}$, $\bar{A}_{33}$) of the contracted 4th order tensor in the principal reference frame after imposing all symmetric and normalization conditions of the exact 4th order tensor, as a scalar function $F_k(\Lambda_1, \Lambda_2)$ of the two largest eigenvalues $(\Lambda_1, \Lambda_2)$ of $\mathrm{a}_{ij}$. Most fitted closures take the form of a $n^{\text{th}}$-order binomial function in $\Lambda_1$ and $\Lambda_2$ (note that $\Lambda_3 = 1 - \Lambda_1 - \Lambda_2$) to represent $F_k$ as

$$\bar{A}_{kk} = F_k(\Lambda_1, \Lambda_2) = f_k^{(n)}(\Lambda_1, \Lambda_2), \quad \Lambda_1 \geq \Lambda_2 \geq \Lambda_3, \qquad k = 1,2,3 \tag{6.102}$$

where, again, no summation is implied by the repeated indices for $\bar{A}_{kk}$. Polynomial order exceeding $n = 4$ fall under the class of eigenvalue based optimal fitting (EBOF) closures. Generally, we can represent the function $f_k^{(n)}$ as a tensor product of a constant coefficient matrix $\mathfrak{C}_{ij}^{(n)}$ and a n^th order permuted bivariate polynomial vector $\mathbb{A}_j^{(n)} = \mathbb{A}_j^{(n)}(\Lambda_1, \Lambda_2)$, i.e.

$$f_k^{(n)}(\Lambda_1, \Lambda_2) = \mathfrak{C}_{kj}^{(n)} \mathbb{A}_j^{(n)} \qquad k = 1,2,3, \quad j = 1 \dots (n+1)(n+2)/2 \tag{6.103}$$

Representations of $\mathfrak{C}_{ij}^{(n)}$ and $\mathbb{A}_j^{(n)}$ for various closure approximations appear in APPENDIX C, **C.2**. The derivative of the components of the orthotropic closure in contracted notation with respect to $\mathrm{a}_{rs}$ are

$$\frac{\partial \bar{A}_{kk}}{\partial \mathrm{a}_{rs}} = \frac{\partial \bar{A}_k}{\partial \mathrm{a}_{rs}} = \mathfrak{C}_{kj}^{(n)} \frac{\partial \mathbb{A}_j^{(n)}}{\partial \mathrm{a}_{rs}} = \mathfrak{C}_{kj}^{(n)} \mathbb{A}'^{(n)}_{jrs} = \mathfrak{C}_{kj}^{(n)} \tilde{\mathbb{A}}_{jl}^{(n)} \Lambda'_{lrs} \qquad k = 1,2,3, \; l = 1,2 \tag{6.104}$$

The $n^{\text{th}}$ order binomial permutation vector $\mathbb{A}_k^{(n)}$ and its derivative coefficient matrix $\tilde{\mathbb{A}}_{kl}^{(n)}$ for the quadratic closure are given from terms of binomial expansion, respectively as

$$\begin{cases} \mathbb{A}_k^{(n)}(\Lambda_1, \Lambda_2) = \Lambda_1^{i-j} \Lambda_2^j \\ \tilde{\mathbb{A}}_{kl}^{(n)} = \dfrac{\partial}{\partial \Lambda_l} \mathbb{A}_k^{(n)} = \begin{cases} (i-j) \cdot \Lambda_1^{i-j-1} \Lambda_2^j & l = 1, \\ j \cdot \Lambda_1^{i-j} \Lambda_2^{j-1} & l = 2 \end{cases} \quad k = 1 + j + i(i+1)/2 \end{cases} \qquad j = 0 \cdots i, \; i = 0 \cdots n \tag{6.105}$$

For a special case of orthotropic fitted closure called rational elliptical (RE) closure by Wetzel and Tucker [312], the scalar function for the 3 independent tensor component is

$$F_k(\Lambda_1, \Lambda_2) = \frac{f_k^{(n)}(\Lambda_1, \Lambda_2)}{f_k^{(m)}(\Lambda_1, \Lambda_2)} \tag{6.106}$$

The derivative of the components of the above with respect to $\mathrm{a}_{rs}$ is

$$\frac{\partial \bar{A}_{kk}}{\partial \mathrm{a}_{rs}} = \frac{1}{[f^{(m)}]^2} \left[ f^{(m)} \frac{\partial f^{(n)}}{\partial \mathrm{a}_{rs}} - f^{(n)} \frac{\partial f^{(m)}}{\partial \mathrm{a}_{rs}} \right] \tag{6.107}$$

Again, no summation is implied by the repeated indices in $\bar{A}_{kk}$. From the normalization condition of the 4th order tensor, we obtain for the derivative of $\bar{A}_{kk}$ $(k = 4,5,6)$ as

$$\frac{\partial \bar{A}_{kk}}{\partial \mathrm{a}_{rs}} = ß_{ki}^{-1}\left\{\frac{\partial Ƙ_i}{\partial \mathrm{a}_{rs}} - \frac{\partial \bar{A}_{ii}}{\partial \mathrm{a}_{rs}}\right\},\ \frac{\partial Ƙ_i}{\partial \mathrm{a}_{rs}} = \mathcal{E}_{il}Ƙ'_{lrs},\ \underline{\underline{\mathcal{E}}} = \begin{bmatrix} 1 & 0 & -1 \\ 0 & 1 & -1 \end{bmatrix}^T \quad (6.108)$$

$$i = 1,2,3,\ \ l = 1,2$$

For the partial derivative of the eigenvalues with respect to the components of the 2nd order orientation tensor, kindly refer to APPENDIX C, C.1. EBF closures are computationally more expensive because of the principal axis transformation.

*6.3.3.6 Invariant Based Fitted (IBF) Closure Approximations*. Of the class of IBF closures, the natural (NAT) closure approximation of Verleye et al. [313] was built on the work of Lipscomb et al. [311] and formed the basis for other IBF developments. Verleye et al. [46] developed a general expression for the fully symmetric 4th order tensor in terms of invariants of the 2nd order tensor, the identity tensor and fitted coefficients which were determined through a least square fitting process. The NAT closure assumed the absence of fiber-fiber interaction and infinitely long fiber geometry. The closure is exact based on the foregoing assumptions; however, it has been reported to possess singularities for axisymmetric orientation states.

The Invariant based optimal fitting (IBOF) closure developed by Chung and Kwon [304] was an extension to the NAT closure development, however the independent coefficients are derived from regression analysis based on actual data obtained from DFC for multiple flow types (similar to the procedure used for EBF closures). In the contracted form, the 4th order tensor $\mathrm{a}_{ijkl}$ based on symmetry properties is given as

$$\underline{\underline{A}} = \begin{bmatrix} A_{11} & A_{12} & A_{13} & A_{14} & A_{15} & A_{16} \\ & A_{22} & A_{23} & A_{24} & A_{24} & A_{26} \\ & & A_{33} & A_{34} & A_{35} & A_{36} \\ & & & A_{44} & A_{45} & A_{46} \\ & & & & A_{55} & A_{56} \\ \ldots Sym & & & & & A_{66} \end{bmatrix} \tag{6.109}$$

which conforms to the following the symmetry requirement

$$\begin{aligned} A_{44} &= A_{23}, \quad A_{45} = A_{36}, \quad A_{46} = A_{25} \\ A_{55} &= A_{13}, \quad A_{56} = A_{14}, \quad A_{66} = A_{12} \end{aligned} \tag{6.110}$$

and normalization condition

$$\sum_{n=1}^{3} A_{nm} = \mathrm{a}_m, \qquad \mathrm{a}_m = \mathrm{a}_{ij}, \qquad m = \begin{cases} i = j & i = j \\ 9 - i - j & i \neq j \end{cases} \tag{6.111}$$

It follows that under these conditions, the relationship between $\mathrm{a}_{ijkl}$ and $\mathrm{a}_{ij}$ can be written as.

$$\begin{aligned} A_{11} + A_{12} + A_{13} &= \mathrm{a}_{11}, \quad A_{12} + A_{22} + A_{23} = \mathrm{a}_{22} \\ A_{13} + A_{23} + A_{33} &= \mathrm{a}_{33}, \quad A_{14} + A_{24} + A_{34} = \mathrm{a}_{23} \\ A_{15} + A_{25} + A_{35} &= \mathrm{a}_{13}, \quad A_{16} + A_{26} + A_{36} = \mathrm{a}_{12} \end{aligned} \tag{6.112}$$

Taking partial derivatives of Equations (6.110) and (6.111) we obtain the following

$$\frac{\partial A_{mn}}{\partial \mathrm{a}_{rs}} = \frac{\partial A_{ij}}{\partial \mathrm{a}_{rs}}, \qquad and \ \sum_{n=1}^{3} \frac{\partial A_{nm}}{\partial \mathrm{a}_{rs}} = \frac{\partial \mathrm{a}_m}{\partial \mathrm{a}_{rs}} \tag{6.113}$$

There are thus only 9 independent components for the 4$^{th}$ order tensor. The IBOF is developed in terms of the fully symmetric 4$^{th}$ order expansion of $\mathrm{a}_{ijkl}$ as a combination of the 2$^{nd}$ order tensor $\mathrm{a}_{ij}$ and identity tensor $\delta_{kl}$ where the Cayley-Hamilton theory is used to obtain

$$\begin{aligned} \mathrm{a}_{ijkl} = \beta_1 \mathbb{S}(\delta_{ij}\delta_{kl}) + \beta_2 \mathbb{S}(\delta_{ij}\mathrm{a}_{kl}) + \beta_3 \mathbb{S}(\mathrm{a}_{ij}\mathrm{a}_{kl}) + \beta_4 \mathbb{S}(\delta_{ij}\mathrm{a}_{km}\mathrm{a}_{ml}) \\ + \beta_5 \mathbb{S}(\mathrm{a}_{ij}\mathrm{a}_{km}\mathrm{a}_{ml}) + \beta_6 \mathbb{S}(\mathrm{a}_{im}\mathrm{a}_{mj}\mathrm{a}_{kn}\mathrm{a}_{nl}) \end{aligned} \tag{6.114}$$

In the above, the $\mathbb{S}$ operator represents the symmetric permutation expansion of its argument, for example,

$$\begin{aligned}\mathbb{S}(\mathbb{T}_{ijkl}) = \frac{1}{24}\big[&\mathbb{T}_{ijkl} + \mathbb{T}_{ijlk} + \mathbb{T}_{ikjl} + \mathbb{T}_{iklj} + \mathbb{T}_{iljk} + \mathbb{T}_{ilkj} + \mathbb{T}_{jikl} \\ &+ \mathbb{T}_{jilk} + \mathbb{T}_{jkil} + \mathbb{T}_{jkli} + \mathbb{T}_{jlik} + \mathbb{T}_{jlki} + \mathbb{T}_{kijl} + \mathbb{T}_{kilj} \\ &+ \mathbb{T}_{kjil} + \mathbb{T}_{kjli} + \mathbb{T}_{klij} + +\mathbb{T}_{klji} + \mathbb{T}_{lijk} + \mathbb{T}_{likj} + \mathbb{T}_{ljik} \\ &+ \mathbb{T}_{ljki} + \mathbb{T}_{lkij} + \mathbb{T}_{lkji}\big]\end{aligned} \tag{6.115}$$

To incorporate the IBOF closure in our NR solution approach, we differentiate $\mathrm{a}_{ijkl}$with respect to $\mathrm{a}_{rs}$ to obtain

$$\begin{aligned}\frac{\partial}{\partial \mathrm{a}_{rs}}\{\mathrm{a}_{ijkl}\} = \Big[&\frac{\partial \beta_1}{\partial \mathrm{a}_{rs}}\mathbb{S}(\delta_{ij}\delta_{kl}) + \frac{\partial \beta_2}{\partial \mathrm{a}_{rs}}\mathbb{S}(\delta_{ij}\mathrm{a}_{kl}) + \frac{\partial \beta_3}{\partial \mathrm{a}_{rs}}\mathbb{S}(\mathrm{a}_{ij}\mathrm{a}_{kl}) \\ &+ \frac{\partial \beta_4}{\partial \mathrm{a}_{rs}}\mathbb{S}(\delta_{ij}\mathrm{a}_{km}\mathrm{a}_{ml}) + \frac{\partial \beta_5}{\partial \mathrm{a}_{rs}}\mathbb{S}(\mathrm{a}_{ij}\mathrm{a}_{km}\mathrm{a}_{ml}) \\ &+ \frac{\partial \beta_6}{\partial \mathrm{a}_{rs}}\mathbb{S}(\mathrm{a}_{im}\mathrm{a}_{mj}\mathrm{a}_{kn}\mathrm{a}_{nl})\Big] + \cdots \\ &+ \Big[\beta_2\mathbb{S}(\delta_{ij}\delta_{kr}\delta_{ls}) + \beta_3\{\mathbb{S}(\delta_{ir}\delta_{js}\mathrm{a}_{kl}) + \mathbb{S}(\mathrm{a}_{ij}\delta_{kr}\delta_{ls})\} \\ &+ \beta_4\{\mathbb{S}(\delta_{ij}\delta_{kr}\delta_{ms}\mathrm{a}_{ml}) + \mathbb{S}(\delta_{ij}\mathrm{a}_{km}\delta_{mr}\delta_{ls})\} \\ &+ \beta_5\{\mathbb{S}(\delta_{ir}\delta_{js}\mathrm{a}_{km}\mathrm{a}_{ml}) + \mathbb{S}(\mathrm{a}_{ij}\delta_{kr}\delta_{ms}\mathrm{a}_{ml}) \\ &+ \mathbb{S}(\mathrm{a}_{ij}\mathrm{a}_{km}\delta_{mr}\delta_{ls})\} \\ &+ \beta_6\{\mathbb{S}(\delta_{ir}\delta_{ms}\mathrm{a}_{mj}\mathrm{a}_{kn}\mathrm{a}_{nl}) + \mathbb{S}(\mathrm{a}_{im}\delta_{mr}\delta_{js}\mathrm{a}_{kn}\mathrm{a}_{nl}) \\ &+ \mathbb{S}(\mathrm{a}_{im}\mathrm{a}_{mj}\delta_{kr}\delta_{ns}\mathrm{a}_{nl}) + \mathbb{S}(\mathrm{a}_{im}\mathrm{a}_{mj}\mathrm{a}_{kn}\delta_{nr}\delta_{ls})\}\Big]\end{aligned} \tag{6.116}$$

The $\beta_i$ coefficients are expressed as functions of the second and third invariants (II and III, respectively) of $\mathrm{a}_{ij}$. Employing Equations (6.114) and (6.115), and applying the Cayley-Hamilton theorem, it can be shown that only 3 independent coefficients remain. It follows that the expressions for the IBOF dependent coefficients $(\beta_1, \beta_2, \beta_5)$ are

$$\begin{aligned}\beta_1 = \frac{3}{5}\Big[&-\frac{1}{7} + \frac{1}{5}\beta_3\left(\frac{1}{7} + \frac{4}{7}\mathrm{II} + \frac{8}{3}\mathrm{III}\right) - \beta_4\left(\frac{1}{5} - \frac{8}{15}\mathrm{II} - \frac{14}{15}\mathrm{III}\right) \\ &- \beta_6\left(\frac{1}{35} - \frac{4}{35}\mathrm{II} - \frac{24}{105}\mathrm{III} + \frac{16}{15}\mathrm{II}\,\mathrm{III} + \frac{8}{35}\mathrm{II}^2\right)\Big]\end{aligned} \tag{6.117}$$

$$\beta_2 = \frac{6}{7}\left[1 - \frac{1}{5}\beta_3(1 + 4\mathrm{II}) + \frac{7}{5}\beta_4\left(\frac{1}{6} - \mathrm{II}\right) - \beta_6\left(-\frac{1}{5} + \frac{4}{5}\mathrm{II} + \frac{2}{3}\mathrm{III} - \frac{8}{5}\mathrm{II}^2\right)\right]$$

$$\beta_5 = -\frac{4}{5}\beta_3 - \frac{7}{5}\beta_4 - \frac{6}{5}\beta_6\left(1 - \frac{4}{3}\mathrm{II}\right)$$

which can be differentiated with respect to $\mathrm{a}_{rs}$ to yield

$$\begin{aligned}
\frac{\partial\beta_1}{\partial\mathrm{a}_{rs}} &= \frac{3}{5}\left[\frac{1}{5}\frac{\partial\beta_3}{\partial\mathrm{a}_{rs}}\left(\frac{1}{7} + \frac{4}{7}\mathrm{II} + \frac{8}{3}\mathrm{III}\right) - \frac{\partial\beta_4}{\partial\mathrm{a}_{rs}}\left(\frac{1}{5} - \frac{8}{15}\mathrm{II} - \frac{14}{15}\mathrm{III}\right) + \cdots \right. \\
&\qquad \left. - \frac{\partial\beta_6}{\partial\mathrm{a}_{rs}}\left(\frac{1}{35} - \frac{4}{35}\mathrm{II} - \frac{24}{105}\mathrm{III} + \frac{16}{15}\mathrm{II}\,\mathrm{III} + \frac{8}{35}\mathrm{II}^2\right)\right] + \cdots \\
&\qquad + \frac{3}{5}\left[\left[\frac{4}{35}\beta_3 - \beta_4 + \beta_6\left(\frac{4}{35} - \frac{16}{35}\mathrm{II} - \frac{16}{15}\mathrm{III}\right)\right]\frac{\partial\mathrm{II}}{\partial\mathrm{a}_{rs}} \right. \\
&\qquad \left. + \left[\frac{1}{5}\frac{8}{3}\beta_3 + \frac{14}{15}\beta_4 + \beta_6\left(\frac{24}{105} - \frac{16}{15}\mathrm{II}\right)\right]\frac{\partial\mathrm{III}}{\partial\mathrm{a}_{rs}}\right]
\end{aligned}$$

$$\begin{aligned}
\frac{\partial\beta_2}{\partial\mathrm{a}_{rs}} &= \frac{6}{7}\left[-\frac{1}{5}\frac{\partial\beta_3}{\partial\mathrm{a}_{rs}}(1 + 4\mathrm{II}) + \frac{7}{5}\frac{\partial\beta_4}{\partial\mathrm{a}_{rs}}\left(\frac{1}{6} - \mathrm{II}\right) \right. \\
&\qquad \left. - \frac{\partial\beta_6}{\partial\mathrm{a}_{rs}}\left(-\frac{1}{5} + \frac{4}{5}\mathrm{II} + \frac{2}{3}\mathrm{III} - \frac{8}{5}\mathrm{II}^2\right)\right] \\
&\qquad + \frac{6}{7}\left[-\frac{1}{5}[4\beta_3 + 7\beta_4 + \beta_6(4 - 16\mathrm{II})]\frac{\partial\mathrm{II}}{\partial\mathrm{a}_{rs}} - \frac{2}{3}\beta_6\frac{\partial\mathrm{III}}{\partial\mathrm{a}_{rs}}\right]
\end{aligned} \tag{6.118}$$

$$\frac{\partial\beta_5}{\partial\mathrm{a}_{rs}} = -\frac{4}{5}\frac{\partial\beta_3}{\partial\mathrm{a}_{rs}} - \frac{7}{5}\frac{\partial\beta_4}{\partial\mathrm{a}_{rs}} - \frac{6}{5}\frac{\partial\beta_6}{\partial\mathrm{a}_{rs}}\left(1 - \frac{4}{3}\mathrm{II}\right) + \frac{8}{5}\beta_6\frac{\partial\mathrm{II}}{\partial\mathrm{a}_{rs}}$$

The independent coefficients $(\beta_3, \beta_4, \beta_6)$ in Chung et al. [304] are obtained from a $5^{\text{th}}$ order binomial fitted function in terms of II and III as

$$\beta_m = \sum_{i=0}^{5}\sum_{j=0}^{i} a_k^m \cdot \mathrm{II}^{i-j}\mathrm{III}^j\,, \qquad k = j + i(i+1)/2\,, \qquad m = 1,2,3 \tag{6.119}$$

where the coefficients of the binomial terms appear in APPENDIX C, C.2.2. The non-unity invariants of $\mathrm{a}_2$ are given as

$$\mathrm{II} = \Lambda_1\Lambda_2 + \Lambda_2\Lambda_3 + \Lambda_3\Lambda_1, \qquad \mathrm{III} = \Lambda_1\Lambda_2\Lambda_3 \tag{6.120}$$

It follows that derivative of the independent coefficients in Equation 120 with respect to the components $\mathrm{a}_{rs}$are

$$\frac{\partial \beta_{\boldsymbol{m}}}{\partial \mathrm{a}_{rs}} = \sum_{i=0}^{5} \sum_{j=0}^{i} a_k^m \left\{ (i-j) \cdot \mathrm{II}^{i-j-1} \mathrm{III}^{j} \frac{\partial \mathrm{II}}{\partial \mathrm{a}_{rs}} + j \cdot \mathrm{II}^{i-j-1} \mathrm{III}^{j-1} \frac{\partial \mathrm{III}}{\partial \mathrm{a}_{rs}} \right\} \tag{6.121}$$

where,

$$\begin{aligned} \frac{\partial \mathrm{II}}{\partial \mathrm{a}_{rs}} &= (\Lambda_2 + \Lambda_3) \frac{\partial \Lambda_1}{\partial \mathrm{a}_{rs}} + (\Lambda_1 + \Lambda_3) \frac{\partial \Lambda_2}{\partial \mathrm{a}_{rs}} + (\Lambda_1 + \Lambda_2) \frac{\partial \Lambda_3}{\partial \mathrm{a}_{rs}} \\ \frac{\partial \mathrm{III}}{\partial \mathrm{a}_{rs}} &= (\Lambda_2 \Lambda_3) \frac{\partial \Lambda_1}{\partial \mathrm{a}_{rs}} + (\Lambda_1 \Lambda_3) \frac{\partial \Lambda_2}{\partial \mathrm{a}_{rs}} + (\Lambda_1 \Lambda_2) \frac{\partial \Lambda_3}{\partial \mathrm{a}_{rs}} \end{aligned} \tag{6.122}$$

### *6.4 Error Estimate*

In the numerical examples to follow, we assess the performance of the Newton-Raphson (NR) method to accurately predict the steady-state values of $\mathrm{a}_{ij}$ based on the relative absolute error between results of our NR approach and that of a reference method, in this case the explicit, single-step, Dormand-Prince 4th and 5th order Runge-Kutta adaptive-stepping method (RK45) with a set relative and absolute tolerance of $10^{-12}$ which ensures the solutions are accurate up to 12 digits. We define the percentage error as

$$err_1 = \frac{\mathrm{a}_{mn}^{NR} - \mathrm{a}_{mn}^{RK4}}{\mathrm{a}_{mn}^{RK4}} \times 100\% \tag{6.123}$$

The NR iteration is said to converge to a unique solution, and the iteration is terminated once the NR error tolerance $err_{NR} = \left\| \underline{\Sigma}_{ij}^{NR} \right\|_2$ falls below some predetermined tolerance value, $err_d = 10^{-6}$, where $\underline{\Sigma}_{ij}^{NR}$ is given by Equation (6.7).

### *6.5 Results and Discussion*

Example calculations are presented below to illustrate and validate our NR based approach for computing steady-state values of $a_{ij}$ under various selected flow conditions. Prior to providing these examples, we first validate the explicit analytical expressions for the Jacobian introduced in Equation 4 for the various fiber orientation tensor evolution models and closures presented above. Results computed at steady state using our Newton Raphson approach are validated with computed results obtained using MATLAB's inbuilt explicit, adaptive-stepping, Dormand-Prince 4$^{th}$ and 5$^{th}$ order, Runge-Kutta ode45 (RK45) solver (MathWorks, Natick, MA, USA). We note that the RK45 basis for validating our NR results are likewise numerical approximations, and as such, the relative error metric provided only serves as an indication of the agreement between these two outcomes. The validation of the NR approach is performed using three different orientation tensor model case studies which include (a) an anisotropic rotary diffusion (ARD) tensor model (b) a slow orientation kinetic (SOK) tensor model, and (c) a composite/coupled tensor model. Subsequent sections are dedicated to assessing the performance of the NR method using various 4$^{th}$ order orientation tensor closure approximations and various homogenous flows. Besides the relative error tolerance, other metrics for assessing the NR performance in the example calculations below include the rate of numerical convergence and the number of NR iterations to converge.

#### *6.5.1 Validation of exact Jacobians based on Finite Difference Approximation*

Exact Jacobians $\underline{\underline{J}}^{exact}$ defined through the exact derivative expressions provided in Section 2 are validated here using approximate values of the Jacobian $\underline{\underline{J}}^{FD}$ calculated with the $O(h^4)$ central finite difference method. The validity of our computed exact Jacobians is assessed here based on the error defined by the Euclidean norm given as

$$err_2 = \left\| \underline{\underline{J}}^{exact} - \underline{\underline{J}}^{FD} \right\|_2 \qquad (6.124)$$

Values of $\underline{\underline{J}}^{FD}$ are computed in this work using the higher order central difference finite difference approximation $CT_4$ written as

$$J_{mnij}^{FD} = \frac{-\Sigma_{mn}(\mathrm{a}_{ij} + 2\delta\mathrm{a}_{ij}) + 8\Sigma_{mn}(\mathrm{a}_{ij} + \delta\mathrm{a}_{ij}) - 8\Sigma_{mn}(\mathrm{a}_{ij} - \delta\mathrm{a}_{ij}) + \Sigma_{mn}(\mathrm{a}_{ij} - 2\delta\mathrm{a}_{ij})}{12\delta\mathrm{a}_{ij}} \qquad (6.125)$$

$$\ldots + O(\delta^4)$$

where $\delta\mathrm{a}_{ij}$ is the finite perturbation of the $ij$-component of the 2nd order orientation tensor. A perturbation size of $\delta = 10^{-6}$ is applied individually to each component of $\mathrm{a}_{ij}$ to define $\delta\mathrm{a}_{ij}$. We also consider the sensitivity of the result to the finite difference type including the $n^{\text{th}}$ order backward ($BW_n$), forward ($FW_n$) and central ($CT_n$) differences, (see e.g., [276] for finite difference formulas). The model parameters used here for the various 2nd order fiber orientation tensor models considered for this validation exercise appear in Table 6.5 [278].

Table 6.5. Case Study 1 parameters for the FT, Dz, iARD, pARD, WPT, MRD and PT models [278]

| | $C_I$ | ARD Parameters | | | | |
|---|---|---|---|---|---|---|
| $FT$ | 0.0311 | - | | | | |
| $Dz$ | 0.0258 | $D_z = 0.051, \hat{n} = [0 \quad 0 \quad 1]$ | | | | |
| $iARD$ | 0.0562 | $C_M = 09977$ | | | | |
| $pARD$ | 0.0169 | $\Omega = 0.9868$ | | | | |
| $WPT$ | 0.0504 | $w = 0.9950$ | | | | |
| $MRD$ | 0.0198 | $D_1 = 1.0,\ D_2 = 0.7946,\ D_3 = 0.0120$ | | | | |
| $PT$ | - | $b_1$ | $b_2$ | $b_3$ | $b_4$ | $b_5$ |

| | | | | |
|---|---|---|---|---|
| $1.924 \times 10^{-4}$ | $5.839 \times 10^{-3}$ | 0.04 | $1.168 \times 10^{-5}$ | 0 |

We assume for this validation exercise a 'randomly' generated orientation state $\underline{\underline{a}}^0$ given as

$$\underline{\underline{a}}^0 = \begin{bmatrix} 0.0622 & 0.0765 & 0.0398 \\ 0.0765 & 0.5521 & 0.0186 \\ 0.0398 & 0.0186 & 0.3857 \end{bmatrix} \quad (6.126)$$

Values of $err_2$ from Eqn. (6.124) appear in Table 6.6 to Table 6.9 for the various 2nd order orientation tensor models, identified in the left most column, and 4th order closure approximations, as identified in the table headers. The results show that with a perturbation size of $\delta = 10^{-6}$, the $err_2$ magnitude values are below $10^{-6}$ which ensures the accuracy of our exact Jacobian derivations and numerical implementations over the various 2nd order orientation tensor models and 4th order closures approximations considered above. The $err_2$ magnitude values appear lower for the Hinch and Leal closure approximations (cf. Table 6.6), on the order of $10^{-9} - 10^{-8}$ as compared to the EBF and IBF closures (cf.

Table 6.7), which is on the order of $10^{-8}$. Alternatively, the $err_2$ magnitude values for the IBOF and EBOF closures (cf.

Table 6.9) are relatively higher on the order of about $10^{-7}$. In Figure 6.2a , the $err_2$ is shown to decrease monotonically with increasing perturbation size, showing the optimum result for $\delta = 10^{-3}$, below which $err_2$ increases monotonically. Additionally, Figure 6.2b shows that the $err_2$ depends on the finite difference type and decreases with increasing order of approximation. Values of $err_2$ are relatively higher with lower order finite difference approximations (i.e., the 1st order backward and forward difference

$BW_1$ $and$ $FW_1$, respectively) and lower with higher order approximations (i.e. the 4th order central difference $CT_4$).

Table 6.6. Computed error $(\times 10^{-8})$ for various orientation tensor models using Hinch and Leal closure approximations.

| | $HYB_1$ | $HYB_2$ | ISO | LIN | QDR | SF2 | $HL_1$ | $HL_2$ |
|---|---|---|---|---|---|---|---|---|
| AT | 0.6436 | 0.9385 | 0.2220 | 0.4188 | 0.2691 | 2.0949 | 0.9618 | 4.3940 |
| PT | 0.8088 | 0.7549 | 0.5837 | 0.5003 | 0.4244 | 1.6776 | 0.8241 | 3.4809 |
| iARD | 0.5737 | 1.2712 | 0.3444 | 0.6336 | 0.5728 | 0.5100 | 0.8774 | 1.6148 |
| pARD | 0.7169 | 0.5475 | 0.2722 | 0.2438 | 0.4155 | 1.4805 | 0.9818 | 3.6185 |
| WPT | 0.8563 | 1.0386 | 0.3773 | 0.2525 | 0.2926 | 1.4543 | 0.9632 | 3.5284 |
| Dz | 0.5899 | 0.8248 | 0.2373 | 0.5484 | 0.3233 | 0.7137 | 1.0594 | 2.7732 |
| NEM | 0.6490 | 0.9306 | 0.4012 | 0.4314 | 0.1612 | 2.1012 | 0.9846 | 4.4062 |
| pARD-RSC | 1.0030 | 1.3343 | 1.3062 | 1.0506 | 1.3699 | 0.5378 | 1.3441 | 1.6482 |
| iARD-RPR | 0.5645 | 0.6900 | 0.3478 | 0.3687 | 0.5222 | 1.0731 | 1.0324 | 2.0512 |

Table 6.7. Computed error $(\times 10^{-7})$ for various orientation tensor models using IBF and EBF closure approximations.

| | $NAT_1$ | $NAT_2$ | ORS | ORT | ORW | ORW3 |
|---|---|---|---|---|---|---|
| AT | 0.3557 | 0.5812 | 0.4573 | 0.3351 | 0.5994 | 0.6496 |
| PT | 0.3076 | 0.3946 | 0.3327 | 0.2517 | 0.5693 | 0.5284 |
| iARD | 0.2512 | 0.2663 | 0.2129 | 0.1934 | 0.4762 | 0.4213 |
| pARD | 0.2975 | 0.3956 | 0.3216 | 0.2479 | 0.5354 | 0.5229 |
| WPT | 0.3040 | 0.4068 | 0.3430 | 0.2622 | 0.5720 | 0.5612 |
| Dz | 0.3532 | 0.3276 | 0.2805 | 0.2920 | 0.6704 | 0.6010 |
| NEM | 0.3649 | 0.5869 | 0.4615 | 0.3388 | 0.6062 | 0.6520 |
| pARD-RSC | 0.2529 | 0.2800 | 0.2074 | 0.1802 | 0.4772 | 0.4155 |
| iARD-RPR | 0.1964 | 0.1709 | 0.1601 | 0.1404 | 0.3687 | 0.2820 |

Table 6.8. Computed error $(\times 10^{-7})$ for various orientation tensor models using IBOF and EBOF closure approximations.

| | IBOF | WTZ | LAR32 | VST | FFLAR4 | LAR4 |
|---|---|---|---|---|---|---|
| AT | 6.1748 | 4.3147 | 5.0800 | 3.1567 | 4.3188 | 4.3101 |
| PT | 5.6437 | 4.0286 | 4.7162 | 2.9443 | 4.0435 | 3.9967 |
| iARD | 4.4942 | 3.0776 | 3.6782 | 2.2662 | 3.0115 | 3.0665 |
| pARD | 5.5458 | 3.8741 | 4.5415 | 2.8548 | 3.8500 | 3.8818 |
| WPT | 5.6412 | 4.0391 | 4.7234 | 2.9350 | 4.0486 | 3.9851 |
| Dz | 6.4226 | 4.8010 | 5.5454 | 3.3924 | 4.8016 | 4.6691 |
| NEM | 6.1978 | 4.3254 | 5.0936 | 3.1653 | 4.3271 | 4.3182 |
| pARD-RSC | 4.1821 | 2.9401 | 3.4878 | 2.1358 | 2.9236 | 2.9070 |
| iARD-RPR | 3.4882 | 2.4509 | 2.9082 | 1.7359 | 2.3590 | 2.4110 |

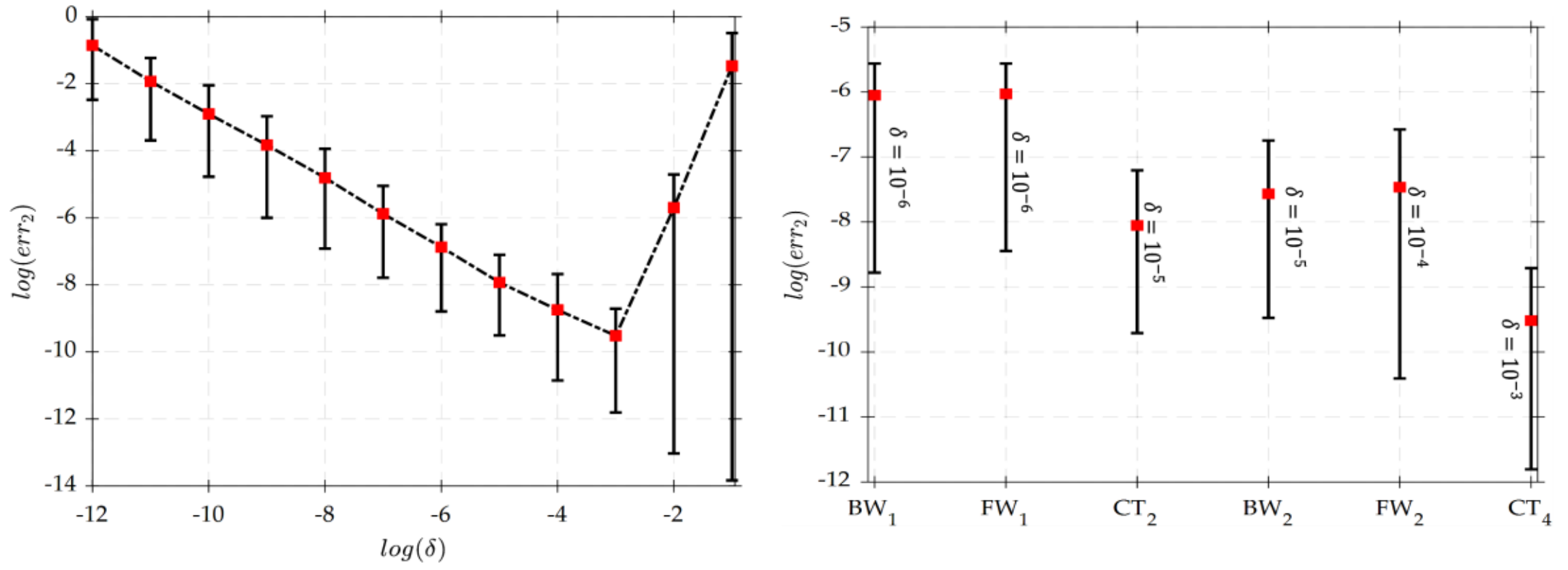

Figure 6.2. Finite difference error $err_2$ for various orientation models and closures (a) over perturbation size δ for the 4th order central difference approximation and for (b) finite difference type and order of approximation at their corresponding optimum perturbation size δ.

### *6.5.2 Comparison of the NR Method and the Dormand-Prince Runge-Kutta (RK45) Method*

In this section, computed results of the steady state orientation states obtained for various orientation evolution models and closure approximations using our Newton Raphson algorithm are compared to those obtained using the RK45 method. The RK45 method is implemented to provide transient solutions to the initial-value orientation ODE (cf. Eqn. (6.7)). A sufficiently small time step is selected for the RK45 simulations to minimize truncation errors [276] and the computations are performed until steady state is achieved, defined where the Euclidean norm of the orientation tensor material derivative is less than $10^{-6}$, i.e. at $err_{RK45}(t) = \left\| \Sigma_{ij}^{RK45}(t) \right\|_2$, $\Sigma_{ij}^{RK45}(t) = \partial a_{ij}^{RK45}/\partial t$.

Three sample cases are included here, the first set of models are based on the study by Falvoro et al. [278], and two other sets are based on the study by Tseng et al. [284]. These cases are used to assess the performance of the NR method relative to the RK4 method for steady state predictions based on various 2nd order orientation tensor models. The first case includes various anisotropic rotary diffusion (ARD) tensor models presented in [278] that account for fiber-fiber and fiber-matrix interactions. The second case considers slow orientation kinetics (SOK) tensor-based models such as the SRF, RSC, and RPR that correct the over-prediction inaccuracies of the transient fiber orientation response computed with the AT model. These models account for the effect of both deformation and diffusion of the fibers and matrix on the transient fibers response [284]. The third case is based on coupled tensor models involving a combination of an ARD model and an SOK model that has been shown to yield improved orientation tensor accuracy including the iARD-RPR, and pARD-RPR which have been incorporated into Moldex3D (CoreTech System Co. of Taiwan) [293] and iARD-RSC models incorporated into AutoDesk MoldFlow® (Autodesk Inc., San Francisco, CA, USA). The models in the latter case study are typically used in injection molding simulations specifically for long-fiber concentrated suspensions [22], [284]. The EBOF closure approximation of Verweyst [302] is employed for all analysis.

**Case 1:** The first study considers various anisotropic rotary diffusion (ARD) orientation tensor models with model parameters appearing in Table 6.5. The initial orientation state for the RK45 simulations is

$$\underline{\underline{a}}^0 = \begin{bmatrix} 0.30 & 0.00 & 0.00 \\ 0.00 & 0.60 & 0.10 \\ 0.00 & 0.10 & 0.10 \end{bmatrix} \tag{6.127}$$

which is the same as that used as the initial guess in the NR method calculations. The initial guess value is selected such that the target steady state orientation is physical with non-negative eigenvalues in all cases. This is required for some tensor models that can exhibit numerical convergence instability such as the PT model [278], [284]. One purpose of this first case study is to show that the NR method adequately predicts the steady state orientation irrespective of the choice and complexity of the rotary diffusion model or transient simulation behavior. The RK45 calculated transient responses for select components of $a_{ij}$ appear in Figure 6.3.

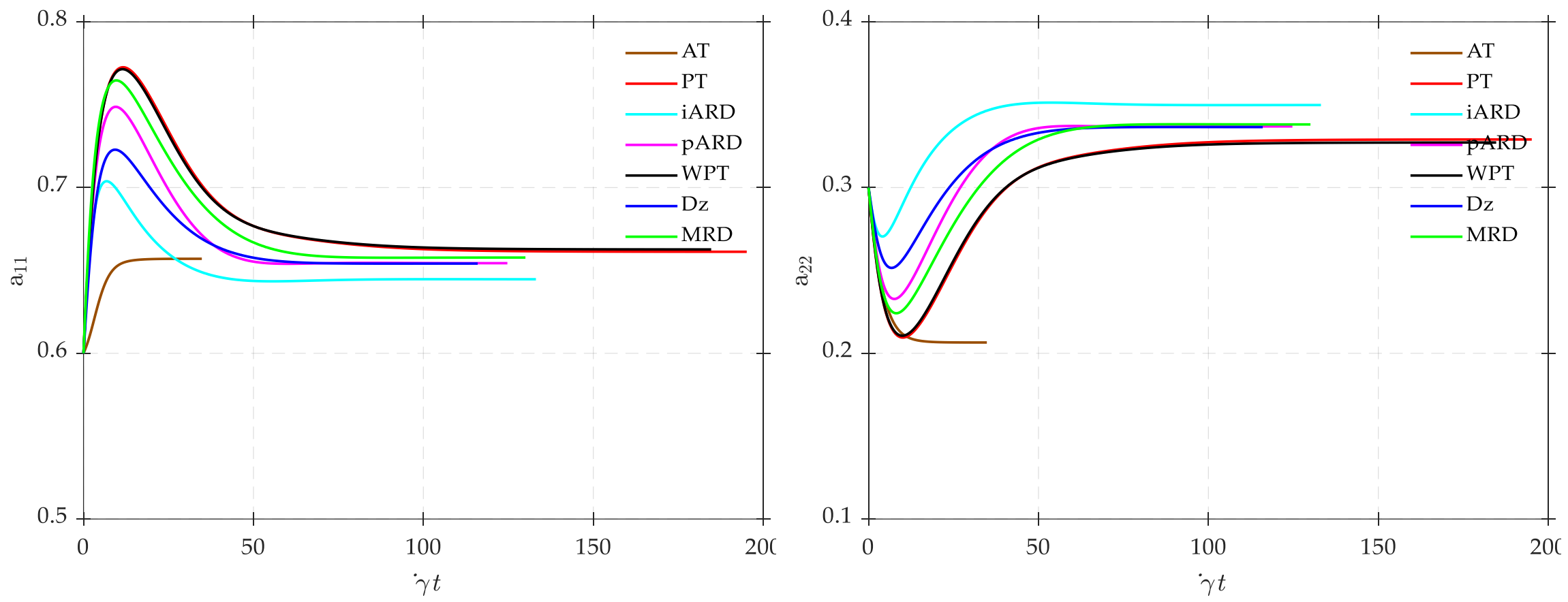

**(a)** **(b)**

Figure 6.3. Time evolution of the 2nd order orientation tensor for calibrated FT, PT, iARD, pARD, WPT, Dz and MRD models for (a) $a_{11}$ component (b) $a_{22}$ component.

The computed responses in Figure 6.3a and b show varying amplitudes of initial overshoot of $a_{11}$ and initial trough of $a_{22}$ depending on orientation model that does not, however, affect the final steady state orientation.

Table 6.9 includes values of $err_1$ (cf. Equation (6.123)) which show that the NR based steady state orientation predictions agree well with RK4 values indicated by $err_1 < 0.005\%$ for both diagonal and off-diagonal tensor components. Although the NR method does not provide information of the transient orientation history such as the initial overshoot and/or the initial trough, steady state orientations appear to be independent of the transient behavior of the orientation tensor model.

Table 6.9. Error estimates of the $a_{11}$, $a_{22}$ and $a_{13}$ steady-state orientation tensor component values for FT, Dz, iARD, pARD, WPT, MRD and PT models

| | $a_{11}$ | $a_{22}$ | $a_{13}$ |
|---|---|---|---|
| AT | 0.0015 | 0.0003 | 0.0000 |
| PT | 0.0046 | 0.0021 | 0.0031 |
| iARD | 0.0009 | 0.0006 | 0.0000 |
| pARD | 0.0006 | 0.0003 | 0.0032 |
| WPT | 0.0043 | 0.0020 | 0.0030 |
| Dz | 0.0021 | 0.0011 | 0.0021 |
| MRD | 0.0015 | 0.0008 | 0.0046 |

The NR convergence history for the various diffusion tensor models appears in Figure 6.4 which indicate a relatively higher rate of convergence with a quadratic nature for the fundamental AT isotropic rotary diffusion (IRD) model requiring only 5 iterations to convergence compared to the PT, WPT, Dz, ARD models which all required 7 iterations to converge. The improved and principal ARD models (iARD and pARD) show the slowest convergence rate, requiring 8 iterations to converge. All models, however, exhibit stable convergence with a monotonically decreasing NR residual norm. With the RK45 method, the orientation tensor reached steady state faster for the AT model, at a time fraction of $\dot{\gamma}t = 35\text{s}$ compared to the iARD, pARD, Dz, MRD which required an average of about $\dot{\gamma}t = 126\text{s}$ to converge. The PT and WPT models were the slowest, requiring at least $\dot{\gamma}t = 185\text{s}$ for the orientation to approach steady state. Unlike the AT, PT, and WPT models, the iARD, pARD, Dz, and MRD model transient responses had undulations which may be an effect of the numerical solver due to the stiffness of the ODE.

In terms of computational efficiency, computation time was obtained for all NR and RK5 simulations using a x64-based, Intel(R) Core(TM) i5-8365U CPU with a 1.90 GHz processing speed and a 16.0 GB RAM, operating Windows 11 Pro. Average computation time for each technique appears in Table 6.10. Overall, the NR technique is seen to be on average 7.4 x faster than the RK45 method for the error tolerance given in

Table 6.9.

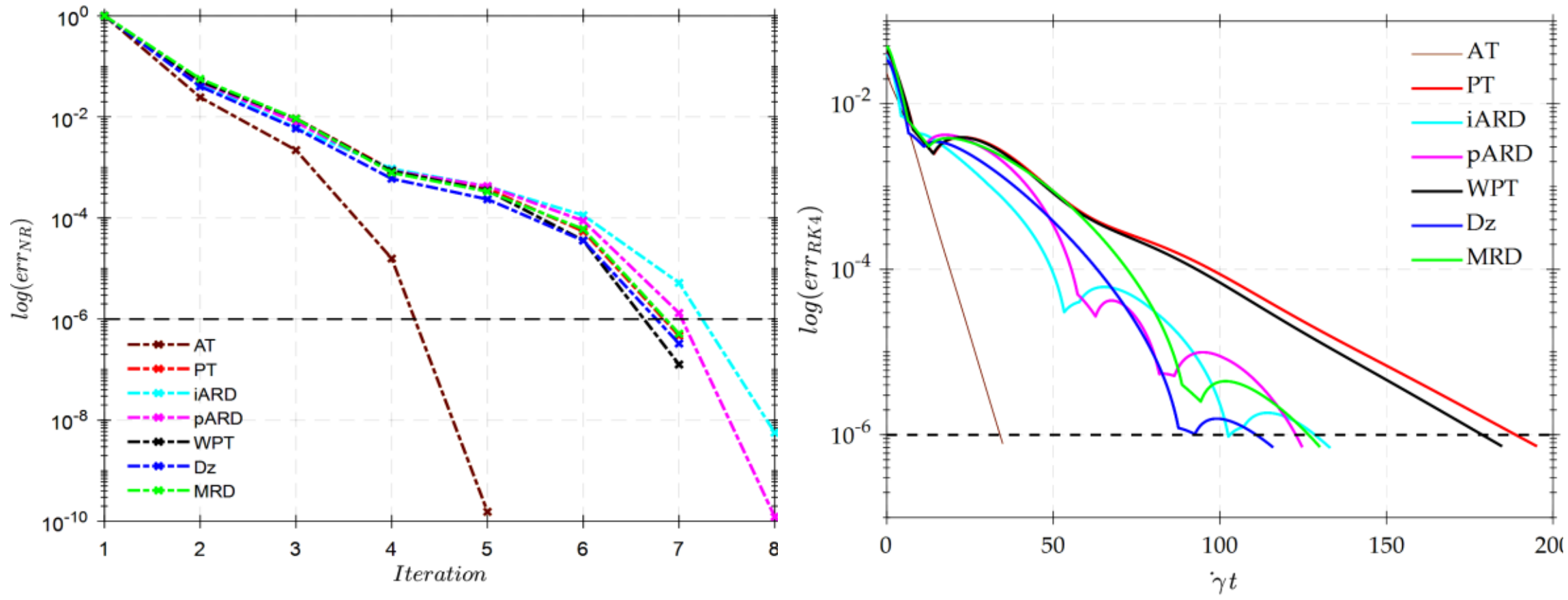

(a) (b)

Figure 6.4. Values of the residual norm for the (a) NR method (b) RK45 method; for the AT, PT, iARD, pARD, WPT, Dz and MRD models.

Table 6.10. Average processor time taken (sec) to compute the steady state orientation values for both numerical techniques and for the various diffusion models.

| | AT | PT | iARD | pARD | WPT | Dz | MRD |
|---|---|---|---|---|---|---|---|
| NR | 0.7210 | 0.4075 | 0.5916 | 0.3784 | 0.3038 | 0.4615 | 0.7908 |
| RK45 | 3.5829 | 3.9775 | 3.0067 | 2.6441 | 2.7791 | 3.4486 | 6.7519 |

**Case 2:** Unlike the first case study, Case 2 investigates the behavior of the NR method for SOK tensor models that involves corrections to slow down the transient fiber orientation kinematic response as compared to the AT model, which includes the SRF, RSC and RPR models. Although the various model corrections have no effect on the final steady state solution, the effect of the SOK model complexity on NR convergence is shown. The model parameters used for this case study are taken from Tseng et al. [284] and appear in

Table 6.11.

Table 6.11. Case Study 2 parameters for the FT, SRF, RSC and RPR models [284].

| | AT | SRF | RSC | RPR |
|---|---|---|---|---|
| $C_I$ | 0.01 | 0.01 | 0.01 | 0.01 |
| ꞔ | – | 0.1 | 0.1 | – |
| $\alpha$ | – | – | – | 0.9 |
| $\beta$ | – | – | – | 0 |

The starting orientation for the RK45 analysis and the initial guess $\underline{\underline{a}}^0$ used for the Newton Raphson method is the same as that used in case study 1 (cf. Eqn. (6.127)). Two flow fields $L_1$ and $L_2$ are considered:

- $L_1$: Simple shear flow in the 1-2 plane, $L_{12} = \dot{\gamma}$.
- $L_2$: Balanced shear/planar-elongation flow, simple shear in 1-2 plane superimposed on planar elongation in 1-2 plane. $L_{11} = -\dot{\varepsilon}, L_{22} = \dot{\varepsilon}, L_{12} = \dot{\gamma}$ given $\dot{\gamma}/\dot{\varepsilon} = 10$.

The time evolution of the components of the 2nd order orientation tensor based on the RK45 method are shown in Figure 6.5. below. The percentage error between NR and RK45 steady state values are presented in Table 6.12.

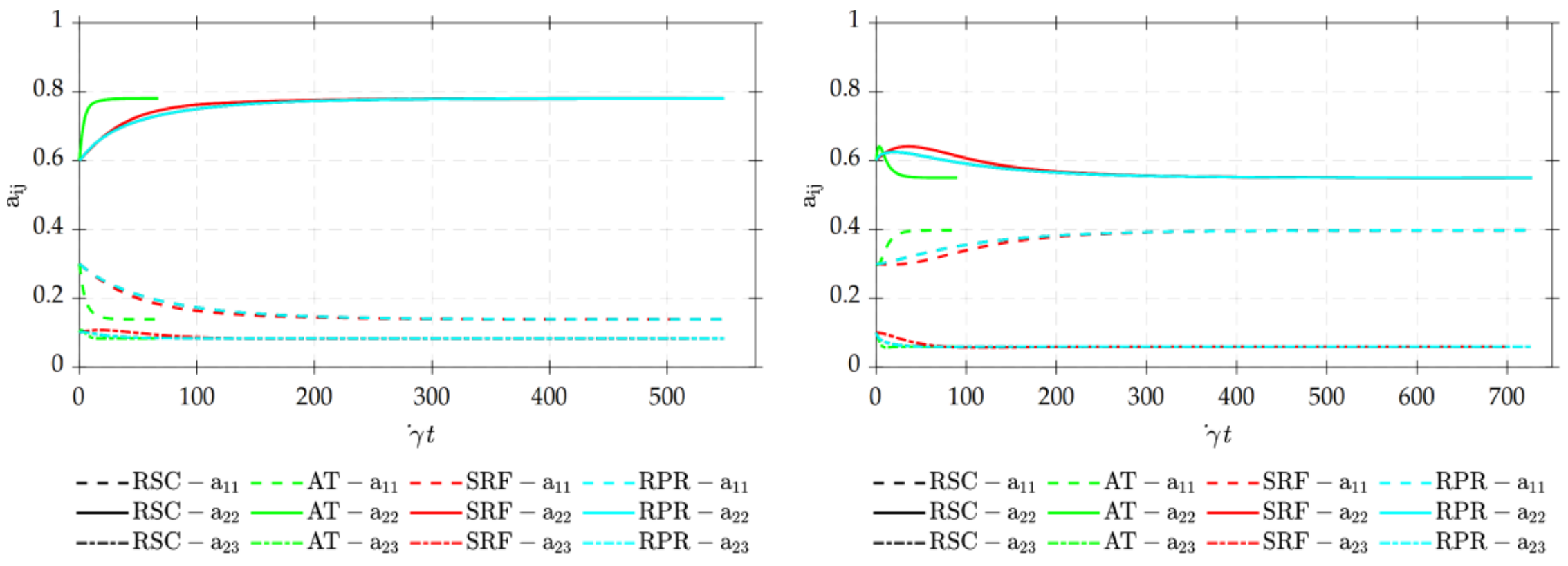

(a) (b)

Figure 6.5. Time evolution of the $a_{11}, a_{22}$ and $a_{13}$ components of the 2nd order orientation tensor for calibrated AT, SRF, RSC and FT-RPR models for (a) simple shear flow, and (b) shearing/stretching combination flow.

Results show a high level in accuracy in prediction based on the NR method. with a maximum error tolerance of only $err_1 < 0.04\%$ among all tensor components for all models. The AT and SRF models exhibited relatively faster convergence rate requiring only 5 iterations to converge for both flow types (cf. Figure 6.6) compared to the RSC and RPR models which required an extra iteration to converge. The RSC and RPR models both require complex eigenvalue and eigenvector spectral decompositions which may influence the convergence rate. Moreover, the derivatives of the eigenvectors as shown in Appendix A may impact the convergence rate of these models. The convergence rate was seen to be independent of both flow types considered.

Table 6.12. Calculated $a_{11}, a_{22}$ and $a_{13}$ steady-state error values for RSC, AT, SRF, and RPR models and for the 2 flow fields $L_1$ and $L_2$.

| | $L_1$ | | | $L_2$ | | |
|---|---|---|---|---|---|---|
| | $a_{11}$ | $a_{22}$ | $a_{13}$ | $a_{11}$ | $a_{22}$ | $a_{13}$ |
| RSC | 0.0374 | 0.0058 | 0.0000 | 0.0178 | 0.0122 | 0.0017 |
| FT | 0.0036 | 0.0005 | 0.0000 | 0.0015 | 0.0011 | 0.0000 |
| SRF | 0.0374 | 0.0055 | 0.0024 | 0.0158 | 0.0109 | 0.0050 |
| RPR | 0.0374 | 0.0058 | 0.0000 | 0.0178 | 0.0122 | 0.0017 |

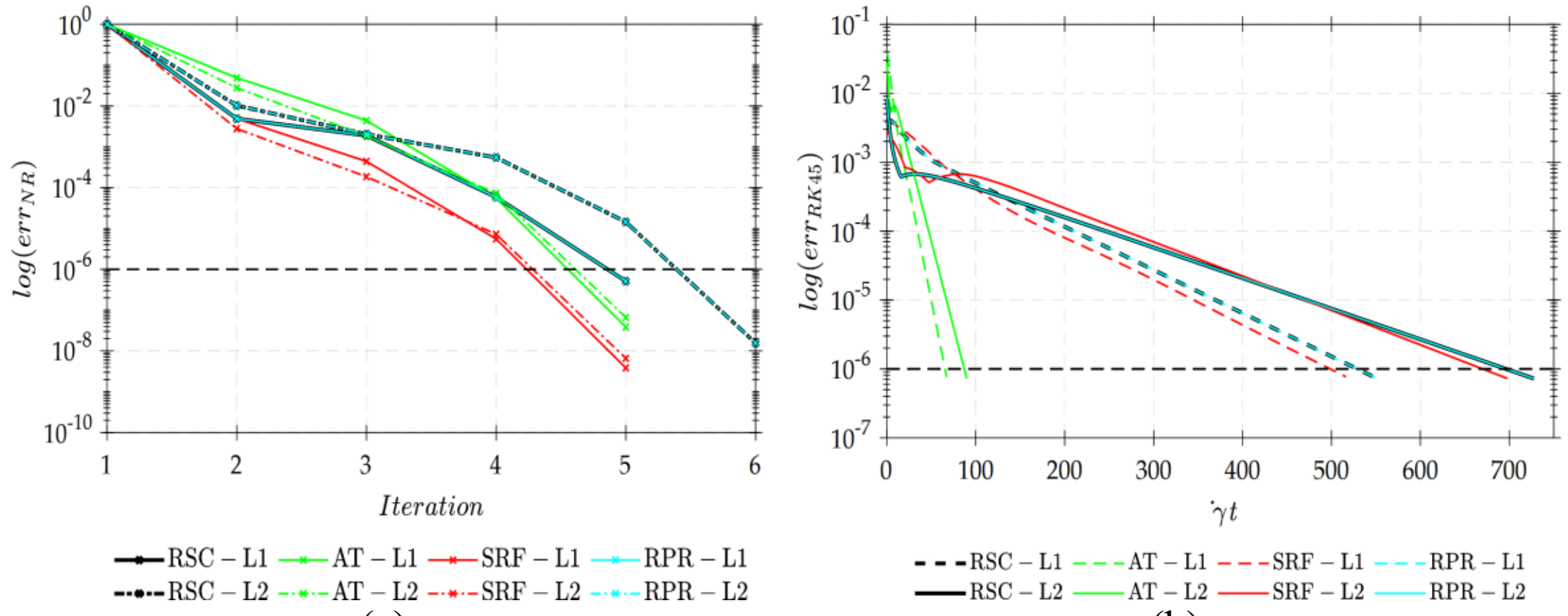

(a) (b)

Figure 6.6. History of the (a) NR residual norm (b) RK45 absolute maximum material derivative tensor component; for the FT, RSC, SRF, and RPR model for the 2 different flow fields (L1 and L2).

Figure 6.7a and b present the NR residual norm for the AT and RSC models and for different initial orientation states obtained by varying $a_{22}^0$ (i.e. $0. \leq a_{22}^0 \leq 0.7$), while retaining the initial states of other $a_{ij}^0$ components. As $a_{22}^0$ increases, the initial orientation states approach the final steady state orientation, thus less iterations are required for the solutions to converge. The AT converged as expected for all $a_{22}^0$ values considered (cf. Figure 6.7a), however the RSC model was very sensitive to the $a_{22}^0$ values (cf. Figure 6.7b). Only the initial orientation states with $a_{22}^0 = 0..49, 0.56, 0.63, 0.70$ converged to physical solutions as shown in Figure 6.7b), while the other $a_{22}^0$ values yielded non-physical solutions.

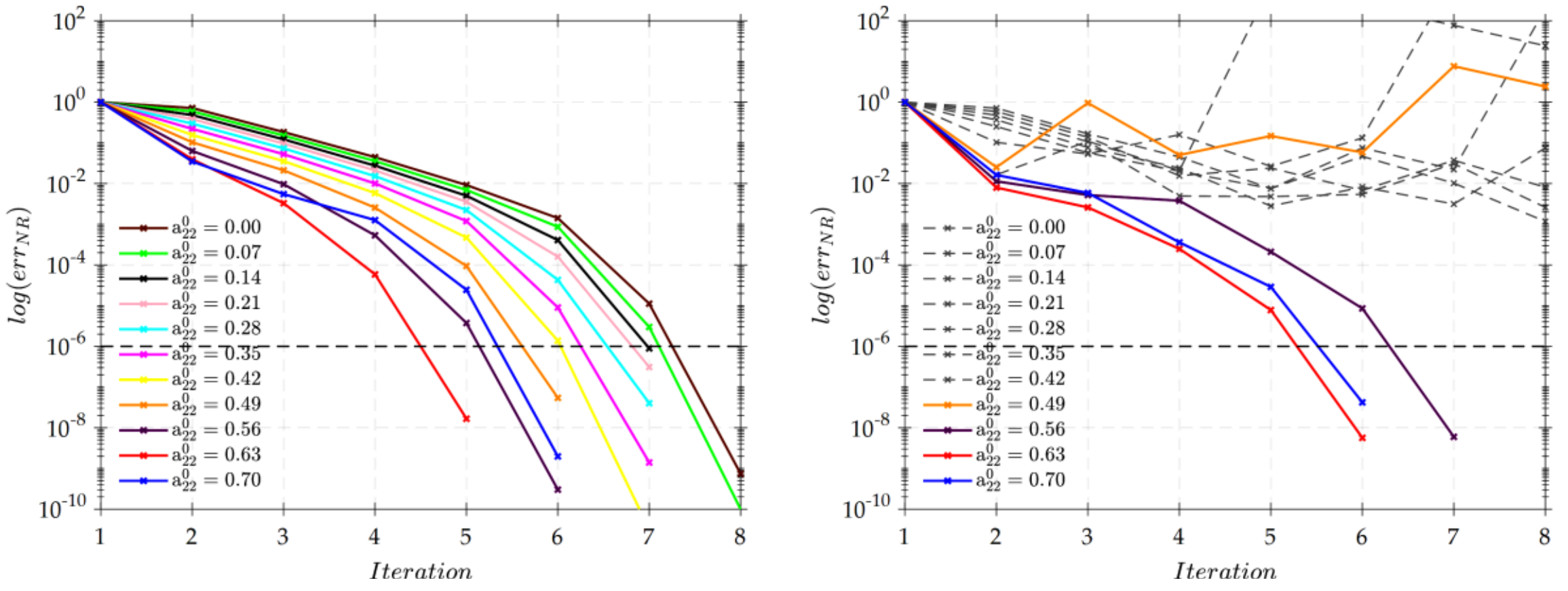

(a) (b)

Figure 6.7. NR residual norm for different initial orientation states, $a_{ij}^0$ and for (a) AT (b) RSC, orientation tensor models.

For the purpose of visualization, we present typical 3D orientation distribution function (ODF) (cf. Figure 6.8) recovered from the deviatoric forms of the average steady state 2nd and 4th order fiber orientation tensors obtained from the NR solutions for two fiber interaction coefficients, ($C_I$ = 0.01 and $C_I$ = 1.0) using the AT model and based on the method of Onat and Leckie [19].

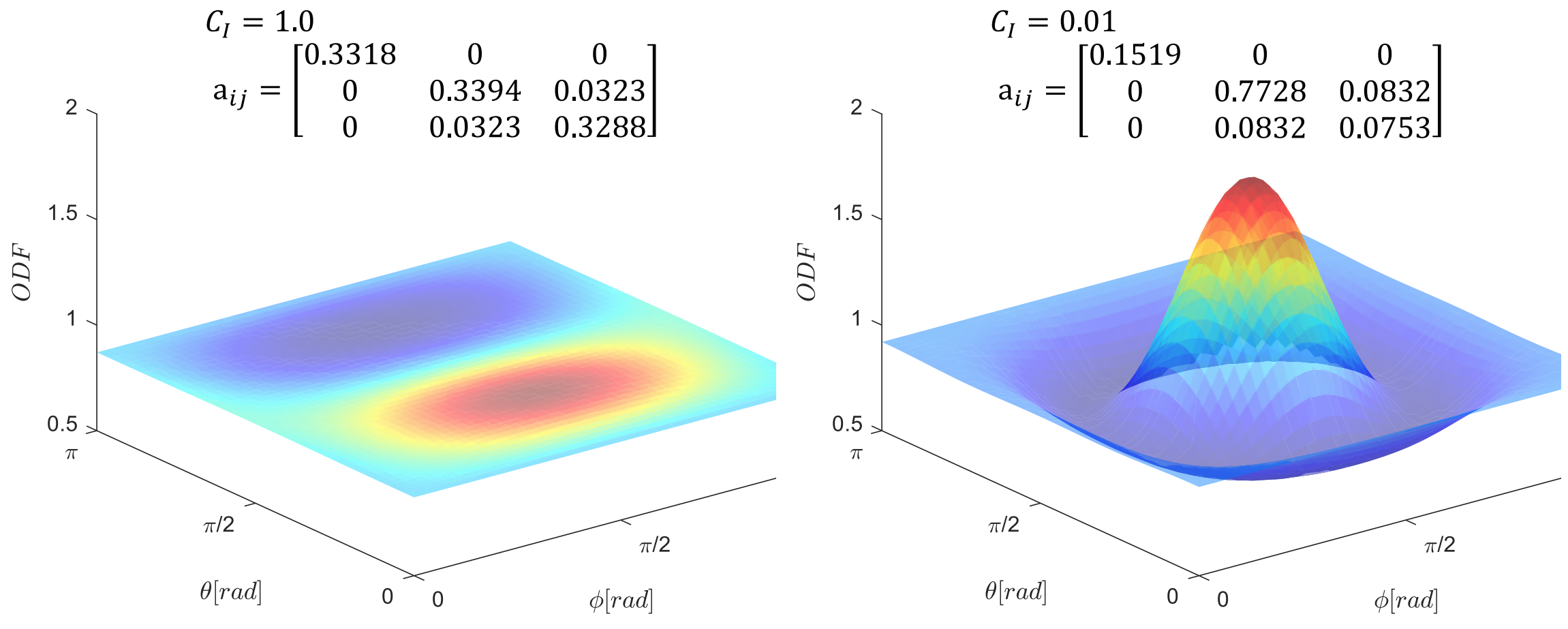

(a) (b)

Figure 6.8. Reconstruction of the orientation distribution function (ODF) for the steady-state orientation for fiber interaction coefficient of (a) CI=1.0, and (b) CI=0.01.

**Case 3:** The third case study investigates the performance of the NR method to coupled *iARD-RPR* and *pARD-RPR* orientation tensor models and include model parameters to simulate three different polymer composites materials in simple shear flow with $\dot{\gamma} = 1s^{-1}$. The model parameters used for the three coupled models and three polymer composites (i.e., M1: 40 wt. % glass-fiber/PP, M2: 31 wt. % carbon-fiber/PP and M3: 40 wt. % glass-fiber/nylon) considered in this case study appear in Table 6.13 and Table 6.14. We assume the same RK45 initial orientation state and NR initial guess as in case study 1 (cf. Eqn. (6.127)). The simulated RK45 results for the coupled orientation models and polymer composites considered here appear in Figure 6.9a - c. The percentage error between the NR and RK45 steady state values are given in Table 6.15. The results show negligible discrepancy in the calculated steady state solutions, generally yielding an error $err_1$ below 0.3% for all tensor components. Figure 6.10a, however, shows that the NR convergence rate is dependent on the model and composite material composition. The M2 (31 wt. % carbon-fiber/PP) with lower fiber content had faster convergence rate for all three coupled tensor models compared to the M1 and M3 composites with higher fiber content which had slower convergence rate. The coupled models showed slight instability through the initial NR iterations, especially for the iARD-RPR model with M1. The RK45 convergence profiles (cf. Figure 6.10b) is seen to have similar behavior as the NR convergence profiles in terms of the order of convergence for the various models and material types.

Table 6.13. ARD-RSC Parameters [284] for (a) M1 - 40 wt. % glass-fiber/PP (b) M2 - 31 wt. % carbon-fiber/PP (c) M3 - 40 wt. % glass-fiber/nylon

| | M1 | M2 | M3 |
|---|---|---|---|
| ƙ | 1/30 | 1/30 | 1/20 |
| $b_1$ | $3.842 \times 10^{-4}$ | $3.728 \times 10^{-3}$ | $4.643 \times 10^{-4}$ |
| $b_2$ | $-1.786 \times 10^{-3}$ | $-1.695 \times 10^{-2}$ | $-6.169 \times 10^{-4}$ |
| $b_3$ | $5.250 \times 10^{-2}$ | $1.750 \times 10^{-1}$ | $1.900 \times 10^{-2}$ |
| $b_4$ | $1.168 \times 10^{-5}$ | $-3.367 \times 10^{-3}$ | $9.650 \times 10^{-4}$ |
| $b_5$ | $-5.000 \times 10^{-4}$ | $-1.000 \times 10^{-2}$ | $7.000 \times 10^{-4}$ |

Table 6.14. Coupled *iARD-RPR* and *pARD-RPR* Model Parameters [284] for (a) model M1: 40 wt. % glass-fiber/PP (b) model M2: 31 wt. % carbon-fiber/PP and (c) model M3: 40 wt. % glass-fiber/nylon

| | M1 | M2 | M3 |
|---|---|---|---|
| $C_I$ | 0.0165 | 0.0630 | 0.0060 |
| $C_M$ | 0.9990 | 1.0100 | 0.9000 |
| $\Omega$ | 0.9880 | 0.9650 | 0.9000 |
| $\alpha$ | 0.9650 | 0.9650 | 0.9500 |
| $\beta$ | 0.0000 | 0.0000 | 0.0000 |

Table 6.15. Calculated errors for $a_{11}, a_{22}$ and $a_{13}$ steady-state values for iARD-RPR, iARD-RSC, pARD-RPR models and materials M1, M2 and M3.

| | iARD-RPR | | | pARD-RPR | | | iARD-RSC | | |
|---|---|---|---|---|---|---|---|---|---|
| | M1 | M2 | M3 | M1 | M2 | M3 | M1 | M2 | M3 |
| $a_{11}$ | 0.1039 | 0.0757 | 0.0876 | 0.0878 | 0.0278 | 0.0971 | 0.1107 | 0.0395 | 0.3231 |
| $a_{22}$ | 0.0550 | 0.0255 | 0.0219 | 0.0444 | 0.0131 | 0.0245 | 0.0553 | 0.0202 | 0.0678 |
| $a_{13}$ | 0.1286 | 0.1703 | 0.0401 | 0.0656 | 0.0270 | 0.0415 | 0.0872 | 0.0450 | 0.0235 |

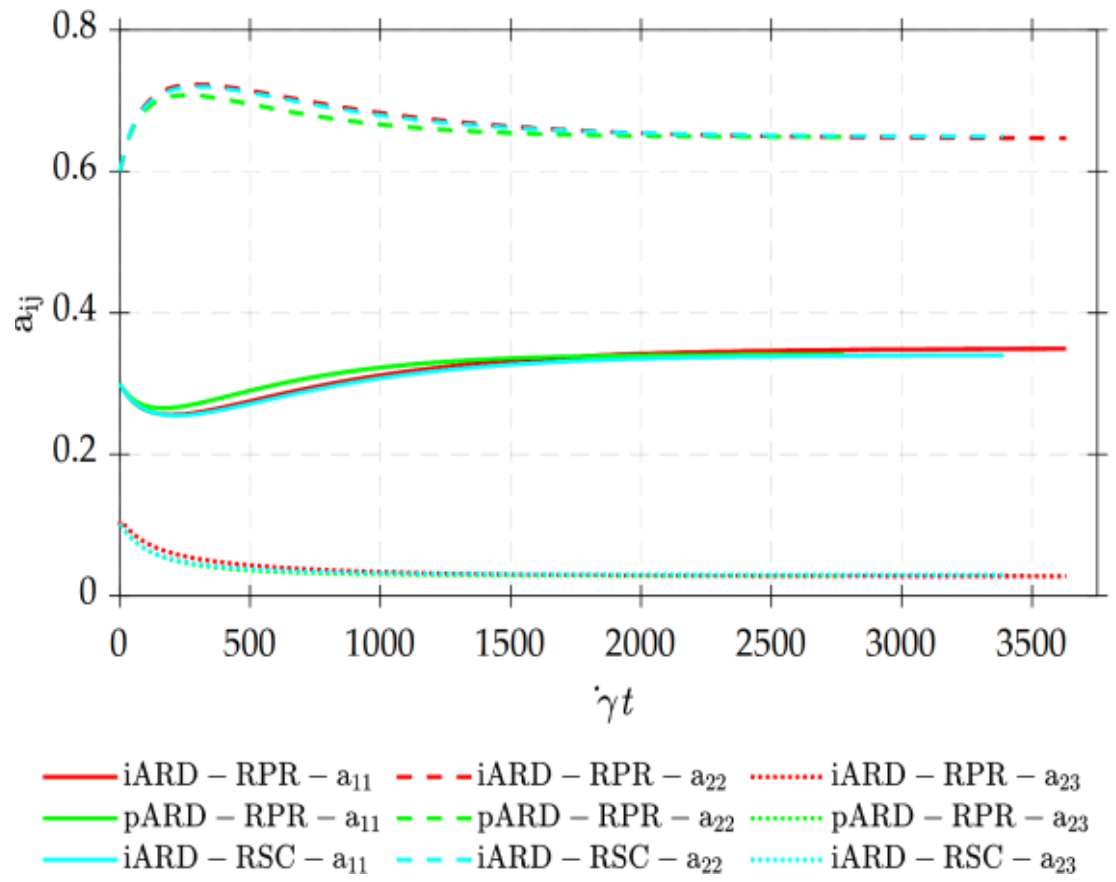

(a)

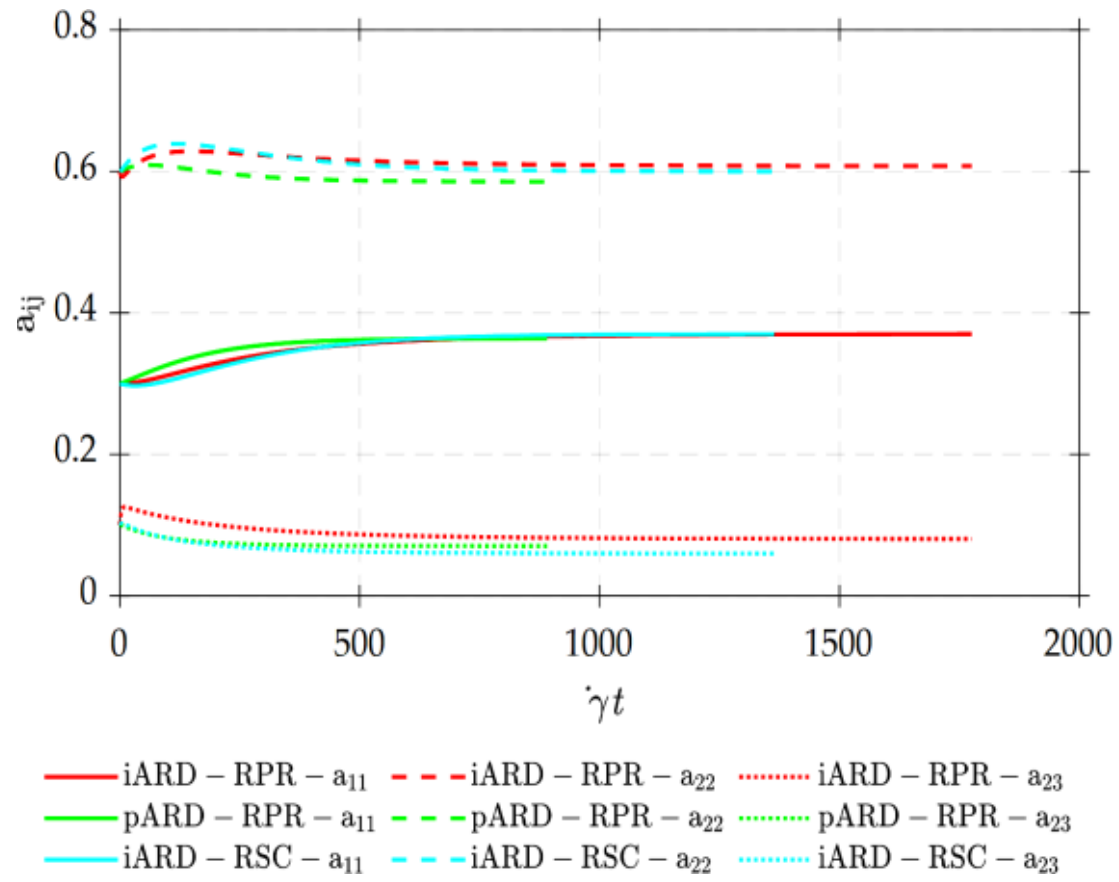

(b)

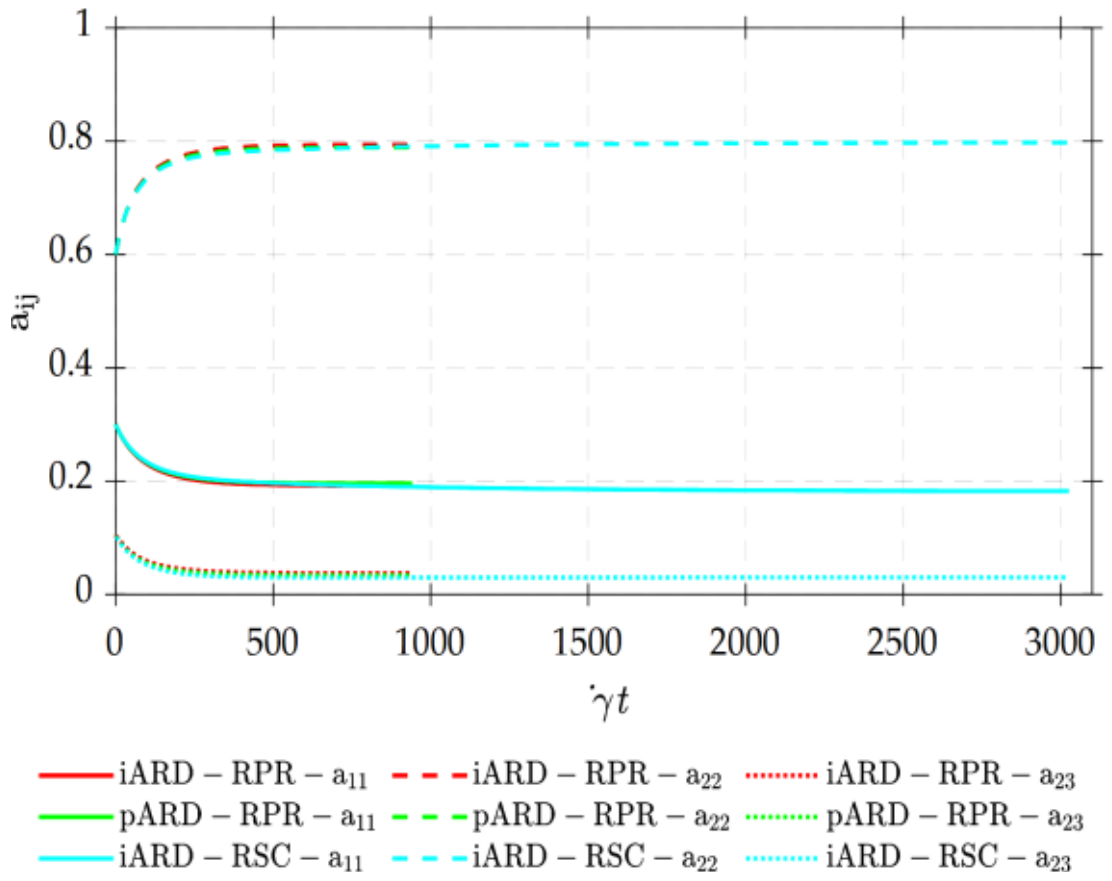

(c)

Figure 6.9. Time evolution of $a_{11}, a_{22}$ and $a_{13}$ for the iARD-RPR, iARD-RSC, pARD-RPR models using materials (a) M1 (b) M2 (c) M3.

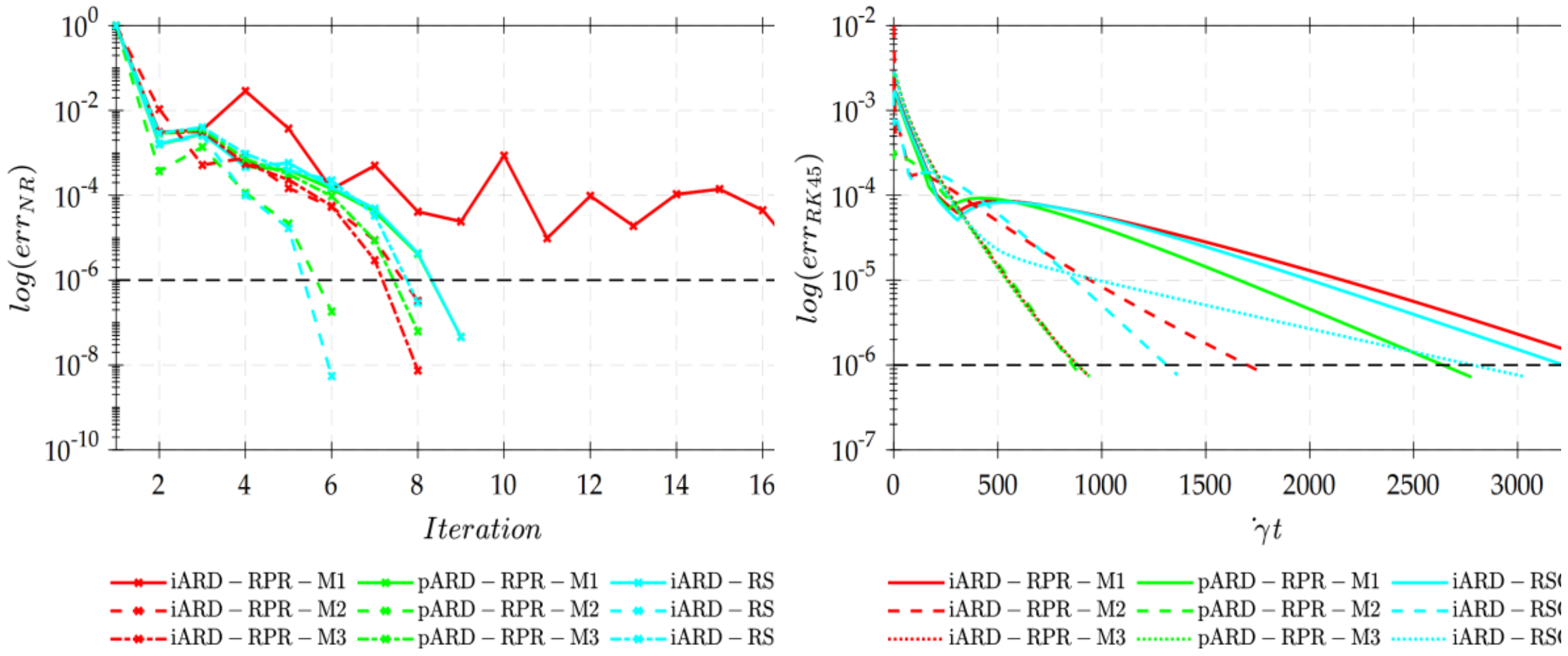

(a) (b)

Figure 6.10. Iteration history of the (a) NR residual norm (b) RK45 residual norm, for the various coupled models and polymer composites.

### *6.5.3 Performance of the NR method using various Closure Approximations*

The performance of the NR method in computing steady state orientation tensor component values for various closure approximations of the 4th order orientation tensor in terms of accuracy and stability is assessed in this section. For this assessment, we consider the AT model with a $C_I = 0.01$.The same initial orientation state for the RK45 method and the NR method initial guess used in the case 1 study of the preceding section (cf. Eqn.

(6.127)) is also employed and simple shear flow with $\dot{\gamma} = 1s^{-1}$ is used for all analysis. The first set of results are presented for Hinch and Leal closures in Section 6.3.3.4 and fitted closures in Section 6.3.3.5 and 6.3.3.6.

Figure 6.11a - d shows that transient response of the $a_{11}$ and $a_{12}$ from RK45 analysis through steady state. These results show that the choice of closure approximation affects the time taken to reach a steady state orientation. The model with the QDR and $HYB_2$ closures reached steady-state faster than other models while models with the LIN, ISO and SF2 closures required more time to reach steady-state. From Table 6.16 the NR method results tend to agree well with RK45 reference values, having a maximum $err_1 = 0.004\%$. Likewise, NR simulations using the fitted closure approximations performed well with a good accuracy in predicted results (cf. Table 6.16. ).

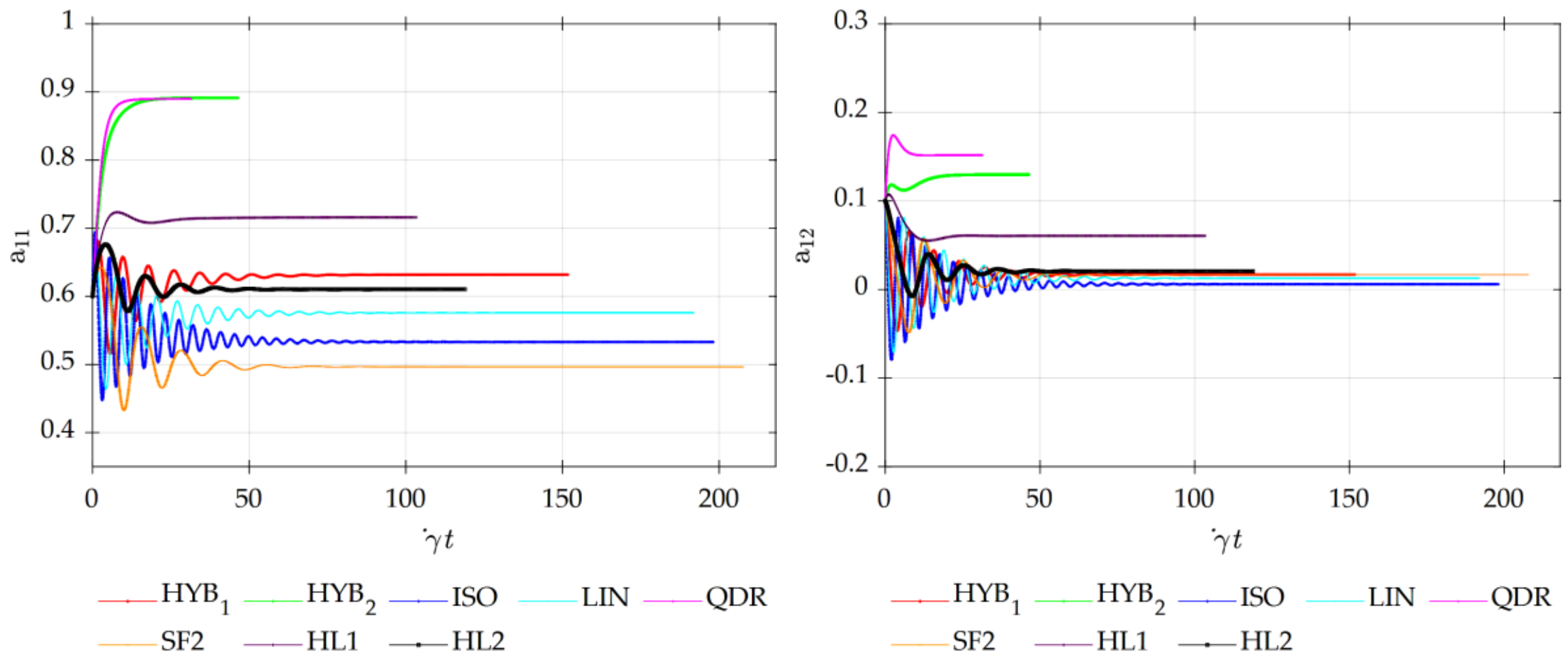

(a) (b)

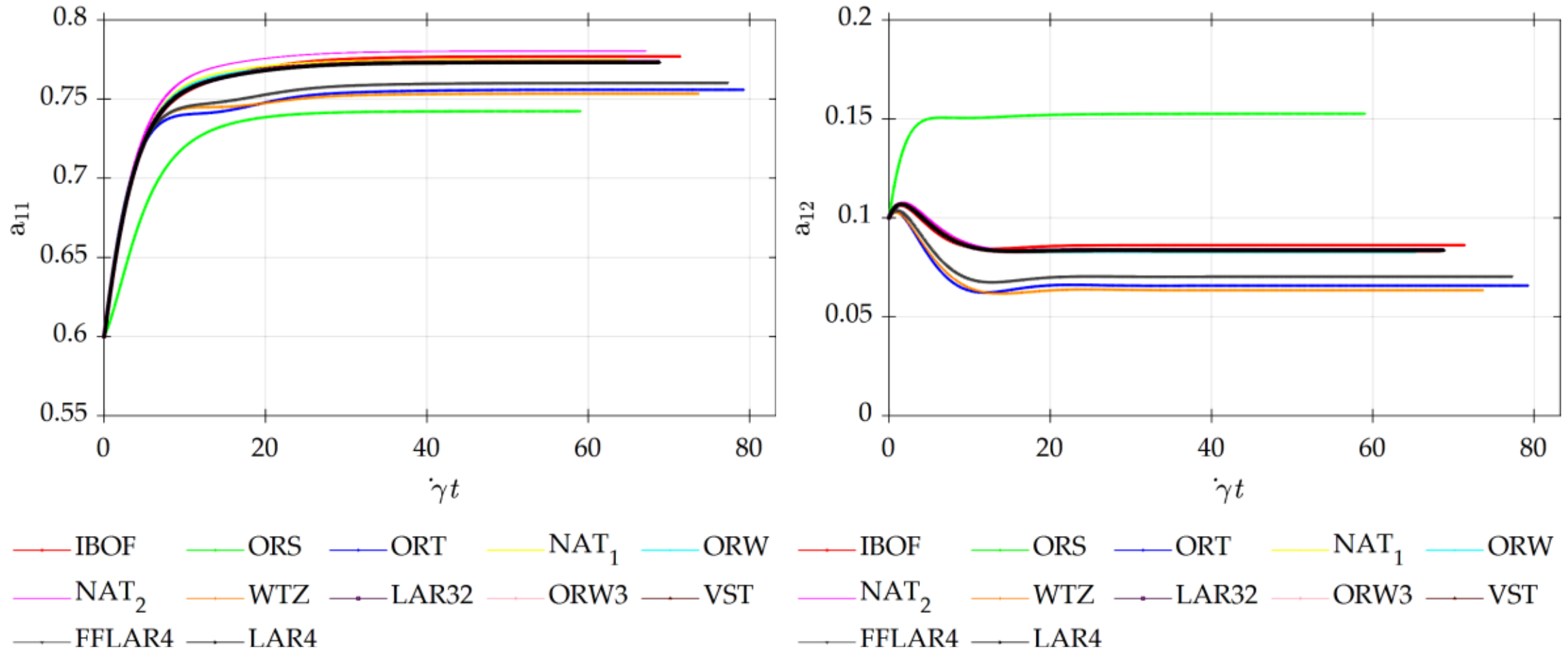

(c) (d)

Figure 6.11. Transient simulation results for (a) $a_{11}$ and (b) $a_{12}$ for the various Hinch and Leal closure approximations, and (c) $a_{11}$ and (b) $a_{12}$ for the fitted closures.

Figure 6.13a shows that the LIN and ISO closure had the fastest convergence rate with a quadratic nature, requiring only 3 iterations to converge followed by the HL1, HL2, QDR and HYB1 closures which required 6 iterations to converge. The invariant based HYB2 closure required an extra iteration to converge while the SF2 closure had the slowest convergence rate, requiring up to 10 iterations to converge. Surprisingly, all the fitted closure approximations had similar convergence behavior with only 5 iterations needed to converge except the IBOF closure that required an extra iteration to converge (cf. Figure 6.13b). Conversely, RK45 convergence using the Hinch and Leal closures (cf. Figure 6.13a) has a rather distinct behavior with the LIN, ISO and SF2 closures having the slowest convergence rate and the $HYB_2$ and QDR closures having the fastest convergence rate. Similar to the NR results that used fitted closures, the RK45 simulations converged over a narrower range (cf. Figure 6.13b) as compared to the Hinch and Leal closures.

Table 6.16. Calculated Error for $a_{11}, a_{22}$ and $a_{12}$ steady-state values for (a) Hinch and Leal closure approximations and (b) fitted closures approximations.

| | (a) | | | | (b) | | |
|---|---|---|---|---|---|---|---|
| | $a_{11}$ | $a_{22}$ | $a_{12}$ | | $a_{11}$ | $a_{22}$ | $a_{12}$ |
| $HYB_1$ | 0.0042 | 0.0016 | 0.0000 | IBOF | 0.0068 | 0.0000 | 0.0000 |
| $HYB_2$ | 0.0000 | 0.0000 | 0.0000 | ORS | 0.0000 | 0.0000 | 0.0000 |
| ISO | 0.0000 | 0.0000 | 0.0000 | ORT | 0.0061 | 0.0013 | 0.0000 |
| LIN | 0.0000 | 0.0000 | 0.0000 | $NAT_1$ | 0.0000 | 0.0000 | 0.0000 |
| QDR | 0.0000 | 0.0000 | 0.0000 | ORW | 0.0067 | 0.0000 | 0.0000 |
| SF2 | 0.0055 | 0.0000 | 0.0000 | $NAT_2$ | 0.0072 | 0.0000 | 0.0000 |
| HL1 | 0.0042 | 0.0014 | 0.0000 | WTZ | 0.0060 | 0.0000 | 0.0000 |
| HL2 | 0.0038 | 0.0000 | 0.0000 | LAR32 | 0.0066 | 0.0000 | 0.0000 |
| | | | | ORW3 | 0.0066 | 0.0000 | 0.0000 |
| | | | | VST | 0.0066 | 0.0000 | 0.0000 |
| | | | | FFLAR4 | 0.0062 | 0.0013 | 0.0000 |
| | | | | LAR4 | 0.0066 | 0.0000 | 0.0000 |

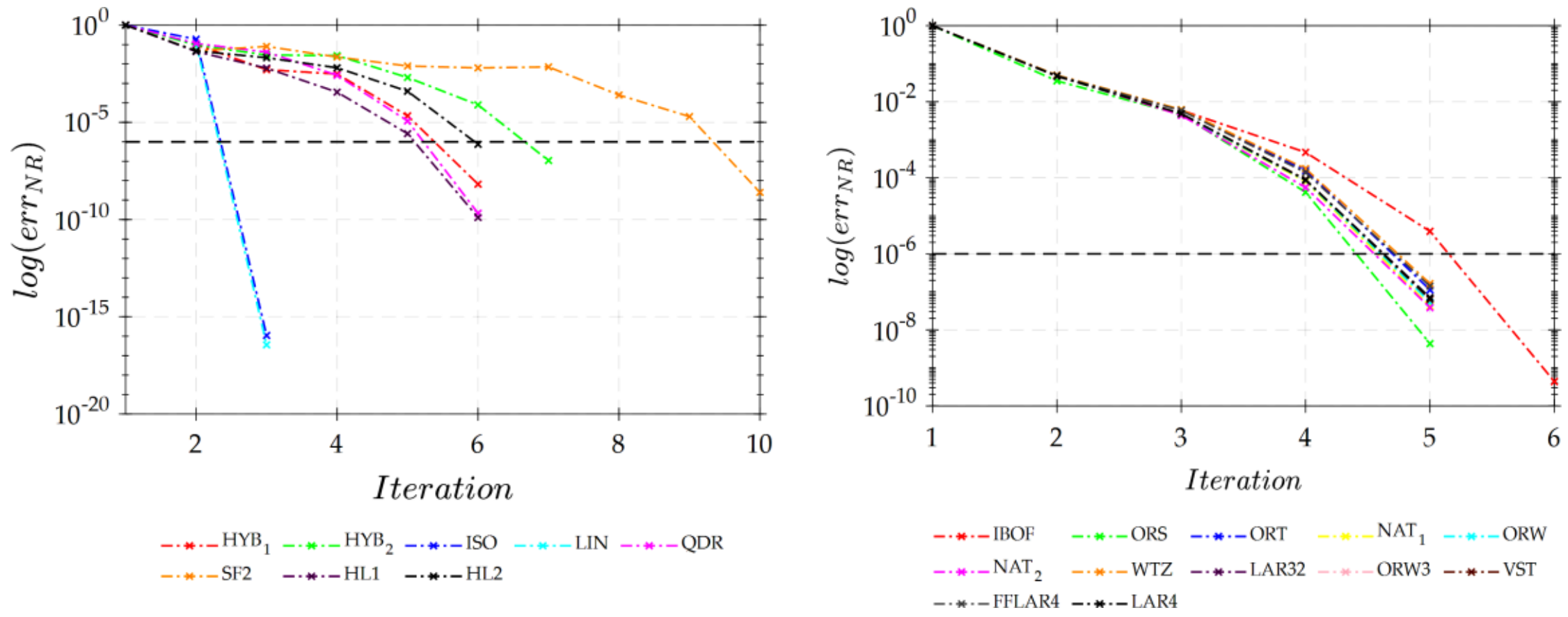

(a) (b)

Figure 6.12. Iteration history of the NR residual norm for the (a) Hinch and Leal closures and (b) polynomial fitted closure approximations.

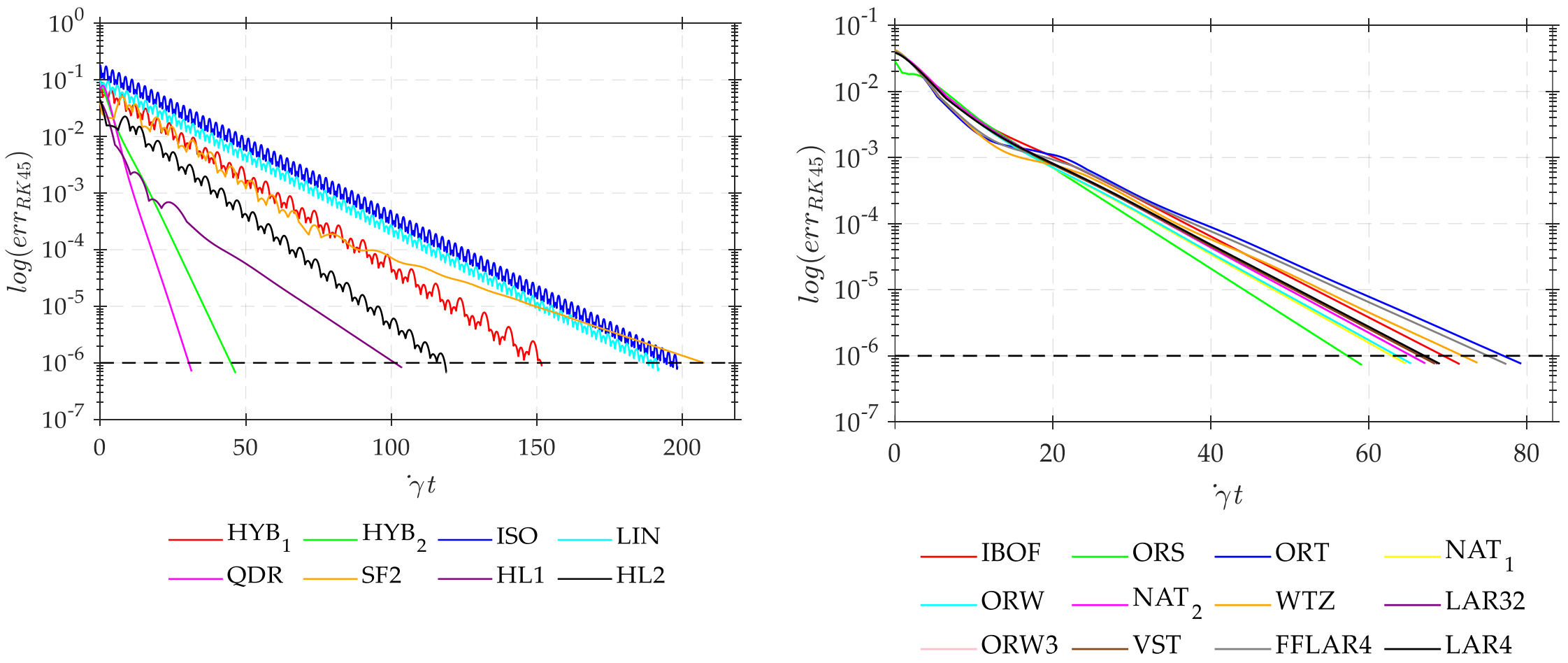

(c) (d)

Figure 6.13. Iteration history of the RK45 residual norm for the (a) Hinch and Leal closures and, (b) polynomial fitted closure approximations.

### *6.5.4 Homogenous Flow Considerations*

In this section we employ six homogenous flows to ensure the stability of our NR method beyond simple shear flow. The following flow conditions taken from [233] were considered:

- SS: Simple shear flow in the 2-3 plane, i.e. $L_{23} = \dot{\gamma}$
- SUA: Two stretching/shearing flow, including simple shear in 2-3 plane superimposed with uniaxial elongation in 3-direction, i.e. $L_{11} = -\dot{\varepsilon},\ L_{22} = -\dot{\varepsilon},\ L_{33} = 2\dot{\varepsilon}\ , L_{23} = \dot{\gamma}$. Two cases are considered, including balanced shear/stretch with $\dot{\gamma}/\dot{\varepsilon} = 10$ and dominant stretch with $\dot{\gamma}/\dot{\varepsilon} = 1$
- UA: Uniaxial elongation flow in the 3 - direction, i.e. $L_{11} = L_{22} = -\dot{\varepsilon},\ L_{33} = 2\dot{\varepsilon}$
- BA: Biaxial elongation (BA) flow in the 2-3 plane, i.e. $L_{11} = -2\dot{\varepsilon},\ L_{22} = L_{33} = \dot{\varepsilon}$
- PST: Two shear/planar-elongation flow, including simple shear in 2-3 plane superimposed on planar elongation in 1-3 plane, $L_{11} = -\dot{\varepsilon},\ L_{33} = \dot{\varepsilon},\ L_{23} = \dot{\gamma}$. Two cases are considered including balanced shear-planar elongation with $\dot{\gamma}/\dot{\varepsilon} = 10$ and dominant planar elongation with $\dot{\gamma}/\dot{\varepsilon} = 1$
- SBA: Balanced shear/bi-axial elongation flow, including simple shear in 2-3 plane superimposed on biaxial elongation, i.e., $L_{11} = -2\dot{\varepsilon}, L_{22} = L_{33} = \dot{\varepsilon}, L_{23} = \dot{\gamma}$. Two cases are considered for the shear to extension rate, $\dot{\gamma}/\dot{\varepsilon} = 1$ and $\dot{\gamma}/\dot{\varepsilon} = 10$

The natural orthotropic closures, i.e. exact midpoint fit - $NAT_1$ was employed for this study. The initial orientation state for the RK4 simulations and the initial guess $\underline{\underline{a}}^0$ for the NR iterations is

$$\underline{\underline{a}}^0 = \begin{bmatrix} 0.40 & 0.00 & 0.00 \\ 0.00 & 0.00 & 0.10 \\ 0.00 & 0.10 & 0.45 \end{bmatrix} \tag{6.128}$$

The NR method provided exceptional prediction accuracy of the orientation tensor components for all flow conditions considered where the percentage error for all orientation tensor components were zero to four decimals (i.e. $err_1 = 0.0000\%$). Figure 6.14a and b show transient results for $a_{33}$ and $a_{23}$ from RK45 simulations, respectively. The NR iteration history for the same (cf. Figure 6.14c) shows that higher shear to extension rate ratio improves the NR rate of convergence. The SS, $PST_2$, $SUA_2$, and $SBA_2$ flows converged faster than other flow types. Conversely, the flow types with the highest extension to shear rate ratio (UA and BA) had a slower rate of convergence. Alternatively, the RK45 solution (cf. Figure 6.14d) showed the higher extension to shear rate ratio enhanced the convergence rate, contrary to what was observed for the NR convergence history. The behavior of both NR and RK45 numerical techniques under the various flow and model considerations can provide useful insight when selecting a suitable numerical technique for solving a particular problem.

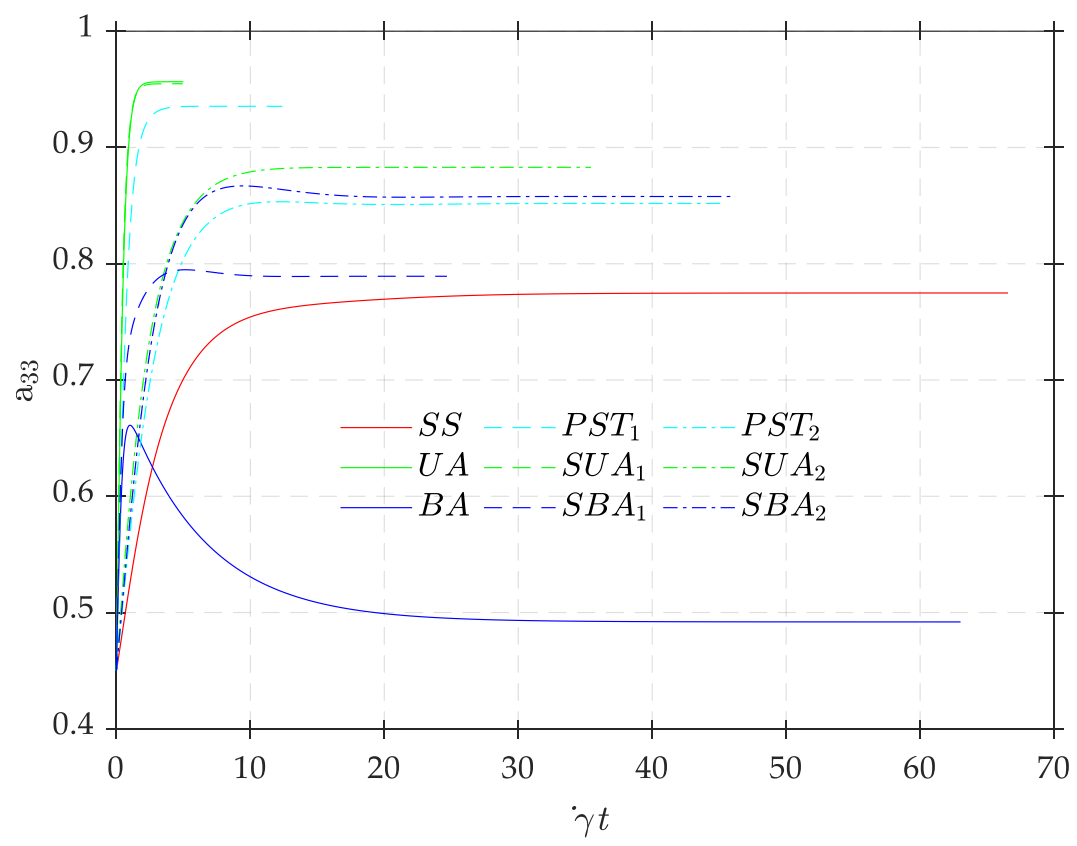

(a)

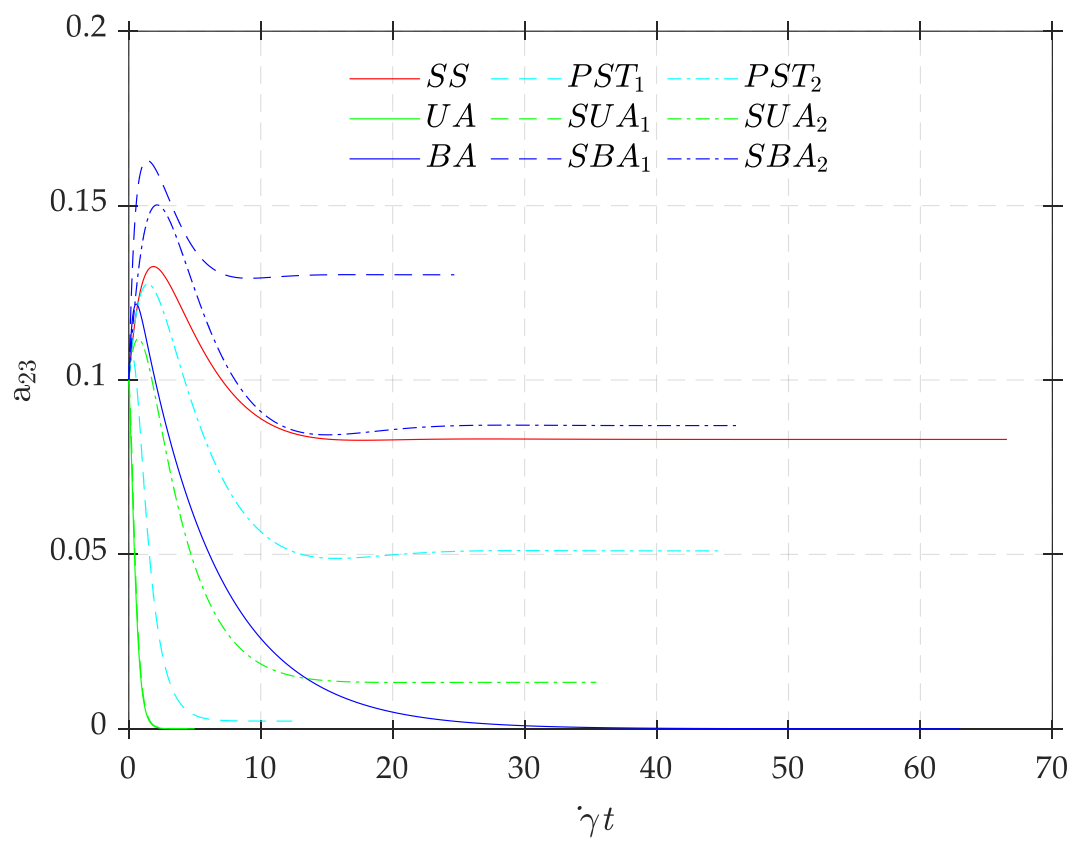

(b)

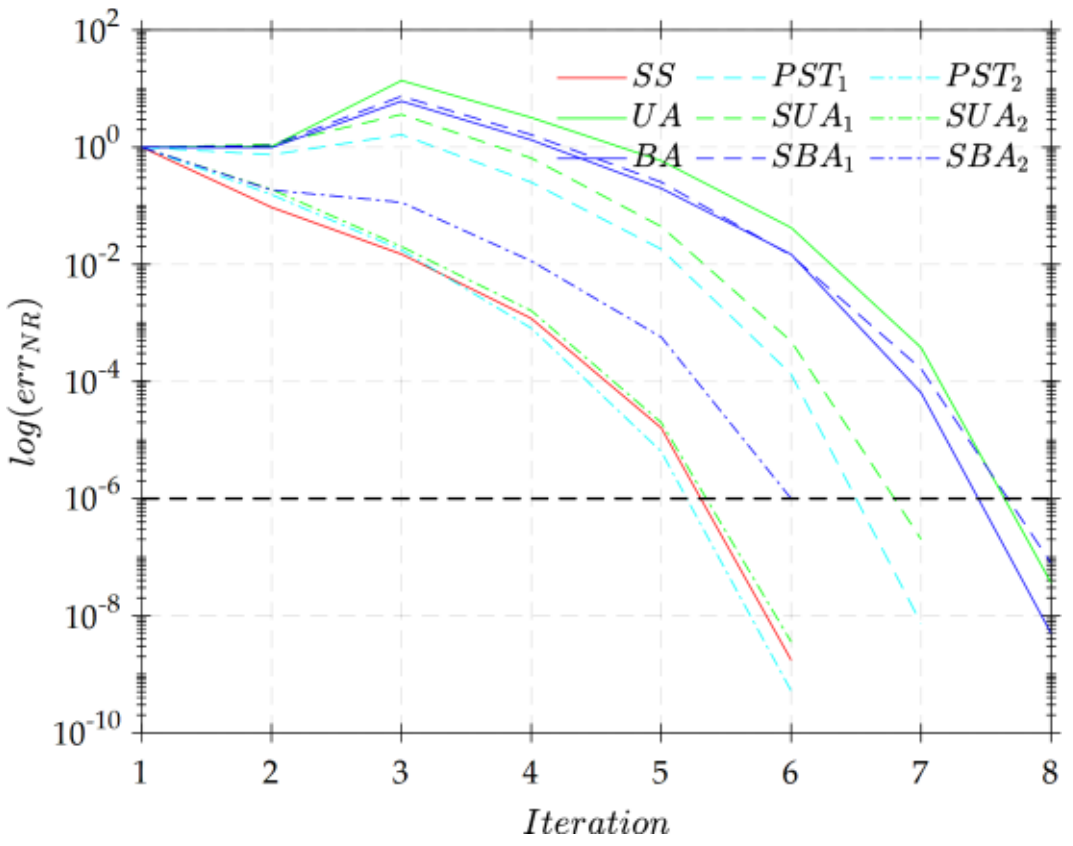

(c)

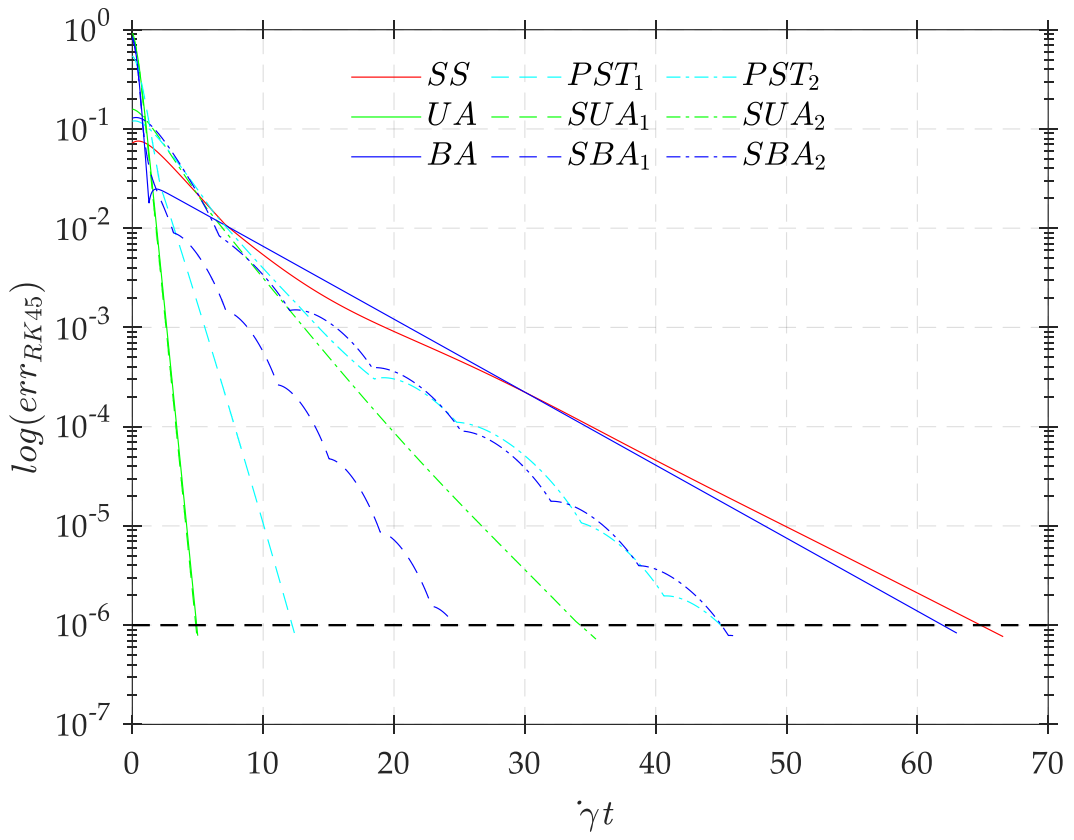

(d)

Figure 6.14. Transient responses of the 2nd order orientation tensor evolution for (a) component $a_{33}$ and (b) component $a_{23}$, residual norm history profile for the (c) NR method (d) RK45 method. Results are presented for the various flow considerations.

## *6.6 Conclusion*

A Newton-Raphson (NR) method has been derived and implemented for computing steady state 2nd - order fiber orientation tensors in homogeneous flows using exact 4th order Jacobians. Analytical expressions are provided for the Jacobians which are derived from partial derivatives of 2nd order fiber orientation tensor material derivatives taken with respect to the 2nd order fiber orientation tensor itself. Various macroscopic fiber orientation moment-tensor models and closure approximations of the 4th order fiber orientation tensor are included in the derivations, and the performance of the NR method in various homogenous flows have been studied. The computed exact Jacobians were validated with finite different calculations yielding an absolute error ranging from the order of $10^{-9} - 10^{-7}$. Results presented here show that the NR method accurately predicts steady state orientations for all tensor models, closure approximations and flow types considered in this paper. The NR iteration history, however, appeared to be sensitive to the selected tensor evolution model. The ARD models showed a slower NR convergence rate as compared to the IRD models. Also, the NR iteration history was seen to be sensitive to the material properties of the underlying composite material. The lower fiber content composite had the fastest NR convergence rate as compared to higher fiber content composites. The NR convergence also depended on the choice of $4^{th}$ order closure approximation. The linear (LIN) and isotropic (ISO) closures yielded the fastest NR convergence rate while all fitted closures had similar convergence behavior. The strong flow Hinch and Leal closure (SF2) exhibited the slowest NR iteration convergence rate. Additionally, the NR convergence was seen to be sensitive to the flow type and extension to shear rate ratio. Lower extension to shear rate ratio with higher shear rate dominance was observed to increase the NR convergence rate. Like any typical application of the NR root finding method, a good initial guess of the steady state orientation is required to avoid non-physical values. The NR method was observed to be comparatively faster as compared to the RK45 method for computing steady state values of the second order orientation tensor in all examples considered.

# CHAPTER SEVEN

## 2D Multi-Scale Extrusion-Deposition Polymer Composite Melt Flow Process Simulation

Sections of this chapter are taken from: Awenlimobor, A., Smith, D.E. and Wang, Z., 2024. Simulation of fiber-induced melt pressure fluctuations within large scale polymer composite deposition beads. Additive Manufacturing, 80, p.103980.

The phenomenon of heterogenous micro-void segregation at the interface between fiber and matrix during polymer melt flow processing has been shown to be significantly influenced by the local surrounding fluid pressure [9], [12], [13], [14], [15], [16], [17]. Note that in the moisture induced void nucleation mechanism [9], [12], [13], [14], [15] and the restrained volumetric shrinkage mechanism [16], [17], the onset of void nucleation occurs once the local fluid pressure drops below a critical value. Once micro-voids nucleate, their growth is driven primarily by the pressure difference between the micro-void internal pressure and external pressure in the surrounding fluid [5], [11]. Simulating the local pressure distribution around the fiber's surface during polymer processing can provide useful insight into the underlying mechanisms responsible for void formation especially at the tips of fiber where they are observed to mostly segregate. [57], [235].

Fiber suspension simulation, particularly those performed for polymer composite melt extrusion-deposition processes, have almost exclusively focused on fiber orientation and spatial distribution within the microstructure. However, little attention has been given to micro-void formation and evolution during polymer composite extrusion-deposition process or to understanding how the suspended fibers influence micro-void development. The flow of polymer-melt through the nozzle during typical EDAM processing is characterized by a complex combination of shear and extensional flows that are dependent on temperature, the viscoelastic polymer melts rheology and the geometry of the extruder nozzle. High shear rates tend to occur on the nozzle walls and the flow is more uniaxial elongation at the nozzle centreline [140], [314]. The shear dominant flow condition has been shown in [57], [235] to be responsible for creating pressure extremes at the fiber surface where heterogenous void nucleation likely occurs. The main objective of this chapter is to present a computational approach aimed at understanding mechanisms that may promote moisture/volatile induced micro-void nucleation on or near suspended fibers within the bead microstructure produced by polymer extrusion-deposition process using a multiscale modelling methodology. While our approach would be applicable to both filaments based FFF and LSAM systems and other extrusion-based processes, our focus here is on the large-scale polymer composite deposition. In the macroscale model, we develop a two-dimensional (2D) planar flow model for predicting the global flow-field and fiber orientation distribution within the polymer melt during the extrusion-deposition process. Then a micro-scale model is developed following the approach of Chapter Five and presented in Zhang et al. [230], [234], [265] which is based on Jeffery's model assumptions for suspended particles [21]. We simulate the evolution of a single ellipsoidal fiber along streamlines of the polymer melt flow through the nozzle and onto the print platform utilizing the field responses (velocity, velocity gradients and pressure) obtained from the macroscale model which defines boundary conditions in the micro-model. Then, a single fiber's translational and rotational velocities are computed by zeroing the net hydrodynamic forces and torques on the fiber's surface where its orientation and evolution along the flow path are updated based on an explicit iterative numerical algorithm which incorporates velocities and pressures from the macro-model. The micro-model is validated by comparing results of fiber motion and pressure distribution on the fiber surface with Jeffery's analytical model equations [21] for the motion of a single particle suspended in purely viscous shear flow. We account for rotary diffusivity due to short-range fiber-fiber interaction in the micro-model FEA simulation by determining an effective fluid domain size that mitigates Jeffery's rotation to match that predicted by the Advani-Tucker fiber orientation evolution equation. We also consider the fiber's evolution along various flow paths based on a given set of random initial fiber conditions to determine pressure bounds on the fiber surface across the melt flow. The pressure distribution on the fiber's surface as it travels along streamlines through the LSAM nozzle and onto the print bed, particularly within the regions of die swell at the nozzle exit, provides insight into a potential mechanism that could promote micro-void formation within printed beads. Knowledge of the relationship between process operating parameters and void formation and evolution can be used to control the quality of printed parts [5], [40].

### *7.1 Multiscale Model Development*

A multiscale modelling approach is developed in this work to better understand micro-void initiation within the beads printed with the LSAM extrusion-deposition process. The computational method here includes a macro-scale model which is used to calculate velocities and pressure along streamlines from the polymer melt flow solution in the extrusion-deposition process, and a micro-scale model which simulates the motion of a single rigid ellipsoidal particle based on the fluid flow solution along the macro-model streamlines. Our approach is a one-way coupling where computed velocities and pressures calculated along macro-model streamlines serve as inputs to define boundary conditions in the micro-model. A Newtonian fluid is assumed in both models. The material properties of the polymer melt employed in this study are taken from Heller et al. [23] and Wang et al. [24] which include a density of $1154\ kgm^{-3}$ and kinematic viscosity of $817 Pa \cdot s$ (i.e., 13% by weight carbon fiber filled ABS at 230ºC with a shear rate of $100\ s^{-1}$). In all of the discussion to follow, a 'fiber' is a rigid two-dimensional ellipsoidal solid having an aspect ratio of $r_e = И_1/И_2$ where $И_1$ and $И_2$ are the lengths of the major and minor ellipsoidal axes.

### *7.2 Macroscale Model - Planar deposition flow simulation*

A typical extrusion-deposition process of fiber filled polymer through a LSAM extrusion nozzle and the subsequent single bead deposition on a translating substrate is shown in Figure 3.1. The internal nozzle geometry used in this study is based on the Strangpresse (Strangpresse, LLC, Youngstown, Ohio, USA) Model 19 LSAM single screw extruder nozzle where an annotated schematic representation of its internal nozzle geometry appears in Figure 3.1. The 2D planar flow domain consists of the internal nozzle geometry region and a single bead layer deposited on the substrate that translates laterally with respect to the nozzle. (cf. Figure 3.1a). The FEM formulation is briefly described here where additional modelling details of planar deposition flow can be found in Zhang, et al. [24].

The governing equations of mass and momentum conservation for polymer melt flow within the nozzle and the printed bead are defined by Stokes's equation (eqns. *(5.86)*-*(5.89)*) based on the assumptions of no inertia in the fluid, the polymer melt is a creeping flow with a low Reynolds number (i.e., Re<<1), and the polymer melt is an isothermal, incompressible, Newtonian fluid [24]. Note that eqn. *(5.89)* does not include the influence of fiber orientation on the deviatoric stress. The ANSYS Polyflow (Ansys, Canonsburg,

PA, USA) commercial software is used for the macro-model polymer melt flow extrusion-deposition analysis. Figure 7.1a illustrates the quasi-steady fluid domain and boundary conditions for the 2D polymer melt flow model. Using data from Heller et al. [23] and Wang et al. [24], [315], the average normal velocity of 24mm/s is prescribed at the nozzle inlet $\Gamma_1$, and the velocity of the moving substrate and deposited material is 101.6mm/s in the positive x-direction which is imposed on $\Gamma_4$ and $\Gamma_5$. A no slip boundary condition is imposed on the nozzle inner wall $\Gamma_2$ and a free-surface boundary condition is prescribed on the exposed surface $\Gamma_3$ of the deposited material. Figure 7.1b shows computed velocity streamlines that form between the nozzle inlet $\Gamma_1$ and the bead flow exit $\Gamma_5$. Also shown in Figure 2b are feature streamlines 4, 10, and 18 in addition to zones of interest 1, 2, and 3 to be discussed below.

For the non-Newtonian simulation, a shear-thinning fluid with a power law index of $n \approx 0.45$ and a consistency coefficient of $m \approx 10^4 Pa \cdot s^n$ is used. As $n$ approaches 1, the viscosity approaches the Newtonian value equal to the consistency coefficient corresponding to a shear-rate of unity. The computed streamlines and resulting velocity profile distribution across sections of the nozzle for the non-Newtonian studies in comparison to results of the Newtonian analysis [24], [315] are presented in Figure 7.1b. While the velocity profiles of the Newtonian analysis are parabolic in shape, the profiles of the non-Newtonian analysis are somewhat hyperbolic shaped with a velocity plateau towards the center tending towards a plug flow velocity distribution.

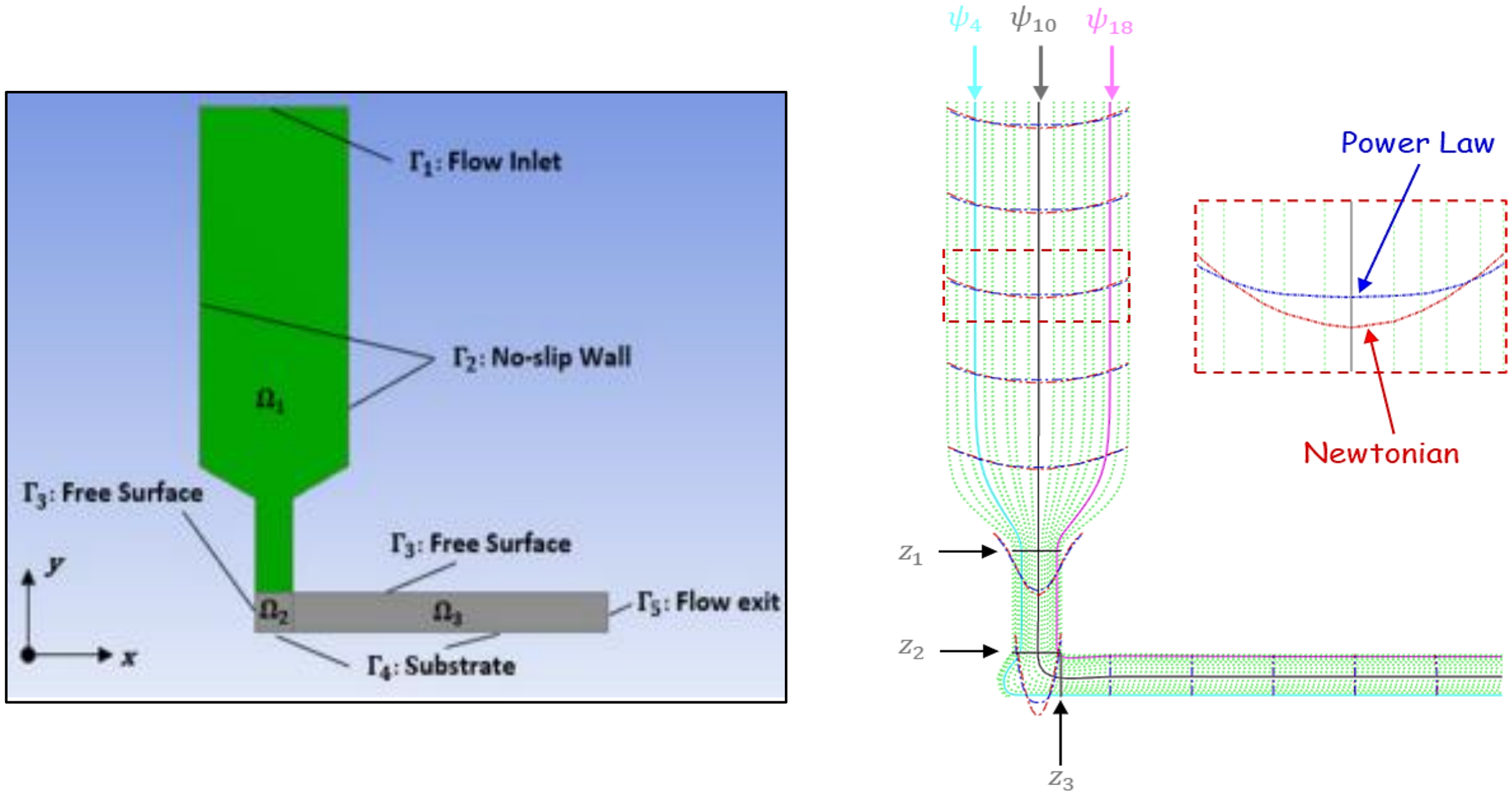

(a) (b)

Figure 7.1: 2D Planar extrusion-deposition flow model a) fluid domain and boundary conditions, b) velocity streamlines of the polymer flow through the nozzle with feature streamlines highlighted.

### *7.3 Microscale Model - Single fiber motion simulation*

Simulation of a rigid ellipsoidal fiber motion along streamlines of the polymeric melt flow is performed in this work using a custom FEM code developed in MATLAB (MathWorks, Natick, MA, USA). The single fiber micromodel is governed by Stokes's assumption of negligible inertia and negligible thermal effects and includes an isotropic homogenous Newtonian fluid that is the same as that used in the extrusion-deposition macro-model described above. Our algorithm for the micro-model simulation of a single 2D rigid ellipsoidal particle is derived from the work in Zhang et al. [230], [234], [265]. The flow domain for the 2D single fiber micro-model appears in axes (cf. Figure 5.3a) where we assume no slip occurs on the fiber surface and there is no flux across the fiber surface. Velocity, velocity gradient and pressure computed along streamlines of the extrusion-deposition macro-model described above are used to prescribe boundary conditions on the micro-model flow boundaries as a function of time. To impose these values in the micro-model, three essential boundary conditions are prescribed with respect to the fiber's local coordinate axes (cf. Figure 5.3b). FEM solutions are obtained by applying the essential boundary conditions to a fixed mesh which is rotated with the local fiber axes. Rotating the model in this manner significantly reduces computation time by maintaining a constant FEA system matrix, avoiding the need of remeshing the domain and/or recalculating the system matrix and its decomposed form at each iteration time step. The far-field velocities on the fluid domain boundary $\dot{X}_i^{BC1}$ (cf. Figure 5.3b) of the micro-model are defined from the streamline velocities $\dot{X}_i^{\psi}$ and velocity gradients $L_{ij}^{\psi}$ based on eqn. *(5.90)* which are obtained from the macro-model velocity solution at each time *t* of the single fiber evolution solution. Likewise, the prescribed pressure $p_{BC2}$ is defined according to eqn. *(5.94)* on a far-field node BC2 located on the fluid domain surface where its value is computed from the macro-model streamline pressure $p_{\psi}$. The prescribed velocities $\dot{X}_i^{BC3}$ on the fiber's surface are obtained in the usual manner according to the equation of rigid body motion (cf. eqn. *(5.93)*). The micro-model formulations are non-linear modifications to the model development by Zhang et. al. [230], [234], [265] and the governing equations are the same Stokes equations for mass and momentum conservation used in the macro-model given in eqns. *(5.86)-(5.89)*, based on the same assumption of isothermal, incompressible, homogenous viscous flow with a non-Newtonian power-law fluid definition. The microscale model development for the single fiber motion along the streamlines of the GNF polymer melt flow has been provided in detail in Chapter Five of this dissertation**.** Similar to the Newtonian analysis [57], [235], the instantaneous velocities, velocity gradients and pressure of the streamline data obtained from velocity solutions of the GNF macro-model analysis are used to derive the far-field fluid domain boundary conditions.

### *7.4 Non-Dilute Fiber Suspension Motion*

Jeffery's model assumes a Newtonian fluid and is valid for dilute suspension where fibers possess a relatively large radius of influence with neighboring fibers and contribute independently to the dissipation of energy in the form of a modified isotropic effective fluid viscosity $\mu^*$ for the suspension, such that $\mu^* = \mu\left(1 + \kappa\vartheta_f\right)$ [21], where κ is the modification factor dependent on the particles dimension which has been accounted for in our extrusion-deposition macro-model appearing above and $\vartheta_f$ is the volume fraction of the ellipsoidal fiber in the suspension. However, for semi-dilute and concentrated suspensions, there exists some degree of stochasticity in an individual fiber's behavior due to momentum diffusion and fiber-fiber interactions as the distance between neighboring particles becomes small relative to its size. In this case, neighboring fibers would introduce some degree of disturbance in a particle's surrounding fluid. As a result, particle-particle interaction necessitates a coupling effect between fibers. In other words, interactions between fibers reduce the effective radii of influence between near neighbors, the proximity of which results in an increased energy dissipation within each fiber's sphere of influence [22], [265].

As the fiber volume fraction and/or aspect ratio increases, collision of particles creates momentum transfer between colliding particles. Kugler et. al [22] classified fiber-fiber interaction into long-range and short-range hydrodynamic interaction, the latter of which can be further sub-divided into short range lubrication regimes, direct mechanical contact and a transition regime. As a result of momentum diffusion, the fibers eventually assume a steady state orientation that depends on the initial condition in accordance with the indeterminacy described by Jeffery. Folgar and Tucker [261] extended Jeffery's analysis by accounting for a collection of interacting suspended particles by incorporating a rotary diffusion term $D_r$. The rotary diffusion term $D_r$ is defined in terms of the scalar magnitude of deformation tensor $\dot{\gamma}$ according to $D_r = C_I\dot{\gamma}$, where $C_I$ is the interaction coefficient which is an empirical constant. Kugler et. al [22] gives a review of existing orientation models that accounts inter-particle interaction such as nematic model, anisotropic and mold flow rotary diffusion model, retarding principal rate model, etc.

To capture fiber-fiber interactions in our single fiber model, we develop a relation between the Folgar-Tucker interaction coefficient $C_I$ and the effective radius of influence in our micro-model (cf. Figure 7.2). Firstly, we determine a relation between the stall angle of the fiber and the interaction coefficient $C_I$ based on equation of change of the 2$^{nd}$ order orientation tensor by Advani and Tucker [19]. Here the stall angle is the fiber angle at which rotary motion ceases which has been found to be a function of the micro-model flow domain size (see e.g., Zhang et al. [265]). Subsequently we obtained a relation between the flow domain size and the fiber stall angle through a series of micro-model FEA simulations with fluid boundary domain BC1 of different sizes. As $\mathbb{m}$ decreases, the ends of the fiber become nearer to the prescribed boundary BC1 such that the velocity field near the fiber tips hydrodynamically interacts with the flow adjacent to BC1. The prescribed boundary creates a flow disturbance as viewed from the fiber in a manner similar to that which would be expected by neighboring fibers in a semi-concentrated flow. We then determine the relationship between the steady-state orientation tensor and the interaction coefficient $C_I$ for a given ellipsoidal aspect ratio. A relationship between $C_I$ and the micro-model flow domain size is then established by equating fiber stall angle in the micro-model to the direction of the first eigenvector of the fiber orientation tensor at steady state. This approach provides a means to approximately account for the effect of fiber-fiber interaction in the FEA simulation of the single fiber evolution along streamlines for a given interaction coefficient.

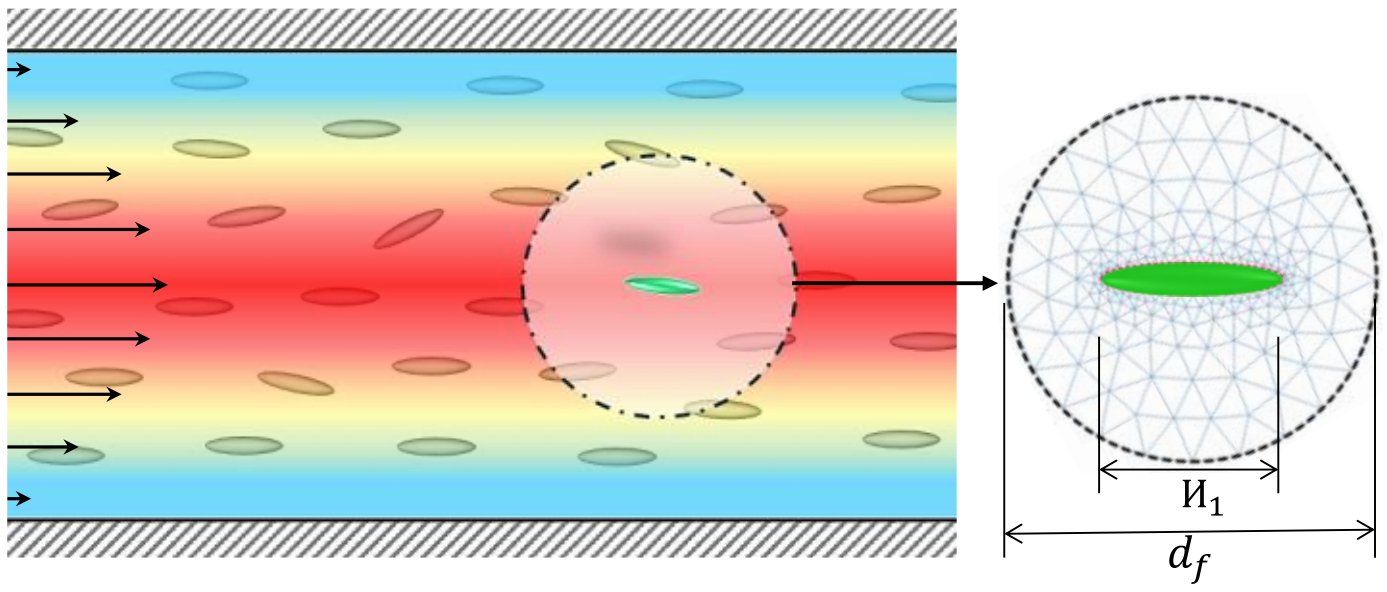

Figure 7.2: Schematic depicting the effective domain of influence around a single particle due to inter-particle hydrodynamic interaction.

Observations of the experimentally determined steady state orientation [181] show that the fibers tend to align with streamlines of the flow field irrespective of the initial conditions, contrary to Jeffery's idealization where suspended particles continue to rotate in simple shear. Saffman [181] shows that non-Newtonian properties of the fluid, not considered by Jeffery, is responsible for a stall in the tumbling motion. Other factors not accounted for in Jeffery's model that adds to the indeterminacy of a particle's motion include the flexural tendency of the particle which would depend on its inherent elastic property, aspect ratio, fluid rheology of the medium and interacting flow field. Moreover, the fibers may eventually break when subject to severe interacting forces, however, fiber flexibility is beyond the scope of our work.

### 7.5 Determining Effective Fluid Domain Size

To quantify the effect of fiber-fiber interactions with our single fiber model, we first establish a relationship between a suspension's interaction coefficient $C_I$ [19] and the stall angle in our single fiber FEA micro-model. The steady state orientation tensor values that correspond to a particular interaction coefficient can be determined from the Advani-Tucker 2nd order orientation tensor equation of change given as

$$\dot{\mathrm{a}}_{ij} = \frac{1}{2}\left(\Xi_{im}\mathrm{a}_{mj} - \mathrm{a}_{im}\Xi_{mj}\right) + \frac{\kappa}{2}\left(\Gamma_{im}\mathrm{a}_{mj} + \mathrm{a}_{im}\Gamma_{mj} - 2\mathrm{a}_{ijkl}\Gamma_{kl}\right) + 2D_r\left(\delta_{ij} - \alpha\mathrm{a}_{ij}\right) \qquad (7.1)$$

where, $\mathrm{a}_{ij}$ and $\mathrm{a}_{ijkl}$ are the 2nd and 4th order fiber's orientation tensors, respectively, $\kappa$ is the shape parameter defined above, $\Gamma_{ij}$ is the strain rate tensor given as $\Gamma_{ij} = \left[L_{ij} + L_{ji}\right]$, $\Xi_{ij}$ is the vorticity tensor given as $\Xi_{ij} = \left[L_{ij} - L_{ji}\right]$ and $\alpha$ is a dimension factor (i.e., $\alpha = 3$ for 3D orientation and $\alpha = 2$ for 2D planar orientation). In the above, the fourth-order orientation tensor $\mathrm{a}_{ijkl}$ is computed from $\mathrm{a}_{ij}$ using a closure approximation as is common in polymer composite suspension simulations. We employ the orthotropic fitted closure of Verweyst et al. [316] in all the calculations to follow. The symmetry properties of the orientation tensors require that $\mathrm{a}_{ij} = \mathrm{a}_{ji}$ and $\mathrm{a}_{ijkl} = \mathrm{a}_{jikl} = \mathrm{a}_{kijl} = \mathrm{a}_{lijk} = \mathrm{a}_{klij}$. The normalization condition also requires that $\mathrm{a}_{ii} = 1$ and $\mathrm{a}_{ijkk} = \mathrm{a}_{ij}$ where repeated indices imply summation in the usual manner here and in the following. We determine the steady state 2nd order orientation tensor that results in zero rate of change, i.e., $\dot{\mathrm{a}}_{ij} = \mathbf{0}$ via a Newton Raphson iteration scheme given as

$$\mathrm{a}_{ij}{}^{+} = \mathrm{a}_{ij}{}^{-} - J_{mnij}{}^{-} \backslash \Sigma_{mn}{}^{-} \qquad (7.2)$$

where the residual $\Sigma_{mn} = \dot{\mathrm{a}}_{mn}$ is

$$\Sigma_{mn} = \frac{1}{2}(\Xi_{mk}\mathrm{a}_{kn} - \mathrm{a}_{mk}\Xi_{kn}) + \frac{\kappa}{2}(\Gamma_{mk}\mathrm{a}_{kn} - \mathrm{a}_{mk}\Gamma_{kn} - 2\Gamma_{kl}\mathrm{a}_{mnkl}) + 2D_r(\delta_{mn} - \alpha\mathrm{a}_{mn}) \tag{7.3}$$

and the Jacobian $J_{mnij}$ is obtained by differentiating the residual with-respect-to components of the 2nd order orientation tensor $\mathrm{a}_{ij}$ as.

$$J_{mnij} = \frac{\partial\Sigma_{mn}}{\partial\mathrm{a}_{ij}} = \frac{1}{2}\left(\Xi_{mk}\frac{\partial\mathrm{a}_{kn}}{\partial\mathrm{a}_{ij}} - \frac{\partial\mathrm{a}_{mk}}{\partial\mathrm{a}_{ij}}\Xi_{kn}\right) + \frac{\kappa}{2}\left(\Gamma_{mk}\frac{\partial\mathrm{a}_{kn}}{\partial\mathrm{a}_{ij}} + \frac{\partial\mathrm{a}_{mk}}{\partial\mathrm{a}_{ij}}\Gamma_{kn} - 2\Gamma_{kl}\frac{\partial\mathrm{a}_{mnkl}}{\partial\mathrm{a}_{ij}}\right) - 2D_r\alpha\frac{\partial\mathrm{a}_{mn}}{\partial\mathrm{a}_{ij}} \tag{7.4}$$

The derivative of the 2nd order orientation tensor with respect to its individual components is simply

$$\frac{\partial\mathrm{a}_{rs}}{\partial\mathrm{a}_{mn}} = \delta_{rm}\delta_{sn} \tag{7.5}$$

where $\delta_{ij}$ is the Kronecker delta. Derivatives of $\mathrm{a}_{ijkl}$ with respect to $\mathrm{a}_{ij}$ are provided elsewhere for various closures approximations that are commonly used with eqn. (7.4) (cf. Awenlimobor and Smith [317], to appear). We define a preferred direction of orientation as the principal direction of the steady state $\mathrm{a}_{ij}$ computed from the *n^th^* eigenvector of $\mathrm{a}_{ij}(\Phi_{mn})$ corresponding to the maximum eigenvalue $\Lambda_n$ which is obtained from

$$\underline{\underline{\Phi}} : \mathbb{A}_{ij} = \Phi_{ki}\mathrm{a}_{kn}\Phi_{nj}, \qquad \Lambda_k = \mathbb{A}_{kk}, \qquad \underline{\Lambda} : \epsilon_{ijk}\left[\mathrm{a}_{ij} - \Lambda_n\delta_{ij}\right] = 0 \tag{7.6}$$

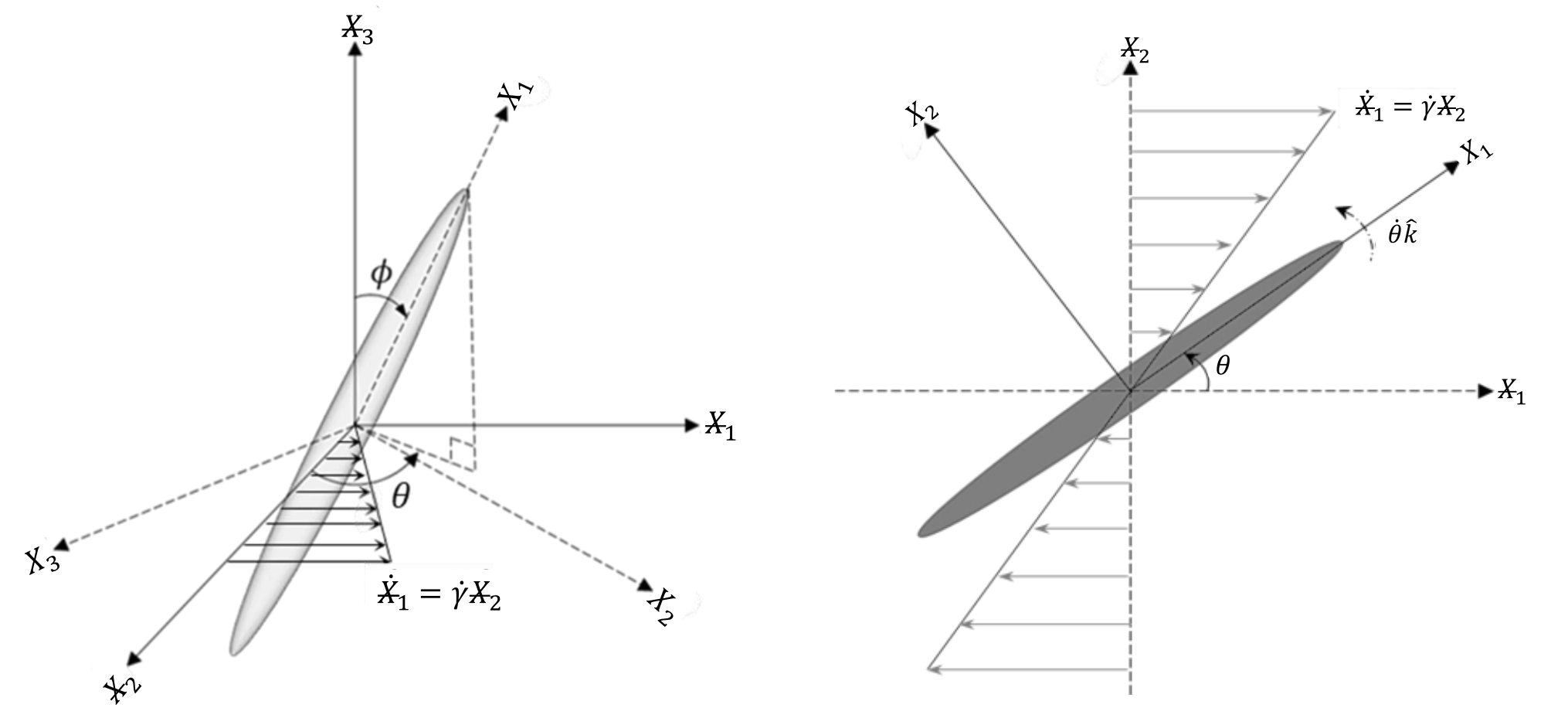

(a) (b)

Figure 7.3: Fiber Orientation Angles: a) 3D coordinates used in fiber orientation tensor equations and b) 2D coordinates used in single fiber motion simulations.

Consider planar simple shear flow having $\dot{X}_1 = \dot{\gamma}X_2$ and $\dot{X}_2 = \dot{X}_3 = 0$ (cf. Figure 7.3b) with a fiber at $\phi = 90°$ rotating in the $X_1 - X_2$-plane. For this flow field, the in-plane steady state orientation angle $\theta$ was evaluated using eqn. (7.1) through (7.6) for various values of $C_I$ and for different closure approximations as given in [317]. Alternatively, a series of FEA simulations were performed for an ellipsoidal fiber rotating through a modified Jeffery's orbit in simple shear for various fluid boundary domain sizes (cf. Figure 5.3). A corresponding pair of FEA simulation and orientation tensor evaluations were performed using the same fiber geometry and shear rate. Values of stall angle were then compared. Results of stall angle as a function of micro-model domain size factor $\mathbb{mm} = d_f/2\text{И}_1$ (where $d_f$ is the diameter of the micromodel flow domain) and $C_I$ appear in Figure 7.4.

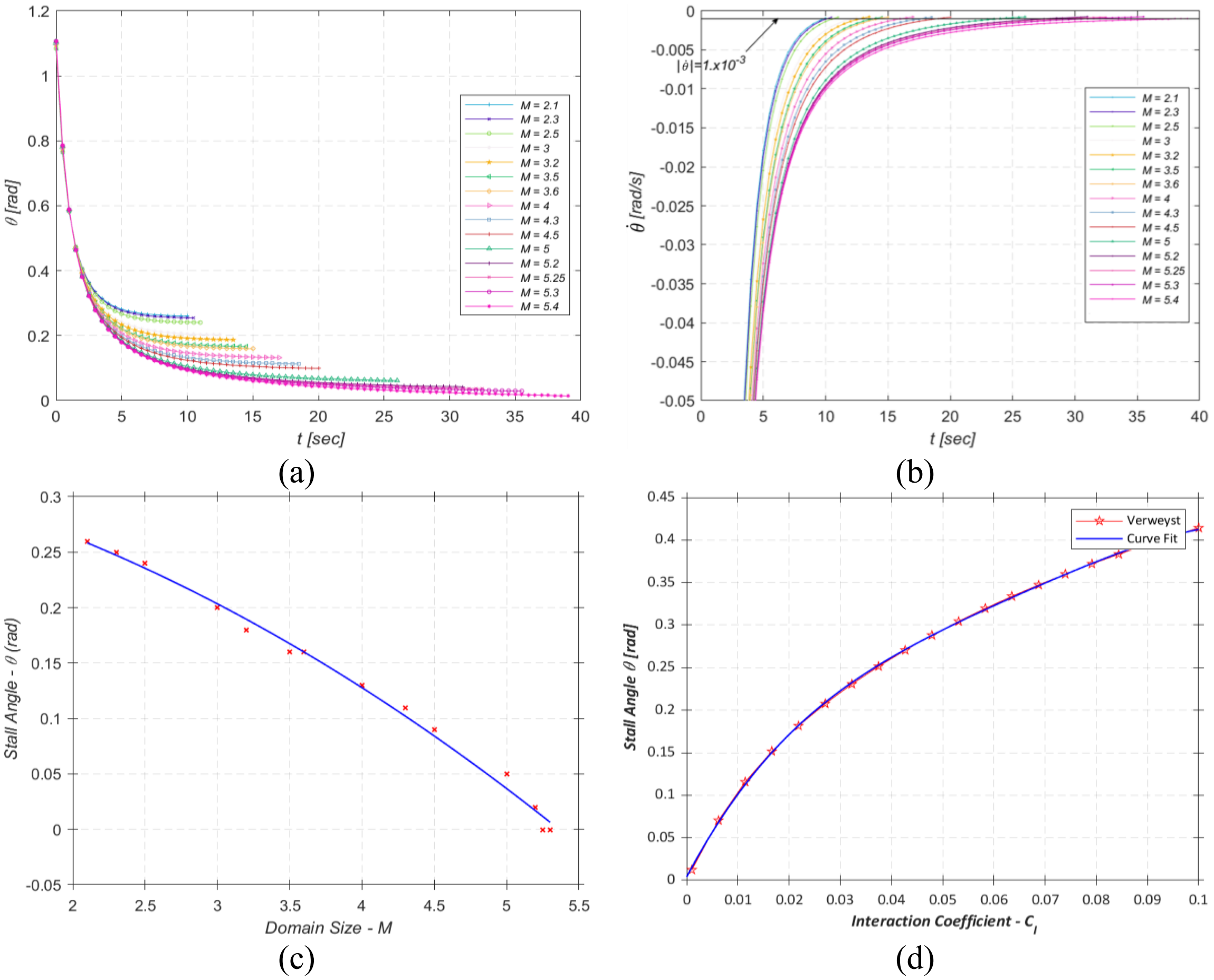

Figure 7.4: Single fiber angular motion and preferred orientation results for varying domain size $\mathbb{m}$ a) fiber orientation angle $\theta$ through its stall angle b) fiber angular velocity $\dot{\theta}$ simulated through a stall angular velocity tolerance of $|\dot{\theta}| = 1.\times 10^{-3}\ 1/s$ c) relationship between fiber stall angle and domain size factor $\mathbb{m}$ from FEA analysis, d) relationship between fiber steady state angle $\theta$ and interaction coefficient $C_I$ (Aspect ratio $r_e = 6$).

The influence of domain size appearing in Figure 7.4c shows a nearly linear relationship between the fiber stall angle and domain size from the micro-model simulations, given by eqn. (7.7) below.

$$\theta = 0.33839 - 0.022\mathbb{m} - 0.0077\mathbb{m}^2 \qquad (7.7)$$

Additionally, results of the orientation angle computed from the eigenvectors of the steady state orientation tensor $\mathbf{a_2}$ show nonlinear relationship between stall angle and interaction coefficient (cf. Figure 7.4d) which can be represented as

$$\theta = \pi/2 - 1.57 + 11.4C_I - 183.5C_I^2 + 1773.4C_I^3 - 6680.1C_I^4 \quad (7.8)$$

Combining results from Figure 7.4c&d, we obtain a relationship between the fluid boundary domain size in our single fiber micro-model and $C_I$ given as (cf. Figure 7.5)

$$\mathbb{m} = -1.4285 + \sqrt{45.89 - 1.48 \times 10^3 C_I + 2.38 \times 10^4 C_I^2 - 2.30 \times 10^5 C_I^3 + 8.68 \times 10^5 C_I^4} \quad (7.9)$$

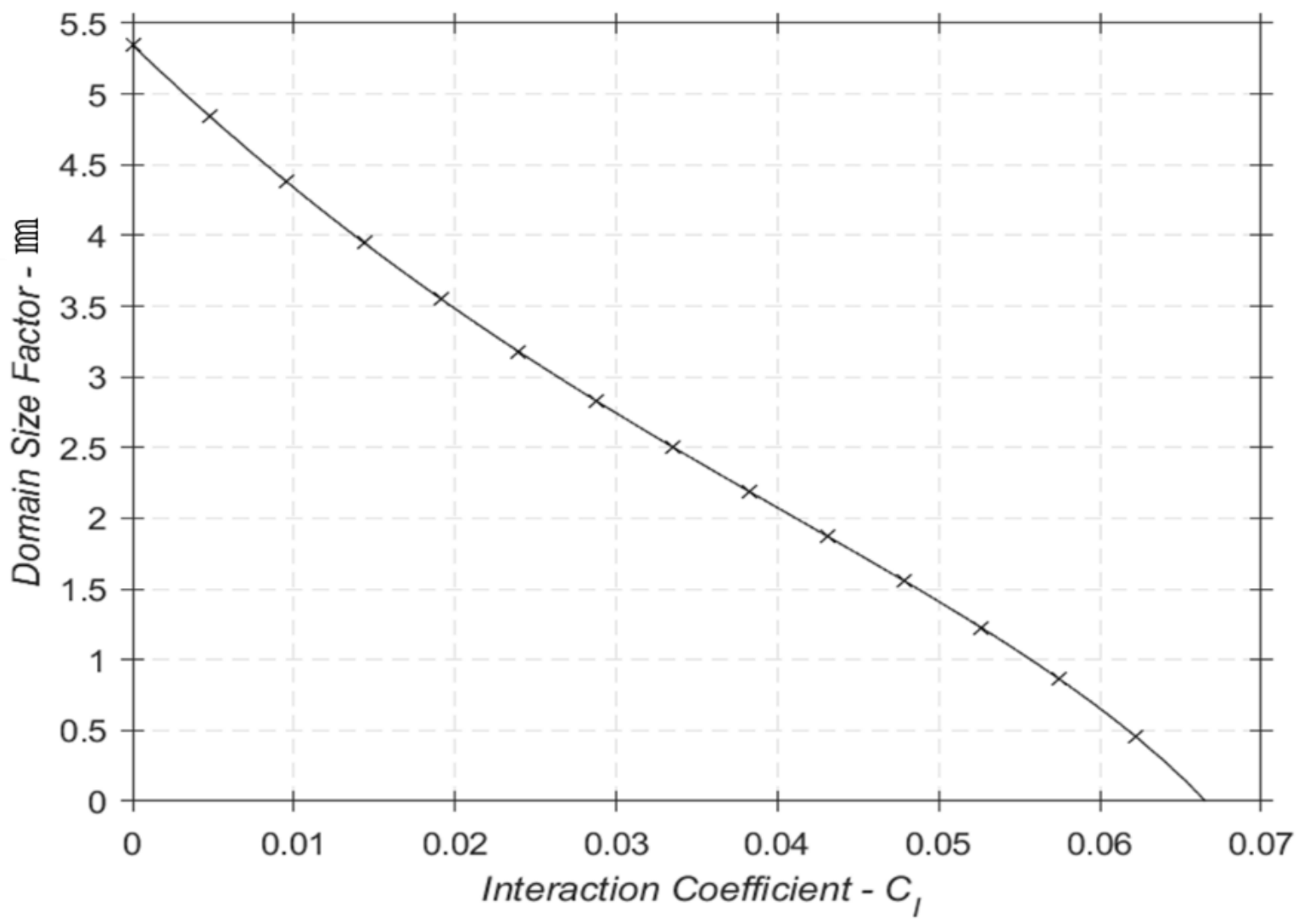

Figure 7.5: Fitted relationship between domain size factor $\mathbb{m}$ vs interaction coefficient $C_I$.

Assuming an ellipsoidal fiber aspect ratio $r_e = 6$ which corresponds to a shape parameter $\kappa = 0.9459$ and given a volume fraction $\vartheta_f = 8.4\%$ by volume (13% by weight) CF/ABS polymer composite, we obtain an interaction coefficient of $C_I = 0.0128$ using Bay's correlation that relates $C_I$ to $\vartheta_f$ and $r_e$ [279]. It follows from Equation (7.9), that the effective domain size based on our $C_I$ is $\mathbb{m} = 4.08$ (~4.0) which we have used in the Newtonian simulations. Given that fiber suspensions are classified into 3 concentration regimes based on $v_f$ and $r_e$ as [62], [277] our simulations are within or nearly within the concentrated regime for the suspension where $C_I = .0128$ and $\mathbb{m} = 4$ are used in the results section below.

## 7.6 Results and Discussion

### 7.6.1 Multi-Scale Newtonian Melt Flow Simulation

The result of the velocity magnitude $|\underline{\dot{X}}|$ and scalar magnitude of deformation tensor $\dot{\gamma}$ from the macroscale Newtonian analysis appear in Figure 7.6a & b, respectively. Computed velocities in Figure 3a show an increase in velocity magnitude from the edge of the nozzle to its center as expected. It follows that material along streamlines near the edges of the nozzle have a higher extrusion-deposition time compared to those closer to the center. The velocity contours (see for example, Figure 5 and 6 in Ref. [24]) show a parabolic velocity distribution across transverse sections of extruder nozzle except near the entrance and exit of the straight capillary portion of the nozzle. Melt flow in these transition regions is characterized by sharp transitions of velocity and velocity gradients along the inside wall of the extruder nozzle. Upon deposition onto the print bed, the melt flow attains a uniform velocity throughout the bead material where all stresses reduce to zero.

The plot of velocity gradient in Figure 7.7 shows unusually high values occurring at the sharp corners of the flow field due to singularities in the velocity solution where the polymer melt flow transitions from a no-slip to a free surface boundary condition, which we attribute a posteriori to be responsible for unexpected behavior of the fiber's motion along streamlines close to these locations. In this figure, as well as in all of the micro-model results, $\dot{X}_1$ and $\dot{X}_2$ are the components of the velocity vector $\underline{\dot{X}}$ in the $X_1$- and $X_2$ -directions, respectively. We see from Figure 7.7 that the velocity gradient component - $L_{12}$ dominates near the nozzle exit and is seen to increase in magnitude when moving outward from the center streamline towards those near the edge of the nozzle.

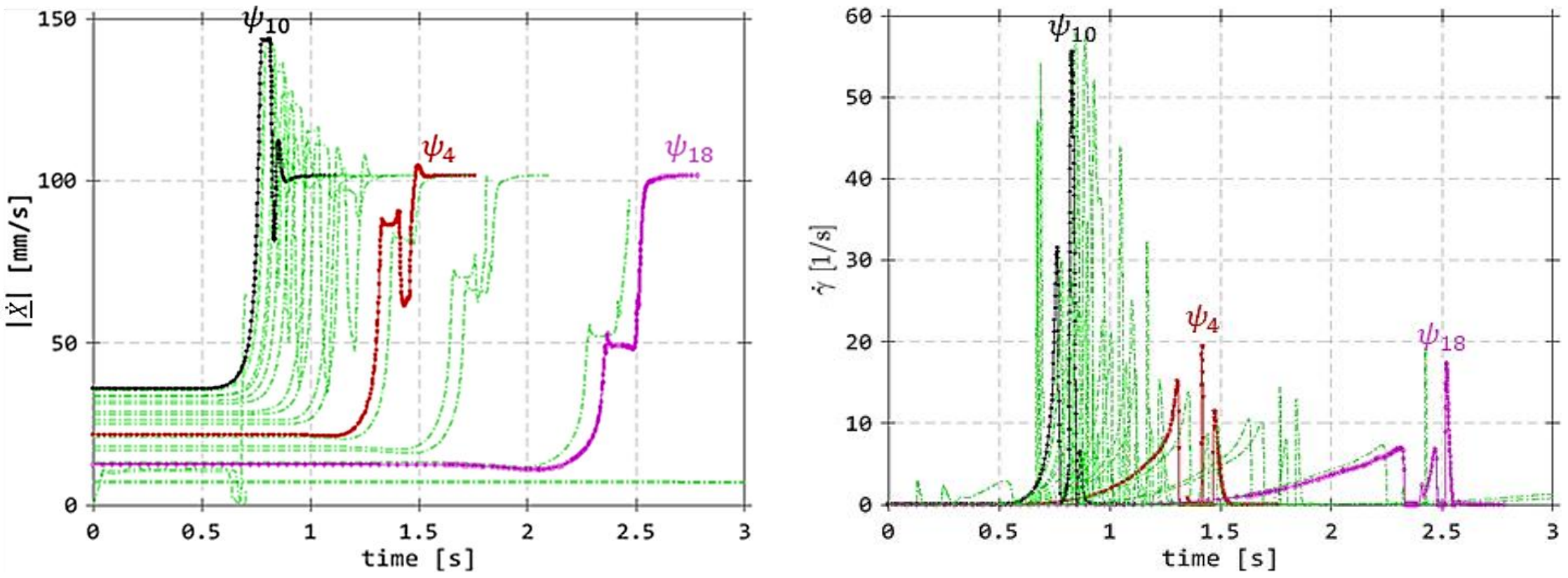

(a) (b)

Figure 7.6: a) Velocity magnitude $|\dot{\underline{X}}|$ b) scalar magnitude of second order deformation tensor for various streamlines with feature streamlines highlighted [5].

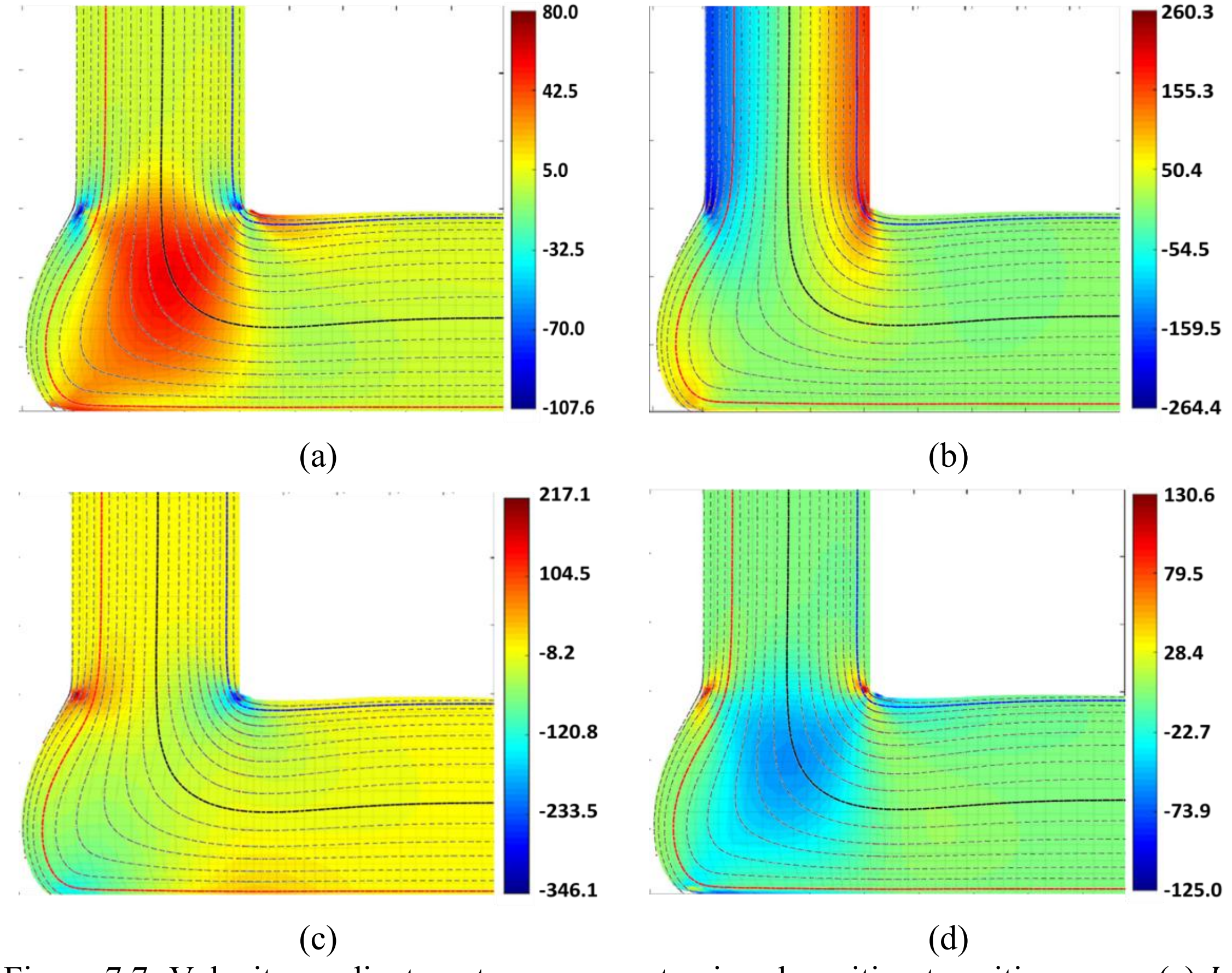

(a) (b)

(c) (d)

Figure 7.7: Velocity gradient contours near extrusion-deposition transition zones (a) $L_{11}$, (b) $L_{12}$, (c) $L_{21}$, (d) $L_{22}$ . The units of the global velocity gradients are $s^{-1}$.

[5] $\psi_n$ here refers to streamline identifier (n) and starts at 1 from the left edge of the nozzle increasing transversely to a maximum number of 22 at the right edge of the nozzle.

Chapter Five describes the ability of our micro-model to reproduce Jeffery's result for single fiber motion, and the determination of an effective single ellipsoidal fluid domain size that approximates the effect of short-range fiber interaction in simple shear flow. All simulations included here use a fiber half-length of $a = 42\mu m$ and an ellipsoidal aspect ratio of $r_e = 6$ which corresponds to a cylindrical geometric aspect ratio of $r_c = 7.66$ using Equation (2.21) in Zhang [265]. Here we limit our discussion to results along streamlines $\psi_4$, $\psi_{10}$, and $\psi_{18}$ to capture effects along the lower, middle, and upper sections of the bead, respectively (cf. Figure 7.1b). The following simulations incorporate velocity, velocity gradients, and pressure computed in the 2D planar extrusion-deposition macro-model to define far field boundary conditions BC1 and BC2 in the single fiber micro-model. To assess the effect of initial conditions in the single fiber analysis, we run multiple simulations, each with its own initial fiber angle $\theta_0$ over a range of $-{}^{\pi}/_{2} \leq \theta_0 \leq {}^{\pi}/_{2}$ in increments of ${}^{\pi}/_{12}$. Simulating fiber motion over this range of initial angles and on various streamlines provides a comprehensive assessment of possible fiber responses and corresponding location where they occur across the extruder nozzle. To better display streamline results, subsequent figures presented in this section have been annotated to show three interest regions of the nozzle geometry appearing in Figure 7.1b which includes:

(i) Zone 1: The entrance to the small capillary section of the nozzle at the point where the polymer- melt just exits the convergent zone.

(ii) Zone 2: The exit from the nozzle where the polymer leaves the nozzle and enters the region of die swell, and the external pressure drops to atmospheric condition.

(iii) Zone 3: The exit of die swell region below and to the side of the nozzle exit where the deposited material has made a complete $90^0$ turn onto the translating bed below and attains a near uniform velocity equal to the print speed.

We consider the simulation of fibers in a concentrated suspension with $C_I = 0.0128$ using the reduced single fiber domain approach with $\mathbb{m}$ = 4 in the micro-model as described above. For each fiber motion simulation result (i.e., a fiber moving along a specific streamline with a designated initial angle), the overall minimum and maximum fiber surface pressure is calculated and the difference between the streamline pressure and overall minimum and maximum fiber surface pressures are noted. In addition, the corresponding coordinate locations where the minimum and maximum fiber surface pressures occur within extrusion-deposition flow are identified. Figure 7.8 shows a typical fiber surface pressure result along streamline $\psi_{10}$ (starting at the centerline of the nozzle inlet) for a concentrated suspension where distinct extremes of minimum and maximum pressures identified as $\Delta P_{min}$ and $\Delta P_{max}$, respectively, are plotted as a function time along with the streamline pressure from the macro-model. The first extreme pressure location, denoted here as Loc. 1, and the second extreme location, denoted as Loc. 2, appear in the pressure history for all streamlines and $\theta_0$ with varying degrees of intensity and at slightly different locations as shown below. Note that the position along the streamline for Loc. 1 and Loc. 2 will occur at different locations depending on the streamline and initial fiber angle.

The initial extreme in minimum fiber surface pressure at Loc. 1 is observed to occur just prior to the entrance of the nozzle capillary section (i.e., zone 1) while the second pressure drop at Loc. 2 occurs within the die swell region between zones 2 & 3. Only at the latter extreme fiber location does the absolute local minimum pressure on the fiber surface drop to a value that is below zero atmosphere (reaching -0.4$MPa$ in the simulation appearing in Figure 7.8). This low pressure extreme is expected to provide a favorable condition for void nucleation to occur based on prior related research [9], [12], [13], [14], [15], [16], [17]. A closer inspection of the fiber's surface pressure distribution at this location shows that the peak sites occur at the fiber's tips (cf. Figure 7.8d) which is typical of all simulations presented in this work.

To gain a better understanding of the effect of streamline location on the fiber response during its motion through the extrusion-deposition flow in the concentrated regime, we present results of time-varying profiles for three select streamlines, one near the left edge - $\psi_4$, the center streamline - $\psi_{10}$, and one at the far-right edge $\psi_{18}$ (cf. Figure 7.1b), each with a range of initial fiber orientation as specified above. The computed results show that the fiber surface extreme pressures on the outer streamlines ($\psi_4$ and $\psi_{18}$) are less sensitive to initial fiber orientation over the entire deposition time as compared to the center streamline $\psi_{10}$ where the initial fiber angle has much more pronounced effect on the characteristic pressure peak values.

The results of the fiber orientation relative to the streamline direction presented in Figure 7.9 shows that the particle eventually tends to align with the streamlines of the flow irrespective of its initial starting angle and the degree of fiber alignment increases from the center streamline ($\psi_{10}$) to streamlines closer to edges of the nozzle ($\psi_4$ and $\psi_{18}$). The asymmetry in the results of the orientation for edge streamlines $\psi_4$ & $\psi_{18}$ shown in Figure 7.9a & c, respectively, signifies that fibers on these streamlines undergoes uneven rotation prior to flow alignment depending on the degree and direction of initial misalignment relative to the prevailing vortex direction ($\omega$) of the undisturbed flow which in turn depends on the relative positioning of the streamline with respect to the centerline.

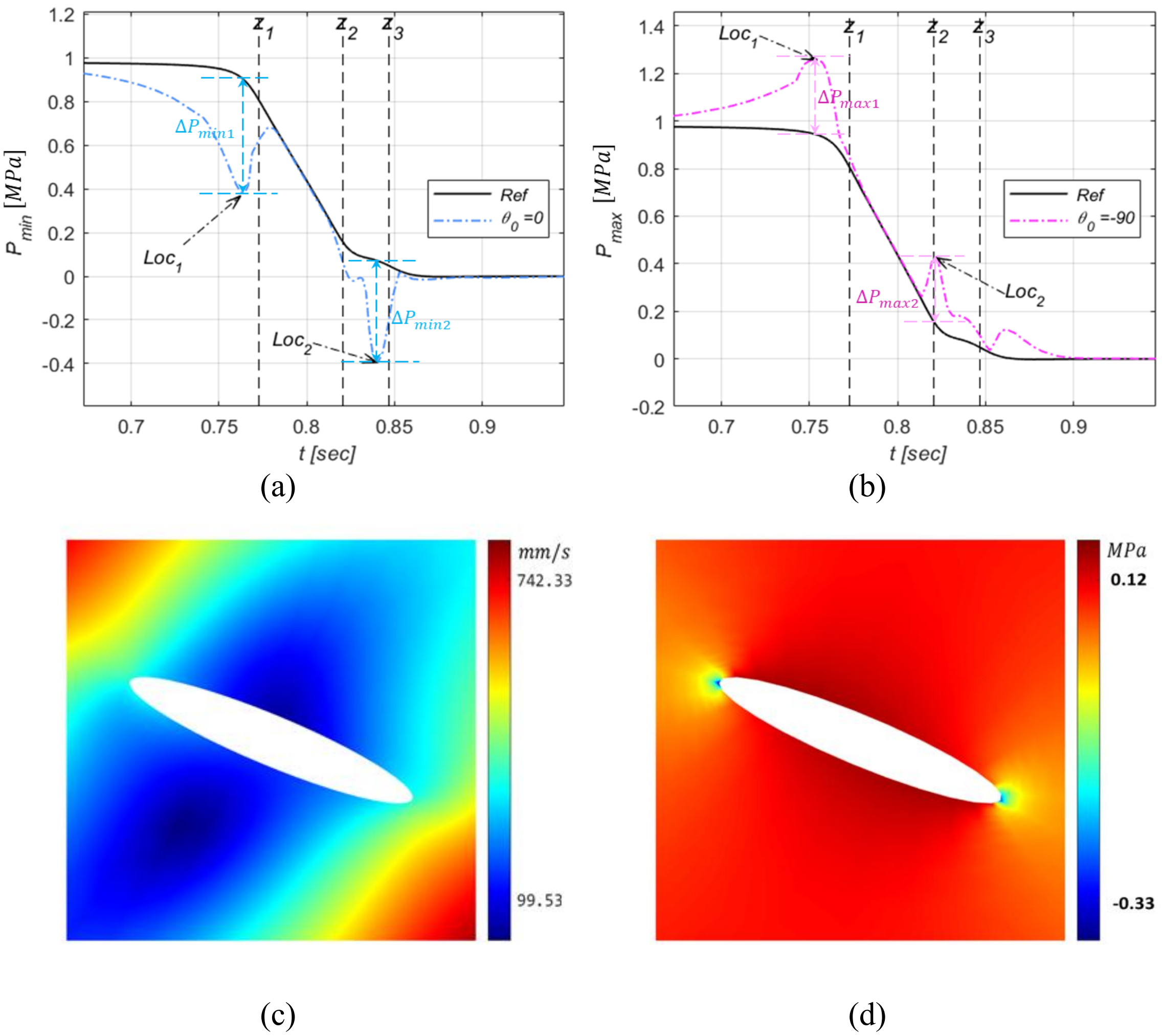

Figure 7.8: Selected computed results along center streamline $\psi_{10}$ for the concentrated suspension ($C_I = 0.0128$ and $\mathbb{m} = 4$). Shown are the fiber's surface (a) minimum pressure ($\theta^0 = 0^0$) (b) maximum pressure ($\theta^0 = -90^0$) at peak locations (Loc. 1 & Loc. 2). Contour plot at the first location (Loc. 1) of minimum pressure drop showing (c) Velocity magnitude (d) Pressure near the fiber.

To better depict the fiber rotation span for fibers initially inclined unfavorably with the flow, the orientation transient profiles have been vectorially added to $\pi$ considering the fiber has no preferred ends (i.e., $\theta(t) = -\theta(t) - \pi,\ \theta_0 < 0,\ \omega > 0$ for streamline $\psi_4$ and $\theta(t) = -\theta(t) + \pi,\ \theta_0 > 0,\ \omega < 0$ for streamline $\psi_{18}$). Alternatively, the fiber motion on the outer streamlines is more sensitive to the initial fiber orientation and possesses some degree of asymmetry with respect to the initial angle. This is due to the relatively high velocity gradients for streamlines closer to the nozzle edge as compared to the center streamline. Moreover, the transition time in the die swell region between zones 2 and 3 increases with streamline location from the right-hand edge to the left-hand edge due to correspondingly larger radius of curvature (cf. Figure 7.1b). Streamline 18 has a sharp 90$^{o}$ turn with negligible dwell time in the die swell region as zones 2 and 3 almost nearly overlaps unlike streamline 4 and 10 which experiences relatively higher dwell in the die swell region as the polymer melt gradually approaches the deposition plate surface.

For subsequent simulation results, we consider a range of initial fiber orientation and report the computed overall minimum and maximum pressure difference with respect to the streamline pressure across the nozzle at the important extreme pressure locations (i.e., Loc. 1 and 2). In addition, we report the corresponding spatial positions where the minimum and maximum pressure extremes occur within extrusion-deposition flow for each of the various streamlines across the nozzle section. Lastly, we report the fiber's orientation relative to the streamline direction at three interest zones of the nozzle (zones 1-3).

Calculated results in Figure 7.10 show that the extreme pressures on center streamlines are more sensitive to initial fiber angle than that for the outer streamlines. We observe a drop in average minimum pressure of -0.5MPa at the first extreme occurrence (Loc. 1) which is almost uniform across all streamlines within the nozzle.

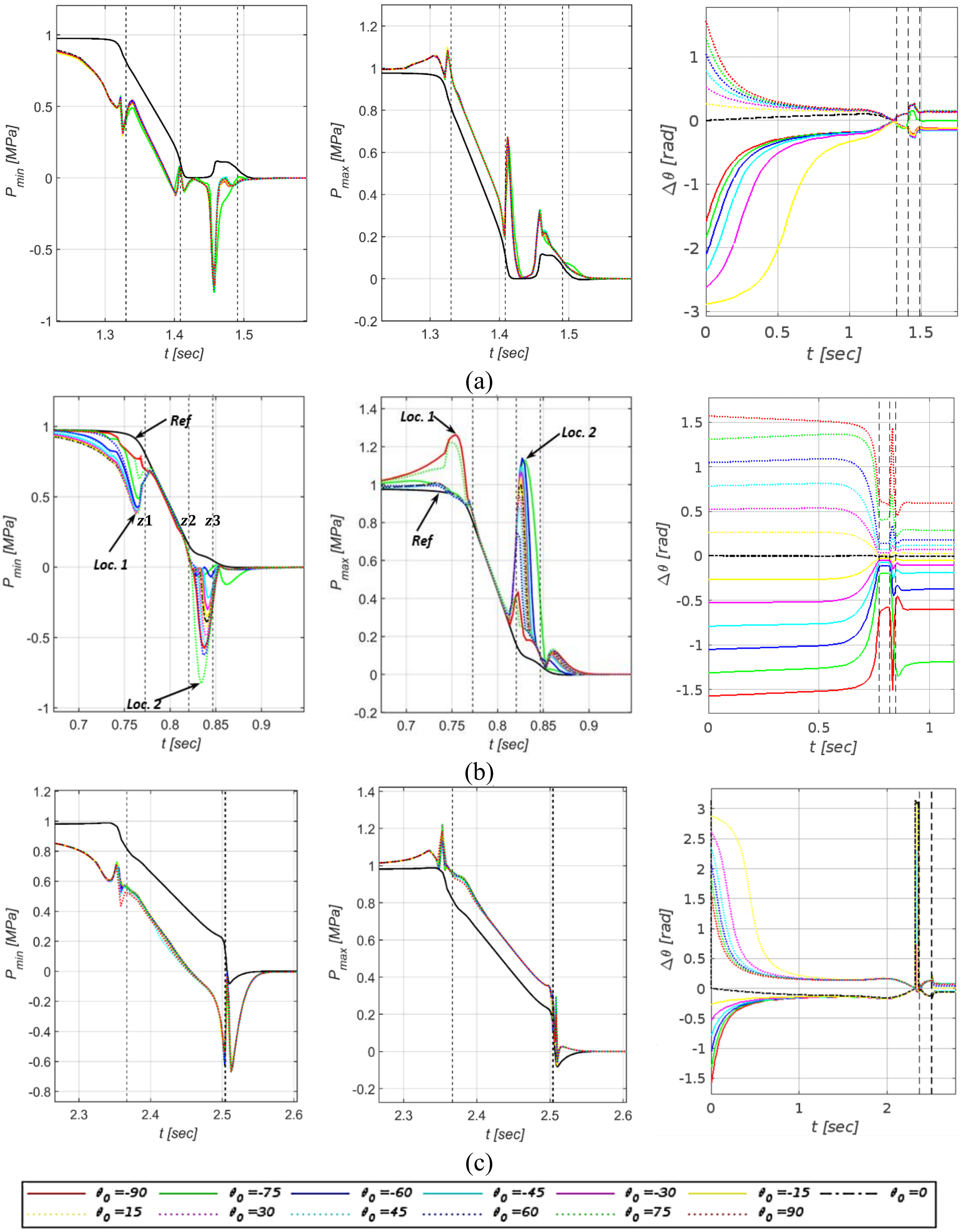

Figure 7.9: Quasi-transient profile plots of the fiber minimum pressure, maximum pressure and relative orientation angle, including various initial fiber angles for selected streamlines a) streamline-4 b) streamline-10 and c) streamline-18 ($C_I = 0.0128$ and $\mathbb{m} = 4$).

Alternatively, the second average pressure extreme occurrence (Loc. 2) has a minimum streamline pressure of -0.8MPa at the left edge streamline and -0.1MPa at the right edge (cf. Figure 7.10b). The spatial position where the first extreme in the minimum pressure drop occurs across the nozzle is seen to be well-above the entrance to the straight nozzle capillary (zone 1) but at the second pressure extreme location, the mean minimum extreme pressure occurs across the die swell region of the flow as shown in Figure 7.11b. This would indicate that the likelihood of void nucleation decreases from the bottom to the upper free surface of the bead. The average extreme maximum pressure at the first peak location (Loc. 1) across streamlines of the nozzle just before zone 1 is seen to be generally less severe than pressure values at the second peak location (Loc. 2), and the mean extreme pressure magnitudes decline asymmetrically with a trough-like appearance from streamlines closer to the edges towards the centerline (cf. Figure 7.10c). The opposite behavior is observed at the second extreme site (Loc. 2) where there is an unsymmetrical rise in the mean extreme pressure magnitude from the edges to the centerline in a crest-like manner (cf. Figure 7.10d), and the spatial position where this occurs is seen just after the nozzle exit, about .5mm beneath zone 2 almost nearly evenly across the flow (cf. Figure 7.11d). This behavior may be attributed to the relatively high shear rates at the wall just before exiting the nozzle compared to the center streamline which transitions abruptly at the edges.

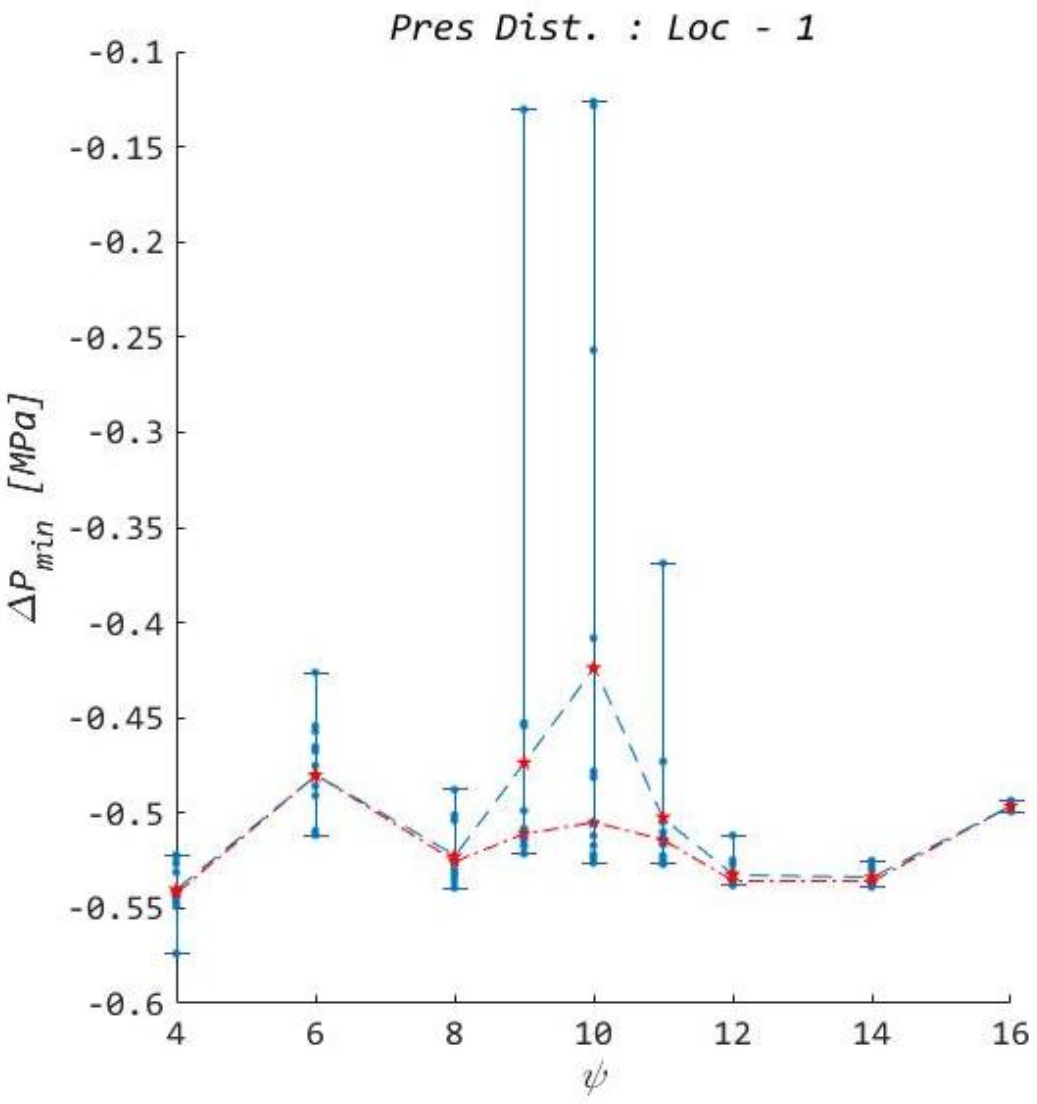

(a)

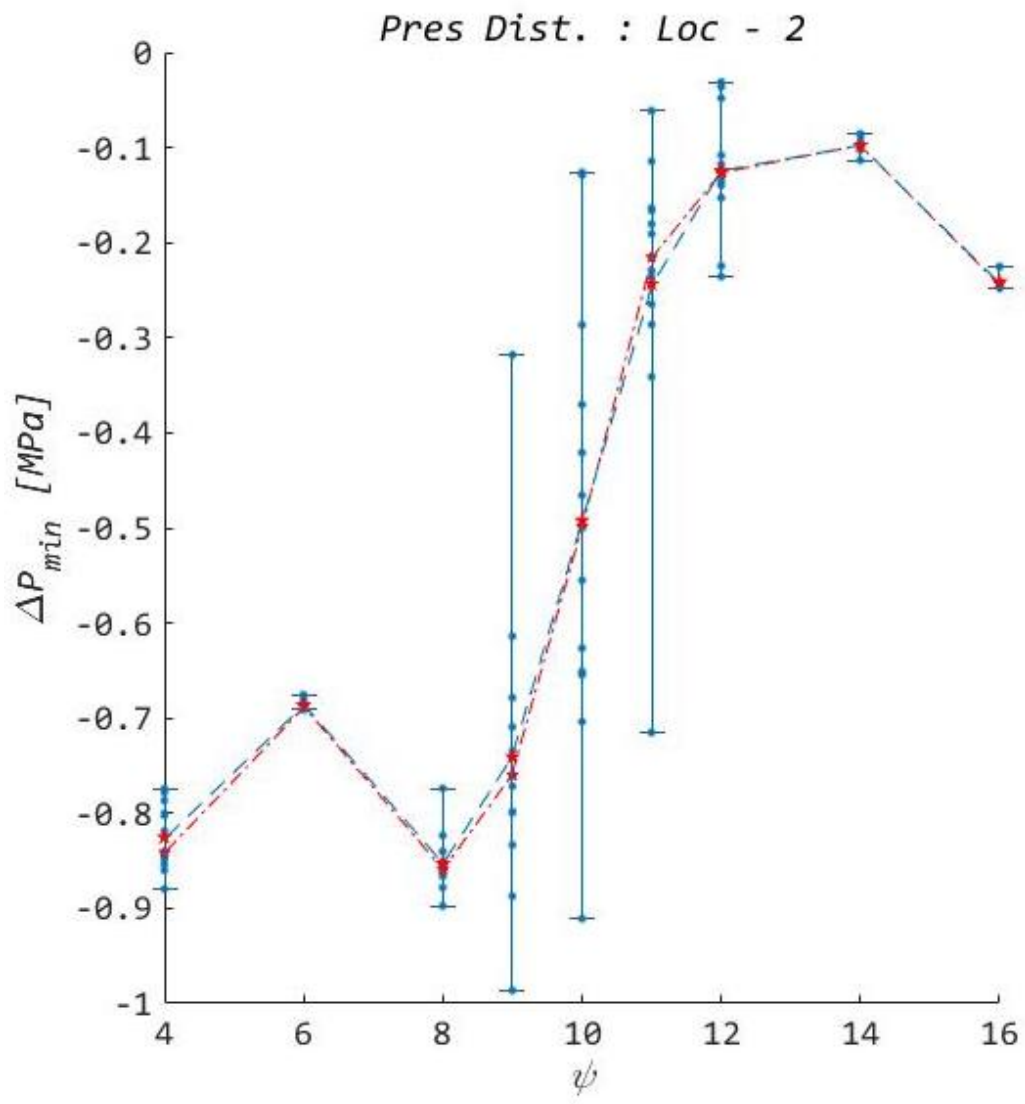

(b)

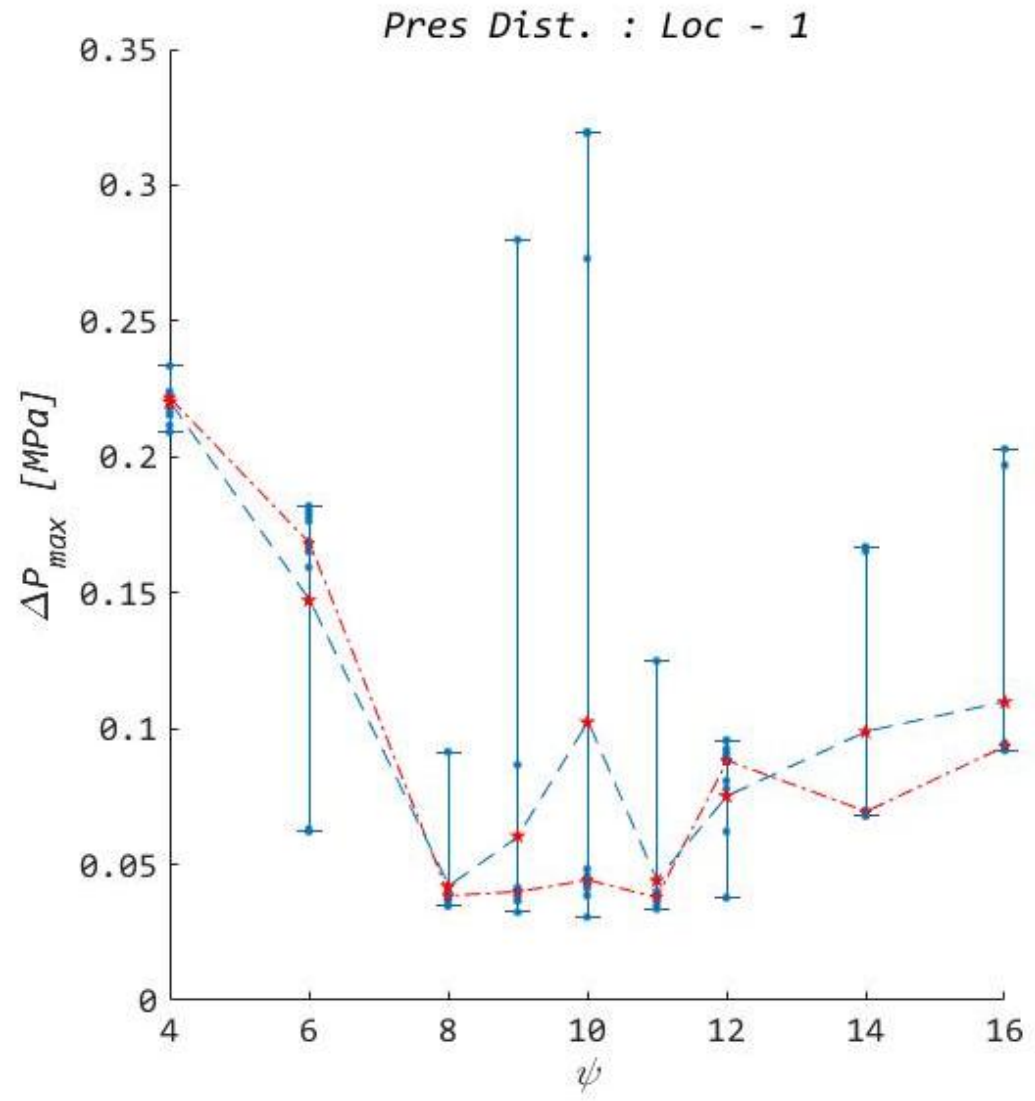

(c)

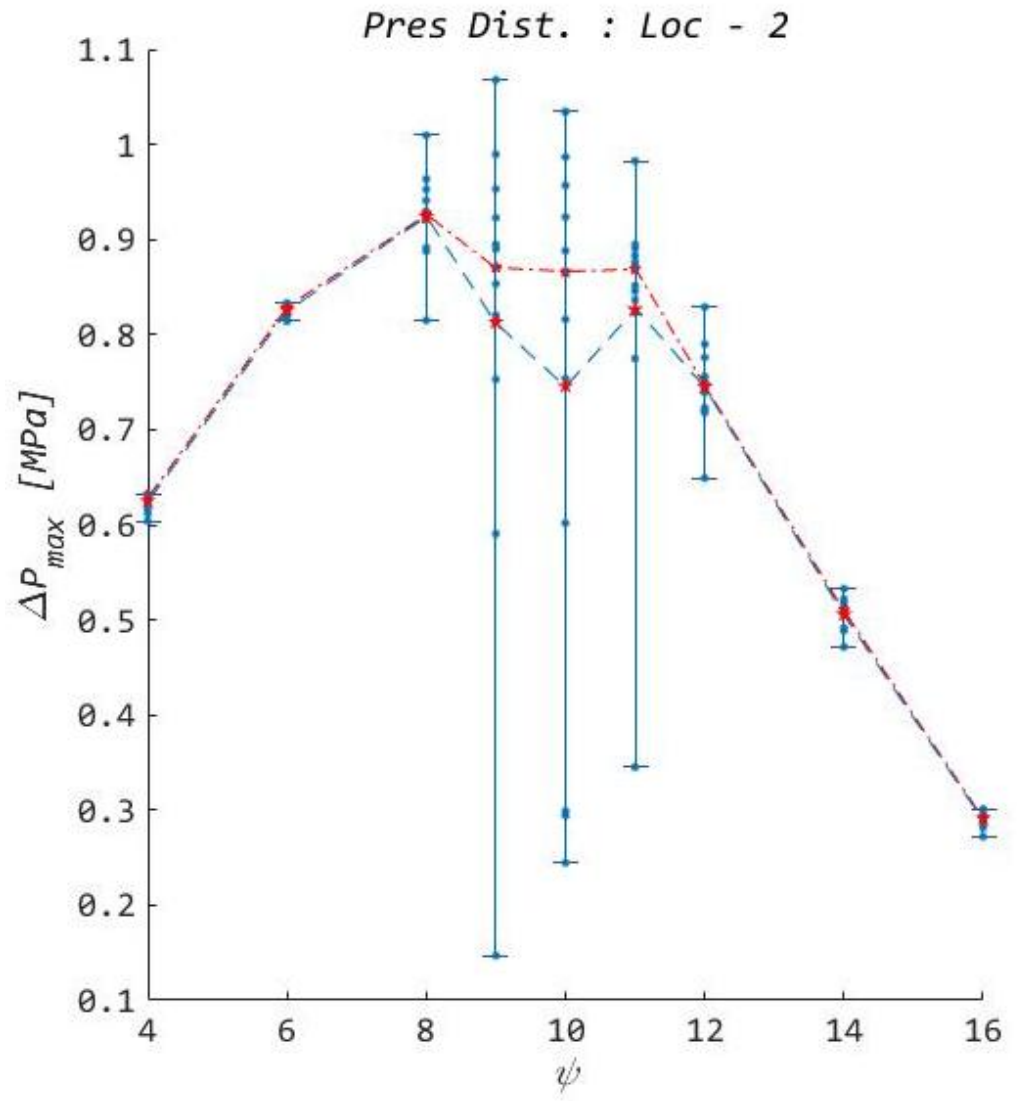

(d)

Figure 7.10: Overall pressure extremes on the fiber surface over the complete period of deposition (the blue trendline represents the mean and the red trendline is the median): (a) overall minimum at Loc. 1 (b) overall minimum at the Loc. 2 (c) overall maximum at Loc. 1 (d) overall maximum at Loc. 2.

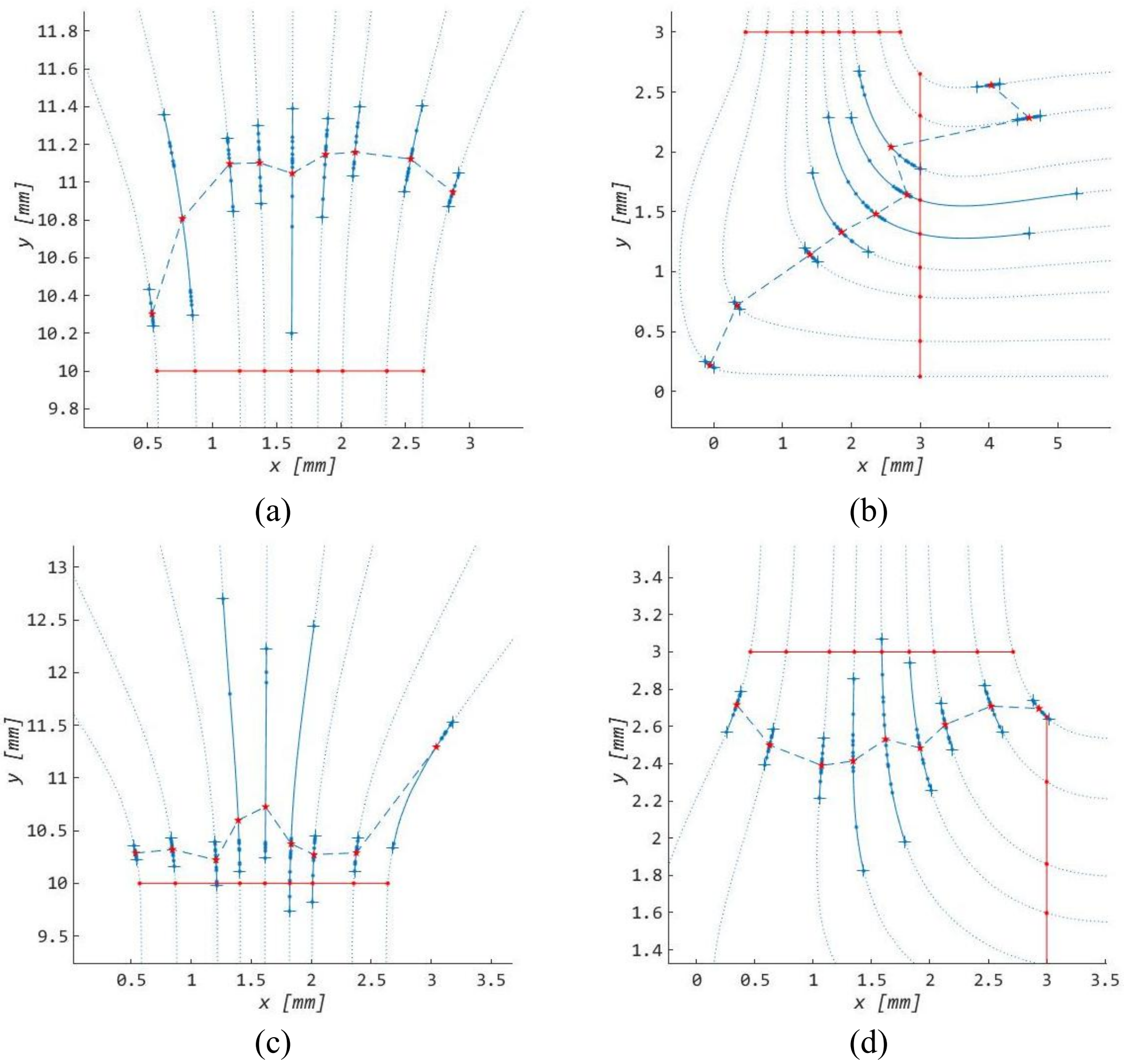

Figure 7.11: Distribution of location within the nozzle where the pressure extremes on the fiber surface occurs over the complete period of deposition and for all computed streamlines: (a) overall minimum at Loc. 1 (b) overall minimum at the Loc. 2 (c) overall maximum at Loc. 1 (d) overall maximum at Loc. 2.

The result of the fiber's orientation distribution relative to the streamline direction at the 3 regions of interest shows that the fiber is almost nearly aligned with the streamlines of the flow across the nozzle section and the degree of alignment increases towards the edge of the nozzle as we observe from Figure 7.12a-c. This is consistent with the conclusion of Saffman [181] who observed that the fibers tend to align with the flow. The error bounds of the fiber's orientation across the nozzle due to the variation of initial fiber angle in all three locations are also similar.

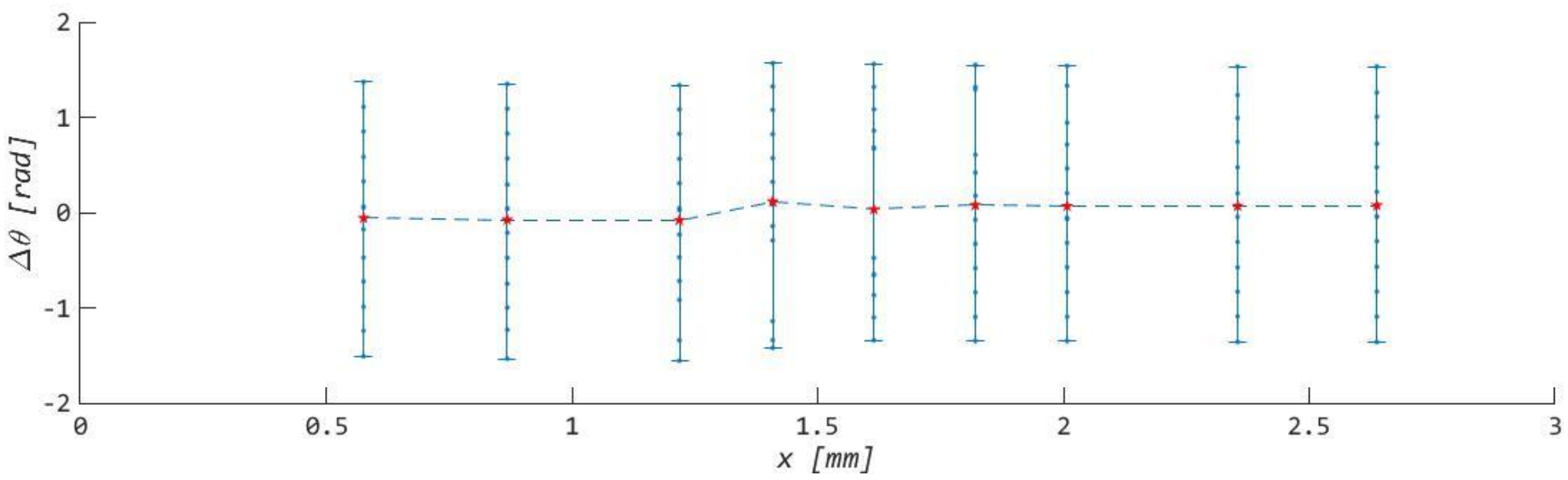

(a)

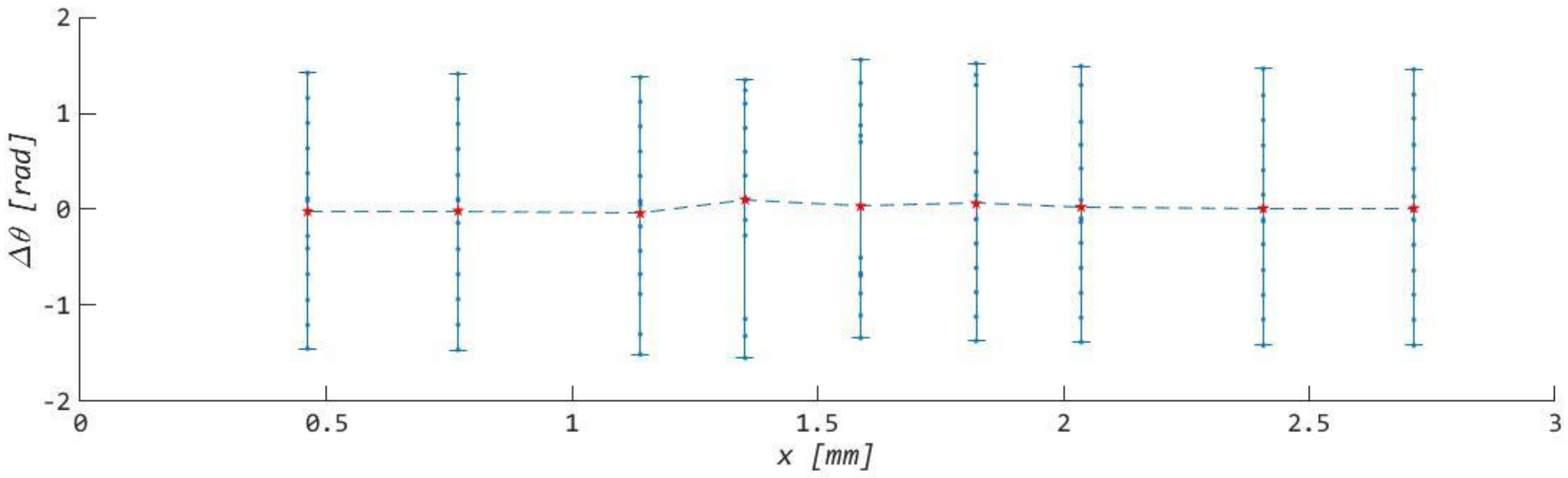

(b)

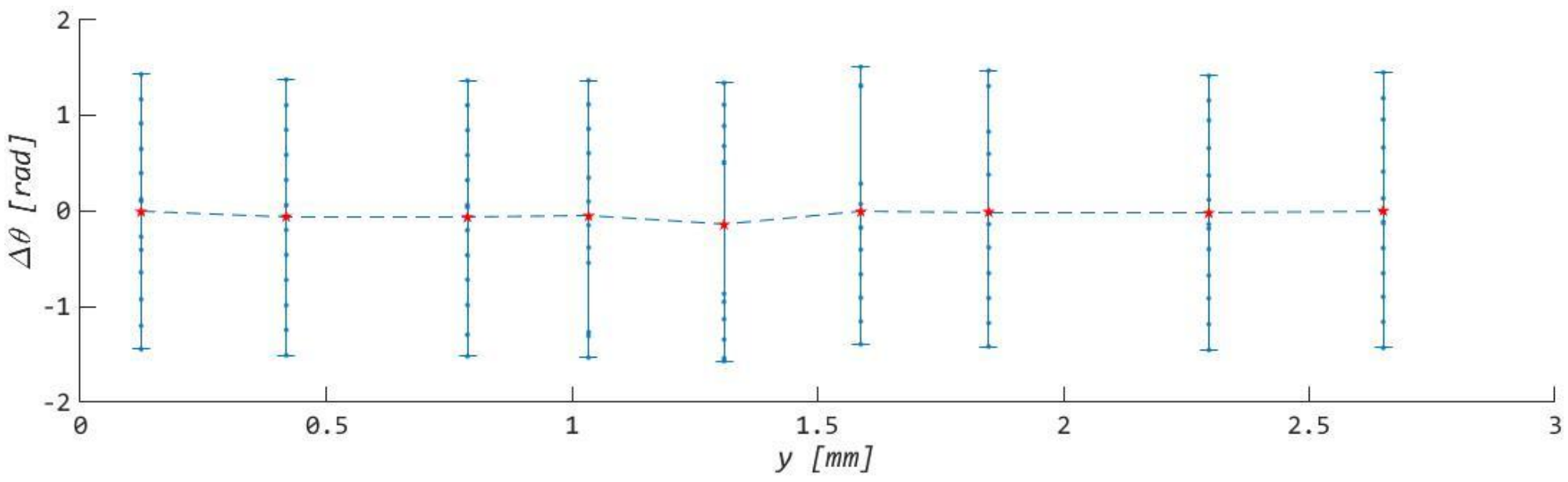

(c)

Figure 7.12: Distribution of fiber orientation angle at the region of interest within the extruder nozzle: (a) Zone 1 (b) Zone 2 (c) Zone 3.

### *7.6.2 Multi-Scale Non-Newtonian Melt Flow Simulation*

Computed results of the velocity magnitudes and shear-rates from the non-Newtonian macroscale analysis are shown in Figure 7.13a & b together with the results from the Newtonian simulation for comparison. The results show relatively higher velocity and shear rate magnitudes for streamlines at the nozzle edges ($\psi_4$ & $\psi_{18}$) and lower values towards the centerline ($\psi_{10}$) for the shear-thinning fluid compared to the Newtonian fluid. Correspondingly, the deposition times are relatively shorter for streamlines closer to the nozzle edge and relatively longer for streamlines closer to the centerline for the shear-thinning fluid compared to the Newtonian fluid. Likewise, the pressure-drop and pressure gradients across the nozzle are less severe for the shear-thinning fluid compared to the Newtonian fluid [232].

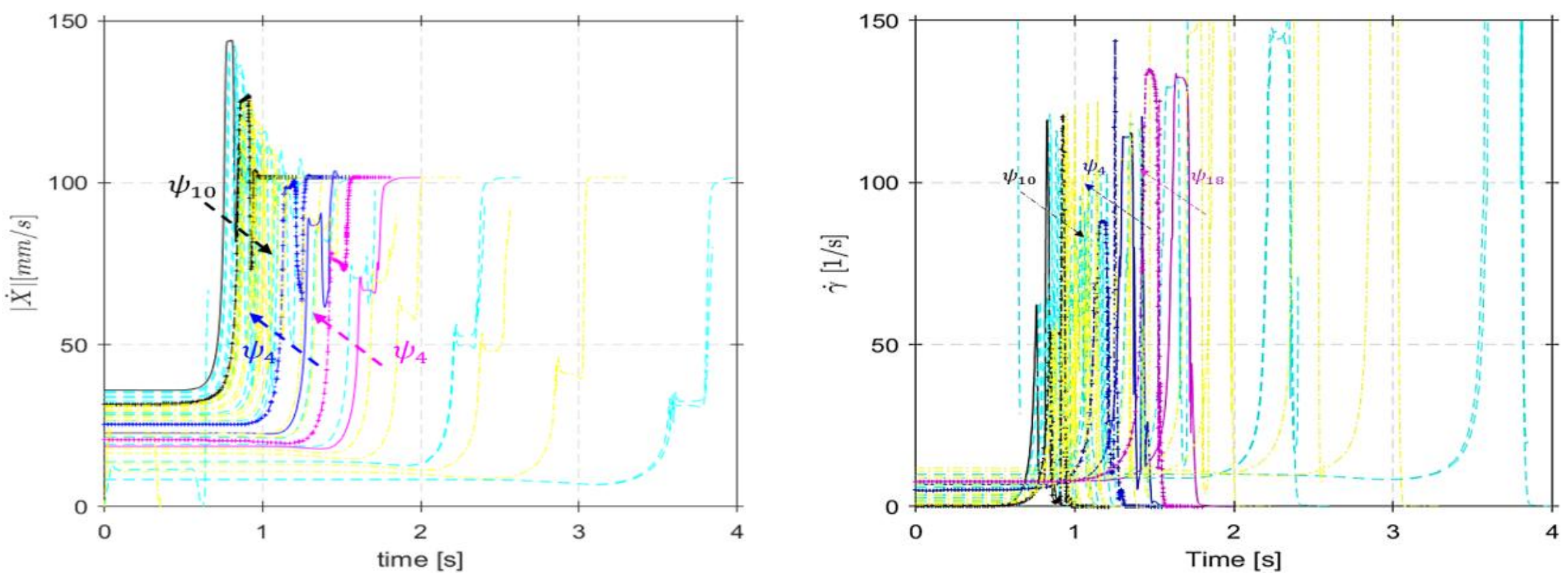

(a) (b)

Figure 7.13: showing a) relationship between shear-thinning fluid viscosity and flow shear-rate. Also shown are time-varying profiles along streamline $\psi_4$ (blue), $\psi_{10}$ (black) and $\psi_{18}$ (pink) for the both Newtonian (continuous line) and non-Newtonian (dotted line) analysis results for b) velocity magnitude c) shear-rate scalar magnitude, and d) pressure distribution.

We present the results of the rigid ellipsoidal fiber's motion and surface limit pressure evolution along streamlines of the extrusion-deposition flow for dilute fiber suspension with shear-thinning fluid rheology based on the micro-model non-Newtonian analysis. The results are presented for three (3) feature streamlines i.e. streamline $\psi_4$ closer the left edge of the nozzle, streamline $\psi_{10}$ at the nozzle center and streamline $\psi_{18}$ at the right edge of the nozzle (cf. Figure 7.1b).

Like the 2D Jeffery studies, we see from Figure 7.14a-c that the fibers angular velocities are unaffected by the shear-thinning fluid rheology irrespective of the non-uniform velocity gradients that characterizes the extrusion-deposition flow-field especially at the nozzle edges and a Newtonian analysis is sufficient to predict fiber's motion. This is evident from the overlap of the angular velocity profiles for all flow-behavior indices considered. On streamline $\psi_4$, the fiber experiences a spin reversal upon exiting the nozzle within the region of die swell due to counter-rotation in the 90° bend that opposes the local shear-vorticity direction at the left inner wall of the straight capillary before returning to steady state during bed deposition. On streamline $\psi_{10}$, the fiber's motion is steady within the straight capillary due to the uniform flow-field at the center streamline however the angular velocity peaks within the die-swell region due to the change in flow direction. On streamline $\psi_{18}$ the fiber experiences two (2) significant peaks in the angular velocity along the flowpath. The first peak occurs as a result of the severe velocity gradient at the right edge of the nozzle while the latter occurs due to abrupt change in flow direction at the sharp notch where the polymer exits the nozzle. Although we expect the particle dynamics would be influenced by the shear-thinning fluid rheology in a 3D simulation based on our studies in Chapter Five, our primary focus here is the particle's surface pressure distribution which our 2D GNF FEA model has been shown to be sufficient for understanding the shear-thinning effect on the pressure response in Chapter Five. Figure 7.14d-e shows that the shear-thinning fluid rheology reduces the magnitude of the fibers surface pressure peaks as the flow behavior index is reduced. The implication of this is that we expect lower probability of void nucleation with higher void formation times for fiber suspension with strong shear-thinning fluid characteristics than for weakly non-Newtonian fiber suspension. The magnitude of minimum pressure drops on the fiber surface are observed to be significantly higher on edge streamlines ($\psi_4, \& \, \psi_{18}$) compared to the center streamline ($\psi_{10}$). The net pressure extremes with respect to the instantaneous streamline pressure are observed to be higher at the second peak location for streamlines ($\psi_4$, & $\psi_{10}$) except on streamline $\psi_{18}$ where the net pressure magnitude is seen to be higher at the first peak location.

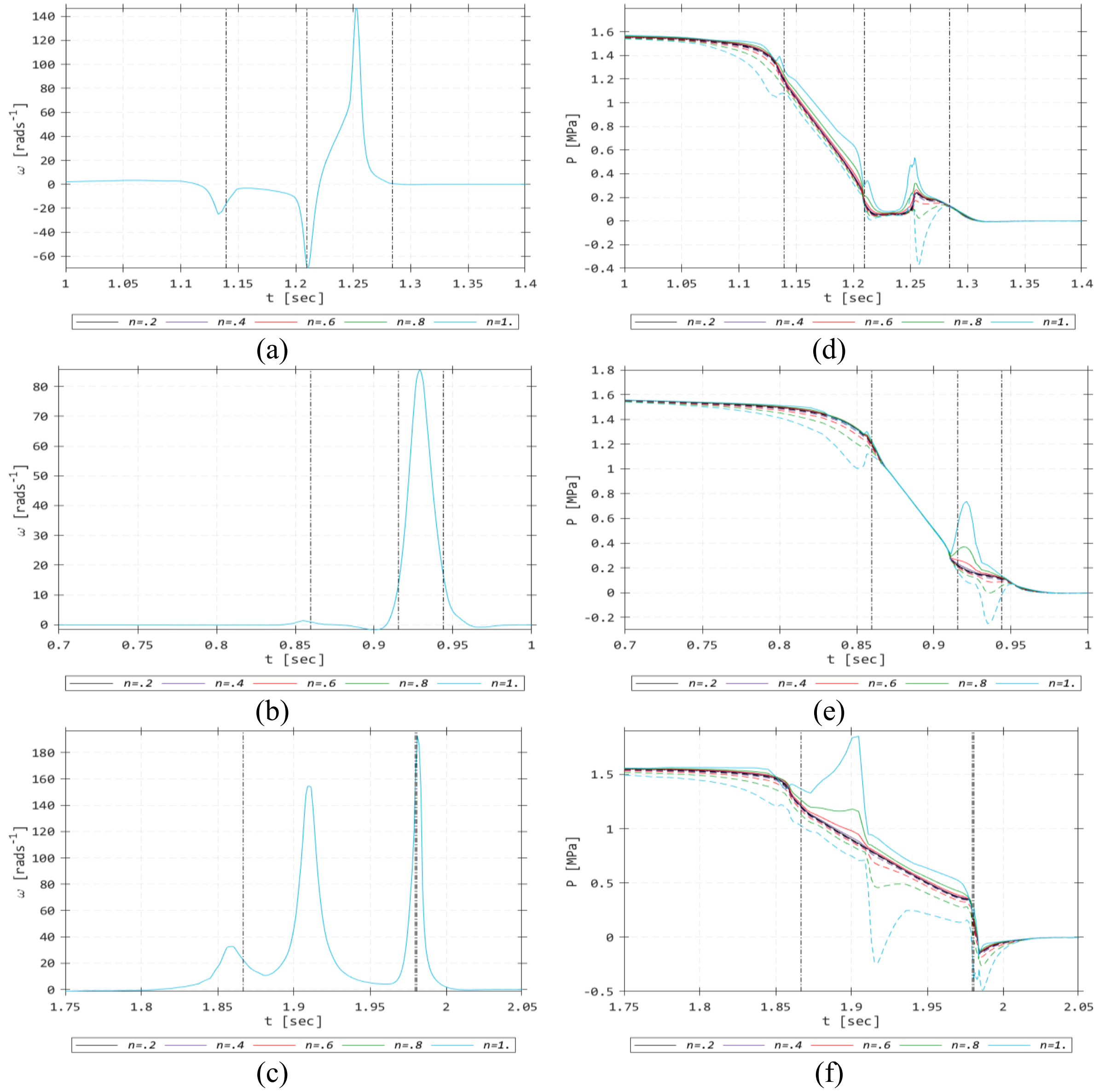

Figure 7.14: Figure showing the time evolution of the net fiber orientation relative to the streamline direction for (a) streamline $\psi_4$, (b) streamline $\psi_{10}$ and (c) streamline $\psi_{18}$. Also shown is the time evolution of the extreme pressure distribution on the fibers surface for (d) streamline $\psi_4$, (e) streamline $\psi_{10}$ and (f) streamline $\psi_{18}$. Results are presented for flow behavior index ranging from $n = 0.2 - 1.0$.

Figure 7.15 shows that like the Newtonian case, the peak sites of minimum pressure drop are observed to occur at the fiber's tips when they do occur. Moreover, the deposition times at which the peak pressure magnitudes occur are only slightly modified by the shear-thinning fiber suspension. For the edge streamlines at the second peak location of minimum pressure drop, the time of occurrence are slightly shifted downstream the extrusion-deposition flow while for the center streamline, the time of occurrence is slightly shifted upstream the flow. As such the orientation angle at which the second peak minimum pressure drop on the fibers surface occurs is slightly modified.

The observed minimum tip surface pressure computed in the simulation above provides fundamental insight into the occurrence of significant tip-voids within CF/ABS EDAM polymer composites as presented in Table 3.2. The occurrence of extreme minimum surface pressure at the tips of a suspended rotating particle potentially explains the high-volume fraction of micro-voids that form at particle ends. Further, the negative fiber tip pressures in the shear dominated flow regions of the EDAM nozzle correspond directly with the observed larger micro-voids in regions of the printed bead specimen close to the bead edges (i.e. ROI-III), as compared to the regions closer to the bead center that encountered a high degree of stretching flow during processing (i.e. ROI-II), shown in Table 3.3 and Figure 3.13. Previous sensitivity studies presented in Chapter Five revealed various factors that influences the pressure distribution on the fiber surface in CF/ABS EDAM which includes the fluid viscosity $\mu$, shear rate $\dot{\gamma}$, and particle aspect ratio $r_e$.

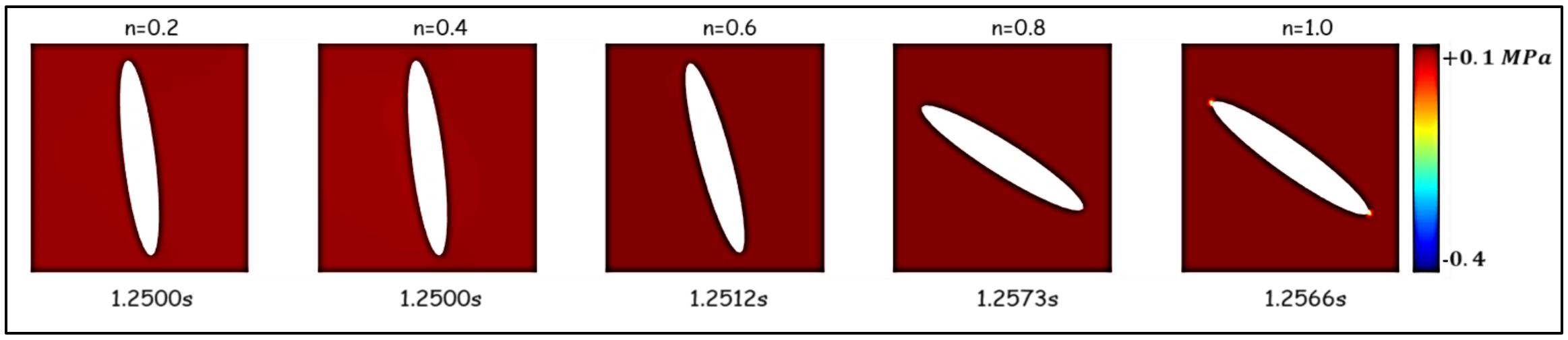

(a)

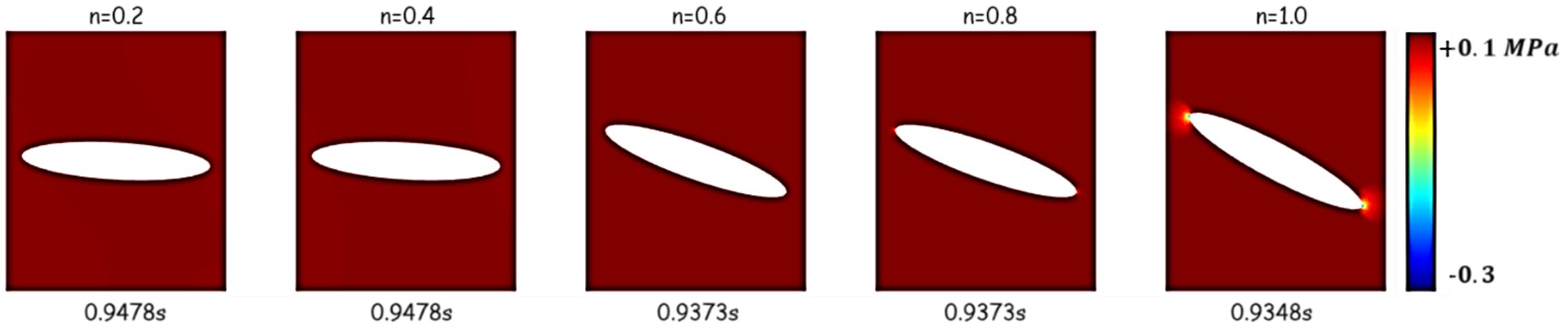

(b)

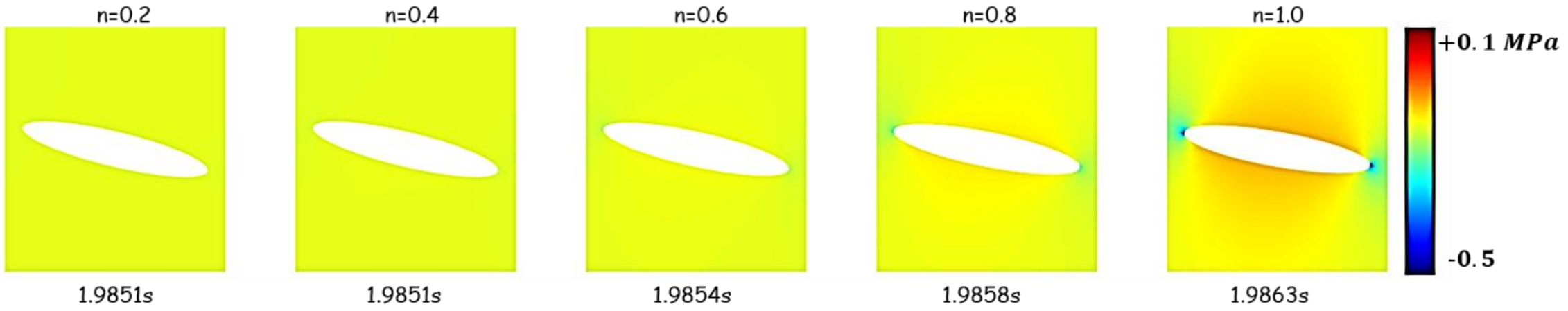

(c)

Figure 7.15: Figure showing the pressure distribution around the vicinity of the fiber at the second peak location of minimum pressure drop on the fibers surface for different flow behavior index ranging from $n = 0.2 - 1.0$ and (a) streamline $\psi_4$, (b) streamline $\psi_{10}$ and (c) streamline $\psi_{18}$.

As we have shown in our non-Newtonian studies, the shear-thinning rheology has no impact on the fibers motion in the 2D simulations. However, for the 3D simulations (cf. Chapter Five), the particle's motion is seen to be affected by the shear-thinning fluid rheology. In both cases, the shear-thinning fluid rheology influences the fiber surface pressure extremes which decreases as the power law index is reduced. Overall, our multiscale simulation shows extreme low pressures at the tips of suspended particles as they travel down streamlines of the polymer melt flow and our μ-CT scan results (cf.

CHAPTER THREE) indicating that a large majority of micro-voids occur at fiber tips in CF/ABS EDAM samples together provide unique insight into a potential mechanism for micro-void nucleation with short fiber polymer composites.

### *7.7 Conclusion*

A computational multiscale FEA methodology has been developed to study the behavior of suspended rigid ellipsoidal fibers during polymer composite melt extrusion-deposition flow through an LSAM nozzle. Sensitivity analysis based on Jeffery's model assumption reveals a direct correlation between the extreme pressures on the fiber surface with its geometry aspect ratio and the rheological properties of the flow (shear rate and viscosity) and these pressure extremes are observed to occur at the fiber's tips. Further, extreme minimum pressures are shown to occur at the fiber tips as the fiber rotates into alignment with the principal direction of the flow. Results of the extrusion-deposition multi-scale analysis that considers the effect of rotary diffusion due to short-range fiber interaction reveals a dependence of the severity and sensitivity of the fibers extreme pressures to streamline location and the initial fiber orientation. In addition, the effect of increasing fiber concentration and aspect ratio increases the magnitude of the pressure extremes on the fiber surface. In the extrusion-deposition flow, a significant minimum pressure extreme occurs on the fiber surface at the entrance to the straight capillary section and across the die swell region immediately outside of the extruder nozzle which indicates an increased likelihood for micro-voids initiation at fiber ends in these regions. Results indicate that we would expect a higher probability of micro-voids formation closer to the plate than the free surface. Results also confirm a high degree of fiber alignment in the extruded bead. The effect of shear-thinning is seen to decrease the fiber surface pressure extremes with decreasing power-law index. Based on the classical nucleation theory, we expect lower probability of void nucleation and higher micro-void formation times for strongly non-Newtonian fiber suspension and vice versa. The non-uniform velocity gradient that characterizes the LSAM nozzle extrusion-deposition flow does not influence the observed effect of the shear-thinning fluid rheology on the fiber dynamics or fiber responses and the peak sites of minimum pressure drop occurs at the fiber's tips as we observed in the Newtonian studies. However, the time interval and corresponding fiber orientation angle at which the peak pressure magnitudes occur are slightly modified in the shear-thinning simulations.

# CHAPTER EIGHT

## Conclusion and Future Work

### *8.1 Conclusion*

The development of computational tools that predict the microstructure and material behavior of large area additively manufactured polymer composite parts can help in controlling the printing parameters and process conditions to optimize the complex microstructure of the beads and resulting properties and performance of the printed part. Previous computational efforts aimed at predicting the microstructure have mainly focused on descriptors such as fiber orientation and distribution within prints. However, very little or no effort has been made to understand and predict the development of voids within the bead microstructure which are known to significantly impair the quality and performance of the printed components. The current research developed and applied a computational approach to investigate underlying mechanisms responsible for the formation of micro-voids within print beads, the various factors that may influence their development and assess the impact of these micro-voids on the resulting effective properties of prints. The various investigations and research outcomes presented in chapters of this dissertation are summarized below.

Firstly, 3D microstructural characterization of a 13% CF/ABS EDAM printed bead specimen using X-ray μ-CT image acquisition and analysis technique was performed in Chapter Three primarily to investigate micro-void formation within the printed bead with respect to various microstructural metrics including the fractions of the various micro-constituent phases and micro-void features, the distribution of micro-voids size and sphericity, and the distribution of the fiber orientation. The results revealed a high-volume fraction of micro-voids (~11% on average) where contained within the bead microstructure, of which more than 90% of the micro-voids by volume formed at the tips of the carbon fiber reinforcement. The heterogenous micro-voids that nucleated at the fiber/matrix interface were on average larger in size and less spherical in shape than the homogenous mode micro-voids isolated within the polymer matrix. Additionally, bead regions with relatively high degrees of fiber alignment in the print direction were found to have less interstitial homogenous micro-voids likely due to the relatively high fiber packing density, which promoted relatively higher micro-void segregation at the fiber terminations as compared to regions with more random fiber orientation distribution. These observations of favorable fiber segregation at the tips of suspended particles, especially in regions with highly aligned fibers, are supported by previous literature [3], [5]. The homogenous mode classical nucleation theory may explain the relatively small sized and highly spherical voids observed to nucleate within the polymer matrix while the heterogenous mode voids with larger size and irregular structure that segregate at the tips of suspended fibers are likely promoted by the low pressure regions at the fibers tips that acts as sinks that draw bubbles to it as well as provide, favorable sites for heterogenous mode void nucleation leading to bubble coalescence/void growth.  The effective material properties of a particular bead specimen are to a large extent dependent on its inherent microstructural configuration. Our novel contribution here is the quantification and characterization of micro-voids that nucleate particularly at the fiber terminations within EDAM SFRP composites which have only previously been assessed from a qualitative perspective in literature.

The subsequent chapter (Chapter 4), sought to evaluate the effective thermo-mechanical properties (including elastic constants, coefficient of thermal expansion, and thermal conductivity) of the 13% CF/ABS bead specimen based on a numerical FEA homogenization approach using voxel based realistic periodic RVE's generated form the actual X-ray μ-CT data. The study involved determination of suitable RVEs given a dispersion error tolerance of 5% in computed effective properties and the numerical results were found to be comparable to the analytical estimates based on Mori-Tanaka's mean-field homogenization approach. Overall, the inherent micro-voids were found to negatively impact the evaluated effective properties of the studied bead region (ROI-II), about 21% decrease in the calculated effective moduli, 4% decrease in the effective coefficient of thermal expansion and 12% decrease in the effective thermal conductivity. Linear regression analysis revealed that the computed effective quantities correlated with the fiber volume fraction and degree of fiber alignment in the print direction across the bead specimen. While the effective modulus and thermal conductivity were observed to vary directly with the fiber volume fraction and degree of fiber alignment with the print direction, the effective thermal expansion coefficient was observed to vary inversely with these microstructural parameters. Additionally, the numerical study revealed relatively higher values of the computed effective modulus and thermal conductivity at ROI's closer to the bead's edge and free surface with relatively higher volume fraction and degree of fiber alignment with the print direction as compared to central ROI's with less compact fiber structure and more randomly oriented fiber distribution. Our unique contribution lies in the creation of realistic 3D X-ray μ-CT based Representative Volume Elements (RVEs) to evaluate the effects of porosities on the effective properties of Short Fiber Reinforced

Polymer (SFRP) composites through Finite Element Analysis (FEA). Previously, assessments were conducted either numerically using deterministic RVEs or analytically with mean-field homogenization methods. However, these methods are less accurate due to their limitations in capturing the geometric peculiarities of inclusions, such as irregular particle morphology and characteristics, as well as the spatial variations in the distribution of microstructural features. Evidently, the presence of porosities within the bead microstructure was shown to result in significant property losses, and as such, understanding the underlying mechanisms responsible for the development of these micro-voids is crucial which the rest of the dissertation was dedicated to.

In the introductory section, we presented a hypothetical basis for studying the distribution of the local pressure around the surface of suspended particles as the primary variable that influences the development of process-induced micro-void within polymer composite beads. Our hypothesis stemmed from the theoretical model development of the most known mechanisms of void nucleation in polymeric liquids found in numerous literature [9], [12], [13], [14], [15], [16], [17] which was seen to be highly dependent on the occurrence of negatively low localized pressure within the polymer melt during material processing. Because micro-voids are localized phenomenon occurring on the microscale level at the order of the smallest dimension of a fiber particle, a multiscale computational modeling approach involving coupling between a macro-scale model that predicts the global flow-field state and a micro-scale model that predicts localized flow-field state was necessary. In Chapter Five (5) we presented the model development of a non-linear finite element analysis (FEA) based micro-scale simulation that considered a generalized Newtonian fluid (GNF) viscosity model to study the effects of shear-thinning fluid rheology in combination with a host of other factors including the particle aspect ratio and initial particle orientation on the particle behavior and flow-field response of a single particle viscous suspension as a starting point for our investigation. The study was based on a special class of homogenous flows that characterizes typical flow conditions found melt flow regions of the of an extrusion nozzle during polymer composite additive manufacturing processing. In the Newtonian asymmetric homogenous flows, the particle's tendency to stall was found to be dependent on the shear-to-extension rate ratio and increased shear dominance resulted in increased flow symmetry and tendency for continued periodic particle tumbling motion. However, in the axisymmetric flows, there is no tendency for the particle to stall irrespective of the magnitude of shear-to-extension rate ratio. Likewise, the results reveal distinct peaks in the pressure extreme transient profiles as the particle tumbles continuously in the shear-dominant flows, which occur at the particle tips and at particle orientation positions that coincide with the principal flow directions. Sensitivity studies revealed that the ellipsoidal particle's orbital peak surface pressure extreme magnitudes decreased exponentially with increasing particle curvature radius or conversely increases exponentially with the particle's geometric aspect ratio and asymptotes as the geometric shape parameter approaches unity. In reality, suspended fiber particles used to reinforce polymers are cylindrical shaped with irregular end conditions. Moreover, the cylindrical particle shape allows for exclusive investigation of the individual contributions of the edge curvature radius and geometric aspect ratio effects on the particle surface pressure response which was impossible with ellipsoidal shaped particles having both geometric parameters coupled. The results showed that the edge curvature radius had significantly greater influence on the particle's surface pressure extreme magnitude compared to its geometric aspect ratio. Additional sensitivity studies for Jeffery's type motion showed that the initial particle's azimuth orientation angle determined the particular tumbling orbit and the magnitude of the surface pressure fluctuations at the particle's tip. The particle surface pressure extreme magnitudes are observed to decrease, and the surface location of pressure extreme occurrence deviates further from the particle's tip location with decreasing orbital constant. The orbital peak particle tip pressure magnitude is seen to approximately obey a linear law with the polar location on the orbit across spectrum of degenerate Jeffery's orbit.

Because the thermoplastic polymer melt behavior is highly non-Newtonian in nature and the reinforcing fiber particles increase the shear-thinning behavior of the polymer melt, it was important to consider the effect of shear-thinning melt rheology on the behavior of the suspended particles in the various homogenous flow-fields. In the 2D studies, the particle's dynamics were observed to be unaffected by the shear-thinning rheology which was not the case in the 3D studies. The particle's motion was observed to be retarded by the shear thinning fluid rheology under axisymmetric flow conditions. Under asymmetric homogenous flow conditions, the cessation of the particle's motion was found to be dependent on the shear-thinning fluid rheology in addition to the Trouton ratio. In both 2D and 3D dimensional spaces however, irrespective of the flow type and Trouton ratio, higher fluid shear-thinning intensity resulted in a reduction in the magnitude of the particle surface pressure distribution due to an associated decrease in the effective viscosity of the fluid around the particle surface. The orbital locations where the peak surface pressure extreme magnitudes occurred were however unaffected by the shear-thinning fluid rheology. For Jeffery's type motion, the particles tumbling orbit were observed to be modified by the shear-thinning fluid rheology and to a greater extent with decreasing orbital constant. Jeffery's orbits were found to either dilate or constrict depending on the initial polar orientation of the particle. Moreover, the lowering of the particle's surface pressure extreme magnitudes by the shear-thinning fluid effect was exacerbated by increasing orbital constant or widening of Jeffery's orbit. Additionally, the effect of fluid shear thinning on the particle's motion was found to initially increase with increasing geometric aspect ratio until a critical point where the effects begin to diminish with the aspect ratio. However, the lowering of the pressure extreme magnitude on the particle surface by the fluid shear-thinning effect was observed to be continuously intensified with increasing aspect ratio. The investigation carried out in Chapter Five was aimed at understanding the effect of various factors and process conditions on the surface pressure distribution of suspended particles which we previously identified a primary variable that influences the development of micro-voids within printed polymer composite beads. Previous studies have primarily examined particle motion in viscous suspensions, focusing largely on linear shear flow. However, there has been limited attention on understanding the development of the flow field around the particle during its motion, while considering factors like particle shape, end effects and shear-thinning rheology, etc. on the particles response and flow-field development which are crucial for understanding complex processes in physical rheological systems. Additionally, the existing research that explores the pressure field around a particle is mostly based on the analysis of flow around a stationary particle. To the best of the author's knowledge, these studies do not consider the impact of the particle's dynamics on velocity and pressure distribution, which is a significant knowledge gap that the current chapter addresses. Although the chapter study assumed single particle dynamics in dilute suspension without considering the effect of momentum diffusion due to short range inter-particle hydrodynamic interactions which is a phenomenon commonly found in highly filled short fiber polymer composite melt flow additive manufacturing process. To account for rotary diffusion due to inter-fiber interaction, we developed a novel two-step numerical approach that correlates the suspensions coefficient of interaction with the effective fluid domain size used in our single fiber model, a method which was discussed in detail in Chapter Seven. A step in the numerical approach involved establishing a relationship between the rotary diffusion interaction coefficient and the steady state fiber orientation using any of the advection-diffusion fiber orientation tensor evolution models and a numerical root finding method, the Newton-Raphson algorithm in our case which led to the study carried out in Chapter Seven.

Traditionally, the steady state 2$^{nd}$ order fiber orientation has been computed using time evolution numerical IVP ODE techniques like the popular 4$^{th}$ order Runge-Kutta (RK4) or predictor-corrector methods. However, Chapter Seven presents a Newton-Rapson (NR) method for determining the steady state 2$^{nd}$ order fiber orientation tensor using exact 4$^{th}$ order Jacobian obtained from partial derivatives of 2$^{nd}$ order fiber orientation tensor material derivative with respect to the 2$^{nd}$ order fiber orientation tensor which is the novel contribution of this chapter. The comprehensive study considered various macroscopic fiber orientation moment-tensor models and various closure approximations of the 4$^{th}$ order fiber orientation tensor and the performance of the NR method in different homogenous flows. The stability of the NR method was found to depend on the flow type and characteristics, likewise the choice of closure approximations used for approximating the 4$^{th}$ order fiber orientation tensor. Results showed that the NR method performed best with the natural orthotropic (NAT) closure and the invariant based orthotropic fitted (IBOF) closure approximations for complex flows. Moreover, the derived exact Jacobians are popularly used in coupled flow/fiber orientation tensor finite element models [315], [316]. Although obtaining exact Jacobian of the 2$^{nd}$ order orientation tensor equation of change involving derivatives of the 4$^{th}$ order fiber orientation tensor, can be very difficult to model, however the method was found to be computationally more efficient compared to a higher order finite difference approximations of matching accuracy which often gives rise to numerical instability and various numerical errors when dealing with relatively small quantities.

As was previously noted, the preliminary sensitivity studies conducted in Chapter Five on single particle motion in viscous homogenous flow assumed dilute particle suspension that neglects the effect of inter-fiber hydrodynamic interactions which typically should not be ignored when considering highly loaded fiber polymeric suspension. Moreover, the actual polymer melt flow-field is inherently complex in nature consisting of non-uniform and spatially varying velocity gradients across the computationally flow domain. Chapter Seven sought to study the behavior of suspended rigid ellipsoidal fibers during polymer extrusion-deposition flow process and more accurately predict the flow-field development using a multiscale computational modelling technique. The chapter was aimed at understanding underlying pressure-based mechanisms that may promote heterogenous mode micro-void nucleation at the interface of suspended fibers within the bead microstructure during polymer composite processing. The macroscale model development provided in detail in Wang et al. [24], [315] was used to simulate the 2D planar deposition polymer melt flow process to predict the global flow-field and fiber orientation distribution within the computational flow domain. The microscale model development which was presented in Chapter Five was used to simulate the evolution of a single ellipsoidal fiber along streamlines of the polymer melt extrusion-deposition flow process utilizing the field responses (velocity, velocity gradients and pressure) obtained from the macroscale model to extrapolate the boundary conditions on the micro-model. The chapter also presented a novel approach to capture the effect of momentum diffusion due to short range hydrodynamic inter-fiber interactions in the single fiber microscale model by determining an effective fluid domain size that results in an equivalent steady state fiber orientation angle as would be predicted by the Advani-Tucker 2$^{nd}$ order fiber orientation evolution equation of change. Additionally, the study considered the fiber's evolution along various streamlines across the nozzle based on a given set of random initial fiber conditions to determine pressure bounds on the fiber surface across the flow. The multiscale analysis results showed that the extreme pressure magnitudes on the fiber surface were exacerbated by considering the effect of fiber-fiber interaction and the severity and sensitivity of the pressure magnitudes depended on the streamline location and the initial fiber orientation. The minimum fiber surface pressure extremes were observed to drop considerably at the entrance to the straight capillary section and across the die swell region immediately outside of the extruder nozzle which indicated an increased likelihood for micro-voids initiation at fiber ends in these regions. In the die-swell region, our results showed higher local pressure dips closer to the print bed which increased gradually towards the free surface indicating a higher probability of micro-voids formation closer to the plate than the free surface. As with the homogenous flow analysis conducted in Chapter Five, the effect of shear-thinning fluid rheology in the polymer extrusion-deposition melt flow is seen to decrease the fiber surface pressure extremes and to a greater degree with increasing non-linearity. The non-uniform melt flow velocity field does not alter the effect of the shear-thinning fluid rheology on the fiber motion, nor does it modify the location on the fiber surface where the peak pressure extreme occurs along the flow-path although the deposition time and associated fiber orientation angle at which the peak pressure magnitudes occurred were observed to be slightly modified by the shear-thinning fluid. The research presented in Chapter Seven marks a pioneering effort in using a multiscale FEA-based modeling approach to simulate particle motion along the streamlines of the EDAM polymer melt deposition flow process. The goal is to investigate the flow-induced mechanisms that might contribute to micro-void formation on the surfaces of suspended particles by analyzing the localized pressure distribution on the particles' surfaces. As experimentally observed in numerous literature [3], [5], [18], [48] and from Chapter Three, the void content in pure polymers are negligible despite being hygroscopic in nature, however the void content was observed to vary directly with the concentration of fiber fillers. It should be noted that the inherent moisture/volatiles species contained within the polymeric material are not void in themselves but sources of voids. These observations are strongly supported from our simulation results and from formation mechanisms presented in the introduction sections. The absence of local effects in the uniform pressure distribution profile of pure polymer melt flows from our macroscale simulation which is typically above the atmospheric pressure may explain the insignificant level of void contents, however from the results of our multiscale simulation of fiber suspension flow, we observe significant localized pressure drop on the surface of the fibers well below the global fluid flow pressure and atmospheric pressure due to the fiber dynamics and which may explain the increased void levels observed in the final bead microstructure from experiments.

The foregoing discussion on the study of rigid spheroidal particle's motion along streamlines of EDAM polymer composite deposition flow provide insight into potential micro-void mitigation strategies that could exploit the fluid rheological behavior to improve component part quality. Based on the proposed micro-void nucleation mechanism, a potential way for mitigating their formation would leverage factors that reduce the pressure intensity at the fibers tip. One potential way of controlling these micro-voids formation would involve suitable rheological adjustment to reduce the local extreme pressure fluctuations on the fibers surface. On one hand, increasing the shear-thinning intensity may help control the void formations, however increased shear-thinning may increase the likelihood of multiphase flow segregation within nozzle and the create more anisotropy in the microstructure of the printed composite. Our findings from previous chapters also reveal that low curvature radius at the fiber ends is a more relevant parameter that results in exacerbated pressure extremes compared to the fiber length. Our simulation results to this regard agrees with our experimental observation of very high content of heterogenous micro-void nucleation at the tip of fibers compared to their formation elsewhere on the particles surface. Accordingly, proper fiber surface finishing that reduces abrupt changes in fiber geometry and the possible pressure singularities at these sharp transitions is a possible way to mitigate the micro-void formation. Additionally, our simulation results show that the pressure extremes on the fiber surface across the nozzle are exacerbated at the die-swell region of the nozzle exit where the polymer melt flow makes a sharp 90° turn upon deposition unto the substrate. This agrees with the experimental observations of Yang et al. [18]. that found the micro-void content to be highest at the die-swell region compared to other polymer melt flow regions. As such, determining an optimal nozzle tilt angle design to control the flow angle at the nozzle exit may help reduce micro-void formation. A recent simulation-based investigation conducted by Guo et al. [256] showed that adjusting the nozzle tilt angle effectively modifies the flow-field pressure and velocity distribution in the die-swell region and the resulting shear rates and viscosity distribution within this region which could potential help control the micro-void formation within this region.

### *8.2 Future Work*

A future direction to our simulation effort would capture effects typically found in the actual structure of the fiber suspension such as the actual fiber geometry imported from the μ-CT data, the inter-particle and intra-particle hydrodynamic forces, and Brownian effects etc. which all potentially affect the resulting pressure distribution on the fibers surface. Moreover, our current simulation has neglected the visco-elastic polymer melt solidification behavior during deposition which is expected to significantly affect the pressure distribution especially at the die swell region of the nozzle exit where fiber surface local pressure extremes in flow analysis and the micro-void contents reported in literature are seen to be very high. Additionally, the current research is based on the assumption of isothermal polymer melt flow process, however the heat transfer process during post-deposition bead cooling necessitates the consideration of the conservation of energy equation in predicting the process state variables particularly the pressure distribution, which implies temperature dependency of the process parameters like viscosity, density, thermal conductivity, specific heat capacity etc.

Although our simulation efforts have revealed underlying process-induced mechanisms that may be responsible for the formation of voids within the microstructure of EDAM printed polymer composite beads, however there is yet need for a comprehensive and reliable computational model that realistically predicts micro-void formation, growth and their characteristics and that directly correlates print parameters and process conditions to the experimentally observed void distribution and characteristics within the microstructure of the printed beads which is potential future research opportunity.

APPENDICES

# APPENDIX A

## The Eshelby's (Strain Concentration) Elasticity Tensor

The strain concentration tensor or Eshelby's tensor ($\Pi_{ijkl}$) that appears in the Mori Tanaka's model for predicting the homogenized elasticity tensor of the short fiber reinforced composite material is given in this section. Given

$$\eta_m = (1-\nu_m)^{-1},\ \chi_r = (1-a_r^2)^{-1},\ \omega_m = (1+a_r) + a_r \begin{cases} (-\chi_r)^{1.5}\cos^{-1} a_r & a_r < 1 \\ -\chi_r{}^{1.5}\cosh^{-1} a_r & a_r > 1 \end{cases} \quad \text{(A. 1)}$$

Then for spheroidal inclusions the non-zero components of $\Pi_{ijkl}$ are given as:

$$\Pi_{1111} = \Pi_{1111} = .50\eta_m[-\chi_r + (4 - 2\nu_m + 3\chi_r)(1-\omega_m)]$$

$$\Pi_{1122} = \Pi_{1133} = .25\eta_m[\ \chi_r - (2 - 4\nu_m + 3\chi_r)(1-\omega_m)]$$

$$\Pi_{2222} = \Pi_{3333} = .25\eta_m[\ 1.5(1+\chi_r)\ + (\ 1 - 2\nu_m - 2.25\chi_r)\omega_m]$$

$$\Pi_{2233} = \Pi_{3322} = .25\eta_m[\ 0.5(1+\chi_r)\ + (-1 + 2\nu_m - 0.75\chi_r)\omega_m] \quad \text{(A. 2)}$$

$$\Pi_{2211} = \Pi_{3311} = .25\eta_m[-2.0(1+\chi_r)\ + (\ 2 + 2\nu_m + 3.00\chi_r)\omega_m]$$

$$\Pi_{2323} = \Pi_{3232} = .25\eta_m[\ 0.5(1+\chi_r)\ + (\ 1 - 2\nu_m - 0.75\chi_r)\omega_m]$$

$$\Pi_{1212} = \Pi_{1313} = .25\eta_m[-2.(\nu_m + \chi_r) + (\ 1 +\ \nu_m + 3.00\chi_r)\omega_m]$$

For spherical shaped inclusions, the non-zero components of the Eshelby tensor are given as

$$\Pi_{1111} = \Pi_{2222} = \Pi_{3333} = \eta_m[7 - 5\nu_m]/15$$

$$\Pi_{1122} = \Pi_{1133} = \Pi_{2211} = \Pi_{3311} = \Pi_{2233} = \Pi_{3322} = \eta_m[-1 + 5\nu_m]/15 \quad \text{(A. 3)}$$

$$\Pi_{1212} = \Pi_{1313} = \Pi_{2323} = \eta_m[4 - 5\nu_m]/15$$

# APPENDIX B

## B.1 Definition of Constants in Jeffery's Equation

The expressions of the components of the variable vector $\underline{\chi}$ and coefficient tensors $\underline{\Lambda}^{I}$, $\underline{\underline{\Lambda}}^{II}$ & $\underline{\underline{\Lambda}}^{III}$ that appear in the definition of the Jeffery's velocity and pressure are defined in eqns. (B. *1*) -(B. *5*) below. For the variable vector $\underline{\chi}$ the components are given as

$$\chi_1 = \text{Ч}_1' X_2 X_3, \qquad \chi_2 = \text{Ч}_2' X_3 X_1, \qquad \chi_3 = \text{Ч}_3' X_1 X_2 \tag{B. 1}$$

The components in $\underline{\Lambda}^{I}$ vector is likewise given as

$$R = -\frac{\Gamma_{23}}{\text{Ч}_{1_0}''}, \qquad S = -\frac{\Gamma_{13}}{\text{Ч}_{2_0}''}, \qquad T = -\frac{\Gamma_{12}}{\text{Ч}_{3_0}''} \tag{B. 2}$$

The components in $\underline{\underline{\Lambda}}^{II}$ tensor is given as

$$U = 2[\text{И}_2^2 B - \text{И}_3^2 C], \qquad V = 2[\text{И}_3^2 C - \text{И}_1^2 A], \qquad W = 2[\text{И}_1^2 A - \text{И}_2^2 B] \tag{B. 3}$$

where the coefficients $A, B, C$ in eqn. (B. 3) above are also components of tensor $\underline{\underline{\Lambda}}^{III}$ containing the stresslets and torque acting on the rigid ellipsoidal particle suspended in linear ambient flow-field [266]. given in eqn. (B. 4) below

$$\begin{aligned}
A &= \frac{1}{6}\left\{\frac{2\text{Ч}_{1_0}''\Gamma_{11} - \text{Ч}_{2_0}''\Gamma_{22} - \text{Ч}_{3_0}''\Gamma_{33}}{\text{Ч}_{2_0}''\text{Ч}_{3_0}'' + \text{Ч}_{3_0}''\text{Ч}_{1_0}'' + \text{Ч}_{1_0}''\text{Ч}_{2_0}''}\right\}, \qquad F = \frac{\text{Ч}_{2_0}\Gamma_{23} - \text{И}_3^2\text{Ч}_{1_0}'(\Xi_1 - \dot{\Psi}_1)}{2\text{Ч}_{1_0}'(\text{И}_2^2\text{Ч}_{2_0} + \text{И}_3^2\text{Ч}_{3_0})}, \\
F' &= \frac{\text{Ч}_{3_0}\Gamma_{23} + \text{И}_2^2\text{Ч}_{1_0}'(\Xi_1 - \dot{\Psi}_1)}{2\text{Ч}_{1_0}'(\text{И}_2^2\text{Ч}_{2_0} + \text{И}_3^2\text{Ч}_{3_0})} \\
B &= \frac{1}{6}\left\{\frac{2\text{Ч}_{2_0}''\Gamma_{22} - \text{Ч}_{3_0}''\Gamma_{33} - \text{Ч}_{1_0}''\Gamma_{11}}{\text{Ч}_{2_0}''\text{Ч}_{3_0}'' + \text{Ч}_{3_0}''\text{Ч}_{1_0}'' + \text{Ч}_{1_0}''\text{Ч}_{2_0}''}\right\}, \qquad G = \frac{\text{Ч}_{3_0}\Gamma_{13} - \text{И}_1^2\text{Ч}_{2_0}'(\Xi_2 - \dot{\Psi}_2)}{2\text{Ч}_{2_0}'(\text{И}_3^2\text{Ч}_{3_0} + \text{И}_1^2\text{Ч}_{1_0})}, \\
G' &= \frac{\text{Ч}_{1_0}\Gamma_{13} + \text{И}_3^2\text{Ч}_{2_0}'(\Xi_2 - \dot{\Psi}_2)}{2\text{Ч}_{2_0}'(\text{И}_3^2\text{Ч}_{3_0} + \text{И}_1^2\text{Ч}_{1_0})}
\end{aligned} \tag{B. 4}$$

$$C = \frac{1}{6}\left\{\frac{2Ч''_{3_0}\Gamma_{33} - Ч''_{1_0}\Gamma_{11} - Ч''_{2_0}\Gamma_{22}}{Ч''_{2_0}Ч''_{3_0} + Ч''_{3_0}Ч''_{1_0} + Ч''_{1_0}Ч''_{2_0}}\right\}, \qquad H = \frac{Ч_{1_0}\Gamma_{12} - И_2^2Ч'_{3_0}(\Xi_3 - \dot{\Psi}_3)}{2Ч'_{3_0}(И_1^2Ч_{1_0} + И_2^2Ч_{2_0})},$$

$$H' = \frac{Ч_{2_0}\Gamma_{12} + И_1^2Ч'_{3_0}(\Xi_3 - \dot{\Psi}_3)}{2Ч'_{3_0}(И_1^2Ч_{1_0} + И_2^2Ч_{2_0})}$$

The integral constants $Ч_j$ and their symmetric forms are defined as [21]

$$Ч_j = \int_{\lambda}^{\infty} \frac{1}{\Delta}\frac{d\lambda}{(И_j^2 + \lambda)}, \qquad Ч'_j = \int_{\lambda}^{\infty} \frac{1}{\Delta^3}(И_j^2 + \lambda)d\lambda, \qquad Ч''_j = \int_{\lambda}^{\infty} \frac{1}{\Delta^3}(И_j^2 + \lambda)\lambda d\lambda \tag{B. 5}$$

A constant $Ч_j$ subscripted with 0, i.e. $Ч_{j_0}$ implies that the lower limit of integration $\lambda = 0$.

## B.2 Two-Dimensional (2D) Reduced Form of Jeffery's Equation

The 2D contracted form of the Jeffery's pressure and velocity [21] can be expressed by eqns. (B. 6) & (B. 7) respectively given as

$$p = p_0 + 2\mu_1\Lambda_{ij}^{V}\nabla_{X_i}\nabla_{X_j}\Omega \tag{B. 6}$$

$$\dot{X}_i = \dot{X}_i^{\infty} + \Lambda_{ij}^{IV}\nabla_{X_j}\chi_3 + \Lambda_{jk}^{V}X_k\nabla_{X_i}\nabla_{X_j}\Omega - \Lambda_{ij}^{V}\nabla_{X_j}\Omega \tag{B. 7}$$

where $p_0$ is the constant mean pressure at a distance from the ellipsoid, $\dot{X}_i$ are the velocity components at arbitrary position $\underline{X} = [X_1 \quad X_2]^T$ and $\dot{X}_i^{\infty}$ is the velocity of the undisturbed fluid at $\underline{X}$ given as

$$\dot{X}_i^{\infty} = L_{ij}X_j \tag{B. 8}$$

coefficient matrices $\underline{\underline{\Lambda}}^{IV}$ and $\underline{\underline{\Lambda}}^{V}$ and the gradient operators $\underline{\nabla}_X$ are respectively given as

$$\underline{\underline{\Lambda}}^{IV} = \begin{bmatrix} Y & W \\ -W & Y \end{bmatrix}, \qquad \underline{\underline{\Lambda}}^{V} = \begin{bmatrix} A & H \\ H' & B \end{bmatrix}, \qquad \underline{\nabla}_X = [\partial/\partial X_1 \quad \partial/\partial X_2]^T$$

The 2D strain deformation tensor in the local fiber reference frame $L_{ij}$ is decomposed in the usual way to obtain the 2D symmetric component $\Gamma_{ij}$ and anti-symmetric components $\Xi_{ij}$ according to

$$L_{ij} = \Gamma_{ij} + \Xi_{ij} \tag{B. 9}$$

$$\Gamma_{ij} = L_{ij} + L_{ji}, \qquad \Xi_{ij} = \left[L_{ij} + L_{ji}\right] = \Xi_3^{-1}Б_{ij}, \qquad Б_{ij} = i - j \tag{B. 10}$$

The 2D Laplace Function $\Omega$ that appears in eqns. (B. 6) & (B. 7) is defined as

$$\Omega = \int_{\lambda}^{\infty} \frac{1}{\Delta}\left\{\sum_{j=1}^{2} \frac{X_j^2}{И_j^2 + \lambda} - 1\right\} d\lambda \tag{B. 11}$$

where

$$\Delta^2 = \prod_{j=1}^{2}\left(И_j^2 + \lambda\right), \qquad and, \qquad \lambda : \sum_{j=1}^{2} \frac{X_j^2}{И_j^2 + \lambda} = 1 \tag{B. 12}$$

At the fiber's surface where $\lambda = 0$, the field velocity must equal the fiber's surface velocity assuming no slip at the fiber's surface, i.e.

$$\dot{X}_i^p = \dot{X}_i \,\Big|_{\lambda=0} = \dot{\Psi}_3 Б_{ij} X_j \tag{B. 13}$$

The constants that appear in $\underline{\underline{\Lambda}}^{IV}$, $\underline{\underline{\Lambda}}^{V}$ above are thus obtained as

$$A = -B = \frac{\Gamma_{11}}{4Ч''_{3_0}}, \quad H = H' = \frac{1}{2}\left[\frac{\Xi_3 - \dot{\Psi}_3}{Ч_{1_0} - Ч_{2_0}}\right], \quad Y = -\frac{\Gamma_{12}}{Ч'_{3_0}}, \quad W \\ = 2(И_1^2 + И_2^2)A \tag{B. 14}$$

where $Ч_j, Ч'_j, Ч''_j$ & $\chi_3$ retain their usual definition given in eqn. (B. *1*) & (B. *5*). above.

The fibers angular velocity is derived as

$$\dot{\Psi}_3 = \Xi_3 + M_{33}D_3, \qquad D_3 = \Gamma_{12} \tag{B. 15}$$

### B.3 Flow-Regimes in Typical EDAM Nozzle

Polymer composite melt flow through the nozzle in typical EDAM polymer composite processing is characterized by complex combination of shear and extensional deformation rate components that are dependent on the viscoelastic polymer melt rheology and the geometry of the extrusion nozzle. The flow condition at the nozzle wall is pure shear and at the nozzle centreline is pure uniaxial elongation (cf. Figure B. *1*) [140], [314]. Away from the convergent zone in the lubrication zone defined by the clearance between the screw edge and the nozzle walls, the flow is predominantly shear dominant while close to the centreline and near the entrance of the nozzle where the flow undergoes acceleration due to geometric constriction, the flow is dominated by extensional rate, and at the vortices created near the notch edges with sharp transitions due to elastic instabilities, the flow is mainly rotational [140]. The flow contraction region consists of a complex combination of the various flow categories with varying dominance.

A simple metric used to classify the flow regimes is based on a flow parameter $\bar{\nu}$ given by [140].

$$\bar{\nu} = \frac{\dot{\gamma}_c + j\omega_c}{\dot{\gamma}_c - j\omega_c} \tag{B. 16}$$

where $\dot{\gamma}_c$ is the magnitude of deformation rate tensors defined as $\dot{\gamma}_c = \sqrt{2\Gamma_{ij}\Gamma_{ji}}$ and $\omega_c$ is the magnitude of the vorticity tensor given as $\omega_c = \sqrt{2\ \Xi_{ij}\ \Xi_{ji}}$. The flow is pure shear when $\bar{\nu} = 0$, pure elongational when $\bar{\nu} = 1$, and purely rotational when $\bar{\nu} = -1$. Typical flow patterns within the convergent zone results in $\bar{\nu}$ lying between $-1 \leq \bar{\nu} \leq 1$.

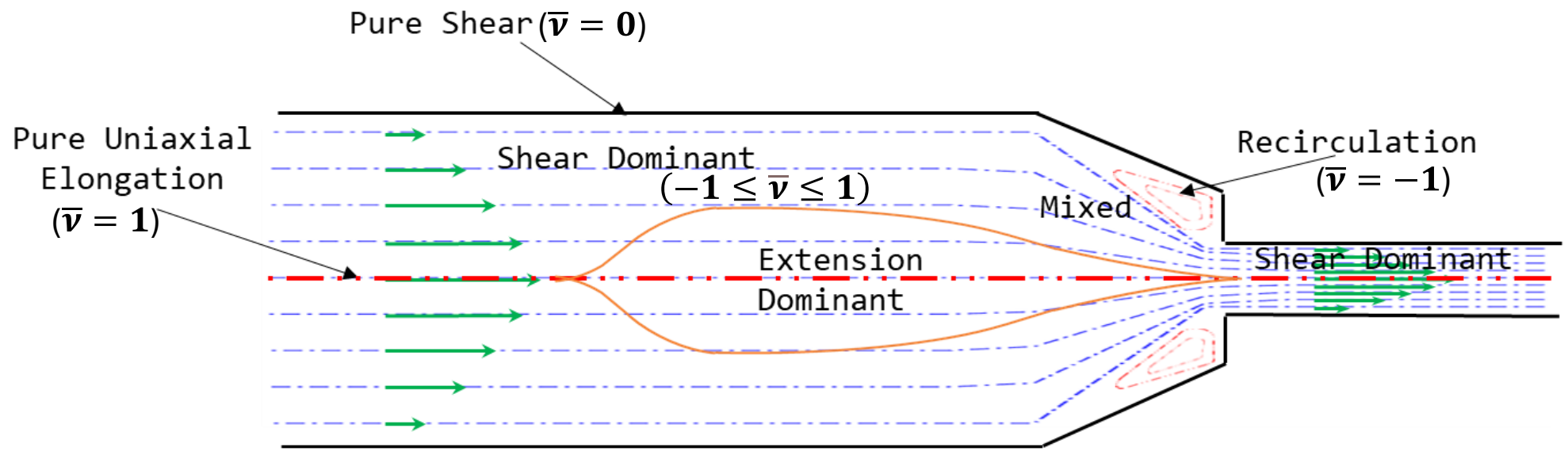

Figure B. 1 Schematic showing flow regimes within a typical EDAM nozzle during polymer processing.

## B.4 Obtaining Particle Stall Orientation Angles in Newtonian Homogenous Flows

The particle stall angles under favorable conditions in general class of homogenous flows can be obtained using the tensorial representation for the particle orientation of an axisymmetric ellipsoidal particle in viscous suspension with velocity gradient $\underline{\underline{L}}$ developed by Dinh et al. [258] based on Jeffery's model assumptions and is given as

$$\dot{\rho}_i = -\Xi_{ij}\rho_j + \kappa\left(\Gamma_{ij}\rho_j - \Gamma_{kl}\rho_k\rho_l\rho_i\right) \quad \text{(B. 17)}$$

where $\underline{\rho}$ is the particle orientation defined by the vector:

$$\underline{\rho} = \left[\cos\theta \quad \sin\theta\sin\phi \quad \sin\theta\cos\phi\right]^T \quad \text{(B. 18)}$$

The Euler angles and angular velocities can be backtracked from the rate of the orientation vectors $\underline{\dot{\rho}}$ thus:

| | |
|---|---|
| $\phi = \tan^{-1}\dfrac{\rho_2}{\rho_3},\ \ \theta = \cos^{-1}\rho_1,\ \ \dot{\phi} = \dfrac{\dot{\rho}_3}{\rho_3}\left[\dfrac{\dot{\rho}_2}{\dot{\rho}_3} - \dfrac{\rho_2}{\rho_3}\right]\left[1 + \dfrac{{\rho_2}^2}{\rho_3}\right]^{-1}$ <br> $\dot{\theta} = -\dot{\rho}_1(1-\rho_1^2)^{-1/2}$ | (B. 19) |

Considering the normalization condition, the independent components of the particle orientation at stall can likewise be obtained via the Newton-Raphson numerical iterative process according to eqn. (B. *20*) below

$$\underline{\rho_s^+} = \underline{\rho_s^-} - {\underline{\underline{J}}_{\Theta_2}}^{-1}\underline{\dot{\Theta}^\rho} \quad \text{(B. 20)}$$

where $\underline{\rho_s} = [\rho_2^s \quad \rho_3^s]^T$, $\sum_{\forall j}\rho_j = 1$, $\underline{\dot{\Theta}^\rho} = [\dot{\phi} \quad \dot{\theta}]^T$, and the components of the Jacobian $\underline{\underline{J}}_{\Theta_2}$ are explicitly defined in eqns. (B. *21*) - (B. *24*) below

$$\underline{\underline{J}}_{\Theta_2,11} = \frac{1}{\rho_3}\left\{J_{\rho,21} - J_{\rho,31}\frac{\rho_2}{\rho_3} - \frac{\dot{\rho}_3}{\rho_3}\left[1 + 2\left(\frac{\dot{\rho}_2}{\dot{\rho}_3} - \frac{\rho_2}{\rho_3}\right)\left(\frac{\rho_2}{\rho_3} + \frac{\rho_3}{\rho_2}\right)^{-1}\right]\right\}\left[1 + \frac{{\rho_2}^2}{\rho_3}\right]^{-1} \quad \text{(B. 21)}$$

$$\underline{\underline{J}}_{\Theta_2,12} = \frac{1}{\rho_3}\left\{J_{\rho,22} - J_{\rho,32}\frac{\rho_2}{\rho_3} - \frac{\dot{\rho}_3}{\rho_3}\frac{\dot{\rho}_2}{\dot{\rho}_3} + 2\frac{\dot{\rho}_3}{\rho_3}\frac{\rho_2}{\rho_3}\left[1 + \left(\frac{\dot{\rho}_2}{\dot{\rho}_3} - \frac{\rho_2}{\rho_3}\right)\left(\frac{\rho_2}{\rho_3} + \frac{\rho_3}{\rho_2}\right)^{-1}\right]\right\}\left[1 + \frac{{\rho_2}^2}{\rho_3}\right]^{-1} \quad \text{(B. 22)}$$

$$\underline{\underline{J}}_{\Theta_2,21} = \left[-J_{\rho,11}(1-\rho_1^2) + \dot{\rho}_1\right](1-\rho_1^2)^{-3/2} \quad \text{(B. 23)}$$

$$\underline{\underline{J}}_{\Theta_2,22} = \left[-J_{\rho,12}(1-\rho_1^2) + \dot{\rho}_1\right](1-\rho_1^2)^{-3/2} \quad \text{(B. 24)}$$

and the tensor $\underline{\underline{J}}_\rho$ is computed from eqn. (B. *25*) below

$$\underline{\underline{J}}_\rho = -\underline{\underline{\Xi}}\,\underline{\underline{\Omega}} + \kappa\left[\underline{\underline{\Gamma}} - \underline{\rho}\underline{\rho}^T\left(\underline{\underline{\Gamma}} + \underline{\underline{\Gamma}}^T\right) - \left(\underline{\rho}^T\underline{\underline{\Gamma}}\underline{\rho}\right)\underline{\underline{I}}\right]\underline{\underline{\Omega}}, \qquad \underline{\underline{\Omega}}^T = \begin{bmatrix} -1 & 1 & 0 \\ -1 & 0 & 1 \end{bmatrix} \quad \text{(B. 25)}$$

## B.5 Principal Flow Directions

The principal flow directions can be obtained by spectral decomposition of the symmetric part of the velocity gradient tensor $\underline{\underline{\Gamma}}$ . The respective eigenvectors $\underline{\Phi}^k$ are the principal flow directions. i.e.

$$\underline{\underline{\Phi}}\ \Big|\ \Gamma_{mn} = \Phi_{mk}\Lambda_{kl}\Phi_{nl}, \qquad \Lambda_{kl} = \begin{cases} \lambda_k & k = l \\ 0 & k \neq l \end{cases}, \qquad \underline{\underline{\Phi}}\ \Big|\ \underline{\underline{\Gamma}} = \underline{\underline{\Phi}}\,\underline{\underline{\Lambda}}\,\underline{\underline{\Phi}}^T \quad \text{(B. 26)}$$

Considering the in-plane homogenous flow velocity gradient of eqn. *(5.70)*, the principal flow directions in the shear plane irrespective of coordinate reference frame are obtained as

$$\tan\phi_p = \frac{\dot{\varepsilon}_2 - \dot{\varepsilon}_3}{\dot{\gamma}} \pm \sqrt{\frac{\dot{\varepsilon}_2 - \dot{\varepsilon}_3}{\dot{\gamma}}^2 + 1} \quad \text{(B. 27)}$$

As would be seen from the simulation results, for a particle tumbling in the flow shear-plane, the particle orientation at the location of minimum pressure extreme on particle's surface corresponds to position of particle alignment with one of the principal flow directions in the flow shear plane. i.e.

$$\underline{\rho}\Big|_{p=p_{max}} = \underline{\Phi}^k \quad \theta = 0 \qquad \text{(B. 28)}$$

Hence the peak pressure occurs at an instant $t_p$ such that $\phi(t_p) = \phi_p$.

### B.6 Optimization of Jeffery's equation for Center Gated Disk Flow

The center gated disk axisymmetric flow finds application in fiber orientation modelling in injection molding system and involves a pressure-gradient that drives the flow of fluid between two parallel plates such that the flow diverges radially outwardly from the inlet gate [318]. The velocity solutions are developed from the lubrication approximation based on the assumption of Newtonian fluid property and constant flow rate with no temperature gradient and the solutions are valid at radial distance $X_{0_r}$ much greater than the gap thickness $\mathbb{h}$ (i.e., $X_{0_r}/\mathbb{h} \gg 1$) [265]. The velocity profiles are fully developed at the disk inlet with reduced curvature at greater radial distances. The velocity gradient varies with time and is characterized by combined spatially dependent shearing and planar elongation components. Analytical and numerical solutions for accurately predicting the fiber orientation currently exists and have been well developed by various researchers [265], [293], [318].

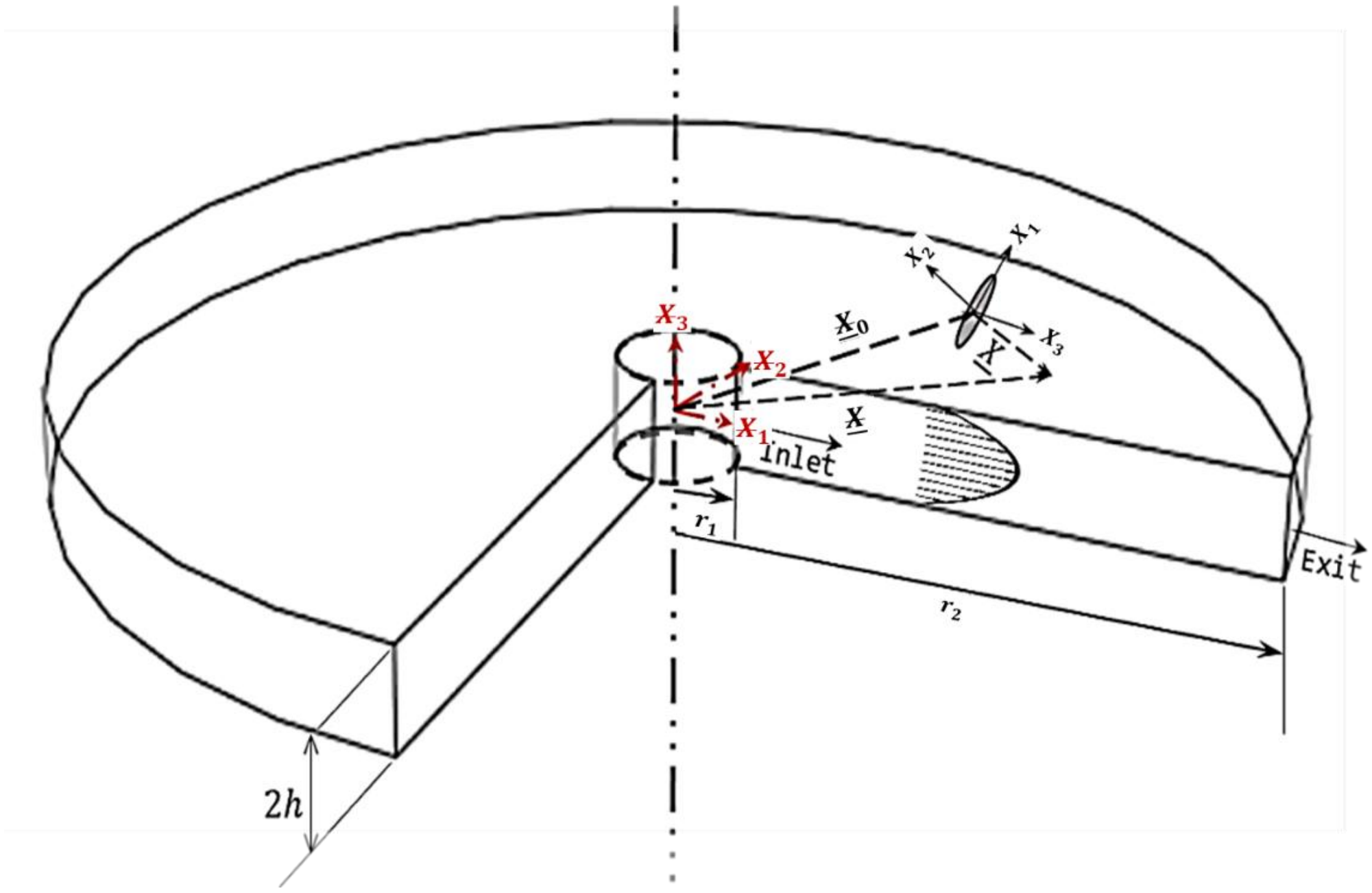

Figure B. 2 Schematic representation of a center gated disk with relevant annotations

Here we utilize Jeffery's model development to study the pressure distribution at the fiber's interface based on this flow condition that may provide insight into micro-void formation injection molding processing of polymer composites. For this flow type, the Jeffery' pressure are not only dependent on the spatial position vector $\underline{X}$ in the local fiber reference frame relative to its center and the fiber orientation angles $\underline{\Theta}$, but also the position vector of the fiber centroid $\underline{\mathcal{X}}_0$ from the global origin at the gate entry is an additional unknown. We thus need to find the partial derivative of the pressure $p$ with respect to $\underline{\mathcal{X}}_0$ in addition to $\underline{X}$ and $\underline{\Theta}$. In the global reference frame, the absolute position vector of an arbitrary point away from the fiber surface is given as

$$\mathcal{X}_j = \mathcal{X}_{0_j} + Z_{X_{jk}} X_k \tag{B. 29}$$

where the spatial quantities have been normalized with the gap height $\mathbb{h}$. The normalized velocity components at $\underline{\mathcal{X}}_0$ for this flow type is given in cylindrical coordinates as

$$\dot{\mathcal{X}}_{0_r} = \frac{1}{\mathcal{X}_{0_r}}\left[1 - \mathcal{X}_{0_z}^2\right], \qquad \dot{\mathcal{X}}_{0_\theta} = \dot{\mathcal{X}}_{0_3} = 0 \tag{B. 30}$$

where $X_{0_r}$ and $X_{0_3}$ are the normalized radial and vertical distance and $\mathcal{X}_{0_r} = \sqrt{\mathcal{X}_{0_1}^2 + \mathcal{X}_{0_2}^2}$. The actual non dimensionless velocity components are $\dot{\mathcal{X}}'_{0_r} = \bar{\mathbb{u}}\,\dot{\mathcal{X}}_{0_r}$, $\dot{\mathcal{X}}'_{0_\theta} = \dot{\mathcal{X}}'_{0_3} = 0$, where $\bar{\mathbb{u}} = 3\mathbb{Q}/8\pi\mathbb{h}^2$ is the average velocity. The non-zero components of the velocity gradient and its partial derivative with respect to the radial distance in terms of the normalized variables is thus given as [318].

$$\mathcal{L}_{11} = -\mathcal{L}_{22} = \frac{\partial \dot{\mathcal{X}}_{0_r}}{\partial \mathcal{X}_{0_r}} = -\frac{\dot{\mathcal{X}}_{0_r}}{\mathcal{X}_{0_r}}, \qquad \mathcal{L}_{13} = \frac{\partial \dot{\mathcal{X}}_{0_r}}{\partial \mathcal{X}_{0_3}} = -2\frac{\mathcal{X}_{0_3}}{\mathcal{X}_{0_r}} \tag{B. 31}$$

The non-dimensionless equivalent of velocity gradient $\mathcal{L}_{ij}$ in eqn. (B. *31*) above can be obtained from $\mathcal{L}'_{ij} = \bar{\mathbb{u}}\,/\mathbb{h}\,\mathcal{L}_{ij}$. Additional optimization constraints to those given in eqns. (5.20) - (5.21) for this flow type would include.

$$r_1 \le \mathcal{X}_{0_r} \le r_2, \qquad 0 \le \mathcal{X}_{0_3} \le 1 \tag{B. 32}$$

The velocity gradient $\mathcal{L}$ for this flow type has been provided in eqn. (B. *31*). The first derivative of the non-zero component of the of the velocity gradient $\mathcal{L}$ with respect to the global position vector $\underline{X}_0$ is given as

$$\underline{\nabla}_{\underline{X}_0}\mathcal{L}_{11} = -\underline{\nabla}_{\underline{X}_0}\mathcal{L}_{22} = 2\frac{\dot{\mathcal{X}}_{0_r}}{\mathcal{X}_{0_r}^3}\underline{\mathcal{X}}_0, \qquad \underline{\nabla}_{\underline{X}_0}\mathcal{L}_{13} = 2\frac{\mathcal{X}_{0_3}}{\mathcal{X}_{0_r}^3}\begin{bmatrix} \mathcal{X}_{0_1} \\ \mathcal{X}_{0_2} \\ -\mathcal{X}_{0_r}^2\mathcal{X}_{0_3}^{-1} \end{bmatrix} \tag{B. 33}$$

Likewise, the second derivatives of the non-zero component of the of the velocity gradient $\mathcal{L}$ with respect to the global position vector $\underline{X}_0$ is given as

$$\underline{\nabla}^2_{\underline{X}_0} Ł_{11} = -\underline{\nabla}^2_{\underline{X}_0} Ł_{22}$$

$$= -2\frac{1}{Ӿ_{0_r}^5}\begin{bmatrix} \dot{Ӿ}_{0_r}\left(3Ӿ_{0_1}^2 - Ӿ_{0_2}^2\right) & 4\dot{Ӿ}_{0_r}Ӿ_{0_1}Ӿ_{0_2} & 2Ӿ_{0_r}Ӿ_{0_1}Ӿ_{0_3} \\ & \dot{Ӿ}_{0_r}\left(3Ӿ_{0_2}^2 - Ӿ_{0_1}^2\right) & 2Ӿ_{0_r}Ӿ_{0_2}Ӿ_{0_3} \\ & & -Ӿ_{0_r}^3 \end{bmatrix} \tag{B. 34}$$

$$\underline{\nabla}^2_{\underline{X}_0} Ł_{13} = -2\frac{1}{Ӿ_{0_r}^5}\begin{bmatrix} \left(2Ӿ_{0_1}^2 - Ӿ_{0_2}^2\right)Ӿ_{0_3} & 3Ӿ_{0_1}Ӿ_{0_2}Ӿ_{0_3} & -Ӿ_{0_r}^2Ӿ_{0_1} \\ & \left(2Ӿ_{0_2}^2 - Ӿ_{0_1}^2\right)Ӿ_{0_3} & -Ӿ_{0_r}^2Ӿ_{0_2} \\ & & 0 \end{bmatrix}$$

The non-dimensionless equivalent of the above derivatives in eqn. (B. *33*) & (B. *34*) above are given as

$$\underline{\nabla}_{\underline{X}_0} Ł'_{ij} = \frac{\bar{ш}}{\mathbb{h}^2}\underline{\nabla}_{\underline{X}_0} Ł_{ij}, \qquad \underline{\nabla}^2_{\underline{X}_0} Ł'_{ij} = \frac{\bar{ш}}{\mathbb{h}^3}\underline{\nabla}^2_{\underline{X}_0} Ł_{ij} \tag{B. 35}$$

The Jeffery's pressure equation in terms of the independent variable vectors of differentiation $\underline{X}_0$, $\underline{X}$ and $\underline{\Theta}$ for this flow type as

$$p\left(\underline{Ӿ}_0, \underline{X}, \underline{\Theta}\right) = p_0 + \mathbb{h}^{-1}\varpi_j\left(\underline{Ӿ}_0, \underline{\Theta}\right)Д_j\left(\underline{X}\right) \tag{B. 36}$$

The gradient vector of the Jeffery's pressure with respect to $\underline{X}_0$, $\underline{X}$ and $\underline{\Theta}$ is thus given as

$$\underline{\nabla}p = \mathbb{h}^{-1}\begin{bmatrix} \underline{\nabla}_{\underline{X}_0} \\ \underline{\nabla}_{\underline{X}} \\ \underline{\nabla}_{\underline{\Theta}} \end{bmatrix} p \tag{B. 37}$$

where

$$\nabla_{X_{0_i}}p = Д_k\nabla_{X_{0_i}}\varpi_k, \qquad \nabla_{X_i}p = \varpi_k\nabla_{X_i}Д_k, \qquad \nabla_{\Theta_i}p = Д_k\nabla_{\Theta_i}\varpi_k \tag{B. 38}$$

and the corresponding hessian is given as

$$\underline{\underline{\nabla}}^2 p = \mathbb{h}^{-1}\begin{bmatrix} \underline{\nabla}_{\underline{X}_0}\underline{\nabla}^T_{\underline{X}_0} & \underline{\nabla}_{\underline{X}_0}\underline{\nabla}^T_{\underline{X}} & \underline{\nabla}_{\underline{X}_0}\underline{\nabla}^T_{\underline{\Theta}} \\ \underline{\nabla}_{\underline{X}}\underline{\nabla}^T_{\underline{X}_0} & \underline{\nabla}_{\underline{X}}\underline{\nabla}^T_{\underline{X}} & \underline{\nabla}_{\underline{X}}\underline{\nabla}^T_{\underline{\Theta}} \\ \underline{\nabla}_{\underline{\Theta}}\underline{\nabla}^T_{\underline{X}_0} & \underline{\nabla}_{\underline{\Theta}}\underline{\nabla}^T_{\underline{X}} & \underline{\nabla}_{\underline{\Theta}}\underline{\nabla}^T_{\underline{\Theta}} \end{bmatrix} p \tag{B. 39}$$

By exploiting the symmetric nature of the hessian, the only six (6) relevant components of $\underline{\underline{\nabla}}^2 p$ are given as

$$\nabla_{X_{0_i}}\nabla_{X_{0_j}}p = Д_k\nabla_{X_{0_i}}\nabla_{X_{0_j}}\varpi_k,\ \ \nabla_{X_i}\nabla_{X_j}p = \varpi_k\nabla_{X_i}\nabla_{X_j}Д_k,\ \ \nabla_{\Theta_i}\nabla_{\Theta_j}p = Д_k\nabla_{\Theta_i}\nabla_{\Theta_j}\varpi_k$$

$$\nabla_{X_{0_i}}\nabla_{X_j}p = \left[\nabla_{X_{0_i}}\varpi_k\right]\left[\nabla_{X_j}Д_k\right],\ \ \nabla_{X_{0_i}}\nabla_{\Theta_j}p = Д_k\nabla_{X_{0_i}}\nabla_{\Theta_j}\varpi_k,\ \ \nabla_{X_i}\nabla_{\Theta_j}p = \left[\nabla_{X_i}Д_k\right]\left[\nabla_{\Theta_j}\varpi_k\right] \tag{B. 40}$$

In the usual manner, the $\underline{\Theta}$ & $\underline{X}_0$ derivative operators are distributive over the non-constant elements and sub-elements of $\underline{\varpi}$. For conciseness, we can write a general expression for permutations of $m^{th}$ order $\underline{\Theta}$ − derivative and $n^{th}$ order $\underline{X}_0$ − derivative of $\underline{\varpi}$ as

$$\underline{\nabla}_{\underline{\Theta}}^{m}\underline{\nabla}_{\underline{X}_0}^{n}\underline{\varpi} = \left[\underline{\nabla}_{\underline{\Theta}}^{m}\underline{\nabla}_{\underline{X}_0}^{n}A \quad \underline{\nabla}_{\underline{\Theta}}^{m}\underline{\nabla}_{\underline{X}_0}^{n}B \quad \underline{\nabla}_{\underline{\Theta}}^{m}\underline{\nabla}_{\underline{X}_0}^{n}C \quad \underline{\nabla}_{\underline{\Theta}}^{m}\underline{\nabla}_{\underline{X}_0}^{n}(F+F') \quad \underline{\nabla}_{\underline{\Theta}}^{m}\underline{\nabla}_{\underline{X}_0}^{n}(G+G') \quad \underline{\nabla}_{\underline{\Theta}}^{m}\underline{\nabla}_{\underline{X}_0}^{n}(H\cdot\right. \tag{B. 41}$$

where

$$\underline{\nabla}_{\underline{\Theta}}^{m} = \nabla_{\Theta_i}^{(1)}\nabla_{\Theta_j}^{(2)}\nabla_{\Theta_k}^{(3)}\cdots\nabla_{\Theta_r}^{(p)}\cdots\nabla_{\Theta_s}^{(m)}, \qquad i,j,k,r,s = 1,2,3 \tag{B. 42}$$

$\underline{\nabla}_{\underline{\Theta}}^{(p)}$ is $p^{th}$ instance gradient operation for the $m^{th}$ order gradient operator $\underline{\nabla}_{\underline{\Theta}}^{m}$ for instance,

$$\underline{\nabla}_{\underline{\Theta}}^{2} = \nabla_{\Theta_j}\nabla_{\Theta_k}, \qquad \underline{\nabla}_{\underline{X}_0}^{2} = \nabla_{X_{0_j}}\nabla_{X_{0_k}}, \qquad \underline{\nabla}_{\underline{\Theta}}\underline{\nabla}_{\underline{X}_0} = \nabla_{\Theta_j}\nabla_{X_{0_k}}, \qquad j,k = 1,2,3 \tag{B. 43}$$

Typical partial derivatives for the components of $\underline{Q}$ are given in eqn. (B. *44*) below, and implicit in their expressions are the definition of other component derivatives.

$$\underline{\nabla}_{\underline{\Theta}}^{m}\underline{\nabla}_{\underline{X}_0}^{n}A = \frac{1}{6}\left\{\frac{2Ч''_{1_0}\underline{\nabla}_{\underline{\Theta}}^{m}\underline{\nabla}_{\underline{X}_0}^{n}\Gamma_{11} - Ч''_{2_0}\underline{\nabla}_{\underline{\Theta}}^{m}\underline{\nabla}_{\underline{X}_0}^{n}\Gamma_{22} - Ч''_{3_0}\underline{\nabla}_{\underline{\Theta}}^{m}\underline{\nabla}_{\underline{X}_0}^{n}\Gamma_{33}}{Ч''_{2_0}Ч''_{3_0} + Ч''_{3_0}Ч''_{1_0} + Ч''_{1_0}Ч''_{2_0}}\right\}$$
$$\underline{\nabla}_{\underline{\Theta}}^{m}\underline{\nabla}_{\underline{X}_0}^{n}F = \frac{Ч_{2_0}\underline{\nabla}_{\underline{\Theta}}^{m}\underline{\nabla}_{\underline{X}_0}^{n}\Gamma_{23} - И_3^2Ч'_{1_0}\left(\underline{\nabla}_{\underline{\Theta}}^{m}\underline{\nabla}_{\underline{X}_0}^{n}\Xi_1 - \underline{\nabla}_{\underline{\Theta}}^{m}\underline{\nabla}_{\underline{X}_0}^{n}\dot{\Psi}_1\right)}{2Ч'_{1_0}\left(И_2^2Ч_{2_0} + И_3^2Ч_{3_0}\right)} \tag{B. 44}$$

The derivative of the tensors $\Gamma_{ij}$ and $\Xi_{ij}$ can be derived from the derivative of the symmetric and antisymmetric decomposition of $L_{ij}$ in the usual manner according to.

$$\langle \underline{\nabla}_{\underline{\Theta}}^{m} \underline{\nabla}_{\underline{X}_0}^{n} \Gamma_{ij} \ , \ \underline{\nabla}_{\underline{\Theta}}^{m} \underline{\nabla}_{\underline{X}_0}^{n} \Xi_{ij} \rangle = \frac{1}{2} \left[ \underline{\nabla}_{\underline{\Theta}}^{m} \underline{\nabla}_{\underline{X}_0}^{n} L_{ij} \pm \underline{\nabla}_{\underline{\Theta}}^{m} \underline{\nabla}_{\underline{X}_0}^{n} L_{ji} \right] \tag{B. 45}$$

where the velocity gradient in the local fibers reference frame $L_{ij}$ is obtained in the usual manner from the global definition $Ⱡ_{ij}$ through the transformation operation according to eqn. *(5.9)* given as $L_{ij} = Z_{X_{mi}} Ⱡ_{mn} \, Z_{X_{nj}}$. We require the first and second partial derivatives of $L_{ij}$ with respect to the components of $\underline{X}_0$ and $\underline{\Theta}$ . The first derivatives in indicial notation are given as

$$\nabla_{X_{0_k}} L_{ij} = Z_{X_{mi}} \nabla_{X_{0_k}} Ⱡ_{mn} \, Z_{X_{nj}}$$
$$\nabla_{\Theta_k} L_{ij} = \nabla_{\Theta_k} Z_{X_{mi}} Ⱡ_{mn} \, Z_{X_{nj}} + Z_{X_{mi}} Ⱡ_{mn} \, \nabla_{\Theta_k} Z_{X_{nj}} \tag{B. 46}$$

Likewise, the second derivative of $L_{ij}$ with respect to the components of $\underline{X}_0$ and $\underline{\Theta}$ in like manner are given as

$$\nabla_{X_{0_r}} \nabla_{X_{0_s}} L_{ij} = Z_{X_{mi}} \nabla_{X_{0_r}} \nabla_{X_{0_s}} Ⱡ_{mn} \, Z_{X_{nj}}$$
$$\nabla_{\Theta_r} \nabla_{\Theta_s} L_{ij} = \nabla_{\Theta_r} \nabla_{\Theta_s} Z_{X_{mi}} Ⱡ_{mn} \, Z_{X_{nj}} + Z_{X_{mi}} Ⱡ_{mn} \, \nabla_{\Theta_r} \nabla_{\Theta_s} Z_{X_{nj}}$$
$$+ \nabla_{\Theta_r} Z_{X_{mi}} Ⱡ_{mn} \, \nabla_{\Theta_s} Z_{X_{nj}} + \nabla_{\Theta_s} Z_{X_{mi}} Ⱡ_{mn} \nabla_{\Theta_r} Z_{X_{nj}} \tag{B. 47}$$
$$\nabla_{\Theta_r} \nabla_{X_{0_s}} L_{ij} = Z_{X_{mi}} \nabla_{\Theta_r} \nabla_{X_{0_s}} Ⱡ_{mn} \, Z_{X_{nj}} + \nabla_{\Theta_r} Z_{X_{mi}} \nabla_{X_{0_s}} Ⱡ_{mn} \, Z_{X_{nj}}$$
$$+ Z_{X_{mi}} \nabla_{X_{0_s}} Ⱡ_{mn} \, \nabla_{\Theta_r} Z_{X_{nj}}$$

The higher order $\underline{\Theta}$ & $\underline{X}_0$ derivatives of the angular velocity $\underline{\dot{\Psi}}$ found in eqn. (B. *44*) can be derived in similar fashion to eqn. (B. *42*) above given as

$$\underline{\nabla}_{\underline{\Theta}}^{m} \underline{\nabla}_{\underline{X}_0}^{n} \dot{\Psi}_j = \underline{\nabla}_{\underline{\Theta}}^{m} \underline{\nabla}_{\underline{X}_0}^{n} \Xi_j + M_{jk} \underline{\nabla}_{\underline{\Theta}}^{m} \underline{\nabla}_{\underline{X}_0}^{n} D_k \tag{B. 48}$$

In the usual manner, the $\underline{\Theta}$ & $\underline{X}_0$ derivatives are distributive over the individual components of the tensors $\Xi_j$, and $D_j$, for instance

$$\underline{\nabla}^m_{\underline{\Theta}}\underline{\nabla}^n_{\underline{X}_0}\Xi_1 = \underline{\nabla}^m_{\underline{\Theta}}\underline{\nabla}^n_{\underline{X}_0}\xi, \qquad \underline{\nabla}^m_{\underline{\Theta}}\underline{\nabla}^n_{\underline{X}_0}D_1 = \underline{\nabla}^m_{\Theta}\underline{\nabla}^n_{\underline{X}_0}\Gamma_{23} \tag{B. 49}$$

The derivatives of the Laplace function $\Omega$ used to obtain the gradient and hessian of $\underline{Д}$ where previously derived in terms of the actual quantities. Recall the dimensionless form for the Jeffery's pressure $\bar{p}$ given in eqn. (5.126). For this flow, $\dot{\gamma}_c$ is not constant but depends on the independent differentiable variables. As such it must be considered when obtaining derivatives of the dimensionless pressure. The first and second derivatives of $\bar{p}$ is given in indicial notation as

$$\nabla_k\bar{p} = \nabla_k(\dot{\gamma}_c^{-1})\frac{p - p_0}{\mu} + \dot{\gamma}_c^{-1}\frac{\nabla_k p}{\mu}, \qquad \dot{\gamma}_c = \sqrt{2\Gamma_{ij}\Gamma_{ji}} \tag{B. 50}$$

$$\nabla_m\nabla_n\bar{p} = \nabla_m\nabla_n(\dot{\gamma}_c^{-1})\frac{p - p_0}{\mu} + \nabla_m(\dot{\gamma}_c^{-1})\frac{\nabla_n p}{\mu} + \nabla_n(\dot{\gamma}_c^{-1})\frac{\nabla_m p}{\mu} + \dot{\gamma}_c^{-1}\frac{\nabla_m\nabla_n p}{\mu} \tag{B. 51}$$

where

$$\nabla_k(\dot{\gamma}_c^{-1}) = -2\dot{\gamma}_c^{-3}\nabla_k\Gamma_{ij}\Gamma_{ji}, \qquad \Gamma_{ij} = \Gamma_{ji} \tag{B. 52}$$

$$\begin{aligned}\nabla_m\nabla_n(\dot{\gamma}_c^{-1}) = {} & 12\dot{\gamma}_c^{-5}\left[\Gamma_{ij}\nabla_m\Gamma_{ji}\right]\left[\Gamma_{pq}\nabla_n\Gamma_{qp}\right]\cdots \\ & - 2\dot{\gamma}_c^{-3}\left[\Gamma_{ij}\nabla_m\nabla_n\Gamma_{ji} + \nabla_m\Gamma_{ij}\nabla_n\Gamma_{ji} + \nabla_n\Gamma_{ij}\nabla_m\Gamma_{ji} + \Gamma_{ij}\nabla_n\nabla_m\Gamma_{ji}\right]\end{aligned} \tag{B. 53}$$

The same validation exercise of the gradient and hessian for this flow type using finite difference such as that described in Appendix IV is carried out. Assuming the same orientation state $\underline{\Theta}^i$ and relative spatial position $\mathbb{h}^{-1}\underline{X}^i$ as provided in (5.64) for a unity $\mathbb{h}$ value and given an arbitrary fiber centroidal position $\underline{X}_0^i = [12 \quad 9 \quad 0.75]^T$, We obtain the following non-zero component of the velocity gradient as $Ł_{11} = -Ł_{22} = -0.011575$, $Ł_{13} = 0.25053$. Subsequently, the pressure and the gradient and hessian error estimates based on definition of eqn. (5.63) are obtained as $ç^{(1)} = 1.0763 \times 10^{-15}$, $ç^{(2)} = 3.9542 \times 10^{-13}$.

# APPENDIX C

## C.1 Eigenvalues and Eigenvector Derivatives

The eigenvalue for a matrix $\underline{\underline{a}}$ is often given in terms of the eigen-values $\Lambda^k$ and corresponding eigen-vectors $\underline{\Phi}^k$ as [283]

$$\left[\underline{\underline{a}} - \Lambda_k \underline{\underline{I}}\right] \underline{\Phi}^k = \underline{F}^k \tag{C. 1}$$

For the homogeneous system $\underline{F}^k = \mathbf{0}$, and requiring $\underline{\Phi}^k \neq 0$ to yield non-trivial solutions, Eqn. A1 becomes $\left|\underline{\underline{a}} - \Lambda_k \underline{\underline{I}}\right| = 0$ where the vertical bars indicate the matrix determinant. The resulting system matrix $[a_{mn} - \Lambda^k \delta_{mn}]$ is known to be rank deficient by order 1, requiring that we impose a normalization condition on the eigen-vectors $\underline{\Phi}^k$ which can be defined through a scalar function $G^k(\underline{\Phi}^k) = 0$. The eigenvectors are thus obtained by replacing the *n^th^* row of the residual column vector $\Sigma_i^k = (a_{ij} - \Lambda^k \delta_{ij})\Phi_j^k - \underline{F}_i^k$ with $G^k(\underline{\Phi}^k) = 0$ and solving for $\underline{\Phi}^k$ from the equation $\underline{\Sigma}^k(\underline{\Phi}^k) = 0$ through any iterative algorithm or explicit solver. Here we employ Newton-Raphson's method to obtain $\underline{\Phi}^k$ such that:

$$\underline{\Phi}^{k^+} = \underline{\Phi}^{k^-} - \underline{\underline{J}}^{k^{-1}} \Sigma^k \tag{C. 2}$$

$$\Sigma_i^k = \begin{cases} S_{ij}^k \Phi_j^k - \underline{F}_i^k & i \neq n \\ G^k(\underline{\Phi}^k) & i = n \end{cases} \tag{C. 3}$$

$$J_{ij}^k = \frac{\partial \Sigma_i^k}{\partial \Phi_j^k} = \begin{cases} S_{ij}^k & i \neq n \\ \dfrac{\partial G^k}{\partial \Phi_j^k} & i = n \end{cases} \tag{C. 4}$$

where

$$S_{ij}^k = a_{ij} - \Lambda^k \delta_{ij} \tag{C. 5}$$

The derivative of the eigenvalues with respect to components $\mathrm{a}_{ij}$ can thus be obtained by differentiating Eqn. (C. 3) assuming symmetry of system matrix, i.e., $S_{ij}^{k} = S_{ji}^{k}$ such that:

$$\Phi_i^k \frac{\partial S_{ij}^k}{\partial \mathrm{a}_{rs}} \Phi_j^k + \underline{F_i}^k \frac{\partial \Phi_i^k}{\partial \mathrm{a}_{rs}} - \Phi_i^k \frac{\partial F_i^k}{\partial \mathrm{a}_{rs}} = 0 \qquad \text{(C. 6)}$$

$$\Phi_i^k S_{ij}^k = S_{ji}^k \Phi_j^k = S_{ij}^k \Phi_j^k = \underline{F_i}^k \qquad \text{(C. 7)}$$

where:

$$\frac{\partial S_{ij}^k}{\partial \mathrm{a}_{rs}} = \delta_{ir}\delta_{js} - \frac{\partial \Lambda^k}{\partial \mathrm{a}_{rs}} \delta_{ij} \qquad \text{(C. 8)}$$

Assuming $\underline{F_i}^k = 0$, we obtain

$$\frac{\partial \Lambda^k}{\partial \mathrm{a}_{rs}} = \frac{1}{\left(\Phi_m^k \Phi_m^k\right)} \{\Phi_r^k \Phi_s^k\} \qquad \text{(C. 9)}$$

Consequently, we can obtain the corresponding derivatives $\partial \Phi_i^k / \partial \mathrm{a}_{rs}$ given $\partial \Lambda^k / \partial \mathrm{a}_{rs}$ from

$$S_{ij}^k \frac{\partial \Phi_i^k}{\partial \mathrm{a}_{rs}} = -\frac{\partial S_{ji}^k}{\partial \mathrm{a}_{rs}} \Phi_i^k \qquad \text{(C. 10)}$$

Recalling the system matrix $\left[\underline{\underline{\mathrm{a}}} - \Lambda^k \underline{\underline{I}}\right]$ is inherently singular, we can substitute an *n^th^* row of the above equation adopting a normalization technique as [283].

$$\frac{\partial G^k}{\partial \Phi_j^k} \frac{\partial \Phi_j^k}{\partial \mathrm{a}_{rs}} = \frac{\partial G^k}{\partial \mathrm{a}_{rs}} \qquad \text{(C. 11)}$$

The modified differential equation can be recast as given in equation (C.12) below allowing the inversion of the modified system matrix to solve

$$J_{ij}^k \frac{\partial \Phi_i^k}{\partial \mathrm{a}_{rs}} = \frac{\partial Q_i^k}{\partial \mathrm{a}_{rs}} \qquad \text{(C. 12)}$$

where:

$$\frac{\partial Q_i^k}{\partial \mathrm{a}_{rs}} = \begin{cases} -\dfrac{\partial S_{ij}^k}{\partial \mathrm{a}_{rs}} \Phi_j^k & i \neq n \\ -\dfrac{\partial G^k}{\partial \mathrm{a}_{rs}} & i = n \end{cases} \tag{C. 13}$$

Smith et. al [283] presents three common normalization techniques given as

1. Mass Normalization

$$G^k\left(\underline{\Phi}^k\right) = \Phi_j^k \Phi_j^k - 1, \qquad \frac{\partial G^k}{\partial \Phi_j^k} = 2\Phi_j^k, \qquad \frac{\partial G^k}{\partial \mathrm{a}_{rs}} = 0 \tag{C. 14}$$

2. Preassigning an *m^th^* Component of $\underline{\Phi}^k$

$$G^k\left(\underline{\Phi}^k\right) = \delta_{mj} \Phi_j^k - \alpha, \qquad \frac{\partial G^k}{\partial \Phi_j^k} = \delta_{mj}, \qquad \frac{\partial G^k}{\partial \mathrm{a}_{rs}} = 0 \tag{C. 15}$$

3. Predefining the Euclidean norm of $\underline{\Phi}^k$

$$G^k\left(\underline{\Phi}^k\right) = \sqrt{\Phi_j^k \Phi_j^k} - \beta, \qquad \frac{\partial G^k}{\partial \Phi_j^k} = \Phi_j^k, \qquad \frac{\partial G^k}{\partial \mathrm{a}_{rs}} = 0 \tag{C. 16}$$

A more direct and efficient approach by Nelson [319] which utilizes mass normalization technique is given below:

$$\frac{\partial \Phi_i^k}{\partial \mathrm{a}_{rs}} = V_i^k + c^k \Phi_i^k, \qquad c^k = -\Phi_i^k V_i^k, \qquad V_i^k = \left\{\underline{\underline{SP}}^{k^{-1}}\right\}_{ij} \frac{\partial QP_j^k}{\partial \mathrm{a}_{rs}} \tag{C. 17}$$

Given a pre-selected fixed index – *m and* noting in eqns. (C. 18) - (C. 19) below that repeated indices do not imply a summation,

$$SP_{ij}^k = \left(K_{ij} - \lambda^k \delta_{ij}\right)(1 - \delta_{mi})\left(1 - \delta_{mj}\right) + \delta_{mi}\delta_{mj} \tag{C. 18}$$

$$\frac{\partial QP_i^k}{\partial \mathrm{a}_{rs}} = -\frac{\partial S_{ij}^k}{\partial \mathrm{a}_{rs}} \Phi_j^k (1 - \delta_{mi}) \tag{C. 19}$$

## C.2 Optimal Fitted Closure Approximation Constants/Coefficients

### *C.2.1 Eigenvalue Based Optimal Fitting Closure (EBOF) Approximation*

In CHAPTER SIX, we presented the generalized formulations for the eigenvalue based orthotropic binomial fitted closures approximation which is written in terms of a constant coefficient matrix $\mathfrak{C}_{ij}^{(n)}$ and an $n^{\text{th}}$ order permuted bivariate polynomial vector $\mathbb{A}_j^{(n)}$ with $(n+1)(n+2)/2$ number of parameters (cf. Equations (6.103)-(6.104) (6.105)). Depending on the polynomial order, $(n)$, the various constants coefficient matrix $\underline{\underline{\mathfrak{C}}}^{(n)}$ for the closures considered in this paper taken from Kuzmin [38] are presented below.

C.2.1.1 *Linear Orthotropic Fitted Closure* $(n = 1)$. For the orthotropic linear closure, the constant coefficient matrix $\underline{\underline{\mathfrak{C}}}^{(1)}$ is given as

$$\underline{\underline{\mathfrak{C}}}^{(1)} = \frac{1}{7}\begin{bmatrix} -3/5 & 6 & 0 \\ -3/5 & 0 & 6 \\ 27/5 & -6 & -6 \end{bmatrix}$$

and for the orthotropic fitted closure by Cintra and Tucker (cf. [267]), the constant coefficient matrix $\underline{\underline{\mathfrak{C}}}^{(1)}$ is given as

$$\underline{\underline{\mathfrak{C}}}^{(1)} = \begin{bmatrix} -0.15 & 1.15 & -0.10 \\ -0.15 & 0.15 & 0.90 \\ 0.60 & -0.60 & -0.60 \end{bmatrix}$$

*C.2.1.2 Quadratic Orthotropic Fitted Closures* $(n = 2)$. The simple general orthotropic quadratic closure has constant coefficient matrix $\underline{\underline{\mathfrak{C}}}^{(2)}$ given as

$$\underline{\underline{\mathfrak{C}}}^{(2)} = \begin{bmatrix} 0 & 0 & 0 & 1 & 0 & 0 \\ 0 & 0 & 0 & 0 & 0 & 1 \\ 1 & -2 & -2 & 1 & 2 & 1 \end{bmatrix}$$

The orthotropic natural closure - exact midpoint fit [251] has constant coefficient matrix $\underline{\underline{\mathfrak{C}}}^{(2)}$ given as

$$\underline{\underline{\mathfrak{C}}}^{(2)} = \begin{bmatrix} 0.0708 & 0.3236 & -0.3776 & 0.6056 & 0.4124 & 0.3068 \\ 0.0708 & -0.2792 & 0.2252 & 0.2084 & 0.4124 & 0.7040 \\ 1.1880 & -2.0136 & -2.1264 & 0.8256 & 1.7640 & 0.9384 \end{bmatrix}$$

The ORF independent coefficients are derived from a 2nd order polynomial fit of the principal axis data obtained from DFC via a minimization technique. For the orthotropic fitted closure by Cintra and Tucker (ORF) (cf. [267]), the constant coefficient matrix $\underline{\underline{\mathfrak{C}}}^{(2)}$ is given as

$$\underline{\underline{\mathfrak{C}}}^{(2)} = \begin{bmatrix} 0.060964 & 0.371243 & -0.369160 & 0.555301 & 0.371218 & 0.318266 \\ 0.124711 & -0.389402 & 0.086169 & 0.258844 & 0.544992 & 0.796080 \\ 1.228982 & -2.054116 & -2.260574 & 0.821548 & 1.819756 & 1.053907 \end{bmatrix}$$

The ORF had better performance compared to non-fitted closure approximations, however, the ORF behaved poorly for flows with very low interaction coefficients and sometimes gave non-physical oscillations like the behavior of the Hinch and Leal closure (HL2) in same condition. Though the ORL behaves better for low interaction coefficient in simple shear flow yet overpredicts the flow direction orientation tensor and is unstable for radial diverging flows. Chung and Kwon [320], improved the ORF and developed the 2nd order ORW closure for wide interaction coefficients that is stable for all flow conditions using new flow data from distribution function calculation (DFC). The improved orthotropic fitted closure (ORW2) by Chung and Kwon (cf. [320]), has constant coefficient matrix $\underline{\underline{\mathfrak{C}}}^{(2)}$ given as

$$\underline{\underline{\mathfrak{C}}}^{(2)} = \begin{bmatrix} 0.070055 & 0.339376 & -0.396796 & 0.590331 & 0.411944 & 0.333693 \\ 0.115177 & -0.368267 & 0.094820 & 0.252880 & 0.535224 & 0.800181 \\ 1.249811 & -2.148297 & -2.290157 & 0.898521 & 1.934914 & 1.044147 \end{bmatrix}$$

Kuzmin [251] presents details on derivations of some orthotropic fitted closures via a numerical bottom top approach.

*C.2.1.3 Cubic Orthotropic Fitted Closures* $(n = 3)$. Recently higher order polynomial fitted closures were developed for improved accuracy. The orthotropic natural closure - extended quadratic fit though of cubic order is essentially quadratic.

1. <u>Non-rational Fitted Closure</u>

The constant coefficient matrix $\underline{\underline{\mathfrak{C}}}^{(3)}$ for this closure approximation is given as

$$\underline{\underline{\mathfrak{C}}}^{(3)} = \begin{bmatrix} 0 & 0.5 & 0 & 0.5 & -0.6 & 0 & 0 & 0.6 & 0.6 & 0 \\ 0 & 0 & 0.5 & 0 & -0.6 & 0.5 & 0 & 0.6 & 0.6 & 0 \\ 1 & -1.5 & -1.5 & 0.5 & 0.4 & 0.5 & 0 & 0.6 & 0.6 & 0 \end{bmatrix}$$

Chung and Kwon [320], also extended the 2nd order ORW to develop 3rd order polynomial ORW3 closure using new flow data from orientation distribution function calculation (DFC). For the improved 3rd order orthotropic fitted closure ORW3 by Chung and Kwon [320], the constant coefficient matrix is given as

$$\left[\underline{\underline{\mathfrak{C}}}^{(3)}\right]^T = \begin{bmatrix} -0.1480648093 & -0.2106349673 & 0.4868019601 \\ 0.8084618453 & 0.9092350296 & 0.5776328438 \\ 0.7765597096 & 1.1104441966 & 0.4605743789 \\ 0.3722003446 & -1.2840654776 & -2.2462007509 \\ -1.7366749542 & -2.5375632310 & -4.8900459209 \\ -1.3431772379 & 0.1260059291 & -1.9088154281 \\ -0.0324756095 & 0.5856304774 & 1.1817992322 \\ 0.8895946393 & 1.9988098293 & 4.0544348937 \\ 1.7367571741 & 1.4863151577 & 3.8542602127 \\ 0.6631716575 & -0.0756740034 & 0.9512305286 \end{bmatrix}$$

Eigenvalue-based optimal fitting (EBOF) closures which are extensions to the ORF but with higher order polynomial fits were developed for improved accuracy. These include the rational elliptical (RE) closure by Wetzel [312] and the 4th order polynomial fitted closure by Verweyst [316] (ORT).

2. Rational Fitted Closure

The rational elliptical (RE) closure developed by Wetzel [41] is a higher order extension to the ORF using Carlson elliptical integrals. The rational ellipsoid fitted closure has two permutation vectors, the denominator being one order less than the numerator, i.e., $m = n - 1$ which is of cubic order, $n = 3$. The corresponding constant coefficient matrix for the Wetzel numerator $(n = 3)$ is [312]

$$\left[\underline{\underline{\mathfrak{C}}}^{(3)}\right]^T = \begin{bmatrix} 0.1433751825 & 0.1433751825 & 0.9685744898 \\ -0.6566650339 & -0.5209453949 & -2.5526857671 \\ -0.5106016916 & -0.6463213306 & -2.5756669706 \\ 3.5295952199 & 0.6031924921 & 2.2044050704 \\ 4.4349137241 & 2.3303190917 & 4.4520903005 \\ 0.1229618909 & 5.1539592511 & 2.2485545147 \\ -2.9144388828 & -0.2256222796 & -0.6202937932 \\ -5.5556896198 & -1.6481269200 & -1.8811803355 \\ -2.8284365891 & -5.4494528976 & -1.9023485762 \\ 0.2292109036 & -3.7461520908 & -0.6414620339 \end{bmatrix}$$

And for the denominator $(m = n - 1 = 2)$

$$\left[\underline{\underline{\mathfrak{C}}}^{(2)}\right]^T = \begin{bmatrix} 1.0000000000 & 1.0000000000 & 1.0000000000 \\ 0.7257989503 & 0.6916858207 & -1.2134964928 \\ 3.0941511876 & 3.1282643172 & -1.2128608265 \\ -1.6239324646 & -1.5898193351 & 0.2393747647 \\ -4.7303686308 & -4.7303686308 & 0.6004510415 \\ -3.1742364608 & -3.2083495904 & 0.2162486576 \end{bmatrix}$$

Mullens et al. [42] developed several high order polynomials fitted closures for short fiber composites and introduced the time derivative fitted closures. For the LAR32 closure by Mullens [275] the corresponding constant coefficient matrix for the numerator $(n = 3)$ is

$$\left[\underline{\underline{\mathfrak{C}}}^{(3)}\right]^T = \begin{vmatrix} 0.087602233 & 0.156805152 & 1.072423739 \\ 0.028205550 & -0.577818864 & -2.803554028 \\ -0.426784335 & -0.514280920 & -2.661576129 \\ 1.274677110 & 0.684250887 & 2.389379765 \\ 0.876469059 & 2.132305029 & 4.566728489 \\ 0.602031647 & 3.454835266 & 2.097523143 \\ -1.066583115 & -0.263237143 & -0.658248930 \end{vmatrix}$$

$$\begin{vmatrix} -1.918931146 & -1.614122610 & -1.904704744 \\ -0.934291306 & -4.005261132 & -1.754978355 \\ -0.262854903 & -2.228133231 & -0.508282668 \end{vmatrix}$$

and for the denominator ($m = n - 1 = 2$)

$$\left[\underline{\underline{\mathfrak{C}}}^{(2)}\right]^T = \begin{vmatrix} 1.000000000 & 1.000000000 & 1.000000000 \\ -0.244001948 & 0.365652907 & -1.068512526 \\ -0.574150861 & 1.385725477 & -0.771356469 \\ -0.432097367 & -1.359687152 & 0.067386858 \\ -0.895226091 & -2.866357848 & 0.206908269 \\ -0.462709527 & -1.518996192 & -0.248999874 \end{vmatrix}$$

3. Quartic Orthotropic Fitted Closures ($n = 4$)

The constant coefficient matrix $\underline{\underline{\mathfrak{C}}}^{(4)}$ based on regression analysis by Verweyst [316] developed from Carlson elliptic integrals.

$$\left[\underline{\underline{\mathfrak{C}}}^{(4)}\right]^T = \begin{bmatrix} 0.6363 & 0.6363 & 2.7405 \\ -1.8727 & -3.3153 & -9.1220 \\ -4.4797 & -3.0371 & -12.2571 \\ 11.9590 & 11.8273 & 34.3199 \\ 3.8446 & 6.8815 & 13.8295 \\ 11.3421 & 8.4368 & 25.8685 \\ -10.9583 & -15.9121 & -37.7029 \\ -20.7278 & -15.1516 & -50.2756 \\ -2.1162 & -6.4873 & -10.8802 \\ -12.3876 & -8.6389 & -26.9637 \\ 9.8160 & 9.3252 & 27.3347 \\ 3.4790 & 7.7468 & 15.2651 \\ 11.7493 & 7.4815 & 26.1135 \\ 0.5080 & 2.2848 & 3.4321 \\ 4.8837 & 3.5977 & 10.6117 \end{bmatrix}$$

The constant coefficient matrix $\underline{\underline{\mathfrak{C}}}'$ based on regression analysis for the FFLAR4 closure by Mullens [275]

$$\begin{bmatrix} 0.678225884 & 0.748226727 & 3.167356369 \\ -3.834359034 & -4.249612053 & -13.288266400 \\ -2.664862865 & -2.987266447 & -11.680179330 \end{bmatrix}$$

$$\left[\underline{\underline{\mathfrak{C}}}^{(4)}\right]^T =$$

$$\left[\begin{array}{rrr}
9.746185193 & 8.641488072 & 23.788431340 \\
14.209962670 & 14.938209410 & 43.700607680 \\
2.700369681 & 5.974489008 & 17.383121430 \\
-8.013024236 & -7.521216405 & -19.959054610 \\
-22.447252700 & -21.757217160 & -58.354308000 \\
-13.078649640 & -15.798676320 & -49.513705640 \\
-0.125467689 & -3.616551654 & -11.755525930 \\
2.417857515 & 2.376441613 & 6.291273472 \\
10.563248410 & 10.222185780 & 25.844317920 \\
12.689484570 & 12.640352670 & 35.425354130 \\
2.487386515 & 4.788201652 & 18.226443930 \\
-0.328195677 & 1.056519961 & 2.925785795
\end{array}\right]$$

The constant coefficient matrix $\underline{\underline{\mathfrak{C}}}^{(4)}$ based on regression analysis for the LAR4 closure by Mullens [275] is

$$\left[\underline{\underline{\mathfrak{C}}}^{(4)}\right]^T = \left[\begin{array}{rrr}
0.813175172 & 1.768619587 & 4.525066937 \\
-3.065410883 & -9.826017151 & -19.259137620 \\
-4.659333003 & -6.484058476 & -17.650178090 \\
6.329870878 & 19.986994700 & 33.901239610 \\
14.747639770 & 28.905936750 & 61.543979540 \\
9.739797775 & 10.759963010 & 28.467355970 \\
-4.216519964 & -17.715409270 & -27.768082700 \\
-15.922240910 & -40.492387100 & -76.738638810 \\
-20.818571900 & -27.442217500 & -68.977583290 \\
-8.993993112 & -7.230748101 & -22.399036130 \\
1.138888034 & 5.785725498 & 8.600822308 \\
5.834142985 & 18.709047480 & 32.480679940 \\
11.470974520 & 19.729631240 & 43.875135630 \\
9.874209286 & 8.882877701 & 26.928320210 \\
3.100457733 & 2.224834058 & 7.101978254
\end{array}\right]$$

### *C.2.2 Invariant Based Optimal Fitting Closure (IBOF) Approximation*

The unknown independent coefficients of the binomial expansion for the six parameters in the IBOF closure representation based on regression fitting by Chung et al. [304] of actual flow data obtained from the orientation distribution function closure considering different flow types like EBF closures are given in Table C. *1* below.

Table C. 1: 5th order binomial fitting coefficients for the IBOF closure approximation

| $k\backslash m$ | *1* | *2* | *3* |
|---|---|---|---|
| *0* | 2.49409081657860E+01 | -4.97217790110754E-01 | 2.34146291570999E+01 |
| *1* | -4.35101153160329E+02 | 2.34980797511405E+01 | -4.12048043372534E+02 |
| *2* | 7.03443657916476E+03 | 1.53965820593506E+02 | 5.73259594331015E+03 |
| *3* | 3.72389335663877E+03 | -3.91044251397838E+02 | 3.19553200392089E+03 |
| *4* | -1.33931929894245E+05 | -2.13755248785646E+03 | -6.05006113515592E+04 |
| *5* | 8.23995187366106E+05 | 1.52772950743819E+05 | -4.85212803064813E+04 |
| *6* | -1.59392396237307E+04 | 2.96004865275814E+03 | -1.10656935176569E+04 |
| *7* | 8.80683515327916E+05 | -4.00138947092812E+03 | -4.77173740017567E+04 |
| *8* | -9.91630690741981E+06 | -1.85949305922308E+06 | 5.99066486689836E+06 |
| *9* | 8.00970026849796E+06 | 2.47717810054366E+06 | -4.60543580680696E+07 |
| *10* | 3.22219416256417E+04 | -1.04092072189767E+04 | 1.28967058686204E+04 |
| *11* | -2.37010458689252E+06 | 1.01013983339062E+05 | 2.03042960322874E+06 |
| *12* | 3.79010599355267E+07 | 7.32341494213578E+06 | -5.56606156734835E+07 |
| *13* | -3.37010820273821E+07 | -1.47919027644202E+07 | 5.67424911007837E+08 |
| *14* | -2.57258805870567E+08 | -6.35149929624336E+07 | -1.52752854956514E+09 |
| *15* | -2.32153488525298E+04 | 1.38088690964946E+04 | 4.66767581292985E+03 |
| *16* | 2.14419090344474E+06 | -2.47435106210237E+05 | -4.99321746092534E+06 |
| *17* | -4.49275591851490E+07 | -9.02980378929272E+06 | 1.32124828143333E+08 |
| *18* | -2.13133920223355E+07 | 7.24969796807399E+06 | -1.62359994620983E+09 |
| *19* | 1.57076702372204E+09 | 4.87093452892595E+08 | 7.92526849882218E+09 |
| *20* | -3.95769398304473E+09 | -1.60162178614234E+09 | -1.28050778279459E+10 |